\documentclass[12pt,a4paper,amstex]{book}
\usepackage{lipsum}
\usepackage{amsmath}
\usepackage{setspace}
\usepackage{subfigure}
\usepackage{appendix}
\usepackage{float}
\usepackage{wrapfig}
\usepackage{textcomp}
\usepackage[small]{caption}
\usepackage{graphics}
\usepackage{psfig}
\usepackage{epsfig}
\usepackage{array}
\usepackage{dcolumn}
\usepackage{asp2010}
\usepackage{epstopdf}
\usepackage{longtable}
\usepackage{amssymb}
\usepackage{bm}
\usepackage{comment}
\usepackage{multicol}
\usepackage{multirow}  
\usepackage{chngcntr}
\usepackage{grffile}
\counterwithin{figure}{chapter}
\counterwithin{table}{chapter}
\usepackage[nottoc]{tocbibind}
\usepackage{tabularx}
\graphicspath{{Images/}{Images/}{Images/}}

\newcolumntype{.}{D{.}{.}{3.2}}

\newcommand{\psra}{PSR J1738$-$2330}
\newcommand{\psrb}{PSR J1752+2359}
\newcommand{\pa}{J1738$-$2330}
\newcommand{\pb}{J1752+2359}
\newcommand{\pdot}{$\dot{P}$}
\newcommand{\ppdot}{$P-\dot{P}$}
\newcommand{\polyco}{{\itshape polyco}}
\newcommand{\edotb}{\boldsymbol{E\cdotp{B}}=0}
\newcommand{\edotbnot}{\boldsymbol{E\cdotp{B}}\neq{0}}
\newcolumntype{L}{>{\centering\arraybackslash}m{3cm}}
\numberwithin{section}{chapter}
\def\deg{^{\circ}}
\def\pdot{\dot{P}}
\def\ppdot{P-\pdot}
\newcommand{\footnoteremember}[2]{
\footnote{#2}
\newcounter{#1}
\setcounter{#1}{\value{footnote}}
}
\newcommand{\footnoterecall}[1]{
\footnotemark[\value{#1}]
}
\renewcommand{\appendixname}{Appendix}
\usepackage[left=1.20in,right=1.20in,top=1.5in,bottom=0.5in, paperheight=11in,paperwidth=8.5in]{geometry}
\begin{document}

\clearpage
\pagenumbering{roman}
%\begin{titlepage}
\thispagestyle{empty}

\begin{center}
{\Huge \bf On the absence of pulses from pulsars}
\end{center}

\vskip 3.5cm
\begin{center}
{\Large A thesis submitted to the}\\
\vskip 0.5 cm
{\large \bf Tata Institute of Fundamental Research, Mumbai}\\
\vskip 0.3 cm
{\large \bf for the degree of Doctor of Philosophy}\\
\vskip 0.3 cm
{\Large (in Physics)}
\vskip 3cm
{\Large by}
\vskip 0.5 cm
{\LARGE \bf Vishal Gajjar}
\vskip 1 cm
{\Large \bf National Centre for Radio Astrophysics}\\
\vskip 0.1 cm
{\LARGE Tata Institute of Fundamental Research}\\
\vskip 0.15 cm
{\LARGE Pune University Campus}\\
\vskip 0.1 cm
{\LARGE Pune -- 411 007}\\
\vskip 0.1 cm
{\LARGE India}
\vskip 1 cm
{{\large \bf e-mail:} {\Large \it vishalg@berkeley.edu}}
\vskip 1 cm
{\LARGE \bf April, 2014}
\end{center}
\thispagestyle{empty}
   \setcounter{secnumdepth}{3}
   \setcounter{tocdepth}{5} 
   \frontmatter
%  \begin{KeepFromToc}
%  \input{Synopsis/Dedication}
%  \clearpage\null\newpage

   \tableofcontents      
%  \end{KeepFromToc}
   \raggedbottom
%  \clearpage\null\newpage
%  \input{Synopsis/Acknow_v2}
%  \clearpage\null\newpage
%  \input{Synopsis/Declaration} 
  % \clearpage\null\newpage
  % \input{synopsis_arxiv} 
     \listoffigures
   \listoftables
%  \clearpage\null\newpage
%  \addtocontents{toc}{\setcounter{tocdepth}{3}}
 
   \mainmatter 
   \makeatletter\@openrightfalse\makeatother

\chapter*{Synopsis} 
\noindent\rule{16cm}{1pt} \\
{\bf\large Thesis title : On the absence of pulses from pulsars  \\
 Vishal Gajjar \\
 Supervisor : Dr. Bhal Chandra Joshi \\}
\noindent\rule{16cm}{1pt}

Pulsars are one of the most fascinating objects in the known Universe. The very nature of 
their regular pulsation led to their discovery in the early days. They  
were the first to provide sound evidence about the existence of the neutron stars. 
In the last 47 years since the discovery of these objects, extensive 
research has been carried out to investigate the origin of the pulsar radiation. 
There have been many attempts to scrutinize the bright radio emission seen from 
pulsars by utilizing numerous observed phenomena. One of such phenomena is 
the rapid shape-changes in the observed single pulses in spite of their average 
profile showing remarkable stability for observations separated by decades. 
The sporadic nature of the single pulses provides important clues regarding the chaotic 
nature of the pulsar radiation. Thus, enormous amount of research 
have been focused on the modelling of the single pulses. 
There are different types of single pulse phenomena seen in 
the radio pulsar \emph{viz.}  drifting, nulling and mode-changing, 
which also shows interdependence in certain cases. Drifting 
and mode-changing has been modelled to some extent by many 
previous studies. However, nulling is one of the  
unexplained phenomena seen in the radio pulsars. 
The absence of emission in the individual single pulses 
has defied satisfactory explanation since its discovery 
44 years ago. 

Nulling in the radio pulsars has been reported by \cite{bac70} 
for the first time in four pulsars. Since its discovery, around 109 
pulsars were reported to show prominent nulling behaviour in their 
single pulses. Nulling in pulsars has been historically quantified 
as the fraction of observed null pulses, also known as the \emph{nulling fraction} (NF). 
In order to find the operating mechanism that causes only a few pulsars 
to exhibit noticeable nulling behaviour, NFs were 
compared with many pulsar parameters by various groups 
\cite[]{rit76,ran86,big92a,viv95,wmj07}.
However, no strong correlation between the NF and any pulsar parameter 
has been reported to date. \cite{rit76} has reported that if the pulsar
period is long, it is likely to have high nulling fraction compared to short
period pulsars. As the period of the pulsar is directly related to its 
characteristic age, \cite{rit76} speculated that pulsars 
die with increasing fraction of nulls in them. 
Contrary to that, \cite{ran86} presented an alternative hypothesis 
of correlation with the profile classes, where no dependence 
was noticed between NF and age of the pulsar in a similar profile
morphological class. It was extensively shown 
in a study of around 60 pulsars, that the nulling fraction is likely
to be less than 1\% in pulsars exhibiting core component, while, it is more
likely to have higher nulling fraction in pulsars exhibiting conal profile \cite[]{ran86}. 
In a separate study by \cite{big92a}, NF was confirmed to show correlation 
with the profile classes along with a few weak correlations. However, 
\cite{wmj07} has shown that nulling does not show correlation with 
profile morphological classes as profiles with almost all classes 
show similar nulling behaviour. Moreover, \cite{wmj07} suggested that nulling 
is an extreme form of mode-changing phenomenon. 
Thus, there is no common agreement between different 
studies regarding the true nature of any correlation. 

The degree and form of pulse nulling varies from 
one pulsar to another. On one hand, there are 
pulsars such as PSR B0826$-$34 \cite[]{dll+79}
which null most of the time, and PSR J1752+2359 
\cite[]{lwf+04}, which exhibits no radio emission for 
3 to 4 minutes. In contrast, pulsars such as PSR B0809+74 
show a small degree of nulling \cite[]{la83}. 
Pulse nulling is frequent in pulsars such as PSR B1112+50, while it is very 
sporadic in PSR B1642$-$03 \cite[]{rit76}. 
Moreover, the pulsed emission abruptly 
declines by more than two orders of 
magnitudes during these nulls \cite[]{la83,vj97}, 
which are as yet not well understood. While 
nulling pulsars have been known for last three 
decades, the recently discovered new class of sources, 
such as Rotating Radio transients 
\cite[]{mll+06,km11,bjb+12} and 
intermittent pulsars \cite[]{klo+06,crc+12,llm+12}, 
also show a behaviour similar to classical nullers 
and are increasingly believed to be nulling 
pulsars \cite[]{bjb+12}, indicating that 
nulling occurs in a significant fraction of pulsar population.

Pulsar emission at different radio frequencies, 
originates at different locations in the pulsar magnetosphere \cite[]{kom70}. 
There are very few long simultaneous 
observations of nulling pulsars reported so far 
in the literature. In a simultaneous single pulse study 
of two pulsars, PSRs B0329+54 and B1133+16 at 327 and 2695 MHz, 
\cite{bs78} showed highly correlated pulse energy 
fluctuations. Simultaneous observations of 
PSR B0809+74 for about 350 pulses 
indicated that only 6 out of 9 nulls were simultaneous 
at 102 and 408 MHz \citep{dls+84}. About half of nulls 
were reported to occur simultaneously at 325, 610, 1400 
and 4850 MHz for PSR B1133+16 \citep{bgk+07}. In contrast, 
simultaneous nulls were reported at 303 and 610 MHz for 
PSR B0826$-$34 \citep{bgg08}. 
It is not clear if nulling represents a global 
failure of pulse radiation or is due to a shift in pulsar 
beam manifesting as lack of emission at the 
given observation frequency due to the geometry 
of pulse emission. 

In light of all these previous investigations, 
long, sensitive, and preferably simultaneous 
observations at multiple frequencies 
of a carefully selected sample of pulsars are 
motivated. Thus, the aim of this thesis is to quantify, model  
and compare nulling behaviour between different classes of pulsars  
to scrutinize the true nature of the nulling phenomenon. 

In this thesis, main results on three main investigations 
are discussed with the necessary background for each of them. 
In Chapter 1, basic introduction 
to the pulsars is provided. Details regarding the 
structure of the pulsar surroundings and origin of 
the radio emission are briefly derived in Chapter 2. 
Observations conducted for each investigation, are 
summarised with necessary details in Chapter 3. 
Further three chapters discuss results on 
the individual study. Chapter 4 presents results 
on the survey of nulling pulsar conducted using the GMRT. 
A comparison study between two high nulling fraction pulsars 
are discussed in Chapter 5. Unique simultaneous multi-frequency 
observations of two pulsars are summarised in Chapter 6. 
Chapter 7 presents summary of all the obtained results 
in the thesis along with a comprehensive view 
about the nulling behaviour suggested by our observations. 

The chapter wise summary of this thesis is given below. 
First three chapters provide necessary background, 
along with the details regarding various observations. 

\section*{\underline{\underline{Introduction}}}
Introduction to pulsars, including their discovery and peculiar 
observed properties, are discussed in \emph{Chapter 1}. 
A simple pulsar toy model is presented to explain the 
pulsating nature of this source. Radio waves 
from the pulsar, travelling through the interstellar medium, 
undergo numerous propagation related effects. A few of these effects 
can reduce the pulse energy to zero level. These effects 
could imitate intrinsic phenomenon like pulsar nulling. 
Thus, they are briefly explained to highlight their true nature. 
Different observed pulsar parameters and their stability 
on various time-scales are also discussed in the chapter. 
Various single pulse phenomena reported in the thesis are 
briefly introduced with examples. 

\emph{Chapter 2} aims to provide necessary background 
about the currently known radio emission mechanism physics of the pulsars. 
It starts with a discussion on the structure of the neutron star along with the 
properties and composition of its surface, which were used to 
build the standard pulsar emission model proposed by \cite{rs75}. 
Origin of the coherent radio emission due to the growth in the 
two stream instability is briefly reviewed. This chapter 
also summarises, in detail, all previous studies conducted to investigate 
pulsar nulling phenomena. Different models proposed over the 
years to explain the mode-changing phenomenon are also listed  
in this chapter to relate them to pulse nulling in extreme 
conditions. Towards the end, primary incentives for the 
work reported in this thesis are listed in order 
to assess these models. 

Details regarding all observations, reported in this 
thesis, are discussed with full details in \emph{Chapter 3}. 
Justification regarding the selections of sources 
and observing frequencies for different objectives are 
elaborated in this chapter. 
In this thesis, observations from 
four different telescopes has been reported \emph{viz.} 
the GMRT, the WSRT, the Arecibo telescope and the Effelsberg 
radio telescope. Details about all these telescopes are 
highlighted in this chapter including various local setting 
which were executed during the time of observations. 
Details about initial and basic analysis procedures 
followed for all our observations in this thesis work 
are also elaborated. These include, (a) obtaining single pulses 
from different observatories, (b) eliminating radio frequency 
interferences related effects, (c) estimation of the NF 
and (d) obtaining the null and burst length histograms from 
the separated null and burst pulses. A novel approach to isolate 
weak burst pulses among the null pulses is also introduced 
here for the first time. 

% \section*{\underline{\underline{Survey of nulling pulsars using the GMRT}}\\
% (Gajjar, V., Joshi B. C. and Kramer M., 2012, MNRAS, 424, 1197)}
\section*{Survey of nulling pulsars using the GMRT \\
\underline{\underline{(Gajjar, V., Joshi B. C. and Kramer M., 2012, MNRAS, 424, 1197)}}}

In recent years Parkes Multibeam Survey (PKSMB) has discovered many new
pulsars. Several of them show nulling behaviour in their discovery plots. 
We have carefully looked through the discovery plots of many pulsars, out of 
which, 5 promising candidates were selected for longer observations 
with the GMRT. In \emph{Chapter 4}, the nulling behaviour 
of 15 pulsars, which include 5 PKSMB pulsars with no previously 
reported nulling behaviour, with the estimates on 
their NFs is reported. For four of these 15 pulsars, 
only an upper/lower limit was previously reported. 
The estimates of reduction in the pulsed emission is 
also presented for the first time in 11 pulsars. 
NF value for individual profile component is also 
presented for two pulsars in the sample \emph{viz.}  
PSRs B2111+46 and B2020+28. Possible mode changing 
behaviour is suggested by these observations 
for PSR J1725--4043, but this needs to be confirmed with
more sensitive observations. An interesting 
quasi-periodic nulling behaviour for PSR J1738--2330 is also reported. 

We find that the nulling patterns differ between 
PSRs B0809+74, B0818--13, B0835--41 and B2021+51, even though
they have similar NF of around 1\%. We showed this by comparing 
the null length and burst length distributions using a 
two sample Kolmogorov-Smirnov test. Our results 
confirm that NF probably does not capture the full detail 
of the nulling behaviour of a pulsar. Thus, it can be 
speculated that, due to this, earlier attempts failed to correlate 
NF with various pulsar parameters.  

We carried out the Wald-Wolfowitz runs test for randomness 
to 8 more pulsars. We find that this test indicates that 
occurrence of nulling, when individual pulses are considered, is 
non-random or exhibits correlation across periods. 
This correlation groups pulses in null and burst states, 
which was also noted by \cite{rr09}. 
However, the durations of the null and the burst states shown 
to be modelled by a stochastic Poisson point process suggesting that 
these transitions occur at random. 
Thus, the underlying physical process for nulls in the 8 pulsars  
studied appears to be random in nature producing 
nulls and bursts with unpredictable durations. 
Moreover, modelling of the null length and the burst length distributions
provided typical nulling and burst timescale (i.e. $\tau_n$ and $\tau_b$ respectively),  
which were derived for 8 pulsars for the first time to the best of our knowledge. 

\section*{On the long nulls of PSRs J1738$-$2330 and J1752+2359\\
\underline{\underline{(Gajjar, V., Joshi B. C., and Geoffrey W. 2014a, MNRAS,  439, 221)}}}
A detailed study of pulse energy modulation in two pulsars, 
\psra\ and \psrb, with similar NF has been presented in \emph{Chapter 5}. 
The NFs were estimated to be 85$\pm$2\% and 81\% 
for \psra\ and \psrb\ respectively. The aim of this study was to investigate 
similarity and differences between two high NF pulsars. 

Both the pulsars exhibit similar bunching of the burst pulses 
classified as the \emph{bright phases}, which are separated by long 
null phases. A similar quasi-periodic switching between 
these two state is observed for both the pulsars. 
However, using a new technique of pair-correlation-function (PCF), 
we have reported significant differences between these two pulsars. 
The PCF for \psra\ indicates that the mechanism responsible 
for bright phases is governed by two quasi-periodic 
processes with periodicities of 170 and 270 pulses, 
On the other hand, the nulling pattern of \psrb\ is dominated 
by 540 pulse quasi-periodicity, which jitters from 490 to 
595 pulses. These processes are not strictly periodic, 
but retain a memory longer than 2000 pulses for \psra, while 
the memory of \psrb's periodic  structure is retained for only 
about 1000 periods. 

Towards the end of each bright phase of \psrb, an exponential 
decline in the pulse energy is reported. 
\psra\ also shows similar exponential decay 
along with a flickering emission characterized by short frequent 
nulls towards the end of each bright phase. 
We modelled the bright pulse energy decay for 
\psrb\ and estimated their average 
lengths. A strong anti-correlation 
was found between the peak energy and 
the decay time for all bright phases in this pulsar. 
We demonstrated that the area under each bright phase is similar, 
suggesting that the energy release during all such events is 
approximately constant for \psrb.  
The first bright phase pulse profile and the last bright phase 
pulse profile show striking differences between both the pulsars, 
hinting differences in transition from null phase to burst phase and 
vice-versa between the two pulsars. We report, for the first time, 
peculiar weak burst pulses during the long null phases of \psrb, 
which we call as inter-burst pulses (IBPs). 
The IBPs are similar to emission seen from RRATs. 
The occurrence rate of the IBPs is random and uncorrelated with 
the preceding or following bright phase parameters. 
The polarization profiles for \psrb\ obtained from observations 
with the Arecibo telescope are reported for the first time.  
The total intensity and the circular polarization profiles 
of IBPs are slightly shifted towards the leading side 
compared to the conventional integrated profile, 
indicating a change in the emission region. 
No such pulses were seen in the long null phases of 
\psra. Lastly, we do not observe any GPs in 
our long observations unlike such pulses being 
reported at low frequencies in \psrb\ \cite[]{EK05}.
Hence, these results confirm that even though these 
two pulsar have similar but significantly high 
NFs, they show very different nulling behaviour.

\section*{Simultaneous multi-frequency study of pulse nulling behaviour in two pulsars \\
\underline{\underline{(Gajjar, V., Joshi B. C., Kramer M., Smith R. and Karuppusamy R., 2014b, ApJ, 797, 18)}}}
A detailed study on the simultaneous occurrence of the nulling phenomena
in two pulsars, PSRs B0809+74 and B2319+60, is reported in \emph{Chapter 6}. 
The observations were conducted simultaneously at four different 
frequencies, 313, 607, 1380 and 4850 MHz, from three different 
telescopes \emph{viz.} the GMRT, the WSRT and the Effelsberg radio telescope. 
The overlap time for each pulsar was around 6 hours 
between different observatories. 

We obtained single pulses at each frequency and the on-pulse 
energies were compared across all four frequencies, which showed 
remarkable similarity in the pulse energy fluctuations 
for both the pulsars. To quantify these similarities, we obtained the NF 
at each frequency which were found to be consistence within the error bars, 
at all four frequencies for both the pulsars. 
Similarly, the one-bit sequences were also compared 
using the contingency table analysis. To measure 
the statistical significance of the contingency tables, 
we measured the Cramer-V and the uncertainty coefficient 
for each pair of frequencies. For PSR B0809+74, 
both the statistical tests showed highly 
significant broadband nulling behaviour. For PSR B2319+60, 
the significance was marginally lower for the pairs 
involving 4850 MHz. 

We also scrutinize all the nonconcurrent 
pulses (i.e. pulses which did not show similar emission states 
across different frequencies) for both the pulsars to investigate their true nature. 
For PSR B0809+74, we found that out of 12 nonconcurrent 
pulses, 7 (about 58\%) occurred at the transition point where emission 
state is switching from null to burst (or vice-verse). 
Similarly for PSR B2319+60, we found that out of 158 
nonconcurrent pulses, 82 (about 52\%) occurred at the transition point where 
emission state is switching. Thus, both pulsars showed remarkable 
similarity in the overall broadband behaviour, also accounting the 
fraction of nonconcurrent pulses at the transition points. 

A slight difference can be seen for the \emph{exclusive} 
pulse profiles, pulses which exclusively occurred at a single 
observing frequency, between both the pulsars. PSR B0809+74 
showed significantly narrow pulses (except at 4850 MHz) 
aligning the overall pulse profile, while PSR B2319+60 showed significantly 
weak pulses which are shifted towards the leading edge compared to the overall 
pulse profile. However, a strong claim can not be made regarding 
their true shapes, due to their small numbers. 
We also compared the null length and the burst length distributions across all 
observed frequencies which showed similar distribution with high 
significance ($>$99\%) for both the pulsars. These results clearly suggest 
that nulling is truly a broadband phenomena. It favours models invoking 
magnetospheric changes on a global scale compared to 
local geometric effects as a likely cause of nulling in these pulsars. 

\section*{\underline{\underline{Conclusions}}} 
Final summary of all obtained results are listed in \emph{Chapter 7} and their 
implications are discussed. We have also combined the obtained results 
on three different aspects of the nulling behaviour. 
In light of these results, following interpretations 
about the nulling phenomena can be suggested. 
\begin{enumerate}
 \item NF is not an ideal parameter to quantify nulling behaviour, confirmed firmly 
  by comparing low NF pulsars as well as high NF pulsars.  
 \item Nulling occurs randomly with unpredictable length durations. 
  Quasi-periodicities seen in the high NF pulsars can also be explained 
  by the Markov models with a forcing function.  
 \item Nulling is an extreme form of mode-changing phenomena which 
  occurs on a global magnetospheric scale. 
 \item Geometric reasons are less favoured as a likely cause of nulling phenomena 
  due to the randomness and broadband behaviour reported in this thesis.  
\end{enumerate}
These results can also be extended to 
explain the peculiar emission behaviour seen in the 
intermittent pulsars and the RRATs. 
Future work motivated by these observations are listed 
towards the end of this chapter.

\section*{List of publications}
\subsection*{Refereed Journal}
\begin{itemize}
 \item Gajjar, V., Joshi B. C. and Kramer M., 2012, \newline
       \emph{Survey of nulling pulsars using the Giant Meterwave Radio Telescope}, \newline
       MNRAS, 424, 1197
       
 \item Gajjar, V., Joshi B. C., and Geoffrey W. 2014a, \newline
       \emph{On the long nulls of PSRs 11738-2330 and J1752+2359}, \newline
       MNRAS, 439, 221

 \item Gajjar, V., Joshi B. C., Kramer M., Smith R. and Karuppusamy R., 2014b, \newline
       \emph{Frequency independent quenching of pulsed emission}, \newline
       ApJ, 797, 18       
\end{itemize}

\subsection*{Proceedings}
\begin{itemize}
 \item Gajjar, V., Joshi B. C. and Kramer M., 2009, \newline
       \emph{Peculiar nulling in PSR J1738-2330}, \newline
       ASP conference series, 407, 304
       
 \item Gajjar, V., Joshi B. C. and Kramer M., 2012, \newline
       \emph{Broadband nulling behaviour of PSR B2319+60}, \newline
       ASP conference proceeding, 466, 79
       
 \item Gajjar, V., Joshi B. C. and Kramer M., 2013, \newline
       \emph{A survey of nulling pulsars using the Giant Meterwave Radio Telescope}, \newline
       Proceedings of IAU Symposium, 291, 385
       
 \item Gajjar, V., Joshi B. C. and Kramer M., 2011, \newline
       \emph{A survey of nulling pulsars using GMRT}, \newline
       Astronomical Society of of India Conference Series, 3, 118
\end{itemize}

\chapter{Introduction}
{\itshape Pulsating stars (aka Pulsars)} are one of the most exotic objects in the known  
Universe due to their unique nature. Pulsars are highly magnetised 
rotating neutron stars. They emit beams of radiation in a concentrated emission cone 
which sweeps the surrounding sky in a light-house manner. When this beam of radiation crosses 
the line-of-sight towards the observer, a pulse can be detected. 
This train of pulses has a unique period associated 
with it, which is the period of rotation of the pulsar. 
Pulsars are known to emit radiation from radio to gamma-rays 
frequency regime, although from widely spaced emitting regions 
with different emission mechanisms. However, this thesis particularly 
focuses on the study in the radio frequency regime only. 

\section{Pulsar Discovery}
The first pulsar was discovered serendipitously 
in the late 1960s \cite[]{hbp+68}. Jocelyn Bell, 
a graduate student from the Cambridge University, and her supervisor Anthony Hewish,  
were conducting experiments to study the interplanetary scintillations. 
During these investigations, a train of pulses was detected. 
In an interesting account by \cite{bell77}, the first ever recorded 
signal from a pulsar was noticed as ``a bit of \emph{scruff}" on a chart recorder. 
The observed train of pulses, with 1.33 second periodicity, were first thought to be 
terrestrial signal. However, regular appearance of this signal following the sidereal time, 
eliminated the possibility of terrestrial origin (the source was subsequently named  
CP 1919 which is now known as pulsar B1919+21).  
All different possibilities were considered, including a possible beacon from 
an extra-terrestrial intelligence civilization. However, after a few months of this   
discovery, couple of more such sources were found at different locations in the sky, 
making it clear that the recorded phenomena has a natural origin. 
The peculiar property of such short periodicity allowed \cite{hbp+68} 
to conclude that, compact objects such as white dwarfs and neutron stars 
could be the possible source. The suggestion that pulsars are rotating neutron 
stars was also provided by \cite{gold68}. This was indeed 
confirmed later when a pulsar with 33 millisecond periodicity was discovered 
powering the Crab nebula, which is a supernovae remnant \cite[]{pac68}. 

The existence of neutron star in a stellar life cycle 
was first predicted by \cite{bz34}. However, a possible 
detection of any signal from the neutron stars was never expected. 
Hence, the discovery of radio signals from pulsars became the single most 
significant event to prove the existence of neutron stars. 
Currently, more than 2000 pulsars are known which exhibit a plethora 
of observed phenomena. Pulsars are fantastic 
laboratory to test many physical and astrophysical problems. 
They were extensively used to study the interstellar medium.  
Pulsars are also proposed to be an important probe in the gravitational 
wave detection [see \cite{joshi13} for a review]. 
% 
% \section{Radio emission properties}
\section{Pulsar Toy model}
\label{pulsar_toy_sect}
\begin{figure}[!h]
 \centering
 \includegraphics[width=1.2 in,height=5 in,angle=-90,bb=0 0 202 720]{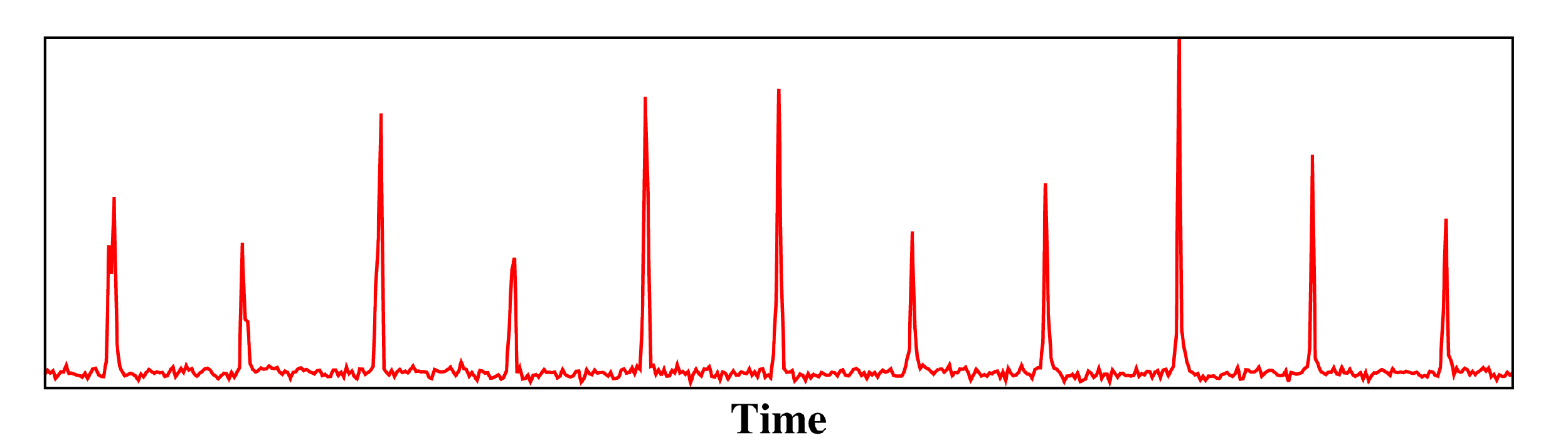}
 \caption[Pulse train from PSR B0809+74]{An observed train of pulses from PSR B0809+74 
 (here \emph{PSR} stands for Pulsating Source of Radio and the number 
 followed by it presents right-ascension and declination of the source)}
 % Pulse_train_for_thesis.eps: 0x0 pixel, 300dpi, 0.00x0.00 cm, bb=352 50 554 770
 \label{pulse_train}
\end{figure}

\begin{figure*}[!h]
 \begin{center}
   \centering
%    \hspace*{1.2in}
%   \includegraphics[bb=0 0 536 539]{pulsar_toy_gimp.jpg}
 % pulsar_toy_gimp.jpg: 536x539 pixel, 72dpi, 18.91x19.01 cm, bb=0 0 536 539
  \includegraphics[width=3 in,height=4 in,angle=0,bb=0 0 449 640]{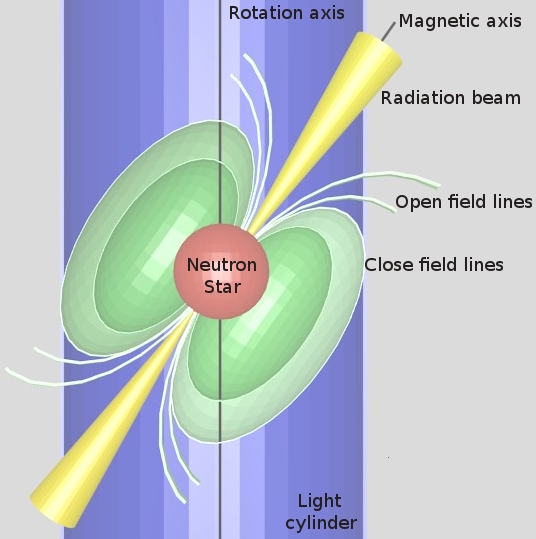}
 % pulsar_toy_gimp.jpg: 449x540 pixel, 72dpi, 15.84x19.05 cm, bb=0 0 449 540
  \caption[Pulsar toy model]{The schematic diagram of a radio pulsar to illustrate various regions around the neutron star (not to scale). 
  The magnetic axis is misaligned with the rotation axis, which gives the peculiar pulsating properties to each pulsar.
  The field lines enclosing inside the light cylinder are classified the close field line region while 
  the field lines which cross the light cylinder belong to open field line region 
  (cartoon using the free 3D modeling package : {\itshape blender} \url{www.blender.org}).}
  \label{toy_model}
 \end{center}
\end{figure*}

Pulsar radio emission have been studied extensively to scrutinize 
the emission mechanism. The radio emission was found to be 
highly polarized. In an interesting study by \cite{rc69}, 
pulsar emission was shown to be originating near the magnetic 
field lines. The rotating vector model suggests, pulsar has a co-rotating 
radiation beam aligned with its magnetic axis. The angle between 
the rotation axis and the magnetic axis, also known as $\alpha$, 
produces pulsed radio emission (shown in Figure \ref{pulse_train}) 
every time the radiation beam sweeps the line-of-sight of the earth \cite[]{rc69}. 
This picture of the pulsar toy model is shown in Figure \ref{toy_model}. 

 
The charged particles, bound to the magnetic field lines, are forced to corotate with the neutron star. 
However, as the distance from the neutron star rotation axis increases, the linear velocity of the rotation 
of the charge particles also increases up to speed of light. Beyond this boundary, 
the particles can not continue to corotate with the star. This boundary, also known 
as the \emph{light cylinder}, divides the pulsar magnetosphere in two different regions, 
namely the open field lines and the close field lines as 
illustrated in Figure \ref{toy_model}. The region of open field lines 
allows the charge particles to accelerate and escape the magnetosphere. The 
radio emission is generated inside this open field line regions. 
The details regarding the production of the radio emission is given in Section \ref{radio_emission_sect}. 

\section{Propagation of radio waves}
The radio waves generated by the pulsar pass through the Galactic interstellar medium (ISM)
before reaching the observer. During their travel, the signal experiences multiple 
propagation effects, such as dispersion, scattering and scintillation, discussed below. 
Some of these effects can be reverted to retrieve the original signal. 

\subsection{Pulse dispersion}
\begin{figure}[h]
 \centering
 \includegraphics[width=3 in,height=4 in,angle=-90,bb=0 0 504 720]{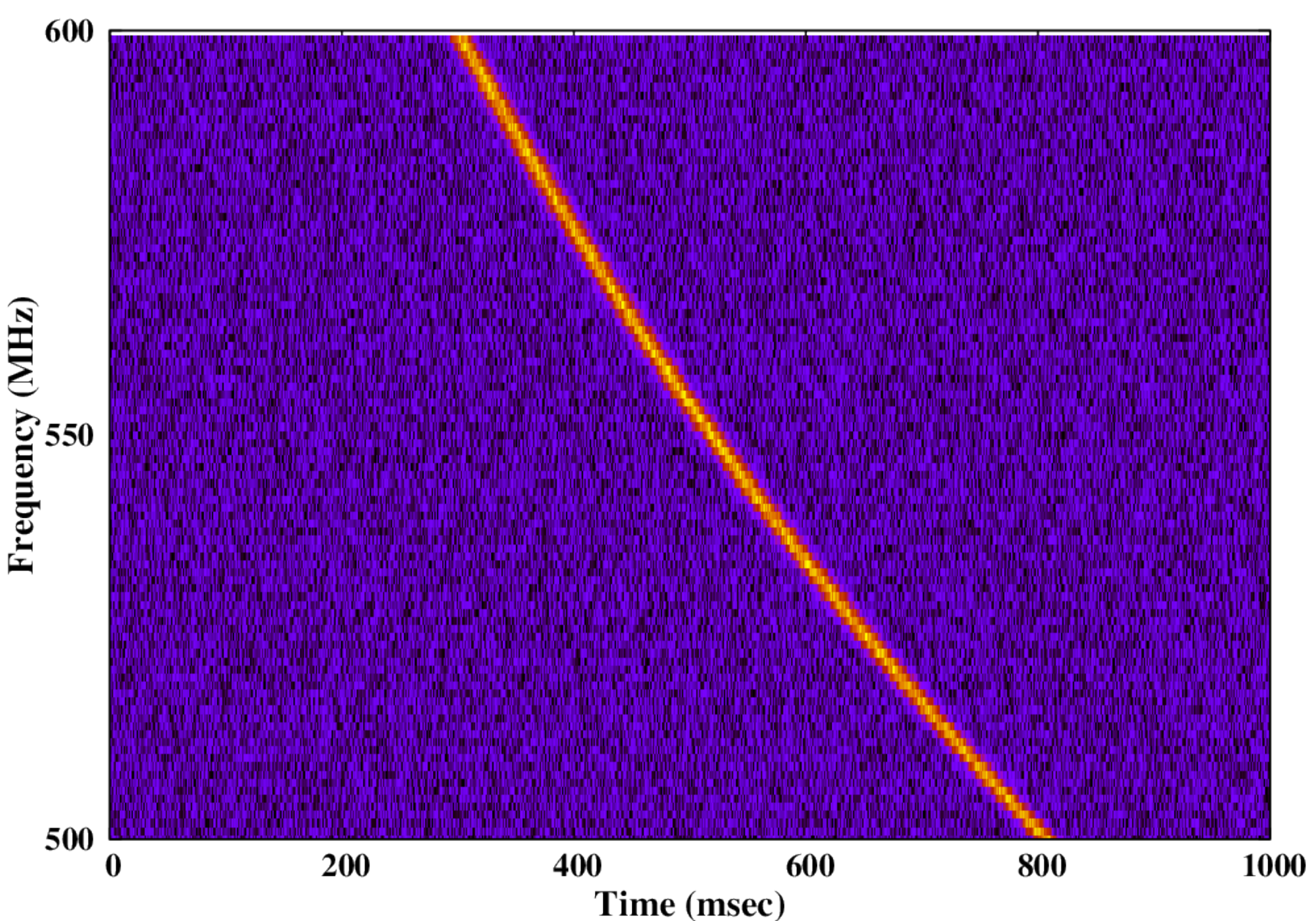}
 % Dispersion_plot.eps: 0x0 pixel, 300dpi, 0.00x0.00 cm, bb=50 50 554 770
 \caption[Pulse dispersion]{Pulse dispersion shown for a simulated pulsar, with period of around 1 sec and DM of 100 pc cm$^{-3}$. 
 The expected delay across the observed bandwidth of 100 MHz at 550 MHz is around half the pulsar period.}
 \label{dispersion}
\end{figure}

The ionised component of the interstellar medium disperses the passing signal by introducing a frequency dependent 
delay. Pulse at higher frequencies experiences relatively shorter delay compared to 
pulse at lower frequencies. The delay in pulse arrival times 
across a finite bandwidth can be seen in Figure \ref{dispersion}. 
Quantitatively, the delay, $\bigtriangleup{\tau}$ in arrival times between a high frequency, $f_{high}$, 
and a low frequency, $f_{low}$, can be given as \cite[]{handbook}, 
\begin{equation}
 \bigtriangleup{\tau} ~ = ~ 4.15 ~ \times ~ \bigg[\bigg(\frac{f_{low}}{GHz}\bigg)^{-2}~ - ~\bigg(\frac{f_{high}}{GHz}\bigg)^{-2} \bigg] ~ 
 \times ~ \bigg(\frac{DM}{pc~cm^{-3} }\bigg)~ msec. 
\end{equation}
Here the dispersion measure (DM), which is an integrated column density of electrons, 
can be calculated as, 
\begin{equation}
 DM ~ = ~ \int_{0}^{d} {n_e} dl ~~ pc~{cm^{-3}}.  
\end{equation}
Here, d is the distance to the pulsar and $n_e$ is the electron density of the ISM. For 
all the known pulsars, the DM can be measured very accurately by comparing the pulse delay 
between higher and lower frequency channels across the observed bandwidth (as seen in Figure \ref{dispersion}). 
The model of the electron density, known as NE2001, across various line-of-sights has been approximately 
measured from independent estimates \cite{NE2001}. Thus, DM of a pulsar also provides a good measure 
of its distance with $\sim$30\% uncertainty.  

The dispersion effects can easily be removed from the DM of the pulsar. 
The procedure to remove the dispersion smearing 
is known as the dedispersion. There are two different ways in which the dispersion 
can be reverted to retrieve the original signal (a) Incoherent dedispersion 
and (b) Coherent dedispersion. During the coherent dedispersion, a model 
of the ISM needs to be adapted to revert the phase delays. This procedure 
is generally carried out during the observation time itself. However, it is 
also possible to coherently dedisperse the recorded data as well. 
While, for the case of  incoherent dedispersion, signals are delay 
compensated for individual frequency channels.  

\subsection{Pulse scattering}
\label{pulse_scattering_sect}
\begin{figure}[h!]
 \centering
%  \hspace*{-1.5in}
 \includegraphics[width=4.5 in,height=1.7 in,angle=0,bb=0 0 505 275]{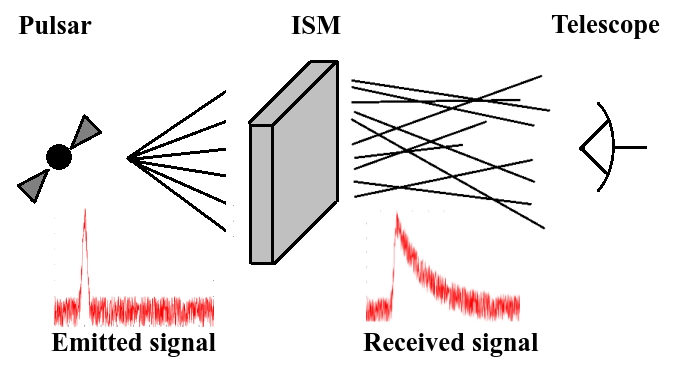}
 % Scattering_picture.jpeg: 673x366 pixel, 96dpi, 17.81x9.68 cm, bb=0 0 505 275
 \caption[Pulse scattering]{A simplified model of the ISM screen approximation. The pulsar signal undergoes 
 multipath scattering inside the ISM. The signals with different path lengths, refracted from 
 different ISM regions, arrive at the telescope with varying delay time-scales. This effect  
 causes the pulse to grow broader with an exponential tail at the trailing edge. [Figure adapted 
 from an original version in \cite{lor08}]}
 \label{scattering_picture}
\end{figure}

Scattering of the pulsar signals manifest as an exponential tail 
at the trailing edge of the received pulse. The irregularities in the 
electron density of the ISM results in a one-sided broadening 
of the pulse. To scrutinize the scattering phenomena, the intermediate ISM can be presented 
as multiple patches with varying refractive index in a thin screen. 
These patches perturbs the phase of the impinging radio wave-front, causing 
a scatter of the resultant signal. A simplified screen approximation 
of the ISM is shown in Figure \ref{scattering_picture}, in which 
the multipath scattering can be seen at the observing end. Signals 
received by the observer from different patches arrive slightly  
later compared to signals that travelled unperturbed. Hence, an exponential 
tail can be observed once the leading phase of the pulse arrives. 

As can be seen from Figure \ref{scattering_picture}, 
the signal to noise ratio (S/N) reduces substantially due 
to the scattering. If the scattering tail decay time-scales 
matches the pulsar period, pulse detection becomes a daunting task. 
The relation between the exponential decay time-scale, 
pulsar distance (d) and observing frequency (f) 
can be presented as \cite[]{handbook}, 
\begin{equation}
 \tau_{s} ~ \propto ~ d^2 ~ f^{-4}. 
\end{equation}
Lower frequencies show relatively longer tails due to the above given relation. 
Hence, most of the pulsar searches are conducted above 1 GHz. 
The ISM is also not uniform in different directions. The Galactic 
plan and specially the Galactic center are regions where the electron density highest 
and the line-of-sight passes through a longer path of electrons, 
making search for pulsars a more challenging task in these regions. 

\subsection{Pulse scintillation}
\label{scintillation_sect}
\begin{figure}[h!]
 \centering
 \includegraphics[width=3 in,height=5 in,angle=-90,bb=0 0 504 720]{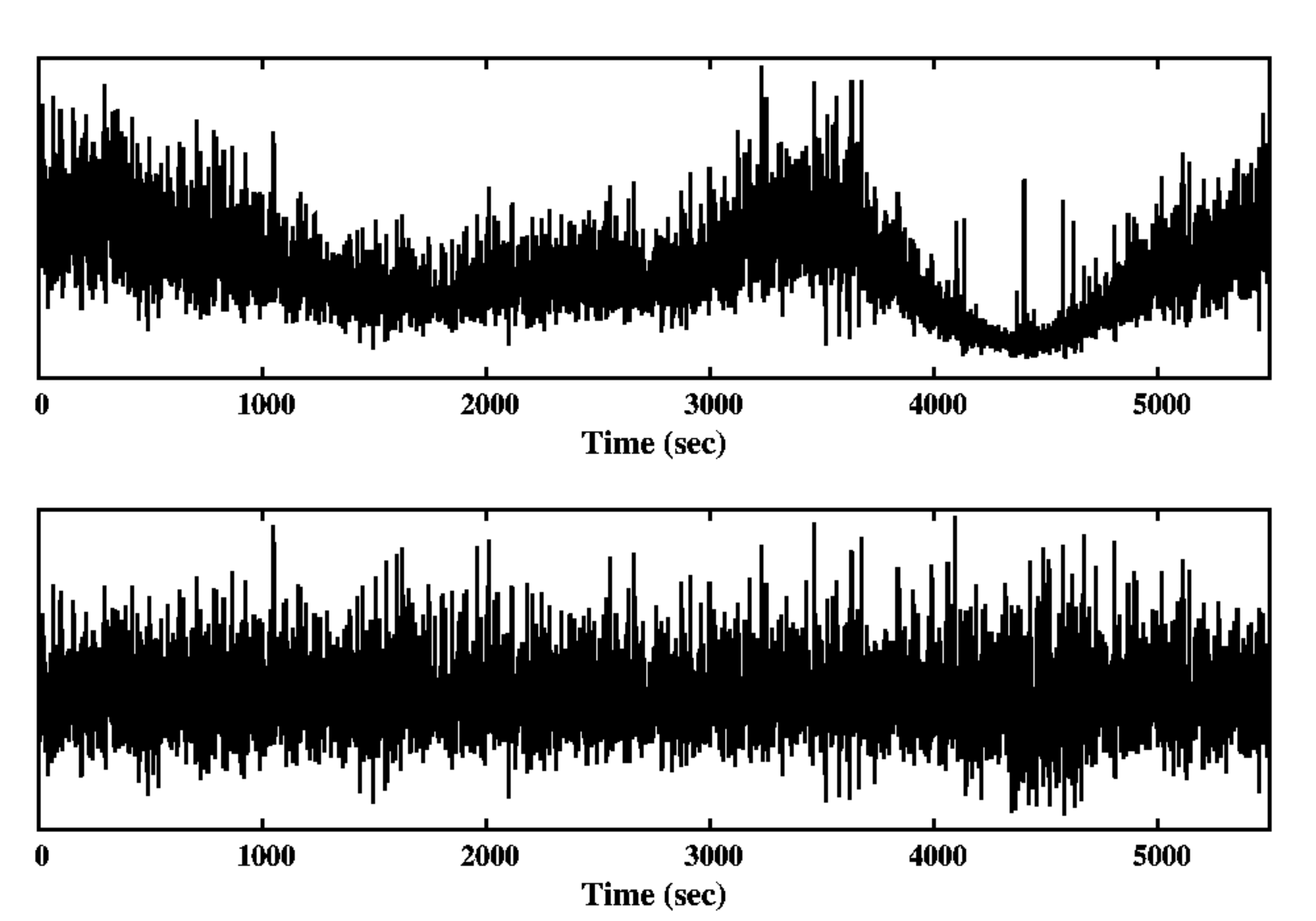}
 % ISS_plot_forThesis.eps: 0x0 pixel, 300dpi, 0.00x0.00 cm, bb=50 50 554 770
 \caption[The interstellar scintillation seen for the pulse peak intensities]
 {The interstellar scintillation seen for the pulse peak intensities. The top panel 
 shows the peak pulse intensity as a function of time. Note the gradual variations due to 
 the changing interference pattern from the turbulent ISM. The bottom panel shows peak pulse 
 energy after removal of the interstellar scintillation normalizing by the block averages.} 
 \label{ISS_plot}
\end{figure}

Closely similar to the scattering phenomena, single pulses from the pulsars 
undergo large pulse intensity fluctuations. This effect is similar to the 
optical twinkling of stars caused by the Earth's atmosphere. 
Figure \ref{pulse_train} shows varying pulse intensities,  
a combination of intrinsic and scintillation effects seen for all known 
pulsars. It was first reported by \cite{lr68}. 
The pulse scintillation is caused by the highly turbulent and inhomogeneous ISM \cite[]{sch68}. 
Different ISM patches, as mentioned in the Section \ref{pulse_scattering_sect}, 
also cause an interference pattern at the observer's plane. Due to the 
relative motion between the observer, the pulsar and the ISM patches, 
the interference pattern also moves across the observer's plane. 
The relative velocity shifts the enhanced and the reduced intensity 
regions to cause changes in the pulse intensity. The time-scale 
of this intensity fluctuation depends upon the aggregated relative 
velocity \cite[]{sch68}. Scintillation can 
also reduce the observed pulse intensity to go below 
the detection threshold, making the single pulse study a challenging task. 
\cite{sch68} has suggested that pulse scintillation are correlated only 
over a limited range of frequencies, also known as the \emph{scintillation 
bandwidth} ($\bigtriangleup{f}$) related to the observing frequency ($f$) as, 
$\bigtriangleup{f}{~\propto~}f^{4}$. This matched well with 
the early observations of ten pulsars by \cite{ric69}. 
The correlation bandwidth for different line-of-sights can be calculated using 
Galactic electron density model \cite[]{NE2001}. Hence, by using a relatively 
larger bandwidth than $\bigtriangleup{f}$, scintillation effects can be reduced.
 
Figure \ref{ISS_plot} shows only the peak pulse energy of around 5000 consecutive burst pulses, 
of PSR B0809+74 observed from the Giant Meterwave Radio Telescope (GMRT) 
at 325 MHz with 33 MHz bandwidth. The gradual fall and rise in the pulse 
intensity is clearly evident. For an unbiased treatment of the individual pulses,
pulse intensities can be averaged in consecutive blocks and each pulse can be 
normalized with their respective block average. 
The effective peak pulse intensities, normalized by their respective block average to 
remove the interplanetary scintillation effect, is also shown in Figure \ref{ISS_plot}.  

\section{Observed pulsar parameters} 
\subsection{Integrated profile} 
\begin{figure}[h!]
 \centering
 \includegraphics[width=5 in,height=3.2 in,angle=0,bb=0 0 350 252]{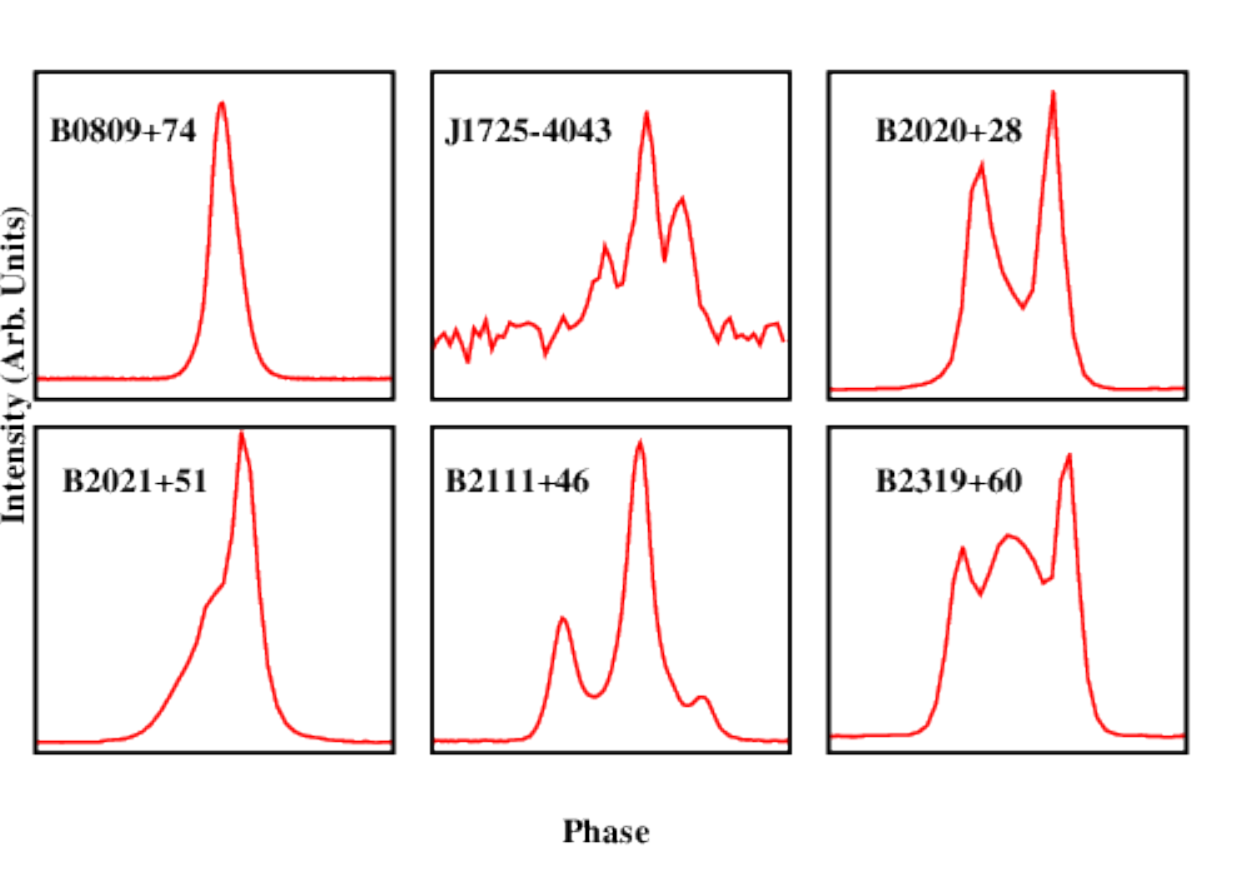}
 % Profiles.eps: 0x0 pixel, 300dpi, 0.00x0.00 cm, bb=50 50 410 302
 \caption[Integrated profiles]{Integrated profiles of six pulsars, observed from the GMRT 
 at 325 MHz. Note the differences in the shape and structure between various profiles.}
 \label{integrated_profile_example}
\end{figure}

Pulsar exhibits varying types of fluctuations in the shape of the 
individual pulses. However, after adding many hundreds or a few 
thousand pulses, a stable shape can be obtained which is known 
as the \emph{integrated profile} of the pulsar. 
The profile of a pulsar remains constant, at a given observing frequency, 
irrespective of the epoch of observations. The integrated profile is like a unique 
signature of the individual pulsar and it shows wide variety of shapes and structures across all known 
pulsars. A few such examples of integrated profiles, to highlight their different shapes, 
are shown in Figure \ref{integrated_profile_example}. The remarkable stability of the integrated 
profile suggest a stable emission process for each pulsar. Hence, the study 
of the integrated profiles is important to map the radio 
emitting regions. However, a few pulsars show switch between different 
integrated profiles, a phenomena known as \emph{mode-changing}, 
which tends to occur for a few hundred to a few thousand pulses (see Section \ref{mode-changing-sect} for details). 
There are also effects like geodetic precession which causes a gradual change 
in profile shapes, although number of pulsars, in which this effect is seen, are very few. 

\subsection{Period and slowdown}
The period of the pulsar, which corresponds to its rotation, is one of the 
pulsar parameter measured with high accuracy. The period does show a gradual slow down due 
to the loss in the rotational kinetic energy of the neutron star with time. 
However, this slowdown of the period is only one part in 10$^{15}$ second for 
normal pulsars. The slowdown of the pulsar, presented as $\pdot$ (= dP/dt), can be measured 
accurately to predict the period of the pulsar at each epoch of observations. 
As the gradual changes in the slowdown rate is related to the amount of energy release by the 
neutron star, young pulsars show higher slowdown rate compared to older pulsars. 
The rate of loss of rotation kinetic energy, the 
spin-down luminosity, can be presented as \cite[]{handbook}, 
\begin{equation}
 \dot{E}~=~ -\frac{dE_{rot}}{dt} ~=~ -\frac{d(I\varOmega^2/2)}{dt} ~=~ -I\varOmega\dot{\varOmega} ~=~ \frac{4\pi^2I\dot{P}}{P^{3}}.  
\end{equation}
Here, $I$ is the moment of inertia and $\varOmega~=~2\pi/P$ is the rotation angular frequency. 
For a normal pulsar with $P$ of 1 sec, $\pdot$ around 10$^{-15}$ s/s and $I$ of the order of $10^{45} ~ g ~ cm^2$ 
[assuming spherical shape with canonical values of mass around 1.4$M_{\odot}$\footnote{$M_{\odot}$ = 1 Solar Mass} 
and radius of 10 km \cite[]{handbook}], the loss of rotational energy is of the order of 10$^{31}$ ergs/sec. 
However, the radio luminosity of a pulsar is a significantly small fraction of this 
energy as most of the energy of the pulsar goes in the emission at X-rays and $\gamma$-rays frequencies 
along with the energy carried away by the particle winds and magnetic dipole emission. 

\begin{figure}[h!]
 \centering
 \includegraphics[width=3 in,height=4 in,angle=-90,bb=50 50 554 770]{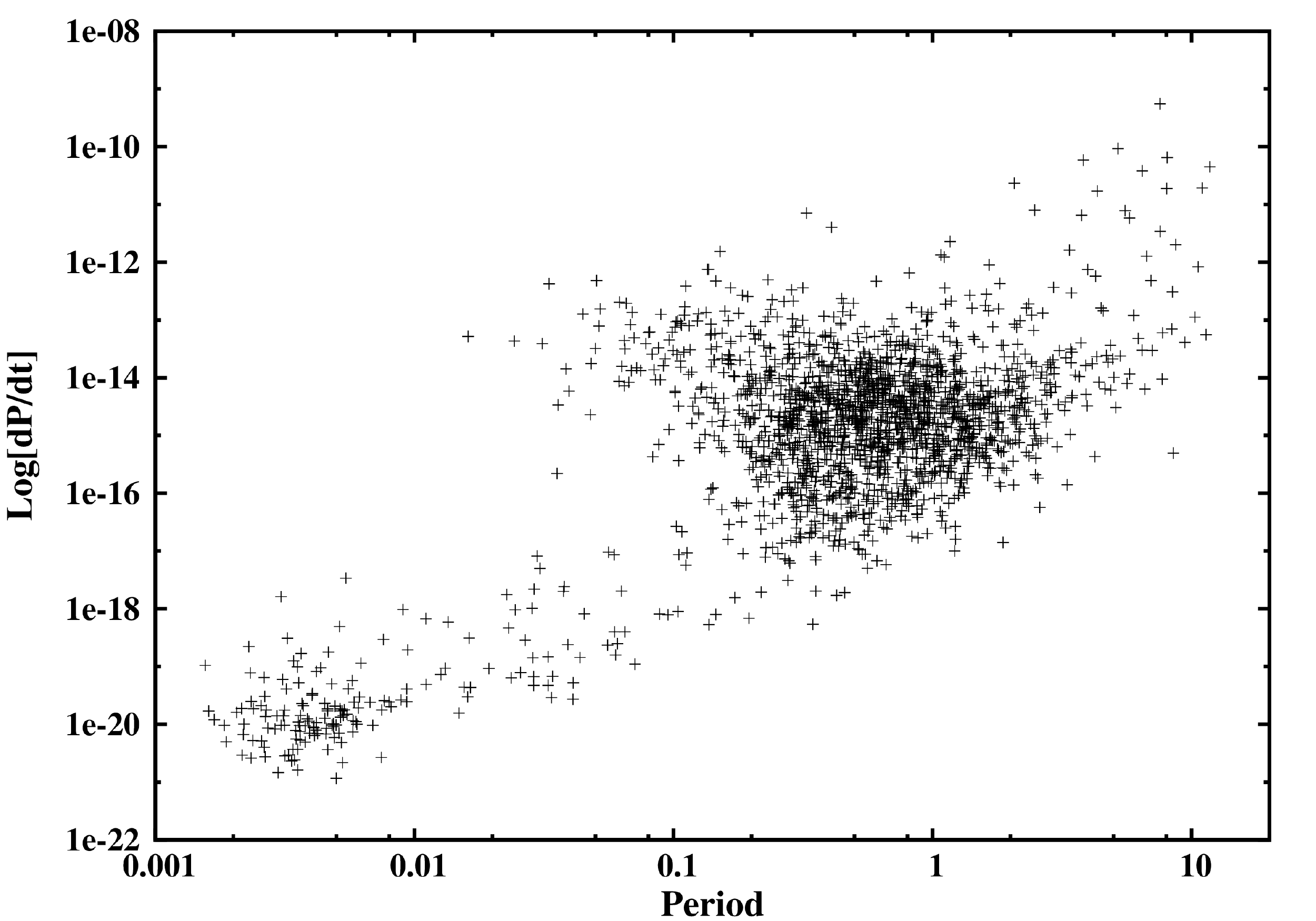}
 % All_PSR_data_p_pdot.eps: 0x0 pixel, 300dpi, 0.00x0.00 cm, bb=50 50 554 770
 \caption[$\ppdot$ diagram of around 2000 pulsars]{$\ppdot$ diagram of around 2000 pulsars. 
 Two cluster of points represent the normal and the millisecond pulsars population at their respective period range.
 A few points, extending from the normal pulsars population towards the upper right corner are the 
 neutron stars discovered at higher energies. [The data were obtained from the 
 \url{www.atnf.csiro.au/research/pulsar/psrcat} \cite[]{mhth05}]}
 \label{ppdot}
\end{figure}
$P$ and $\pdot$ are unique parameters which can also be used to calculate other 
parameters such as the characteristic age and the magnetic field.  
Pulsar exhibits wide range of periods, starting from the fastest known pulsar 
PSR J1748$-$2446ad with period of around 1.3 msec \cite[]{hrs+06} to the slowest known pulsar 
PSR J2144$-$3933\footnote{Few high energy sources were found to show 
longer periods compared to this pulsar.} with period of around 8.5 sec \cite[]{mld+96}. 
Figure \ref{ppdot} shows the $P-\pdot$ diagram, obtained from around 2000 pulsars. 
The plot shows scatter of points with two loosely bound clusters with a bridge 
joining them. The clusters correspond to two main groups, 
which seem to show slightly different observational properties. 
Pulsars with period less than 30 msec belong to a special class 
known as the \emph{millisecond pulsars}. They correspond to old 
neutron stars which have been 'recycled' by phases of mass accretion 
from the binary companion, to radiate once again. The big cluster at relatively 
slow rotation rate, with periods between 0.1 to 1 sec, are the normal pulsars. 
In this thesis, radio emission properties of only the normal pulsars have been discussed. 

\subsection{Pulse Polarization}
\label{pulse_polrization_intro_sect}
As mentioned in Section \ref{pulsar_toy_sect}, pulsar signals are highly polarized. 
The plane of the linear polarization point towards the orientation of the 
magnetic field lines at the point of emission. Hence, pulse polarization measurements are useful to scrutinize 
the geometry of the radio emitting region. Using the radio observations, four Stokes 
parameters, I, Q, U and V can be obtained, where I is the total intensity of 
the signal, while V is the circular polarization intensity.  
The linear polarization intensity, obtained from Q and U profiles 
as L ($=\sqrt{Q^2+U^2}$), is also of great importance to investigate. 
Along with these parameters, it is also of interest to study the position angle (PA) of 
linear polarization (Equation \ref{PA_eq}) as a function of pulse phase. 
\begin{equation}
PA~=~\frac{1}{2}{tan^{-1}\bigg(\frac{U}{Q}\bigg)} 
\label{PA_eq}
\end{equation}
The error in the PA can be measured as \cite[]{mli04},   
\begin{equation}
 \bigtriangleup{PA}~=~\frac{\sqrt{(U\times\bigtriangleup{U})^2~+~(Q\times\bigtriangleup{Q})^2}}{2\times{L^2}}. 
\end{equation}
Here, $\bigtriangleup{U}$ and $\bigtriangleup{Q}$ are the root mean square deviations  
from Stokes U and Q average profiles respectively. 

\begin{figure}[!h]
 \centering
 \subfigure[]{
 \centering
 \includegraphics[width=2.5 in,height=2.5 in,angle=0,bb=0 0 350 252]{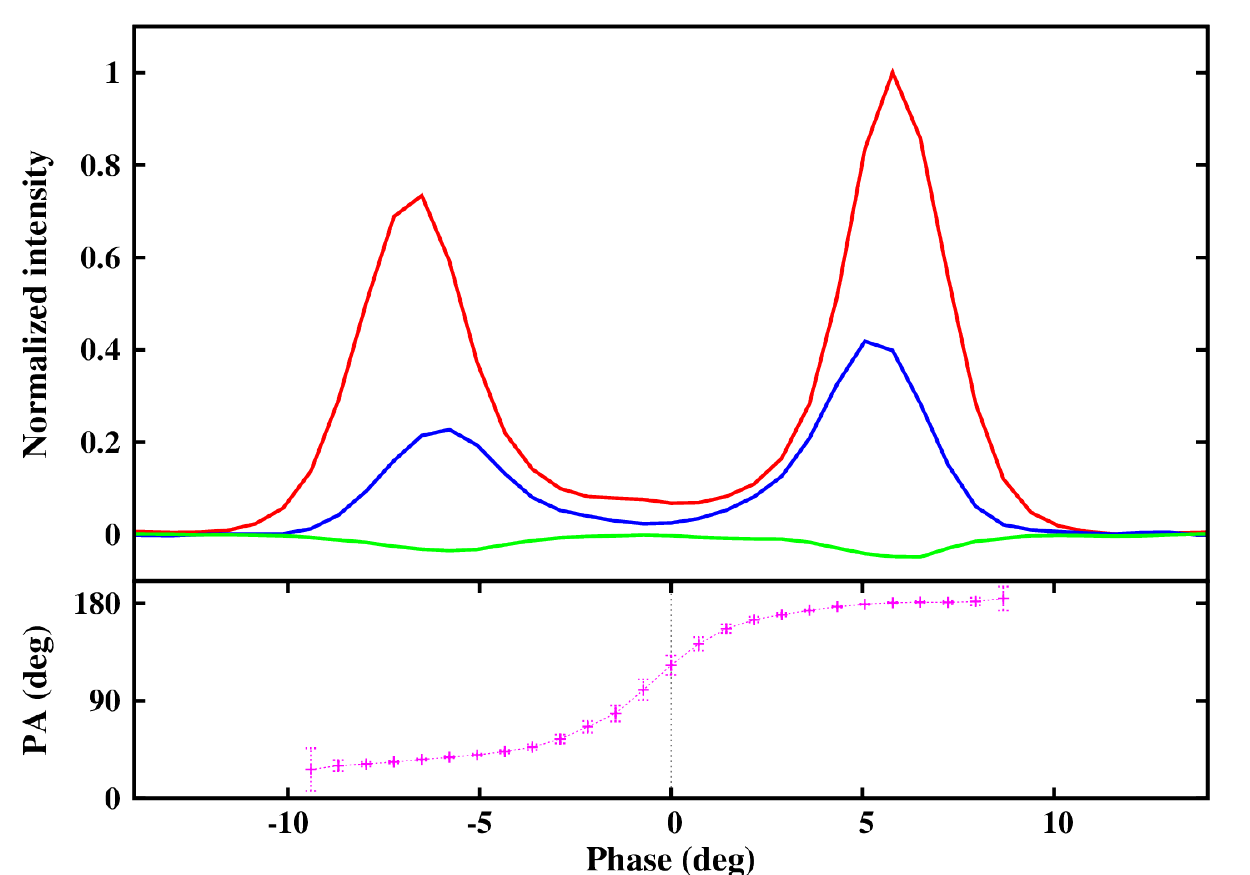}
 % B0525_PAswing_forThesis.eps: 0x0 pixel, 300dpi, 0.00x0.00 cm, bb=50 50 410 302
 \label{pa_swing}
 }
 \subfigure[]{
  \centering
 \includegraphics[width=2.5 in,height=1.8 in,angle=0,bb=0 0 281 242]{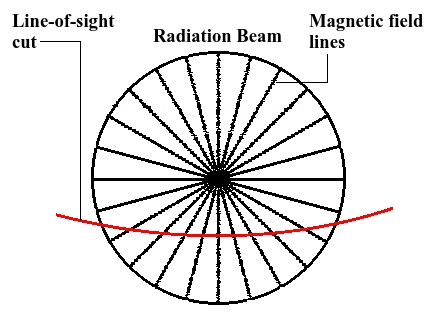}
 % rotating_vector_los.jpg: 374x323 pixel, 96dpi, 9.90x8.55 cm, bb=0 0 281 242
 \label{rotatig_vector}
 }
 \caption[Polarization profile with the rotating vector model]{
 Polarization profiles with the rotating vector model.
 (a) The panel shows different polarization profiles for PSR B0525+21 observed 
 at 1.4 GHz [Archival data obtained from the EPN pulsar network\footnotemark[2]; 
 original observations by \cite{gl98}]. The total intensity is shown with red line 
 while the linear and circular polarization profiles are shown with 
 blue and green solid lines. The bottom panel shows the PA (magenta line) across the 
 pulse longitude with the measured error bars. The characteristic $S$-shaped 
 swing is clearly evident in the bottom panel. (b) The panel shows the top 
 view of the radiation beam, with magnetic field lines emanating around the 
 magnetic pole. The red line-of-sight cut is shown to trace the PA swing as 
 it passes through magnetic field lines of changing orientation, 
 highlighting the rotating vector model proposed by \cite{rc69}.}
 \label{rotating_vector_fig}
\end{figure}

\cite{rc69} has demonstrated that the emission in the radio regime originates near the magnetic pole, 
using the PA changes across the pulse profile. 
In the proposed Rotating-vector model, 
\cite{rc69} demonstrated the expected shape of the PA angle changes as 
characteristic {\bf$S$}-shape curve by simple geometric arguments. 
The expected polarization angle swing, as the line-of-sight cross the magnetic field lines, 
matched perfectly well with the observed PA swing for many pulsars. 
It should be noted that a few pulsar also exhibit an abrupt 90$^\circ$ swing in the 
otherwise smooth {\bf$S$}-shape curve. These kind of changes tend to occur due to 
the presence of two different orthogonal polarization modes. The change in their dominance 
can cause such abrupt swing of 90$^\circ$. Figure \ref{rotating_vector_fig} shows an example 
of different polarization profiles obtained for 
PSR B0525+21\footnote[2]{www.jb.man.ac.uk/research/pulsar/Resources/epn}, 
along with the {\bf$S$}-shaped PA swing across the pulse profile. 
In principle, the observed PA swing can also help to derive 
other pulsar parameters, such as $\alpha$ and $\beta$ using the geometry. 
Here, $\beta$ (also known as the impact angle) is an angle between 
the magnetic pole and the center of the line-of-sight cut on the emission beam 
measured away from the rotation axis. However, a degeneracy can occur between 
an outer line-of-sight cut (positive $\beta$) and an inner line-of-sight 
cut (negative $\beta$) for limited range of active pulse longitude \cite[]{nv82}.  

As the pulsar signal passes through the interstellar medium, 
it also undergoes Faraday rotation, which is a frequency dependent 
rotation of the polarization angle. The effect can cause artificial changes 
in the PA, although these changes are large and can easily be distinguished 
from the intrinsic PA changes. However, a careful calibration 
is required using an artificial source before obtaining the four 
Stokes parameters. 

\section{Single pulse phenomena}
\subsection{Giant Pulses}
Pulsar in the Crab nebula was one of the first source discovered through 
its extremely strong burst pulses \cite[]{sr68}. These pulses, 
also known as the \emph{Giant pulses}, have peak intensities 
up to 1000 times higher than normal individual pulses. Detail study of 
these Giant pulses from the Crab pulsar has revealed that they 
are superposition of extremely narrow nanosecond duration structures \cite[]{hkw+03}. 
Currently, giant pulses are known to occur in around 10 pulsars. 
They are mostly reported in the millisecond pulsars \cite[]{kt00,rj01,J04}. 
They have also been reported in three regular pulsars 
\cite[]{ke04,EK05}, only at lower frequencies of 100 MHz, which remain to be verified 
at higher frequencies. In the earlier studies \cite[]{sr68,kt00,J04}, 
it was reported that giant pulses have significantly smaller 
pulse widths compared to average pulses. The expected brightness 
temperature of these pulses reached up to 10$^{37}$ K, making them  
the brightest source in the known Universe \cite[]{cbh+04}. 
They also tend to occur near the edge (either near the trailing edge or near the leading edge)
of the pulse profile. The emission mechanism behind the production 
of giant pulses, with this peculiar nature, still remain unidentified. 
\subsection{Drifting}
\label{drifting_intro_sect}
\begin{figure}[h!]
 \centering
 \includegraphics[width=4 in,height=4 in,angle=0,bb=0 0 562 487]{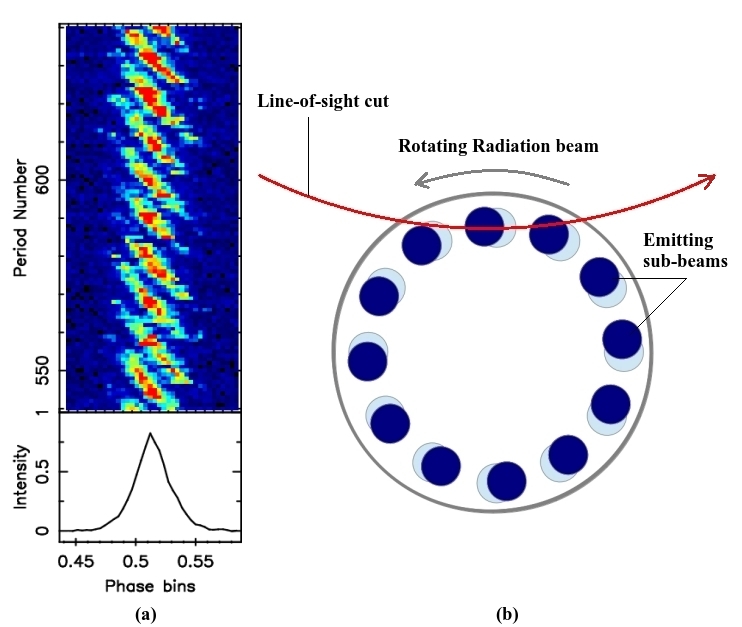}
 % Drifting_beam_combine.jpg: 749x649 pixel, 96dpi, 19.82x17.17 cm, bb=0 0 562 487
 \caption[Example of drifting subpulses in PSR B0809+74]{An example of drifting subpulses in PSR B0809+74. 
 (a) A stack of around 100 pulses are shown in the top panel
 with the average profile in the bottom panel, obtained from the GMRT observations at 325 MHz.
 Individual pulse intensities are shown with a color ramp from blue to red. Note the varying structure 
 of individual pulses with two (sometimes three) subpulses. The drifting towards the leading side 
 is clearly evident in the consecutive pulse sequence. (b) The rotating sub-beams carousal 
 with a line-of-sight cut is shown to demonstrate the drifting of subpulses each time the line-of-sight 
 cuts the radiation beam during each period. Two sets of sub-beams are shown (light blue and dark blue) to show 
 the displacement of sub-beams during a period interval.}
 \label{drifting_beam}
\end{figure}

A few pulsars tend to show individual pulses with multiple peaks (i.e. subpulses), 
although the integrated profiles are smooth with single component in many cases. 
For such pulsars, contiguous pulses are reported to show marching of these 
subpulses across the pulse longitude, also known as the \emph{drifting}. 
The drifting of the subpulses was first reported by \cite{dc68}. 
Figure \ref{drifting_beam}(a) shows sequence of observed pulses 
from PSR B0809+74. The drifting of the subpulses is clearly evident here 
with a periodicity of around 11 periods after which the pattern repeats itself 
(see Section \ref{rotating_carousal_sect} for more details regarding the drifting periodicities).
\cite{rs75} have suggested the cause of the periodicity due to a rotating 
carousal of sub-beams within a hollow radiation beam (as shown in Figure \ref{drifting_beam}(b)). 
Each time the line-of-sight passes across the radiation beam, it encounters 
a slightly different set of arrangements of the sub-beams because of their  
rotation around the magnetic pole. As these sub-beams are uniformly distributed, the 
patten repeats itself after certain number of periods as each sub-beam is replaced by an 
adjacent sub-beam. \cite{dr01} has proposed a method to determine the number of 
sub-beams in the rotating carousal. 

The periodicity of the drifting subpulses is defined as $P_3$ while 
the separation between the subpulses in the individual pulse 
is defined as $P_2$. A significant correlation was reported 
between the $P_3$ and pulsar age \cite[]{wol80,ran86}. 
However, in the largest survey conducted on the drifting pulsars 
\cite[]{wes06,wse07,wel07}, it was concluded that 
no such correlation exist. Drifting of subpulses can occur in both directions, 
from leading side to trailing side and vice-versa. 
\cite{ran86} has reported that pulsars do not have any preferred sense 
of drifting as equal number of pulsars were found exhibiting 
different drifting directions. \cite{wes06,wse07} has also concluded 
that drifting is an intrinsic property of emission mechanism. 
However, for a few pulsars it is not possible to detect classical 
drifting behaviour because their subpulses are disordered. 
As pulsar gets older, the sub-beams structure becomes more and more organized to 
produce detectable subpulse drifting pattern \cite[]{wes06,wse07}. 

\subsection{Mode-changes}
\label{mode-changing-sect}
\begin{figure}[!h]
 \begin{center}
 \subfigure[]{
 \centering
 \includegraphics[width=2 in,height=2.5 in,angle=-90,bb=50 50 554 770]{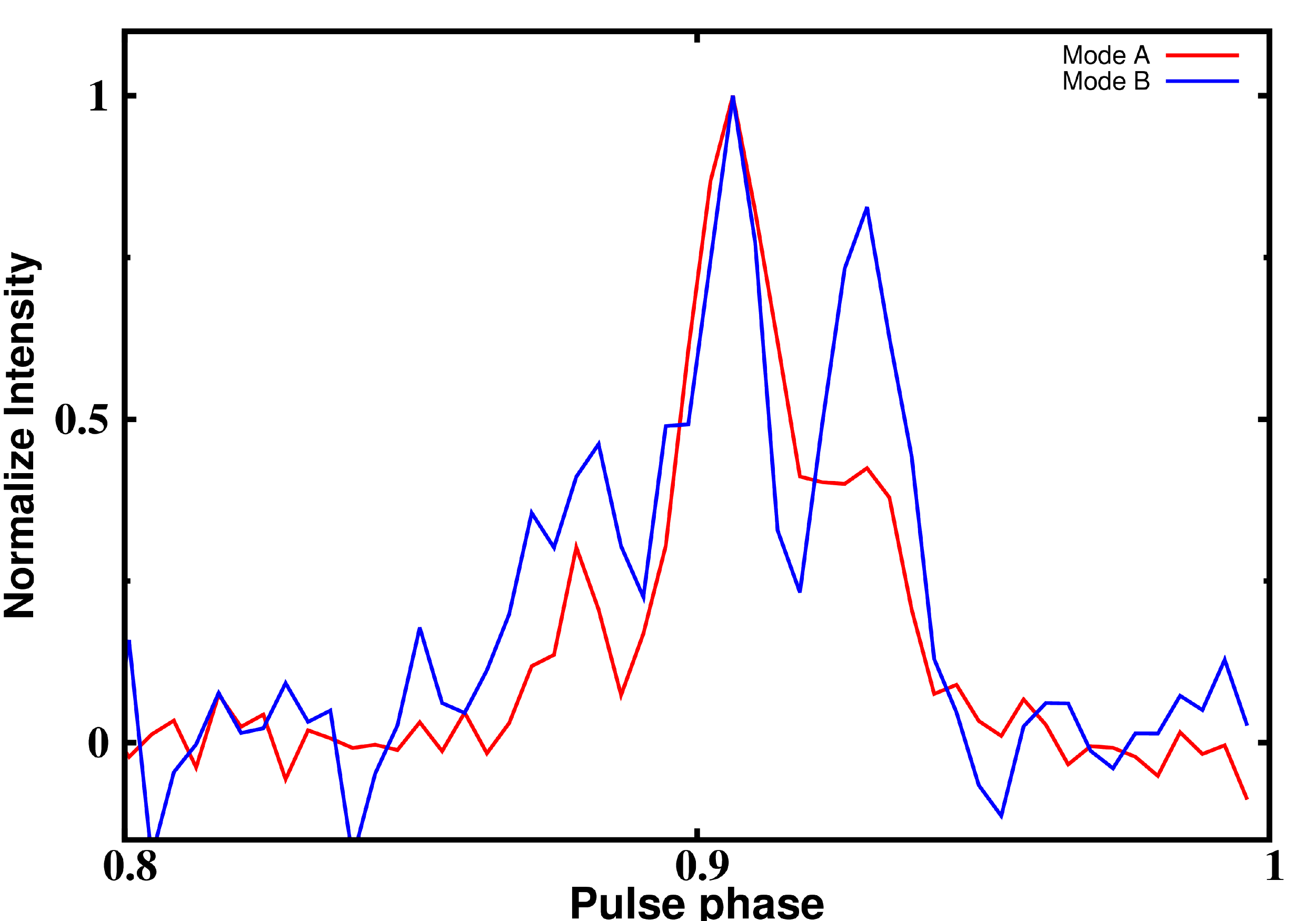}
 % Both_mode_forThesis.eps: 0x0 pixel, 300dpi, 0.00x0.00 cm, bb=50 50 554 770
 }
 \subfigure[]{
 \centering
 \includegraphics[width=2 in,height=2.5 in,angle=-90,bb=50 50 554 770]{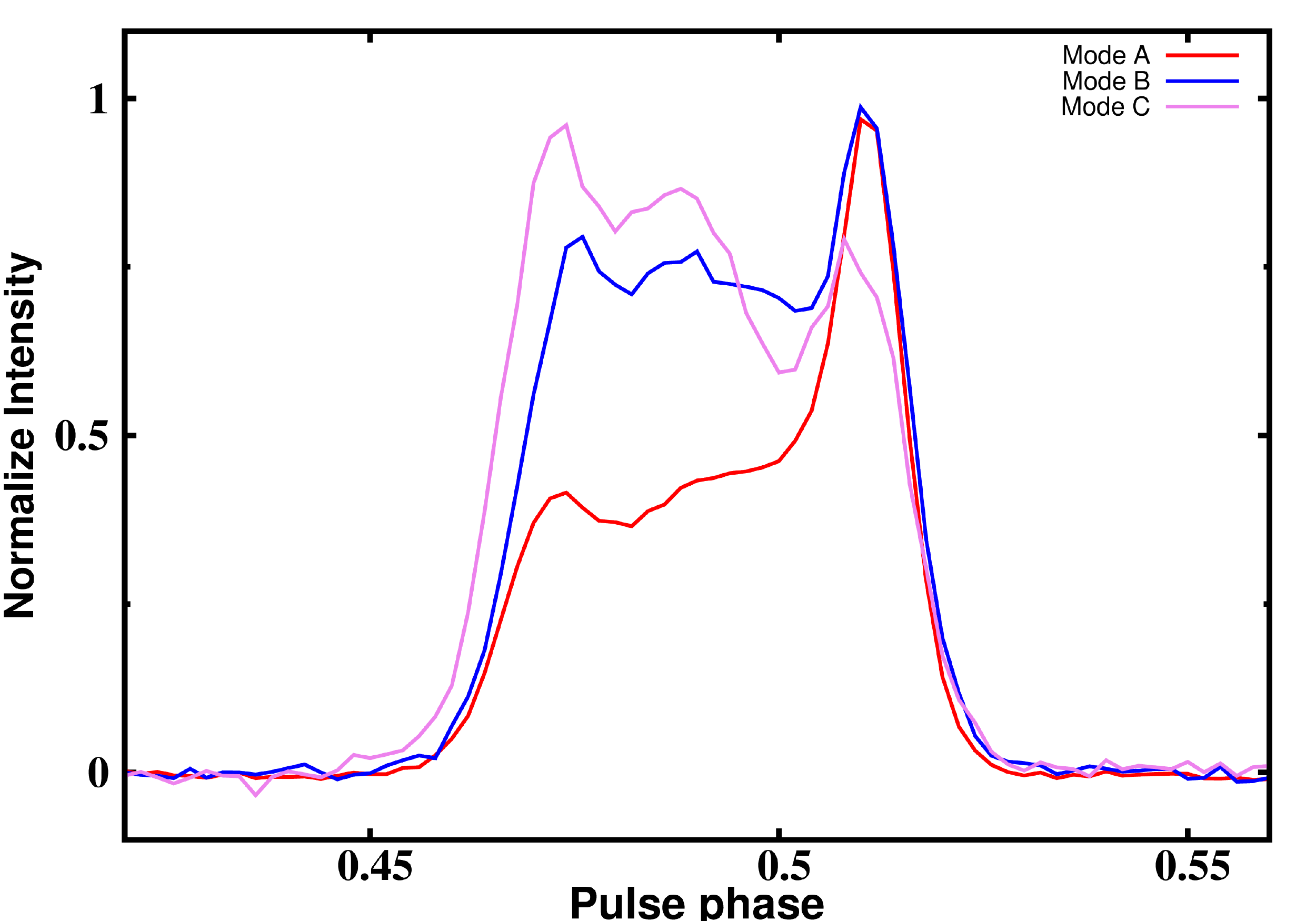}
 % Mode_profiles_forThesis.eps: 0x0 pixel, 300dpi, 0.00x0.00 cm, bb=50 50 554 770
 }
 \caption[Two examples of mode-changing pulsars]{Two examples of mode-changing pulsars PSRs (a) J1725$-$4043 (b) B2319+60, 
 observed from the GMRT at 325 MHz. Note the clear differences in the profiles for different modes.}
 \label{mode-changing-fig}
 \end{center}
\end{figure}

The integrated profile of a pulsar is known to be one of the stable parameter 
and in most cases, characterising cuts across the radiation beam.
However, a few pulsars display switching between different integrated profiles. 
This phenomena is known as the \emph{mode-changing}. In all mode-changing pulsars, 
one of the profile mode is more favoured over the other. This can be distinguished 
by measuring the amount of time pulsar spends during each profile mode. 

Figure \ref{mode-changing-fig} shows two examples of mode-changing pulsars. 
PSR J1725$-$4043 shows two different modes with different intensities in the 
trailing component. However, PSR B2319+60 shows three different modes 
with significantly different profiles \cite[]{wf81}. The transition between 
different modes are rather sudden for both these pulsars. 
In many pulsars, the modes are classified as the normal mode/s 
(for more frequent modes) and abnormal mode/s (for non-frequent modes). 
Mode-changes also manifests itself in the form of changing drift rates with 
different modes (see Section \ref{null_dritft_corr_sect}). 
For example, PSR B2319+60 shows slightly different drifting periodicities between  
Mode-A to Mode-B, while the Mode-C shows disordered subpulses \cite[]{wf81}. 

\cite{ran86} has tabulated different mode-changing pulsars and concluded that 
mode-changing is mostly observed in pulsars with multiple component profiles. 
The intensity of the central component was reported to enhance 
during the abnormal modes \cite[]{ran86}. A few pulsars also show 
significant changes in different polarization profiles, along with 
the total intensity profiles, indicating a global magnetospheric change. 
Recently, \cite{wmj07} has enhanced the number of known nulling 
pulsars by identifying mode-changing phenomena in around 6 pulsars. 
 
\subsection{Pulse Nulling}
\begin{figure}[h!]
 \centering
 \includegraphics[width=1.5 in,height=5 in,angle=-90,bb=0 0 202 720]{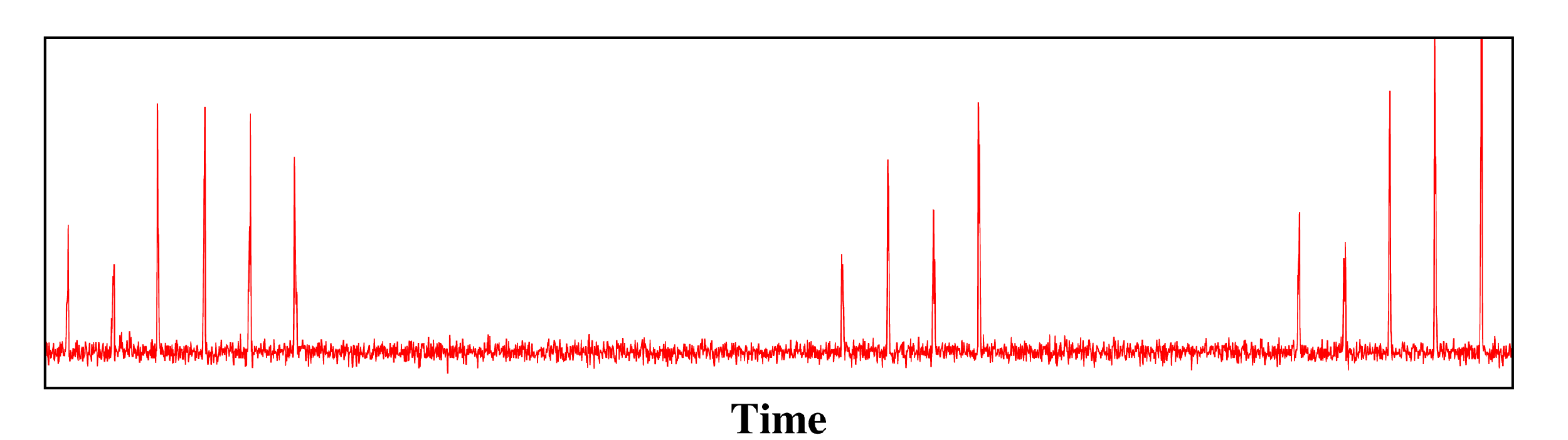}
 % Pulse_train_forThesis.eps: 0x0 pixel, 300dpi, 0.00x0.00 cm, bb=352 50 554 770
 \label{nulling_example}
 \caption[Nulling in single pulses]{A sequence of pulses from PSR B2319+60 observed at 1420 MHz 
 from the Westerbork Synthesis Radio Telescope. The nulling in pulses is clearly evident.}
\end{figure}

The abrupt cessation of pulsed radio emission for 
several pulse periods, exhibited by many pulsars, 
has remained unexplained despite the discovery 
of this phenomenon in many radio pulsars. 
This phenomenon, called \emph{pulse nulling}, was first 
discovered in four pulsars in 1970 \cite[]{bac70}. 
Subsequent studies have revealed pulse nulling 
in about 100 pulsars to date \cite[]{bac70, rit76, big92a, viv95, wmj07}. 

The fraction of pulses with no detectable 
emission is known as the nulling fraction (NF) 
and is a measure of the degree of nulling in 
a pulsar. However, NF does not specify 
the duration of individual nulls, nor does it 
specify how the nulls are spaced in time. 
Nulling can also considered to be an extreme form of mode-changing 
phenomena \cite[]{wmj07}, in which pulsar emission 
goes below the detection threshold. Chapter 2 discusses 
the nulling phenomena in full detail. 

\section{Outline of the thesis}
The aim of this thesis is to investigate the absence of pulses 
in pulsar, a phenomena also know as pulsar nulling. 
The outline of the thesis is as follows. 
The motivation for the work done in thesis 
is listed in Chapter 2. It also discusses the 
emission mechanism as proposed by the standard model. 
A summary on the previous work carried out to investigate the 
pulsar nulling phenomena is also included in Chapter 2. 
Chapter 3 discusses in details all the observations 
carried out for the work that has been reported in thesis. 
Observations from various radio telescopes are discussed 
and compared in the similar chapter. The common analysis techniques deployed 
during most of the observations are also listed 
in Chapter 3. In Chapter 4, details regarding the 
nulling behaviour of various pulsars are listed. 
The comparison has been made between similar NF pulsars 
in order to demonstrate that, the NF does not quantify nulling 
behaviour in full details. Chapter 4 also highlights the fact 
that nulling is unpredictable. Chapter 5 presents a 
comparison between two pulsars with high NFs. It is shown 
that both the pulsars exhibit similar bursting behaviour, 
but strikingly different quasi-periodicities and emission properties. 
Chapter 6 discusses the details regarding the simultaneous observations 
of two nulling pulsars at four different frequencies. 
Chapter 7 summaries the results obtained in three different studies 
along with their implications. The future work to extend the work carried 
out in the thesis is also listed in Chapter 7.  

\chapter{Background}
% \graphicspath{{/home/vishal/My_paper/Thesis/Chapter2/}{/home/vishal/My_paper/Thesis/Chapter2/}{/home/vishal/My_paper/Thesis/Chapter2/}}
\graphicspath{{Images/}{Images/}{Images/}}

This chapter discusses the peculiar single pulse phenomena, 
exhibited by many radio pulsars, known as the pulsar nulling 
in full detail. The basic background regarding different relevant 
works that have been carried out, since the discovery of this 
phenomena, is also discussed with details in this chapter. 
In order to understand the cessation of radio 
emission, the chapter also highlights the standard emission 
mechanism model. Towards the end of the chapter, 
various models proposed over the years to 
explain the cessation of radio emission are discussed. 
The chapter also lists the broad motivations, 
for the work carried out in this thesis. 
\section{Pulsar emission mechanism}
\subsection{Neutron star}
\label{ns_sect}
Neutron stars are born after the violent death of massive stars. 
When a core of a star attains the critical mass limit, 
after the `burning' of hydrogen, the core becomes unstable to support the balance 
between the radiation pressure and the gravitational collapse. 
During this period, a chain of events occurs in which star's core 
collapses in a massive explosion known as the supernova.  
The supernovae mainly ejects the rest of the star in a vast 
energy release impulsively, while collapsing the core further 
to form a neutron star. There are two main types of supernovae events, 
namely type I and type II supernovae. Type II supernovae is associated 
with the isolated massive stars. Type I supernovae 
is classified as a contact binary system of a white dwarf  
and a companion star, in which the companion 
star transfers matter to White Dwarf, there by pushing 
it to the Chandrasekhar limit. To distinguish between the two classes, 
their spectra can be studied. Type II supernovae are hydrogen rich 
as it was still available in the envelop surrounding the core, while 
type I supernovae are hydrogen depleted due to their origin 
from the white dwarfs. Theoretically, type II supernovae are more likely  
to form a neutron star because during the type I supernovae, the 
white dwarf is also likely to get disintegrated rather 
than collapse further to form a neutron star. 

The neutron star avoids further collapse due to 
the \emph{neutron degeneracy pressure}. \cite{ov39} 
analysed the structure of a star consisting 
of degenerate neutron gas. It was shown that degeneracy was 
so complete that density and pressure are more important 
than the temperature. The relationship between the density 
and pressure, also known as the equation of state, have 
been modelled through multiple theories which can be tested 
by the observed relationship between the radius and the mass of 
the neutron star. Hence, these quantities are important to measure 
for a large number of pulsars. Current estimate suggest the 
radius of the neutron star to lie within a small range of  
10.5 to 11.2 km \cite[]{gho07}. The theoretical limit on the neutron star 
masses lies between 0.5 to 2 M$_{\odot}$ \cite[]{gho07}. 
The maximum expected mass of a neutron star is about 3 M$_{\odot}$ \cite[]{lp01}.  
Most observed values of the neutron star masses are around 1.4 M$_{\odot}$ \cite[]{handbook}. 

The structure of the neutron star is divided 
in two different regions, namely the crystalline 
solid crust (about 1 km thick) and liquid 
interior mainly consisting of superfluid neutrons. 
The density at the surface of the neutron 
star is around 10$^6$ g cm$^{-3}$ which 
reaches to 10$^{15}$ g cm$^{-3}$ near the core.  
The constituents of the central region could be further 
exotic state of matter, like mesons or kaons \cite[]{bay91}.  
At the exact core, neutrons are also likely to dissolve 
to form quarks and gluons. The surface of the neutron star 
is relatively less dense (about 10$^9$ orders of magnitude) 
compared to the interior, hence its likely to be made of 
solid crystalline lattice, mainly of iron nuclei with 
a sea of free electrons flowing between them. The iron 
is more likely element to exist because of it's high binding energy. 
The surface is believed to be extremely smooth with structure 
irregularity of only $\sim$5 mm because of the high gravitation 
potential. Neutron stars are also proposed to have very thin layer of 
helium \cite[]{rc72} and hydrogen on the surface.  

\subsection{Magnetosphere of pulsars}
\label{GJ_sect}
Pulsars are known to be one of the highly magnetized objects
in the known Universe. The magnetic field strength of pulsars
ranges from 10$^8$ G to 10$^{14}$ G. The high energy
X-ray sources, known as \emph{Magnetars}, are known to have
the highest magnetic field of 10$^{13}$ to 10$^{14}$ G while
the old millisecond pulsars have low magnetic fields
of around 10$^{8}$ G. The direct measurements
of the high magnetic fields come from the absorption
lines observed in the X-ray spectra \cite[]{wdp+79,wsh83}.
\cite{pac67} has suggested that, the magnetic flux conservation during the 
core compression can cause magnetic fields to reach up to 10$^{12}$ G for the neutron stars.
This is possible to occur if the interior of the neutron star is
highly conductive. Although, the high magnetic field
has very little effect on the overall structure of the neutron star,
it can alter the structure of the lattice on the surface \cite[]{rud74}.
The primary loss of rotational energy from the neutron star is also
due to the radiation caused by the high magnetic field.

The region around the neutron star were first thought to be complete vacuum
due to their origin from a massive explosion which strips the envelop around
the neutron star. However, in reality, it is filled with plasma of different polarities
partitioned by the magnetic field, hence also known as the \emph{pulsar magnetosphere}.
This plasma is accumulated by pulling the charged particles from the surface
of the neutron star. \cite{deu55} was first to suggest this scenario, before the
discovery of pulsars, in which rotating magnetized star would
generate enough electric field to accelerate particle
to substantial energies. Thus, he also suggested that accelerating particles
from these magnetized star could produce cosmic rays,
an idea persist to the present day. \cite{gol68} suggested a model in which
a bunch of particles were proposed to corotate with the pulsar, trapped
in the equatorial magnetic field. However, no explanation was given
regarding how such bunch would remain stable. This model was the first
to propose bunching of particles which influenced future work.
\cite{og69} proposed a model for oblique rotator, where
the magnetic and the rotation axes are not aligned, in which
the loss of rotational energy was associated with the low frequency magnetic dipole
radiation. The non-alignment will cause the magnetic dipole to
radiate pulses. \cite{gj69} were the first
to formalize the magnetosphere around pulsars using a
simple aligned rotator case. They extended the model
proposed by \cite{deu55} for the neutron stars.
Although, the \cite{gj69} model does not describe
a realistic scenario, it was important work because
it demonstrated, using simple electrostatics, that region surrounding the pulsar 
is filled with plasma of two different polarities.
Multiple work for the oblique rotator have been
conducted using numerical simulations to extend the
\cite{gj69} interpretation of the magnetosphere \cite[]{km85,ckf99,spi04}.
The origin of this plasma was proposed as follows, 
for an aligned rotator \cite[]{gj69}.

\begin{figure*}[!h]
 \begin{center}
   \centering
%    \hspace*{1.2in}
%   \includegraphics[bb=0 0 536 539]{pulsar_toy_gimp.jpg}
 % pulsar_toy_gimp.jpg: 536x539 pixel, 72dpi, 18.91x19.01 cm, bb=0 0 536 539
%   \includegraphics[width=5 in,height=3 in,angle=0,bb=14 14 961 555]{pulsar_toy_align_gimp.eps}
  \includegraphics[width=4 in,height=3.5 in,angle=0,bb=0 0 536 539]{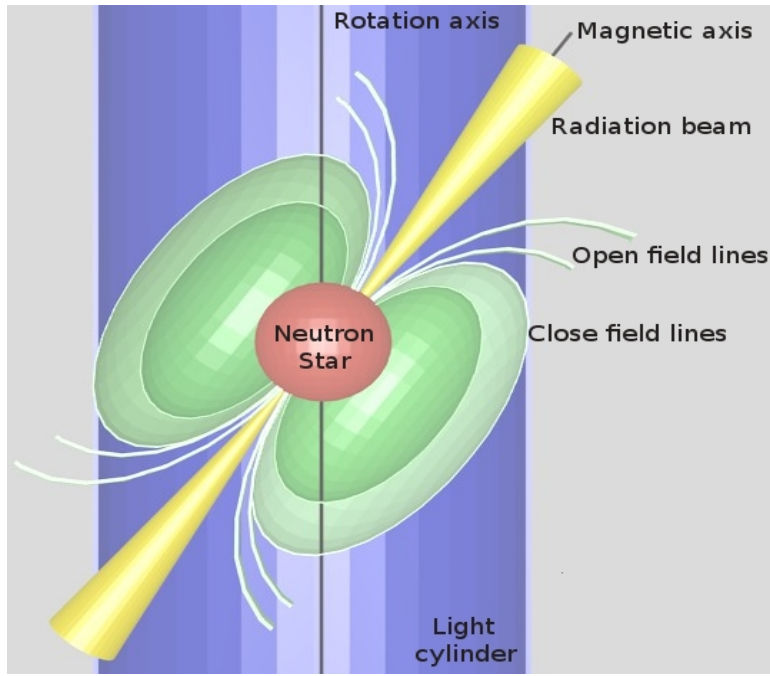}
  \caption[Pulsar toy model with GJ charge density distributions]{Schematic diagram of a radio pulsar 
  (not to scale) to illustrate charge density around the neutron star 
  (in grey color) for an aligned rotator adapted by \cite{gj69}. 
  The yellow beam is the beam of radio emission. The closed 
  field lines are shown with green enclosed curves, while the light-cylinder 
  is shown in background with blue color. The magnetosphere of the 
  pulsar is divided in two parts, with negative charges on top 
  of the polar cap and positive charges around the equatorial region. 
  The boundary dividing the two regions occurs at $cos^{-1}(\pm1/\sqrt{3})$ 
  (cartoon using the free 3D modeling package : {\itshape blender} \url{www.blender.org}).}
  \label{toy_model2}
 \end{center}
\end{figure*}

The induced electric field ({\bf E$_{ind}$}) at a distance (r) due to
the rotation of the neutron star with magnetic field ({\bf B})
would be around,
\begin{equation}
\boldsymbol{E_{ind}} ~ = ~ -(\boldsymbol{\varOmega}~\times~{\bf{r}})~\times~{\bf{B}}/c.
\label{E_eq}
\end{equation}
Here, \boldsymbol{$\varOmega$} is the angular velocity of the pulsar. 
For an aligned rotator, as assumed by \cite{gj69}, 
the rotation and the magnetic axes point 
in the $\boldsymbol{\widehat{z}}$ direction with 
the neutron star at the center. The angular velocity 
in such case can be simplified in polar coordinates as,
\begin{equation}
\boldsymbol{\varOmega}~=~\varOmega{cos\theta}{~\boldsymbol{\widehat{r}}}~-~ \varOmega{sin\theta}{~\boldsymbol{\widehat{\theta}}}. 
\label{omega_eq}
\end{equation}
Similarly, the aligned dipole magnetic field at a distance r, from the center 
of the conducting sphere (neutron star) with radius R, can also be presented as \cite[]{jackson}, 
\begin{equation}
{\bf B} ~= ~ \frac{B_p{R^3}}{r^3}\left(cos{\theta}{~\boldsymbol{\widehat{r}}}~+~\frac{1}{2}{sin\theta}{~\boldsymbol{\widehat{\theta}}}\right). 
\label{dipole_B}
\end{equation}
Here, $B_p$ is the surface magnetic field near the pole, which can also be given as, 
$B_p~=~2\mu/R^3$, where, $\mu$ is the magnetic dipole moment. 
By putting equations \ref{omega_eq} and \ref{dipole_B} into equation \ref{E_eq}, 
we can obtain the induced electric field inside the neutron star ($\boldsymbol{E_{in}}$ for $r<R$) as, 
\begin{equation}
\boldsymbol{E_{in}} = \frac{\varOmega{B_p}{R^3}sin{\theta}}{cr^2}\left(\frac{1}{2}{sin\theta}
{~\boldsymbol{\widehat{r}}}~-~cos{\theta}{~\boldsymbol{\widehat{\theta}}}\right).
\label{Ein_eq}
\end{equation}
As mentioned in Section \ref{ns_sect}, the crust region
is highly conductive for the neutron stars. Thus, the charges
will move and arrange inside the star to cancel the
induced electric field. If outside of the sphere is taken as a vacuum
in the initial conditions, the boundary conditions will allow the estimate of
the electrostatics potential near the surface as, 
\begin{equation}
 \phi_{out}~=~-\frac{B_{p}\varOmega{R^5}}{6cr^3}\big(3cos^{2}\theta~-~ 1\big). 
\end{equation}
The external electric field can be derived by taking a gradient 
of the above mentioned electric potential as,  
\begin{equation}
 \boldsymbol{E_{out}}~=~ -\boldsymbol{\triangledown}\phi_{out}. 
 \label{Eout_eq}
\end{equation}
It can be shown from equations \ref{dipole_B}, \ref{Ein_eq} and \ref{Eout_eq}  
that, $\boldsymbol{E_{out}\cdotp{B}~\neq~0}$ while $\boldsymbol{E_{in}\cdotp{B}~=~0}$.  
Hence, the component of the electric field ($E_{\|}$) aligning the magnetic field 
and perpendicular to the surface at the the polar cap ($r=R$) can be given as \cite[]{handbook,gho07}, 
\begin{equation}
E_{\|} ~ = ~ \frac{\boldsymbol{E_{out}\cdotp{B}}}{B} ~ = ~ ~ - \frac{\varOmega{B_p}R}{c}cos^3{\theta}.   
\label{Epara_eq}
\end{equation}
% 
% The discontinuity in the normal component across stellar 
% surface can be calculated by using the Gauss's law to derive 
% the surface charge density ($\sigma_{s}$) on the star as \cite[]{gho07}, 
% \begin{equation}
%  \sigma_{s} = - ~ \frac{B_p{\varOmega}{R}}{4\pi{c}}~cos^2\theta. 
% \end{equation}
The force exerted by the perpendicular induced
electric field (equation \ref{Epara_eq}) 
on the surface is 
\begin{equation}
 F_{\|}~=~ q E_{\|}.
\end{equation}
For a pulsar with the magnetic field of around 10$^{12}$ G,
the force exerted by the induced electric field exceeds the gravitation
force on the charge particles of the neutron star by many orders of
magnitude (around 10$^9$ for protons and more for electrons). 
Hence, particles get pulled out from the surface of the
neutron star to cancel the electric field component 
(making ${\boldsymbol {E_{out}\cdotp{B}}=0}$).
Moreover, the starting conditions of vacuum around the neutron star 
is no longer valid as it gets filled with the plasma supplied from the surface. 
These charge particles also experience, due to the toroidal 
component of the electric field (E$\perp$ to the magnetic 
field), $\boldsymbol{E\times{B}}$ drift, which force 
them to corotate rigidly with the neutron star. 
However, particles bounded inside the magnetosphere can only corotate 
to the point where their rotational velocity ($\boldsymbol{r\times{\varOmega}}$) 
do not cross the upper bound of the speed of light, c. This boundary is also known 
as the \emph{light-cylinder} as mentioned in the Chapter 1 
and shown Figure \ref{toy_model2}, reproduce here with aligned axes. 
The radius of the light-cylinder (R$_c$) is given by, 
\begin{equation}
 R_c~=~\frac{c}{\varOmega}. 
 \label{rc_eq}
\end{equation}
For a pulsar with a period around 1 sec, the light-cylinder 
radius is around 5$\times{10^{9}}$ cm. 
As mentioned in Section \ref{pulsar_toy_sect}, the light-cylinder divides 
the magnetic field lines in two separate regions, 
closed field lines and open field lines. 
% Particles travelling in the open field 
% line escapes the magnetosphere and it creates a potential different 
% across these field lines, on top of the polar caps. 
% 
% For these field lines the particle depletion again creates 
% a non-zero electric field as $\boldsymbol{E\cdotp{B}}\neq0$ 
% (replacing $\boldsymbol{E_{out}}$ with $\boldsymbol{E}$ for simplicity).  
At any instance the magnetosphere acquires certain 
charge distribution, by pulling out particles from the surface, 
to cancel the induced electric field 
outside the neutron star to keep $\edotb$ 
(replacing $\boldsymbol{E_{out}}$ with $\boldsymbol{E}$ for simplicity).   
The charge distribution, required for the corotation 
and validity of the force free conditions (for r $\ll$ $R_c$), 
is known as the Goldreich-Julian (GJ henceforth) charge 
distribution and can also be expressed as \cite[]{gj69}, 
\begin{equation}
\sigma_{GJ}~ = ~\frac{{\boldsymbol{\triangledown\cdot{E}}}}{4\pi}~=~-\frac{\boldsymbol{{\varOmega~}\cdotp{B}}}{2\pi{c}}.
% \frac{1}{\left(1~ - ~ \left(r/{R_L}\right)^2{sin^2{\theta}}\right)}	
% ~ \frac{{\bf \triangledown\cdot{E}}}{4\pi} ~ = ~-\frac{{\bf \varOmega\cdot{B}}}{2\pi{c}}\frac{1}
% {1~-~(\varOmega{r}/c)^2{sin^2{\theta}}}                                                            
\end{equation}
Using equations, \ref{omega_eq} and \ref{dipole_B}, the charge 
density of the particles in the magnetosphere can be estimated  
as, 
\begin{equation}
 n_{GJ} ~ = ~ \frac{\sigma_{GJ}}{e} ~ = ~-~\frac{\varOmega{B_s}R^3}{4\pi{c e r^3}}\left(3cos^2\theta~-~1\right).
 \label{ngj_eq}
\end{equation}
The $n_{GJ}$ is known as the Goldreich-Julian charge density 
and it plays essential role in building the pulsar emission model. 
The charge density, at the polar cap for a pulsar with period of around 
1 sec and magnetic field of $10^{12}$ G, is around 
$7~\times~10^{10}$ cm$^{-3}$. 
As can be seen from equation \ref{ngj_eq}, the magnetosphere of the 
pulsar consist of charges with two different polarities, with 
a boundary at $cos^{-1}({\pm1/\sqrt{3}})$. 
Figure \ref{toy_model2} shows the distribution 
of the different polarity charge particles 
for an aligned rotator as assumed by \cite{gj69} and \cite{aro81}. 
\cite{rs75} assumed an anti-parallel alignment of rotation 
and magnetic axes, hence giving opposite polarity 
of charge distributions from the one shown in 
Figure \ref{toy_model2}. 
In both cases, the charge separated magnetosphere 
maintains $\edotb$ in all regions. 
% \subsection{Sparking and Pairs creation}

\subsection{Sparking}
\label{spark_sect}
In the open field line region, the particles are not bounded.  
Thus, travelling through the open field lines, 
they escape the magnetosphere.
This creates depletion of charge particles on top 
of the polar cap region, causing a non-zero 
electric field with $\edotbnot$. 
The potential drop keeps growing on top of 
the polar cap as particles depart. 
Part of the magnetosphere, with sufficient charge particles 
to keep $\edotb$, pulls away from the surface 
of the neutron star at polar cap and a gap 
forms, which is the origin of the radio emission according to  
the standard model [\cite{rs75}; also referred to as RS model henceforth]. 
This gap is also known as the {\itshape inner} gap and it occurs 
on either poles due to quadratic nature of the electric field. 
The radius of the inner gap ($r_p$) extends 
from the magnetic field axis at the center 
to the last open field line. The RS model argues 
that iron nuclei (ion) has higher bounding energy and 
hence they can not be pulled out from the surface 
of the neutron star, while electrons can be pulled out 
easily from the surface as their binding energies are much lower.
This is known as the binding energy problem. If the binding energy 
of the ion decreases, they can continuously flow from the 
surface and no such gap will grow. Consequences of such event has 
important implication in different pulsar nulling models (see Section \ref{intrinsic_effects_sect})
\begin{figure}[h!]
 \centering
 \includegraphics[width=3.5 in,height=3 in,angle=0,bb=14 14 320 293]{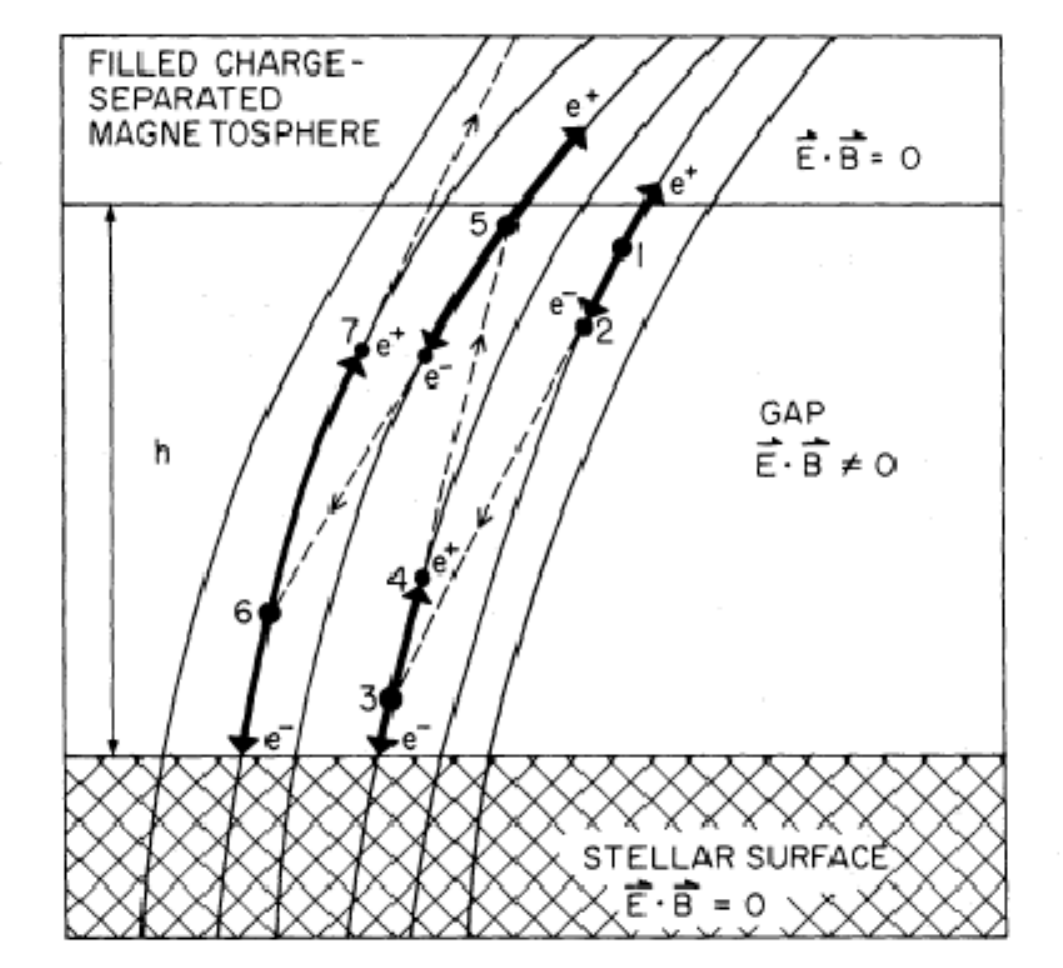}
 % Sparking_rs.eps: 0x0 pixel, 300dpi, 0.00x0.00 cm, bb=14 14 320 293
 \caption[A cross section of the polar cap]{A cross section of the polar gap. Adopted here from the \cite{rs75} to 
 demonstrate the breakdown by an avalanche of primary pair particles.}
 \label{sparking_gap}
\end{figure}

For the inner gap with height h, the potential 
across the gap ($\bigtriangleup{V}$) can be given in terms of the surface 
magnetic field $B_s$ and charge density $\sigma_e$ as \cite[]{rs75,gho07}, 
\begin{equation}
 \bigtriangleup{V}~=~\frac{\varOmega{B_s}}{c}h^2 ~ = ~ 2\pi{n_{GJ}}h^2 
 \label{delta_V_eq}
\end{equation}
As the gap between the neutron star surface and the magnetosphere ($h$) 
grows, the potential drop also increases ($\propto{h^2}$). 
At the instance when the gap 
height reaches to a limit where its equivalent to 
the polar cap radius ($r_p~=~h$), the gap attains 
maximum potential difference, suggested by the RS model 
as, 
\begin{equation}
 \bigtriangleup{V_{max}}~=~\frac{\varOmega\Phi}{2\pi{c}} ~ \sim ~ 10^{12} B_{12}/P^2 ~~ volts.
 \label{vmax_eq}
\end{equation}
Here, $B_{12}$ is the surface magnetic field in units of $10^{12}$ G and  
$\Phi$ is the total magnetic open field line flux.  
However, before it acquires this limit, an avalanche of electron-positron pairs 
will discharge the gap potential. The discharge occurs due 
to the interaction of high energy background photons ($\gamma$-rays) with the high 
magnetic field lines on top of the polar cap.
Under the influence of high magnetic field, a $\gamma$-ray photon splits 
into an electron-positron ($e^{-}~-~e^+$) pair.  
The gap potential then accelerates both these particles to relativistic 
velocities along the curved magnetic field lines in different directions.
Depending upon the direction of the electric field (given model), 
positron or electron escapes the gap while the other falls back and hits 
the polar cap region. According to the RS model, positrons were speculated 
to escape the polar gap while the electrons travel back towards the surface 
(as shown in Figure \ref{sparking_gap}). 
The energy of these particles, accelerate to relativistic velocities, 
is around $E_e ~\sim ~ e(\bigtriangleup{V})~\sim~10^{11}B_{12}h_{3}^{2}/P(s)$ eV. 
Here, $h_3$ is the gap height in units of $10^3$ cm, while $P(s)$ is the pulsar period 
in seconds. The particle with this energy, moving in the 
curvature field lines, again emits high energy photon which 
in turn, again creates another electron-positron pair and so on 
[see \cite{stu71,rs75,aro83b} and Figure \ref{sparking_gap}]. 
\cite{dh86} have also suggested an alternative 
mechanism to produce pairs by an \emph{inverse Compton scattering} 
the low-energy thermal photons by the accelerated charged particles. 
The low-energy photons may come from the surface through 
simple blackbody radiation due to the high neutron stars temperatures. 
It is not very clear which of these processes 
drives the pair creation but either of them create an avalanche of pairs, 
within $10^4$ cm from the surface to discharge the gap potential. 
This process repeats as the particles depart from the open field line region 
and gap again starts to grow. \cite{stu71} was the first one to suggest 
the above mentioned model of pairs production through 
\emph{curvature radiation} and continuous ejection 
of particle from the magnetic poles at a controlled rate in pulsars. 
The localized spots on the polar cap, where the magnetic field lines that caused 
the breakdown are anchored, are known as \emph{``sparks"}. Thus, 
the above discussed phenomena is also known as \emph{sparking} \cite[]{rs75}. 
\begin{figure}[h!]
 \centering
 \includegraphics[width=3 in,height=4 in,angle=-90,bb=50 50 554 770]{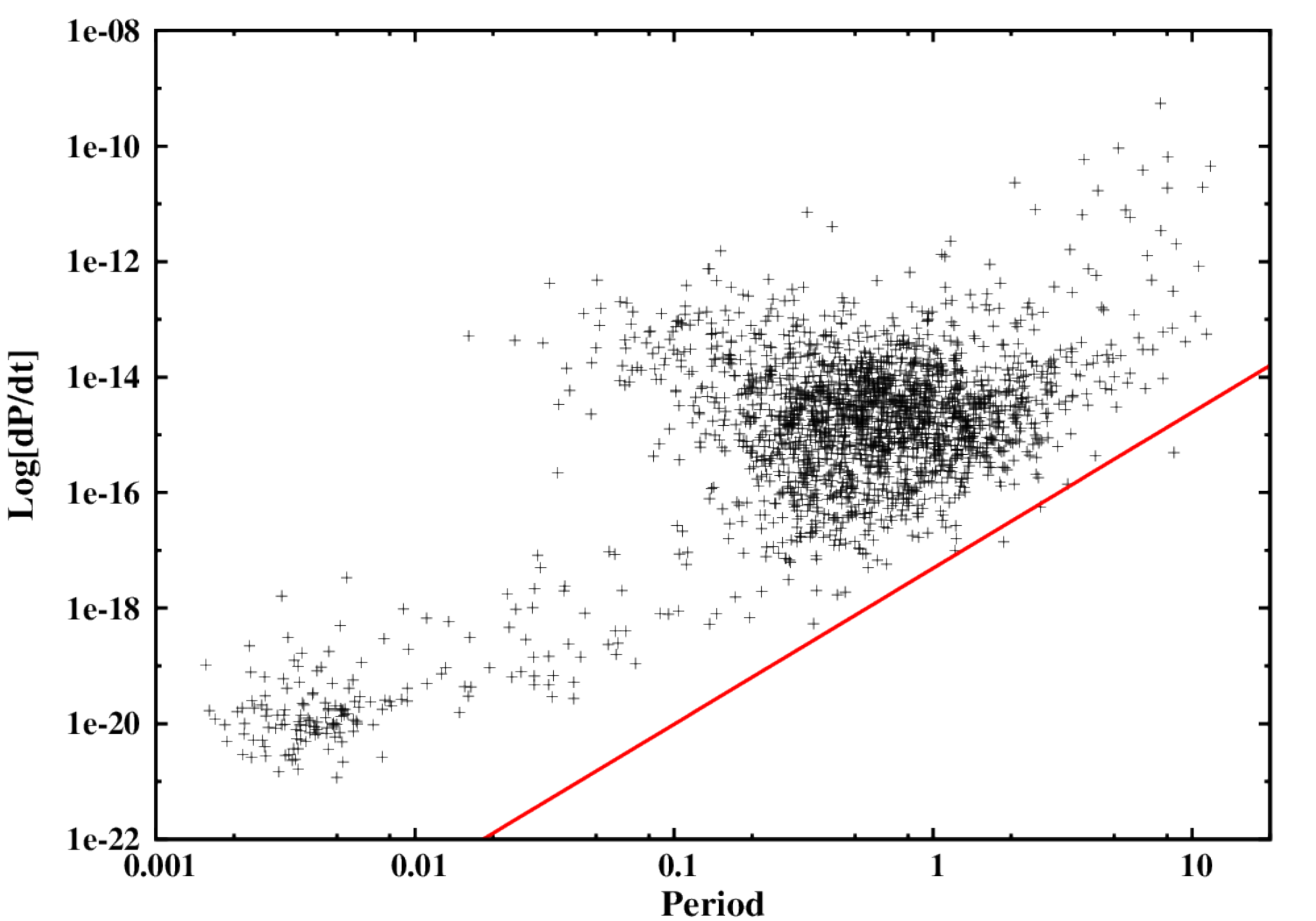}
 % All_PSR_data_p_pdot_wdeath.eps: 0x0 pixel, 300dpi, 0.00x0.00 cm, bb=50 50 554 770
 \caption[$P-\dot{P}$ diagram of all known pulsars]{$P-\dot{P}$ diagram of around 2000 pulsars. 
 The red solid line is the death line, beyond which necessary conditions 
 for pair creation does not exist. This line was derived from the 
 relation given in \cite{cr93}. 
 [The data were obtained from the \emph{www.atnf.csiro.au/research/pulsar/psrcat} \cite[]{mhth05}]}
 \label{ppdot_wdeath_fig}
\end{figure}

\cite{stu71} and \cite{rs75} discussed the necessary conditions for the 
sparking. As pulsar gets older, it's rotation period gets slower and slower 
and hence the maximum achievable polar gap potential 
($V_{max}$ shown in equation \ref{vmax_eq}) drops down. 
For these pulsars. the gap potential is not sufficient 
to create sparking and hence the necessary conditions 
for radio emission, discussed below, will cease. 
This is the death of a pulsar! Figure \ref{ppdot_wdeath_fig} 
shows the $P-\dot{P}$ diagram, also discussed in Chapter 1. 
Its clearly evident that, very few pulsars (only two or three) 
exist beyond the theoretically estimated \emph{death line} 
using the standard dipole approximation \cite[]{cr93}.
According to the standard model, pulsars are born in the upper left 
corner and they progress slowly to the island of normal pulsars. 
They eventually evolve to go beyond the death line into the 
what is known as the pulsar graveyard at the lower right corner 
in the $P-\dot{P}$ diagram.  

\subsection{Radio emission}
\label{radio_emission_sect}
Extensive research have been focused on developing the emission mechanism model which can 
give rise to observed phenomena in radio pulsars. The mechanism to generate the radio waves, 
with the brightness temperature of around $10^{25}~ - ~10^{31}~~K$, 
is of main interest in pulsar phenomenology. It became clear in the early days 
that such mechanism has to be of \emph{coherent} nature, 
as non of the incoherent phenomena can account 
for such bright radiation. If there are N particles 
emitting coherently, the resulting emission would enhance by   
$N^2$ times the emission from a single particle during the coherent mechanism, 
while it only enhances by $N$ times during the incoherent mechanism. 
The coherent emission models can broadly be divided into 
three main groups of theories, \emph{viz.} (a) emission 
by the coherent bunches [see \cite{rs75,cr77,bb77} and references therein], 
(b) maser emission, corresponding to negative absorption 
[see \cite{bla75,mel78,mel92b} and references therein] and 
(c) reactive instability due to an intrinsically growing wave mode 
\cite[]{ab86,bgi88,aps90}. Each of these models have their own numerous 
variants, which is beyond the scope of this thesis to discuss [for a review see \cite{mic91}]. 
Among these theories, the \emph{two-stream instability} 
to cause bunching is the most favoured one as it has been 
considered by many authors \cite[]{rs75,bb77,cr77,uso87,uu88,mgp00,uso02} and 
will be discussed further here. Broadly the origin of the two-stream instability 
can be expressed as follows. 

Positrons which escape the gap on top of the polar cap, acquire 
relativistic velocities with Lorentz factor, $\gamma$, reaching up to 
$10^6$, which are also know as the \emph{primary} particles. 
These high energy primary particles do not get further acceleration 
beyond the gap as they enter the region of the magnetosphere 
where $\edotb$ with $n_{GJ}$ charge density.  
Travelling in the curved magnetic field lines, they 
emit high energy photons through curvature radiation. 
A few of these photons again interact with the magnetic field to produce further pairs of 
$e^{-}~-~e^+$, which are known as the \emph{secondary} particles.  
However, the energies of these secondary particles are much lower 
(with $\gamma$ of around $800$). In the absence of any accelerating 
electric field, both secondary particles pair travel in the outward direction, 
along the curved field line to conserve the momentum. 
These particles are the prime source in three different variants of two-stream 
instability theories which can give rise to the observed radio emission, 
\emph{viz.} (a) interaction between primary and secondary 
particles \cite[]{rs75}, (b) electrostatic interaction between 
electron and positron of the secondary particles \cite[]{cr77}, 
(c) interaction between clouds of secondary plasma \cite[]{uso87,uu88,mgp00}. 

According to the RS model, the most energetic positrons ($\gamma~\sim~10^6$)
radiate away most of their energies within $10^6$cm from the stellar 
surface after leaving the gap. Most of these energies get 
converted to outward moving secondary particle pairs as discussed above. 
It can be shown that curvature radiation from these secondary particles 
falls in the radio regime \cite[]{rs75}. However, incoherent radiation 
from individual secondary particle can not give rise to bright 
radio emission seen in pulsars. To generate coherent 
radiation from these particle, they have to form \emph{bunches}. 
Bunching occurs due to the Coulomb interaction between the relativistic 
secondary particles with the passing ultra-relativistic positrons.  
These positrons, generated at the later stage of the gap discharge,  
when the gap potential has decreased substantially, travel 
much further in the magnetosphere without losing much of 
their energies. Due to their negligible energy loss, 
they catch up and pass through the slow moving secondary particles cloud 
which causes the two-stream instability to grow and enhance 
the curvature radiation from bunches. 
Similar formalism was also suggested by \cite{stu71}. 
However, \cite{bb77} have obtained more precise estimate 
on the development of the bunches and showed that, 
the beams of primary particles passing through the secondary pairs 
do not have enough time to develop two-stream instability. 
These authors have also tried to correct the RS model by invoking 
ion beams, originating from surface along with the 
positrons. 
 
In other models of two-stream instability, \cite{cr77} have suggested 
interaction between electrons and positrons in the secondary particle pairs. 
It was shown by these authors that, when the secondary particles cloud moves  
along the curved magnetic field lines of a rotating magnetosphere, 
a relative streaming between electron and positron causes two-stream 
instability to grow and form bunches. The resultant bunch, travelling 
in the curved field lines, give rise to the observed radio emission. 
Similarly, \cite{uso87} has suggested interaction between two consecutive 
secondary pair clouds with different momenta, which leads to two-stream 
instability at a distance of around $10^8$ cm from the surface.  
At these heights, the high energy particles from a 
lower cloud catch up with the slow moving low energy particles 
from a cloud going ahead of it, which causes the two-stream instability to 
develop. \cite{mgp00} extended this model further by introducing the 
concept of plasma \emph{solitons}, which emit curvature radiation 
in the radio regime. 

In the above mentioned three models, 
the basic assumption is the non-stationary nature of the pair 
creation. However, \cite{as79} and \cite{aro81} have invoked steady 
pair creation, which, according to \cite{uso87}, may be just 
time average values. It is not very clear that which of the 
coherent emission models is operating in pulsars. Observational 
constraints on the emission heights has, however, confirmed the 
origin of emission near 50 stellar radius. At these heights, it is unlikely 
for the plasma instability models to be active as they are expected to 
occur at much higher heights. The other coherent mechanism, 
maser emission, requires much stronger magnetic field and 
thus fails to operate in millisecond pulsars which have magnetic field of 
around $10^8 - 10^9$ G. Hence, many years of extensive 
study has pointed out the bunching mechanism as the most 
viable coherent mechanism operating in rotation powered pulsars. 
However, it is not very clear which of the two-stream instability 
model is responsible for bright radio emission. 

\section{Radiation beam}
\label{radiation_beam_sect}
% Size and shape of the emission beam
The coherent radiation from pulsars, originating from 
one of the above mentioned emission model, give rise to 
conical beam of emission with the magnetic axis as a center
\cite[]{rc69,kom70}. If the magnetic axis is misaligned with 
the rotation axis, each time this beam of emission point 
towards the observer, a pulse can be detected. 
The integrated profile, obtained from around thousand of such pulses, 
represents average shape of the beam for a given line-of-sight cut (see Chapter 1). 
This model seems to be in accord with 
most of the observed phenomena seen in radio pulsars.

A charge particle travelling along the magnetic 
field line emits curvature photon towards the observer 
from a point at which the line-of-sight is tangent with 
the corresponding field line \cite[]{kom70}. 
Thus, the detected curvature photons\footnote{Photon emitted from the curvature radiation.} 
also carry information regarding the orientation of the magnetic field lines 
at the location of their origin. This idea can be extended to speculate that, 
the last open field lines define the angular extent of the radiation beam 
and hence the width of the integrated profile also. 
As the open field line region is inversely related to the period 
of the pulsar (see equation \ref{rc_eq}, where $R_c$ is proportional to period 
and thus inversely related to the angular extent of the open field line region), 
fast spinning pulsars tend to have wider profiles, which has been 
confirmed by various observations. 
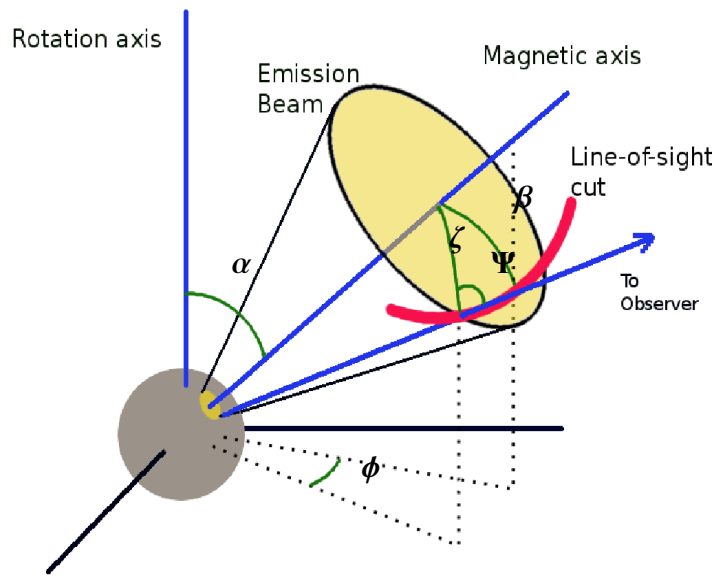
\begin{figure}[h!]
 \centering
%  \includegraphics[width=4 in,height=3.3 in,angle=0,bb=14 14 815 615]{geometry_beam.eps}
 % geometry_beam.eps: 0x0 pixel, 300dpi, 0.00x0.00 cm, bb=14 14 815 615
% \setlength{\unitlength}{0.1\textwidth}
  \begin{picture}(230,230)
  \put(0,0){\includegraphics[width=4 in,height=3.3 in,angle=0,bb=14 14 815 615]{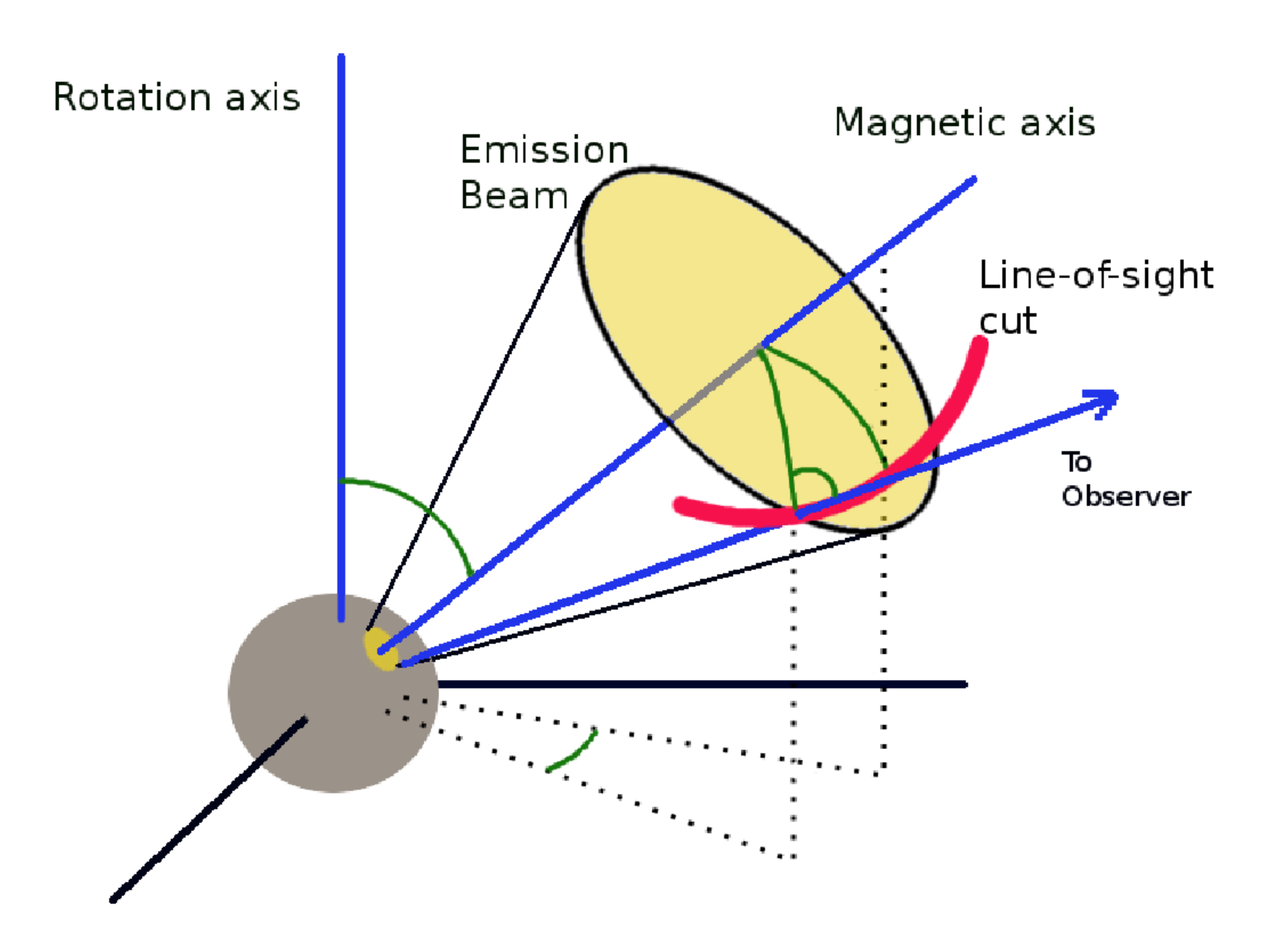}}
  \put(90,120){$\boldsymbol{\alpha}$}
  \put(139,43){$\boldsymbol{\phi}$}
  \put(172,130){$\boldsymbol{\zeta}$}
  \put(188,120){$\boldsymbol{\Psi}$}
  \put(197,145){$\boldsymbol{\beta}$}
 \end{picture}
 \protect\caption[Emission beam geometry]{Emission beam geometry of a pulsar, where the rotation and magnetic axes 
 are inclined with each other by an angle $\alpha$. The line-of-sight is shown with a red solid line which cuts the emission beam 
 by an angle $\beta$ in the plane containing rotation and magnetic axes. The change in the longitude 
 is measure by the angle $\phi$, which is an angle between the plane containing rotation 
 and magnetic axes (zero longitude) with the location of the instantaneous line-of-sight, a \emph{subobserver} point. 
 The angle $\zeta$ is the angle between the subobserver point with the magnetic axis, while the angle 
 $\Psi$ is the angle between the line joining subobserver point and magnetic axis with the line-of-sight trajectory 
 (cartoon using the free 3D modeling package : {\itshape blender} \url{www.blender.org}).}
 \label{geometry_beam}
\end{figure}

The linear polarization of the observed photon, 
depicts the orientation of the magnetic field plane (the plane
containing the field line). This linear polarization in many pulsars show 
peculiar variation across the pulse profiles, as discussed 
extensively in Section \ref{pulse_polrization_intro_sect} 
by introducing the rotating-vector model proposed by \cite{rc69}. 
According to the model, at the origin of radio emission, 
the magnetic field can be assumed to be of purely dipole nature. 
Thus, by modelling the polarization angle changes, one can 
decipher the beam geometry around the line-of-sight cut. 
Figure \ref{geometry_beam} illustrates the emission beam along 
with the misaligned magnetic and rotation axes. 
The angles shown in the diagram has following relationships 
using simple geometry \cite[]{mt77,mic91}. 
\begin{equation}
tan(\Psi) = \frac{sin(\alpha)sin(\phi)}{sin(\zeta)cos(\alpha)~-~cos(\zeta)sin(\alpha)cos(\phi)}
\end{equation}
The measured polarization angle (PA) is the angle $\Psi$ shown in Figure \ref{geometry_beam}. 
Variation in the PA with $\phi$ is a measurable observed pulsar property. However, along with 
the different propagation effects mentioned in Section \ref{pulse_polrization_intro_sect}, 
there are various intrinsic phenomena which causes 
further changes in the polarization angle, such as  
the orthogonal mode changes.  However, this topic is beyond 
the scope of this thesis to discuss further. 

\subsection{Emission height}
\label{emission_hgt_sect}
The pulsars show gradual widening of the integrated profiles 
from higher to lower frequencies. This effect is known as 
the \emph{radius-to-frequency mapping}, which suggest that, 
the emission regions are localized at different 
heights from the neutron star surface for different frequencies. 
Estimation of the emission heights are crucial to scrutinize different 
emission mechanism models, as some of them failed to operate at lower 
heights (as discussed in Section \ref{radio_emission_sect}). 
According to \cite{stu71} the emission region 
is around 1 stellar radii from the surface while \cite{rs75} suggested 
much higher emission heights ($\approx$10-100 stellar radii). 
For pulsars with double profile components, separation between 
both of them were measured at multiple frequencies to derive 
the separation to frequency dependence. \cite{kom70} suggested this 
relation to be around $f^{-1/4}$ by using the curvature of 
the field line arguments. According to the RS model, the emission height at a given 
frequency is dependent upon the plasma density, given in equation \ref{ngj_eq}, 
which decreases with increasing height by a factor of $r^{-3}$. 
Thus, these authors suggested frequency dependence of $f^{-1/3}$ 
for the profile evolution.  Many studies have reported  
the emission heights in number of pulsars \cite[]{cor78,ran83a,lm88,kg98,gg01,mr02a,kg03a}. 
\begin{figure}[h!]
 \centering
 \includegraphics[width=3 in,height=4 in,angle=-90,bb=50 50 554 770]{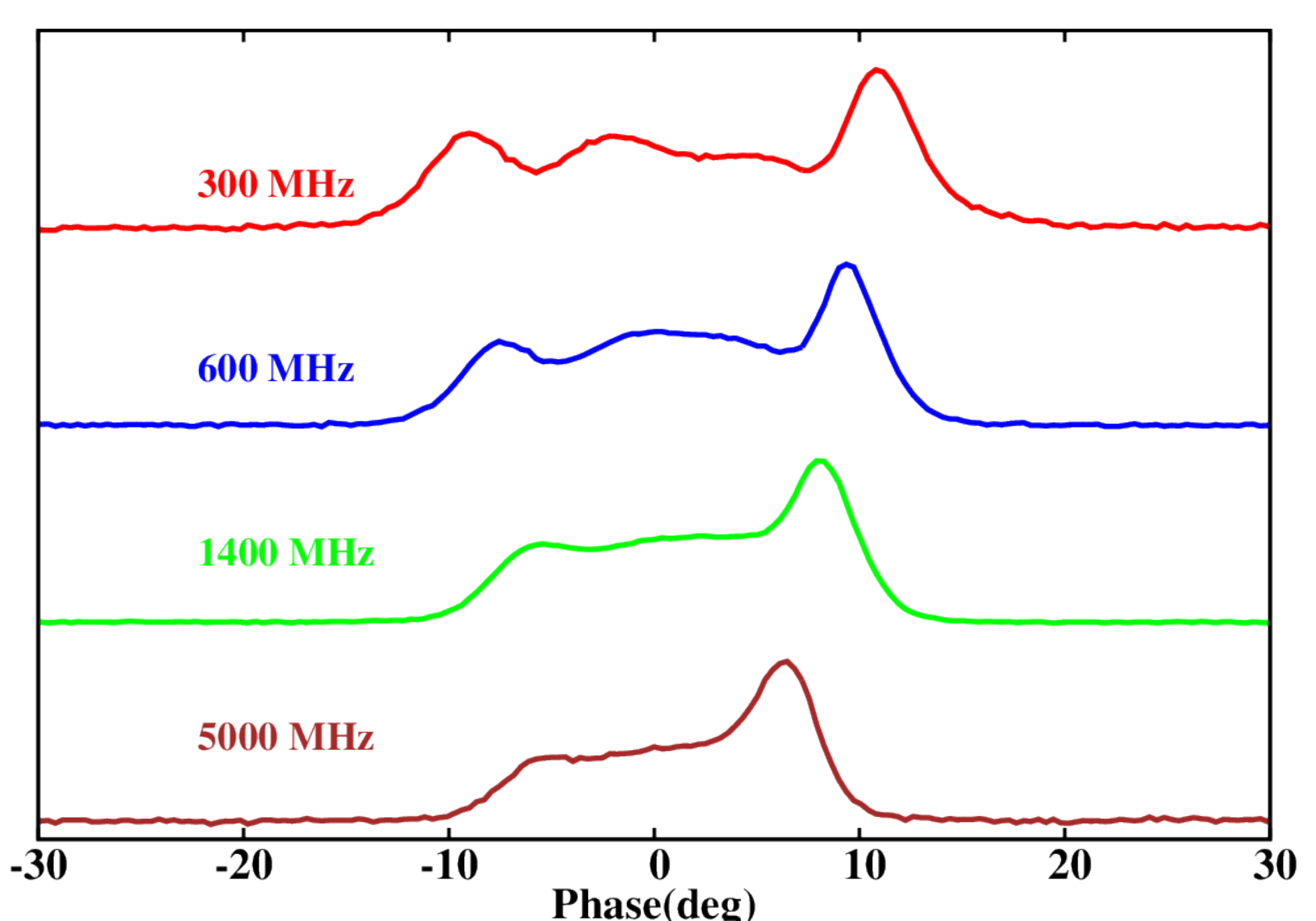}
 % Profiles_for_thesis.eps: 0x0 pixel, 300dpi, 0.00x0.00 cm, bb=50 50 554 770
 \caption[Integrated profiles at four frequency for PSR B2319+60]{Frequency evolution of the integrated profile for PSR B2319+60. 
 The observations were taken at four different frequencies (\emph{viz.} around 300,600,1400 and 5000 MHz) as a part of this thesis work.}
 \label{freq_profile_evolution}
\end{figure}
\begin{figure}[h!]
 \centering
 \includegraphics[width=5 in,height=3 in,angle=0,bb=14 14 975 555]{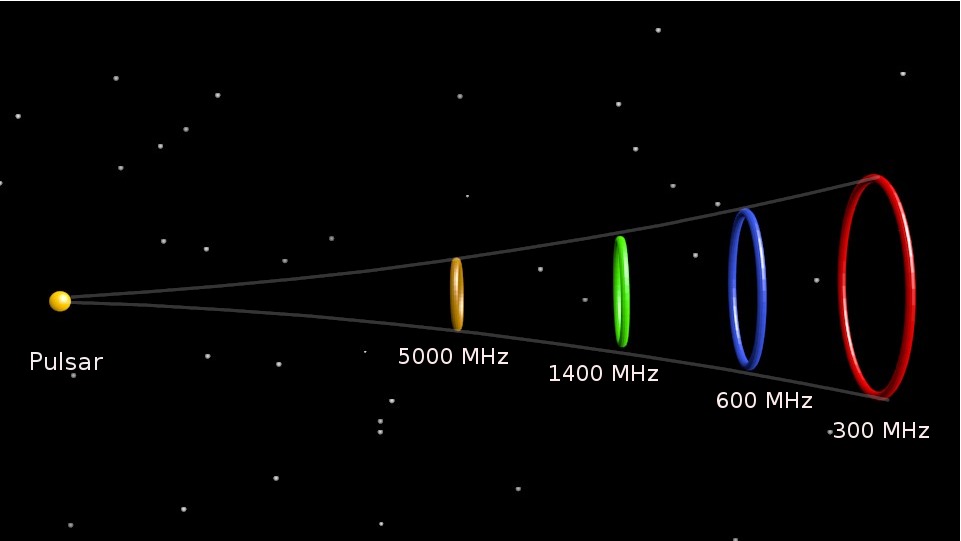}
 % Emission_height_new2_gimp.eps: 0x0 pixel, 300dpi, 0.00x0.00 cm, bb=14 14 975 555
 \caption[Radius to frequency mapping for PSR B2319+60]{Beam of emission with localized emission 
 regions at different frequencies for PSR B2319+60. The height and beam size were derived as 
 explained in the text. The size and height of the emission region are in scale compare 
 to a pulsar with radius of around 10 km. Note that these are only 
 rough estimates and displayed here just to demonstrate the radius-to-frequency mapping 
 (cartoon using the free 3D modeling package : {\itshape blender} \url{www.blender.org}).}
 \label{radiation_beam_height}
\end{figure}

To demonstrate the relative heights and size of the emission beams at 
different frequencies, we compared the integrate profiles 
of PSR B2319+60 at four frequencies \emph{viz.} 300, 600, 1400 and 5000 MHz. 
The profile evolution is clearly evident in Figure \ref{freq_profile_evolution}
with narrow higher frequency profile to wider lower frequency profile. 
To estimate their relative heights, we used the relationship 
suggested by \cite{kg03a} as, 
\begin{equation}
 r_{h} ~ = ~ 400~\pm~80~f_{GHz}^{-0.26\pm0.09}~\dot{P}_{15}^{0.07\pm0.03}~P^{0.30\pm0.05} ~~ km. 
 \label{rh_eq}
\end{equation}
Here, $r_h$ is the emission height at frequency $f$ given in GHz, 
while $\dot{P}_{15}$ is the rate of change of period 
($P$) in units of $10^{-15} ~ s/s$. For PSR B2319+60, 
\cite{wg95} estimated the inclination angle $\alpha$ (31$\deg$) and impact 
angle $\beta$ (4$\deg$), which were used to estimate the radiation beam opening angle ($\rho$)
given as \cite[]{kg03a}, 
\begin{equation}
sin^2\left(\frac{\rho}{2}\right) ~=~ sin^2\left(\frac{W}{4}\right)\cdotp{sin(\alpha)}\cdotp{sin(\alpha+\beta})~+~sin^2\left(\frac{\beta}{2}\right).
\label{beam_angle_eq}
\end{equation}
Here, $W$ is the width of the pulse profile at the corresponding frequency. 
Thus, using equations \ref{rh_eq} and \ref{beam_angle_eq}, the beam shape at each 
of the observing frequency was derived and shown with a relative scale 
in Figure \ref{radiation_beam_height}. The emission height were estimated 
to be around 780, 665, 535 and 385 km at 300, 600, 1400 and 5000 MHz, respectively. 
These heights are only rough estimates and calculated just to demonstrate the radius-to-frequency 
mapping of the emission regions. There are multiple effects that plays crucial role in 
determining these heights which are known as retardation and aberration of the 
emission beam due to the rotation. To estimate the true heights of emission regions, 
one has to correct for these effects which is beyond the scope of this thesis. 

\subsection{Rotating carousal}
\label{rotating_carousal_sect}
\begin{figure}[h!]
 \centering
 \includegraphics[width=5.6 in,height=2.5 in,angle=0,bb=14 14 812 343]{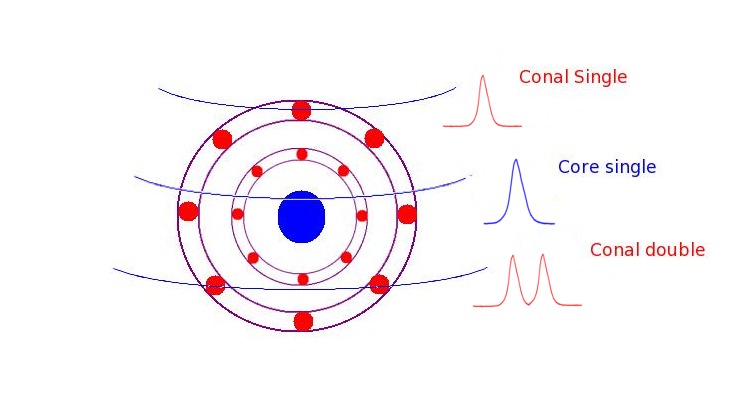}
 % rankin_conal.eps: 0x0 pixel, 300dpi, 0.00x0.00 cm, bb=14 14 812 343
 \caption[Rotating carousal model]{Rotating sub-beam model to 
 classify pulsar profiles according their line-of-sight cuts as suggested 
 by \cite{ran83a}.}
 \label{rankin_conal}
\end{figure}

As mentioned in Section \ref{spark_sect}, the emission from the 
polar cap comes from group of localized regions (\emph{aka.} sparks). 
The primary particle beams generated from these sparks 
are the sources of secondary pair plasma, which give rise to 
radio emission by one of the bunching mechanism. 
Thus, the location of sparks on the polar cap 
also reflects in the structure of the emission beam. 
Due to which, the emission beam does not have uniform illumination but, 
as mentioned in Section \ref{drifting_intro_sect} and shown 
in Figure \ref{rankin_conal}, constitute of small 
patches of emission sub-beams. These sub-beams are not 
stationary as they rotate around the magnetic axis 
due to the following mechanism. 

According to the RS model, the charge particles in the sparks 
do not corotate with the stellar surface (or rotate 
with a slightly different speed) as long as the charge densities 
do not reach $n_{GJ}$ (given in equation \ref{ngj_eq}).  
As the gap discharges, it attains the $n_{GJ}$. 
At this instant, just like the particles inside the neutron 
star (see Section \ref{GJ_sect}), the particles in the spark regions 
also experience $\boldsymbol{E\times{B}}$ drift.   
These particles move with the neutron star 
like a rigid rotator around the magnetic axis. 
Thus, the presence of the gap alters the velocity 
of the rotation of the spark regions. 
This alteration or difference in the velocity of sparks, 
moving around the magnetic axis, mimics a slow drifting of their 
locations. This difference in velocity is given by \cite[]{rs75}, 
\begin{equation}
\bigtriangleup\nu~=~\frac{\bigtriangleup{V}}{B_s{r_{p}}}c.  
\end{equation}
Here, $r_{p}$ is the radius of the polar gap through 
which net positive charge is supplied to the magnetosphere, 
while $\bigtriangleup{V}$ is the potential across the gap 
height given in equation \ref{delta_V_eq}. As these sparks 
give rise to sub-beams in the emission beam, similar drifting 
of sub-beams can also be observed each time the line-of-sight 
cuts the emission cone. The periodicity of drifting is given as the time taken by 
a sub-beam to move to an adjacent sub-beam location. Total time taken 
by a sub-beam to circulate around the polar cap is around, $2\pi{r_p}/{\bigtriangleup{\nu}}$. 
If there are $n$ such sub-beams in the emission beam, the observed 
drifting periodicity ($P_3$) can be given as \cite[]{rs75}, 
\begin{equation}
 P_3 ~=~ 5.6 ~\frac{B_{12}}{n ~ P} ~ ~ sec.
\end{equation}
Here, $B_{12}$ is the magnetic field in units of $10^{12}$ G, and 
$P$ is the pulsar period. It should be noted at this point that, 
along with comprehensive picture of pulsar emission mechanism, the 
major success of \cite{rs75} was in predicting the drifting periodicity. 
However, \cite{vsrr03} has shown that drift velocity is not always correctly
predicted by the RS model. 

\cite{kom70} has estimated the open field line region which 
possess field line with sufficient curvature to 
produce the radio emission. Thus, intensity of 
the radio emission should be higher at the boundary 
of the open field line region, while it should vanish 
at the magnetic pole. Thus, he suggest a hollow cone 
model of the emission beam where emission is further 
localized in a ring of sub-beams as shown in Figure \ref{rankin_conal}.  
According to which, for each line-of-sight cut, a double component profile 
should be observed. However, many pulsars show multi-component 
profiles which was explained by a introducing further concentric circular rings 
of sub-beams around the \emph{core} component \cite[]{bac76,os76a,ran93}. 
\cite{ran83a} classified different profiles in various 
categories depending upon their line-of-sight cuts (also shown
in Figure \ref{rankin_conal}), and proposed rotating \emph{carousal} as 
the cause of drifting seen in pulsars. In later years, 
the carousal beam was modelled for PSR B0943+10 and number 
of sub-beams were calculated \cite[]{dr99,dr01}. This direct modelling 
of the radiation beam, using single pulse observations, made the 
rotating carousal as one of the highly popular model. 
% Thus, the electrons which generate due to the primary pair production inside the gap, 
% will follow a curved path, due the component of the electric field 
% perpendicular to the direction it's motion, and hit the surface in the vicinity 
% of the original spark. This regions has slightly higher 
% temperature due to the back-flowing electron hitting it with relativistic 
% velocities. Hence, it has a higher chance of creating a 
% spark then a nearby region. This effect mimics drifting 
% of the spark regions on top of the polar cap around the 
% magnetic axis due to $\boldsymbol{E\times{B}}$ drift. 

\section{Pulsar nulling}
Among the various single pulse phenomena exhibited 
by the radio pulsars, absence of pulses or pulse nulling 
is the most peculiar one. Pulsar nulling is discussed 
in this section, with necessary details regarding earlier findings,
as it is the core theme of this thesis. 
% The structure of the 
% section is as follows, first the detail description regarding 
% various observational studies are listed, followed by a discussion on 
% various proposed models. 

\subsection{Chronicles of pulsar nulling observations}
\label{chronical_obs_sect}
Pulsar nulling was first reported by {\bfseries \cite{bac70}} 
in four pulsars, PSRs B0834+06, B1133+16, B1237+15 and B1929+10. 
These pulsars were found suddenly to miss one to ten pulses at 
various times. He classified nulled pulses in two categories according 
to their duration and spacing. Type I nulls are the most prominent, with 
a spacing of about fifty pulses. Type II nulls have width of only one or 
two pulses and recur with spacing of three to ten pulses. 
\cite{bac70} claimed the nulled pulses to be periodic. To find the periodicity 
of these null pulses, a sequence of null and burst pulses with 
nulled pulses as 1 s and burst pulses as 0 s was formed. The Fourier transform 
of this one-zero time series was used to look for periodicity 
of occurrence of null pulses. He reported possible periodicity 
of 10.5 P, 2.8 P and 25 P (where P is the period of the pulsar) 
for PSRs B1929+10, B1237+25 and B1133+16, respectively. 
He suggested the origin of nulling due to   
(a) the radiation region of pulsar is undergoing some rapid changes, 
and (b) type II nulls are mostly related to subpulse drifting, as 
marching of subpulse with a spacing greater than the pulse profile 
could periodically cause null pulses. 

In a first ever comprehensive study on the nulling pulsars, 
{\bf \cite{rit76}} reported nulling behaviour of 32 pulsars 
using the Mk IA radio telescope at Jodrell Bank. 
He also introduced a method to quantify occurrence of nulling 
using a quantity called `Nulling Fraction (\emph{NF})', 
which quantifies the fraction of null pulses in the pulse 
energy sequence. His method to estimate the NF for a 
pulsar is discussed in Section \ref{NF_tech_sect}. 
By comparing their NFs, it became possible 
to compare nulling between different pulsars and also to 
look for any possible correlations with pulsar parameters. 
He reported correlation of NF with the period of the pulsar 
and thus, concluded that the pulsars with longer period seem 
to posses high NFs. As the period is related to the 
age of the pulsar ($\propto{\sqrt{P}}$), he concluded that pulsar die with 
increasing fraction of nulling in them. 
An important characteristic reported by \cite{rit76} 
is that, pulsar emission is produced in burst of pulses and 
the length of the burst does not depend upon the age of the pulsar.
However, the separations between these bursts of pulses, also known as null states, 
increases with the age of the pulsar. He also suggested that, if 
subpulse drift is due to drifting of the spark regions 
in the polar cap then inactive or absence of few spark 
regions in the observer's line-of-sight can sometime give 
rise to nulling in some pulsars. Similarly, using the 
Arecibo telescope, single pulse observations were conducted for 
20 pulsars by \cite{bac81}. Out of which, nine showed nulling 
behaviour for which accurate estimation on the NFs were obtained. 

In a further attempt to correlate nulling fraction 
with various pulsar parameters, {\bf\cite{ran86}} 
reported a study of around 60 pulsars. It was 
claimed that nulling is quite common phenomena and 
should occur for half of the known pulsars. 
The pulsars, in the studied sample, were classified based on their 
integrated profile classes defined by \cite{ran83a} (also shown in Figure \ref{rankin_conal}). 
By comparing NFs of pulsars between different classes, 
she concluded that core single pulsars possess small NFs compared 
to other classes. Conal double and multiple component profile classes 
possess larger number of pulsars with high NFs compared to all other classes. 
In an earlier study, \cite{rit76} concluded that pulsar die with increasing NF.
Contrary to that, \cite{ran86} suggested that apparent relation 
of nulling and age is because of profile classes. 
In a given profile class, there is no strong correlation between 
the NF and the pulsar age. Pulsars which possess larger NFs has 
no greater age then those which possess small NFs in the same class. 
\cite{ran83a} suggested that young pulsars usually have profiles with prominent core components, 
thus, they are likely to show small NFs \cite[]{ran86}. 
She also pointed out that, nulling is a global effect 
as it is seen in main pulse as well as in interpulse\footnote{Some pulsars show two 
prominent components, separated by about half the period, 
in their integrated profiles. The more prominent component is 
usually called the main pulse, while the other component is called 
the interpulse. The main pulse and the interpulse are believed to 
originate from widely separated emission regions.}. 
Thus nulling represents \emph{``global failure of the 
condition necessary for emission".}
\cite{wab+86} carried out high sensitive modulation study of around 28 
pulsars, in a similar year, using the Arecibo telescope at 400 MHz. 
These authors concluded that, pulse energy modulation in 
the core component is much lower compared to conal components, 
thus, supporting \cite{ran86} claim of core component having 
smaller NFs. This study also reported NFs in around 20 pulsars 
for the first time. 

In later years, {\bf \cite{big92a}} conducted a correlation study 
of 72 nulling pulsars comparing their NFs with several other parameters. 
He observed around 25 of them at 645 and 843 MHz and 
compared the results with those reported in previous studies. 
On the basis of these observations, he concluded that in about 30\% of 
the studied pulsars of core dominated class, nulling may be detected 
with higher sensitive observations. \cite{big92a} confirmed 
the results of \cite{ran86} about the other profile classes, 
with 40\% of conal single, 65\% of multiple profile pulsars 
and nearly all conal double pulsars have been observed to null. 
He also made similar remarks regarding the correlation of NF-age, 
as no correlation was seen in the same profile class. 
The apparent NF-age correlation is due to differences in the age between 
core single and conal profile classes. One of the apparent correlation 
which was highlighted by \cite{big92a} was about Period-NF. 
Pulsar with longer period seems to possess higher NF. 
This may indicate that pulse emission mechanism is faltering 
for these objects since quantities related to polar cap emission 
mechanisms, such as maximum potential difference across the gap, 
are inversely related to P (see equation \ref{vmax_eq}). 
The other weak correlations which he noticed were with   
the rate of energy loss ($\dot{E}$) and the magnetic field at the light-cylinder ($B_{lc}$). 
However, these correlations can occur due to the Period-NF correlation as 
all of these quantities are highly correlated with the period of the pulsar. 
Unlike other earlier studies, \cite{big92a} reported 
an anti-correlation between NF and $\alpha$, 
thus, NF is large for small $\alpha$. \cite{lm88} and \cite{ran90} both gave 
different methods to estimate $\alpha$. It was suggested by 
\cite{lm88} that as a pulsar gets older, the magnetic axis tends 
to align to the rotation axis. Hence, the anti-correlation between 
NF-$\alpha$ points towards underline NF-age correlation in the 
\cite{lm88} model. On the other hand, $\alpha$ estimated 
by \cite{ran90} does not appear to be related to pulsar age. 
Thus, from her model, NF-$\alpha$ correlation might have an 
independent origin. In any case, this correlation suggests 
nulling to have geometric origin. Moreover, the 
correlation of NF with magnetic field at light cylinder 
suggests that large fields are associated with the increased 
pulsar nulling. However this correlation is not an independent estimate, 
as determined value of B also depends upon $\alpha$ and 
period P which are, according to \cite{big92a}, correlated with NF. 
\cite{lw95} also reported a statistical study of 72 nulling pulsars, 
contesting corrections reported by \cite{big92a}. 
These authors concluded that, NF is more strongly 
correlated with the pulse width, $W$, compared to 
period and $\alpha$ of the pulsar. These authors also 
claimed that NF-$W$ correlation can also 
explain NF-$P$ and NF-$\alpha$ correlation reported by \cite{big92a}. 

The method to estimate the NF, given by \cite{rit76}, was further improved 
by {\bf \cite{viv95}}. With this new method, it became possible to estimate 
NF for the weak pulsars. \cite{viv95} observed around 15 pulsars and 
estimated NFs for 10 of them. Only two of these pulsars in the sample were 
new, while for the other pulsars, NFs values were improved with his 
new technique. He also reported similar results regarding the null 
and burst length as those reported by \cite{rit76}. 
The emission from PSR B0031$-$07 comes in burst of pulses and there are 
two time-scales involved. Short time-scale bursts and nulls come 
in $\leq$ 10 periods, while a long time-scale process can causes 
burst duration of $\approx$ 24 periods and null duration 
$\approx$ 29 periods. \cite{viv95} also reported that there 
were no \emph{truncated} burst pulses before or after the null 
phase, indicating that subpulses are basic units of pulsar emission. 

Parkes multibeam survey has discovered more than half of the known 
pulsars \cite[]{mlc+01,mhl+02,kbm+03,hfs+04,fsk+04,lfl+06}. 
In one of such investigation, \cite{fsk+04} reported 
for the first time, nulling in a binary pulsar, PSR J1744$-$3922. 
Pulsars are also found in binary systems, where the other 
companion could be a white dwarf or a normal star [see \cite{gho07} for a review]. 
The reported pulsar rotates in a tight circular orbit every 4.6 hours. 
The time-scale of the nulling seen in this pulsar are around 
few tens of minutes repeating after every few minutes of burst 
states. They also reported occasions where pulsar is not seen 
during a full 35 minute observing run. All external effects, 
such as scintillation and obscuring binary companion 
have been ruled out by these authors. The potential gap 
values were also calculated for non-nulling and nulling 
pulsars. It was suggested by these authors that nulling pulsars 
tend to have smaller median gap potential values and hence 
they are likely to null more. Similarly, PSR J1744$-$3922 was 
shown to have smaller gap potential which is analogues to 
nulling pulsars. In the similar category of strange nulling pulsars, 
\cite{cl07} reported \emph{signs} of nulling behaviour in an extra 
Galactic pulsar, PSR B0529$-$66, located in the Large Magellanic Cloud.  
However, this was never confirmed. 

Recently, {\bf\cite{wmj07}} carried out observations of 25 newly 
discovered pulsars from the above mentioned Parkes multibeam survey and 
estimated their NFs for the first time. Their study 
increased the total number of nulling pulsars by a substantial amount. 
Moreover, a pulsar with NF of 93\% was also found, which is  
the largest known NF. Their study actually doubled the total number of 
known nulling pulsars, which possess NF above 50\%. 
It was concluded that nulling is a property of old pulsars.  
It was also claimed that there is no correlation of 
NF and profile morphological classes as claimed by \cite{ran86}. 
These authors showed that, almost all classes of 
pulsars show nulling and mulitcomponent profiles tend to 
have higher NFs but such pulsars are old. Thus, their study once again 
supported the NF-age correlation. Along with these nulling pulsars, they 
also reported weak mode in two of their studied pulsar. 
It was also claimed that nulling and mode changing 
are highly correlated, which was also suggested by \cite{ran86}.
Moreover, \cite{wse07} reported tentative evidence 
of nulling in 8 pulsars but no estimates on the NFs were reported. 
Number of nulling pulsars were further enhanced by 
pulsar searches conducted from the Giant Meterwave Radio Telescope \cite[]{jml+09}
and the Parkes telescope \cite[]{bbj+11,bjb+12}.

\subsection{Nulling$-$Drifting/Mode-changes interaction studies}
\label{null_dritft_corr_sect}
\cite{bac70a} reported subpulse drifting in PSR B1929+10 which 
was shown to have null pulses. Sub-pulse drifting is associated with 
nulling in many pulsars, for example PSRs B0809+74, B0818--13 and B0031--07. 
In the early attempts, various studies \cite[]{bac70c,col70a,th71,pag73} 
have reported associated changes in the drift rate with null states. 
\cite{urwe78} have reported unique phase memory of 
of the subpulse location before and after the null states in PSR B0809+74. 
They also concluded that nulling occurs due the quenching of sparks (an intrinsic phenomena) 
and not caused by deflection or absorption of the radiation beam as 
it shows clear phase memory of subpulses across the null length. 
However, \cite{fre83} contested these results by showing 
lack of subpulse phase memory in their observations of PSR B0809+74. 
A more detailed account of this phenomena was reported by \cite{la83} 
in two pulsars, PSRs B0809+74 and B0818--13. These authors showed clear changes 
in the drifting speed before and after the null states with estimate on the 
relaxation time in acquiring regular drifting behaviour directly 
proportional to the length of the null states. Thus, nulling in these 
pulsars were compared with a damped oscillator, which also has a finite 
relaxation time constant. Contrary to \cite{urwe78}, 
these authors reported phase jumps in the subpulse location 
across the nulls. They suggested that these changes are due to 
the gradual speed up of drifting during nulls. In a similar line of investigations, 
\cite{vj97} reported phase memory across nulls in PSR B0031--07, 
where a method to estimate reduction in the pulse energy during the null state was also 
reported for the first time. This quantity, defined as $\eta$, 
can put tight constrain on the possible nulling models. 
In the later study of the similar pulsar, PSR B0031--07, 
\cite{jv00} reported lack of subpulse phase memory 
for longer null lengths. These authors also reported 
a change in the drift rate across nulls, exclusively noted 
to occur during one of the profile mode. These studies suggest 
that, nulling and drifting could be related in their fundamental 
origin or might have a similar sporadic parameter controlling them. 

In a continuing attempt to study interactions between 
nulling and drifting phenomena, \cite{vkr+02} again looked 
at PSR B0809+74 with more sensitive observations. 
These authors classified different drift rates 
as two different emission modes, among which 
one is the normal emission mode while the later 
was classified as the slow drifting mode which 
tends to occur after all null states. 
They also showed increase in the intensity, 
with a shift in the profile towards an earlier longitude 
during the slow drift mode. Thus, these authors  
suggested that mode-changing in PSR B0809+74 
is induced by nulls. In a later study,  
\cite{vsrr03} claimed that post nulls,
the drift rates are similar to those seen at higher frequencies. 
Hence, during the post-null burst state, the emission originates 
from lower heights in the magnetosphere. 
Similarly, \cite{jv04} extended work of \cite{la83} on PSR B0818--13 
with more sensitive observations. These authors also 
concluded analogous behaviour, by reporting 
changes in the subpulse phase memory to speculate that 
during the nulls the drift rate appears to speed up. 

\cite{wf81} reported interesting nulling behaviour in PSR B2319+60, 
which also shows three profile modes, also shown in Figure \ref{mode-changing-fig}. 
These modes were classified as modes A, B and ABN in order of   
decreasing $P_3$ values (increasing drift rate) and their stability. 
The ABN mode showed highly irregular drifting pattern, while mode 
A showed very prominent stable drifting.  They reported a highly 
correlated mode-switching and nulling. Before an onset 
of the null state, the modes are shown to transit 
from A $\rightarrow$ B $\rightarrow$ ABN, depicting an 
stepwise increase in the drift rate. 
In an interesting study of PSR B1944+17, a large nulling fraction 
pulsar, \cite{dchr86}  investigated similar nulling and 
mode-changing interactions. These authors reported 
four different emission modes, mode A and B with 
regular drift rates while mode C with zero drifting. 
The fourth mode, mode D, displays chaotic subpulse phase 
distributions. Changes between these mode were shown to triggered 
by the occurrence of nulls. Contrary to PSR B0809+74, 
this pulsar did not show any phase memory 
across the null lengths. The transition states, 
i.e. null-to-burst (and vice-versa) were shown to have 
different profile shapes indicating a gradual 
change in emission properties before the onset of 
the null state. A slow decrease in the intensity was also noticed which 
was modelled as an increasing temperature, causing gap potential to 
yield lower magnitudes. After a few years, in line of similar studies, 
\cite{rwr05} reported interesting observations of interactions  
between nulling and mode changing in PSR B2303+30. The pulsar exhibited 
two profile modes, classified as mode B and Q. In the mode B 
pulsar showed regular drifting while in mode Q drifting is 
less ordered and it shows wider profile. Nulls were seen to occur 
exclusively during mode Q and also often either at the start or at the 
end of the mode Q. These authors also reported gradual changes 
in the emission from mode B to Q as damped oscillator with 
finite relaxation time, similar to those reported by \cite{la83}. 

Nulling behaviour in an unusual pulsar, PSR B0826--34, with presence of 
emission over the entire pulse longitude, was reported by \cite{dll+79} at 408 MHz. 
It is likely that, pulsar is almost an aligned rotator with a very small $\alpha$ 
and a small viewing angle from the rotation axis. Moreover, the most interesting 
aspect of it's emission is the high NF of around 70\%, making it as one 
of the first known pulsar with long nulls \cite[]{dll+79}. 
In contrast, \cite{elg+05} have reported presence 
of weak emission during the null states, in their observations 
at 1374 MHz. The pulsar also showed regular drifting across 
13 subpulse components. These authors suggested that, 
the line-of-sight always remains inside the emission beam, 
cutting inner and outer conal rings. They proposed a hypothesis 
that, nulls in this pulsar are not real and they occur due to different 
weak emission mode. Recently, \cite{kr10} reported a 
study on an interesting nulling and mode-changing pulsar, 
PSR B1944+17, which was investigated earlier by \cite{dchr86}. 
These authors reported analogous results regarding the presence 
of four drift-modes along with the 70\% NF. 
They report a mixture of short and long nulls, in which 
the short nulls show weak average power. These short nulls 
were classified as \emph{pseudo}-nulls, nulls due to 
the empty line-of-sights (further discussed in Section \ref{periodic_nulling_sect}).

\subsection{Periodic nulling}
\label{periodic_nulling_sect}
Excluding the first reported study of nulling phenomena by \cite{bac70}, 
nulling was thought to be a randomly occuring phenomena and no further studies 
were conducted for for many decades. This view was challenged by later 
investigators. \cite{naf03} reported 
periodic nulling in PSR J1819+1305. \cite{rw08} have 
conducted an extensive study of the same pulsar and  
obtained a better estimate on the nulling periodicity 
of around 57$\pm$1 periods. They concluded an empty line-of-sight
passing between the emitting sub-beams,  as the cause of 
periodic nulling seen in this pulsar. \cite{wr07} reported 
a detail study of PSR B0834+06. In the earlier studies, subpulses 
were considered as a basic units of emission due to the absence 
of truncated profiles or \emph{partial nulls} \cite[]{la83,viv95,vkr+02,jv04}.  
\cite{wr07} reported such partial nulls in their study
as pulses with only one of the profile component active at an instance. 
They reported that nulls occur after ever alternating pulses 
and their total observed numbers are too large for their origin from 
random expectation. In similar line of studies, \cite{hr07}
have reported a periodic feature in the pulse energy sequence of PSR B1133+16. 
A low frequency feature was earlier reported in this pulsar but was 
not confirmed with any orderly subpulse structure. 
\cite{hr07} reported that, this low frequency feature 
is because of the ordered sequence of null and 
burst pulses. The sample of these periodic nulling pulsars 
were further enhanced by \cite{hr09}. In a statistical 
study of the occurrence of null pulses, \cite{rr09} 
showed non-randomness in 14 pulsars using a randomness test 
known as the \emph{Wald–Wolfowitz runs test} \cite[]{ww40}. 
These authors concluded that, this non-randomness point towards 
the quasi-periodic nature of the null pulses, originating from 
the empty line-of-sight cuts between the sub-beams. 
Recently, \cite{hlk+12} carried out observations of 
PSR B1133+16 at 8.35 GHz, and found supporting 
evidence of periodic nulling as reported by \cite{hr07}. 

\subsection{Broadband nulling behaviour}
\label{broadband_intro_sect}
In a simultaneous single pulse study 
of two pulsars, PSRs B0329+54 and B1133+16 at 327 and 2695 MHz, 
\cite{bs78} showed highly correlated pulse energy fluctuations. 
As nulling and pulse energy fluctuations are highly correlated, 
this study suggests that nulling is a broadband phenomena 
within this frequency range. However, later simultaneous 
observations of PSR B0809+74 for about 350 pulses 
indicated that only 6 out of 9 nulls were simultaneous 
at 102 and 408 MHz \cite[]{dls+84}. 
Contrary to that, \cite{big92a} reported indirect evidence of broadband nulling 
behaviour in many pulsars from non simultaneous observations at two 
frequencies \emph{viz.} 645 and 843 MHz. In a first ever attempt 
to investigate broadband aspect of nulling behaviour, 
\cite{bgk+07} reported partially simultaneous nulls 
in PSR B1133+16. These authors conducted simultaneous
observations at four different frequencies, \emph{viz.} 325, 610, 1400 
and 4850 MHz from three different telescopes. 
They reported that, only half of the nulls occur 
simultaneously for PSR B1133+16. Observations at higher 
frequencies showed little nulling compared to observations 
at lower frequencies. It was concluded that, about 5\% of the time 
pulsar showed exclusive burst emission only at the highest observing 
frequency, with clear nulls at lower frequencies. 
The profile of these exclusive pulses were shown to be 
extremely narrow compared to the average profile. The location 
of these pulses were noted to arrive at earlier phase, towards 
the leading edge of the pulse profile. Thus, they were suspected to 
have similar origin as the giant pulses. In contrast, 
simultaneous nulls were reported at 303 and 610 MHz for 
PSR B0826$-$34 \cite[]{bgg08}. 

\raggedbottom
\subsection{Extreme Nullers}
\label{extreme_null_sect}
In the last decade, the view regarding the 
nulling phenomena has completely changed due 
to the discoveries of extreme nullers, which 
are sources exhibiting extreme form of nulling behaviour. 
These sources have been classified in two categories, 
first are the \emph{intermittent} pulsars, 
which null for several days before emerging as 
an active normal pulsar for a few days. 
\cite{klo+06} reported a quasi-periodic nulling 
behaviour in PSR B1931+24, as the first of this class of known sources. 
B1931+24 showed complete absence of emission for around 
25 to 35 days while it remained active for around 5 to 10 days. 
In the high cadence observations, it was observed to transit 
rapidly within 10 sec from an active burst state to a null state. 
One of the worth highlighting characteristic of this source 
is the changes in the $\dot{P}$. During the active state, pulsar displayed 
regular slow down, as expected from a normal pulsar. However, 
during the off-state (null), the slowdown rate decreased, 
suggesting an overall reduction in the energy loss. 
Thus, \cite{klo+06} suggested cessation of pair 
production on top of the polar cap as the cause of 
long off-states seen in this pulsar. In later years, 
three more sources were found. PSR J1841$-$0500
was observed to show off-state for around 580 days \cite[]{crc+12}, 
while PSR J1832+0029 showed off-state for around 650 to 850 days \cite[]{llm+12}. 
Recently, \cite{sjm+13} have also reported a tentative 
detection of intermittent behaviour in PSR J1839+15.  

The second class of extreme nullers are the 
\emph{Rotating Radio Transients} (RRATs), 
discovered during the reprocessing of the Parkes multibeam 
survey data \cite[]{mll+06}. RRATs are pulsars which 
only emit single burst pulses separated by 
long null states lasting several seconds or minutes. 
They were discovered by their dispersion smear
of these strong single pulses seen in their discovery plots. 
In the original discovery paper by \cite{mll+06}, only 11 such 
sources were reported. However, this sample was further 
enhanced to around 85 sources in just 7 years of their discovery 
through various surveys \cite[]{dcm+09,kle+10,bb10,kkl+11,bjb+12,skd+09}. 
Many of these sources are not yet published but a tentative list of them 
can be found at RRATalog (\url{http://astro.phys.wvu.edu/rratalog/}). 
Recent observations by \cite{pmk+11} reported quasi-periodicity 
on the time-scale of minutes to years in six RRATs. \cite{mmr+11} reported 
spectral behaviour of these sources, which is similar to normal pulsars. 
One of the strongest RRAT, PSR J1819--1458 has also been observed 
in optical and X-rays. The optical observations returned 
with negative results \cite[]{dkm+11}, while the X-ray observations 
with \emph{Chandra} and \emph{XMM Newton} provided clear detection \cite[]{rbg+06,mrg+07}. 
However, the emission mechanism and the origin of their peculiar behaviour is 
still remain to be explained by a suitable model [see \cite{km11} for a review].

\subsection{List of nulling pulsars}
In the past 44 years, since the discovery of this phenomena, 
around 109 pulsars were found to show nulling behaviour. 
These pulsars exhibit a variety in their behaviour, 
including interaction with drifting and mode-changing 
as discussed above. Table \ref{table_all_null_psr} lists most of the reported 
nulling pulsars in the literature, including the pulsars discussed in this 
thesis. NF values are listed along with their respective 
errors and corresponding references. 
It can be seen from Table \ref{table_all_null_psr} that, 
pulsar exhibits wide range of NFs, starting 
from as small as 0.0008\% to as large as 93\%. 
These pulsars also exhibit a wide range in their basic 
parameters, thus the operational mechanism of 
nulling in pulsars remain to be identified. 

 \setlength{\LTleft}{-20cm plus -0.5fill}
 \setlength{\LTright}{\LTleft}
 \flushbottom
 \scriptsize
 \setlength{\tabcolsep}{4pt}
 \begin{longtable}{lllllllcl}
 \hline
 \hline 
No. & PSRs   & P        & $\dot{P}$   &  Age     &$B_{surf}$& $\dot{E}$ & NF  & Ref. \\
    &        & (sec)    & ($\times{10^{-15}}$) s $s^{-1}$ & ($\times{10}^6$) Year & ($\times{10}^{12}$) G & ($\times{10}^{32}$) ergs $s^{-1}$  & (\%) \\
\hline
\endhead
1 & B0031-07 & 0.94 & 0.408 & 36.6 & 0.628 & 0.192	& 44 (1)	 & This study\\ 
2 & B0045+33 & 1.21 & 2.35 & 8.19 & 1.71 & 0.516	& 21 (1)	 & \cite{rr09}\\  
3 & B0148-06 & 1.46 & 0.443 & 52.4 & 0.815 & 0.0556	& $\leq$ 5	 & \cite{big92a}\\  
4 & B0149-16 & 0.83 & 1.3 & 10.2 & 1.05 & 0.888	& $\leq$ 2.5	 & \cite{viv95}\\  
5 & B0301+19 & 1.38 & 1.3 & 17 & 1.36 & 0.191	& 10 (5)	 & \cite{ran86}\\  
6 & B0329+54 & 0.71 & 2.05 & 5.53 & 1.22 & 2.22	& $\leq$ 0.5	 & \cite{rit76}\\  
7 & B0450-18 & 0.54 & 5.75 & 1.51 & 1.8 & 13.7	& $\leq$ 0.5	 & \cite{rit76}\\  
8 & B0523+11 & 0.35 & 0.0736 & 76.3 & 0.163 & 0.653	& $\leq$ 0.06	 & \cite{wab+86}\\  
9 & B0525+21 & 3.74 & 40.1 & 1.48 & 12.4 & 0.301	& 25 (5)	 & \cite{rit76}\\  
10 & B0626+24 & 0.47 & 2 & 3.78 & 0.987 & 7.28	& $\leq$ 0.02	 & \cite{wab+86}\\  
11 & B0628-28 & 1.24 & 7.12 & 2.77 & 3.01 & 1.46	& $\leq$ 0.3	 & \cite{big92a}\\  
12 & B0656+14 & 0.38 & 55 & 0.111 & 4.66 & 381	& 12 (4)	 & \cite{wab+86}\\  
13 & B0736-40 & 0.37 & 1.62 & 3.68 & 0.788 & 12.1	& $\leq$ 0.4	 & \cite{big92a}\\  
14 & B0740-28 & 0.16 & 16.8 & 0.157 & 1.69 & 1430	& $\leq$ 0.2	 & \cite{big92a}\\  
15 & B0751+32 & 1.44 & 1.08 & 21.2 & 1.26 & 0.142	& 34 (0.5)	 & \cite{wab+86}\\  
16 & B0809+74 & 1.29 & 0.168 & 122 & 0.472 & 0.0308	& 1.4 (0.02)	 & \cite{rit76}\\  
17 & B0818-13 & 1.23 & 2.11 & 9.32 & 1.63 & 0.438	& 1.01 (0.01)	 & \cite{rit76}\\  
18 & B0820+02 & 0.86 & 0.105 & 131 & 0.304 & 0.0638	& $\leq$ 0.06	 & \cite{wab+86}\\  
19 & B0823+26 & 0.53 & 1.71 & 4.92 & 0.964 & 4.52	& $\leq$ 5	 & \cite{rit76}\\  
20 & B0826-34 & 1.84 & 0.996 & 29.4 & 1.37 & 0.0622	& 75 (35)	 & \cite{dll+79}\\  
21 & B0833-45 & 0.08 & 125 & 0.0113 & 3.38 & 69200	& $\leq$ 0.0008	 & \cite{big92a}\\  
22 & B0834+06 & 1.27 & 6.8 & 2.97 & 2.98 & 1.3	& 7.1 (0.1)	 & \cite{rit76}\\  
23 & B0835-41 & 0.75 & 3.54 & 3.36 & 1.65 & 3.29	& 1.7 (1.2)	 & This study\\ 
24 & B0919+06 & 0.43 & 13.7 & 0.497 & 2.46 & 67.9	& $\leq$ 0.05	 & \cite{wab+86}\\  
25 & B0940-55 & 0.66 & 22.9 & 0.461 & 3.94 & 30.8	& $\leq$ 12.5	 & \cite{big92a}\\  
26 & B0940+16 & 1.08 & 0.0911 & 189 & 0.318 & 0.028	& 8 (3)	 & \cite{wab+86}\\  
27 & B0942-13 & 0.57 & 0.0453 & 200 & 0.163 & 0.0963	& $\leq$ 7	 & \cite{viv95}\\  
28 & B0950+08 & 0.25 & 0.23 & 17.5 & 0.244 & 5.6	& $\leq$ 5	 & \cite{rit76}\\  
29 & J1049-5833 & 2.20 & 4.41 & 7.91 & 3.15 & 0.163	& 47 (3)	 & \cite{wmj07}\\  
30 & B1055-52 & 0.19 & 5.83 & 0.535 & 1.09 & 301	& $\leq$ 11	 & \cite{big92a}\\  
31 & B1112+50 & 1.65 & 2.49 & 10.5 & 2.06 & 0.217	& 64 (6)	 & This study \\
32 & B1133+16 & 1.18 & 3.73 & 5.04 & 2.13 & 0.879	& 15 (2)	 & \cite{rit76}\\  
33 & B1237+25 & 1.38 & 0.96 & 22.8 & 1.17 & 0.143	& 6 (2.5)	 & \cite{rit76}\\  
34 & B1240-64 & 0.38 & 4.5 & 1.37 & 1.34 & 30.3	& $\leq$ 4	 & \cite{big92a}\\  
35 & B1322-66 & 0.54 & 5.31 & 1.62 & 1.72 & 13.1	& 9.1 (3)	 & \cite{wmj07}\\  
36 & B1358-63 & 0.84 & 16.7 & 0.798 & 3.8 & 11	& 1.6 (2)	 & \cite{wmj07}\\  
37 & B1426-66 & 0.78 & 2.77 & 4.49 & 1.49 & 2.26	& $\leq$ 0.05	 & \cite{big92a}\\  
38 & B1451-68 & 0.26 & 0.0983 & 42.5 & 0.163 & 2.12	& $\leq$ 3.3	 & \cite{big92a}\\  
39 & J1502-5653 & 0.53 & 1.83 & 4.64 & 1 & 4.7	& 93 (4)	 & \cite{wmj07}\\  
40 & J1525-5417 & 1.01 & 16.2 & 0.991 & 4.09 & 6.17	& 16 (5)	 & \cite{wmj07}\\  
41 & B1530+27 & 1.12 & 0.78 & 22.9 & 0.948 & 0.216	& 6 (2)	 & \cite{wab+86}\\  
42 & B1530-53 & 1.36 & 1.43 & 15.2 & 1.41 & 0.219	& $\leq$ 0.25	 & \cite{big92a}\\  
43 & B1556-44 & 0.25 & 1.02 & 4 & 0.518 & 23.7	& $\leq$ 0.01	 & \cite{big92a}\\  
44 & B1604-00 & 0.42 & 0.306 & 21.8 & 0.364 & 1.61	& $\leq$ 0.1	 & \cite{big92a}\\  
45 & B1612+07 & 1.20 & 2.36 & 8.1 & 1.71 & 0.53	& $\leq$ 5	 & \cite{wab+86}\\  
46 & J1639-4359 & 0.58 & 0.015 & 621 & 0.095 & 0.0292	& $\leq$ 0.1	 & This study \\
47 & B1641-45 & 0.45 & 20.1 & 0.359 & 3.06 & 84.2	& $\leq$ 0.4	 & \cite{big92a}\\  
48 & B1642-03 & 0.38 & 1.78 & 3.45 & 0.841 & 12.1	& $\leq$ 0.25	 & \cite{rit76}\\  
49 & J1648-4458 & 0.62 & 1.85 & 5.38 & 1.09 & 2.93	& 1.4 (11)	 & \cite{wmj07}\\  
50 & J1649+2533 & 1.01 & 0.559 & 28.8 & 0.763 & 0.211	& $\leq$ 20	 & \cite{rr09}\\  
51 & B1658-37 & 2.45 & 11.1 & 3.49 & 5.29 & 0.297	& 19 (6)	 & This study \\
52 & J1702-4428 & 2.12 & 3.3 & 10.2 & 2.68 & 0.136	& 26 (3)	 & \cite{wmj07}\\  
53 & J1703-4851 & 1.39 & 5.08 & 4.35 & 2.7 & 0.737	& 1.1 (4)	 & \cite{wmj07}\\  
54 & J1715-4034 & 2.07 & 3.01 & 10.9 & 2.53 & 0.134	& $\geq$ 6	 & This study \\
55 & J1725-4043 & 1.46 & 2.79 & 8.32 & 2.05 & 0.35	& $\leq$ 70	 & This study \\
56 & J1727-2739 & 1.29 & 1.1 & 18.6 & 1.21 & 0.201	& 52 (3)	 & \cite{wmj07}\\  
57 & B1727-47 & 0.82 & 164 & 0.0804 & 11.8 & 113	& $\leq$ 0.05	 & \cite{big92a}\\  
58 & J1738-2330 & 1.97 & 8.56 & 3.66 & 4.16 & 0.436	& 85.1 (2.3)	 & This study \\
59 & B1737+13   & 0.80 & 1.45 & 8.77 & 1.09 & 1.11	& $\leq$ 0.02	 & \cite{wab+86}\\  
60 & J1744-3922 & 0.17 & 0.00155 & 1760 & 0.0165 & 0.119	& $\leq$ 75	 & \cite{fsk+04}\\  
61 & B1742-30   & 0.36 & 10.7 & 0.546 & 2 & 84.9	& $\leq$ 17.5	 & \cite{big92a}\\  
62 & B1749-28   & 0.56 & 8.13 & 1.1 & 2.16 & 18	& $\leq$ 0.75	 & \cite{rit76}\\  
63 & J1752+2359 & 0.40 & 0.643 & 10.1 & 0.519 & 3.71	& $\leq$ 89	 & This study \\
64 & B1809-173  & 1.20 & 19.1 & 1 & 4.85 & 4.3	& 5.8 (4)	 & \cite{wmj07}\\  
65 & B1818-04   & 0.59 & 6.33 & 1.5 & 1.97 & 11.7	& $\leq$ 0.25	 & \cite{big92a}\\  
66 & J1820-0509 & 0.33 & 0.932 & 5.73 & 0.567 & 9.59	& 67 (3)	 & \cite{wmj07}\\  
67 & B1821+05   & 0.75 & 0.227 & 52.6 & 0.418 & 0.21	& $\leq$ 0.4	 & \cite{wab+86}\\  
68 & J1831-1223 & 2.85 & 5.47 & 8.27 & 4 & 0.0926	& 4 (1)	 & \cite{wmj07}\\  
69 & J1833-1055 & 0.63 & 0.527 & 19.1 & 0.585 & 0.817	& 7 (2)	 & \cite{wmj07}\\  
70 & B1839+09   & 0.38 & 1.09 & 5.54 & 0.652 & 7.76	& $\leq$ 5	 & \cite{wab+86}\\  
71 & B1842+14   & 0.37 & 1.87 & 3.18 & 0.848 & 14	& $\leq$ 0.15	 & \cite{wab+86}\\  
72 & J1843-0211 & 2.02 & 14.4 & 2.22 & 5.48 & 0.684	& 6 (2)	 & \cite{wmj07}\\  
73 & B1848+12   & 1.20 & 11.5 & 1.66 & 3.77 & 2.6	& $\leq$ 54	 & \cite{rr09}\\  
74 & B1857-26   & 0.61 & 0.205 & 47.4 & 0.358 & 0.352	& 10 (2.5)	 & \cite{rit76}\\  
75 & J1901+0413 & 2.66 & 132 & 0.321 & 18.9 & 2.75	& $\leq$ 6	 & This study \\
76 & B1907+03   & 2.33 & 4.47 & 8.26 & 3.27 & 0.139	& 4 (0.2)	 & \cite{wab+86}\\  
77 & B1911-04   & 0.82 & 4.07 & 3.22 & 1.85 & 2.85	& $\leq$ 0.5	 & \cite{rit76}\\  
78 & J1916+1023 & 1.61 & 0.681 & 37.7 & 1.06 & 0.0634	& 47 (4)	 & \cite{wmj07}\\  
79 & B1917+00   & 1.27 & 7.67 & 2.63 & 3.16 & 1.47	& $\leq$ 0.1	 & \cite{ran86}\\  
80 & B1919+21   & 1.33 & 1.35 & 15.7 & 1.36 & 0.223	& $\leq$ 0.25	 & \cite{rit76}\\  
81 & J1920+1040 & 2.21 & 6.48 & 5.42 & 3.83 & 0.235	& 50 (4)	 & \cite{wmj07}\\  
82 & B1923+04   & 1.07 & 2.46 & 6.92 & 1.64 & 0.783	& $\leq$ 5	 & \cite{wab+86}\\  
83 & B1929+10   & 0.22 & 1.16 & 3.1 & 0.518 & 39.3	& $\leq$ 1	 & \cite{rit76}\\  
84 & B1931+24   & 0.81 & 8.11 & 1.59 & 2.6 & 5.94	& $\leq$ 80	 & \cite{klo+06}\\  
85 & B1933+16   & 0.35 & 6 & 0.947 & 1.48 & 51.3	& $\leq$ 0.06	 & \cite{big92a}\\  
86 & B1942+17   & 1.99 & 0.73 & 43.3 & 1.22 & 0.0362	& $\leq$ 60	 & \cite{lcx02}\\  
87 & B1942-00   & 1.04 & 0.535 & 31 & 0.757 & 0.185	& 21 (1)	 & \cite{wab+86}\\  
88 & B1944+17   & 0.44 & 0.0241 & 290 & 0.104 & 0.111	& 50 (7)	 & \cite{viv95}\\  
89 & B1946+35   & 0.71 & 7.06 & 1.61 & 2.28 & 7.55	& $\leq$ 0.75	 & \cite{rit76}\\  
90 & B2016+28   & 0.55 & 0.148 & 59.7 & 0.291 & 0.337	& $\leq$ 0.25	 & \cite{rit76}\\  
91 & B2020+28   & 0.34 & 1.89 & 2.87 & 0.816 & 18.5	& 0.2 (1.6)	 & This study \\
92 & B2021+51   & 0.52 & 3.06 & 2.74 & 1.29 & 8.16	& 1.4 (0.7)	 & This study \\
93 & B2034+19   & 2.07 & 2.04 & 16.1 & 2.08 & 0.0902	& 44 (4)	 & \cite{hr09}\\  
94 & B2044+15   & 1.13 & 0.182 & 98.9 & 0.461 & 0.0488	& $\leq$ 0.04	 & \cite{wab+86}\\  
95 & B2045-16   & 1.96 & 11 & 2.84 & 4.69 & 0.573	& 10 (2.5)	 & \cite{rit76}\\  
96 & B2053+36   & 0.22 & 0.369 & 9.51 & 0.289 & 13.4	& $\leq$ 0.7	 & \cite{wab+86}\\  
97 & B2110+27   & 1.20 & 2.62 & 7.27 & 1.8 & 0.595	& $\leq$ 30	 & \cite{rr09}\\  
98 & B2111+46   & 1.01 & 0.715 & 22.5 & 0.862 & 0.27	& 21 (4)	 & This study \\
99 & B2113+14   & 0.44 & 0.289 & 24.1 & 0.361 & 1.34	& $\leq$ 1	 & \cite{wab+86}\\  
100 & B2122+13  & 0.69 & 0.768 & 14.3 & 0.739 & 0.907	& $\leq$ 22	 & \cite{rr09}\\  
101 & B2154+40  & 1.52 & 3.43 & 7.04 & 2.32 & 0.382	& 7.5 (2.5)	 & \cite{rit76}\\  
102 & J2208+5500 & 0.93 & 6.99 & 2.12 & 2.58 & 3.39	& $\leq$ 7.5	 & \cite{jml+09}\\  
103 & B2217+47   & 0.53 & 2.77 & 3.09 & 1.23 & 6.99	& $\leq$ 2	 & \cite{rit76}\\  
104 & J2253+1516 & 0.79 & 0.0665 & 189 & 0.232 & 0.0528	& $\leq$ 49	 & \cite{rr09}\\  
105 & B2303+30   & 1.57 & 2.89 & 8.63 & 2.16 & 0.292	& 1 (0.5)	 & \cite{ran86}\\  
106 & B2310+42   & 0.34 & 0.112 & 49.3 & 0.201 & 1.04	& $\leq$ 11	 & \cite{rr09}\\  
107 & B2315+21 	 & 1.44 & 1.05 & 21.9 & 1.24 & 0.137	& 3 (0.5)	 & \cite{wab+86}\\  
108 & B2319+60 	 & 2.25 & 7.04 & 5.08 & 4.03 & 0.242	& 29 (1)	 & This study \\
109 & B2327-20 	 & 1.64 & 4.63 & 5.62 & 2.79 & 0.412	& 12 (1)	 & \cite{big92a}\\ 
\hline 
\hline 
\caption[Table of all known nulling pulsars]{Table of all known nulling pulsars with their basic parameters. 
 The columns give period, $\dot{P}$, age, surface magnetic field ($B_{surf}$), 
 rate of energy loss ($\dot{E}$) and the reported NFs with their respective 
 errors in the parenthesis along with the corresponding references.} \\
\label{table_all_null_psr}
\flushbottom
\end{longtable}
\normalsize
% \newpage
\section{Why do pulsars null?}
\label{null_model_all_sect}
\begin{figure}[h!]
 \centering
 \includegraphics[width=2.7 in,height=4 in,angle=-90,bb=50 50 554 770]{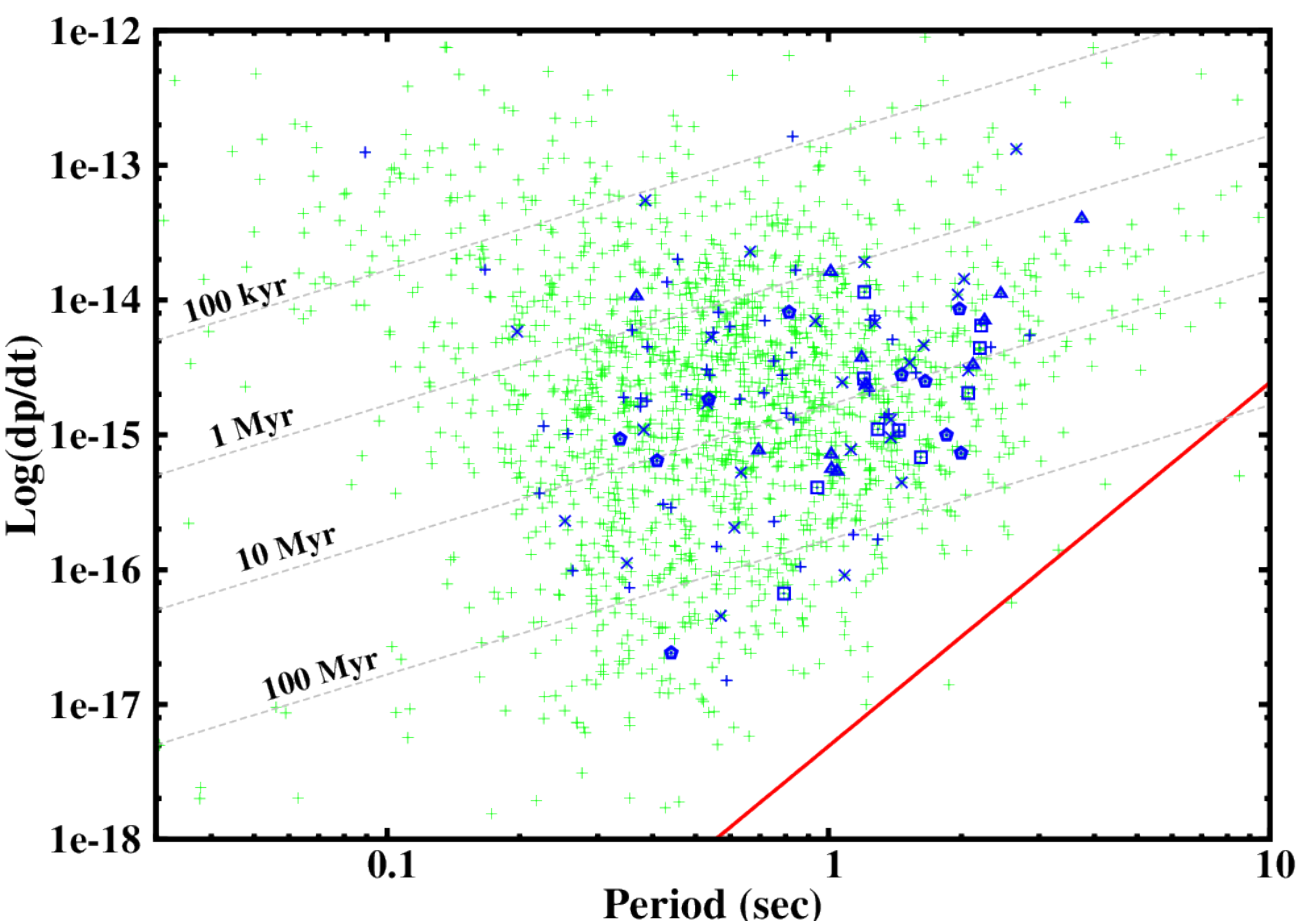}
 % All_Nulling_PSR_p_pdot_NF.eps: 0x0 pixel, 300dpi, 0.00x0.00 cm, bb=50 50 554 770
  \caption[Nulling pulsars in the $P-\dot{P}$ diagram]{Nulling pulsars in the $P-\dot{P}$ diagram with the population of 
 normal pulsars zoomed in. The nulling pulsars are shown with a blue points while their shape 
 represented their corresponding NFs range. These NFs were binned in five groups, to show their relative localization,  
 which are (1) NF$\leq$5\% with pluses, (2) 5\%$<$NF$\leq$15\% with crosses, (3) 15\%$<$NF$\leq$30\% with triangles, (4) 30\%$<$NF$\leq$60\% with 
 squares and (5) NF$>$60\% with pentagons. The green points are the normal pulsars while the red solid line 
 is the speculated death line according to the model given by \cite{cr77}. 
 The characteristic age of the pulsar are also shown  with the dashed lines.}
\label{all_psr_nulling_nf}
\end{figure}

In the early studies, it was argued extensively 
that nulling is a property of older pulsars and they 
tend to lie close to the death line in the $P-\dot{P}$ diagram 
\cite[]{tm77,mt77,wc81,ao82,oa82,gi83,lw95}. 
Hence, its worth testing this hypothesis by estimating 
the positions of nulling pulsars in the $P-\dot{P}$ diagram. 
Figure \ref{all_psr_nulling_nf} shows the distribution of 
all nulling pulsars (listed in Table \ref{table_all_null_psr}). 
Its clearly evident from this distribution, 
in the $P-\dot{P}$ diagram, that they are not 
localized near the death line. Moreover, it should be 
noted, that pulsars with high NFs lie right in the middle 
of the normal pulsar population. Thus, its unlikely 
that nulling presents dying state of a radio pulsar.
Similar conclusions were also suggested by \cite{mic91}.

If nulling is not due to the age of the pulsar, 
then certain mechanism, operating along with the 
emission processes, may cause faltering of 
the radiation beam. There are two different group of theories 
in literature that attempt to explain the cause of nulling. 
First group of theories are more inclined towards 
changes in the geometry of the beam and the line-of-sight, while 
the second group of theories suggest intrinsic effects 
that can cause cessation of the full radiation beam. 
\raggedbottom

\subsection{Geometric effects}
\label{geometric_effects_sect}
In the line of geometric reasoning, 
\cite{aro83c} proposed nutation of the emission beam as a likely cause. 
In Section \ref{periodic_nulling_sect} missing 
line-of-sight as one of the geometric effects was briefly discussed. 
This model was proposed due to the observed periodic 
fluctuations seen in a few nulling pulsars.
According to which, the line-of-sight cuts between 
the radiating sub-beams, and thus missing the emitting regions. 
It is also possible that, only a few radiating sub-beams  
extinguished and when line-of-sight passes 
over them, nulls are observed. 
As the sub-beams are organised in an evenly spaced 
uniform pattern (Figure \ref{rankin_conal}), 
empty line-of-sight cuts or extinguished individual sub-beams 
are more likely to repeat after certain pulsar periods.
\cite{rit76} concluded, for the first time, the geometric 
origin of nulling phenomena due to these extinguishing sub-beams 
in his study. This was attributed as the 
nulling periodicity \cite[]{hr07,hr09,rw08}. 

In further models of geometric origin, \cite{dzg05} suggested 
reversal of emission beam for PSR B1822$-$09 to explain 
the peculiar main pulse and interpulse relation. 
This pulsar exhibit strong main pulse with a \emph{precursor}. 
Precursors are known as the pulse component in front of the main pulse, 
separated by a few degrees in longitude [see \cite{gjk+94} for a review]. 
PSR B1822$-$09 exhibit interesting anti-correlation between this precursor and interpulse, 
as when the precursor goes in the null state the interpulse emission 
appears and vice-verse \cite[]{gjk+94}. Thus, \cite{dzg05} suggested 
that, the interpulse and the precursor are likely to originate from the same 
side of the emission beam. The emission seen during the interpulse is simply 
reversed emission coming from the precursor, from the main emission beam facing away 
from the observer. Thus, during these occasions, emission in the 
precursor beam reverses it's propagation direction. If such action takes place in the main 
beam also then it will give rise to the observed null pulses. However, recent studies 
from \cite{bmr10} showed that during many pulses, the precursor and 
the interpulse are both present in PSR B1822--09. 
Thus, its unlikely that such emission reversal exist for this pulsar. 
  
In other models, \cite{zx06} reported precession of the rotation axis as the cause of 
nulling to explain absence of regular emission from one of the transient 
sources near the Galactic centre  namely, GCRT J1745--3009. However, the 
precession in pulsars are very rare as there are only a few tentative evidence 
\cite[]{wrt89,cgc97,ja01,hh02,wt05}. Moreover, \cite{gle90} 
reported deformation of the emission beam due to the misalignment between the 
rotation and the neutron star symmetry axes. However, it should 
also be noted that, such models are likely to give periodic oscillations 
in the pulse energy with consecutively occurring null and burst states.  
In other studies, \cite{cr04a} reported non-radial oscillation as the 
cause of may single pulse phenomena including nulling, drifting and mode-changing. 
These authors suggest that mode-changing and nulling are due to the 
changes between different oscillation modes. 

\subsection{Intrinsic effects}
\label{intrinsic_effects_sect}
In the RS model, as explained in Section \ref{radio_emission_sect}, 
the coherent radiation is a highly tuned process. Change 
in any of the necessary conditions of coherence may lead to 
stoppage of emission. In the early years, \cite{che81} suggested 
fluctuations in the temperature to invoke changes in the coherence 
conditions as a possible cause of nulling. \cite{dchr86} also reported 
possible increase in the temperature, which can cause ion outflow and extra X-ray flux, 
to decrease the polar cap potential to stop sparking and primary particle flow.  
This rise in temperature can occur due to the 
backflowing electrons, generated during the sparking action, 
hitting the polar cap region with relativistic 
velocities. The heated up spark regions (also known as the hot spots)
are more likely to produce further pairs on top of them in the consecutive 
sparking action (post null). Thus, this model was proposed as the reason 
for the reported subpulse phase memory across the nulls in B0809+74. 
\cite{dchr86} also estimated the null-to-burst (and vice-versa)
transition time-scales of around 2 to 30 milliseconds, 
which is equivalent to the subpulse widths. Thus, it was also suggested 
as an explanation for the lack of partial nulls in B0809+74. 
However, these claims were contested by \cite{fr82} 
and later observations from \cite{fre83}. 
\cite{fr82} showed that, these hot spots will cool too fast to 
persist any memory on the time-scale of null lengths. 
In contrast, they proposed a model of continuous flow 
of charge particle as the cause of nulling. In their model, 
the polar gap potential attains a stationary state in which 
gap discharges roughly at the same rate as it charges, thus 
the flow of primary particle does not stop. 
However, in an non-stationary flow, as discussed 
in Section \ref{radio_emission_sect}, any of the proposed two stream 
instability conditions will not be satisfied to produce bunching. 
Thus, this stationary flow of primary plasma may give rise to 
absence of emission. This polar cap potential can suddenly 
discharge also, retaining normal interrupted sparking to cause 
non-stationary flow to form bunches again. To explain the 
memory across nulls, \cite{fr82} invoked the mechanism of 
\emph{flux tubes} through which the particles are 
transferred to the magnetosphere to form sub-beams. 
The electric field in these flux tubes are much higher 
compare to regions between them, and thus they are 
more likely to produce sparking. During the post-null state also, 
the sparking takes place inside these tubes to retain the 
subpulse phase memory. 

\cite{kmms96} reported a model of nulling in pulsars considering 
a different emission mechanism proposed by \cite{kmm91}. 
According to \cite{kmms96}, the drifting of subpulses are not 
due to the shifting spark regions, but  rotation of plasma wave around the 
magnetosphere. Particles are extracted from the surface to form 
a primary beam distribution. The most energetic particles of this 
beam give rise to secondary plasma and backflowing particles. 
These backflowing particles heat up the surface, which causes broadening of 
the primary beam energy distribution. As this process repeats several times, 
gap potential slowly drops (on much larger time-scale compared to RS sparking). 
Thus, the distribution of the primary plasma moves towards the 
low Lorentz-factors, ultimately switching off the emission mechanism. 
When the gap attains the $n_{GJ}$, it closes to charge again
as pulsar retains its emission state. This phenomena is different 
from the RS model, in which gap discharges 
roughly on the time-scale of a few microseconds. However, in the \cite{kmms96} model, 
the gap discharges on the time-scale of primary beam distribution function. 
Thus, if this process takes more time than the pulsar period, nulls 
can be observed. One of the main argument that can be put forward to 
contradict this model is the speculated emission height. 
This theory suggest generation of radio emission 
around 10 to 20\% height of the light cylinder, which is around 5000 
times stellar radii. However, recent observations have confirmed 
the hight of emission to be around 50 stellar radii 
\cite[]{cor78,ran83a,lm88,kg98,gg01,mr02a,kg03a}. 

As discussed in Section \ref{spark_sect}, the gap discharge can occur 
due to (a) avalanche of pair plasma triggered by the  
photons created from the curvature radiation (CR) or (b) by inverse 
Compton scattering (ICS) of the thermal photons from 
the relativistic charged particles ejected from the surface. 
In the original RS model, the ICS process was assumed to be very 
inefficient mechanism and hence was not considered. 
However, \cite{dh86} suggested that due to the strong magnetic field, 
ICS could be an efficient mechanism. \cite{zq96} argued that 
in case of ICS gap discharge, the surface temperature plays 
an important role. If the temperature is above certain threshold, 
the binding energy of ions will decrease, which can cause 
the ion flow from the surface to extinguish the polar gap 
(as suggested in Section \ref{spark_sect}), and ultimately to cause nulls. 
\cite{zqlh97} proposed that a mixture of CR and ICS discharge 
might be operating with different temperature ranges 
in a same pulsar to explain the observed mode-changing phenomena. 
\cite{zqh97b} extrapolated this model to nulling pulsars, 
specially for PSR B1055--52. They argued that normally 
the gap remains in a thermal ICS mode, which causes 
relatively low gap height but very energetic secondary particles. 
In contrast, the resonant ICS mode has higher gap height and 
low energy secondary plasma. When pulsar switches 
from thermal ICS to resonant ICS mode, due to the temperature fluctuations, 
the secondary pair plasma will not have sufficient energy in this pulsar 
to maintain the coherence conditions, and it exhibits nulls. 

The chemical composition and structure of the neutron star also plays 
an important role in characterising its emission behaviour. 
Section \ref{ns_sect} discusses the constituents of the 
surface materials and also highlights that the surface of the neutron star 
is smooth. However, \cite{vr80} showed that, it is possible for the 
neutron star surface to have irregularities of the order of around 
10 cm. These tiny hills on the polar cap region will have different electric 
potential on top of them. Thus, they can heavily affect the outflow of particles. 
\cite{vr80} showed that this 10 cm hill can cause 10$^3$ times fluctuations 
in the pulse energy, which was proposed as a possible cause 
of brighter and weaker profile modes seen in a few pulsars. 
This model can be extrapolated to suggest that it can 
also cause the radio emission to go below the detection threshold of the 
telescope to produce nulls. 
\begin{figure}[h!]
 \centering
 \includegraphics[width=3.5 in,height=3.7 in,angle=0,bb=14 14 465 431]{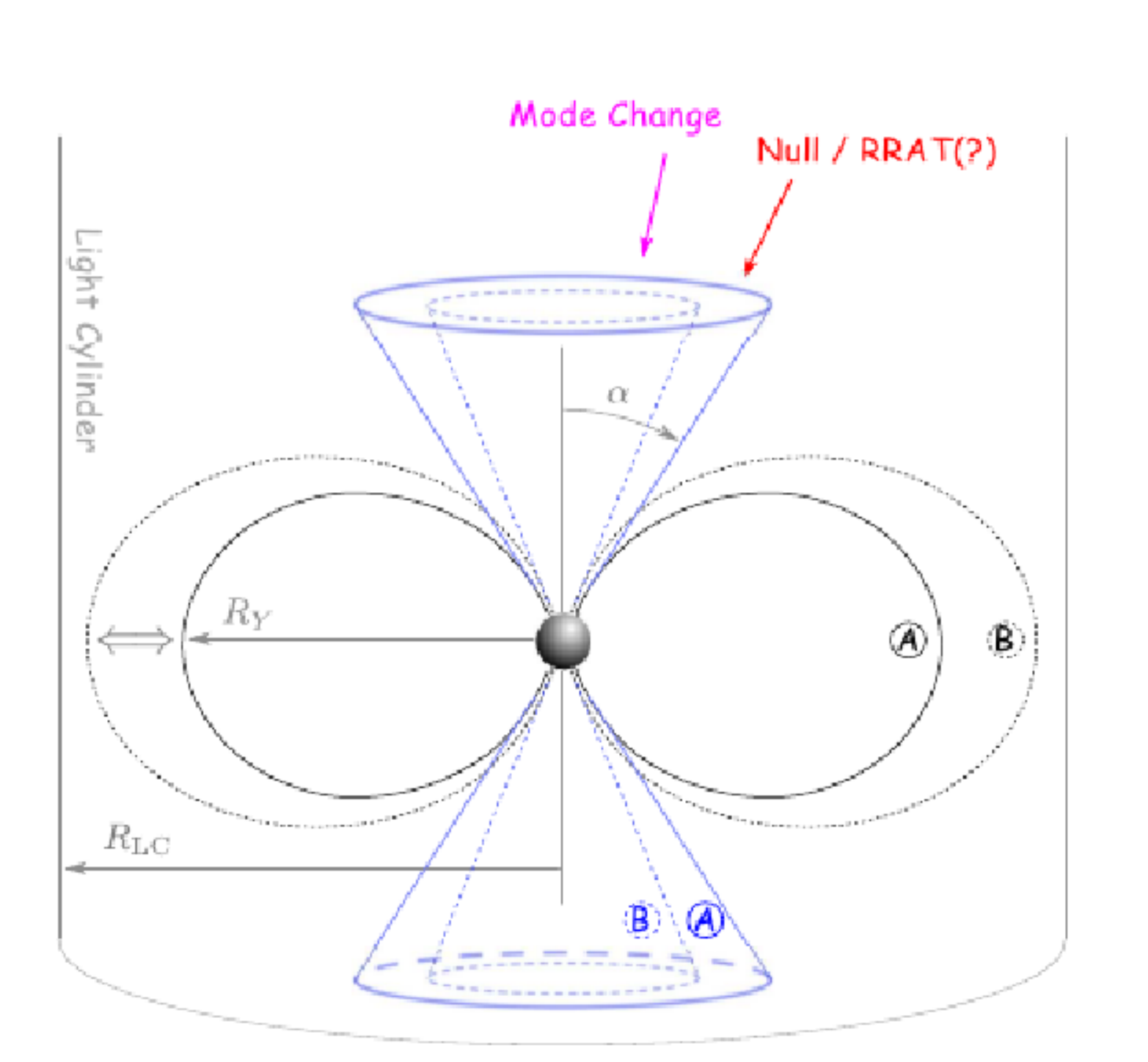}
 % Timokhin.eps: 0x0 pixel, 300dpi, 0.00x0.00 cm, bb=14 14 465 431
 \caption[Nulling/Mode-changing model by \cite{tim10}]{Schematic view of the pulsar magnetosphere adapted 
 here from \cite{tim10}. The two magnetospheric states are indicated with A and B to 
 highlight the differences in their opening angles. Switching between these states can cause 
 pulsar to display mode-changes or nulls depending upon the viewing angle.}
 \label{timokhin_beam}
\end{figure}

In the recent study, \cite{tim10} reported interesting model 
for nulling and mode-changing pulsars\footnote{Ideally, this model should be 
classified along with the other geometric effects, however, due to 
the changes proposed in this model occurs on a global magnetospheric scale, 
it is describe here along with the intrinsic phenomena.}. 
The motivation for his proposed model was to explain 
the $\dot{P}$ variations seen in the intermittent pulsars as 
they suggest an underline phenomena 
on a global magnetospheric scale during the nulls. 
\cite{tim10} proposed two different scenarios which can 
keep the entire magnetosphere in different configurations. 
These include (1) configurations with different close field line 
regions as shown by \cite{tim06} and (2) configurations 
with different current density distributions \cite[]{con05}. 
The magnetosphere of a pulsar can switch between 
different close field line regions or it can switch 
between different current density distributions or both. 
However, the emission beaming angle will change 
for any type of above mentioned state changes, 
to cause nulling as shown in Figure \ref{timokhin_beam}. 
For an outer line-of-sight cut pulsars, the change 
in the beaming angle can cause the entire emission 
to propagate away from the observer and thus nulls will be observed.
For the pulsars with slightly inner cut, such changes will 
allow sampling of different parts of the emission region 
to show different sub-beam structures with different integrated profiles.
\cite{tim10} argued that the difference between the rate 
of energy loss for both types of state changes is similar 
to explain the $\dot{P}$ changes during the nulls in 
the intermittent pulsars. 

\subsection{External effects}
There are also models proposed in various studies which 
are related to magnetospheric changes triggered by external 
influences. \cite{mic91} suggested that nulling might be due to 
accretion of interstellar medium to pulsar magnetosphere. 
In a more recent study, \cite{csh08} reported influence of 
asteroid belt around the pulsars to cause various temporal changes. 
This ``cirumpulsar" debris entering the light cylinder can change the 
conditions for pair productions and coherence, which can 
cause cessation of observed emission. \cite{csh08} argued that the 
time-scale of this phenomena ranges from seconds to months to 
trigger nulling as well as intermittency of radio emission. 

\section{Motivations for this study}
\label{motivation_sect}
Pulsar nulling remained an unexplained phenomena even 
after many decades of investigations. The fundamental 
question, \emph{Why only a few pulsars null and others do not?},  
remain unanswered. Various models have been proposed 
which aim to tackle the problem for individual cases. 
However, a broad overall picture regarding the 
true nature of the nulling phenomena has escaped 
a satisfactory explanation. The reason could be a lack common 
agreement between different studies regarding 
the correlations of NFs with different pulsar parameters. 
For example, \cite{rit76} reported NFs are correlated with 
the age of the pulsar while \cite{ran86} contradicted these 
claims by showing correlations with profile classes. 
These claims were also contested by later studies 
conducted by \cite{wmj07} to show that all profile 
classes of pulsars show similar nulling behaviour. 
Moreover, the degree and form of pulse nulling varies from 
one pulsar to another even though they exhibit similar NFs. 
On one hand, there are high NF pulsars such as PSR B1931+24 
which shows absence of emission for several days \cite[]{klo+06},  while 
there are pulsars such as PSR J1752+2359 \cite[]{lwf+04}, 
which exhibits nulls for only for 3 to 4 minutes, while exhibiting similar NF. 
Thus, it is possible for two pulsars to have similar NF while their 
nulling time-scales could be complete different, 
which raises an important question, \emph{does the NF of a pulsar an ideal way 
to quantify it's behaviour?}

Similarly the broadband aspect of the nulling 
behaviour remains ambiguous due to various contradictory 
studies. For example, a few studies have shown the 
nulling to be a broadband phenomena \cite[]{bs78,bgg08}, 
while a a few other studies have contested these results 
by showing lack of concurrent behaviour \cite[]{dls+84,bgk+07}.
In light of all these previous investigations, 
long, sensitive, and preferably simultaneous 
observations at multiple frequencies of a carefully 
selected sample of pulsars are motivated. 
The main motivations behind the major part of the 
work carried out in this thesis can be listed as follows. 

\begin{enumerate}
 \item The nulling behaviour of similar NF pulsars should be compared to quantify their differences. 
 \vspace{0.2cm}
 \item  Nulling behaviour should be quantified incorporating their time-scales. 
 \vspace{0.2cm}
 \item The periodicities observed in the nulling phenomena by recent observations are intriguing 
       to scrutinize for a large number of pulsars. 
 \vspace{0.2cm}
 \item Broadband aspect of the nulling phenomena needs to be investigated from more number of pulsars with 
       sensitive and simultaneous observations at wide range of frequencies. 
\end{enumerate}

Thus, the aim of this thesis is to quantify, model  
and compare nulling behaviour between different classes of pulsars  
to scrutinize the true nature of the nulling phenomenon at multiple frequencies. 

\clearpage\null\newpage
\chapter{Observations and Analysis}
% \graphicspath{{/home/vishal/My_paper/Thesis/Chapter3/}{/home/vishal/My_paper/Thesis/Chapter3/}{/home/vishal/My_paper/Thesis/Chapter3/}}
\graphicspath{{Images/}{Images/}{Images/}}

\section{Introduction}
This chapter focuses on the details of the observations and the data analysis procedures 
carried out in this thesis. The first part discusses the details regarding 
the observations carried out using the Giant Meterwave Radio Telescope (GMRT), the Westerbork Synthesis Radio Telescope (WSRT), 
the Effelsberg and the Arecibo radio telescopes. In the second part, analysis procedures followed 
for the data reductions have been discussed briefly. 
Table \ref{shedual} lists all the pulsars that were observed as a part of this thesis work 
along with the observing frequencies and the approximate number of observed pulses for these pulsars. 
For two pulsars (PSRs B0809+74 and B2319+60), observations were 
carried out simultaneously at four different frequencies, which includes 
the GMRT at 313 and 607 MHz, the WSRT at 1380 MHz and the Effelsberg at 4850 MHz. 
For PSR J1752+2359, long observations were carried out from the GMRT. Along with 
that, archival data from shorter independent observation from the Arecibo telescope were also investigated. 
The prime motivation behind these observations are listed in Section \ref{motivation_sect}. 
Moreover, a few observations were also conducted on the newly discovered 
pulsars to enhance the number of known nulling pulsars (motivation for 
these observations are mentioned in Sections \ref{new_pulsar_sect} and \ref{chap4_intro_sect}). 
\begin{table}[h!]
\begin{center}
\hfill{}
\newline
\footnotesize
% \scriptsize
\centering
% \hspace{1.1cm}
\begin{tabular}[ht]{||c|c|c|.|.|c|c|c|c|c|c|}
\hline
JName          &  BName     & Period  & \multicolumn{1}{c}{DM}          & \multicolumn{1}{|c|}{Flux at}  &  Frequency of  &  N       \\
               &           &         & \multicolumn{1}{c}{}            & \multicolumn{1}{|c|}{1400 MHz} &  observations   &          \\   
               &           & (sec)   & \multicolumn{1}{c}{(pc cm$^{-3}$)} & \multicolumn{1}{|c|}{(mJy)}    &  (MHz)          & (Pulses) \\
\hline
\hline
% J0034-0721       &  B0031-07   & 0.942950 & 11.3     & 52.0  & 325, 610,   & 12000   \\
% 	         &             &          &          &       & 1380        &         \\   
J0814+7429       &  B0809+74   & 1.292241 & 6.1      & 10.0  & 325, 610,   & 13000  \\
	         &             &          &          &       & 1380, 4850  &        \\   
J0820$-$1350     &  B0818$-$13 & 1.238130 & 40.9     & 7.0   & 610	     & 3400    \\
J0837$-$4135 	 &  B0835$-$41 & 0.751624 & 147.2    & 16.0  & 610  	     & 3400  \\
J1115+5030   	 &  B1112+50   & 1.656439 & 9.2      & 3.0   & 325 	     & 2700  \\
J1639$-$4359$^\dag$ &    $-$      & 0.587559 & 258.9    & 0.92  & 610 	     & 13000 \\
J1701$-$3726$^\dag$ &    $-$      & 2.454609 & 303.4    & 2.9   & 610 	     & 2500  \\
J1715$-$4034$^\dag$ &    $-$      & 2.072153 & 254.0    & 1.60  & 610 	     & 1600  \\
J1725$-$4043$^\dag$ &    $-$      & 1.465071 & 203.0    & 0.34  & 610 	     & 2500  \\
J1738$-$2330$^\dag$ &    $-$      & 1.978847 & 99.3     & 0.48  & 325, 610     & 8000  \\
J1752+2359   	 &    $-$      & 0.409051 & 36.0     & 3.50  & 610	     & 52000 \\	
J1901+0413$^\dag$ &    $-$      & 2.663080 & 352.0    & 1.10  & 610  	     & 2600  \\
J2022+2854   	 &  B2020+28   & 0.343402 & 24.6     & 38.0  & 610	     & 8000  \\
J2022+5154   	 &  B2021+51   & 0.529196 & 22.6     & 27.0  & 610	     & 1400  \\
J2037+1942   	 &  B2034+19   & 2.074377 & 36.0     & $-$   & 610	     & 1600  \\
J2113+4644   	 &  B2111+46   & 1.014685 & 141.3    & 19.0  & 325,610,    & 6200   \\
	         &             &          &          &       & 1380,4850   &        \\
J2321+6024   	 &  B2319+60   & 2.256488 & 94.6     & 12.0  & 325, 610,   & 10000  \\
		 &             &          &          &       & 1380, 4850  &        \\  
\hline
\end{tabular}
\caption[Observed list of pulsars with their basic parameters]{Observed list of pulsars with their basic parameters. 
The columns give, Pulsar name, Period of the pulsar, Dispersion measure, 
Flux of the pulsar reported at 1400 MHz, Frequencies of observations, 
and number of observed pulses. Pulsars marked with a dagger are the newly discovered 
pulsars from the Parkes multibeam pulsar survey.}
\label{shedual}
\label{table_list_of_pulsars}
\end{center}
\end{table}

\section{Selection of observing frequencies}
Observations of the recently discovered pulsars, reported in this thesis work, were 
carried out at 610 MHz with the GMRT. However, observations of many known nulling pulsars 
were carried out at various other frequencies, as listed in Table \ref{table_list_of_pulsars}. 
The GMRT is capable of operating at six different frequencies. The choice of the observing 
frequency depends upon many factors. However, the most important factor 
that drives the selection of observing frequency is the 
source's flux density at that particular frequency. Pulsars are intrinsically 
stronger at lower frequencies. Dependence of the pulsar flux density with the observing 
frequency can be expressed as,
\begin{equation}
S_{ \nu } ~ \propto ~ \nu ^ {- \alpha}. 
\end{equation}
Here, $\alpha$ is the spectral index and S$_\nu$ is the flux density at an observing frequency $\nu$.
Most of the pulsars exhibit negative spectral indices, which makes them 
stronger in flux as one goes to lower and lower radio observing frequencies. 

Only a few pulsars were observed at 325 MHz while rest of the newly discovered pulsars were observed 
at 610 MHz. At frequencies around 325 MHz, the Galactic noise, due to the emission 
from the interaction of the charge particles with the Galactic magnetic fields, 
will dominate the background noise. It would also become much more difficult to phase equalise the antennas 
with larger baselines for longer duration at lower frequencies (see Section \ref{antenna_selection}). Many of the newly discovered 
pulsars have high DM and situated near to the Galactic centre. These pulsars suffer large scattering, 
which reduces their single pulse sensitivity at 325 MHz (as discussed in Section \ref{pulse_scattering_sect} 
and shown in Figure \ref{scattering_picture}). Hence, relatively lower frequency of 
610 MHz was used for most the observations of the newly discovered pulsars to overcome 
these effects as well as to take advantage of the negative spectral indices. 

\section{Study of newly discovered pulsars}
\label{new_pulsar_sect}
The GMRT was used to study some of the recently discovered nulling pulsars in 
this thesis work as it has an advantage of larger 
collecting area along with a wider low frequency coverage. 
In the last few years, several new pulsars have been discovered. 
For example, the Parkes Multibeam Survey (PKSMB) has 
discovered around 800 pulsars in the last decade 
\cite[]{mlc+01,mhl+02,kbm+03,hfs+04,fsk+04,lfl+06}. 
Analysis on the PKSMB data with improved techniques still promises to 
enhance this number. As mentioned in Section \ref{extreme_null_sect}, 
RRATs are one of such example, which were discovered in the reprocessing of the PKSMB 
data \cite[]{mll+06}. Among all of these recently discovered pulsars, 
several of them show interesting single pulse intensity variations. However, many 
of these have not been systematically studied to investigate pulsar nulling. 
Thus, to investigate this further, discovery plots of around 600 pulsars were 
studied carefully in order to find indication of nulling in them.  
Out of this sample, 93 seem to exhibit nulling like behaviour.

The GMRT can only observe pulsars which are located  
higher than $-50^\circ$ in the declination. As all these pulsars were discovered by 
the Parkes telescope, some of them are located deep in the southern sky. Out of 93 selected 
candidates, 13 are located outside the GMRT observable sky range. 
The remaining 80 pulsars were categorised 
according to the pulse intensity variations seen in the discovery plots. 
In some of these pulsars, degree of pulse intensity variations clearly 
suggest nulling. In other cases, it is difficult to make clear judgement about the possible 
nulling because of low S/N. Third criteria for selection of suitable observable candidates, 
is based on the choice of an appropriate observing frequency. 

To estimate, whether a particular pulsar is above the sensitivity limit 
of the GMRT, it is essential to know its flux densities at 325 and 610 MHz. 
Most of the known pulsars exhibit spectral index of around -1.8$\pm$0.2 \cite[]{mkk+00}. 
PKSMB pulsars were discovered at 1420 MHz so for most of our candidates, 
flux densities were reported only at 1420 MHz. A few of these pulsars, among the 
selected candidates, have reported flux densities at 400 MHz. For these pulsars, the 
known spectral indices were used to calculate the flux densities at 325 and 610 MHz. 
While for the remaining pulsars, flux densities at 325 and 610 MHz were 
calculated using their 1420 MHz flux density with an assumed common spectral index of $-$1.8.
Only those pulsars were selected for the final observations, 
which showed reasonably high expected single pulse sensitivity (i.e. expected S/N $>$ 5).

\section{Radio Telescopes and Observations}
\subsection{The GMRT}
\label{gmrt_observations}
% \begin{wrapfigure}{r}{0.5\textwidth}
\begin{figure}[h]
 \begin{center}
 \includegraphics[width=0.68\textwidth,bb=0 0 459 480]{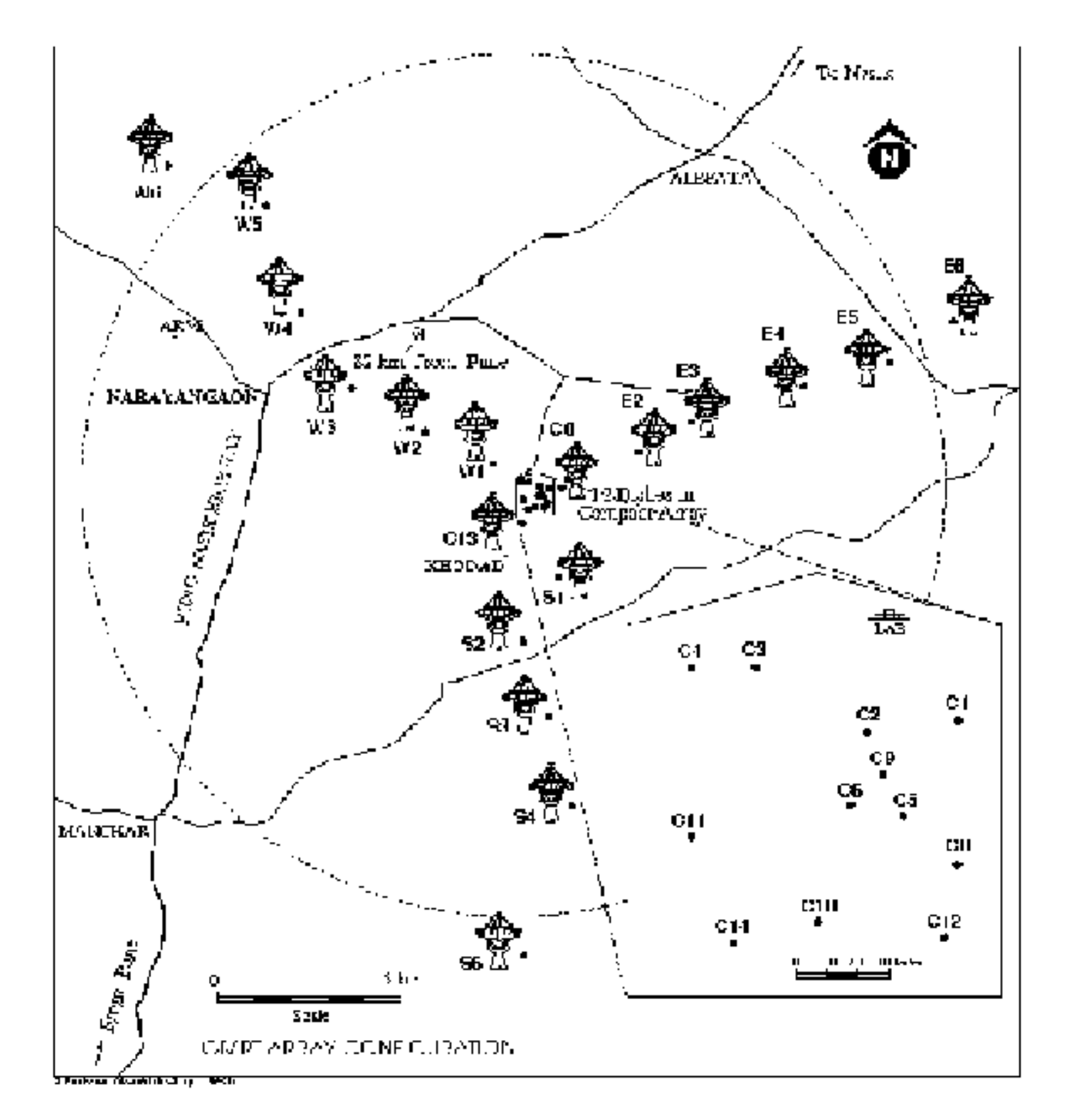}
  \end{center} 
\caption[The GMRT array of 30 dish antennas]{The GMRT array of 30 dish antennas, located at +74$\deg$03$\textquoteright$E and 
+19$\deg$06$\textquoteright$N at Khodad, India (Courtesy : Prem Kumar)}
\label{gmrt}
\end{figure}
% \end{wrapfigure}
The observations with the GMRT were carried out during the cycle 15, 16, 18 and 19 under the project 
name of 15VRG01, 16$\_$048, 18$\_$036, 18$\_$071 and 19$\_$022 between 2008 to 2011. 

The individual dishes of the GMRT are 45-meter in diameter. 
Their reflecting surface forms a paraboloid which allows the parallel 
waves to converge to one common point. If the parallel waves fall on the dish, then the reflected 
waves will still remain in phase at the focal point of the dish which is also known as the \emph{feed}. 
There are five dipole/cylindrical feeds, corresponding to the central frequencies of
150, 233, 327, 610 and 1420 MHz, 
mounted on the four faces of a rotating turret at the focal point.
The feed detects both the approaching polarisation components 
from the sky. The antenna has two axes of rotation to point towards any direction 
in the sky through its azimuth and elevation electric motors. 
The elevation axis has a limit of 17.5$^\circ$ from the horizon. 
The reflecting surface of the dish is a rectangular mesh which 
allows the antenna to operate in comparatively higher wind speeds. 
It also reduces the load on the electric motors which cuts down 
the construction cost by a large margin. Due to this design of 
the reflecting surface, the maximum operating frequency with this dish is around 3 GHz. 
The size of the mesh is also not uniform. The inner one third of the dish 
has a mesh with the size of 10 mm $\times$ 10 mm, the middle one third 
has the size of 15 mm $\times$ 15 mm and the outer one third has the size of 20 mm $\times$ 20 mm. 
The available effective area at 1420 MHz is smaller related to 
the other frequencies due to a comparatively bigger size of the mesh in the outer area, 
which does not act as an efficient reflecting surface at 1420 MHz.  

The inner part of the array is called a central square (CQ). 
The CQ is 1 km$\times$1 km in size and contains 14 antennas. 
The purpose of these antennas is to provide short baselines 
specially for wide field imaging and pulsar observations \cite[]{sak+91}. 
Rest of the 16 antennas are distributed in a $\textquoteleft$Y$\textquoteright$ shaped 
configuration as shown in Figure \ref{gmrt}. The three arms according to their positions from the CQ 
are called the eastern, the western and the southern arms. The eastern and the southern arms 
contain five antennas each while the western arm contains six antennas. 
The longest baseline possible with the GMRT is 25 km long. 
The antennas in the arms make the GMRT a very important instrument for high resolution imaging 
by an unique technique called the \emph{earth rotation aperture synthesis}. 
However, this will not be discussed in this thesis as non of 
the observations used the imaging capabilities of the instrument. 
To study the single pulse behaviour of the nulling pulsars, we used 
the GMRT in a single dish mode which will be discussed in the following 
sections. The GMRT provided very valuable and high sensitivity observations to 
investigate single pulse behaviour with its large collecting area and 
good frequency coverage. 

\subsubsection{Selection of antennas}
\label{antenna_selection}
%\paragraph{}
The GMRT was used in a phased array mode for all the observations 
reported in this thesis. To construct a phase array beam, 
the instrumental phase differences between the array antennas need to be compensated. 
This procedure, called the \emph{phasing}, uses a strong point source to estimate phases of the 
array antennas with respect to a reference antenna. The instantaneous phase of the 
astronomical signal received by a given antenna depends on the signal's path 
through the ionosphere which can exhibit turbulence due to various external factors. 
Thus, antennas which are widely separated (i.e. antenna pair with a long baseline), 
receive signals from completely different ionospheric zones above them. 
At lower frequencies, the ionospheric changes are comparatively rapid for long baselines, 
hence it is not possible to phase equalise such antennas for more than two hours.
The observations at 325 and 610 MHz were carried out with only 
short baseline antennas (CQ + 6 antennas) to keep the respective 
phase fluctuations minimum during a single stretch of two hours long observations. 
Prior to any schedule observations, all fourteen CQ and first two antennas of each 
arm were selected (i.e. total 20 antennas). 
However, during the observations, only 16 to 17 of these antennas 
could be used due to the following reasons concerning their performance. 

To conduct the observations at a particular frequency, 
the feed for the required frequency was rotated and pointed towards the dish. 
For all the selected antennas, a solar attenuator, situated 
inside the feed, was set to 14 dB in order to get rid of the 
excess Galactic background at 325 MHz. 
Before the observations of the targeted source, the pointing offset experiment was carried out 
to estimate the sensitive and the beam pattern for each antenna. 
During the pointing offset experiment, a strong point 
source was selected (either Cassiopeia A or Cygnus A). All selected antenna beams were 
scanned across the source in order to get the beam pattern. The beam pattern 
of each antenna was carefully examined to identify any anomaly. The corresponding 
offset in the pointing for individual antenna was corrected manually. For any known 
strong point source, there are corresponding counts an antenna should provide as an 
output from the calibrated back-end. Hence, the sensitivity of the antenna can be determined 
from the corresponding counts it displayed on the observed strong point source.  
Those antennas which showed degraded sensitivity were excluded before the observations. 
At the start of the observations, it is essential to equalise the default output power 
from all the antennas. Those antennas which even after a manual adjustment 
failed to reach the required power level were also excluded before the observations. 
After carrying out power equalisation, all selected antennas were also 
phase equalised to give minimum phase difference with a reference antenna. 
Only those antennas were selected which showed phase differences 
within acceptable limits ($\pm$ 20$\deg$) across the entire frequency band. 
Phasing is however not possible for a few of the selected 
antennas, as occasionally the true instrumental power is masked 
by the radio frequency interference (RFI). Those antennas, 
for which phasing was not successful, were not included. 
Band shapes of all the antennas were also observed carefully in order to find any significant 
RFI lines in any frequency channel. Those antennas which showed RFI were excluded 
from the observations. Every exclusion of an antenna from the array will degrade the 
sensitivity of the experiment, hence it is not possible to throw away more than 
3 to 4 antennas. In the final selection, one has to find a balance between the degradation 
of the sensitivity and the antenna quality. 
% \subsection{Baseband and Correlator settings}

For initial studies of the newly discovered pulsars and the a few of the known nulling pulsars, 
hardware backend was used (Cycle 15 and 16). For rest of the observations, 
a newly developed software backend was used. 
First the settings for the hardware backend are described and later  
the settings for the software backend will be discussed. 
The I$^{st}$ local oscillator (LO) needs to be set such that the corresponding 
RF gets converted to an intermediate frequency (IF) of 70 MHz with 32 MHz bandwidth. 
For the observations at 325/610 MHz as an observing frequency, I$^{st}$ LO was set at 255/595 MHz respectively. 
These signals were again modulated to the optical wavelengths to 
transfer them through the optical fiber links to the central electronics 
building. The same procedure was applied to both the polarisation outputs. 

\subsubsection{Hardware backend and settings}
In the central electronics building, the optical wavelengths 
were down-converted to obtain the 70 MHz IF for each polarisation.  
The retrieved signal were again down-converted to baseband 
using a IV$^{th}$ LO of 70 MHz in the baseband system. 
This conversion was carried out using a quadrature mixer splitting the overall 32 MHz band
in two 16 MHz sidebands, namely the Upper side band (USB) and the Lower side band (LSB). 
The baseband also has an Automatic level control (ALC), 
which is usually switched off during the pulsar observations for the following reasons. 
The purpose of the ALC is to keep the signal in an optimum range 
(i.e  above quantization error limit but below saturation limit). 
The ALC flattens the changes in the signal passing through it. 
The pulsar signals on the other hand are train of pulses. 
The ALC introduces a distortion to the pulsar signals because of 
the inherent finite time constant. Hence, it is essential 
to keep the ALC off during the pulsar observations. 
The signals were then orderly processed by the analog-to-digital 
converter (ADC), the delay processing units and the FFT engines. 
The FFT unit was configures to take 512 points FFT, which corresponds to 256 channels across 
the observed bandwidth. The output of the FFT is a time multiplexed 
polarisation data with 256 channels in a time-series. The output was 
then transferred to the corresponding USB or the LSB GMRT Array Combiner (GAC) unit. 

GAC first demultiplexes both the polarisation channels. The phased 
array output is produced by taking a sum of the voltages from each antenna and then 
adding them for the final detection \cite[]{pr97}. An additional selection of 
gain values were made inside the GAC. One has to choose the gain values 
with a certain precision, between 1 to 15 with a 3 dB step, because 
the lower gain value will introduce the quantization errors 
and the higher gain value will saturate the output signals. 
For these observations, gain value number 8 was selected.
The GAC permits the users to select their choice of antenna to be 
added in the phased array beam. After considering various 
issues related to individual antennas (as discussed in section \ref{antenna_selection})
final selections were made in the GAC output. 
Same antennas were selected for both the USB and the LSB. 
%\paragraph{}

The GAC-USB is connected to the Phased Array (PA) backend while the GAC-LSB is connected to 
the Polarimeter pulsar backend (PMTR). Both the backends were configured 
to give an integration time of around 1 msec. The output from these backeds are 
also time multiplexed polarisation data similar to output of the FFT units. 
These backends can be configured to provide the Stoke's parameters or the cross and the self power 
terms. However for these observations, only total intensity modes were used for both the USB and the LSB.
In this configuration, both the polarisations were combined together in the backends and 
the total intensity outputs were produced. The pulsar backends also provide intermediate 
GPS bits, which can be used to get a time stamp for the recorded data. 
The general information regarding the start and the stop times, 
in the Coordinated Universal Time (UTC), 
are recorded in a the separate header file along with the 
recorded data file for the observations of each source. 
The USB and the LSB files were recorded separately and then transferred 
to their respective standard SDLT tape units mounted 
on the backend control computers.  

\subsubsection{Software backend and settings}
In 2009, the GMRT was upgraded with more efficient and easy to handle 
backend called the \emph{GMRT Software Backend (GSB)}. 
The GSB provides instantaneous bandwidth of 33.333 MHz
with a high time resolution upto 61 $\mu$sec \cite[]{rgp+00}. 
The GSB consist of a Linux based cluster of 
48 Intel Xeon nodes, which are connected with 11 GB/s gigabit switches. 
For acquisition, 16 nodes are used with 4 channel input to each of them. 
These input receives two polarisation channels from each antenna. All these nodes are synchronized to 
trigger start observations simultaneously with the observatory's Rubidium clock. 
It also carries out the analog-to-digital conversion with an additional gain amplifier at each input of 
the ADC to digitally adjust the gain of each antenna. Additional 16 nodes are used for 
the central processing pipe-line and the rest of 16 nodes are used as a recording cluster 
for the raw voltage dump mode. The basic operational pipe-line of the GSB is similar to the earlier discussed 
hardware backend. The signals from each antenna are corrected for the path length, the geometric delays and 
the instrumental phase differences. On-line FFTs are calculated for each antenna before 
adding them in the phased array beam-former. The beam data then gets recorded 
in the peripheral nodes depending upon the coherent and the incoherent beam requirements.  

The GAC can be configured either in a 16 MHz with 8 bit quantization or in a 32 MHz with 4 bit quantization.
The later one was used for all of our observations. The sampling time used during most of the observations 
was around 128 $\mu$sec with 512 channels across 33.333 MHz band. The power level of each antenna 
were precisely set to required level using the individual antenna gain values. The phase of each antenna 
was also equalize to match with a reference antenna after the FFT operation and before the beam-former. 
The data were then recorded into a separate node (node 50), which is dedicated for the coherent 
beam data recording with the appropriate time stamp files. The recorded data were transferred to 
the standard LTO tape drives for further processing. 
      
\subsection{The WSRT}  
The WSRT was used for the observations of two nulling pulsars, 
which were simultaneously observed at four different frequencies from three different telescopes. 
The observations were carried out between 5 to 17 February 2011 with eight hours per source during its semester 11A. 

% \begin{figure}[h!]
%  \begin{center}
%     \includegraphics[width=0.68\textwidth]{WSRT.eps}
% %  \includegraphics[bb=0 0 1601 1201]{./Images/WSRT.jpg}
%   \end{center} 
% \caption[The WSRT antennas]{Three of the WSRT antennas. Last two antennas can be seen on the rail tracks.
% The WSRT is situated at 6$\deg$36$\textquoteright$15$\textquotedblright$E and 
% 52$\deg$55$\textquoteright$00$\textquotedblright$N in the Netherlands. }
% \label{wsrt}
% \end{figure}

The WSRT is an aperture synthesis radio telescope \cite[]{bh74} which consists of 14 dish antennas, 
each of size 25-meter in diameter on an equatorial mount. Four of the dish antennas are 
placed on the rail tracks, which can be positioned to obtain a suitable baseline coverage. 
These observations were taken with a short spacing configuration of the array 
in order to keep the ionospheric effects minimum. 
The signals are received by the Multi-Frequency Front-End (MFFE) at each antenna. 
The MFFE is designed to receive signals in any of the various available bands 
between 110 MHz to 9 GHz. The front-end also has crossed dipole to receive the linear polarisation for 
most of the frequency band (few bands provide circular polarisation). Signals are then down converted 
to an IFs and transferred to the centrally located IF amplifiers using the coaxial cables to compensate 
for the cable losses. The IFs have the bandwidth of 160 MHz which is split into eight subbands 
of 20 MHz each for every antenna. For pulsar observations, after compensating for various instrumental and geometric delays, 
the digitized signals are added in a tied array added module (TAAM). 

Both pulsars were observed at a central frequency of 1380 MHz with 160 MHz of total bandwidth using 
an upgraded WSRT Pulsar Machine II (PuMa-II). PuMa-II is a fully digital and a flexible cluster of 
44 nodes, which processes instantaneous observing bandwidth of 160 MHz \cite[]{ksv08}. 
PuMa-II receives the phased array output from the TAAM using the optical fibre links. 
It converts the digital optical signals to the electric signals for the baseband recording of the data. 
A Hydrogen maser provides the necessary time-stamps to the recorded samples 
for an accurate time keeping. The data were recorded on the PuMa-II storage cluster 
with the available high time resolution of around 50 nsec for further processing. 
These observations generated few terabytes of data, which were reduced off-line to the standard 
filter-bank formats. Both the polarisations were also added during the off-line processing to get the total intensity. 
After integrating the high time resolution samples, we acquired an effective integration time 
of around 1 msec in the off-line processing to reduce the data volume.  
The off-line processes at the WSRT generated 8 dedispersed time-series files 
from all the subbands.  These data were transferred to a local facility at the National 
% Center of Radio Astrophysics (NCRA), India, using the standard LTO storage tapes for further processing. 

\subsection{The Effelsberg Radio Telescope} 
% \begin{figure}[h!]
%  \begin{center}
%    \includegraphics[width=0.68\textwidth]{eff.eps}
%   \end{center} 
% \caption[The 100-meter Effelsberg radio telescope]{The 100-meter Effelsberg radio telescope. 
% It is located at 6$\deg$52$\textquoteright$58$\textquotedblright$E 
% and 50$\deg$31$\textquoteright$29$\textquotedblright$N in the Germany (Credit: \emph{www.mpifr-bonn.mpg.de}).}
% \label{eff}
% \end{figure}

The Effelsberg Radio Telescope (ERT) was used for the observations of two nulling pulsars 
which were simultaneously observed at four different frequencies from two other telescopes.
It is among the biggest fully steerable single dish radio telescope in the world. 
The antenna has a Gregorian design to receive the signals through a primary dish of 100-meter and 
a secondary dish of 6.5-meter. The primary dish is a paraboloid and the secondary 
dish is an ellipsoid. The ERT is capable of carrying out observations at 
various frequency bands between 333 MHz to 95 GHz. There are various types 
of receivers which can be divided among the prime and the secondary focus 
receivers. The prime focus receivers are flexible to change according to their usage 
while the secondary receivers are permanently mounted on their positions. 

Our observations were carried out at a central frequency of 4850 MHz with 
500 MHz of observing bandwidth in the total intensity mode. 
We used a 6-cm dual horn secondary focus receiver.  
The detected signals were converted to pulses (voltage-to-frequency conversion) 
to reduce the path losses. They were digitally transferred to the receiver 
room and counted in the digital backend. The signals were converted back to 
the analog signals for control and adjustment purposes in the receiver room
\footnote{\url{http://www3.mpifr-bonn.mpg.de/div/electronic/content/receivers/6cm.html}}. 
We used the PSRFFTS search backend for these observations. This
backend was reconfigured to have 128 frequency channels across 500 MHz 
and a dump time of 64 $\mu$sec. This time resolution is adequate to 
correct for the highest observed dispersion smearing in our observed pulsars 
of $\sim$0.1 msec across the 3.9 MHz wide frequency channels. 
The above configuration of the PSRFFTS generated a net data 
rate of 8 MB/s with the output written out in the 32-bit format.
Appropriate observing start time in the Modified Julian Date (MJD) 
were also recorded in the header part of the data files. 
These data were written into the LTO disk and transferred 
to the local facility at NCRA, India, for the off-line processing. 

\subsection{The Arecibo radio telescope}
\label{Arecibo_sect}
% \begin{figure}[h]
% \centering
%  \includegraphics[width=3 in,height=3 in,angle=0,bb=0 0 468 371]{Arecibo.jpg}
%  % Arecibo.jpg: 1950x1545 pixel, 300dpi, 16.51x13.08 cm, bb=0 0 468 371
%  \caption[The Arecibo radio telescope solid dish]{The Arecibo radio telescope solid dish. 
%  The feed is suspended above the fixed reflector by stretched cables from three supporting legs. 
%  It is located at 66$\deg$45$\textquoteright$10$\textquotedblright$W  
% and 18$\deg$20$\textquoteright$39$\textquotedblright$N in the Puerto Rico, USA.
% (Credit : \emph{www.naic.edu})}
%  \label{Arecibo_dish}
% \end{figure}

The Arecibo radio telescope is a single dish radio telescope,
with a diameter of around 305-meter, fixed in a natural crater. 
It is capable of operating at a wide range of frequencies. 
It was used to observe PSR J1752+2359 for around 30 minute 
at 327 MHz on 2006 February 12. The primary mirror 
of the dish is a spherical reflector. The advantage 
of using the spherical surface it to gain tracking 
capabilities for longer observations. The focus of 
the primary reflector is on a line and hence the line feeds 
are used to receive the reflector signals. However, current 
observations made use of a Gregorian feed designed to 
operate from 312 MHz to 342 MHz. The data were recorded with 
a Wideband Arecibo Pulsar Processors (WAPP1). The voltages 
produced by the receivers connected to the orthogonal 
linearly polarised feeds were 3-level sampled. 
The stokes time-series were obtained from the 
auto-correlations and the cross-correlations of these 
voltages. Fourier transform of these correlations 
was used to synthesise 64 channels across a 25 MHz 
bandwidth with about 1 $\mu$s sampling time. 

The single pulse polarisation data were obtained directly at the 
observatory. The recorded time-series were dedispersed and folded to 
256 bins every period using the rotation period 
obtained from the \polyco\footnote{Polynomial coefficients 
of the period and period derivatives for an accurate 
period prediction \url{(www.jb.man.ac.uk/~pulsar/Resources)}. 
For details, see section \ref{single_pulse_technique}}. 
The single pulse data were then converted to the European pulsar network (EPN) 
format \cite[]{ljs+98}. These single pulse polarisation 
data were sent to the local facility at NCRA, India. 

\subsection{Sensitivity of observations}
This section briefly describes the justifications behind the usage 
of the phased array during the observations with the GMRT and the WSRT. 
A comparison of single pulse sensitivity of the observations 
carried out by various telescopes in this thesis work is also 
presented in this section. 

For a single dish, the minimum detectable pulsar signal, 
with a period (P) and a pulse width (W), can be expressed as, 
\begin{equation}
S_{min} ~ = ~ \frac{KT_{sys}}{G \sqrt{{n_p} {\bigtriangleup\nu} {\tau}}}\sqrt{{W}\over{P-W}}. 
\label{smin_sd}
\end{equation}
Where, K is a sensitivity constant, n$_p$ is the number of polarisation, 
$\bigtriangleup \nu$ is the observed bandwidth, $T_{sys}$ is the system 
temprature and $\tau$ in the effective integration time. 
Here the detected signal with an unity S/N is assumed. 
The antenna gain, G, can also be expressed as,
\begin{equation}
 G~=~\dfrac{A_{e}}{2k}. 
\end{equation}
Here A$_e$ is an effective collecting area 
and $k$ is the Boltzmann constant. Hence, the minimum detectable 
signal, by substituting the value of G, can be expressed as, 
\begin{equation}
S_{min} ~ = ~ \frac{2k}{A_{e}} \frac{KT_{sys}}{\sqrt{{n_p} {\bigtriangleup\nu} {\tau}}}\sqrt{{W}\over{P-W}}. 
\label{Smin}
\end{equation}

Above equation expresses the fact that, a bigger antenna with a wider bandwidth 
can provide better sensitivity to detect weak signals. 
The ERT is a single dish antenna with a diameter of around 100-meter. 
The minimum detectable signal of 0.04 mJy, at the observed frequency of 
4850 MHz, can be obtained after appropriate substitutions in equation \ref{Smin} with,  
K = 1, G = 2.8$\times$10$^{26}$ K/Jy, n$_p$ = 2, $\bigtriangleup\nu$ = 500 MHz
and T$_{sys}$ = 9 K\footnote{www3.mpifr-bonn.mpg.de/div/electronic/content/receivers/6cm.html}. 
To obtain the single pulse sensitivity, we used the effective integration time 
($\tau$) similar to the assumed pulsar period of 1 sec with 10\% duty cycle (W = 0.1 sec).

However, the single dish telescopes have a limitation on 
the feasible dish size as it is a mechanical challenge 
to build a fully steerable antenna bigger than $~$100-meter in diameter. 
However, an array of small dish antennas can provide 
necessary collecting area resembling a bigger dish when 
the signals from each of them are combined. There are two way 
in which the signals from different antennas can be combined, 
known as, (1) the Phased Array (PA) and (2) the Incoherent Array (IA). 

With N antennas in the PA mode, where signals from different antennas 
are added in phase, the sensitivity of the observations improves by a factor of 
N compared to an observations from the single dish. 
The minimum detectable flux, with N antennas in the PA mode and both polarisations 
combined (i.e. total intensity), can be expressed as, 
\begin{equation}
S_{min} ~ = ~ \frac{1}{N} \times \frac{KT_{sys}}{G \sqrt{2 {\bigtriangleup\nu} \tau}}\sqrt{{W}\over{P-W}}. 
\label{SminPA}
\end{equation}
In the IA mode, where signals are not added with the necessary phase corrections,  
the sensitivity increases just by $\sqrt{N}$ times. The minimum detectable flux, 
with N antennas in the IA mode, can be expressed as, 
\begin{equation}
S_{min} ~ = ~ \frac{1}{\sqrt{N}} \times \frac{KT_{sys}}{G \sqrt{2 {\bigtriangleup\nu} \tau}}\sqrt{{W}\over{P-W}}.
\label{SminIA}
\end{equation}
From the S$_{min}$ of both the configurations (i.e equation \ref{SminPA} and \ref{SminIA}), 
it can be concluded that, the PA mode is around $\sqrt{N}$ times more sensitive 
than the IA mode. In the PA mode, the size of the final beam 
is inversely proportional to the longest possible baseline in the array.   
While in case of the IA mode, the beam size depends upon the diameter 
of a single dish in the array. Both the configurations have their 
own usefulness depending upon the science goals of the observations. 
The PA mode is useful for targeted observations, where the source positions 
are known with a great accuracy. While in the IA mode, large area of the 
sky can be scanned in a relatively shorter time to search for new pulsars or 
to point towards targets with a large uncertainty in their coordinates. 

For the observations using the GMRT and the WSRT, the PA mode 
was the most suitable choice due its capability in providing higher 
sensitivity to obtain single pulses with high S/N. Table \ref{sensitivity_table} 
lists the minimum detectable flux at different observatories 
with their respective frequencies used during the observations. 
For the GMRT, the minimum detectable flux was obtained after substituting, 
K = 1, G = 3.3$\times$10$^{25}$ K/Jy, $\bigtriangleup \nu$ = 16 MHz, 
$\tau$  = 1 sec, N = 20, T$_{sys}$ = 106 K and 102 K for 325 MHz 
and 610 MHz\footnote{\url{www.gmrt.ncra.tifr.res.in}}respectively, 
into equation \ref{SminPA}. Similarly for the WSRT, the 
minimum detectable flux was obtained after substituting, 
K = 1, G = 8.8$\times$10$^{24}$ K/Jy, $\bigtriangleup \nu$ = 160 MHz, 
$\tau$  = 1 sec, N = 14 and T$_{sys}$ = 27 K\footnote{\url{www.astron.nl}}, 
into equation \ref{SminPA}. For the Arecibo radio telescope, the minimum detectable 
flux was obtained using equation \ref{Smin} with the substitutions  
of, K = 1, G = 25$\times$10$^{26}$ K/Jy, $\bigtriangleup \nu$ = 25 MHz, 
$\tau$  = 1 sec and T$_{sys}$ = 113 K\footnote{\url{www.naic.edu/~astro/RXstatus/327/327greg.shtml}}.
In all four cases, a pulsar of period around 1 sec 
with 10\% duty cycle was assumed. All four observatories 
provided sufficient S/N to detect single pulses with high significance. 
For example, the weakest pulsar in our sample (Table \ref{table_list_of_pulsars}), 
PSR J1725$-$4043, was expected to provide single pulses 
with S/N of around 2 for observations at 610 MHz from the GMRT. 
Similarly, the second weakest pulsar, PSR J1738$-$2330, was expected to provide S/N 
of around 10 for observations at 325 MHz from the GMRT. Moreover, It was possible 
to achieve single pulse S/N of around 25 at 1420 MHz from the WSRT for 
PSR B0809+74, while PSR B2319+60 was expected to provide single pulse S/N of around 26 
at 4850 MHz from the ERT. For these estimations, spectral index of around $-$2, was assumed.

\begin{table}[h!]
\begin{center}
 \begin{tabular}{|c|c|c|c|c|c|}
 \hline
Observatory & Arecibo &\multicolumn{2}{|c|}{GMRT} &  WSRT  & ERT \\
 \hline
Frequency (MHz) & 325  & 325 & 610 & 1380  & 4850 \\
 \hline
S$_{min}$ (mJy) & 0.21 & 0.93 & 0.89 & 0.39 & 0.04 \\
 \hline 
 \end{tabular}
\caption[The minimum detectable flux comparison]{The minimum detectable flux (in units of mJy) for detecting 
single pulses with a unity S/N for all four observatories. The limits are only 
reported for the corresponding frequencies used at the respective 
telescopes.}
\label{sensitivity_table}
\end{center}
\end{table}

\section{Analysis}
This section describes the common analysis procedures followed 
for all our observations in this thesis work. There were different 
techniques deployed to investigate individual science goals, which 
will be discussed in the following chapters. 
However, this section summarises the initial steps in obtaining 
the single pulses from different observatories, estimation of the NF, 
separation of the null and the burst pulses, construction of the 
null-length and the burst-length histograms and fluctuation spectra analysis. 

\subsection{Obtaining single pulses}
Most part of the thesis is based on the statistical analysis 
of the single pulses. Hence, it suffices to highlight the steps 
involved in obtaining these single pulses. 
Numerous C programs were used during the analysis\footnote{partial credit: Dr. B. C. Joshi}. 
Along with that, a publicly available package, SIGPROC\footnote{\url{http://sigproc.sourceforge.net}} 
was also used. The man-made signals are known to be more active at lower 
frequencies. Due to which, analysis of the GMRT 
data required various procedures to remove the RFI at different stages. 
First, the procedure to remove the RFI before acquiring the single pulses,  
from the GMRT raw data, is described.   

\subsubsection{RFI removal from the GMRT raw data}
\begin{figure}[h]
 \centering
 \includegraphics[width=4 in,height=5 in,angle=-90,bb=0 0 697 915]{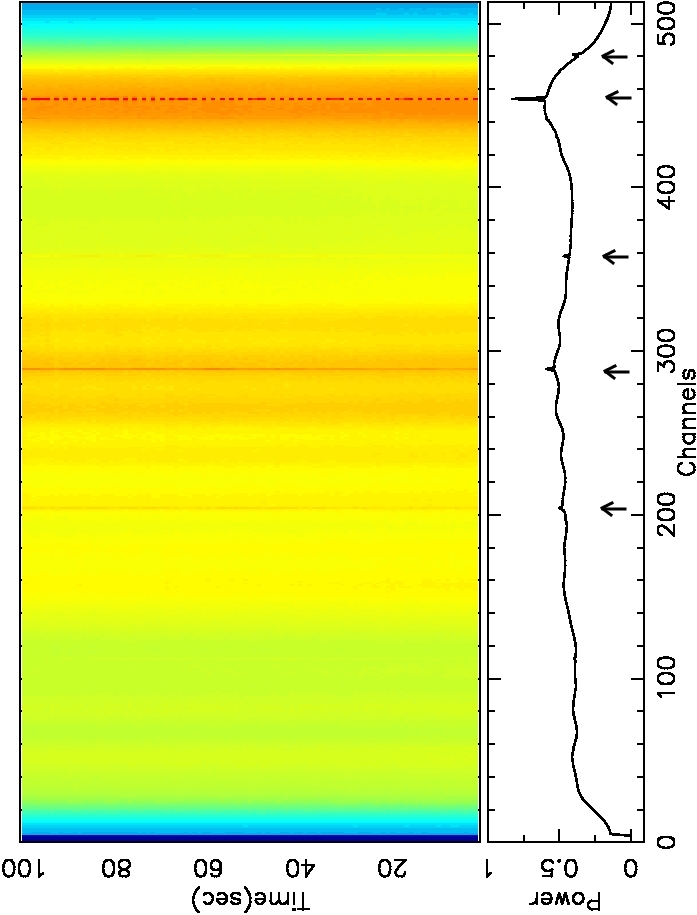}
 % Bandshape_RFI.jpg: 697x915 pixel, 72dpi, 24.59x32.28 cm, bb=0 0 697 915
 \caption[Raw data with the 33.333 MHz bandwidth divided into 512 channels, obtained from the GMRT]
 {Raw data with the 33 MHz bandwidth divided into 512 channels at 325 MHz, obtained from the GMRT.  
 The top panel shows the observed frequency band power in a colour ramp from blue to red.
 The band is shown for around 100 seconds of data. The bottom panel shows 
 the average band obtained after averaging 100 seconds of the observed data. 
 The narrow band RFI lines are clearly visible in both the 
 panels and also tagged with arrows in the bottom panel.}
 \label{GMRT_raw_data_RFI}
\end{figure}

As discussed in section \ref{gmrt_observations}, the GMRT 
data were recorded in the time-multiplexed frequency channels from  
both the backends. The recorded data have bandwidth of 16 MHz or 33.333 MHz 
depending upon the usage of the hardware correlator or the software correlator, respectively 
during the observations. For the 16 MHz band, the data were divided into 256 spectral channels while 
for the 33.333 MHz band, the data were divided into 512 spectral channels. 
The data were taken only from those antennas where much lower level of the RFI 
were noticed before a start of the observations. 
However, it is not possible to exclude large number of antennas due to the 
possible loss in sensitivity. The RFI, which occurs randomly in time, can 
occur for any antenna. It occurs mostly due to the man-made signals picked up by 
the antennas, although it can also be generated in the receiver electronics. 
Hence, a standard way to automatically remove all the RFI can not be adapted. 
In this thesis, mostly a manual identification 
of the RFI is implemented at various stages. 
To remove the RFI affected channels, we obtained the observed frequency band 
from the observations. Figure \ref{GMRT_raw_data_RFI} shows 
an example of 33.333 MHz frequency band from the GMRT observed at 325 MHz. 
The narrow-band RFI, which are concentrated in the individual frequency channels, 
can easily be identified from the average band (as marked by 
the arrows in Figure \ref{GMRT_raw_data_RFI}). These channels 
were tagged and replaced with zeros to create flagged files 
to be used for the remaining part of the analysis.  
Removal of the RFI infected channels reduces the overall sensitivity of 
the observations, however it is useful for a true identification of the nulled 
and the burst pulses. 

\subsubsection{Single pulses from different observatories} 
\label{single_pulse_technique}
The single pulses from different observatories were obtained 
using slightly different techniques. For all of the single 
pulse analysis reported in this thesis, 
the SIGPROC single pulse file format was used. Hence, a conversion is required 
at an initial stage for the different formats of the data from 
different observatories to the selected SIGPROC format. 

The GMRT data were recorded in the 
raw filterbank mode with the header information stored in the separate 
manually written log files. A conversion was required at this stage into 
the standard SIGPROC filterbank format from the raw GMRT filterbank data format. 
For the GMRT data, we used following set of codes to obtain the single pulses. 

%The single pulses were obtained using programs in a publicly available package SIGPROC. 
% It provides number of programs to carry out preliminary steps in 
% pulsar data analysis. The programs which were used in this thesis work 
% are listed below along with a description of their functions. 
\begin{itemize}
\item {\bf Filterbank\footnote{Credit : Dr. B. C. Joshi}} converts the raw GMRT pulsar 
data into a SIGPROC filterbank format. 
The code acquires the necessary details, regarding the observations, by a manual 
entry before the conversion. In the SIGPROC filterbank format, the data are arranged 
in the multiple frequency channels per sample with the observations details in the header. 
\item {\bf Dedisperse\footnoteremember{SIGPROC}{\url{http://sigproc.sourceforge.net}}} combines frequency channels after adjusting 
the dispersion delay between them caused by the interstellar medium. 
This delay was calculated using the DM of the pulsar 
obtained from the ATNF catalogue \cite[]{mhth05}\footnote{\url{http://www.atnf.csiro.au/research/pulsar/psrcat/}}.
The dedisperse time-series consists of samples separated by the effective integration time. 
\item {\bf Fold\footnoterecall{SIGPROC}} is the program which folds the SIGPROC time-series 
data to the given pulsar period. The period can be given manually as one of the input 
arguments or it can be provided as a \polyco\ file. 
The \polyco\ file is generated from a publicly available 
package called \emph{TEMPO}\footnote{\url{http://www.jb.man.ac.uk/~pulsar/Resources/tempo\_usage.txt}}. 
The TEMPO code, calculates the pulse arrival time at a given observatory 
using the known pulsar period and period derivatives and, 
where necessary, one of the several binary models. The \polyco\ file 
is the prediction regarding the pulse phase behaviours in a given 
time range, which are calculated from the input models. The Fold 
program allows the user to choose the output to be produced 
in the single pulse format or in the integrated profile format. 
Fold was also used to bin the time-series in the user defined number of phase bins for 
each single pulses. For most of the pulsars in this thesis, the period 
was divided into 256 phase bins.
\end{itemize}

The WSRT data were obtained as 8 dedisperse time-series 
from the 8 observed subbands for each pulsar. A separate conversion code 
was needed to combine these 8 subbands and form a single time-series in 
the SIGPROC format for the {\bf Fold} program. 
The code {\itshape data${\_}$converter.c}\footnote{I would like to thank Patrick Weltevrede for 
providing me various PuMa off-line analysis packages},
combines these time-series with the respective DM 
to produce a single dedisperse time-series 
in the SIGPROC format with the 160 MHz bandwidth. 
These SIGPROC time-series were folded using the {\bf Fold} 
program, using the \polyco\ files, to obtain the single pulses. 

At the the Effelsberg telescope, the PSRFFTS recorded data in the 
SIGPROC filterbank format only. Hence, these filterbank 
files were directly used to obtain the dedisperse time-series using 
the appropriate DMs. These SIGPROC time-series were also folded using the {\bf Fold}
program, utilising the \polyco\ files, to obtain the single pulses. 

The Arecibo telescope data were converted in the EPN 
format at the observatory. A separate code, {\itshape EPN\_to\_sigproc.c} was 
implemented to convert the observed single pulses with the calibrated polarisations 
into four separate single pulse files corresponding to the four Stoke's parameters, 
I, Q, U and V. 

\subsection{RFI removal from single pulses}
%\paragraph{}
After obtaining the single pulses, NF of the pulsar can be obtained. 
In many occasions, the data show broadband RFI, where the RFI is present for 
short duration in time but over the entire observed band. These RFI mask the true 
intensities of the pulses and hence a judgement on the true nature of the 
pulse, to be either a null or a burst, can not be made. Thus, it is necessary 
to remove these RFI infected single pulses before the NF analysis. 
To carry out this task automatically on the observed single pulse files, 
a code {\itshape rfi\_singlepulse.c} was implemented. 

\begin{figure}[h!]
\centering
\includegraphics[scale=0.5, angle=-90,bb=0 0 492 658]{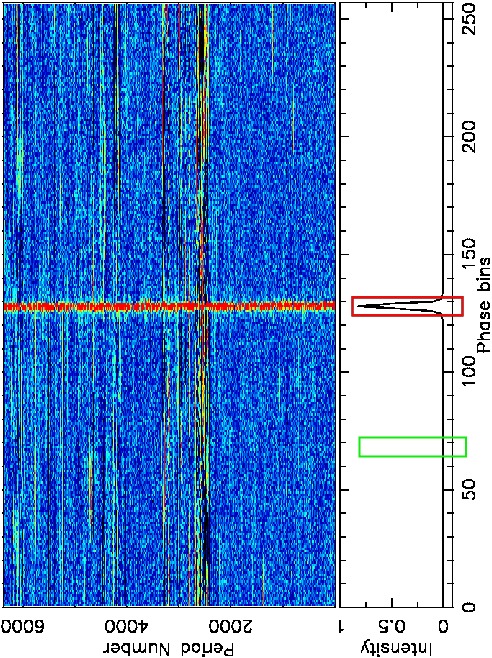}
\caption[Single pulses from PSR B0835$-$41]{Single pulses from PSR B0835$-$41, observed from the GMRT at 610 MHz. The top 
panel shows stack of single pulses in a color ramp intensity from blue to red. The bottom panel 
show the integrated profile, obtained from all the observed pulses. The red square shows the on-pulse 
window while the green square displays the off-pulse window.}
\label{RFI_example}
\end{figure}

\begin{figure}[h!]
 \centering
 \includegraphics[width=3 in,height=5 in,angle=-90,bb=14 14 609 857]{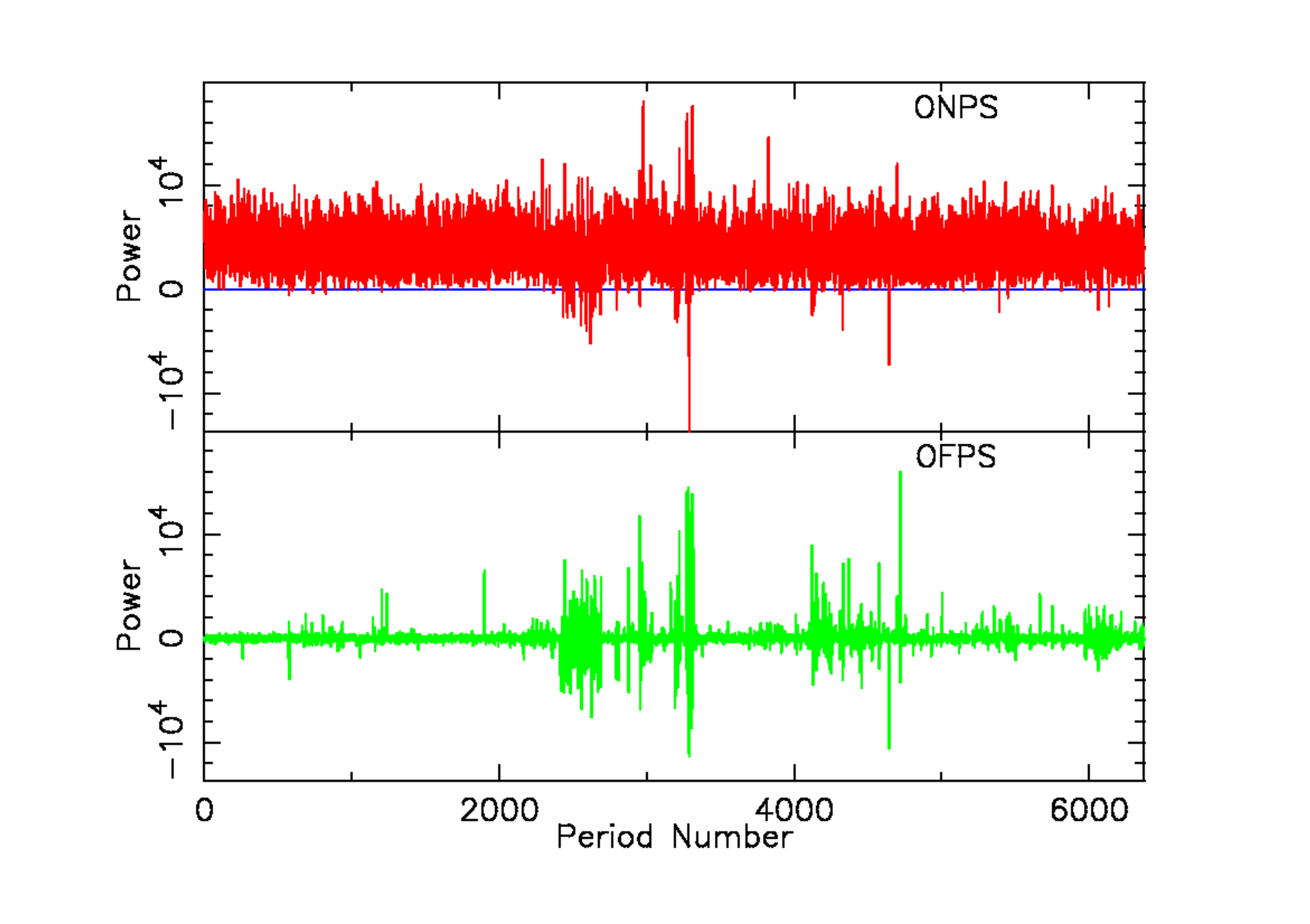}
 % pulseenseq.eps: 0x0 pixel, 300dpi, 0.00x0.00 cm, bb=52 93 482 646
 \caption[The pulse energy sequence obtained from PSR B0835$-$41]{The pulse energy sequence 
 obtained from PSR B0835$-$41. The top panel shows 
 the on-pulse energy sequence while the bottom panel shows the off-pulse energy sequence.}
 \label{pulse_energy_sequence}
\end{figure}

Figure \ref{RFI_example} shows the single pulses obtained from PSR B0835$-$41 
observed from the GMRT at 610 MHz. To identify RFI infected pulses, we selected 
two windows on the phase bins. The on-pulse window (shown with a red square 
in Figure \ref{RFI_example}) bounds the entire pulse region. All the phase bins 
inside this windows were averaged to obtain the on-pulse energy sequence (ONPS). 
Similarly, equal number of phase bins were selected, away from the 
main pulse, to form an off-pulse window. All the phase bins, inside the 
off-pulse window, were added to form the off-pulse energy sequence (OFPS). 
A baseline was selected (from the remaining phase bins) to obtain the mean 
counts in the entire off-pulse region, and subtracted for every period. 
The off-pulse is a region where one does not expect any emission. Hence, 
for the periods, where a baseline has been subtracted, the OFPS should show 
distribution around zero mean energy. Figure \ref{pulse_energy_sequence} 
shows the ONPS and OFPS for PSR B0835$-$41. The presence of RFI 
can be seen in the OFPS, for periods around 2500, where the OFPS 
fluctuations are larger. To identify these periods, a model root-mean-square 
(rms) deviation of the off-pulse mean energy was calculated 
from the periods where such large fluctuations were not present
(for example periods between 1500 to 1600 in Figure \ref{pulse_energy_sequence}). 
A threshold was adapted as 2 to 3 times the model rms to identify all 
the pulses which showed mean off-pulse energy beyond this range. 
All the pulses which were above such threshold were removed from 
the single pulse data. After removal of these pulses, 
the remaining pulses were used for the NF analysis. 
\begin{figure}[h!]
 \centering
  \begin{center}
    \includegraphics[scale=0.5, angle=-90,bb=0 0 595 843]{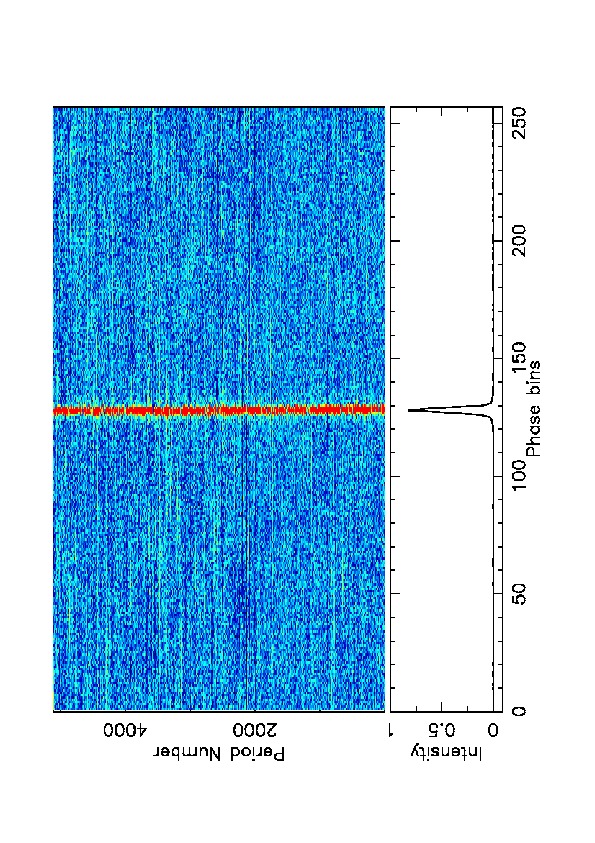}
 % RFI_remov.spdisplay.new.jpg: 595x843 pixel, 72dpi, 20.99x29.74 cm, bb=0 0 595 843
 % RFI_remov.spdisplay.new.eps: 490x657 pixel, 72dpi, 17.29x23.18 cm, bb=0 0 490 657
 \end{center}
 \caption[Single pulses from PSR B0835$-$41 after RFI removal]{Single pulses from 
 PSR B0835$-$41 at 610 MHz similar to Figure \ref{RFI_example}. 
 All the identified RFI pulses, which showed large fluctuations of 
 the mean off-pulse energy, were tagged and removed.}
 \label{RFI_example2}
\end{figure}

For most of the observations, the threshold technique was useful to remove the RFI pulses, 
however for certain observations, the occurrence of RFI were short and did not 
overlap with the off-pulse window. Hence, it was not possible to remove them using 
the above mentioned strict threshold. We carried out a visual inspection 
of the single pulse data to identify such pulses and they 
were removed manually for a few pulsars. 
Figure \ref{RFI_example2} displays a stack of 
single pulses from PSR B0835$-$41 after the removal of 
all RFI infected pulses. The drawback in removing 
the RFI infected pulses is the break in continuation 
of the observed pulse sequence. Hence, these data 
sets were not used to obtain the null length and the burst length statistics. 

\subsection{Nulling fraction estimation} 
\label{NF_tech_sect}
\begin{figure}[h!]
 \centering
  \begin{center}
  \includegraphics[width=5 in,height=2 in,angle=-180,bb=-30 0 499 232]{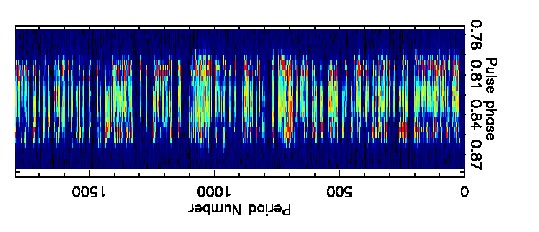}
% J2321+6024.spdisplay.ps: 485x199 pixel, 72dpi, 17.11x7.02 cm, bb=0 0 485 199
 \end{center}
 \caption[A small section of the single pulses observed from PSR B2319+60]{A small section of the 
 single pulses observed from PSR B2319+60 at 610 MHz from the GMRT. The intensities are shown 
 in a color ramp from blue to red. Presence of null pulses can be seen between the burst pulses.}
 \label{spdisplay_b2319}
\end{figure}

\begin{figure}[h!]
 \begin{center}
 \includegraphics[width=3 in,height=5 in,angle=-90,bb=0 80 582 806]{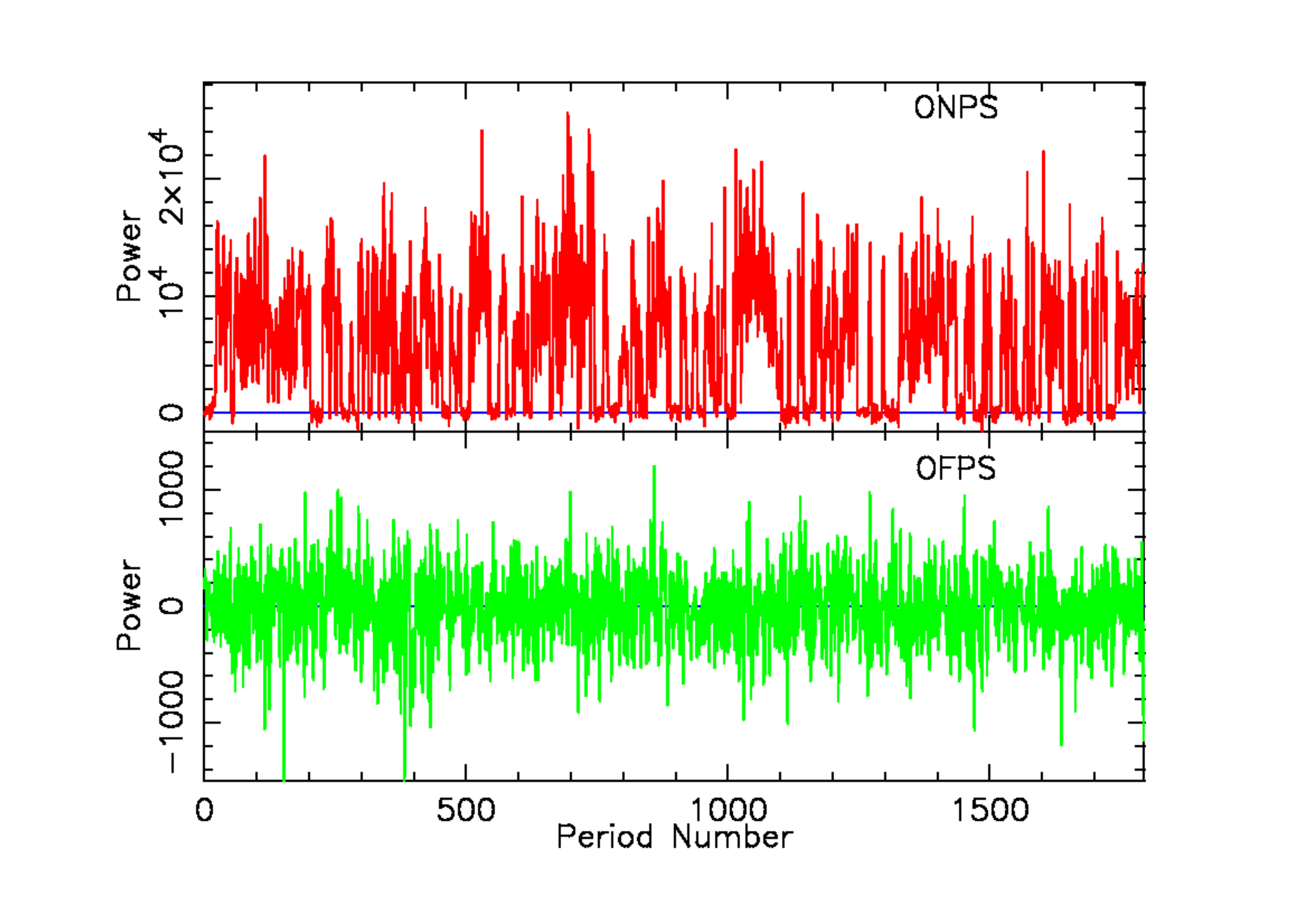}
 \caption[Pulse energy sequence for PSR B2319+60]{The top panel shows the ONPS while the bottom panel displays the OFPS obtained 
 from PSR B2319+60 observed from the GMRT at 610 MHz. Both the panels show power 
 (in arbitrary units) as a function of period number. The ONPS shows pulses reaching the zero pulse 
 energy, which can be classified as null pulses.}
 \label{on_off_energy}
 \end{center}
\end{figure}

In this section, the method to estimate the NF is explained 
in detail. It is similar to the methods used by \cite{rit76} and \cite{viv95} [as mentioned 
in Section \ref{chronical_obs_sect}]. The single pulses observed from the PSR B2319+60 at 610 MHz from the GMRT is shown 
in Figure \ref{spdisplay_b2319}. The on-pulse and the off-pulse windows were selected 
from the integrated profile. The total energy in the on-pulse window and the off-pulse 
window were calculated for each pulse to form the ONPS and the OFPS. 
Figure \ref{on_off_energy} presents the obtained ONPS and the OFPS 
from PSR B2319+60. The presence of nulling can be seen for pulses near 
the zero pulse energy. 

\begin{figure}[h]
 \centering
 \includegraphics[width=7cm,height=10cm,angle=-90,bb=50 50 554 770]{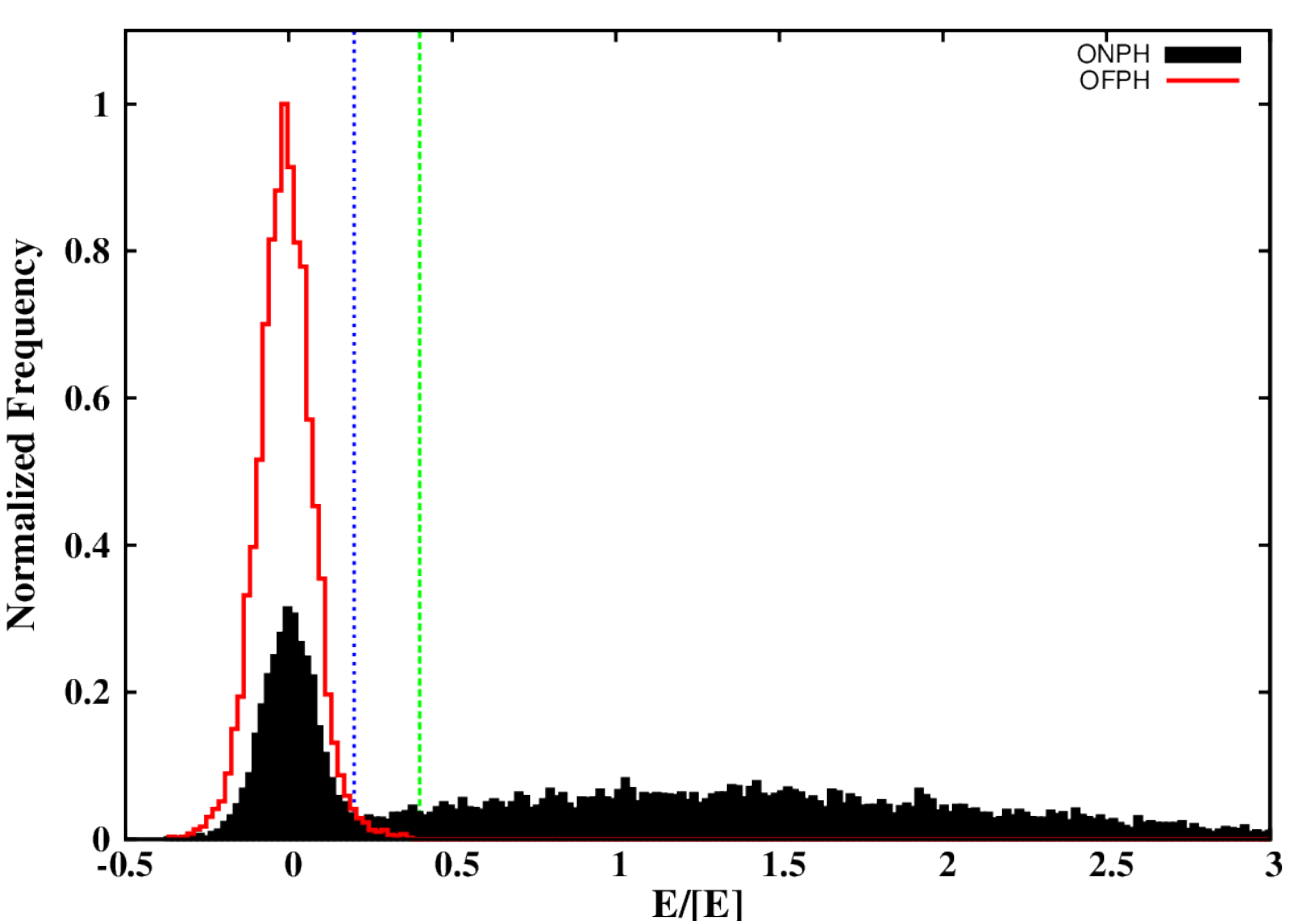}
% histogram_for_thesis.eps: 0x0 pixel, 300dpi, 0.00x0.00 cm, bb=50 50 554 770
 \caption[Normalized pulse energy histograms for PSR B2319+60]{Normalized histograms of the on-pulse and the off-pulse energies as a function of 
 mean pulse energy, [E], for PSR B2319+60. The blue dotted vertical line is a visually selected threshold at the 
 point where null and burst pulse distribution crosses each other (around 0.2 E/[E]) 
 in on-pulse histogram. The green dotted line is a relatively higher threshold (around 0.4 E/[E]) set 
 around the point where the off-pulse energy distribution terminates (see text).}
 \label{hist_comb}
\end{figure}

In a few pulsars, the interstellar scintillation can cause the on-pulse 
energy to fluctuate on a larger scale. The effects of such large 
fluctuations can be addressed by averaging pulse energy 
in blocks of around 200 pulses. This technique is introduced briefly 
in Section \ref{scintillation_sect}. Every pulse in the ONPS and the OFPS 
were normalised by the average energy estimated for the corresponding block of 200 pulses. 
The ONPS and the OFPS were binned to various energy bins depending 
upon the available S/N. The histograms of the on-pulse and the off-pulse energy, 
after a normalisation by the highest number of occurrence, 
are shown in Figure \ref{hist_comb} for PSR B2319+60 (The highest 
number of occurrence count is always at the zero mean energy bin 
of the off-pulse energy histogram for a nulling pulsar). 
It should be noted that, histograms with these normalisation 
do not give probability distribution, however, choice of such 
normalisation makes it easier to discern the NF from the histograms. 

To obtain the NF for each pulsar, modelling of the 
on-pulse energy histogram (ONPH) and the off-pulse 
energy histogram (OFPH) were carried out (using the {\itshape NF.c} code). 
The OFPH represents the telescope noise (a Gaussian random noise), 
the width of which can be used to estimate the rms fluctuations of the data. 
During the period, when the pulsar goes into a null state, 
the on-pulse energy also represents the telescope noise 
similar to the off-pulse energy. The height of the zero centred 
energy in the ONPH provides the fraction of the pulses in the null state. 
If nulling is very prominent, then the ONPH would be indistinguishable 
from the OFPH due to the large fraction of the null pulses. 
Pulsar with a steady power output, tends to show a bimodal 
distribution in the ONPH with two peaks, one at the zero energy bin and 
other at the mean pulse energy bin. Figure \ref{hist_comb} 
shows such bimodal distribution with two peaks in the ONPH for the PSR B2319+60. 
Both the histograms were normalised by the maximum number of occurrence,  
at the zero mean energy bin, from the OFPH. 
As mentioned earlier, such normalisation was adapted 
to discern the NF directly from the plot. The height 
of the ONPH, near the zero mean energy bin, represents the fraction 
of pulses in the ONPH which are similar to the OFPH. 
This fraction represents the observed nulled pulses in the data, 
also known as the NF. A scaled version of the OFPH was 
modelled for the ONPH to estimate this fraction. 
The ONPH is contaminated with noise which has distribution
similar to OFPH distribution. Hence, before subtraction, OFPH has to be 
deconvolved from the ONPH. However, \cite{rit76} has reported that 
such deconvolution does not provide significantly different results. Hence, no such 
deconvolution was carried out in the analysis of our data. 
Before the subtraction, the OFPH was modelled as a zero centred Gaussian 
function, also represented as, 
\begin{equation}
G(x)~ =~ A \times e^{-(x/C)^2}. 
\label{Gauss}
\end{equation}
Here, A is the height of the histogram at zero mean energy, in the 
units of fraction of pulses, which can be identified as A$_{OFPH}$.
The rms noise of the system can be estimated from the width of the 
fitted Gaussian (i.e. C = $\sqrt{2}\times rms$). 
In the ONPH, the zero energy excess represents the nulled pulses. 
The height of which can be estimated by fitting the 
similar Gaussian function given by equation \ref{Gauss}. 
Fitting this function on the ONPH provided the a width of the 
Gaussian function (i.e. C) obtained from OFPH. This parameter 
was kept fixed in the later fitting process on the ONPH. 
Thus, only height was used as a free parameter for fitting 
the ONPH around the zero mean energy bin. This fitting gave an 
estimation of the A$_{ONPH}$. The ratio of the parameters, obtained from the ONPH  
(A$_{ONPH}$) to the one obtained from the OFPH  (A$_{OFPH}$), states the NF of a pulsar as, 
\begin{equation}
NF ~ = ~ \frac{A_{ONPH}}{A_{OFPH}} ~ \times 100\%. 
\label{NF}
\end{equation}
The fitted parameters A$_{ONPH}$ and A$_{OFPH}$ will have 
the corresponding fitting errors on them. 
The error on the NF was obtained using these fitting errors as, 
\begin{equation}
 \bigtriangleup{NF} =  \frac{A_{OFPH}\times (\bigtriangleup{A_{ONPH}}) - A_{ONPH}\times (\bigtriangleup{A_{OFPH}})}{(A_{OFPH}^2)}~ \times 100\%. 
\end{equation}
Here, $\bigtriangleup$NF is the error on the NF while $\bigtriangleup{A_{ONPH}}$ and $\bigtriangleup{A_{OFPH}}$ are 
the fitting errors on the $A_{ONPH}$ and $A_{OFPH}$, respectively. For a few pulsars, the null pulse 
distribution and the burst pulse distribution were not well separated in the ONPH. Hence, fitting the 
Gaussian on these histograms can give large fitting errors.
For these pulsars, where the null and the burst pulse distributions are indistinguishable in the ONPH, 
the estimation of the NF using the above mentioned method leads to an overestimation. 
This happens due to the low S/N burst pulses mixing  
with the null pulse distribution in the ONPH. For pulsars 
with low S/N burst pulses and relatively good amount of null pulses, 
consecutive pulses (around 5 to 10 pulses depending 
upon the single pulse S/N) can be sub-integrated to 
get the bimodal distribution in the ONPH. Estimation of the NF from the sub-integrated 
pulses leads to a underestimation, as it is likely that sub-integration will 
combine few null pulses with the neighbouring burst pulses. Hence, the above mentioned 
methods has to be used with precautions on the individual cases. 

\subsection{Separation of null and burst pulses}
\label{separation_of_null_burst_sect}
The null and the burst pulses need to be separated in order to carry out 
further analysis. There are different ways in which these pulses 
can be separated. We have at large used two different methods 
in various different cases depending upon the available S/N. A simplest method is 
to apply a visual threshold on the ONPH as shown in Figure \ref{hist_comb}. 
This threshold is selected at the location where the null pulse 
and the burst pulse distributions cross each other. Pulses 
below the threshold can be identified as the null pulses, while 
the pulses above the threshold can be tagged as the burst pulses. 
For pulsars with high S/N burst pulses, such threshold can be set very easily. 
However, separation using such threshold can lead to erroneous results for 
the pulsars where the bimodal distributions of ONPH, from the null and the burst 
pulses, are not well separated. For such ONPH, a threshold 
can cause a few null pulses to get mix with the separated burst pulses. 
Similarly, a few low S/N burst pulses can also get mix with the separated null pulses. 
Hence, for such cases we undertook a slightly different approach. 
\begin{figure}[h!]
\begin{center}
 \centering
 \subfigure[]{
 \centering
 \includegraphics[width=2 in,height=5 in ,angle=0,bb=40 0 234 500]{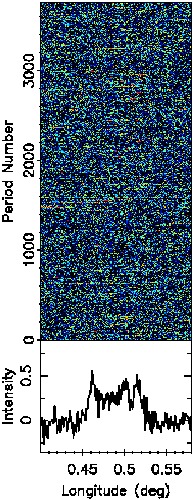}
  \label{nullpulses_after_threshold_raw}
  }
  \subfigure[]{
  \centering
  \includegraphics[width=2 in,height=5 in,angle=0,bb=40 0 250 535]{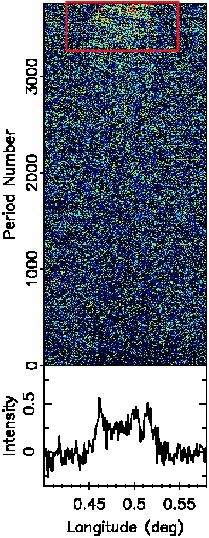}
  % J2321+6024.spdisplay.highthreshold.nullpulses.accending2.eps: 204x553 pixel, 72dpi, 7.20x19.51 cm, bb=0 0 204 553
  \label{nullpulses_after_threshold_accending}
  }
 \caption[Pulses in the acceding order of their pulse energy]{The separated null pulses after applying a slightly 
 higher threshold. The top panels in both the plots show the single pulses while the bottom panels show 
 the integrated profiles. (a) All the separated pulses below the threshold. 
 Note that, it is difficult to identify burst pulses from the single pulses, however 
 a significant integrated profile can be seen, indicating the presence of numerous weak burst pulses. 
 (b) Arranged single pulses in the ascending order of their on-pulse energy. The burst pulses with a 
 relatively higher S/N can be  seen towards the high energy end (inside an indicative red square)}
 \label{nullpulses_after_threshold}
\end{center}
\end{figure}

\begin{figure}[h]
 \centering
 \includegraphics[width=3 in,height=4.5 in,angle=-90,bb=0 0 600 732]{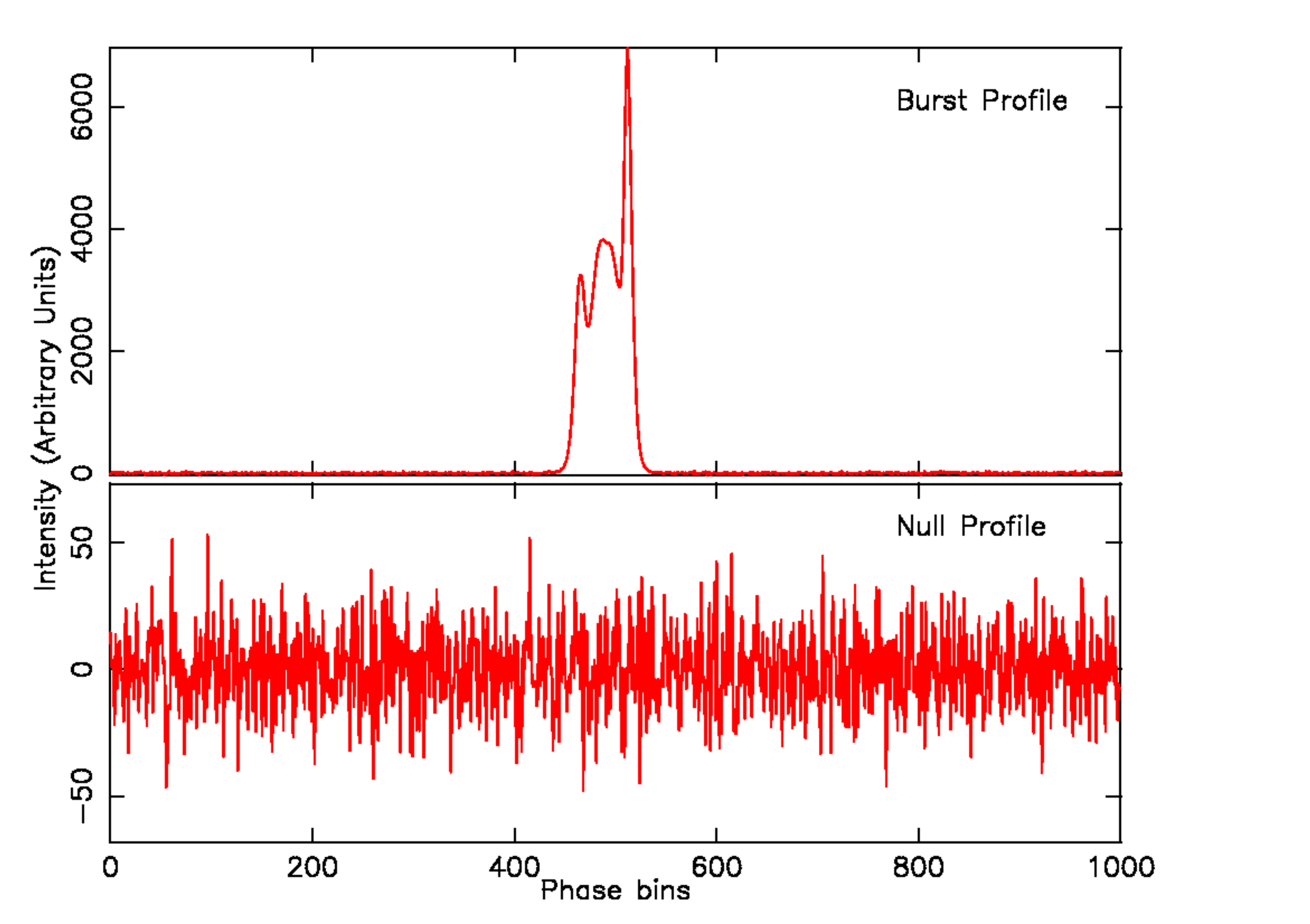}
 % J2321+6024.Null_burst_profile.eps: 537x732 pixel, 72dpi, 18.94x25.82 cm, bb=0 0 537 732
 \caption[The null and the burst profiles for PSR B2319+60]{The null and the burst pulse profiles 
 obtained after final separation of the corresponding pulses 
 for PSR B2319+60 observed from the GMRT at 610 MHz. The null pulse profile does not show any significant 
 emission component.} 
 \label{null_burst_profile}
\end{figure}

A relatively higher threshold was set initially on the ONPH histogram, 
to separated strong burst pulses above the set threshold (around 
0.4 E/[E] in the ONPH of Figure \ref{hist_comb}). 
The set higher threshold obviated any possibility of 
null pulses mixing with the high S/N burst pulses. The pulses 
below such threshold will have a mixture of true null pulses 
and weak burst pulses as shown in Figure \ref{nullpulses_after_threshold}(a).  
They show a significant integrated profile indicating the presence 
of weak burst pulses. To separate these pulses robustly, 
we calculated the on-pulse energy for every period after 
weighting the on-pulse bins, for every period, with the average pulse profile. 
We arranged all these pulses in the acceding order of their on-pulse energy. 
Such arrangement will cause few pulses, with pulse shape 
resembling the average profile and having high on-pulse energy, 
to move towards the high energy end. 
A box can be selected, as shown in Figure \ref{nullpulses_after_threshold}(b),
which encloses all the significant burst pulses at the high energy end. 
To include most of the weak burst pulses, the lower end of the box was moved from the high energy end towards 
the low energy end till the pulses outside the box did not show 
a significant profile (S/N $\sim$ 1). Pulses outside this box were tagged 
as the null pulses while the pulses inside the box were tagged as the burst pulses. 
After this final separation, the integrated profiles obtained from the 
null pulses and the burst pulses are shown in Figure \ref{null_burst_profile}. 

\subsection{Estimation of the reduction in the pulse energy} 
The degree by which the radio emission from a   
nulling pulsar declines during the nulls can be 
obtained from the average null pulse and the 
burst pulse profiles of the pulsar 
for the null and the burst pulses respectively. 
We calculated the reduction parameter, $\eta$, 
for a pulsar in a manner similar 
to that described by \cite{vj97}. 
First, the total energy from the on-pulse 
bins of the burst pulse profile was obtained. 
The rms deviation was obtained from the phase bins inside 
the predefined on-pulse window of the null pulse profile. 
To estimate an upper limit on any possible emission 
inside the on-pulse phase bins, in the null pulse profile, 
three times the corresponding rms estimate was used. 
The ratio between the total on-pulse energy of the 
burst pulse profile and an upper limit on any 
detectable emission from the null pulse profile 
can be defined as, 
\begin{equation}
\eta ~ = ~ \frac {\sum\limits^{N}_{i=1} P_{bpulse} (i) } {3 \times rms_{npulse} }. 
\label{R}
\end{equation}
Here, $P_{bpulse}(i)$ is the intensity of the i$^{th}$ bin inside the on-pulse window 
of the burst pulse profile, rms$_{npulse}$ is the rms estimated over 
the on-pulse window of the null pulse profile and N is the 
total number of bins in the on-pulse window. The error 
on the estimated factor, $\bigtriangleup\eta$, was obtained 
as three times the off-pulse rms of the burst pulse profile. 
The $\eta$ robustly quantifies the reduction in the pulse 
energy during the null state. It can be used to constrain 
various hypothesis regarding the nulling phenomena. 

\subsection{Null length and burst length histograms} 
\label{nlh_blh_intro_sect}
\begin{figure}[h!]
\begin{center}
 \centering
 \includegraphics[width=2.7 in,height=4 in,angle=-90,bb=0 100 554 720]{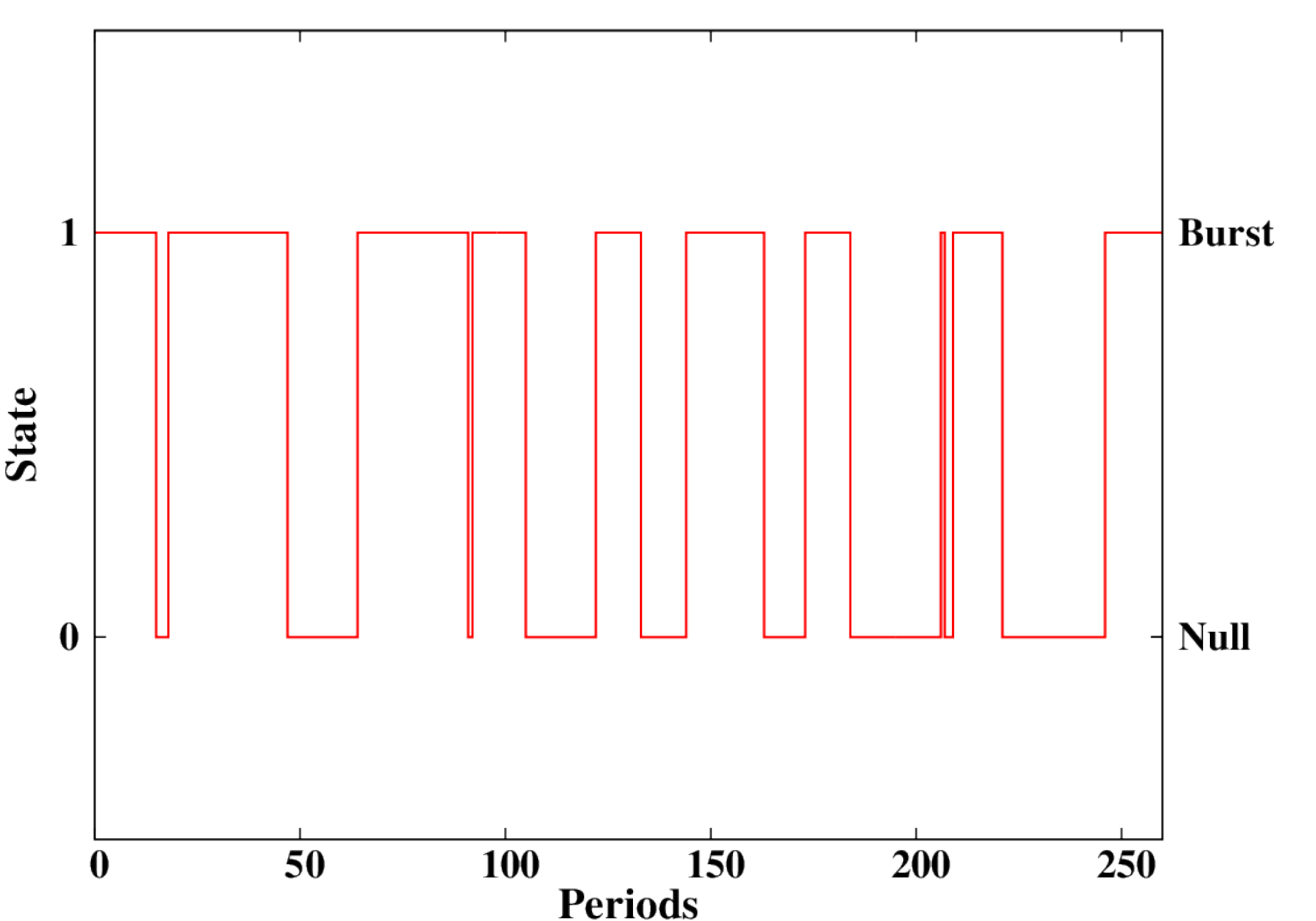}
 % J2321+6024.onezero.eps: 504x720 pixel, 72dpi, 17.78x25.40 cm, bb=0 0 504 720
 \caption[A sequence of pulses with their corresponding identified null/burst state]
 {A sequence of pulses with their corresponding identified 
 null/burst state for PSR B2319+60 observed at 610 MHz. 
 All the null pulses were tagged as zeros and all the 
 burst pulses were tagged as ones.}
 \label{onezero_example}
 \end{center}
\end{figure}

To estimate the distributions of null and burst durations,  
the null length and the burst length histograms were obtained. 
For each pulsar, single pulses were tagged 
with an index number according to their order of occurrence in the observed sequence. 
The null and the burst pulses were separated as explained in Section \ref{separation_of_null_burst_sect}. 
The original index numbers, which remained unaltered after the separation, 
were used to find a continuous sequence of consecutive null and burst pulses. 
Figure \ref{onezero_example} displays a small section 
of the observed data with their identified emission states 
(null or burst). To carry out various fluctuations analysis, 
each null pulse was tagged as zero and each burst pulse was 
tagged as one [as discussed in Section \ref{chronical_obs_sect}, \cite{bac70} introduced 
this technique]. Thus, from the identified state sequence of the pulses, 
a onezero time-series was formed (as shown in Figure \ref{onezero_example}). 
A Fourier transform of such one-zero series can be used 
to scrutinize periodicities associated with the nulling phenomena. 
\begin{figure}[h!]
 \begin{center}
 \includegraphics[width=2.7 in,height=4 in,angle=-90,bb=10 100 554 770]{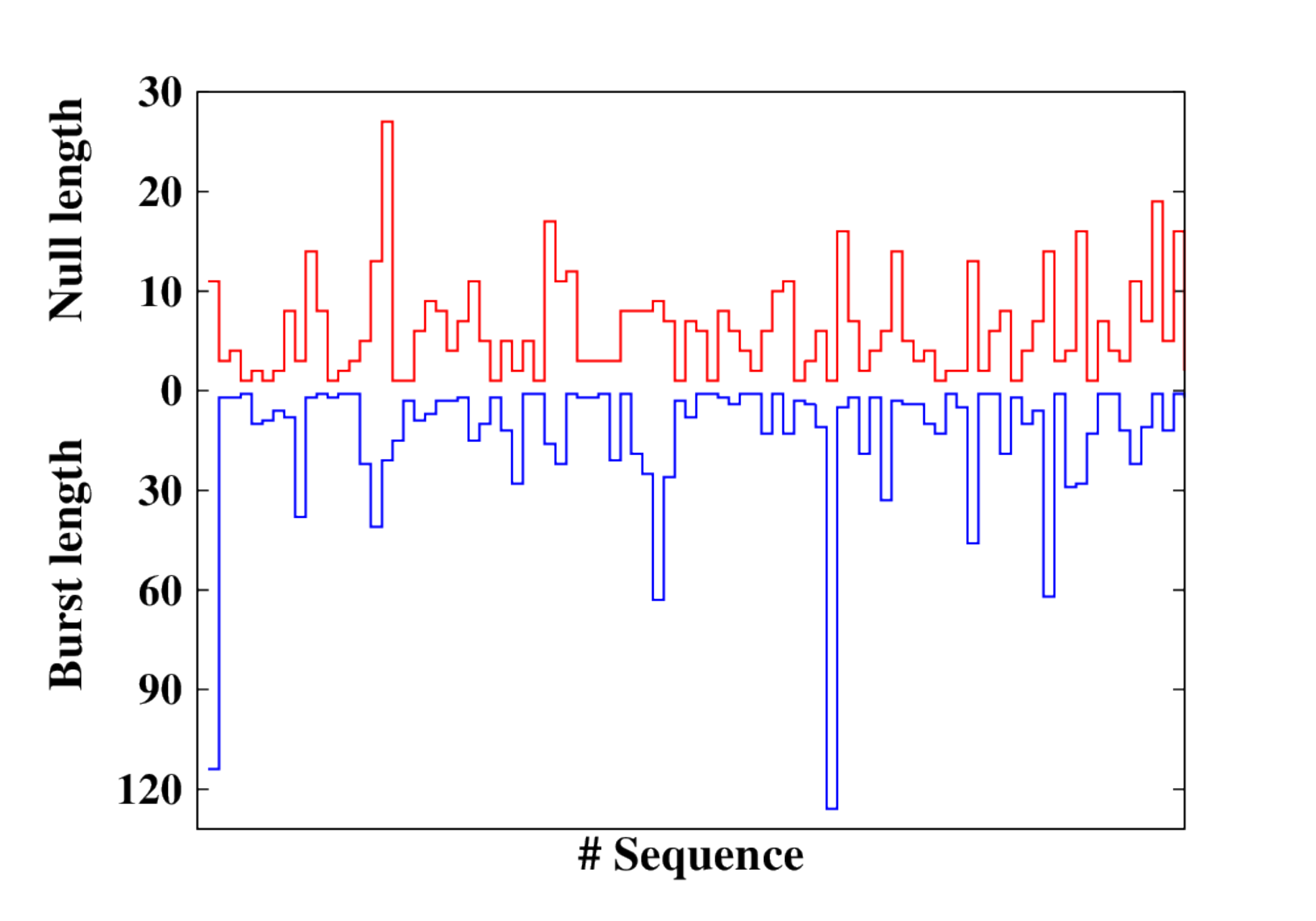}
 % J2321+6024.nbpair.forthesis.eps: 504x720 pixel, 72dpi, 17.78x25.40 cm, bb=0 0 504 720
 \caption[Sequence of consecutive null lengths and burst lengths]
 {A section of the observed null length and the burst length sequences for PSR B2319+60. 
 The red solid line shows the observed null length sequence. 
 The bottom solid blue line shows the length of bursts succeeding 
 the null shown at the top for each step. The ordinate scale for both the 
 sequences are different to display them in a same plot.}
 \label{nblen_seq_example}
 \end{center}
\end{figure}
\begin{figure}[h!]
 \centering
 \begin{center}
  \includegraphics[width=2.7 in,height=4 in,angle=-90,bb=50 100 554 770]{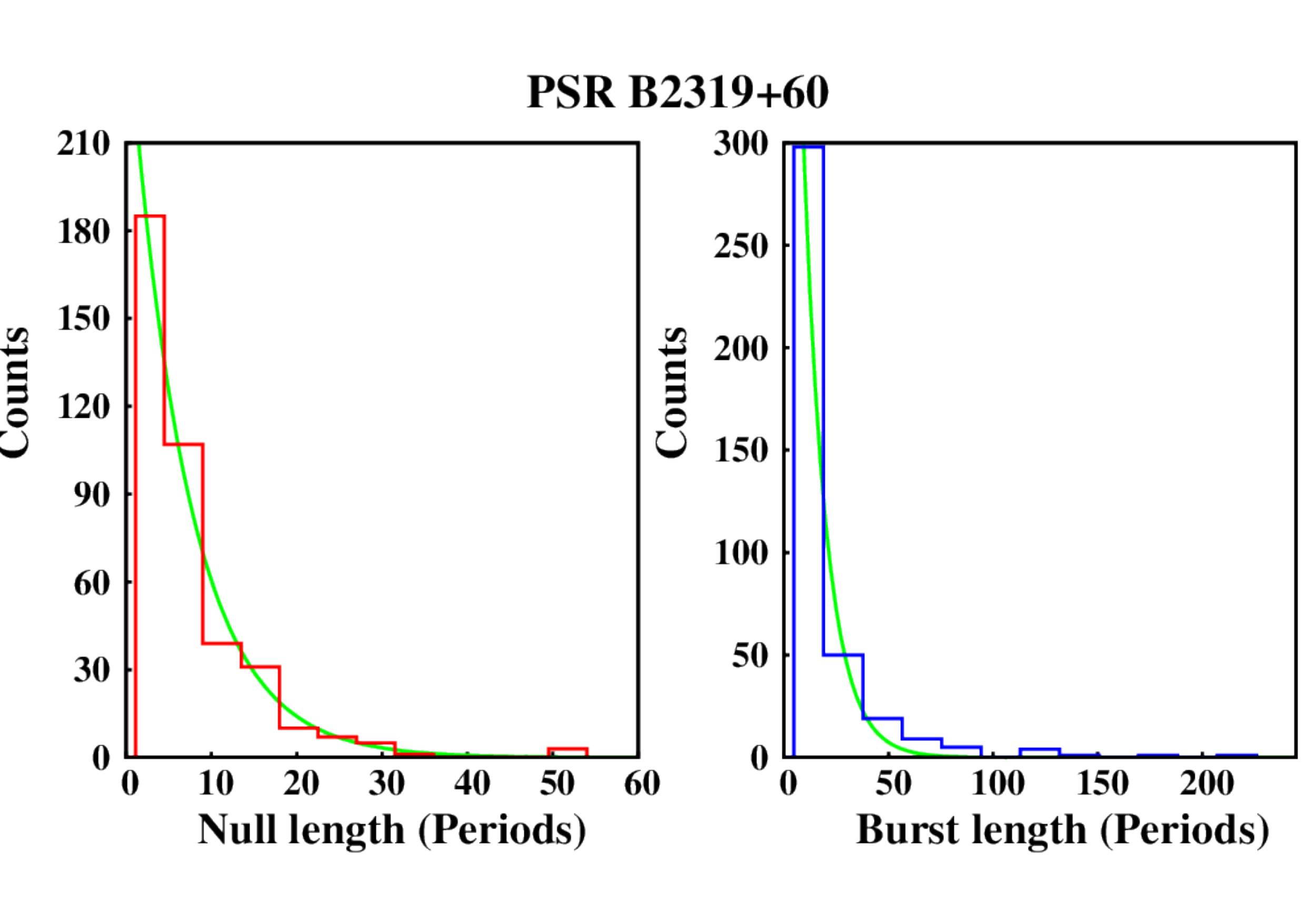}
 % J2321+6024.nbhist.eps: 504x720 pixel, 72dpi, 17.78x25.40 cm, bb=0 0 504 720
 \caption[The observed null length and the burst length histograms]
 {The observed null length and the burst length histograms from PSR B2319+60.
 The exponential fits are shown with a green solid lines. Note the clear exponential 
 decay in the frequency of occurrence from shorter 
 lengths to longer lengths in both histograms.}
 \label{NLH_BLH_example}
 \end{center}
\end{figure}

An individual uninterrupted sequence of null (or burst) pulses 
can be used to count the length of the corresponding null (or burst) state 
in the units of number of pulses. A nulling pulsar switches between 
the null state and the burst state with different time-scales. 
From the identified emission state sequence, such as Figure \ref{onezero_example}, 
a sequence of consecutively occuring null lengths 
and burst length can also be formed.  
Figure \ref{nblen_seq_example} shows a small section of 
consecutive null lengths and burst lengths for PSR B2319+60.
It can be seen from Figure \ref{nblen_seq_example}, 
the pulsar exhibits various lengths of null and burst phases. 
It should be noted that, for the in case of lower S/N, 
identification of true null and burst pulses can become 
difficult task. It can cause large differences in the obtained 
null length and burst length histograms, specially near the short nulls 
short bursts. \cite{cor13} has presented a compahensive study on this aspect. 

These null length and burst length sequences were binned to 
form the null length histogram (NLH) and the burst length histogram (BLH). 
A few pulsars exhibit large fraction of RFI infected pulses, where 
a judgement regarding the pulse emission state can not be made. 
For these pulsars, null lengths and burst lengths,  
which include RFI infected pulses, were entirely excluded. 
Figure \ref{NLH_BLH_example} displays the obtained 
null length and burst length histograms for PSR B2319+60. 
The distributions follow an exponential slope with 
relatively higher number of short null/burst lengths. 
% More detail regarding this peculiar behaviour will be discussed 
% in the next Chapter. 
As discussed in Section \ref{intrinsic_effects_sect}, 
its important to investigate the time-scale of these length distributions to 
scrutinize various nulling mechanism models. 
They are discussed further in Section \ref{sect_expted_time_scale}. 

\subsection{Longitude resolved fluctuation spectra}
\begin{figure}[h!]
 \centering
 \subfigure[]{
  \includegraphics[width=2.4 in,height=1.5 in,angle=-90,bb=14 14 544 220]{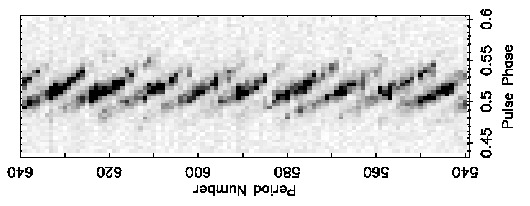}
  % Drift_example.eps: 0x0 pixel, 300dpi, 0.00x0.00 cm, bb=506 193 1 1
 \label{drift_example}
 }
 \centering
 \subfigure[]{
  \includegraphics[width=6 cm,height=6 cm,angle=-90,bb=0 14 584 751]{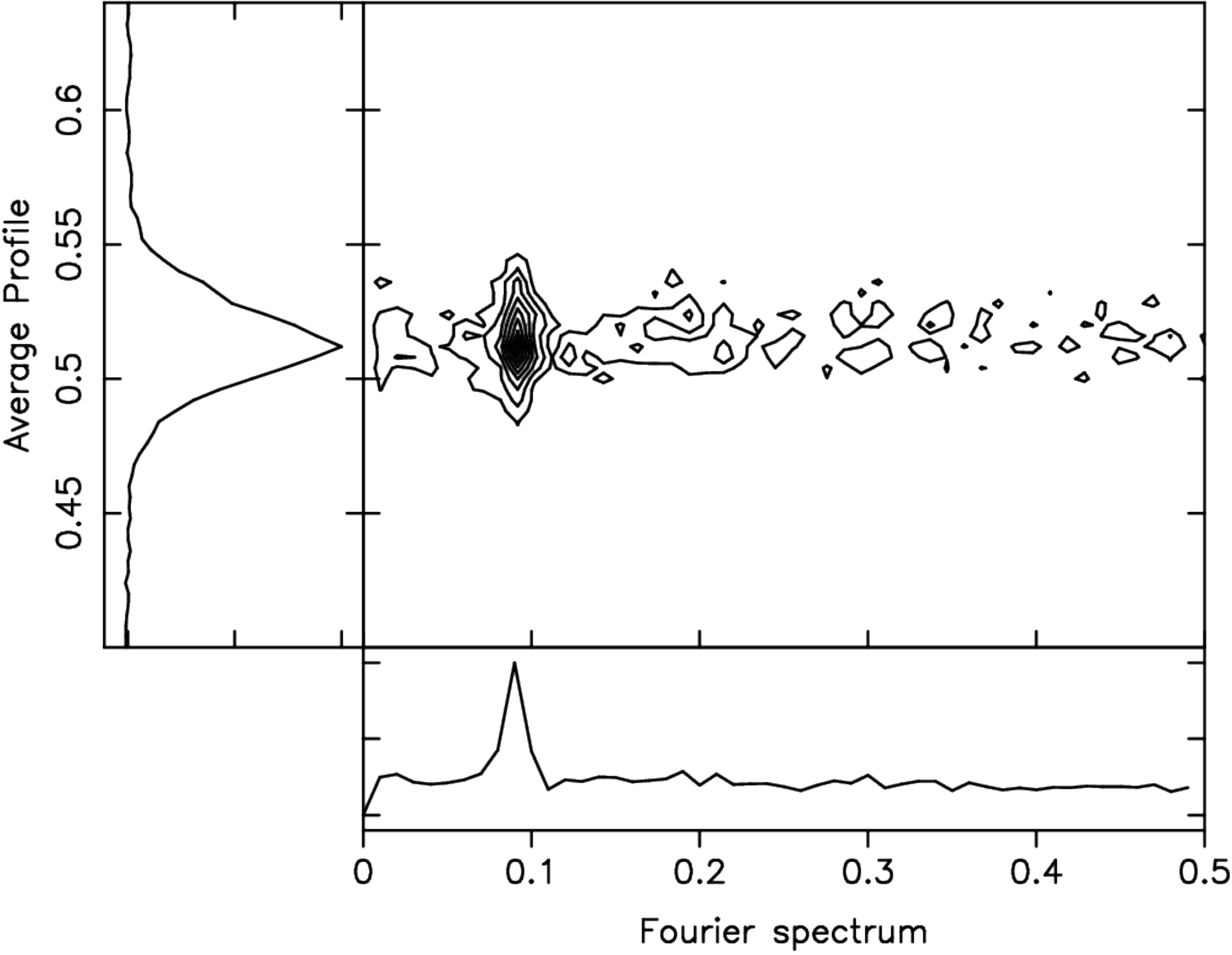}
 % J0814+7429.PMFS_example.eps: 556x719 pixel, 72dpi, 19.61x25.36 cm, bb=0 0 556 719
 \label{prfs_example}
  }
  \caption[A composite plot of the Longitude resolved fluctuation spectra]
  {A composite plot of the Longitude resolved fluctuation spectra. 
  (a) A section of the single pulses observed from PSR B0809+74 at 610 MHz. 
  The drifting of the subpulses is clearly visible. 
  (b) The phase-resolved fluctuation spectra obtained from the 
  single pulses displayed in (a). The center panel shows a contour plot 
  of the spectra over the given phase bin range. The left side panel shows the integrated profile. 
  The bottom panel shows the combined spectrum after averaging the above spectra 
  over the on-pulse longitude. A clear spectral feature with 11 period periodicity 
  is visible in the average spectrum and also in the contour plot.}
  \label{b0809_prfs}
\end{figure}

Many pulsars in our sample do exhibit regular drifting 
behaviour. Section \ref{drifting_intro_sect} introduced 
this phenomena, while a detail discussion regarding 
its origin in presented in Section \ref{rotating_carousal_sect}. 
To identify the drifting periodicities, 
we obtained a Longitude resolved fluctuation spectra (LRFS). 
The LRFS represents a Fourier transform computed 
over each individual phase bin in a contour plot. 
Figure \ref{b0809_prfs}(a) shows a section 
single pulses from a very prominent drifting 
pulsar, PSR B0809+74. To obtain the drifting 
periodicity, the pulsar period was 
divided into 256 phase bins (512 bins in a few cases). 
Pulse-to-pulse fluctuation time-series was obtained 
for the individual phase bin. 
A Fourier transform was carried out for each phase bin 
time-series to identify the fluctuation periodicities. 
The obtained Fourier spectra from all  
the phase bins were displayed in a contour plot to 
highlight the common features over the on-pulse phase bins. 
If a pulsar exhibit regular drifting behaviour, then 
the contour plot will show peak at the corresponding 
periodicity only at the on-pulse phase bins. 
An example of such composite LRFS for PSR B0809+74 is shown in 
Figure \ref{b0809_prfs}(b). The central panel in Figure \ref{b0809_prfs}(b) 
shows the contour plot of the Fourier spectra obtained from all the phase bins. 
The peaks in the contour plot and a strong peak in the average spectrum  
located near 0.09 cycles/period which corresponds 
to drifting periodicity of around 11 periods. 
The drifting periodicity, also clearly visible in 
Figure \ref{b0809_prfs}(a), is consistent with the reported 
periodicity by \cite{la83}. 

\clearpage\null\newpage

\chapter{A survey of nulling pulsars using the GMRT}

\section{Introduction}
\label{chap4_intro_sect}
Pulsar nulling is reported to occur in around 
109 pulsars (see Table \ref{table_all_null_psr}). 
NF, as discussed in Section \ref{NF_tech_sect}, is a widely used 
quantity to measure the degree of nulling in these pulsars. 
However, NF does not specify the duration of 
individual nulls, nor does it 
specify how the nulls are spaced in time. 
Although some attempts of characterising 
patterns in pulse nulling were made in the 
previous studies \cite[]{bac70,rit76,jv04,kr10}, 
not many pulsars have been studied for 
systematic patterns in nulling, partly because 
these require sensitive and long observations. 

Recent discoveries suggest that nulling pulsars 
with similar NF may have different null durations. 
These include intermittent pulsars, 
such as PSR B1931+24 \cite[]{klo+06} 
and PSR J1832+0029 \cite[]{lyn09}, and the RRATs, 
which show no pulsed emission between single burst of emission \cite[]{mll+06,kkl+11}[see Section 
\ref{extreme_null_sect} for details]. These pulsars also show extreme 
degree of nulling similar to a few classical nulling pulsars. 
As mentioned in Section \ref{extreme_null_sect}, 
PSR B1931+24 exhibits radio pulsations for 5 to 10 days 
followed by an absence of pulsations for 25 to 35 days 
\cite[]{klo+06}. If the cessation of radio emission in this 
pulsar is interpreted as a null, it has a NF of about $\sim$72 to 85\% 
similar to PSR J1502$-$5653 \cite[]{wmj07,lem+12}. Yet the latter shows nulls 
with a typical duration of few tens of seconds in contrast 
to a much longer duration for PSR B1931+24. A similar 
conclusion can be drawn by comparing RRATs with classical high 
NF pulsars. While this leads to the expectation that pulsars 
with similar NF may have different nulling time-scales, 
no systematic study of this aspect of nulling is available to 
the best of our knowledge. In this chapter, a modest attempt to 
investigate this is initiated.

Pulse nulling was usually believed to be a random phenomenon  
\cite[]{rit76,big92a}. However, recent studies indicate a 
non-random nulling behaviour for a few classical nulling pulsars 
\cite[]{hjr07,hr09,rr09,kr10} [see Section \ref{periodic_nulling_sect} for details]. 
\cite{rr09} also report random nulling behaviour 
for at least 4 out of 18 pulsars in their sample. Therefore, 
it is not clear if non-randomness in the sense defined in 
\cite{rr09} is seen in most nulling pulsars and 
such a study needs to be extended to more nulling pulsars. 
This issue is investigated in this chapter with a distinct set of 
nulling pulsars.

In this chapter, we present observations of 15 pulsars, 
carried out using the GMRT at 325 and 610 MHz. 
Among these, five were discovered in the  PKSMB survey 
\cite[]{mlc+01,mhl+02,kbm+03,lfl+06}, 
which have no previously reported nulling behaviour. 
Selection of these pulsars is justified in Section \ref{new_pulsar_sect}. 
Rest of the sample consists of well 
known strong nulling pulsars. The observations and analysis techniques 
are similar to the one described in Chapter 3. In Section \ref{null_behavior_sect}, 
nulling behaviour for individual pulsar is discussed along 
with estimates of their NFs and the reduction 
in the pulsed energy during the null state. 
A comparison of the null length and burst 
length distributions for pulsars, which have similar NF, 
is presented in Section \ref{sect_null_comp} and a discussion 
on the randomness of nulls is presented in Section \ref{sect_randomness}. 
The expected time-scales for the null and the burst durations 
are presented in Section \ref{sect_expted_time_scale}. 
Finally, the conclusions are presented in Section \ref{sect_survey_conclusion} 
while the implications of these results are discussed in Section \ref{sect_survey_discussion}. 

\section{Single pulse behaviour of individual pulsars} 
The on-pulse and the off-pulse energy histograms for each observed pulsar 
are shown in Appendix A. The null and the burst pulses were 
separated, using the method discussed in Section \ref{separation_of_null_burst_sect}, 
for each pulsar. The separated null and burst pulses were used to 
construct the null length histograms and the burst length 
histograms, shown in Appendix B. 

\label{null_behavior_sect}
\setcounter{secnumdepth}{0}
\subsection{B0809+74}
% \paragraph{B0809+74}
\label{b0809}
As mentioned in Section \ref{chronical_obs_sect}, this is one of the well studied nulling pulsar. 
Nulling in this pulsar was first reported by \cite{th71}. 
A section of observed single pulses at 325 MHz, is shown in 
Figure \ref{sp_0809_0818_0835}(a), which 
displays a clear null region around period number 440. 
An upper limit on NF for this pulsar was estimated by \cite{rit76}. 
A refined value of the NF was estimated by \cite{la83}
using 40 hrs long observations. \cite{la83} also reported reduction 
in burst pulse energy by $>$ 78 times. This is the only pulsar in the sample 
with a previously reported $\eta$ value. 
The on-pulse and the off-pulse energy histograms 
obtained from our observations are shown in Figure \ref{NF_b0809}. 
The ONPH shows clear bimodal distribution originating due to 
small fraction null pulses and large fraction of strong burst pulses. 
We obtained the NF of 1.0$\pm$0.4\% using these histograms. 
The estimated $\eta$ is 172.0$\pm$0.5, which matches with the previously 
reported lower limit by \cite{la83}. 
Figure \ref{nlh_blh_b0809} shows the NLH and the BLH, constructed 
using 73 null lengths and 73 burst lengths. 
The NLH shows typical null length of around 1 to 3 periods occurring more frequently 
while extending the overall distribution up to 8 periods null. 
The BLH shows exponentially declining distribution of burst lengths up to 800 periods. 
We also carried out LRFS analysis (shown in Figure \ref{prfs_b0809}) to highlight 
the 11 period periodicity, confirming the earlier claims by \cite{la83} and \cite{vkr+02}. 
The pulsar also shows intriguing change in the drift rates before 
and after the null states \cite[]{la83,vkr+02}. 
\begin{figure}[h!]
 \centering
 \subfigure[]{
 \includegraphics[width=4in,height=1.5in,angle=-90,bb=0 0 504 203]{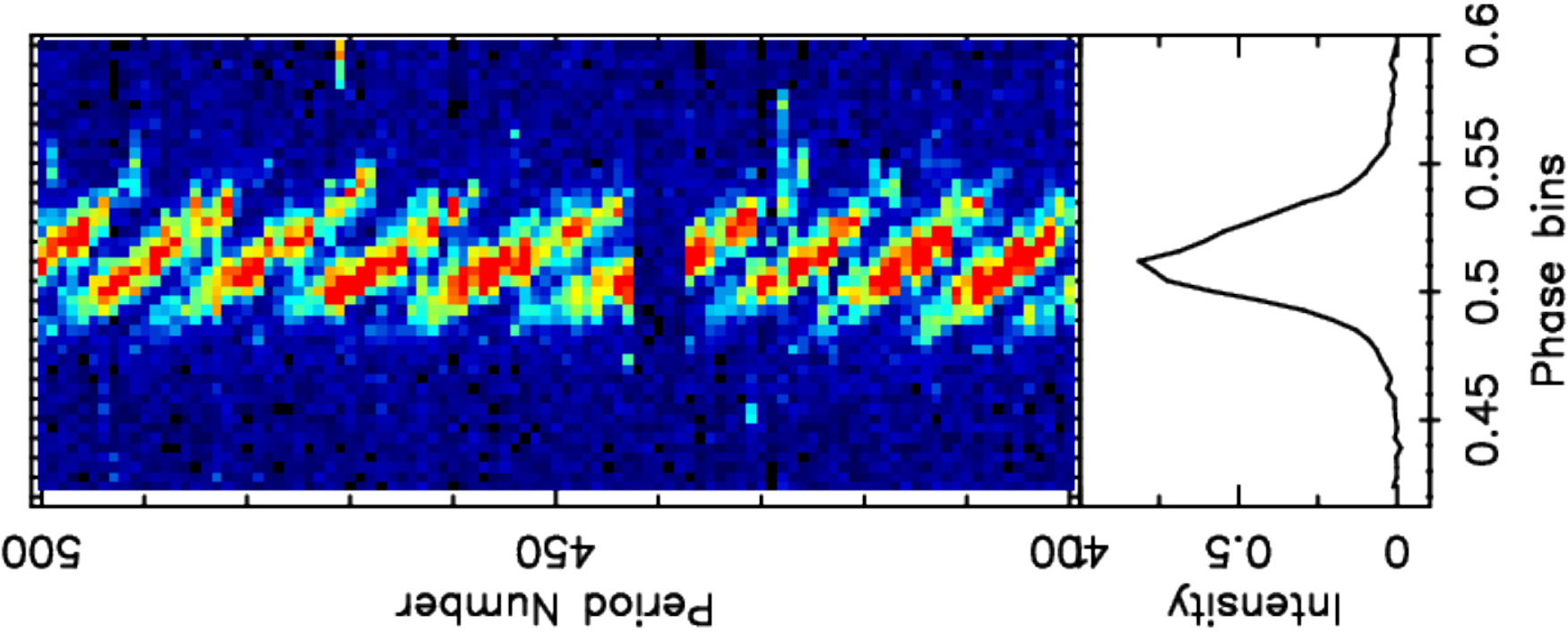}
 % b0809_spdisplay_forThesis.eps: 0x0 pixel, 300dpi, 0.00x0.00 cm, bb=503 202 1 1
 }
 \subfigure[]{
 \includegraphics[width=4 in,height=1.5 in,angle=-90,bb=0 0 504 203]{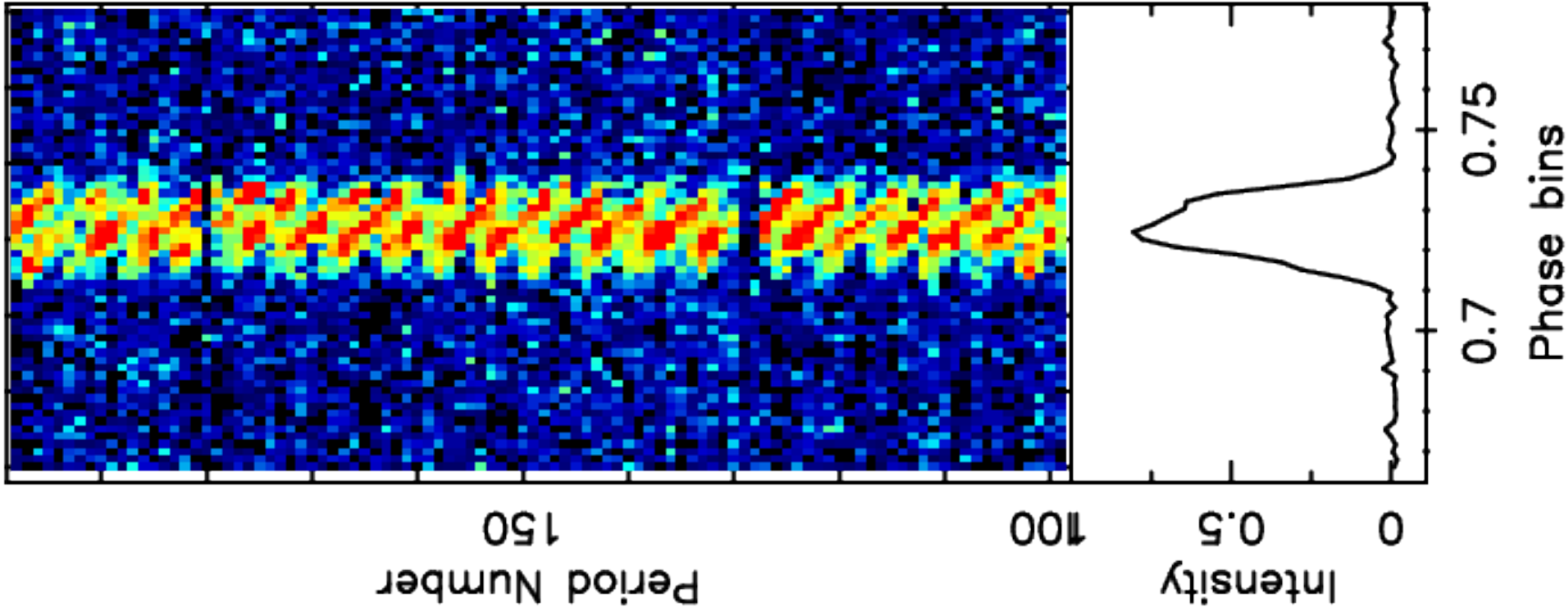}
 % b0818.spdisplay.forThesis.eps: 494x192 pixel, 72dpi, 17.43x6.77 cm, bb=0 0 494 192
 }
 \subfigure[]{
 \includegraphics[width=4 in,height=1.5 in,angle=-90,bb=0 0 504 203]{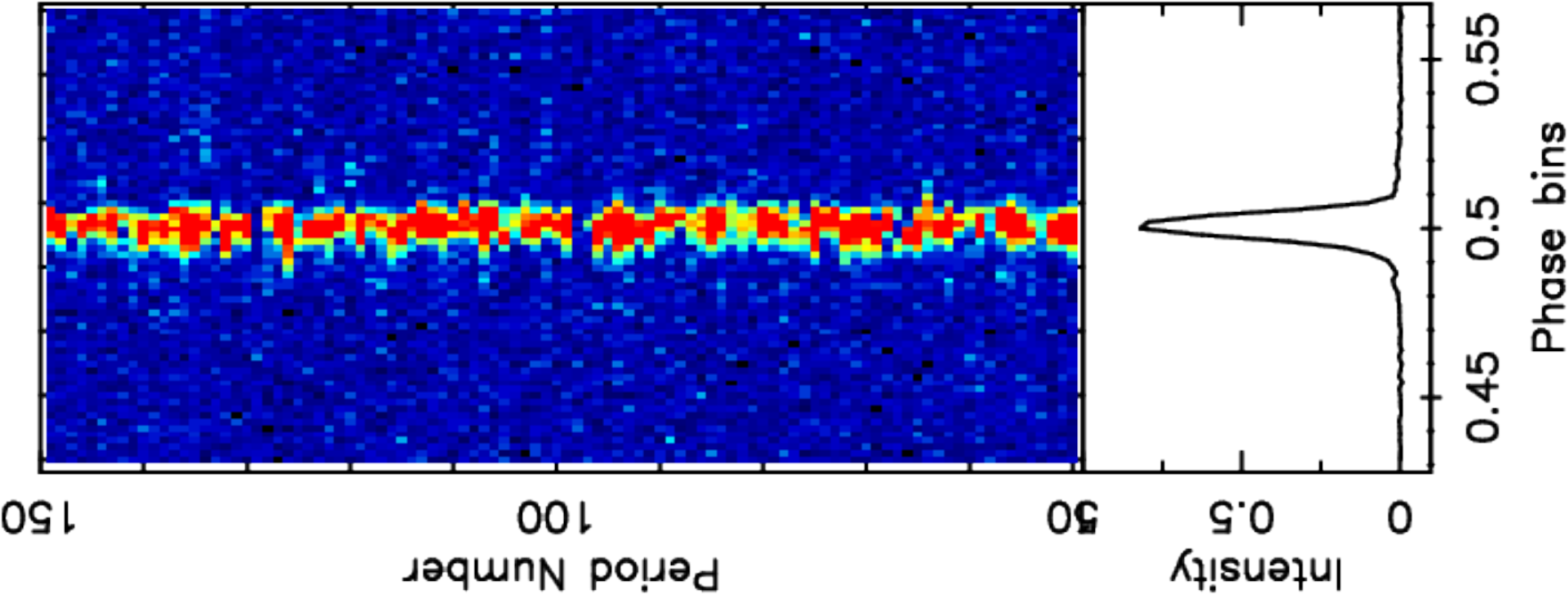}
 % b0818.spdisplay.forThesis.eps: 494x192 pixel, 72dpi, 17.43x6.77 cm, bb=0 0 494 192
 }
 \begin{picture}(0,0)
  \put(-330,0){PSR B0809+74}
  \put(-210,0){PSR B0818$-$13}
  \put(-90,0){PSR B0835$-$41}
 \end{picture}
 \caption[Sections of observed single pulses from three pulsars namely, PSRs B0809+74, B0818$-$13 and B0835$-$41.]
 {Sections of observed single pulses from three pulsars namely, PSRs (a) B0809+74 (b) B0818$-$13 (c) B0835$-$41. 
 The top panel in all sub-plots shows stack of 100 pulses with pulse intensity displayed in a color ramp from 
 blue to red. The bottom panel in each sub-plot presents the respective integrated profile. 
 Presence of null pulses are clearly evident for all three pulsars 
 in addition to clear drifting of subpulses in PSRs B0809+74 and B0818$-$13.}
 \label{sp_0809_0818_0835}
\end{figure}
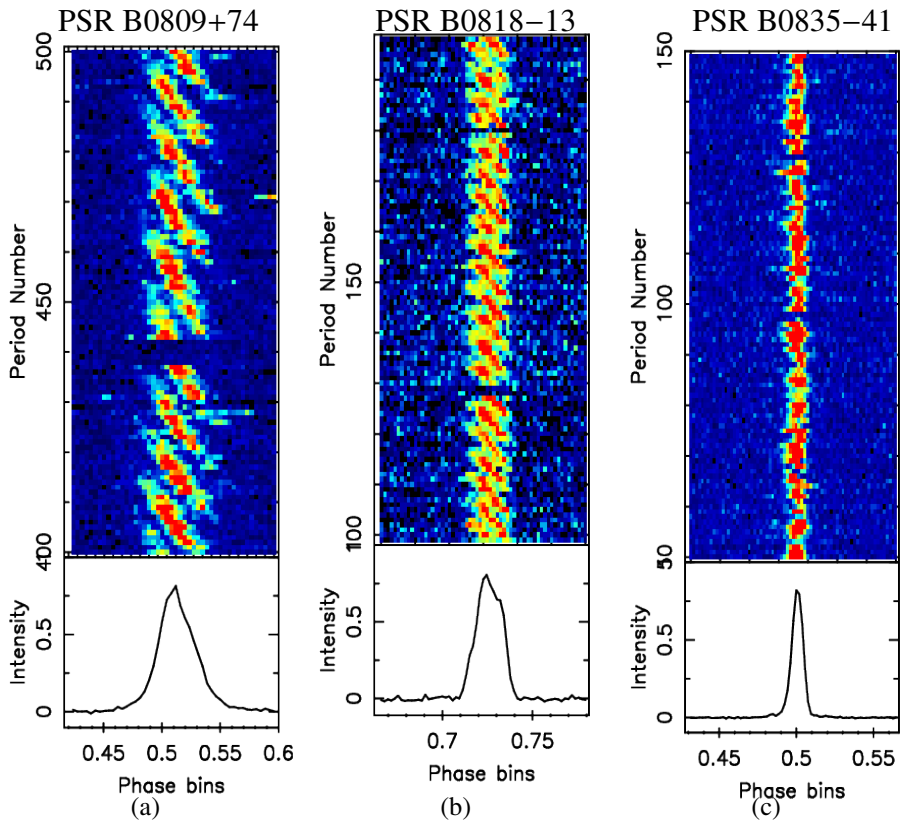
\begin{figure}[!h]
 \centering
 \includegraphics[width=2.7 in,height=3.7 in,angle=-90,bb=0 0 557 719]{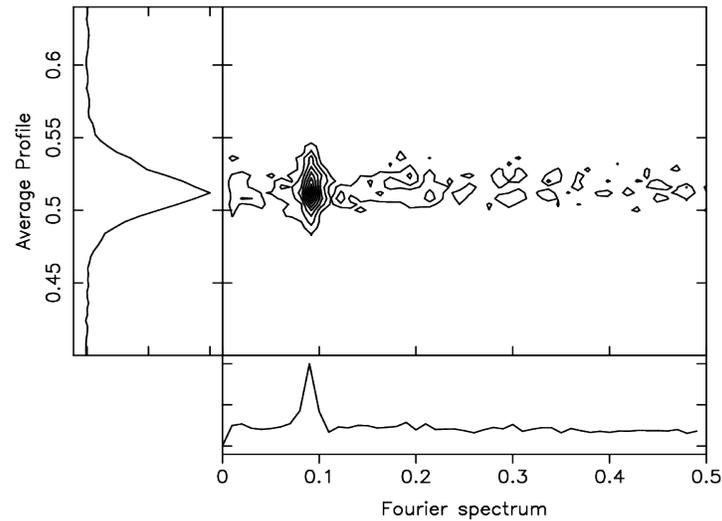}
 % J0814+7429.PMFS_example.eps: 557x719 pixel, 72dpi, 19.65x25.36 cm, bb=0 0 557 719
  \vspace{0.1 cm}
  \caption[Phase-resolved fluctuation spectra obtained from the observed single pulses at 325 MHz 
  of PSR B0809+74]{Phase-resolved fluctuation spectra obtained from the observed single pulses at 325 MHz 
  of PSR B0809+74. The centre panel shows a contour plot of the spectra over 
  the given phase bin range. The left side panel shows the integrated profile. 
  The bottom panel shows the combined spectrum after averaging the above spectra over 
  the on-pulse bins. A clear spectral feature with 11 period periodicity 
  is visible in the average spectrum and also in the contour plot.}
  \label{prfs_b0809}
\end{figure}

\subsection{B0818$-$13}
\label{b0818}
\begin{figure}[h!]
 \centering
 \includegraphics[width=2.7 in,height=3.7 in,angle=-90,bb=0 0 557 719]{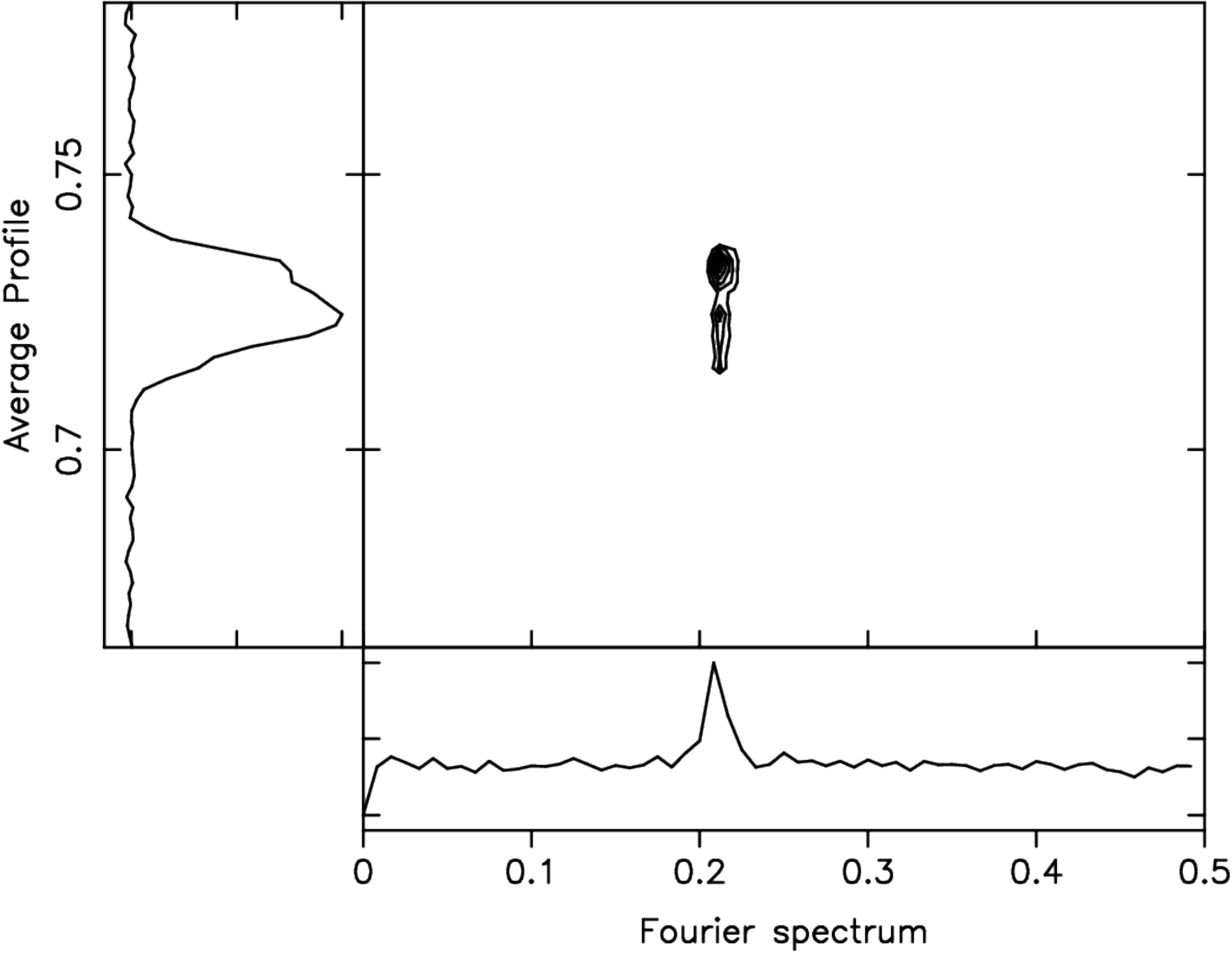}
 % b0818.PRFS.forThesis.eps: 557x719 pixel, 72dpi, 19.65x25.36 cm, bb=0 0 557 719
 \vspace{0.1 cm}
 \caption[Phase-resolved fluctuation spectra obtained from the single pulses of PSR B0818$-$13]
 {Phase-resolved fluctuation spectra obtained from the single pulses of PSR B0818$-$13 
 (see Figure \ref{prfs_b0809} for details regarding various panels). 
 The drifting feature of around 0.21 cycles per period is clearly 
 evident which corresponds to 4.8 period periodicity.}
 \label{prfs_b0818}
\end{figure}
This is one of the well known nulling pulsar. 
A section of observed single pulses at 610 MHz is shown in Figure \ref{sp_0809_0818_0835}(b) 
which shows clear null pulses at period numbers around 130 and 180. 
\cite{la83}, did an extensive study about the change in the drift pattern before 
and after the null states in this pulsar (see Section \ref{chronical_obs_sect}). 
Our data show drifting feature at 0.21 cycles per period (Figure \ref{prfs_b0818}) 
which matches with the earlier reported P$_{3}$ \cite[]{la83,jv04}. The pulse energy histograms 
are shown in Figure \ref{NF_b0818}. Our NF (0.9$\pm$1.8\%) also matched with the earlier 
reported NF (1.01$\pm$0.01\%) within the error bars. 
The new result from our study is the amount of flux reduction during the null states. 
Although this is a relatively stronger pulsar, the estimated $\eta$($\sim$ 4.2) is comparatively small.
The NLH and the BLH are shown in Figure \ref{nlh_blh_b0818}, 
which were constructed using 57 null lengths and 57 burst lengths. 
The NLH shows around 90\% of nulls tend to occur for only single period while the rest 10\% are distributed 
up to 4 periods. The BLH shows smooth exponential decline of the burst lengths distribution 
up to 400 periods.   
\subsection{B0835$-$41}
Nulling in this pulsar was first reported by \cite{big92a}. 
A section of observed single pulses at 610 MHz is shown in Figure \ref{sp_0809_0818_0835}(c), 
which shows regions of short null states around period numbers  
100 and 140. The on-pulse and the off-pulse energy histograms are 
shown in Figure \ref{NF_b0835}. The estimated NF, of around 1.7$\pm$1.2\%, 
is small like in the case of PSR B0818$-$13. 
However, estimated $\eta$ (15.7$\pm$0.2) is comparatively higher. 
The NLH and the BLH are shown in Figure \ref{nlh_blh_b0835}, which were 
constructed using 74 null lengths and 74 burst lengths. 
The NLH shows very interesting null length distribution as more than 95\% of 
the nulls tend to occur only for one period, while the rest 5\% tend to 
occur for two periods. It can be speculated that the 
true nulling time-scale could be shorter than the period of the pulsar and 
the observed NLH presents only a tail of this distribution. 
The BLH shows exponentially declining distribution of burst 
lengths up to 500 periods. 
\subsection{B1112+50}
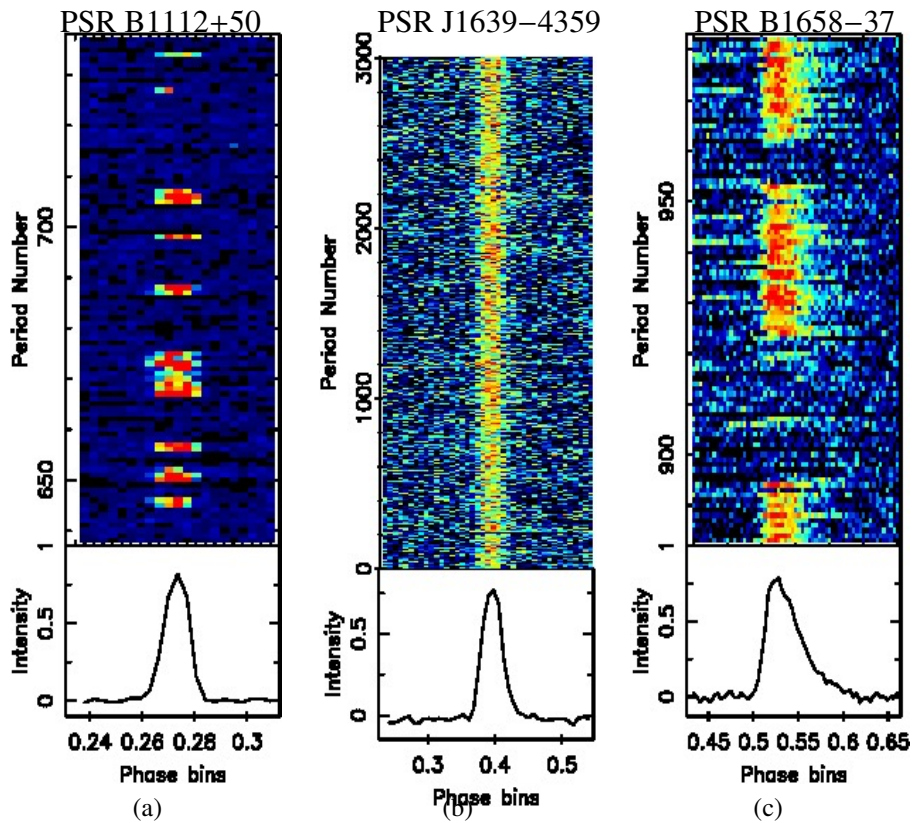
\begin{figure}[h!]
 \centering
 \subfigure[]{
 \includegraphics[width=4in,height=1.5in,angle=-90,bb=0 0 504 203]{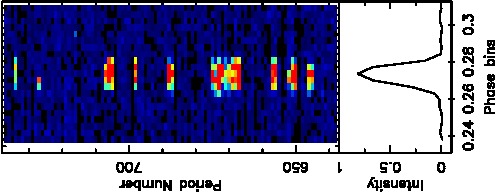}
 % b0809_spdisplay_forThesis.eps: 0x0 pixel, 300dpi, 0.00x0.00 cm, bb=503 202 1 1
 }
 \subfigure[]{
 \includegraphics[width=4in,height=1.5in,angle=-90,bb=0 0 504 203]{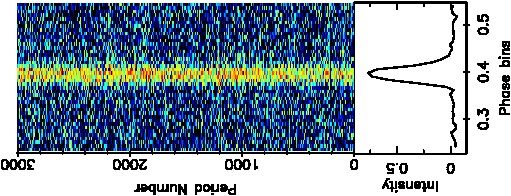}
 % b0809_spdisplay_forThesis.eps: 0x0 pixel, 300dpi, 0.00x0.00 cm, bb=503 202 1 1
 }
 \subfigure[]{
 \includegraphics[width=4in,height=1.5in,angle=-90,bb=0 0 504 203]{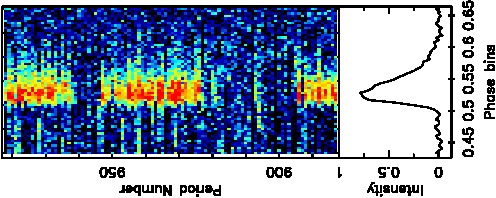}
 % b0809_spdisplay_forThesis.eps: 0x0 pixel, 300dpi, 0.00x0.00 cm, bb=503 202 1 1
 }
  \begin{picture}(0,0)
  \put(-330,0){PSR B1112+50}
  \put(-210,0){PSR J1639$-$4359}
  \put(-90,0){PSR B1658$-$37}
 \end{picture}
 \caption[Modulation of the pulse energy for PSRs B1112+50, J1639$-$4359 and J1701$-$3726]
 {Modulation of the pulse energy for PSRs (a) B1112+50 (b) J1639$-$4359 (c) J1701$-$3726 
 (see Figure \ref{sp_0809_0818_0835} for details regarding panels) 
 observed at 610 MHz. For PSR J1639$-$4359, successive 16 periods 
 were sub-integrated as the single pulses were below the detection level. 
 The plot shows the resulting sub-integrations for a section of the observed data. 
 The period range is kept unaltered for comparison.}
 \label{sp_1112_1639_1701}
\end{figure}

\begin{figure}[h!]
 \centering
 \includegraphics[width=2.7 in,height=3.7 in,angle=-90,bb=0 0 557 719]{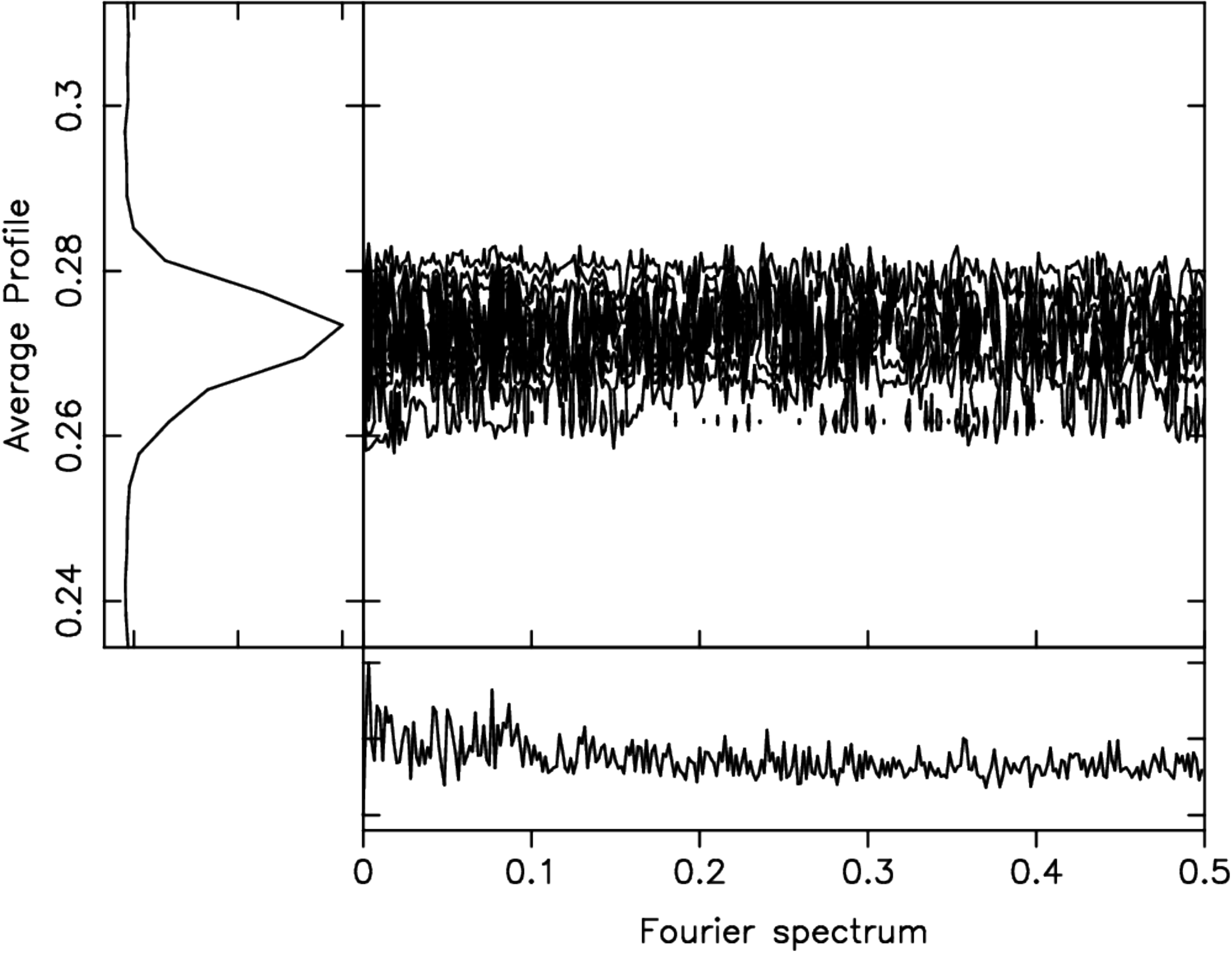}
 % b0818.PRFS.forThesis.eps: 557x719 pixel, 72dpi, 19.65x25.36 cm, bb=0 0 557 719
 \vspace{0.1 cm}
 \caption[Phase-resolved fluctuation spectra for PSR 1112+50]
 {Phase-resolve fluctuation spectra for PSR 1112+50 (see Figure \ref{prfs_b0809} for details regarding 
 various panels). The spectra is featureless with absence of any significant periodicity.}
 \label{prfs_b1112}
\end{figure}

For this pulsar, nulling was first reported by \cite{rit76}. 
This pulsar was reported to have three different profile modes of emission 
at 1420 MHz \cite[]{wsw86}, where it shows two distinct profile components. 
In this survey, observation at 610 MHz showed single profile component.  
The pulsar shows sporadic nature of pulse energy modulations. 
The pulse energy histograms are shown in Figure \ref{NF_b1112}, 
using which the NF of around 64$\pm$6\% (which 
matches with the reported NF of around 60$\pm$5\% by \cite{rit76})
was obtained. For the large fraction of the observed duration, pulsar 
switched rapidly between the null state and the burst state 
within one or two periods. The reduction in the 
pulse energy, $\eta$, during these null states is 44.7$\pm$0.2. 
Due to the intriguing flickering nulls seen in  
Figure \ref{sp_1112_1639_1701}(a), it is interesting to investigate periodicity 
of this pulse energy fluctuations. However, the obtained LRFS, 
shown in Figure \ref{prfs_b1112}, did not 
show any significant spectral feature, rejecting 
any possibility of periodicity in the pulse energy modulation. 
\cite{wse07} also reported no strong spectral feature in their observations at 92-cm. 
Theoretically, it would be challenging to propose any nulling mechanism which 
can cause changes to the pulse energy by such a large fraction (around 45 times) 
in such a short time-scale (one to two periods). 
The overall NLH and the BLH, constructed using 635 null lengths and 635 burst lengths, 
are shown in Figure \ref{nlh_blh_b1112}.  
They show large fraction of short nulls and short bursts with exponentially  
declining null length and burst length distributions up to 50 periods. 
\subsection{J1639$-$4359}
\label{sect_j1639}
This is one of the newly discovered pulsars from the PKSMB survey \cite[]{kbm+03}. 
The single pulses from this pulsar were very weak. 
The single pulse data were averaged over  
successive pulses to form average profiles for every block of 16 
single pulses (sub-integration) to show the pulse energy modulations.
The pulse sequence of around 180 sub-integrated 
blocks (3000 pulses) is shown in Figure \ref{sp_1112_1639_1701}(b). 
As its clearly evident from the sub-integrated pulse plot, 
there are no clear null states. Figure \ref{NF_j1639} shows 
the on-pulse and the off-pulse energy histograms, 
obtained from the sub-integration of successive 16 periods. 
The histograms are presented just to demonstrate 
the absence of null pulses among the sub-integrated pulses.
Estimation of the NF was not possible as averaging 
over consecutive periods resulted in detection of emission for every 
sub-integration giving a NF of zero (as shown in Figure \ref{NF_j1639}). 
However, some of the single pulses in a sub-integration may well have been nulled pulses. 
To estimate an upper limit on the fraction of such pulses for this  
pulsar, we arranged all the single pulses in the ascending order of 
their on-pulse energy. A threshold was moved from the lower energy end towards
the high energy end till the pulses below the threshold did not show  
a significant (S/N $>$ 3) profile component (similar to the method 
discussed in Section \ref{separation_of_null_burst_sect}).  
These pulses, located below the threshold at the lower energy end, were tagged as 
the null pulses. Although, it is likely that these pulses are a mixture of true null pulses 
and weak burst pulses (as the expected significance was comparatively higher with S/N $>$ 3 
compared to such threshold mentioned in Section \ref{separation_of_null_burst_sect}). 
Hence, the fraction of these pulses were used only as an 
upper limit of around 0.1\%, as the true NF for this pulsar 
could be lower than this limit, if more sensitive observations are conducted. 
The NLH and the BLH were not possible to obtain as the pulsar showed weak 
single pulses. 

\subsection{B1658$-$37}
\begin{figure}[h!]
 \begin{center}
 \centering
 \includegraphics[width=3 in,height=4 in,angle=-90,bb=0 120 554 770]{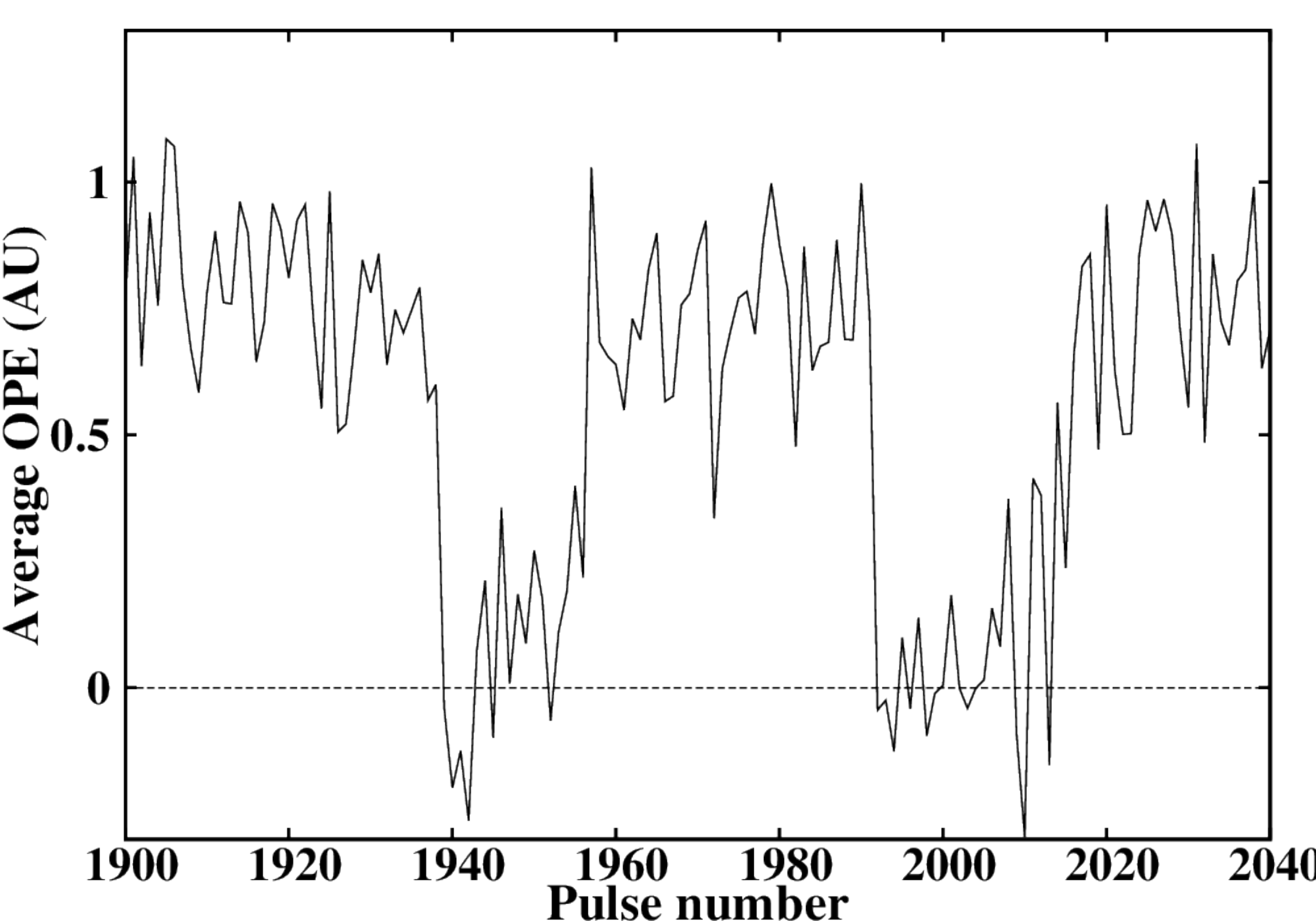}
 % J1701.energyon.eps: 504x720 pixel, 72dpi, 17.78x25.40 cm, bb=0 0 504 720
 \caption[On-pulse energy sequence of around 140 pulses obtained from PSR B1658$-$37 observed 
 at 610 MHz]{On-pulse energy sequence of around 140 pulses obtained from PSR B1658$-$37 observed 
 at 610 MHz. Two consecutive null regions are shown, with the pulse energy reaching zero, around 
 pulse numbers 1940 and 2000. A sudden drop and a gradual rise of 
 the pulse energy is clearly evident before and after the null states respectively.}
 \label{j1701_ope}
 \end{center}
\end{figure}

Nulling in this pulsar was first reported by \cite{wmj07} with 
a lower limit on the NF ($>$14\%). It was also reported to have short nulls. 
However due to lower S/N, \cite{wmj07} sub-integrated 10 pulses. 
Our observations were more sensitive with higher S/N on single pulses [shown 
in Figure \ref{sp_1112_1639_1701}(c)], hence we obtained better estimate 
on the NF (22$\pm$4) compared to earlier reported value. 
The pulse energy histograms used to estimate this NF are shown in Figure \ref{NF_j1701}. 
The null pulses and burst pulses were separated and the obtained 
length histograms are shown in Figure \ref{nlh_blh_b1701}, 
which show exponentially declining null length and burst length distributions. 
The NLH also shows a slight excess of around 10 period nulls. 
Total 73 null lengths and 73 burst lengths were used to obtain these 
length histograms. A careful look at the single pulse sequence reveals interesting 
pulse energy fluctuation with a sudden fall of the pulse energy at the beginning of the null state 
followed by a gradual rise towards the end. Figure \ref{j1701_ope} 
shows an example of two consecutive null states which shows 
a gradual rise of the on-pulse energy towards the end of the 
null state. This behaviour has been reported for the first time in this pulsar. 
\cite{wmj07} reported two different profile modes 
in this pulsar at 1420 MHz. However, scattering caused by 
the observations at a lower frequency hindered identification 
of different profile components. Hence, no profile mode-changing was 
possible to identify from our observations. 
\subsection{J1715$-$4034}
\label{sect_j1715}
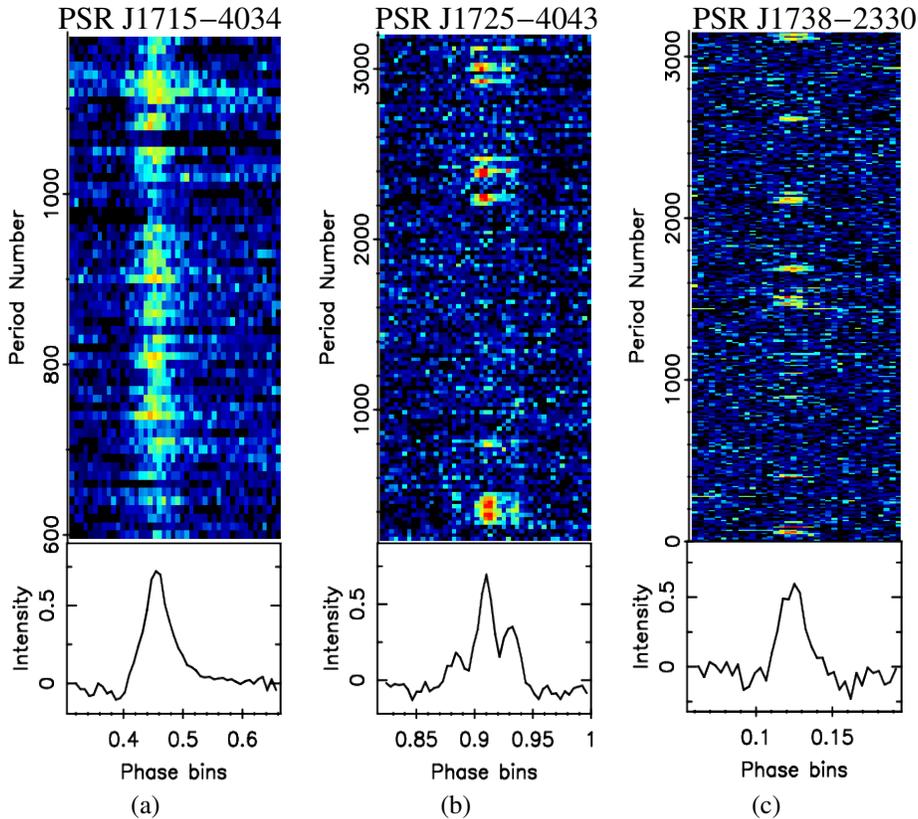
\begin{figure}[h!]
 \centering
 \subfigure[]{
 \includegraphics[width=4in,height=1.5in,angle=-90,bb=0 0 504 203]{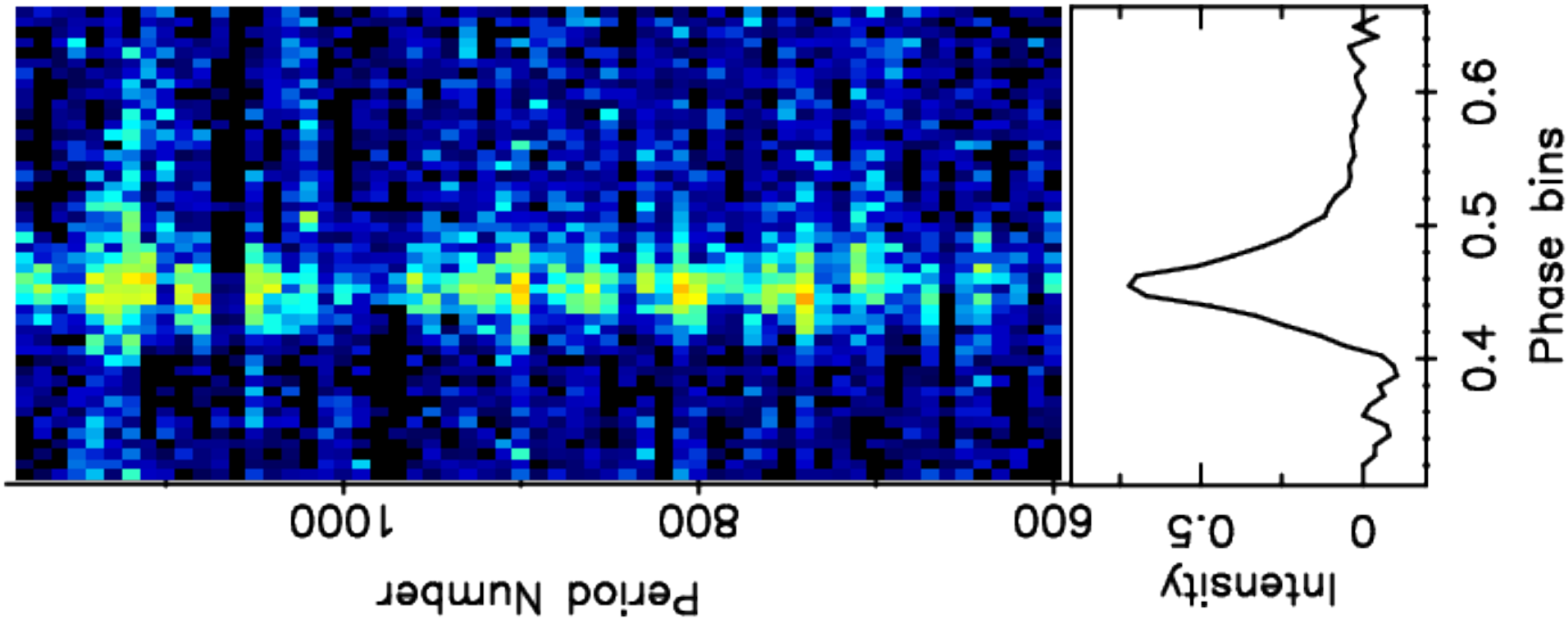}
 % b0809_spdisplay_forThesis.eps: 0x0 pixel, 300dpi, 0.00x0.00 cm, bb=503 202 1 1
 }
 \subfigure[]{
 \includegraphics[width=4in,height=1.5in,angle=-90,bb=0 0 504 203]{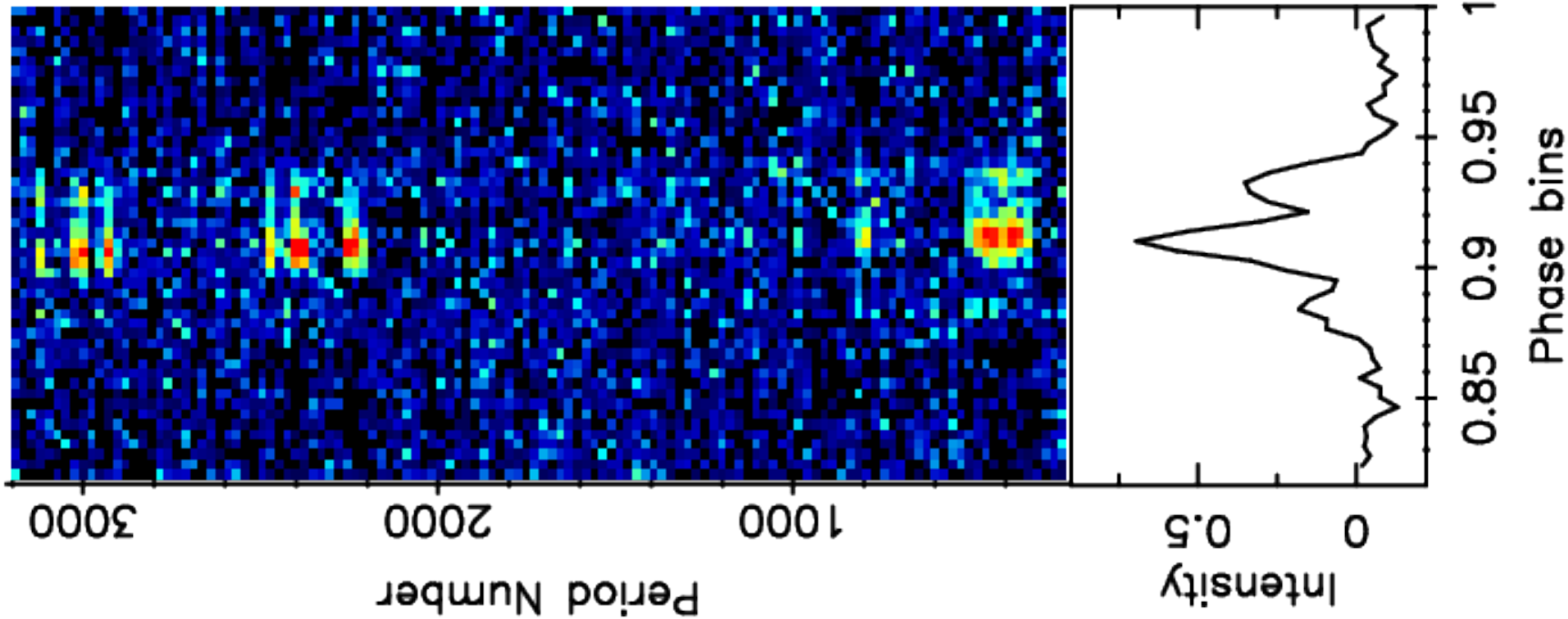}
 % b0809_spdisplay_forThesis.eps: 0x0 pixel, 300dpi, 0.00x0.00 cm, bb=503 202 1 1
 }
 \subfigure[]{
 \includegraphics[width=4in,height=1.5in,angle=-90,bb=0 0 504 203]{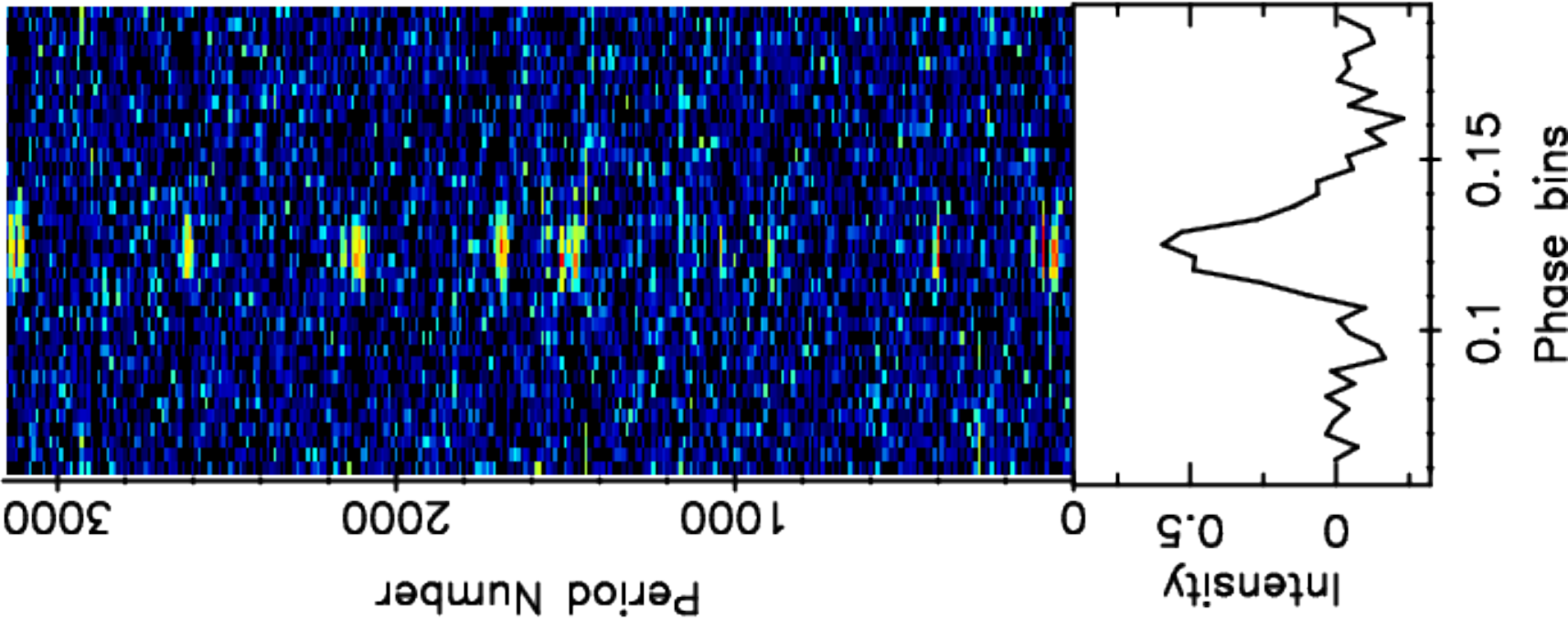}
 % b0809_spdisplay_forThesis.eps: 0x0 pixel, 300dpi, 0.00x0.00 cm, bb=503 202 1 1
 }
 \begin{picture}(0,0)
  \put(-330,0){PSR J1715$-$4034}
  \put(-210,0){PSR J1725$-$4043}
  \put(-90,0){PSR  J1738$-$2330}
 \end{picture}
 \caption[Modulation of the pulse energy for PSRs J1715$-$4034, J1725$-$4043 and J1738$-$2330]
 {Modulation of pulse energy for three pulsars. Successive 10, 24 and 5 periods    
 were averaged as a sub-integration respectively for PSRs (a) J1715$-$4034, (b) 
 J1725$-$4043 and (c) J1738$-$2330. The period numbers are kept unaltered for 
 comparison. PSR J1725$-$4043 shows weak emission during its long null regions  
 (for example periods between 1000 to 2000). PSR J1715$-$4034 shows rapid modulation of the pulse width while 
 PSR J1738$-$2330 shows burst bunches separated by long null states.}
 \label{sp_1715_1725_1738}
\end{figure}

This is one of the pulsar discovered in the PKSMB survey \cite[]{kbm+03} 
and it shows single broad integrated profile of more than 15\% duty 
cycle at 610 MHz with a prominent scattering tail. 
Width of the pulses shows significant modulation as seen in Figure \ref{sp_1715_1725_1738}(a). 
The pulsar shows two profile components at 1420 MHz \cite[]{kbm+03}, which is difficult to identify at 
610 MHz because of the interstellar scattering. Modulation in these components 
can cause this apparent change in the pulse width.
The estimation of the NF, was obtained after sub-integrating successive 10 pulses. 
As discussed in Section \ref{NF_tech_sect}, such sub-integration will cause few 
null pulses to merge with the neighbouring burst pulses. Thus, only 
a lower limit on the NF was possible to obtain using the conventional 
estimation technique. The pulse energy histograms, used in the estimation of the NF ($\geq$ 10\%), 
are shown in Figure \ref{NF_j1715}. The pulsar was too weak to obtain the NLH and the BLH. 

\subsection{J1725$-$4043}
\begin{figure}[h!]
\begin{center}
 \centering
 \includegraphics[height=4in, width=3in, angle=-90,bb=14 104 520 776]{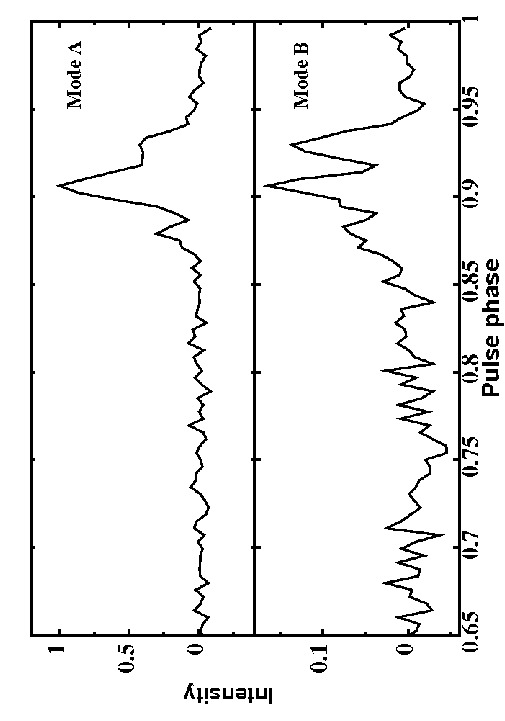}
\caption[Integrated pulse profiles for Mode A and Mode B of PSR J1725$-$4043]
{Integrated pulse profiles for Mode A and Mode B of PSR J1725$-$4043. 
The ordinate is in arbitrary units and was obtained after scaling 
the two profiles by the peak intensity of the Mode A profile. 
The Mode B is around 10 times weaker than Mode A.}
\label{Both_mode}
\end{center}
\end{figure}

This is also one of the pulsar discovered in the PKSMB survey \cite[]{kbm+03} 
with no previously reported nulling behaviour. 
Its integrated profile at 610 MHz  
exhibits three narrow components - one strong central 
component with weak trailing and leading components.
To improve the S/N, 24 successive pulses were sub-integrated. 
A section of these sub-integrations as well as the integrated profile for 
this pulsar is shown in Figure \ref{sp_1715_1725_1738}(b). 
The pulse energy histograms, obtained from the sub-integrated pulses, 
are shown in Figure \ref{NF_j1725}, which shows large fraction of null pulses. 
Visual inspection of the Figure \ref{sp_1715_1725_1738}(b) shows emission 
bunches of 50 to 200 strong pulses, which correspond to the 
normal integrated profile [hereafter referred as Mode A - 
for example periods between 120 to 264 in Figure \ref{sp_1715_1725_1738}(b)]. 
After adding all the sub-integrations during the null states, separated 
by a visual inspection, a weak profile [hereafter referred as Mode B - for 
example periods 264 to 780 in Figure \ref{sp_1715_1725_1738}(b)], different from 
the normal integrated profile (Mode A), is obtained. 
The integrated profiles for the two modes are shown in Figure \ref{Both_mode}.  
Although the two profiles are similar in shape with distinct three components, Mode B profile 
shows relatively stronger trailing component (Figure \ref{Both_mode}). 
Table \ref{j1725snrtable} shows S/N of various peaks from Mode A and Mode B profiles 
to compare their significance. Hence, it appears that the pulsar shows sporadic 
emission with two distinct modes. As the pulses were weak, the NLH and the BLH were 
not possible to obtain.
% 
%%%%%%%%%%%%%%%%%%%%%%%%%%%%%%%%%% j1725 Modes %%%%%%%%%%%%%%%%%%%%%%%%%%%%%%%%%%%%%%%%%%%%%%
%%%%%%%%%%%%%%%%%%%%%%%%%%%%%%%%%%%%%%%%%%%%%%%%%%%%%%%%%%%%%%%%%%%%%%%%%%%%%%%%%%%%%%%%%%%%
\begin{table}[h!]
\begin{center}
\begin{tabular}[h]{|c|c|c|}
\hline
Component & Mode A & Mode B \\
\hline
\hline
Central &  37     &   10  \\
\hline
Trailing & 15     &   9 \\
\hline
\end{tabular}
\caption[Peak S/N for various peaks of Mode A and Mode B profiles]
{Peak S/N for various peaks of Mode A and Mode B profiles. The S/N were 
calculated using the root mean square deviation from the off pulse regions of respective profiles.}
\label{j1725snrtable}
\end{center}
\end{table}

Visual inspection of the single pulses, forming the null 
sub-integrations,  reveals weak individual pulses among nulled pulses. 
Hence, it is difficult to identify null pulses as these could be 
low intensity Mode B pulses. The pulsar spends 30\% of time in Mode A 
emission. The remaining 70\% could be combination of null pulses 
and Mode B emission. Hence, only an upper limit on the NF, 
of around 70\% for this pulsar was possible to obtain. 
\subsection{J1738$-$2330}
\label{j1738_chap4_sect}
\begin{figure}[h!]
 \centering
 \subfigure[]{
 \includegraphics[width=3in,height=1.5in,angle=-90,bb=0 0 504 203]{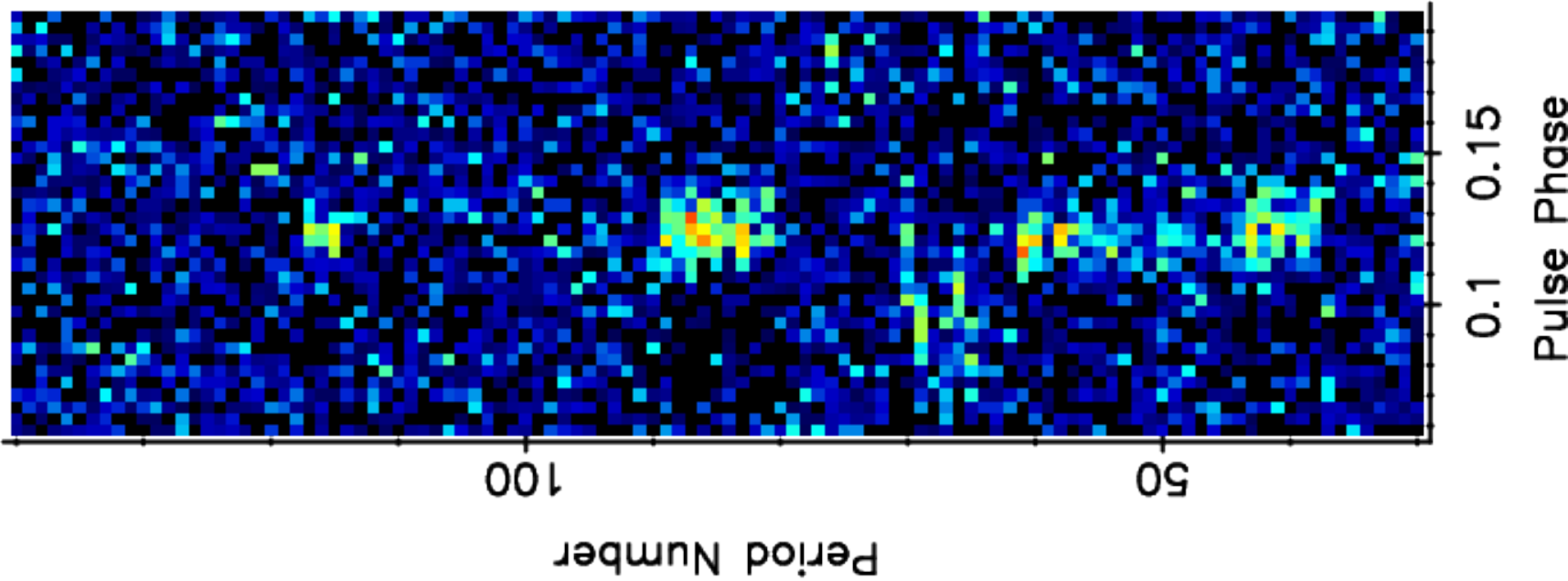}
 % b0809_spdisplay_forThesis.eps: 0x0 pixel, 300dpi, 0.00x0.00 cm, bb=503 202 1 1
 }
 \subfigure[]{
 \includegraphics[width=3in,height=1.5in,angle=-90,bb=0 0 504 203]{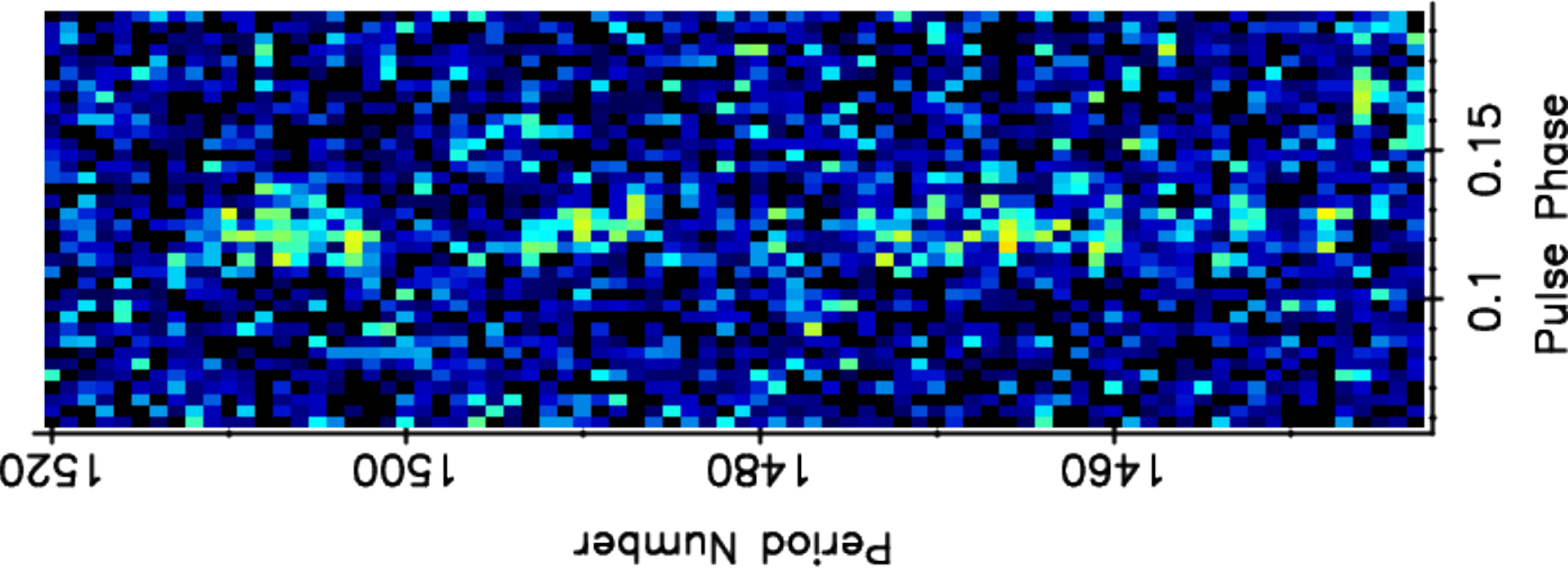}
 % b0809_spdisplay_forThesis.eps: 0x0 pixel, 300dpi, 0.00x0.00 cm, bb=503 202 1 1
 }
 \subfigure[]{
 \includegraphics[width=3in,height=1.5in,angle=-90,bb=0 0 504 203]{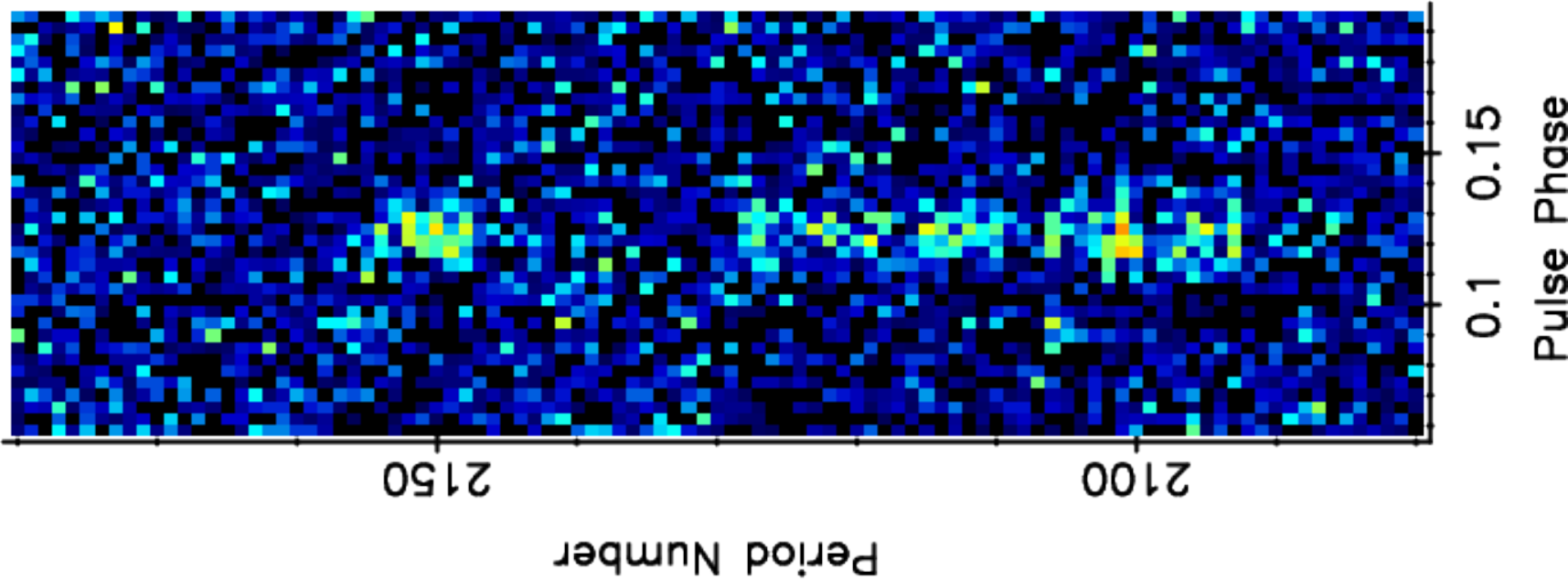}
 % b0809_spdisplay_forThesis.eps: 0x0 pixel, 300dpi, 0.00x0.00 cm, bb=503 202 1 1
 }
 \caption[Three example of burst bunches observed at 325 MHz obtained from PSR J1738$-$2330]
 {Three example of burst bunches observed at 325 MHz obtained from PSR J1738$-$2330. 
 Note the structure of burst bunches with longer to smaller short bunches, 
 separated by short null states, inside each big bunch. Details regarding this peculiar 
 behaviour are further discussed from longer observations in Chapter 5.}
 \label{j1738_bunches}
\end{figure}
\begin{figure}[h!]
 \begin{center}
 \centering
 \includegraphics[width=3in,height=4in,angle=90,bb=0 0 505 685]{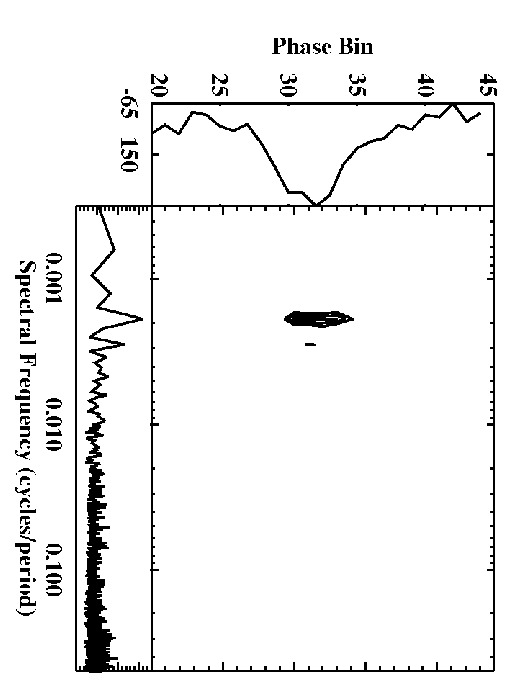}
 \caption[Phase-resolved spectra for PSR J1738$-$2330]
 {Phase-resolved spectra for PSR J1738$-$2330 
 (see Figure \ref{prfs_b0809} for details regarding panels). 
 The spectra in the bottom panel and in the contours 
 are shown with the abscissa plotted in a logarithmic scale 
 upto the Nyquist frequency.}
\label{j1738lrf}
\end{center}
\end{figure}

This pulsar was also discovered in the PKSMB survey \cite[]{lfl+06}. 
We observed this pulsar at 325 MHz using the GMRT. 
We are reporting a unique nulling 
behaviour of this pulsar in this chapter.
The pulsar seems to have quasi-periodic bursts, 
with an average duration of around 50 to 100 periods,  
interspersed with nulls of around 300 to 400 periods. 
This interesting single pulse behaviour in this pulsar is evident 
in Figure \ref{sp_1715_1725_1738}(c), where a plot,  
with 5 successive single pulses integrated, is shown.
A visual inspection shows that it seems to have regular periodic bursts for total   
1200 periods (bursts at period number 1700, 2100, 2600 and 3100 in Figure \ref{sp_1715_1725_1738}(c)). 
Figure \ref{j1738lrf} shows the LRFS for this pulsar. 
It shows quasi-periodicities at nearly 
0.0019 cycles per period and 0.0028 cycles per period, 
which correspond to periodicities of 
approximately 525 and 350 periods, respectively. 
Such large periodicities can only occur from a quasi-periodic modulation of the pulse 
energy. This quasi-periodic behaviour is similar to PSR B1931+24 but with 
much shorter time-scale. Figure \ref{NF_j1738} shows the on-pulse and the off-pulse 
energy histograms from the sub-integrated pulses. Only a lower limit of 
around 69\% was possible to obtain from these data due to the sub-integration. 
As the pulses were weak, the NLH and the BLH were not possible to obtain. 

Close examination of single pulses suggests that the bursts 
typically consist of a sequence of a 20 to 30 period 
bursts followed by 2 short bursts of 5 to 10 pulses. 
These three bursts are separated by shorter nulls of 
about 10 to 25 pulses. The overall burst bunches are 
separated by around 400 periods nulls. Three examples 
of this burst pattern are shown in Figure \ref{j1738_bunches}. 
However, the single pulse S/N was too low to confirm this with high 
significance. A more sensitive observations, motivated from 
this study, was carried out and more details regarding this pulsar's 
nulling behaviour are discussed in Chapter 5. 

\subsection{J1901+0413}
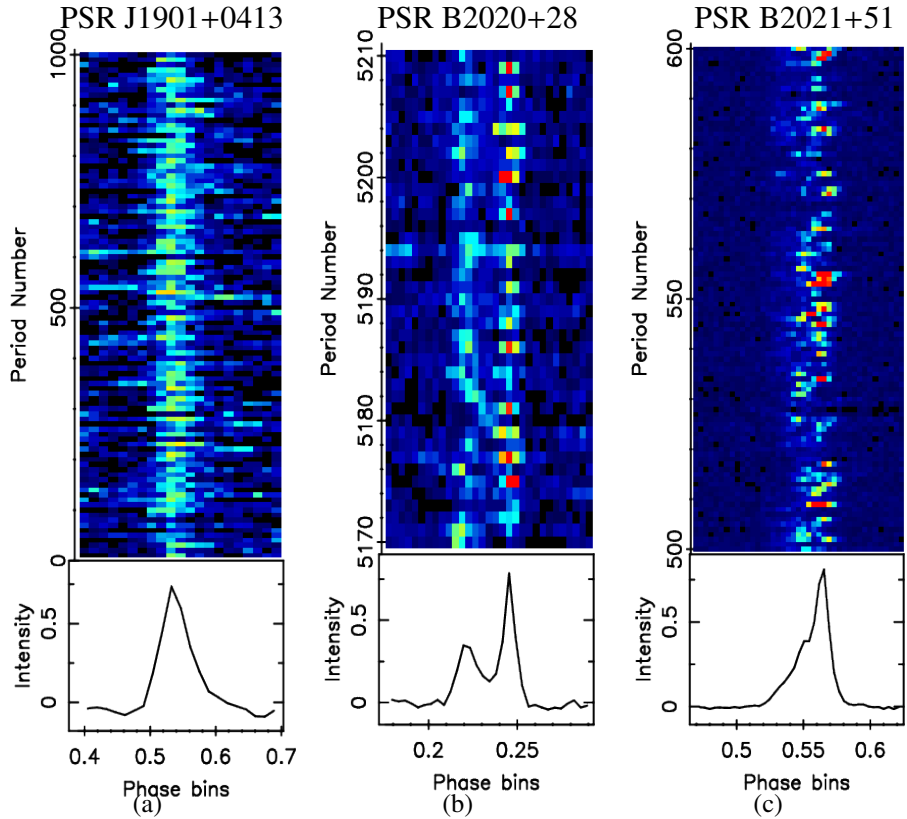
\begin{figure}[h!]
 \centering
 \subfigure[]{
 \includegraphics[width=4in,height=1.5in,angle=-90,bb=0 0 504 203]{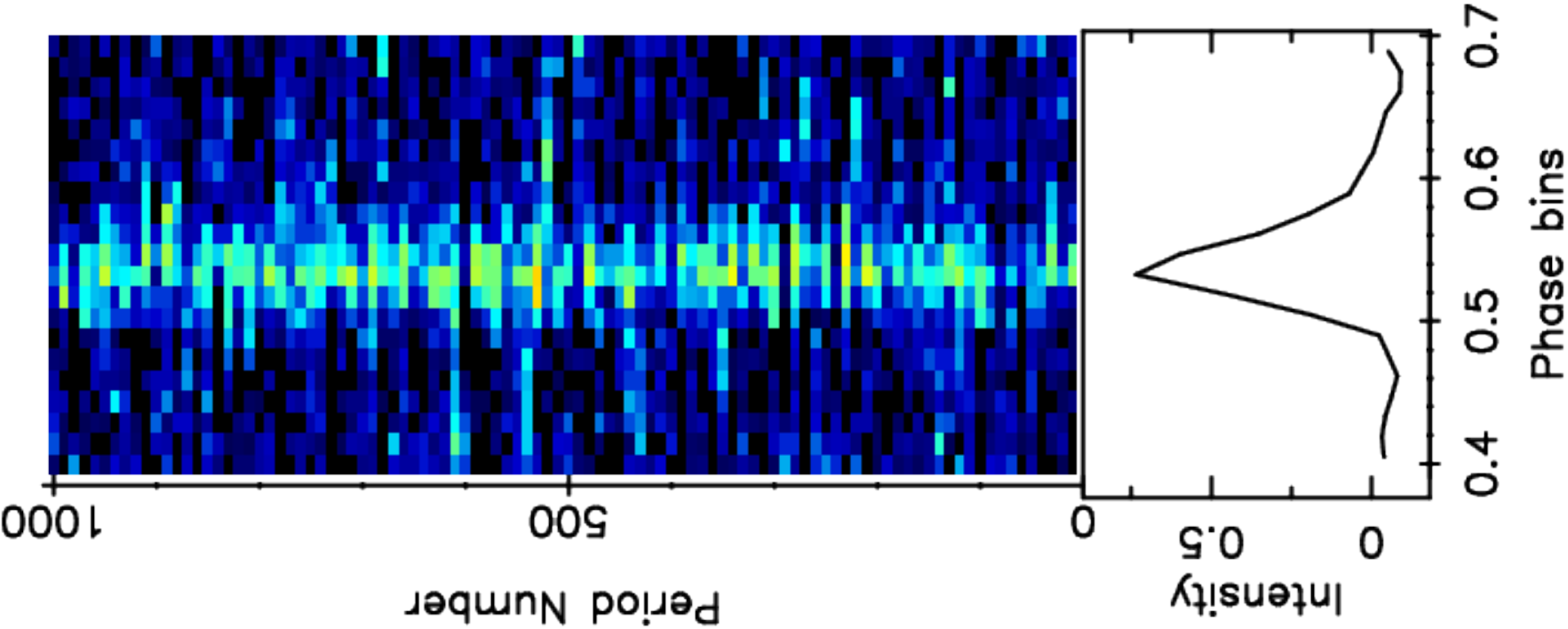}
 % b0809_spdisplay_forThesis.eps: 0x0 pixel, 300dpi, 0.00x0.00 cm, bb=503 202 1 1
 }
 \subfigure[]{
 \includegraphics[width=4in,height=1.5in,angle=-90,bb=0 0 504 203]{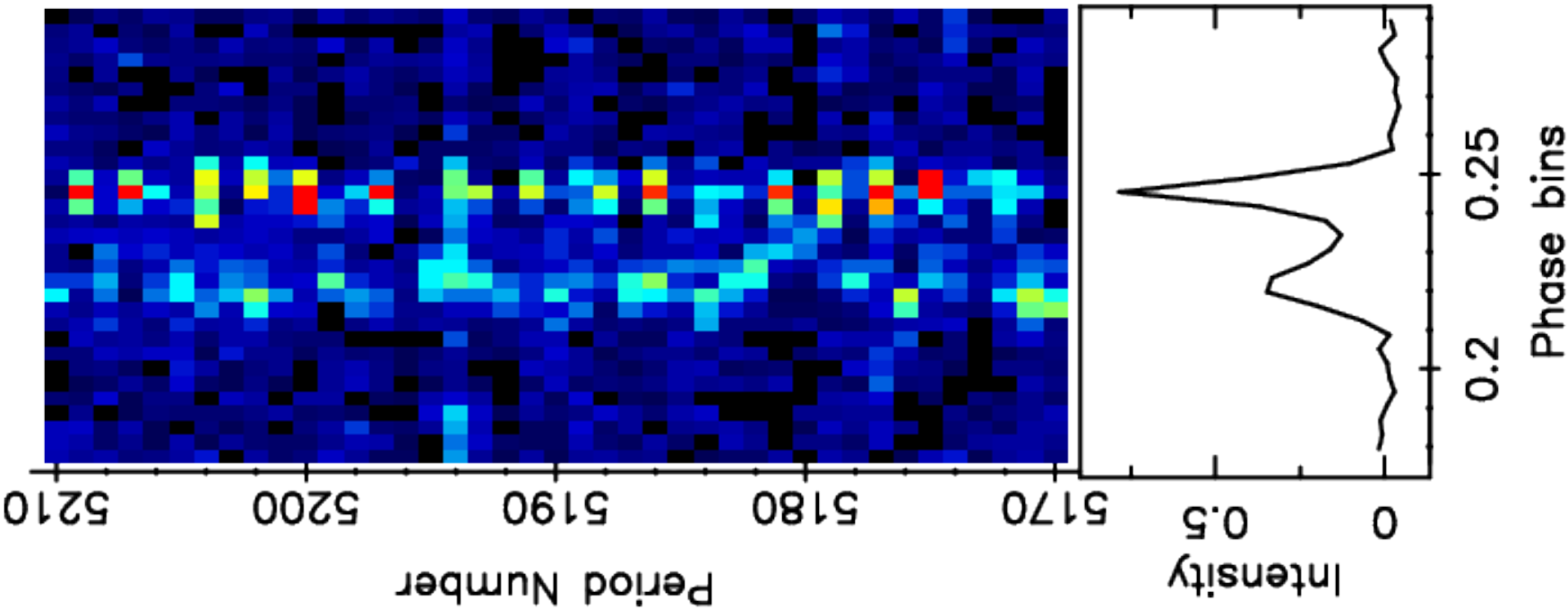}
 % b0809_spdisplay_forThesis.eps: 0x0 pixel, 300dpi, 0.00x0.00 cm, bb=503 202 1 1
 }
 \subfigure[]{
 \includegraphics[width=4in,height=1.5in,angle=-90,bb=0 0 504 203]{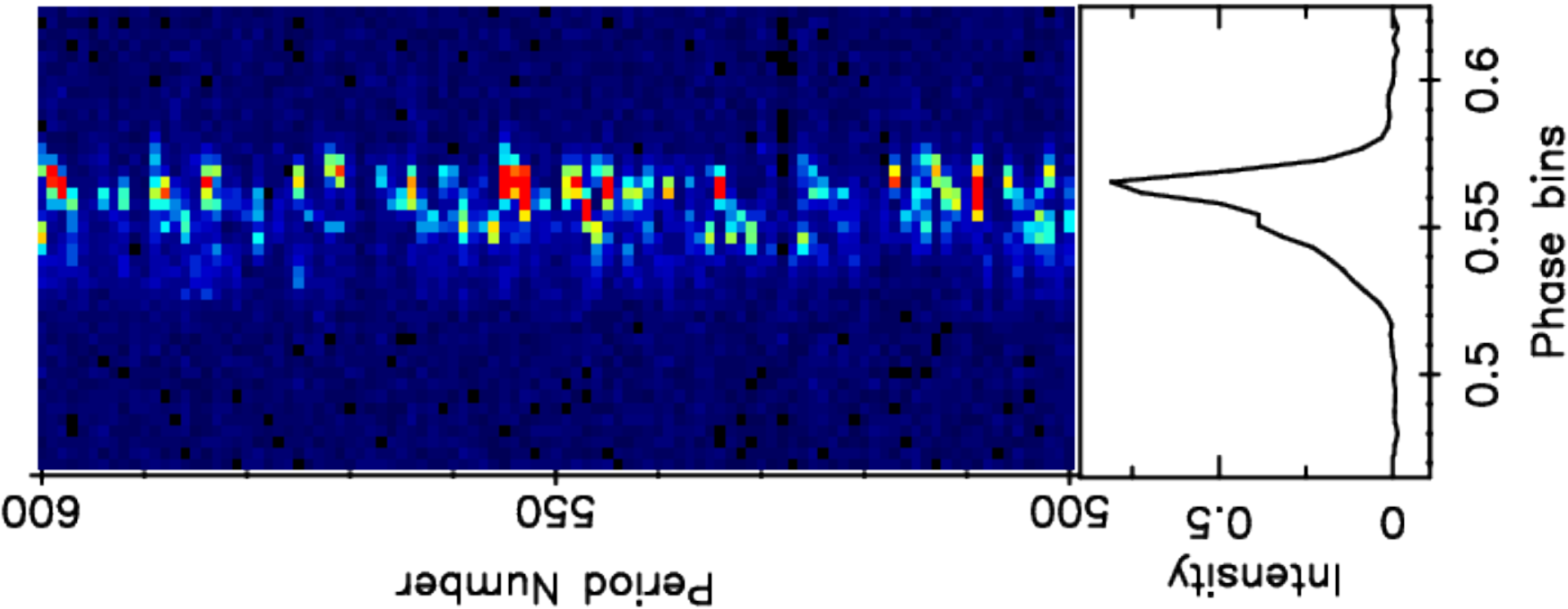}
 % b0809_spdisplay_forThesis.eps: 0x0 pixel, 300dpi, 0.00x0.00 cm, bb=503 202 1 1
 }
 \begin{picture}(0,0)
  \put(-330,0){PSR J1901+0413}
  \put(-210,0){PSR B2020+28}
  \put(-90,0){PSR  B2021+51} 
 \end{picture}
 \caption[Modulation of pulse energy for PSRs J1901+0413, B2020+28 and B2021+51]
 {Modulation of pulse energy for PSRs  (a) J1901+0413, 
 (b) B2020+28 and (c) B2021+51. For PSR J1901+0413, successive 10 periods    
 were sub-integrated to improve the S/N for this display with period numbers 
 unaltered for comparison. Note the even-odd modulation in the trailing component of 
 PSR B2020+28.}
 \label{sp_1901_2020_2021}
\end{figure}
This pulsar was also discovered in the PKSMB survey \cite[]{mhl+02} with 
no previously reported nulling behaviour. 
The pulsar showed single component profile with 15\% duty cycle
with a prominent scattering tail at 610 MHz [shown 
in the bottom panel of Figure \ref{sp_1901_2020_2021}(a)].
The single pulses were weak, hence 10 successive single pulses were 
sub-integrated, as shown in Figure \ref{sp_1901_2020_2021}(a), to display 
the pulse energy modulation. Visual inspection of the single pulses showed 
small amount of nulling, hence estimation of the NF using conventional technique 
was not possible. It was only possible to obtain an upper limit using the similar method 
discussed for PSR J1639$-$4359. We arranged all the single pulses in the 
ascending order of their on-pulse energy. A threshold was moved 
from low to high on-pulse energy end till the pulses below the 
threshold did not show a collective profile with a pulse with 
high significance (S/N $>$ 3). All the pulses below such threshold were tagged as null pulses. 
The fraction of these pulses gave an estimate for the upper limit on 
the NF ($<$ 6\%). As the pulses were weak, the NLH and the BLH were not 
possible to obtain. 

\subsection{B2020+28}
\begin{figure}[h!]
 \centering
 \includegraphics[width=3 in,height=4 in,angle=-90,bb=14 14 565 730]{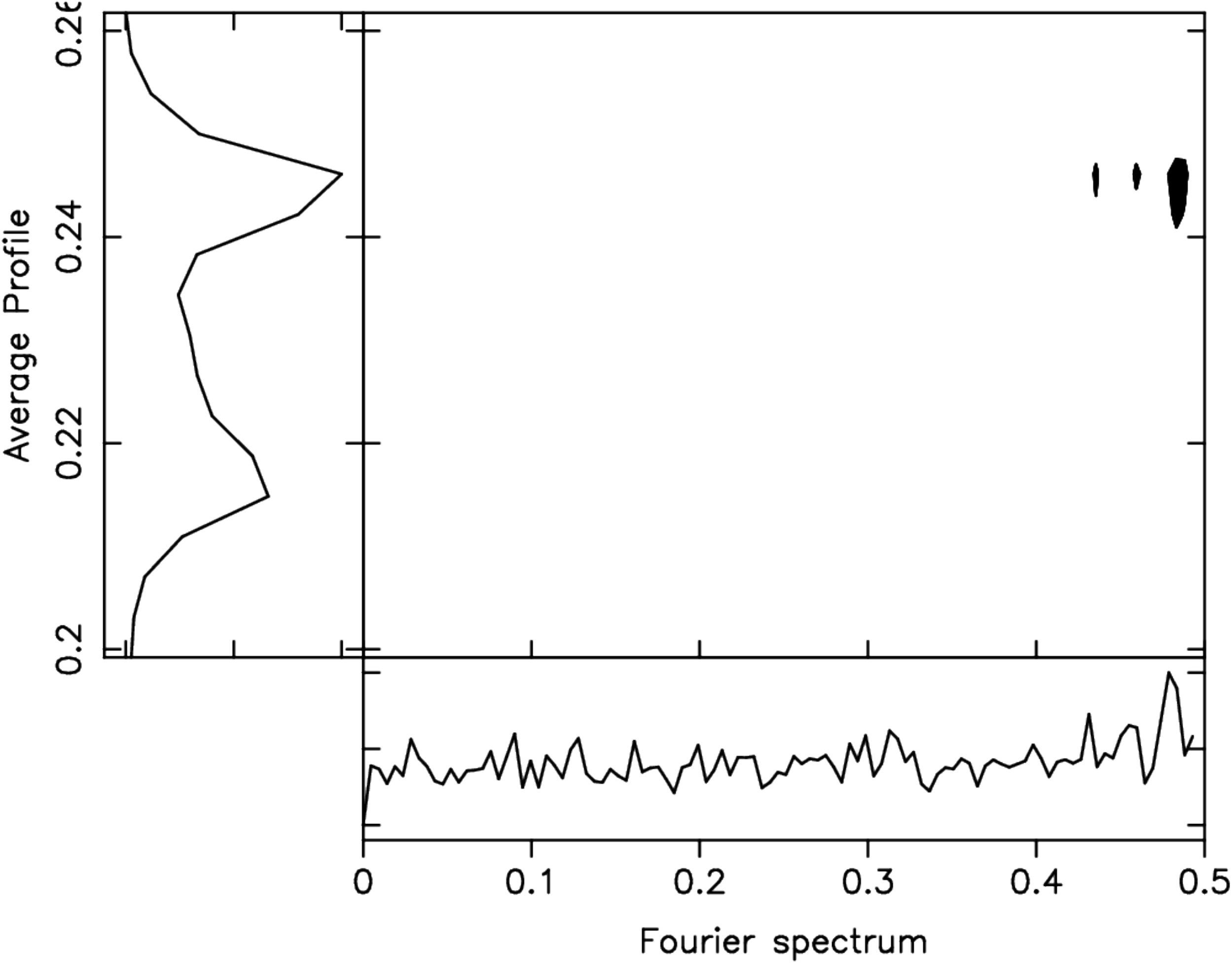}
\caption[LRFS for PSR B2020+28]{LRFS for PSR B2020+28 (see Figure \ref{prfs_b0809} for details regarding panels). 
 The combined spectra shows peak at 0.48 cycles per period, near the alias frequency (Nyquist frequency) 
 which corresponds to periodicity of around 2.1 pulses in the trailing component.}
 \label{prfs_b2020}
\end{figure}

This is one of the well studied pulsar. The pulsar profile shows two strong components 
[shown in the bottom panel of Figure \ref{sp_1901_2020_2021}(b)], 
but this pulsar was classified as triple profile class 
pulsar due the core emission in the saddle region between the two components \cite[]{rsw89}. 
A small section of single pulses is shown in Figure \ref{sp_1901_2020_2021}(b). 
Interestingly, the emission in the individual profile components 
reduces significantly for only for a small fraction of pulses, which is 
different for the two components. A clear even-odd modulation 
pattern can be seen in the trailing component while 
the leading component does not show such rapid fluctuations. 
To scrutinise these modulations, the LRFS was obtained 
(as shown in Figure \ref{prfs_b2020}) which clearly 
shows even-odd modulation in the trailing component. 
Our results matches with the earlier reported behaviour by \cite{nuk+82}. 
The pulsar shows small amount of nulling when the entire pulse longitude 
is considered [see Figure \ref{sp_1901_2020_2021}(b)]. 
The pulse energy histograms are shown in Figure \ref{NF_b2020} which 
shows significantly small number of pulses near the zero pulse energy. 
We estimated NF of 0.2$\pm$1.6\% of for this pulsar. 
Estimation of the reduction in the pulse energy was around 2.5$\pm$0.2. 
The large errors in the NF is due to the fewer number 
of full null pulses in the data. However, there are many pulses 
which show partial nulls (i.e. null in only one of the component). 
Our time resolution was not sufficient to identify the saddle region clearly so we 
were able to estimate the NF for two components only. The 
estimated NF for the leading component is 3.5$\pm$0.8\% 
and $\eta$ is around 9.7.  The estimated NF for the trailing 
component is 9$\pm$1 \% and $\eta$ is around 21, suggesting 
that at least the trailing component shows a nulling behaviour 
similar to a regular nulling pulsar. Both the NFs 
were significantly higher than the overall NF. 
Due to the small number of full null pulses, the NLH and the BLH 
were not possible to obtain. 

\subsection{B2021+51}
This pulsar was reported nulling by \cite{rit76} with only an upper limit on the 
NF($\leq$ 5\%). A small section of observed single pulses is shown 
in Figure \ref{sp_1901_2020_2021}(c). 
As can be seen from the single pulse plot, we were able to obtain sufficient 
S/N for single pulses. The pulse energy histograms are shown in Figure \ref{NF_b2021}. 
We were able to obtain a better constrain on the NF (1.4$\pm$0.7\%). 
The estimated reduction in the pulse energy, $\eta$ is 2.6$\pm$0.2.  
The null and burst length histograms are shown in Figure \ref{nlh_blh_b2021}, 
obtain using 12 null lengths and 12 burst lengths. 
The NLH shows that 60\% of nulls are single period nulls 
while the rest 40\% nulls are double period nulls. 
This nearly equal distribution 
between single and double period nulls can be extrapolated to claim small number of 
shorter nulls (i.e. size smaller than pulsar period) compared to 
large number of such nulls in PSR B0835$-$41. 
Such large number of double period nulls, 
assuming exponential null length distribution seen in most pulsars, 
suggest the pulsar to have few long nulls (i.e. longer than double period) which 
are not seen in our data. BLH histogram shows around 65\% of 
burst are less than 100 periods while the distribution extends up to
800 periods. 
\subsection{B2034+19}
\begin{figure}[h!]
 \centering
 \subfigure[]{
 \includegraphics[width=4in,height=1.5in,angle=-90,bb=0 0 504 203]{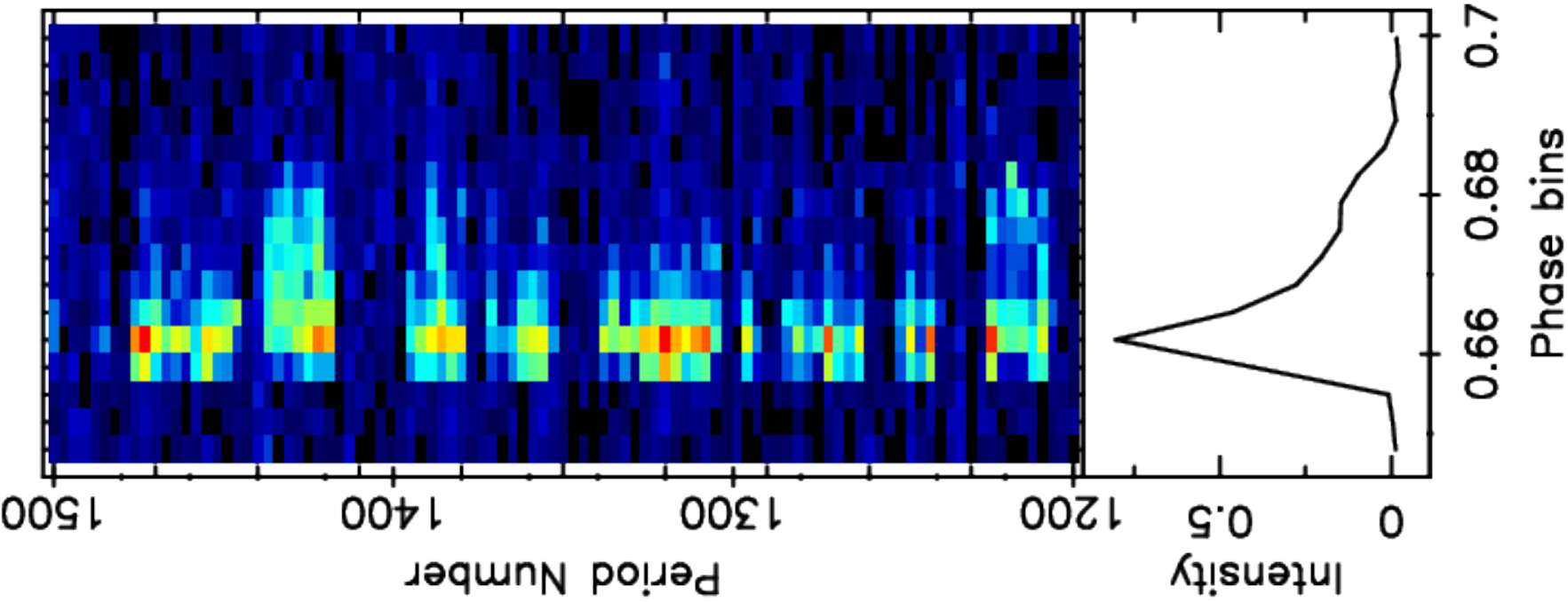}
 % b0809_spdisplay_forThesis.eps: 0x0 pixel, 300dpi, 0.00x0.00 cm, bb=503 202 1 1
 }
 \subfigure[]{
 \includegraphics[width=4in,height=1.5in,angle=-90,bb=0 0 504 203]{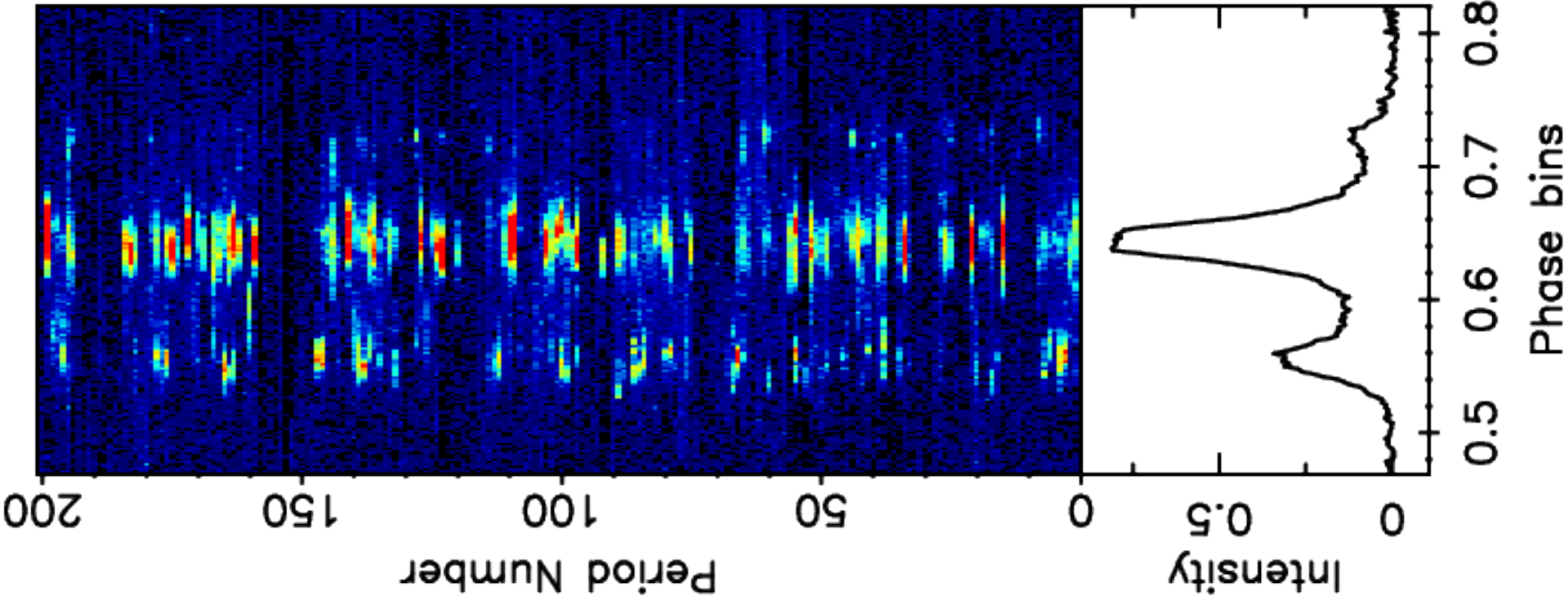}
 % b0809_spdisplay_forThesis.eps: 0x0 pixel, 300dpi, 0.00x0.00 cm, bb=503 202 1 1
 }
 \subfigure[]{
 \includegraphics[width=4in,height=1.5in,angle=-90,bb=0 0 504 203]{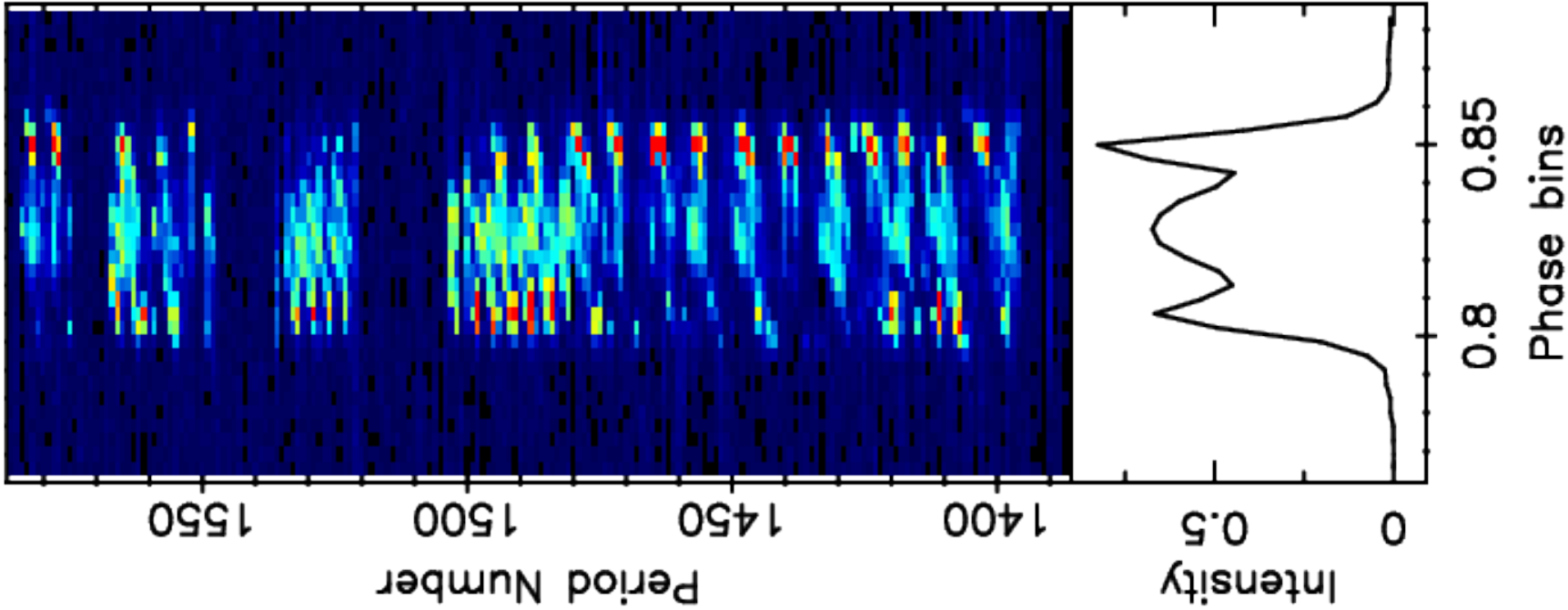}
 % b0809_spdisplay_forThesis.eps: 0x0 pixel, 300dpi, 0.00x0.00 cm, bb=503 202 1 1
 }
 \begin{picture}(0,0)
  \put(-330,0){PSR B2034+19}
  \put(-210,0){PSR B2111+46}
  \put(-90,0){PSR  B2319+60} 
 \end{picture}
 \caption[Modulation of the pulse energy for PSRs B2034+19, B2111+46 and B2319+60]
 {Modulation of the pulse energy for PSRs  (a) B2034+19, (b) B2111+46 and 
 (c) B2319+60. For PSR B2034+19, successive 3 periods were sub-integrated 
 to improve the S/N for this display with period numbers 
 unaltered for comparison. PSR B2111+46 shows interesting modulation pattern in  
 its three components, with different components showing independent occurrences of 
 null events for a few periods. PSR B2319+60 clearly shows change in the  
 drift rate around period number 1480 where it also switches the profile modes.}
 \label{sp_2037_2111_2319}
\end{figure}
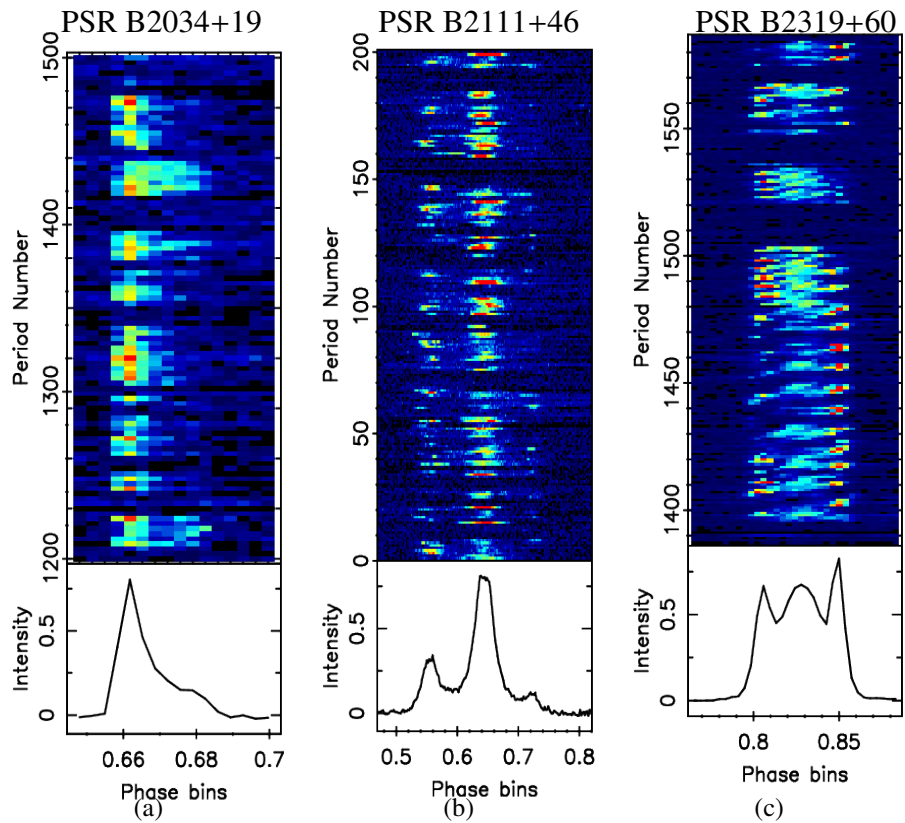
This pulsar was reported nulling  by \cite{hr09}. 
We observed this pulsar at 610 MHz and a section of 
observed pulse sequence is shown in Figure \ref{sp_2037_2111_2319}(a). 
To improve the S/N, successive 3 periods were sub-integrated. 
The pulse energy histograms are shown in Figure \ref{NF_b2037}, obtained 
using these sub-integrated data. Hence, we were able 
to estimate only a lower limit on the NF ($\geq$26\%). 
Although, the estimated NF matches with the earlier reported value within the error bars, 
it is slightly smaller due to our inability to detect short nulls. 
By arranging pulses in the ascending order of their on-pulse energy 
and using the variable threshold method discussed in Section \ref{separation_of_null_burst_sect}, 
we were able to clearly separate null and burst pulses from the single 
pulse data. Using these separated null and burst pulses, with estimated 
$\eta$ of 6.4$\pm$0.1, the obtained null and burst length histograms 
are shown in Figure \ref{nlh_blh_b2037}. Total 336 null lengths 
and 336 burst lengths were used to obtain these length histograms. 
The NLH shows gradual distribution of null lengths up to 10 periods. 
This is small given that pulsar spends more than 26\% of time in the null state. 
The BLH shows more than 50\% of bursts are of size smaller than 3 periods while the 
other half extends up to 50 periods. 
\subsection{B2111+46}
\begin{figure}[h!]
 \centering
 \subfigure[]{
 \includegraphics[width=3.7in,height=2.0in,angle=-90,bb=0 0 504 203]{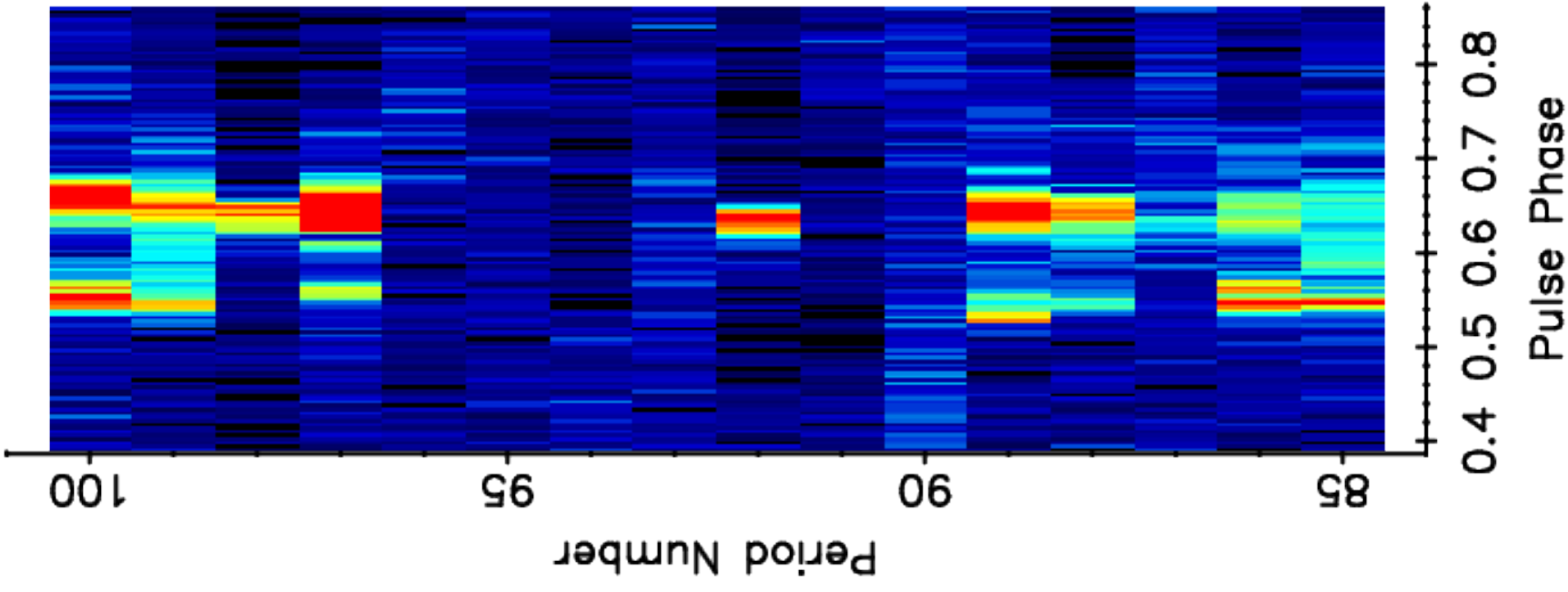}
 % b0809_spdisplay_forThesis.eps: 0x0 pixel, 300dpi, 0.00x0.00 cm, bb=503 202 1 1
 }
 \subfigure[]{
 \includegraphics[width=3.7in,height=2.0in,angle=-90,bb=0 0 504 203]{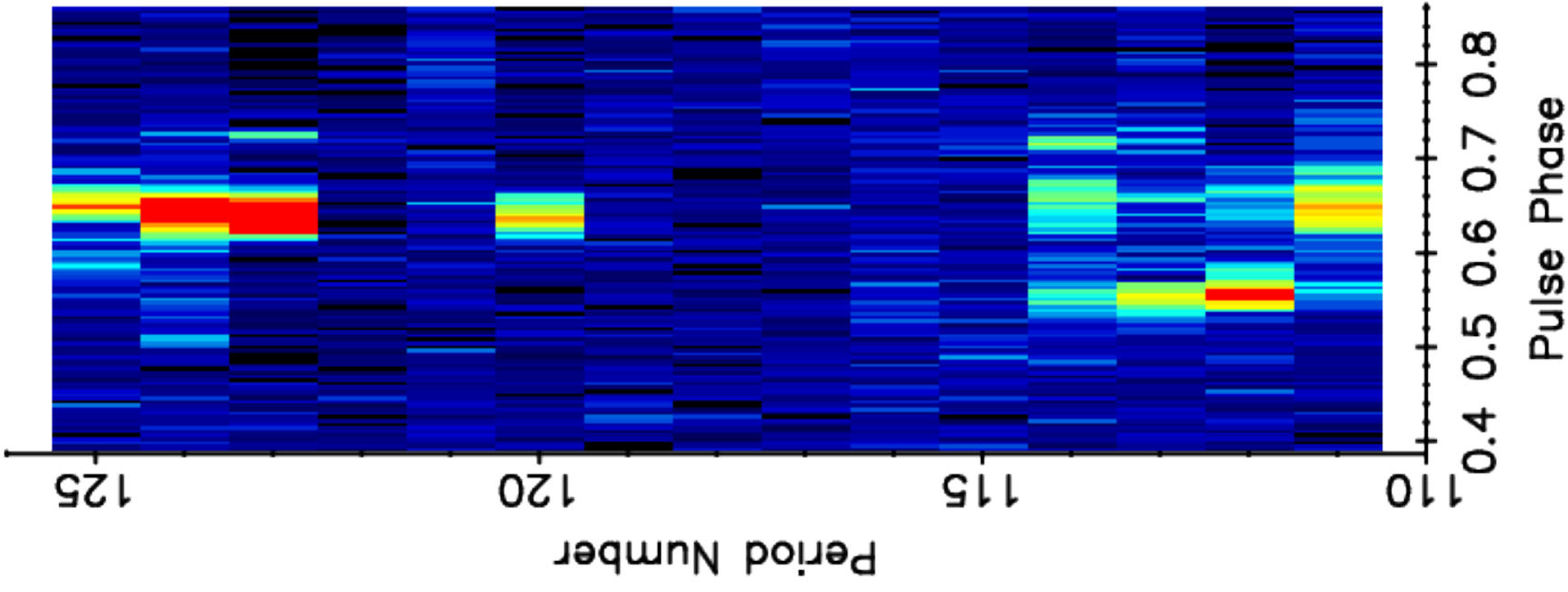}
 % b0809_spdisplay_forThesis.eps: 0x0 pixel, 300dpi, 0.00x0.00 cm, bb=503 202 1 1
 }
 \subfigure[]{
 \includegraphics[width=3.7in,height=2.0in,angle=-90,bb=0 0 504 203]{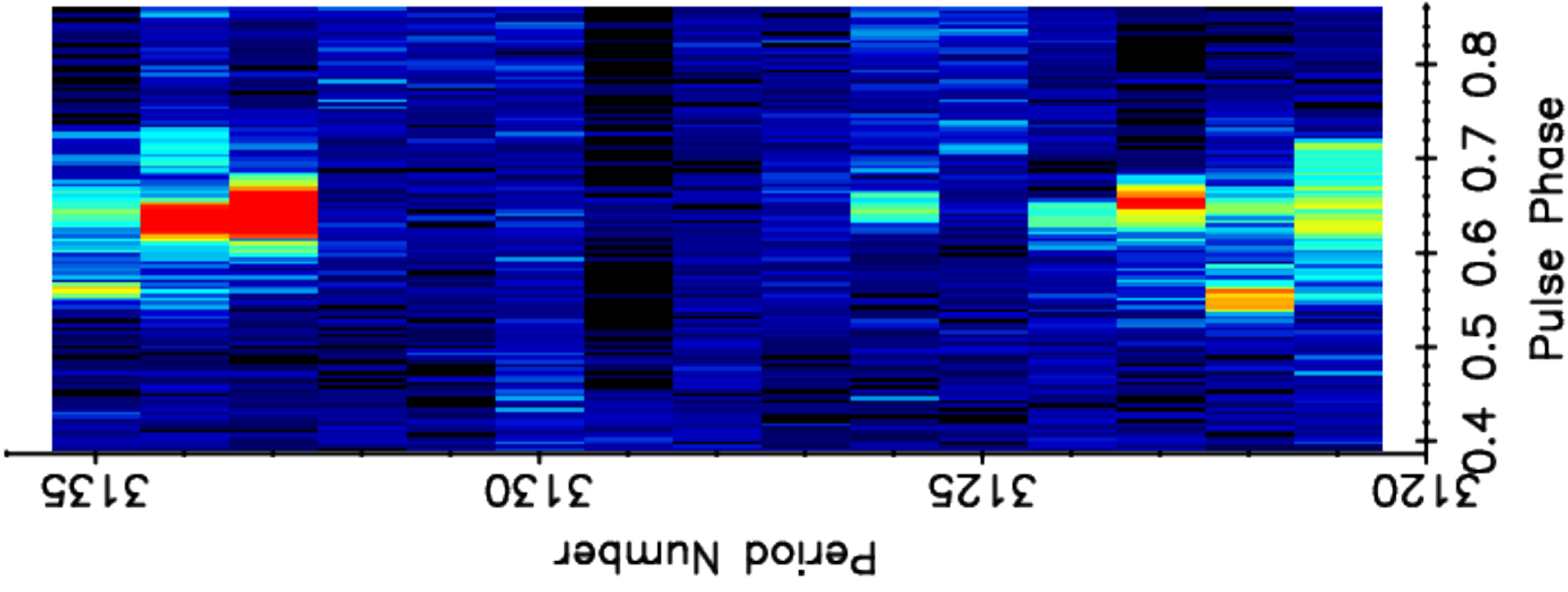}
 % b0809_spdisplay_forThesis.eps: 0x0 pixel, 300dpi, 0.00x0.00 cm, bb=503 202 1 1
 }
 \subfigure[]{
 \includegraphics[width=3.7in,height=2.0in,angle=-90,bb=0 0 504 203]{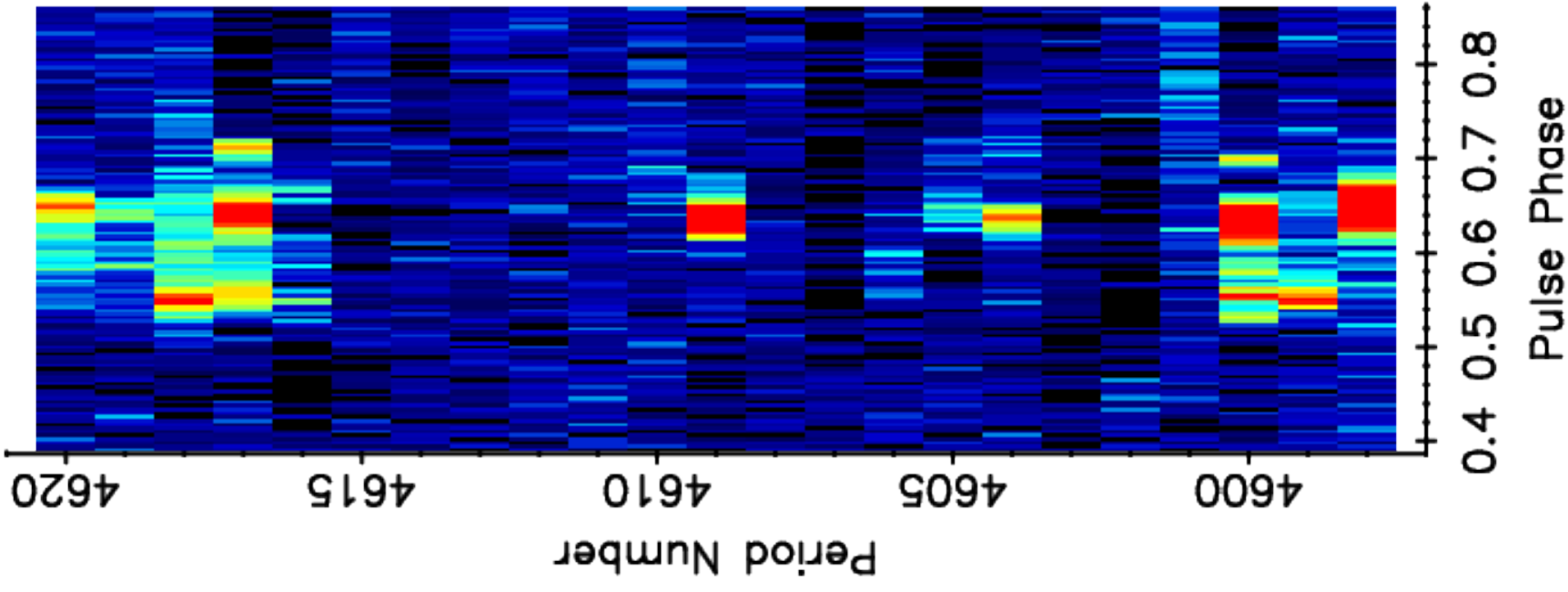}
 % b0809_spdisplay_forThesis.eps: 0x0 pixel, 300dpi, 0.00x0.00 cm, bb=503 202 1 1
 }
 \caption[Four example of long nulls seen in PSR B2111+46]
 {Four example of long nulls seen in PSR B2111+46, observed at 610 MHz. Note the 
 presence of core emission in the individual burst periods inside the long null regions, 
 where both the conal components show clear nulls.}
 \label{longnull_b2111}
\end{figure}
This pulsar is one of the long period well studied pulsar 
with multicomponent profile [bottom panel of Figure \ref{sp_2037_2111_2319}(b)]. 
\cite{rsw89} classified this pulsar as triple component profile class 
pulsar but \cite{zqh+07} reported two additional components in the saddle region connecting 
the core component and the outer components. 
The depolarisation shows swing in the middle component at 600 MHz \cite[]{gl98}, 
hence, the profile is of core multicomponent class. 
A section of the observed single pulses at 610 MHz 
is shown in Figure \ref{sp_2037_2111_2319}(b). 
Our data shows stronger emission 
in the saddle region compared to core and outer conal 
components for many individual pulses. 
Nulling in this pulsar was first reported by \cite{rit76}. 
Figure \ref{NF_b2111} shows the conventional pulse energy histograms 
obtained after considering the entire pulse longitude. 
We measured a slightly higher NF (21$\pm$4\%) compared to
the previously reported estimate by \cite{rit76}.  
Like PSR B2020+28, the fraction of pulses for which emission 
is not detected varies from component to component.  
For the core component, NF was estimated to be of 26$\pm$2\%, 
slightly higher than the NF of the entire pulse. 
The estimated $\eta$ value for this component is 
around 56. For the leading component, NF was found to be 
53$\pm$5\%, almost double the NF for the entire 
pulse. The NLH and the BLH, shown in Figure \ref{nlh_blh_b2111}, 
considering the emission in the entire pulse are 
reported here for the first time for this pulsar. 
The NLH shows 50\% of nulls are single period nulls while other 
50\% are gradually distributed up to 10 periods. 
Total 145 null lengths and 145 burst lengths were used to obtain these 
length histograms. 

Visual inspection of the single pulse data shows many 
long nulls with lengths varying from 8 to 16 periods. 
Many of these long nulls show single burst from the 
core component located inside them. 
Some of the interesting cases of such 
nulls are shown in the Figure \ref{longnull_b2111}.
This behaviour has been reported for the first time in this pulsar 
to the best of our knowledge. It is consistent with the estimated NF of the leading 
component (around 52\%) compared to the NF of the core component (around 26\%) 
as these burst pulses reduce the NF of the core component. 
However, length of our observations were not sufficient to 
claim a statistically significant result of this behaviour. 
Our study motivates longer observations of this pulsar as 
many of the interesting single pulse behaviour has not been 
investigated yet. 
\subsection{B2319+60}
This is also one the long period well studied pulsar. 
Pulsar has three component at 610 MHz. 
It has reported NF of around 25$\pm$5\% 
\cite[]{rit76}. Observed single pulse data show 
that this pulsar displays a regular switching between 
three different profile modes. During two of these modes, 
it exhibits regular drifting behaviour while in the third mode 
no such drifting feature is seen. Figure \ref{sp_2037_2111_2319}(c) 
shows a section of observed single pulses from this pulsar. 
A clear change in the drift rate (transition from slower drift mode 
to faster drift mode) is clearly seen around period number 1480. 
This matches with the reported behaviour for this pulsar at 1 GHz by \cite{wf81}
[also discussed in Section \ref{null_dritft_corr_sect}]. 
More details regarding the broadband nulling behaviour 
of this pulsar are discussed in Chapter 6. 
Figure \ref{NF_b2319} shows the obtained pulse energy histograms.  
The estimated NF, using these histograms, is around 29$\pm$1 \% and 
estimated $\eta$ is around 115.8$\pm$0.4 which is second highest in our sample. 
Figure \ref{nlh_blh_b2319} shows the exponentially declining length 
distributions where the NLH extends up to 30 periods 
while the BLH extends up to 90 periods. Total 225 null lengths and 225 burst lengths 
were used to construct these length histograms. 
% \begin{figure}[h!]
%  \centering
%  \subfigure[]{
%  %  \includegraphics[width=4 in,height=2 in,angle=-270,bb=606 793 1 501]{Drift_example.eps}
%  \includegraphics[width=3.5in,height=1.5in,angle=-90,bb=44 574 546 775]{b2319.modeA.spdisplay.eps}
%  % b0809_spdisplay_forThesis.eps: 0x0 pixel, 300dpi, 0.00x0.00 cm, bb=503 202 1 1
%  }
%  \subfigure[]{
%  %  \includegraphics[width=4 in,height=2 in,angle=-270,bb=606 793 1 501]{Drift_example.eps}
%  \includegraphics[width=3.5in,height=1.5in,angle=-90,bb=44 574 546 775]{b2319.modeB.spdisplay.eps}
%  % b0809_spdisplay_forThesis.eps: 0x0 pixel, 300dpi, 0.00x0.00 cm, bb=503 202 1 1
%  }
%  \subfigure[]{
%  %  \includegraphics[width=4 in,height=2 in,angle=-270,bb=606 793 1 501]{Drift_example.eps}
%  \includegraphics[width=3.5in,height=1.5in,angle=-90,bb=44 574 546 775]{b2319.modeC.spdisplay.eps}
%  % b0809_spdisplay_forThesis.eps: 0x0 pixel, 300dpi, 0.00x0.00 cm, bb=503 202 1 1
%  }
%  \begin{picture}(0,0)
%   \put(-330,0){Mode A}
%   \put(-210,0){Mode B}
%   \put(-90,0){Mode C} 
%  \end{picture}
%  \caption{}
%  \label{sp_2037_2111_2319}
% \end{figure}
% 
\nopagebreak
%%%%%%%%%%%%%%%%%%%%%%%%%%%%%%%%%%%%%%%%%%%% Table of Results %%%%%%%%%%%%%%%%%%%%%%%%%%%%%%%%%
\subsection{Summary of results}
The summary of results on all pulsars is shown in Table \ref{Results}. It also shows 
basic parameters\footnote{\cite{gg74,dtms88,jml+95,lylg95,wmz+01,mhl+02,kbm+03,hlk+04,lfl+06} 
and ATNF Catalogue : \url{http://www.atnf.csiro.au/research/pulsar/psrcat} \cite[]{mhth05}} 
along with the previously reported NF for each pulsar. As it can be seen from Table \ref{Results}, 
our study reported NFs for around 15 pulsars out of which 5 were PKSMB pulsars (marked 
with a dagger in Table \ref{Results}) with no previously reported nulling behaviour. 

\begin{landscape}
% \addtolength{\oddsidemargin}{-.875in}
\hspace{1cm}
\begin{center}
\begin{table}[h!]
\centering
{\small
\footnotesize
% \scriptsize
% \centering
\begin{tabular}[ht]{||c|c|c|c|c|c|c|c|c|c|}
\hline
J2000 	     & B1950 	 & Period  & DM          & S1400 &Obtained& Known&  $\eta$  & Number of  &  N (Sub-integration) \\
Name  	     & Name  	 &         &             &       & NF    & NF    &          & Runs       &                      \\  
 	     &		 & (s)	   & (pc/cm$^3$) & (mJy) & (\%)  & (\%)  &   $-$    & $-$        & $-$			\\ 
\hline
\hline
J0814+7429   &  B0809+74   & 1.292241 & 06.1  & 10.0 & 1.0(0.4) & 1.42(0.02) $^{[1]}$   & 172.0(0.5)   & 246 & 13766 (1)  \\
J0820$-$1350 &  B0818$-$13 & 1.238130 & 40.9  & 7.0  & 0.9(1.8) & 1.01(0.01) $^{[1]}$   & 4.2(0.2)     & 114  & 3341 (1)  \\
J0837$-$4135 &  B0835$-$41 & 0.751624 & 147.2 & 16.0 & 1.7(1.2)   & $\leq$1.2 $^{[2]}$    & 15.7(0.2)  & 148  & 3335 (1)  \\
J1115+5030   &  B1112+50   & 1.656439 & 9.2   & 3.0  & 64(6)     & 60(5) $^{[3]}$       & 44.7(0.2)  & 1270 & 2634 (1)  \\
J1639$-$4359$^\dag$ &    $-$      & 0.587559 & 258.9 & 0.92 & $\leq$0.1  & $-$	                  & $-$         & $-$ & 13034 (1)  \\
J1701$-$3726 &  B1658$-$37 & 2.454609 & 303.4 & 2.9  & 22(4)     & $\geq$14 $^{[5]}$     & 6.4(0.2)   & 146 & 2464 (1)  \\
J1715$-$4034$^\dag$ &    $-$      & 2.072153 & 254.0 & 1.60 & $\geq$10    & $-$                   & $-$         & $-$ & 1591 (10) \\
J1725$-$4043$^\dag$ &    $-$      & 1.465071 & 203.0 & 0.34 & $\leq$70   & $-$		          & $-$         & $-$ & 2481 (24) \\
J1738$-$2330$^\dag$ &	  $-$      & 1.978847 & 99.3  &	0.48 & $\geq$69   & $-$	                  & 5.3(0.3)   & $-$ & 2178 (5)  \\
J1901+0413$^\dag$   &    $-$      & 2.663080 & 352.0 & 1.10 & $\leq$6          & $-$                   & $-$         & $-$ & 2605 (1) \\
J2022+2854   &  B2020+28   & 0.343402 & 24.6  & 38   & 0.2(1.6)  & $\leq$ 3 $^{[3]}$     & 2.5(0.2)   & $-$ & 8039 (1)  \\
J2022+5154   &  B2021+51   & 0.529196 & 22.6  & 27.0 & 1.4(0.7)   & $\leq$5 $^{[3]}$      & 2.6(0.2)   & 24  & 1326 (1)  \\
J2037+1942   &  B2034+19   & 2.074377 & 36.0  & $-$  & $\geq$26   & 44(4) $^{[4]}$       & 6.4(0.1)   & 672 & 1618 (3)  \\
J2113+4644   &  B2111+46   & 1.014685 & 141.3 & 19.0 & 21(4)     & 12.5(2.5) $^{[3]}$   & 14.9(0.3)  & 290 & 6208 (1)  \\
J2321+6024   &  B2319+60   & 2.256488 & 94.6  & 12.0 & 29(1)      & 25(5) $^{[3]}$       & 115.8(0.4) & 450 & 1795 (1)  \\
\hline
\end{tabular}}
\caption[Summary of results from the pulsar nulling survey]
{Parameters for the pulsars, observed in this survey, 
along with the obtained nulling fraction (NF) and reduction 
in the pulse energy during the null state ($\eta$). 
Columns give pulsar name at 2000 and 1950 epochs, period 
(P), dispersion measure (DM), flux density at 
1400 MHz (S1400), NF obtained in this study, NF reported 
previously, estimate of $\eta$ obtained in this study, number of runs 
and the number of pulses used (N) along with the 
number of contiguous pulses integrated (given 
in parentheses) for the analysis. The PKSMB pulsars are 
marked with a dagger. The error bars on obtained values of NF and $\eta$
are indicated after the estimates by the number in the round parentheses and represent 
3 times the standard deviation errors. The references for the previously 
reported NF in Column 7 are as follows: (1) \cite{la83}
(2) \cite{big92a} (3) \cite{rit76} (4) \cite{hr09}
and (5) \cite{wmj07}}. 
\hfill{}
\label{Results}
\end{table}
\end{center}
\end{landscape}
%%%%%%%%%%%%%%%%%%%%%%%%%%%%%%%%%%%%%%%%%%%%%%%%%%%%%%%%%%%%%%%%%%%%%%%%%%%%%%%%%%%%%%%%%%%%%%%%%%%%%%%
% 
\setcounter{secnumdepth}{3}
\section{Comparison of nulling behaviour}
\label{sect_null_comp}
This section discusses comparison of nulling behaviour 
between four nulling pulsars. Each of this pulsar  
has estimated NF of around 1\%. 
% 
% \subsection{Use of NF as a quantifying parameter}
% Last Chapter discussed a method to estimate 
% NF of a pulsar. It was the only parameters 
% which was used as an quantifying parameter 
% for nulling pulsars. Pulsar nulling remained 
% unexplained due to lack of strong correlation 
% between NF and any of the pulsar parameters. 
As summarised in Section \ref{chronical_obs_sect}, over the years various 
attempts have been made to correlate the NF 
with various pulsar parameters. 
\cite{rit76}, who presented a novel approach to quantify   
amount of nulling by the NF, reported a correlation 
between the pulsar period and the NF. \cite{rit76} claimed 
that, pulsars with longer periods tend to null more frequently  
compared to pulsars with smaller periods. 
Pulsar period is also correlated with the 
age of the pulsar. Hence, \cite{rit76} suggested 
that pulsar die with increasing fraction of nulls. 
% \cite{rit76} also suggested that 
% if the emission is produced in bursts, then the 
% increase in pulse nulling as a pulsar 
% ages may be produced by either a decrease in burst length 
% or increase in their separation. 
% 
% \cite{ran86} classified pulsars in various groups 
% acording to the components seen in their profie. 
% By comparing NF of pulsars between different 
% classes, 
\cite{ran86} concluded that core single 
pulsars possess small NF compared to other classes.
% \cite{rit76} concluded that pulsar die with increasing NF. 
Contrary to \cite{rit76}, \cite{ran86} had suggested that apparent relation 
between nulling and pulsar age is due to the profile morphologies. 
In a given profile class, there is no strong correlation 
between the NF and the pulsar age. 
% Pulsars which possess larger NF has 
% no greater age then those which possess small NF in the 
% same class. \cite{ran83} suggested that young pulsars usually 
% have profiles with prominent core components. 
% It was also found by \cite{ran86} that the 
% core dominated pulsars possess small NF. 
% Hence, this may responsible for small NF in young 
% pulsar and large NF in older pulsars, giving 
% rise to an overall correlation of NF with age.
% 
\cite{big92a} reported correlation study between the NF with several 
pulsar parameters, using 72 nulling pulsars. 
One of the highlighted correlation was 
again between the NF and the pulsar period. 
% Pulsar with longer period seems to possess higher NF. 
% This may indicate that pulse 
% emission mechanism is faltering for these objects since 
% quantities related to polar cap emission mechanisms, 
% such as maximum potential difference across the gap, 
% are inversely related to P (see \cite{mic91}, and reference 
% therein). 
% The other correlations which were reported by \cite{big92a} 
% includes spin-down luminosity($\dot{E}$), magnetic field at 
% light cylinder ($~\log B_{lc}$) and angle between the rotation 
% and magnetic axes ($\alpha$). However, these correlations can 
% be the results of the reported NF-period 
% correlation as all these quantities are strongly correlated with the period 
% of the pulsar. 
Contrary to \cite{ran86}, \cite{wmj07} have claimed 
that there is no correlation between the NF and the profile morphological classes. 
% It was observed that almost all classes of 
% the pulsars show nulling. 
Multicomponent profile pulsars tend to have higher NFs 
but such pulsars are old. Thus it can be due to the reported 
NF-age correlation.  
% 
% In summary, the NF$-$age correlation 
% first reported by \cite{rit76} was ruled 
% out by \cite{ran86} as pulsar with similar 
% profile morphology shows no such 
% correlation. However, the NF$-$profile morphology 
% correlation was not supported by \cite{wmj07}
% in the extensive survey of 72 pulsars.  
Hence, there is no common agreement between  
various studies on the true nature 
of any correlation. 
Apart from the weak NF-period correlation, reported by 
many authors, NF do not appear to strongly correlate 
with any of the pulsar parameter. 
Thus, none of the previous studies were able to provide 
a common agreement among various reported correlations with the NF. 
This presents an important question regarding the NF that, 
{\itshape Is the NF an ideal parameter to quantify 
nulling behaviour of a pulsar ?} 
% We tried to justify that NF does not quantify 
% nulling behaviour in full detail in the remaining 
% part of this Chapter. 

\subsection{Nulling behaviour of similar NF pulsars}
\begin{figure}[H]
\addtolength{\oddsidemargin}{-.875in}
 \centering
  \subfigure[]{
  \includegraphics[width=3.8in,height=2in,angle=-90,bb=0 0 504 203]{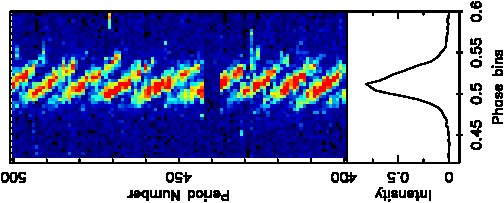}
  % b0809_spdisplay_forThesis.eps: 0x0 pixel, 300dpi, 0.00x0.00 cm, bb=503 202 1 1
  }
  \subfigure[]{
  \includegraphics[width=3.8in,height=2in,angle=-90,bb=0 0 504 203]{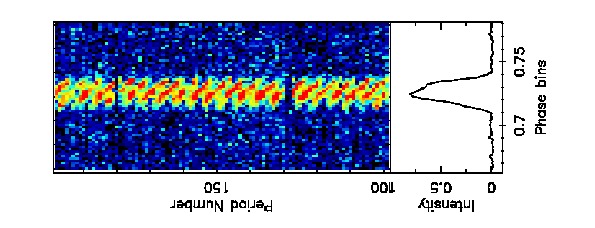}
  % b0818.spdisplay.forThesis.eps: 494x192 pixel, 72dpi, 17.43x6.77 cm, bb=0 0 494 192
  }\\
  \subfigure[]{
  \includegraphics[width=3.8in,height=2in,angle=-90,bb=0 0 504 203]{b0835_spdisplay_forThesis.pdf}
  % b0818.spdisplay.forThesis.eps: 494x192 pixel, 72dpi, 17.43x6.77 cm, bb=0 0 494 192
  }
  \subfigure[]{
  \includegraphics[width=3.8in,height=2in,angle=-90,bb=0 0 504 203]{b2021_spdisplay.pdf}
  % b0809_spdisplay_forThesis.eps: 0x0 pixel, 300dpi, 0.00x0.00 cm, bb=503 202 1 1
  }
  \begin{picture}(0,0)
   \put(-260,292){PSR B0809+74}
   \put(-100,292){PSR B0818$-$13}
   \put(-260,-4){PSR B0835$-$41}
   \put(-100,-5){PSR B2021+51}
  \end{picture}
 \caption[Single pulse plot of four nulling pulsars namely PSRs B0809+74, 
 B0818$-$13, B0837$-$41 and B2021+51]{Single pulse plot of four nulling pulsars namely PSRs 
 (a) B0809+74 (b) B0818$-$13 (c) B0837$-$41 and (d) B2021+51 reproduced 
 again here for a comparison. For each pulsar, 100 observed pulses are shown 
 (see Figure \ref{sp_0809_0818_0835} for details regarding panels).}
 % four_psr_NF_sp.eps: 0x0 pixel, 300dpi, 0.00x0.00 cm, bb=14 14 909 705
 \label{four_psr_NF_sp}
 \normalsize
 \addtolength{\oddsidemargin}{0in}
 \end{figure}

% We have seen in the previous section that 
% pulsar which has high NF do not behave 
% in the similar manner. In this section we 
% will try to justify similar behaviour of nulling 
% pulsars at the other nulling extreme of small NF. 
We have carried out observations of four nulling pulsar 
which exhibit similar NF. These pulsars include 
PSRs B0809+74, B0818$-$13, B0835$-$41 and B2021+51. 
All these pulsars exhibit small amount of nulling 
with estimated NF of around 1\%. The single 
pulse plot of all these four pulsars 
are shown in Figure \ref{four_psr_NF_sp}. 
All these pulsars exhibit sufficiently high S/N 
to detect nulling in the single pulses with high significance.
From Figure \ref{four_psr_NF_sp}, it can be 
seen that although these pulsars have similar 
NF of around 1\%, they exhibit very different single pulse 
behaviour. PSRs B0809+74 and B0818$-$13 show clear drifting 
bands while drifting in PSRs B0835$-$41 and B2021+51 are 
difficult to identify from these plot. 
PSR B0809+74 also has very high duty cycle with 
much wider pulse width compared to other three pulsars.
Table \ref{para_table} lists a few basic pulsar parameters 
of these four pulsars for a comparison.  
PSR B0809+74 is among the oldest (around is 10 times)
compared to the characteristic age of other three pulsars. 
It also shows weak inferred surface magnetic field. 
Although, with these differences in their derived parameters,  
they show similar fraction of nulling. 
\begin{table}[h!]
\begin{center}
 \begin{tabular}[h]{|c|c|c|c|c|}
\hline
PSRs      &     Period         &       Age     &    B$_{surf}$	      &      $\dot{E}$     \\
          &     (sec)          &     (years)   &    (G)             &       (ergs/s)     \\
\hline
\hline
B0809+74    &     1.292241     &    1.22e+08   &    4.72e+11        &  3.08e+30    \\
B0818$-$13  &     1.238130     &    9.32e+06   &    1.63e+12        &  4.38e+31 \\
B0835$-$41  &     0.751624     &    3.36e+06   &    1.65e+12        &  3.29e+32 \\
B2021+51    &     0.529197     &    2.74e+06   &    1.29e+12        &  8.16e+32 \\
\hline
 \end{tabular}
 \caption[A few basic derived pulsar parameters of four nulling pulsars]
 {A few basic derived pulsar parameters \cite[]{hlk+04,wmz+01,hfs+04,mhth05} 
 of four nulling pulsars which show similar NF of around 1\%.}
\label{para_table}
\end{center}
\end{table}
 
\begin{center}
\begin{figure}[h!]
 \centering
 \includegraphics[width=8 cm,height=12 cm,angle=-90,bb=0 0 504 720]{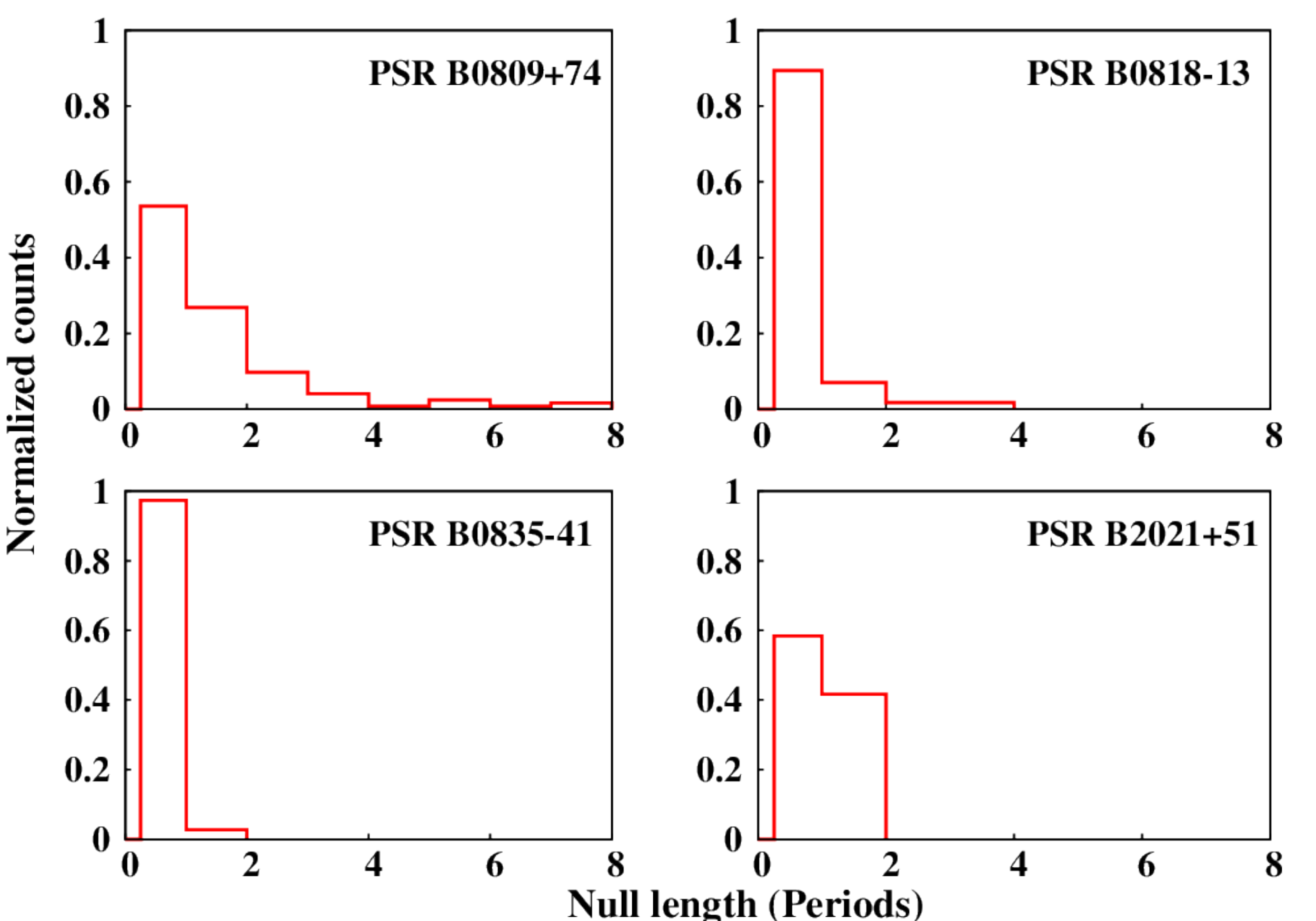}
 \vspace{0.4cm}
 % Four.nlh.eps: 504x720 pixel, 72dpi, 17.78x25.40 cm, bb=0 0 504 720
 \caption[The null length histograms of four nulling pulsars]
 {The null length histograms of four nulling pulsars. Note the relatively larger null lengths 
 for PSRs B0809+74 and B0818$-$13 compared to single and double period nulls in PSRs B0835$-$41 and B2021+51.}
\label{four_nlh}
\end{figure} 
\end{center}
 
\begin{figure}[h!]
\begin{center}
 \centering
 \includegraphics[width=9 cm,height=11 cm,angle=-90,bb=0 0 504 720]{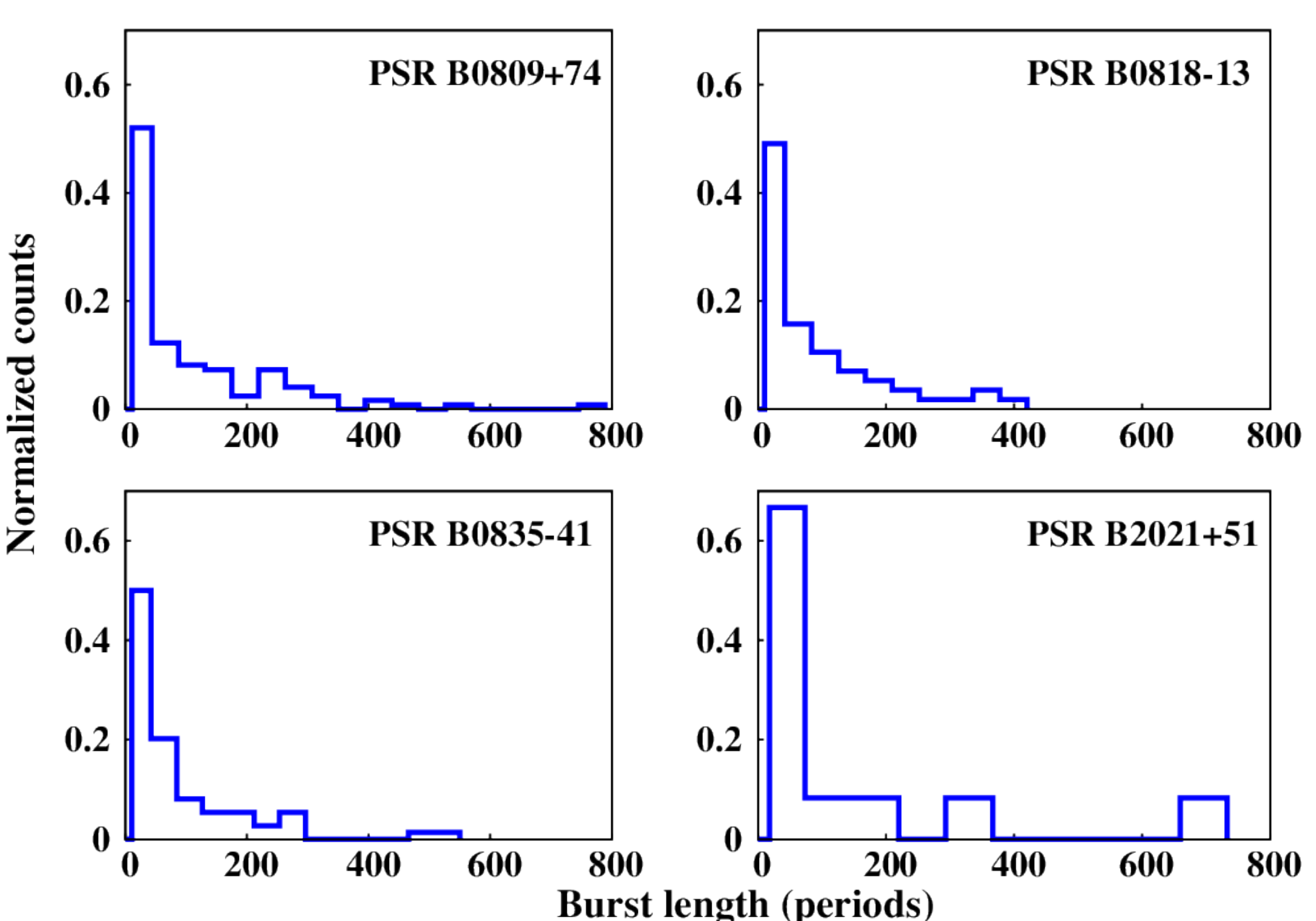}
 % Four.nlh.eps: 504x720 pixel, 72dpi, 17.78x25.40 cm, bb=0 0 504 720
 \caption[The burst length histograms of four nulling pulsars]
 {The burst length histograms of four nulling pulsars. Note the gradual decay of 
 burst lengths seen in all four pulsars.}
\label{four_blh}
\end{center}
\end{figure} 

To compare the nulling pattern seen in these four pulsars, 
we also compared their NLHs and BLHs with 
each other using the the Kolmogrov-smirnov test. 
Figure \ref{four_nlh} shows the NLHs for these four 
pulsars. Similarly, Figure \ref{four_blh} 
shows the BLHs. These histograms, specially the NLHs, 
show significant differences between each other. 
PSRs B0809+74 and B0818$-$13 exhibit null lengths 
of around 4 to 8 periods, while PSRs B0835$-$41 and 
B2021+51 show only single and double period nulls. 
Moreover, the NLH of PSR B0835$-$41 shows around 95\% of 
only single period nulls while less 
than 5\% are two period nulls, which can 
also be seen in Figure \ref{four_psr_NF_sp}.
The NLH of PSR B2021+51 shows 60\% single period 
nulls while the rest 40\% are double period nulls.  
Hence, just from the visual examinations of the NLHs, 
it can be seen that these four pulsar have 
very different nulling time-scales. 
\subsection{KS-test comparison}
To quantify these differences, a two sample Kolmogorov-Smirnov 
(KS) test \cite[]{press} was carried out between the above mentioned four pulsars. 
The KS-test is a non-parametric distribution free test, applicable to unbinned data. 
KS-test provides a statistic, D, which is the maximum difference between the two cumulative 
distribution functions (CDFs). To compare the null length distributions 
of two pulsars, we formed their CDFs from the observed unbinned null length 
sequences. We found D from their CDFs and estimated the rejection probability 
of null hypothesis, which assumes that the measured distributions are drawn from the same underlying 
distribution. The results of these tests are given in Table 
\ref{K-S comparison} for each pair of the four pulsars. 
The null hypothesis is rejected for all pairs with 
high significance (except for the comparison of null lengths for 
PSRs B0818$-$13 and B0835$-$41, where the significance is marginally smaller).
Thus, the nulling patterns differ between each of the four pulsars 
even though they have the same NF. This kind of differences are not only seen in 
pulsars with small NF, but also in pulsars with larger NF. 
For example, PSRs B2319+60 and B2034+19 have NF $\sim$  
30\% but their NLH are different (Figures \ref{nlh_blh_b2319} and \ref{nlh_blh_b2037}).  
A KS-test once again rejects the null hypothesis with high significance. 
Likewise, NLH for PSRs B2111+46 also differ with PSRs B2319+60 and B2034+19. 
However, as these pulsars have slightly different 
NFs, it is difficult to draw a strong conclusion from 
these data. Chapter 5, discusses a comparison between two pulsars namely, PSRs 
J1738$-$2330 and J1752+2359 with NF of around 85\%, where further differences are shown with 
much higher significance. Hence, comparisons of null lengths at different NFs suggest  
that NF does not quantify nulling behaviour in full details.    
\begin{table}[h!]
% \scriptsize
\centering
\begin{tabular}{|l|c|c|c|}
\hline
		& B0818$-$13(57)   & B0835$-$41(74)  & B2021+51(12)    \\ 
\hline
B0809+74(123)  &     99.9 (0.48)   &  99.9 (0.44)    &   97.9 (0.52)   \\ 
% \hline
B0818$-$13(57) &                   &   88.5 (0.2)    &   98.0 (0.46)   \\ 
% \hline
B0835$-$41(74) &                   &                  &  99.1 (0.48)   \\ 
\hline
\end{tabular}
\caption[KS-test statistics from comparison of null length distributions]
{KS-test statistics from comparison of null length distributions between  
four pulsars with similar NF of 1\%. 
The number in the parentheses beside the pulsar names are the number 
of null sequences used for comparison.
The number given for various pairs are the significance of 
null hypothesis rejection, which assumes that the samples 
are drawn from the same distribution. The respective 
D values are given in the parentheses beside the rejection 
probabilities.}
\label{K-S comparison}
\end{table}
\section{Do nulls occur randomly ?}
\label{sec5}
\label{sect_randomness}
Previous studies \cite[]{hjr07,rr09,hr09,kr10}
indicated that nulling may not be random. To test the above 
premise, non-randomness tests were carried out on our data 
for 8 pulsars where it was possible to obtain the NLH 
and the BLH. 

If the null pulses of a pulsar are characterised by 
an independent identically distributed (iid) 
random variable, for which NF represents the proportion 
statistics, then one can Monte-Carlo simulate synthetic 
data sets using a random number generator \cite[]{press}. 
We simulated around 10,000 random one-zero time series of the same   
length as that of the sequence of the observed pulses for each pulsar 
with a given NF. The distributions of null and burst lengths were  
derived from the synthetic data sets. If the underlying 
distribution of observed null lengths does not differ from the 
the simulated distribution with high significance, then it can be concluded that 
the observed nulls are sampled from a  
distribution characterising such an iid random variable, 
for which NF represents the proportion statistics. 
The above premise can be tested by carrying 
out a one sample KS-test \cite[]{press}. 
As explained in the earlier section, 
KS-test provides D statistic, which is the maximum deviation between the two CDFs. 
For our test, one CDF was obtained from a simulated null sequence 
while the other CDF was obtained from the observed null sequence. 
As usual, the test was carried out on the unbinned data. 
The D statistic from this comparison was averaged over all 
10,000 simulated sequences.

Table \ref{randmoness_table} summarises significance level of rejection for the null 
hypothesis, which assumes that the two data sets are drawn 
from the same underlying distribution (or the observed nulls 
are drawn from a random distribution). Apart from PSRs B0818$-$13 
and B2021+51 (where the significance is marginally lower - 
$>$ 82\%), the null hypothesis is rejected at high 
significance for the rest of the pulsars.

A stronger test is Wald$-$Wolfowitz statistical runs$-$test 
\cite[]{ww40}. A dichotomous data set, such as 
the nulling pattern, can be represented by a series of length 
n consisting of n$_1$ 1s (i.e. burst pulses) and n$_2$ 0s (i.e. null pulses), 
with each contiguous series of 1 or 0 defined as a run, r 
(i.e. number of runs given in Column 9 of Table \ref{Results}). 
In order to quantify the degree to which the runs are likely to 
represent a non-random sequence, we calculated Z, defined as, 
\begin{equation}
Z = \frac{r - E(R)}{\sqrt{Var(R)}}.
\label{Zeq}
\end{equation}
Where, the mean of the random variable, R, is given by,
\begin{equation}
E(R) = 1  + \frac{2 n_1 n_2}{n_1 + n_2}.
\end{equation}
The variance of R is given by,
\begin{equation}
Var(R) = \frac{2 n_1 n_2 (2 n_1 n_2 - n_1 - n_2)}{(n_1 + n_2)^2(n_1 + n_2 -1)}.
\end{equation}
Sampling distribution of Z asymptotically tends to a standard 
normal distribution in case of large n with a zero mean and unity 
standard deviation. Therefore, Z will be close to 
zero for a random sequence and the value of Z, derived from R, can be used 
to test the hypothesis that the given sequence is random in a 
distribution free manner. Note that a sequence judged 
random by the runs test indicates that each observation in 
a sequence of binary events is independent of its 
predecessor.
\begin{center}
\begin{table}[h!]
% \scriptsize
\centering
\begin{tabular}[ht]{|l|c|c|c|c|}
\hline
PSRs       &  NF(\%)    &  KS test & Z &  Runs test   \\
\hline
B0809+74   &  1.42      & 99.9 & -38.29  &   99.9   \\
B0818$-$13 &  1.01      & 85.4 & -10.52  &   99.9   \\
B0835$-$41 &  1.7       & 99.3 & -2.77   &   92.1   \\
B1112+50   &  64        & 99.9 & -22.37  &   99.9   \\
B2021+51   &  1.4       & 82.2 & -15.72  &   99.9   \\
B2034+19   &  $\geq$26  & 99.9 & -12.53  &   99.9   \\
B2111+46   &  21        & 99.9 & -17.50  &   99.9   \\
B2319+60   &  29        & 99.9 & -43.10  &   99.9   \\
\hline
\end{tabular}
 \caption[Summary of the  two randomness tests for eight pulsars]
 {Summary of the  two randomness tests, described in the text, for eight pulsars. The null hypothesis assumes that 
 the nulls occur randomly. Rejection of the above hypothesis with the rejection  significance from KS statistic and 
 runs test statistic are given in the Column 3 and 5 respectively. NF 
 (reproduced from Table \ref{Results}) are given for comparison} 
\label{randmoness_table}
\normalsize
\end{table}
\end{center}

This statistic was calculated for 8 pulsars for which the NLH and the BLH 
are presented in Figures \ref{nlh_blh_b0809} to \ref{nlh_blh_b2319}
(except for PSR B1658$-$37). The results are given in Table \ref{randmoness_table}. 
The observed values of Z for all 8 pulsars (except for PSR B0835$-$41) were large and hence  
the null hypothesis is rejected with more than 95 \% significance. 
Even for PSR B0835$-$41, the rejection significance is 92 \%, 
although this is not as significant as the other pulsars. 
Interestingly, Z is negative for all 8 pulsars suggesting that 
the null and burst pulses tend to occur in groups. 

The preceding two tests confirm that a null (or a burst), 
in the null-burst sequence for the 8 pulsars studied, 
is not independent of the state of the pulse preceding it. 
In other words, individual nulls (bursts) are correlated across 
several periods. However, these tests place no constraint on the randomness 
of the duration of nulls (bursts). A careful examination of 
data suggests that the interval between two 
{\it transition events}, defined as a 
transition from a null to burst and vice-verse, 
does not depend on the duration of previous nulls or 
bursts and appears to be randomly distributed.

This random behaviour of the null (burst) duration 
is also supported by the following arguments. 
When the complete pulse sequence is divided into 
several subintervals, consisting of equal number of periods 
(typically 200 pulses), the count of the number of such 
transitions (events) is distributed as a Poisson distribution 
for all pulsars in our sample.  
Likewise, the interval between two transitions is distributed 
as an exponential distribution as its evident in the NLHs and BLHs 
shown in Appendix B. Lastly, featureless spectra 
are obtained from the sequence of null (burst) durations 
indicating no correlations between these durations. 
Thus, it appears that the duration of nulls and bursts can 
be considered as a random variable, at least over the time-scale 
spanned by our data for our sample of pulsars. These claims 
are further supported by the remarkable matching of the 
null length and burst length distributions with 
the stochastic Poisson point process presented in 
the next section. 

In summary, the Wald$-$Wolfowitz runs tests imply non-randomness 
(i.e. correlation in the one-zero sequence, derived from the pulse 
sequence, across periods) in nulling in the sense that the absence (or presence) 
of emission in a given pulse is not independent of the state of 
the pulse preceding it, hinting a memory of the previous state.
However, the duration of the null and burst states
and the  time instants of these transitions appear to be random.
Hence, these pulsars produce nulls and bursts 
with unpredictable durations.

\section{Expected time-scale for nulls and bursts}
\label{sec5a}
\label{sect_expted_time_scale}
The nature of random variable, characterizing the null 
(burst) duration, is investigated further in this section, 
to also obtain the expected time-scale for nulls (bursts). 
The NLHs and the BLHs in Figures \ref{nlh_blh_b0809} to \ref{nlh_blh_b2319} 
suggest that the null and the burst durations are distributed 
as an exponential distribution, which characterizes a stochastic 
Poisson point process. The CDF, F(x), of a Poisson point 
process is given by \cite[]{pp02}, 
\begin{equation}
F(x)  = 1  - \exp{(-x/\tau)}. 
\label{poiscdf}
\end{equation}
Where, $\tau$ represents a characteristic 
time-scale of the stochastic process. A least square  
fit to this simple model  provides the 
characteristic null and burst 
time-scales ($\tau_n$ and $\tau_b$ respectively;
equation \ref{poiscdf}). 

Figures \ref{cdf_b0809} to \ref{cdf_b2319} show 
the CDFs corresponding to NLHs and BLHs for eight pulsars 
along with the least square fits to the proposed Poisson 
point process CDFs given in equation \ref{poiscdf}. 
These fits show remarkable similarity between the observed 
CDFs and the proposed CDF from the Poisson point process (equation \ref{poiscdf}).  
These fits suggest that the interval between one transition from null 
to burst state (and vice-verse) to another transition 
from burst to null state appears to be modelled well by a 
Poisson point process for these pulsars. 
It should be noted that, the model given in 
equation \ref{poiscdf} need not be unique and other models 
may fit the CDFs equally well [See \cite{viv95,wwr+12,cor13}]. 
However, we use this model (a) as this is the simplest model suggested 
by our data, and (b) we did not have sufficient data 
to try more complicated models.  

The characteristic null and burst 
time-scales ($\tau_n$ and $\tau_b$ respectively; 
equation \ref{poiscdf}) and the uncertainties 
on these parameters, obtained from these fits,  
are listed in Table \ref{tabtau}. No fits were carried 
out for the CDF of nulls for PSRs B0835$-$41 and B2021+51 
as only two points were available for the fit.
The fitted model was checked by carrying out a two-sample KS-test 
in the following manner. First a pulse sequence, consisting 
of a million pulses, was simulated using the parameter 
$\tau$ obtained in these fits. Then, the NLH and the BLH 
were obtained for this simulated pulse sequence, 
which provides much larger sample of nulls and bursts 
than the observed sequence. A two-sample KS-test was 
carried out on the NLH and the BLH, obtained from the observed 
sequence and the simulated pulse sequence. The significance 
level of rejection for the null hypothesis, which assumes, 
in this case, that the two distributions are different,   
is given in Column 4 and 6 of Table \ref{tabtau} for the null and the burst durations, 
respectively. They are also indicated in Figures \ref{cdf_b0809} to \ref{cdf_b2319}. 
The null hypothesis is rejected with high 
significance for null duration in all six pulsars, 
for which the fit was carried out. Apart from PSRs B2034+19  
and B2319+70 the null hypothesis is rejected with high 
significance for the burst duration for the other six pulsars.
The lower significance of rejection 
of null hypothesis for the burst duration in the above listed two 
pulsars may be due to the use of the simple model given by 
equation \ref{poiscdf}.
%%%%%%%%%%%%%%%%%%%%%%%%%%%%%%%%%%%%%%%%%%%%%%%%%%%%%%%%%%%%%%%%%%%%%%%%%%%%%%%%
% Fitted CDF figure
%%%%%%%%%%%%%%%%%%%%%%%%%%%%%%%%%%%%%%%%%%%%%%%%%%%%%%%%%%%%%%%%%%%%%%%%%%%%%%%%
\begin{comment}
\begin{figure}
 \centering
  \psfig{figure=nulcdf2111.ps,width=2.5in,angle=90}
\caption{CDF of null length (top plot) and burst length (bottom plot) 
distributions for PSR B2111+46 
(solid line) is shown alongwith the best fit Poisson point process 
model (dashed line), given by equation \ref{poiscdf} }
\label{cdfall}
\end{figure}
\end{comment}
%%%%%%%%%%%%%%%%%%%%%%%%%%%%%%%%%%%%%%%%%%%%%%%%%%%%%%%%%%%%%%%%%%%%%%%%%%%%%%%%

%$$$$$$$$$$$$$$$$$$$$$$$$$$$$$$$$$$$$$$$$$$$$$$$$$$$$$$$$$$$$$$$$$$$$$$$$$$$$$
\begin{center}
\begin{table}
% \scriptsize
\centering
\begin{tabular}[ht]{|l|c|c|c|c|c|}
\hline
PSRs       &  Period   &   $\tau_n$   &  KS-prob & $\tau_b$ & KS-prob  \\
           &   (s)     &    (s)       &    \%    &   (s)    &   \%     \\
\hline
B0809+74   & 1.29  &    1.9 (0.3)    & 99    & 176 (9)   & 98    \\
B0818$-$13 & 1.24  &    0.7 (0.3)    & 99    & 84 (9)    & 78    \\
B0835$-$41 & 0.75  &    -            & -     & 44 (4.5)  & 74    \\
B1112+50   & 1.66  &    4.8 (0.1)    & 99    & 4.3 (0.8) & 88    \\
B2021+51   & 0.53  &    -            & -     & 21 (10)   & 95    \\
B2034+19   & 2.07  &  2.6 (0.2)      & 99    & 11 (2)    & 22    \\
B2111+46   & 1.02  &   1.6 (0.1)     & 99    & 8.7 (0.5) & 99    \\
B2319+60   & 2.26  &   11 (1)        & 94    & 23 (2)    & 33    \\
\hline
\end{tabular}
\normalsize
\caption[The characteristic null ($\tau_n$) and burst ($\tau_b$) time-scale 
for the eight pulsars]{The characteristic null ($\tau_n$) and burst ($\tau_b$) time-scale 
for the eight pulsars in Figures \ref{nlh_blh_b0809} to \ref{nlh_blh_b2319} 
obtained from a least squares fit to the CDF 
of these pulsars to a Poisson point process. 
These fits are shown in Figures \ref{cdf_b0809} to \ref{cdf_b2319}. 
These time-scales have been expressed in seconds (i.e. column 3 and 5) 
after multiplying the fitted parameter (shown in Figures \ref{cdf_b0809} to \ref{cdf_b2319}) 
$\tau$ in equation \ref{poiscdf} with the period of the pulsar. The numbers 
in the parentheses are the corresponding errors on the time-scales, 
obtained from the least squares fits. 
The Kolmogorov-Smirnov probability of rejection for the null hypothesis, 
which assumes that the two distributions are different, are given in 
column 4 and 6 for the null and burst durations respectively.} 
\label{tabtau}
\end{table}
\end{center}

\section{Conclusions}
\label{sec7}
\label{sect_survey_conclusion}
The nulling behaviour of 15 pulsars, out of which 5 were 
PKSMB pulsars with no previously reported nulling behaviour, 
is presented in this chapter with estimates of their NFs. 
For four of these 15 pulsars, only an upper/lower limit was previously 
reported. The estimates of reduction in the pulsed emission 
is also presented for the first time in 11 pulsars. NF value for 
individual profile component is also presented for two pulsars in the 
sample, namely PSRs B2111+46 and B2020+28. 
Interesting pulse energy fluctuations around the null region 
is reported in PSR B1658$-$37. Possible mode changing behaviour 
is suggested by these observations for PSR J1725$-$4043, 
but this needs to be confirmed  with 
more sensitive observations. An interesting quasi-periodic 
nulling behaviour for PSR J1738$-$2330 is also reported. 
We also reported intriguing single period core emission during the 
long nulls in PSR B2111+46. We find that the nulling patterns differ between  
PSRs B0809+74, B0818$-$13, B0835$-$41 and B2021+51, 
even though they have similar NF of around 1\%. 

The null and burst pulses in 8 pulsars in our sample appear 
to be grouped and seem to occur in a correlated way, when 
individual periods are considered. However, 
the interval between  transitions from the null to the  burst states
(and vice-verse) appears to represent a Poisson point process. The typical null 
and burst time-scales for these pulsars have been obtained 
for the first time to the best of our knowledge.  

\section{Discussion}
\label{sect_survey_discussion}
The estimates for the factor, $\eta$, by which the pulsed emission 
reduces during the nulls for 11 pulsars were presented for the 
first time in this study (Table \ref{Results}). Although the physical process,  
which causes nulling, is not yet understood, it could be due to a 
loss of coherence in the plasma generating the radio emission or 
due to geometric reasons as summarized in Section \ref{null_model_all_sect}. 
In the former case, $\eta$ provides a constraint on the process responsible 
for this loss of coherence. Our estimates provide lower limits for different pulsars as 
this estimate is limited by the available S/N. Nevertheless, 
reduction by two orders of magnitude is seen in at least two 
pulsars. If nulling is caused by a shift in the radio beam due to 
global changes in magnetosphere, $\eta$ provides a constraint 
on the low level emission and will depend on the orientation 
to the line-of-sight and the morphology of the beam during the 
null. 

Our results confirm that NF probably does not capture 
the full detail of the nulling behaviour of a pulsar. 
We find that the pattern of nulling can be quite different 
for classical nulling pulsars with similar NFs. Estimates for 
typical nulling time-scales, $\tau_n$, for 6 pulsars in our sample were 
obtained for the first time. For 2 of these with a NF of about 
1\%, $\tau_n$ varies by a factor of three. In particular, 
the typical nulling time-scale for PSR B1112+50  
(NF $\sim$ 65\%) is about 2 s, more than 6 orders of magnitude 
less than that for the intermittent pulsar 
PSR B1931+24 (NF $\sim$ 75\%). In the \cite{rs75} model, the 
pulsar emission is related to relativistic pair plasma generated due 
to high accelerating electric potential in the polar cap. 
As mentioned in Section \ref{intrinsic_effects_sect}, 
changes in this relativistic plasma flow \cite[]{fr82,lhk+10}, 
probably caused by changes in the polar cap 
potential, have been proposed as the underlying cause 
for a cessation of emission during a null. The typical nulling 
time-scale, $\tau_n$ (and burst time-scale $\tau_b$) provides a 
characteristic duration for such a quasi-stable state, which is 
similar to a profile mode-change. An interesting possibility 
may be to relate this to the polar cap potential. In any case, 
any plausible model for nulling needs to account for 
the range of null durations for pulsars in our sample 
and relate it to a physical parameter and/or 
magnetospheric conditions in the pulsar magnetosphere.

We have extended the Wald-Wolfowitz runs test for randomness 
to 8 more pulsars. Results for 15 pulsars were  published 
in previous studies \cite[]{rr09,kr10}. 
Our sample  has no overlap with  the sample presented 
by these authors. Like these authors, we find that this test indicates that 
occurrence of nulling, when individual pulses are considered, is 
non-random or exhibits correlation across periods. 
Unlike these authors, all 8 pulsars in our sample show such a behaviour. 
This correlation groups pulses in null and burst states, which was also 
noted by \cite{rr09}. However, the durations of the null and the burst states seem 
to be modelled by a stochastic Poisson point process suggesting that 
these transitions occur at random. 
Thus, the underlying physical process for nulls in the 8 pulsars  
studied appears to be random in nature producing 
nulls and bursts with unpredictable durations. 
More details regarding this behaviour are further 
discussed in Section \ref{on_randomness_sect}. 

Lastly, our estimates of NFs in Table \ref{Results} are consistent with those published earlier 
for PSRs B0809+74, B0818$-$13, B0835$-$41, B1112+50 and 
B2319+60 \cite[]{la83,big92a,rit76}, 
which indicates that NF are consistent over a 
time-scale of about 30 years. 

In Section \ref{null_model_all_sect}, it was highlighted that 
nulling is an open question even after 
40 years since it was first reported. Recent studies 
suggests that both nulling and  profile mode changes probably 
represent a global reorganization of pulsar magnetosphere 
probably accompanied by changes in the 
spin-down rate of these pulsars \cite[]{klo+06,lhk+10}. 
Interestingly, such quasi-stable states of magnetosphere 
have recently been proposed, based on MHD calculations, 
to explain the release of magnetic energy implied by high energy bursts 
in soft-gamma ray repeaters \cite[]{ckf99,con05,tim10}. 
Global changes in magnetospheric state is likely to be 
manifested in changes in radio emission regardless of the 
frequency of observations. This study motivates simultaneous multi-frequency observations 
of pulsars with nulling, as they will be useful to study these 
changes and constrain such magnetospheric models. 
Chapter 6 discusses similar observations conducted on two pulsars. 
\pagebreak
\clearpage\null\newpage

\chapter[On the long nulls of PSRs J1738$-$2330 and J1752+2359]{On the long nulls of PSRs J1738$-$2330 and J1752+2359}
\graphicspath{{Images/}{Images/}{Images/}}

\section{Introduction}
Nulling pulsars exhibit a variety of NFs, as shown in Table \ref{table_all_null_psr}, 
and one might hope that this apparently fundamental parameter could 
be used to characterise further aspects of the pulsar's behaviour. 
However, as demonstrated in Section \ref{sect_null_comp}, our study 
dashed this hope and concluded that pulsars with similar low NFs ($\sim$ 1 \%) 
are not necessarily similar in their nulling pattern. 
In this chapter, we set out to see if 
this result still holds even for pulsars with large 
NF ($>$ 80 \%). We compare and contrast two high-NF 
pulsars and assess to what extent their nulling patterns 
follow a common statistical rule.
 
PSR J1752+2359 was discovered in a high Galactic latitude 
survey with the Arecibo telescope \cite[]{fcwa95}. Subsequent 
timing observations indicated interesting single pulse behaviour 
with bursts up to 100 pulses, separated by 
nulls of about 500 pulses \cite[]{lwf+04}, giving an NF of about 80 \%.
This study also noted an intriguing exponential decrease in the pulse 
energy during a burst, a feature seen in very few nulling pulsars \cite[]{rw08,bgg10,lem+12}
to the best of our knowledge. The second, PSR J1738$-$2330, 
as mentioned in Section \ref{j1738_chap4_sect}, was discovered in the PKSMB survey \cite[]{lfl+06}. 
As discussed in Section \ref{j1738_chap4_sect} and shown in Figures \ref{j1738lrf} and \ref{NF_j1738}, 
\psra\ shows quasi-periodic bursts interspersed with long nulls 
at 325 MHz with a lower limit of 69\% on its NF. 
Table \ref{paratable} highlights some of 
the basic parameters for both the studied pulsars for a comparison. 
Note that \psra\ is approximately five times slower and three times 
younger than \psrb. 
\begin{table}[h!]
{\centering 
 \begin{center}
 \begin{tabular}[ht]{l c c c c c}
 \hline 
 \multirow{3}{*}{PSRs} &    &                    &           &         \\
%  \cline{2-10} \\
              &    P    &  DM 		           & Age       & B$_{s}$  \\
              &  (sec)  &  (pc$\cdot$cm$^{-3}$)   & (MYr) & ($\times$10$^{12}$G) \\              
 \hline 
            &         &                         &           &        &    $~~~~~$     \\
 J1738$-$2330 &  1.98   & 99.3                    & 3.6       & 4.16	 &    $~~~~~$     \\
 J1752+2359   &  0.41   & 36.0                    & 10.1      & 0.52  	 &    $~~~~~$     \\ 
              &         &                         &           &        &    $~~~~~$     \\
 \hline
 \end{tabular}
 \end{center} 
 \caption[The basic parameters and the observations details for the two observed pulsars]
 {The basic parameters \cite[]{lfl+06,lwf+04,mhth05} and the observations details for the two observed pulsars. 
  Columns give pulsar name at 2000 epoch, period (P), dispersion measure (DM),
  characteristic age ($\times$10$^6$ Yr) and surface magnetic field (B$_s$).}
 \label{paratable} 
 }
\end{table}

In previous studies of these pulsars the typical duration 
for single pulse observations was 1-2 hours. Such short observations, particularly 
for pulsars with periods around 1 s, do not yield 
sufficient nulls and bursts for a satisfactory comparison 
of their statistical properties. In this study 
much longer observations were undertaken to obtain 
a large sample of nulls and bursts in each pulsar. 

The observations of both pulsars were carried 
out with the GMRT at 325 MHz, using the 
GMRT software backend (details of which are given in Chapter 3). 
The off-line analysis procedures are also similar to 
those discussed in Chapter 3 to obtain the NFs 
and to construct the NLHs and BLHs. 
For \psrb\ around 30 minutes of archival full polar 
data (Rankin, private communication), 
observed at 327 MHz from the Arecibo telescope 
was also analysed to confirm the presence of weak 
burst pulses during the null states and to compare 
the linear and the circular polarization profiles. 
Details regarding the data reduction 
and the polarization calibration are similar 
to those discussed by \cite{rwb13} (also see Chapter 3 
for details regarding the Arecibo telescope and 
off-line data analysis). 

The nulling patterns and their quasi-periodicities are discussed  
in Sections \ref{nulls} and \ref{quasiperiod}. The variations in 
the burst pulse energy and the modelling thereof is described in Section \ref{BBB_patten}. 
The emission behaviour during null-to-burst transitions and vice-versa are 
discussed in Section \ref{first_and_last_bbb_section}. Unusual emission 
behaviour present in the null sequences of \psrb\ is analysed in  
Section \ref{emission_in_null}. Section \ref{Arecibo} discusses 
the polarization profiles of \psrb. Section \ref{gps} examines the 
presence of Giant pulses in \psrb. The results of this study and their 
implications are summarized in Sections \ref{conclusion} and 
\ref{discussion}, respectively. 
% Supporting material on pair correlation function and 
% modeling used in the paper is provided in Appendix \ref{apppcf} 

\section{Null-burst statistics}
\label{nulls}

\begin{figure}[h!]
  \centering
   \subfigure[]{ 
% %  \includegraphics[width=4.2 in, height=2 in,angle=-90, b=14 14 212 495]{j1738spdisplay.new.eps} 
     \includegraphics[width=2 in,height=4.5 in,angle=0,bb=0 0 201 497]{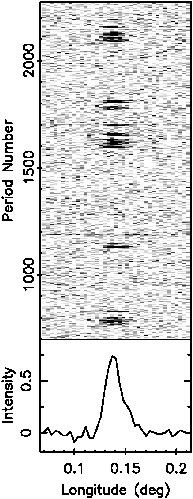}
% % j1752spdisplay.eps: 0x0 pixel, 300dpi, 0.00x0.00 cm, bb=0 0 201 497
    \label{j1738spdisplay}
    }
    \subfigure[]{
    \includegraphics[width=2 in,height=4.5 in,angle=0,bb=0 0 201 497]{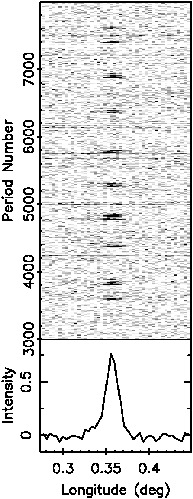}
    \label{j1752spdisplay}
    }
%    \begin{picture}(0,0)
%     \centering 
%     \put(-260,5){PSR J1738$-$2330}
%     \put(-105,5){PSR J1752+2359}
%     \end{picture}
  \caption[Single pulse sequences for both pulsars in grey-scale intensities 
  observed at 325 MHz with the GMRT]{Single pulse sequences for both pulsars in grey-scale intensities 
  observed at 325 MHz with the GMRT. 
  (a) A sequence of around 1500 consecutive 
  pulses from \pa. Short nulls, towards the end of 
  the bright phases, can be seen for this pulsar.  
  (b) A consecutive sequence of around 5000 pulses 
  from \pb. The quasi-periodic pattern of occurrence 
  for bright phase is clearly evident.}
  \label{spdisplay}
\end{figure}

Small sections of the observed single pulses 
are shown in Figure \ref{spdisplay} 
for both the pulsars. Figure \ref{spdisplay}(a) shows 
around 1500 pulses for \pa, while Figure \ref{spdisplay}(b) 
shows a section of around 5000 pulses for \pb. 
Both the pulsars display unique nulling behaviour with a large 
fraction of null pulses. The arrangement of burst and null pulses in the single 
pulse sequence for the two pulsars showcase both similarities and differences. 
The single pulse sequences of Figure \ref{spdisplay} show the burst pulses of both
pulsars clustering together in groups, which we will refer to as the {\it bright phases}
interspersed with long null phases (the {\it inter-burst} or {\it off-phases}),
giving a quasi-periodic effect. 

\begin{figure}[h!]
 \centering
 \subfigure[]{
 \includegraphics[width=3.1 in,height=4.0 in,angle=-90,bb=30 25 588 768]{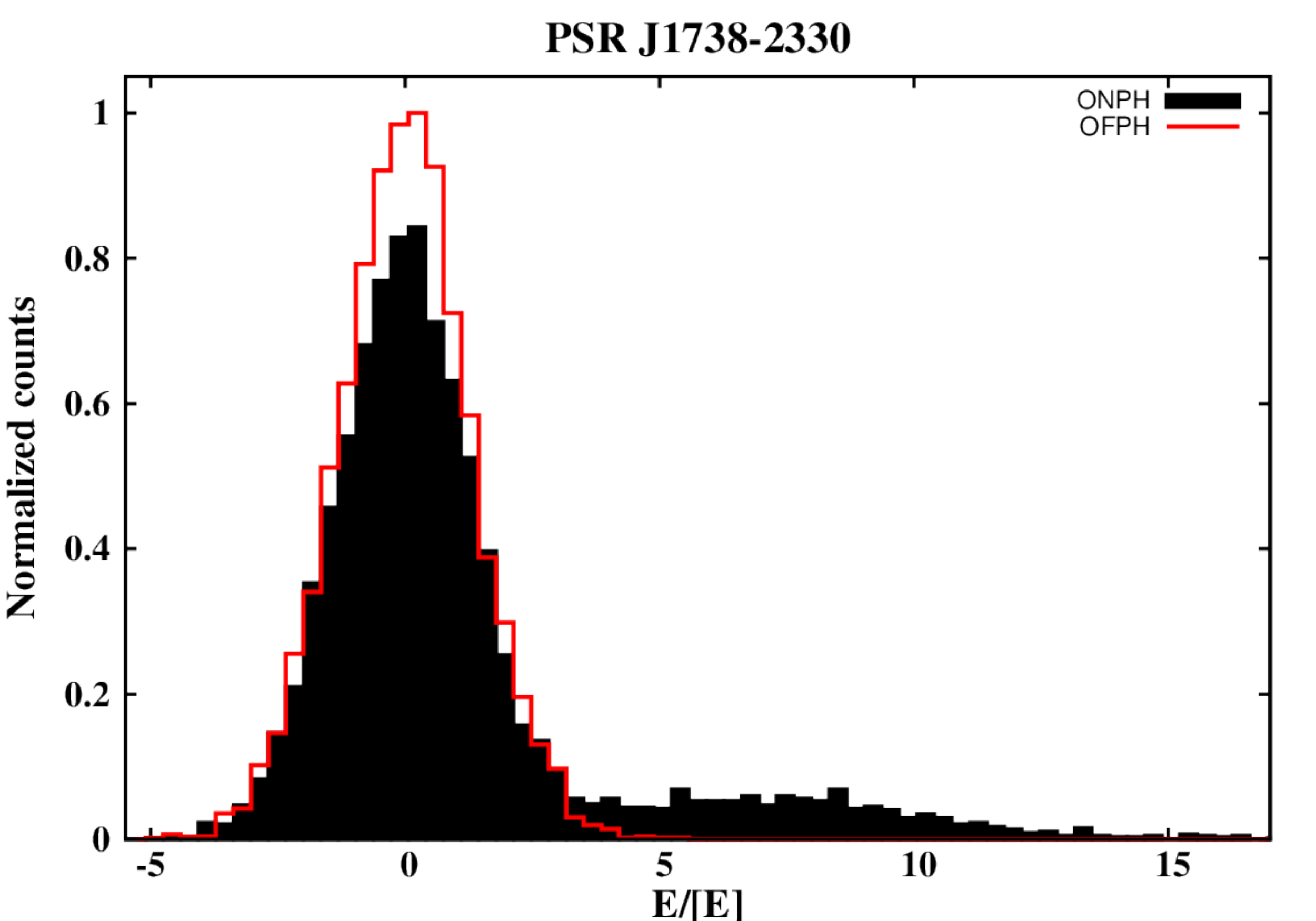}
 % j1738histogram.eps: 0x0 pixel, 300dpi, 0.00x0.00 cm, bb=0 0 734 537
 \label{j1738_hist}
 }
 \centering
 \subfigure[]{
 \includegraphics[width=3.1 in,height=4.0 in,angle=-90,bb=30 25 588 768]{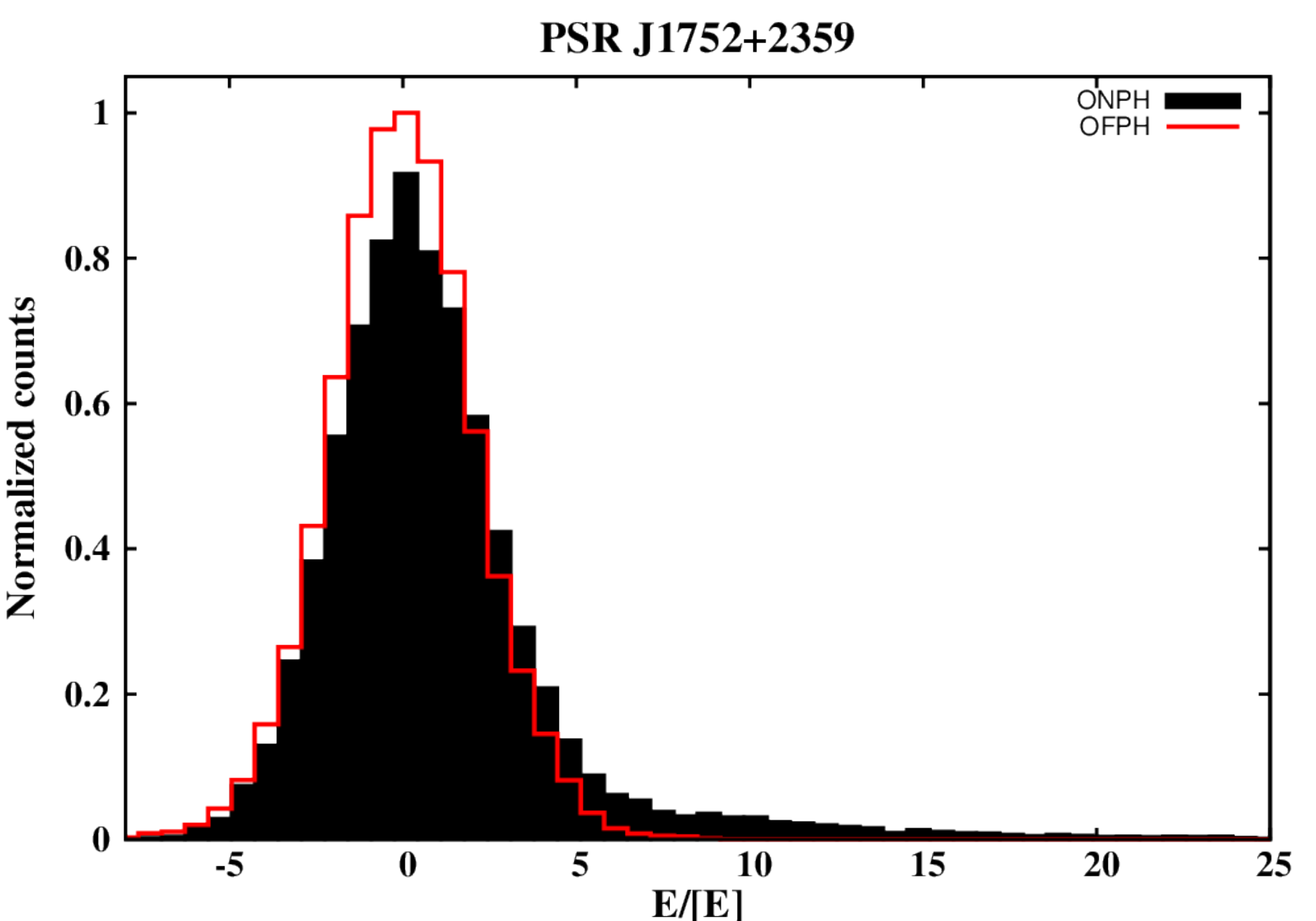}
 % j1738histogram.eps: 0x0 pixel, 300dpi, 0.00x0.00 cm, bb=0 0 734 537
 \label{j1752_hist}
 }
 \begin{picture}(0,0)(0,0)

  \put(-70,180){\scriptsize Freq : 325 MHz}
  \put(-70,170){\scriptsize N : 6200 pulses}
  \put(-70,160){\scriptsize NF : 85.1$\pm$2.3\%}
  
  \put(-70,-60){\scriptsize Freq : 325 MHz}
  \put(-70,-70){\scriptsize N : 21000 pulses}
  \put(-70,-80){\scriptsize NF : $\leq$89\%}
 
 \end{picture}
 \caption[The on-pulse and the off-pulse energy 
 histograms for two pulsars]{The on-pulse (filled curve) and the off-pulse energy 
 (red solid line) histograms for PSRs (a) J1738$-$2330 and 
 (b) J1752+2359 obtained with the GMRT observations at 325 MHz. 
 Both pulsars show large fraction of null pulses. \pa\ shows
 clear bimodal burst pulse distribution compared to 
 smooth on-pulse intensity distribution seen for \pb.}
\label{hist_both_fig}
\end{figure}
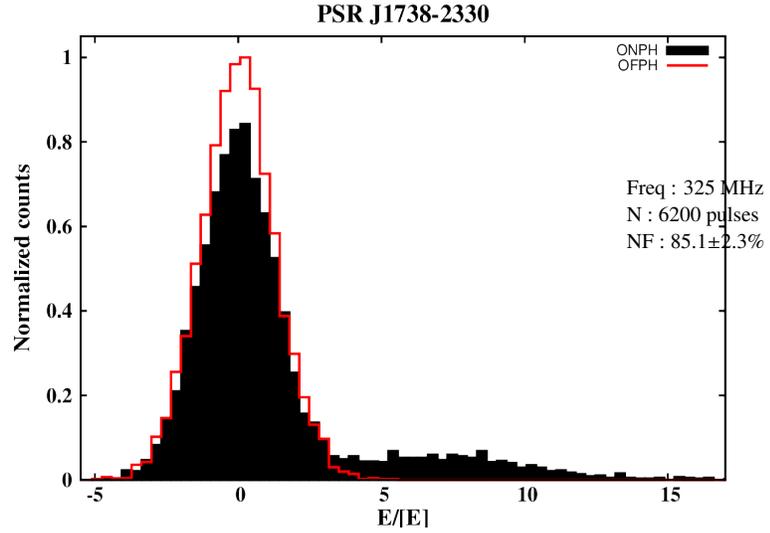
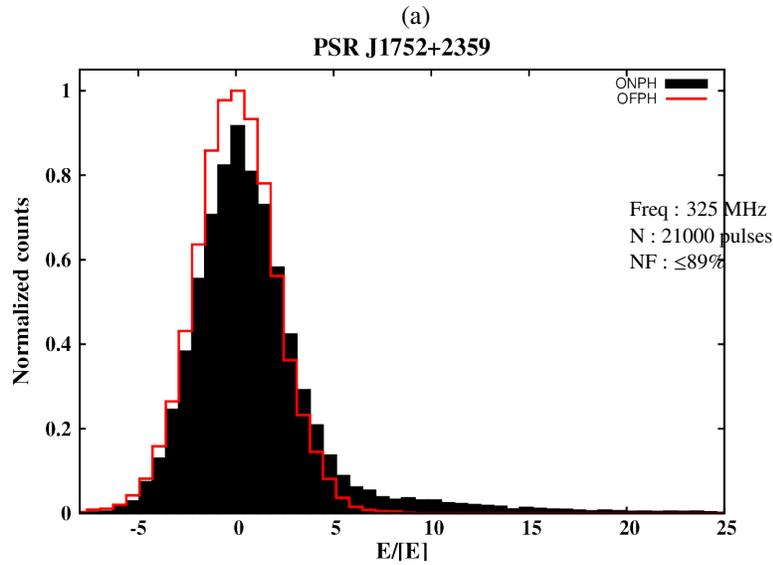
\begin{figure}
 \centering
 \subfigure{
  \centering
  \includegraphics[width=4.5 in,height=3 in,angle=0,bb=0 60 360 312]{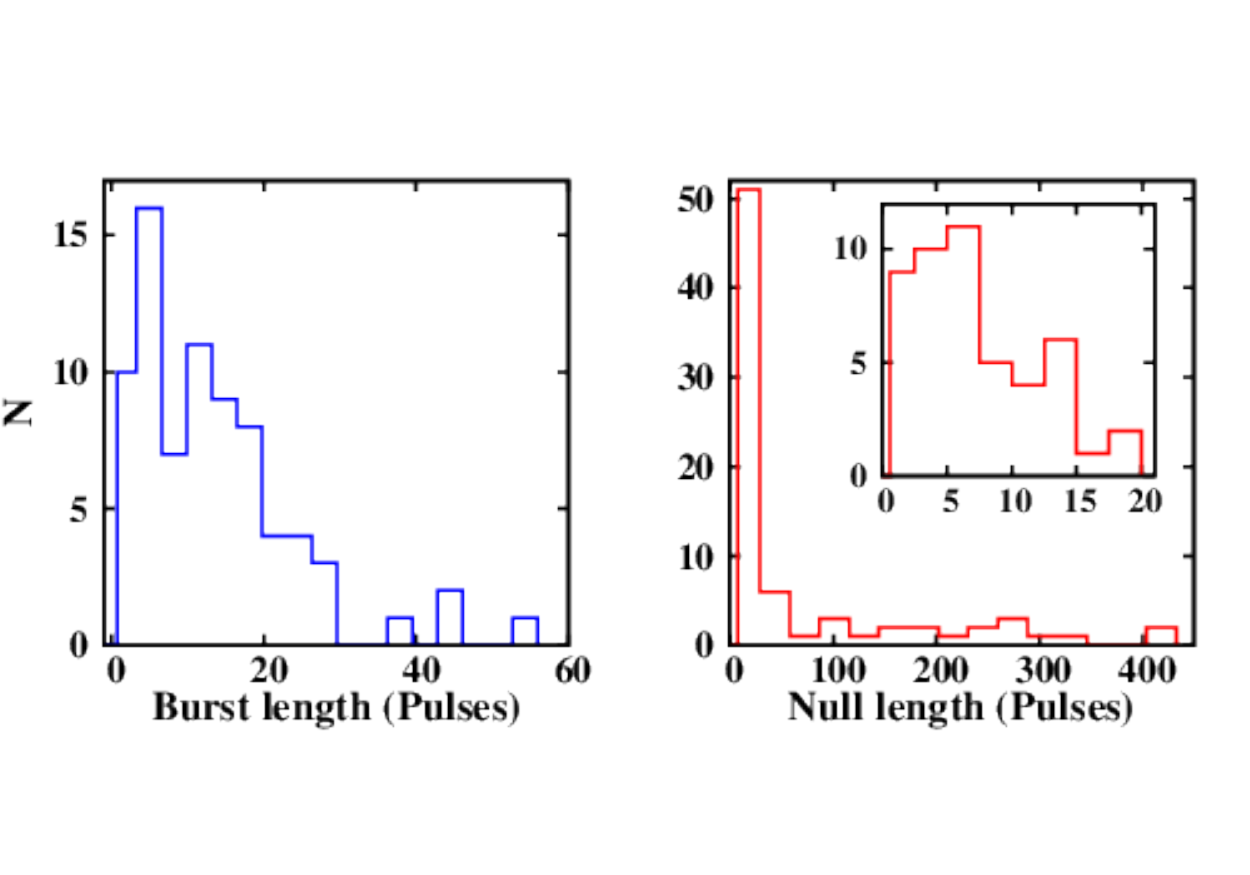}
 % J1738.nbhist.combine.eps: 360x252 pixel, 72dpi, 12.70x8.89 cm, bb=0 0 360 252
%  \label{nbhist_j1738}
 }
 \subfigure{
 \centering
 \includegraphics[width=4.5 in,height=3 in,angle=0,bb=0 0 360 252]{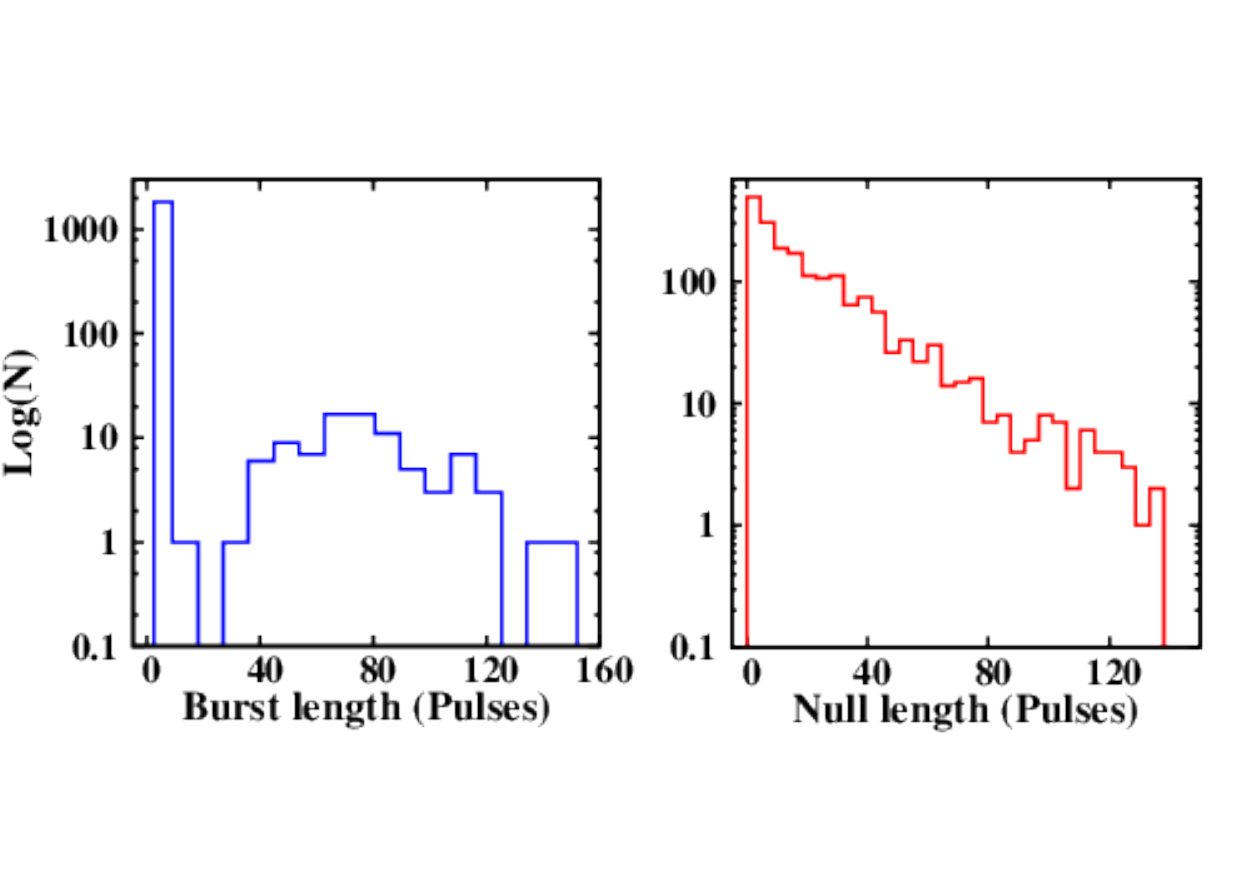}
% %  \includegraphics[width=3 in,height=2 in,angle=0,bb=70 70 360 272]{test.eps}
%  % J1752_nbhist_bbb_tau_range_inter_combine.eps: 0x0 pixel, 300dpi, 0.00x0.00 cm, bb=50 50 410 302
% \label{nbhist_j1752}
 \begin{picture}(0,0)
  \centering 
  \put(-200,380){PSR J1738$-$2330}
  \put(-200,180){PSR J1752+2359}
  \put(-250,200){(a)}
  \put(-80,200){(b)}
  \put(-250,20){(c)}
  \put(-80,20){(d)}
 \end{picture}
  }
 \caption[The conventional BLHs and NLHs for PSRs \pa\ and \pb]
 {The conventional BLHs and NLHs for PSRs \pa\  and \pb\ 
obtained from the GMRT observations. The NLH (b) for \pa\ is shown with an inset plot depicting 
the distributions of short nulls which is very similar to the distribution of the bursts in (a), 
indicating that the high NF of this pulsar arises from an excess of long nulls. The NLH (d) and 
the BLH (c) for \pb\ are shown with the measured counts on a log-scale to bring out the details 
for longer bursts and nulls. Note the large number of single bursts in (c), whose random occurrence 
among the long nulls generates an exponential distribution in the NLH (d). Thus the two pulsars have 
very different null distributions despite their similar NFs.}
\label{nbhist_both}
\end{figure}

The energies in a window centred on the pulse and 
a window with equal number of samples away from the 
pulse were obtained from the single pulse sequence. 
The histograms for these on-pulse and off-pulse energies  
are shown in Figures \ref{hist_both_fig}(a) and 
\ref{hist_both_fig}(b) respectively for PSRs \pa\  
and \pb. The distributions around zero 
mean pulse energy in the ONPH 
indicate the large fraction of null pulses 
in both the pulsars. The NFs were estimated using 
the method discussed in Chapter 3 [also see \cite{rit76}]. 
All the RFI affected pulses were removed during 
the estimation of NFs. 
  
A NF of around 85.1$\pm$2.3\% was estimated for \pa\   
from the bimodal distribution in the ONPH. 
\pb\ shows large number of weak pulses and its 
ONPH is not bimodal. Hence, we can only 
estimate an upper limit of around $<$89\% for the NF 
as some fraction of weak burst pulses will  
be included in the null pulse distribution. 
A method to separate these individual weak burst pulses 
from the null pulses is discussed in Section \ref{emission_in_null}, 
which is similar to technique discussed in Section \ref{separation_of_null_burst_sect}. 
After this separation, the fraction of null pulses are around 81\% for \pb. 
A variation up to 20 to 25 times the mean pulse energy 
is visible in the ONPH for both the pulsars. 
As both the pulsars spend about 85\% 
of time in the null state, the estimated mean pulse 
energies also get reduced by the same fraction for 
both of them. 

\begin{figure}[h!]
 \centering
 \includegraphics[width=4 in,height=3 in,angle=0,bb=0 0 360 252]{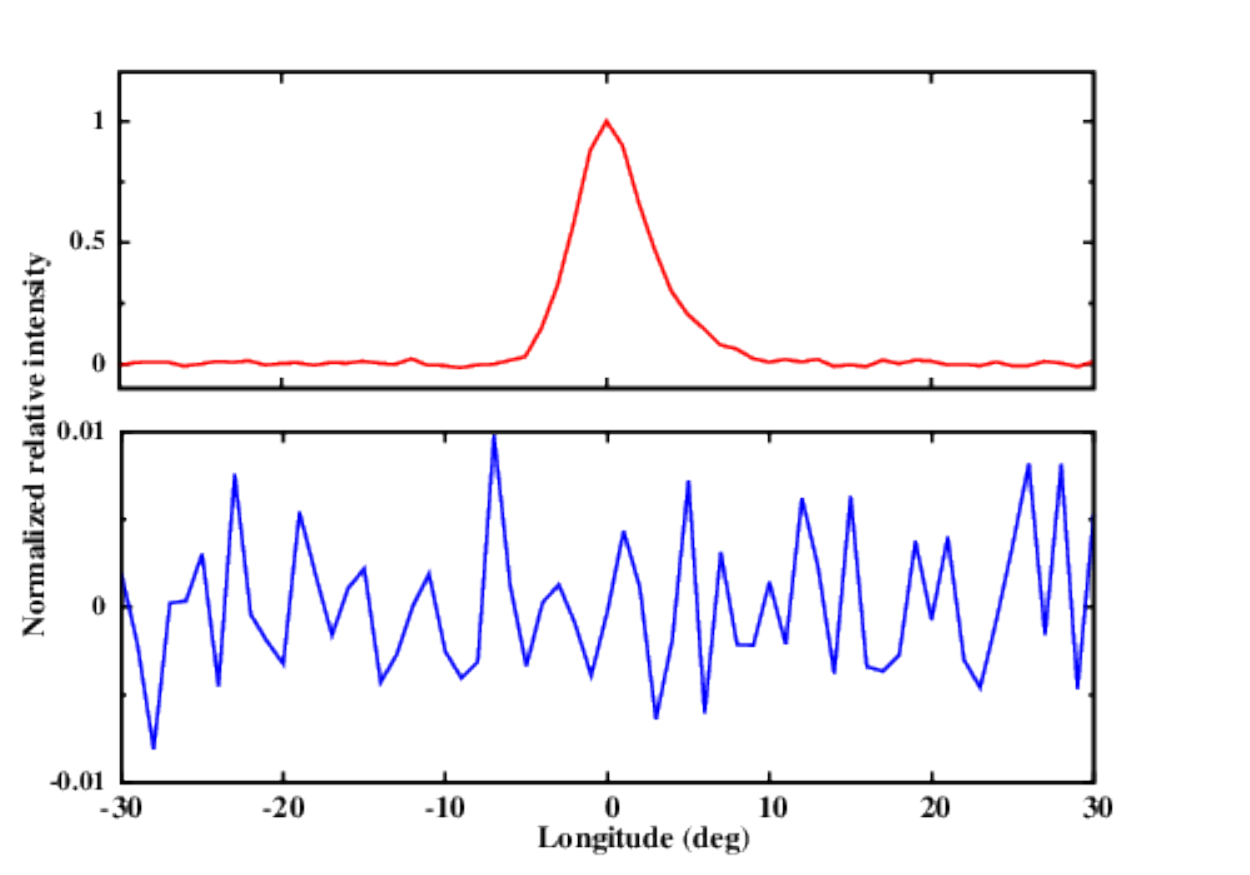}
 % J1752_null_burst_profile.eps: 0x0 pixel, 300dpi, 0.00x0.00 cm, bb=50 50 410 302
 \caption[Separated null and burst pulse profile from \psra]
 {Separated null and burst pulse profile from \psra. The top panel shows 
 the integrated profile from all the burst pulses (solid red line), while the bottom 
 panel shows the integrated profile from all the separated null pulses (solid blue line). Both the 
 profiles were normalized by the peak of the burst pulse profile.}
 \label{j1738_null_burst_profile}
\end{figure}

However, behind this general similarity we find a number of important differences 
between the two pulsars, and these can be seen in the burst and null length histograms 
of each pulsar (Figure \ref{nbhist_both}). In \pa, 
the lengths of the bursts [see Figure \ref{nbhist_both}(a)] 
are much shorter than those typical of \pb\ [shown in Figure \ref{nbhist_both}(c)] 
because in its bright phases, the bursts of \pa\ are equally mixed 
with short nulls, as can be seen from the inset 
of short nulls in Figure \ref{nbhist_both}(b). 
Together this results in bright phases of up to 60$-$80 pulses in 
length, with the nulls predominating towards the 
end (see Section \ref{BBB_patten}). The high NF of this 
pulsar then arises through the presence of a long tail of long 
nulls in the histogram of Figure \ref{nbhist_both}(b). 
All the designated null pulses of this pulsar, whether 
occurring within or between bright phases, were integrated 
and found to show no profile of significance. 
Figure \ref{j1738_null_burst_profile} shows the 
integrated profiles from all separated null and burst pulses for 
a comparison, which clearly shows absence of any emission during the 
nulls. 

By contrast, in \psrb\ the bright phases consist of sustained non-null pulses 
[see Figure \ref{nbhist_both}(c)], typically of 70-80 pulses. 
What is unusual about this pulsar is the very large number of 
isolated burst pulses which occur in the inter-burst phases. 
%Although the total number of burst {\it pulses} is dominated by those participating 
%in burst phases, the total number of {\it bursts} of any length is dominated by the single bursts. 
These are not evident in Figure \ref{spdisplay} and can only be found by a careful 
inspection of the sequences. We designate them as inter-burst pulses (IBPs) and 
as shown in Section \ref{emission_in_null} that they appear at random between the 
bright phases, and maybe throughout the entire emission of \pb. 
As a result, the apparently long nulls of this pulsar become subdivided 
in a random way giving rise to the exponential distribution strikingly 
visible in the NLH [in Figure \ref{nbhist_both}(d)]. 
No such effect is seen in \pa.     

One consequence of the burst-null mix in the bright phases of \pa\  and the
IBPs in \pb\ is that the conventional burst and null length histograms of both
pulsars in Figure \ref{nbhist_both} show no evidence of the quasi-periodic
behaviour of the bright and off-phases despite it being very clear in the pulse
sequences shown in Figure \ref{spdisplay}. To overcome this, we carried out a visual
inspection of the single pulses of both pulsars and identified appropriate
bright phases and their separation (the separation being defined as the
number of pulses between the first pulses of two consecutive bright phases). 

\begin{figure}[h!]
 \centering
  \vspace{0.2in}
  \subfigure[]{
  \centering
   \includegraphics[width=2.1 in,height=2 in,bb=0 0 350 260]{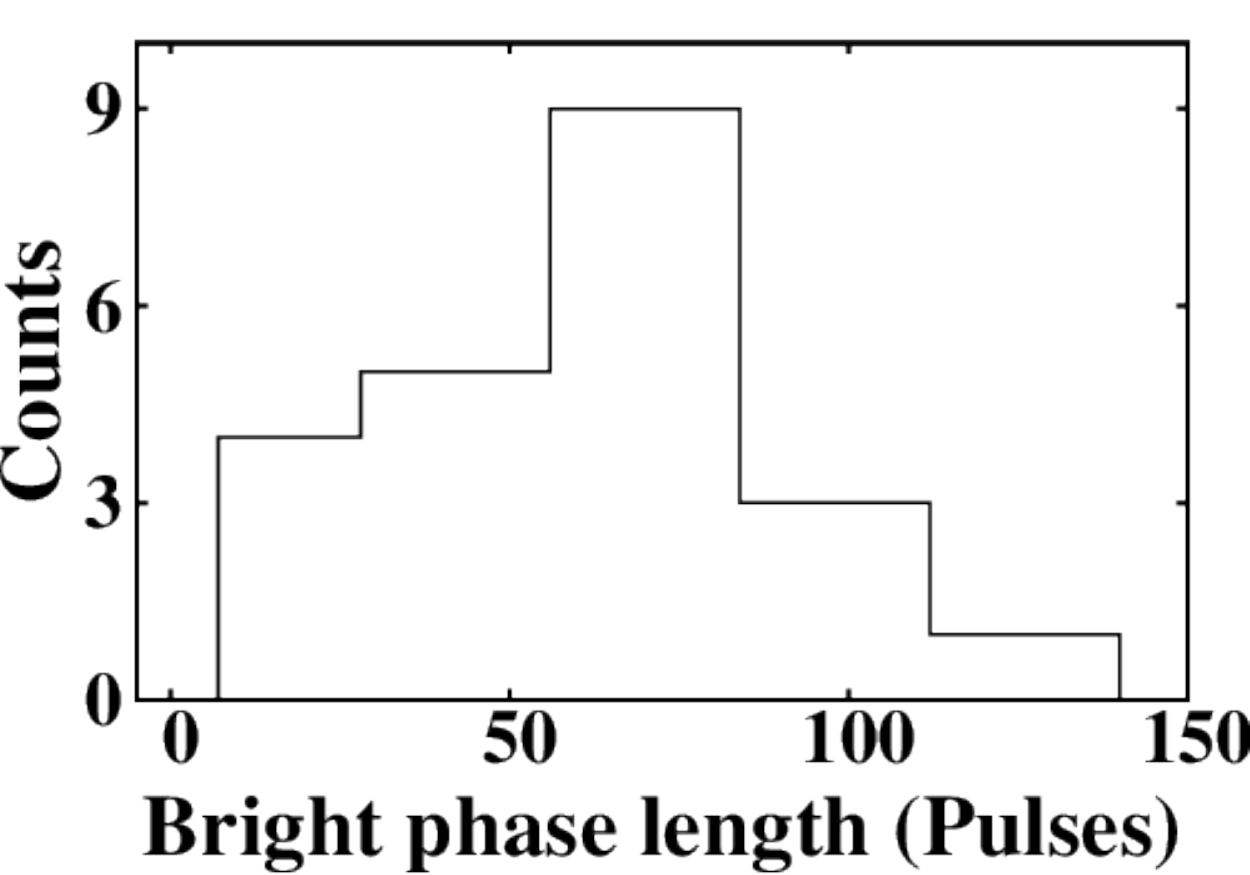}
  % BBB_length_hist_j1738.eps: 0x0 pixel, 300dpi, 0.00x0.00 cm, bb=50 50 410 302
  \label{j1738BBB_length}
  }
  \subfigure[]{
  \centering
  \includegraphics[width=2.1 in,height=2 in,bb=0 0 350 260]{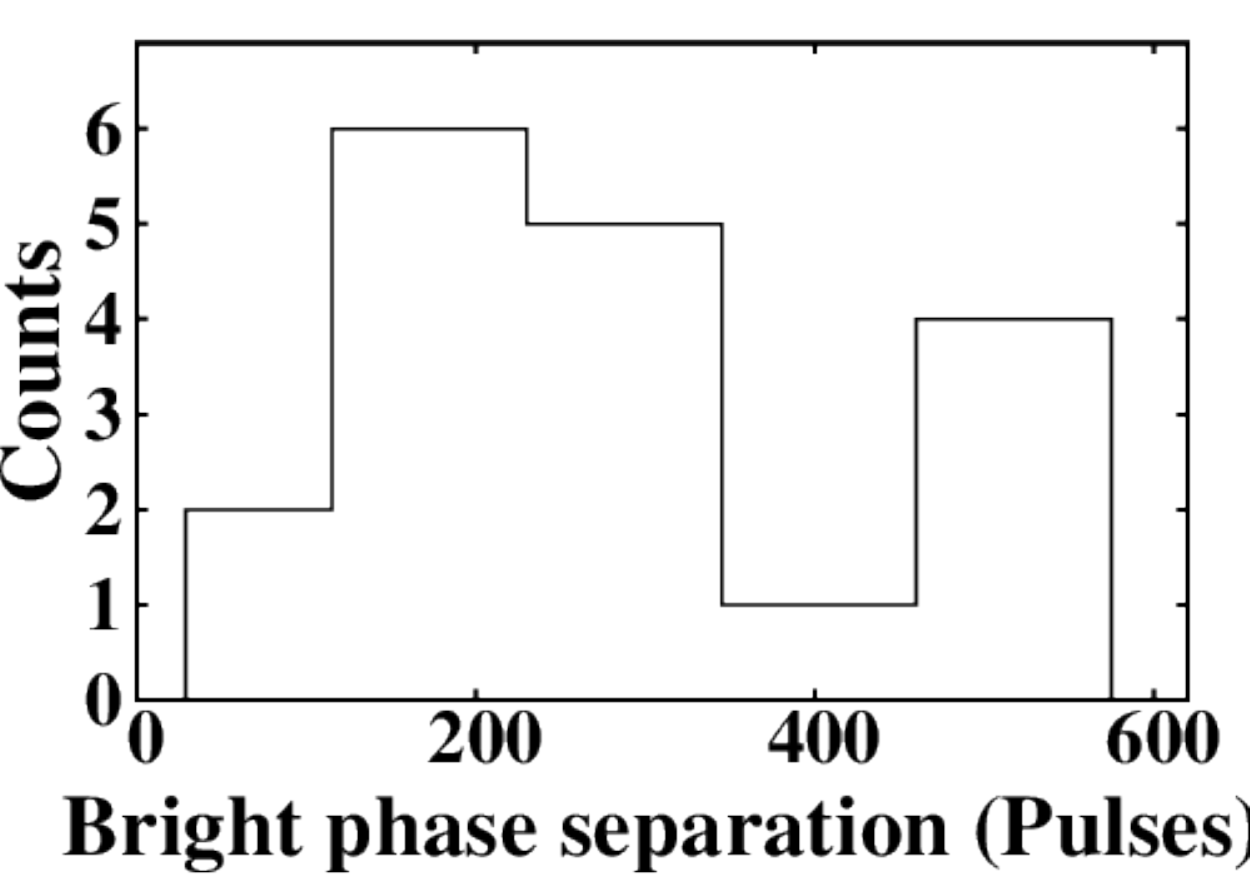}
  % BBB_First_period_gap_hist_j1738.eps: 0x0 pixel, 300dpi, 0.00x0.00 cm, bb=50 50 410 302
  \label{j1738BBB_gap}
  }
  \begin{picture}(0,0)
  \centering 
  \put(-200,150){PSR J1738$-$2330}
  \put(-200,-40){PSR J1752+2359}
  \end{picture}
  \subfigure[]{ 
  \includegraphics[width=2.1in,height=2 in,angle=0,bb=0 0 350 260]{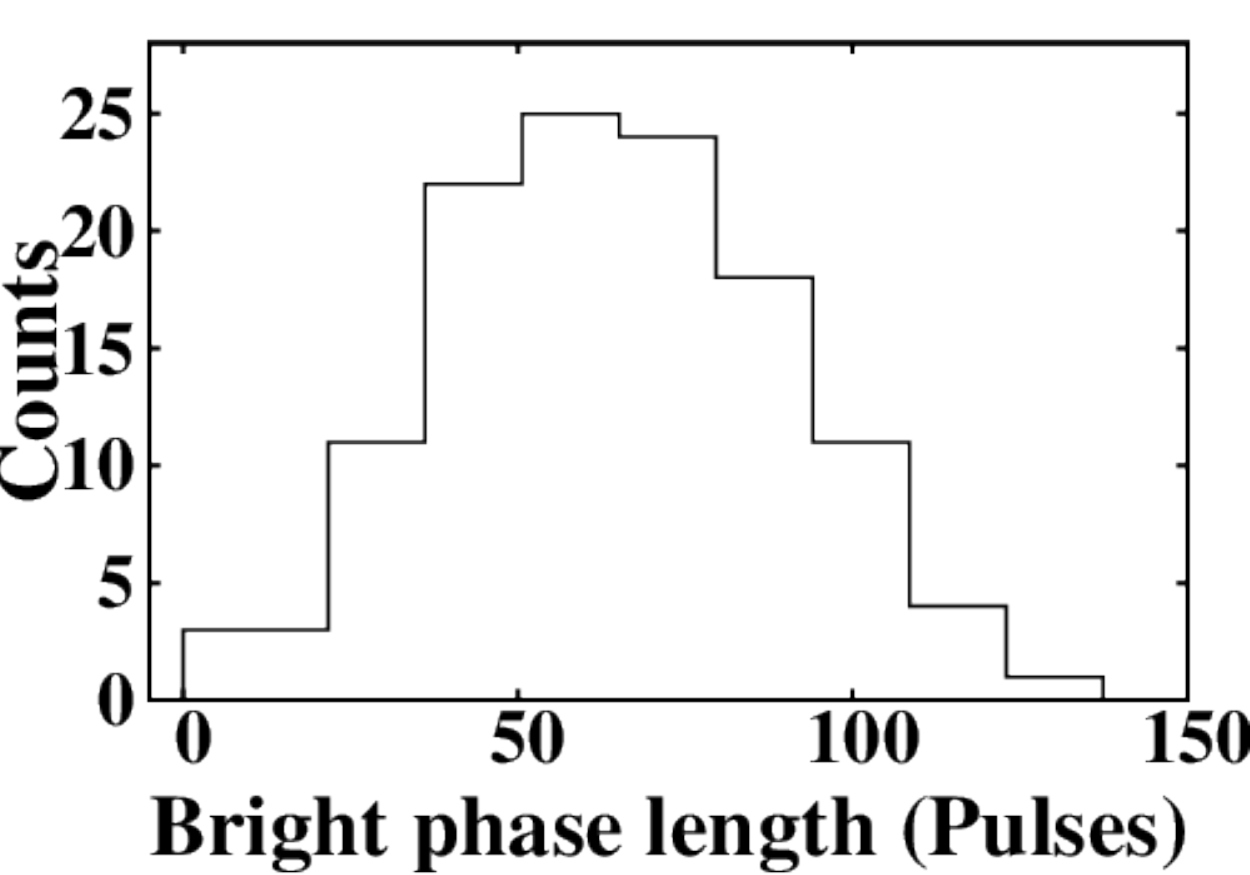}
  % BBB_length_hist.eps: 0x0 pixel, 300dpi, 0.00x0.00 cm, bb=59 52 504 720
  \label{j1752BBB_length}
  }
  \centering
  \subfigure[]{
  \includegraphics[width=2.1in,height=2 in,angle=0,bb=0 0 350 260]{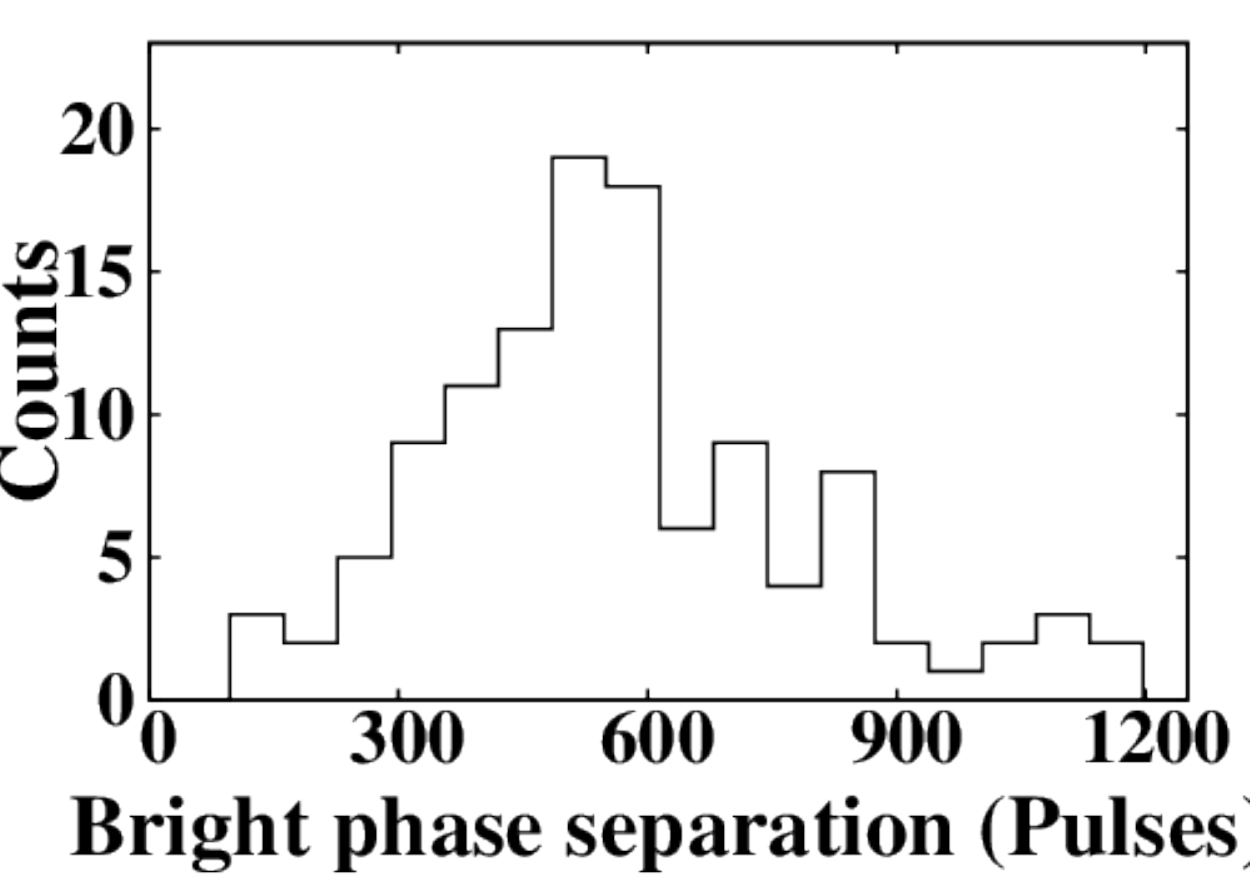}
  % BBB_sepration_from_start_combine_hist.eps: 0x0 pixel, 300dpi, 0.00x0.00 cm, bb=59 52 504 720
  \label{j1752BBB_gap}
  }
\caption[The histograms for the length of the visually identified bright phases]
{The histograms for the length of the visually identified bright phases for 
PSRs (a) \pa\ (c) \pb\ obtained from the 325 MHz observations at the GMRT. 
The histograms of separation between the first  
pulse of successive bright phases for PSRs (b) \pa\ (d) \pb\ are also shown.  
The distributions for bright phase length are similar, while 
those for bright phase separation are very different}
\label{BBB_lengh_gap_fig}
\end{figure}
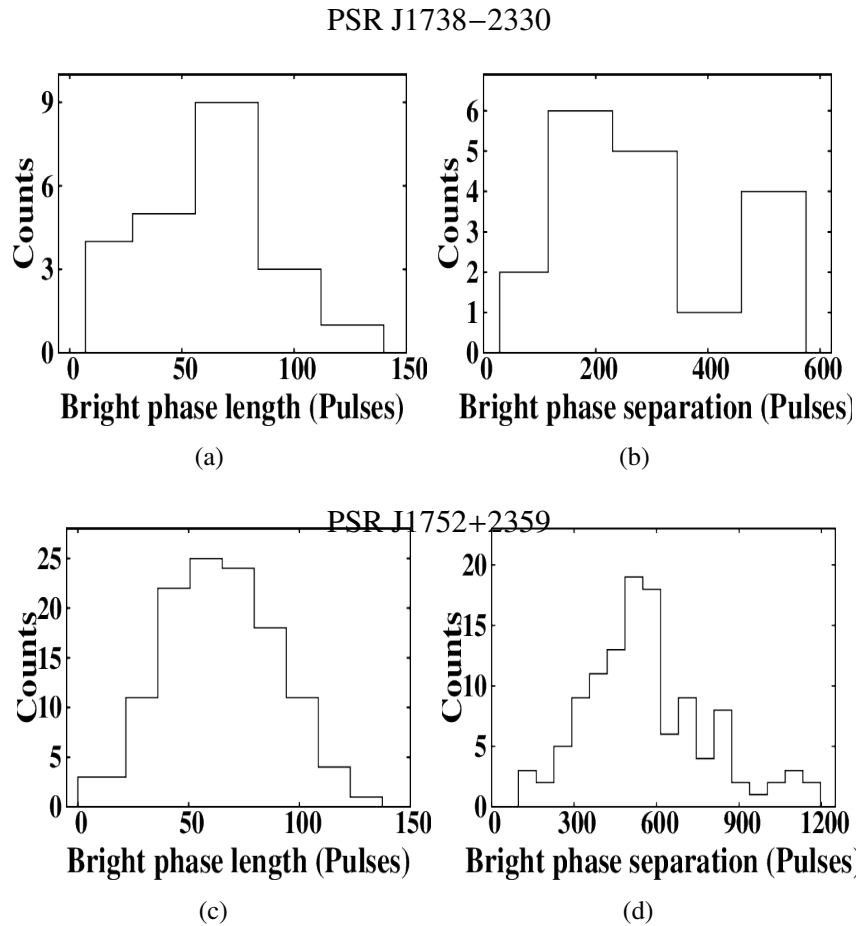

In the case of \pa, 21 bright phases were identified and a histogram of these
is shown in Figure \ref{BBB_lengh_gap_fig}(a). This distribution shows a peak at around
50 to 70 pulsar periods with a spread of around 40 pulsar periods, a result
which might be expected from combining the short bursts and nulls of Figures
\ref{nbhist_both}(a) and \ref{nbhist_both}(b). Likewise, a histogram of the 
separations between the first pulse of two successive bright phases is 
shown in Figure \ref{BBB_lengh_gap_fig}(b). This has a surprising bimodal character with peaks around
170 and 500 pulses. Lengths of around 500 pulses cannot be formed by combining a
typical burst length from Figure \ref{nbhist_both}(a) and the longest null
length from Figure \ref{nbhist_both}(b),
where the maximum is 400 pulses. We can therefore deduce that the longest
inter-burst phases must be interrupted at some point by very short (and maybe
weak) bursts which were rejected as burst pulses. This explains the longer
off-phase stretches seen in this pulsar's sequence of Figure \ref{spdisplay} and
foreshadows our discussion of this pulsar's quasi-periodic patterns in Section
\ref{quasiperiod}. However, strong claim can not be made on the bi-modality of the 
bright phase separation due to their small numbers. 

For \pb, the identification of bright phases was a more difficult task, since
it had to take into account the intrusion of IBPs in the
off-pulse phases and the fact that the precise end point of a fading bright phase
was sometimes difficult to fix (see Section \ref{BBB_patten}). We were able to
identify around 123 bright phases from a visual check of the single pulse
sequences in the 8-hour observations. The bright phase lengths of Figure
\ref{BBB_lengh_gap_fig}(c) show much the same distribution as the burst lengths of
Figure \ref{nbhist_both}(c) (apart from the single-pulse bursts). 
We find a prominent peak at around 60 pulsar periods with 
a spread of around 40 pulsar periods. The distributions of 
bright phase lengths of \pa\ and \pb\ are similar, but note that the former has a smaller 
number of bright phases. The bright phase separations of \pb\ also formed a
broad distribution around a central peak. We measured 120 examples and the
histogram is shown in Figure \ref{BBB_lengh_gap_fig}(d). The peak is
at about 570 pulses with separations ranging from 150 to 
1200 pulsar periods. The wide range of separations indicates that the nulling 
pattern is not strictly periodic, as can be seen in Figure \ref{spdisplay}, and is
discussed in the next Section.

\section{Quasi-periodicity of pulse energy modulation} 
\label{quasiperiod}
\label{qperiodsection}

In the previous section we elucidated the basic statistics of the null-burst
distributions in both pulsars. Both have bright phases whose lengths are
approximately normally (or possibly lognormally) distributed over lengths with a
similar number of pulses [Figures \ref{BBB_lengh_gap_fig}(a) and
\ref{BBB_lengh_gap_fig}(b)], but their respective separations follow very different
statistics [Figures \ref{BBB_lengh_gap_fig}(b) and \ref{BBB_lengh_gap_fig}(d)]. In the case of
\pa\  the separations of the bright phases have a bimodal distribution,
suggesting that the pulsar must sometimes `skip' a burst [Figure\ref{BBB_lengh_gap_fig}(b)], 
giving an exceptionally long off-phase. In \pb, the burst
separations cover a very wide range from 100 up to 1000 pulses 
[Figure \ref{BBB_lengh_gap_fig}(d)]. These features are apparent in Figure \ref{spdisplay}, 
with the sequence of \pa\ including a very long off-phase, and that of \pb\
producing a quasi-periodic effect despite the varying separations.  

\begin{figure}[h!]
  \centering
 \subfigure[]{
 \includegraphics[width=4.2 in,height=2.7 in,angle=0,bb=0 0 360 252]{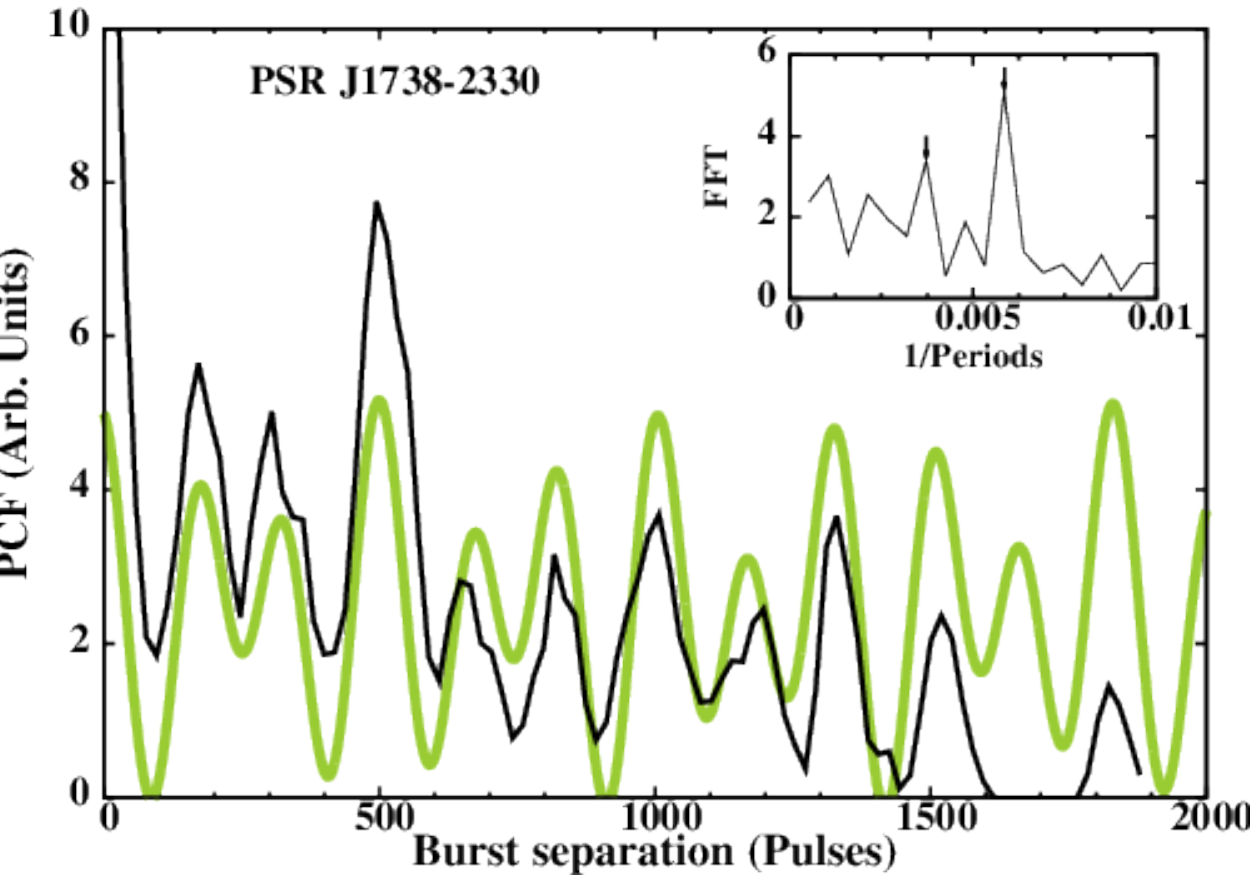}
 % Hist_PCF_J1738.eps: 0x0 pixel, 300dpi, 0.00x0.00 cm, bb=50 50 410 302
 }
 \centering
 \subfigure[]{
 \includegraphics[width=4.2 in,height=2.7 in,angle=0,bb=0 0 360 252]{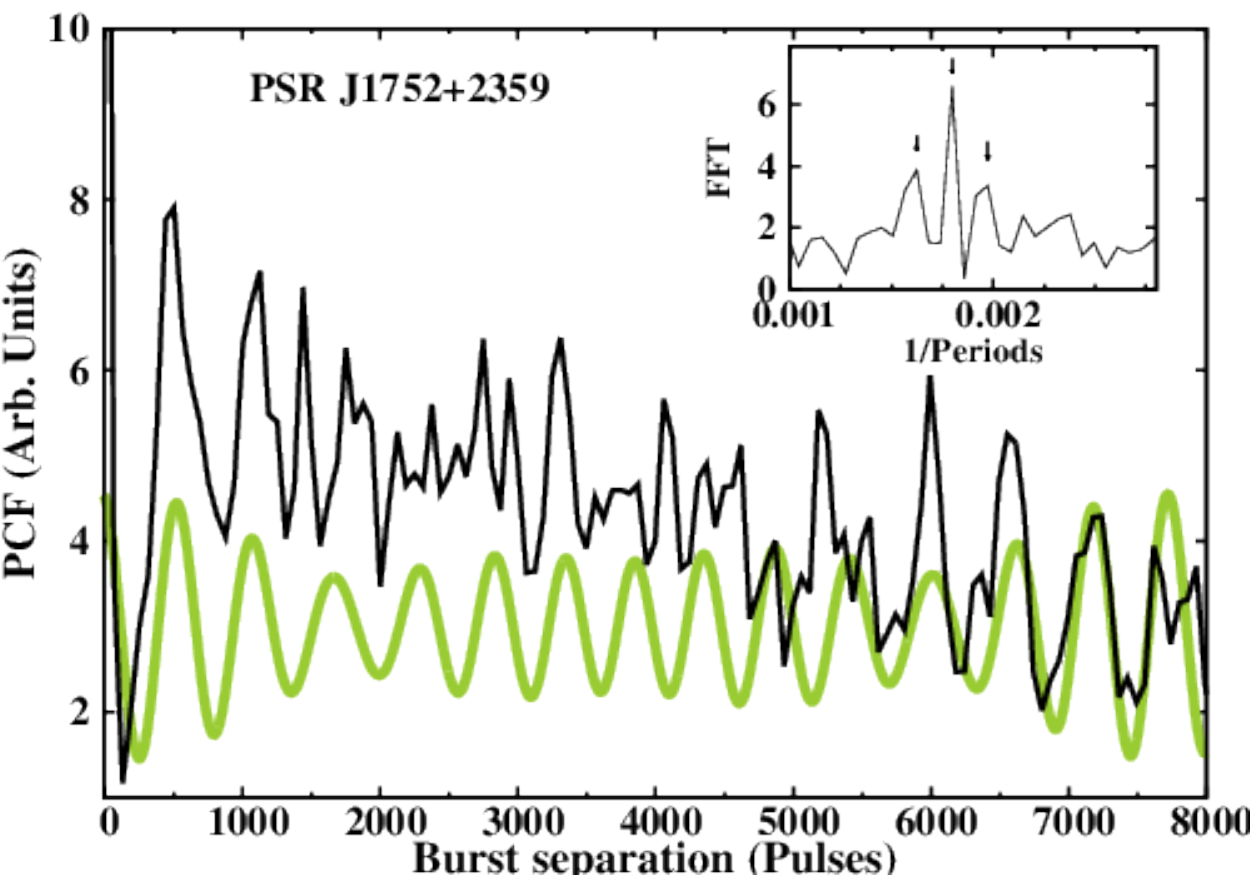}
 % Hist_PCF_J1752.eps: 0x0 pixel, 300dpi, 0.00x0.00 cm, bb=50 50 410 302
 }
 \caption[The Pair Correlation Function for PSRs \pa\ and \pb]
 {The Pair Correlation Function for PSRs (a) \pa\ and 
 (b) \pb. The PCF is calculated as a function of measured separation between the observed burst 
  pulses in the units of pulsar periods. The inset figures show the Fourier spectra 
  of the corresponding PCF fluctuations (significance of these spectral features 
  are 4$\sigma$ and 8$\sigma$ for \pa\ and 3$\sigma$, 3.5$\sigma$ and 6.5$\sigma$ for \pb). 
  The sine-waves after combining the corresponding 
  periodicities are overlaid along with the PCFs in grey colour for comparison. Note the reduced 
  amplitude of the PCFs for large separations due to the finite 
  length of the observed data. \psra\ shows longer quasi-periodicity coherence time-scale than that for for \psrb. 
  See Section \ref{qperiodsection} and Appendix D for more details.}
 \label{pcfplots}
\end{figure}

To probe deeper into the nature of ``quasi-periodicity" in our two pulsars, 
we require a suitable tool. Power spectra are a common device [as discussed in 
Section \ref{null_behavior_sect} for various pulsars and also by \cite{hjr07}], 
but in pulsars whose burst pulses form clusters such spectra are very much
dominated by red noise due to the observed jitter in cluster separation. In the
present context, a more useful procedure is to form a Pair Correlation Function
(PCF) for each pulsar. This is simply a histogram of all burst-to-burst
separations, whether successive or not, and can be utilized to find the
coherence time over which any quasi-periodicity is maintained. A formal
description of PCFs can be found in Appendix D. 

The PCF for \pa\ in Figure \ref{pcfplots}(a) is formed from adding the
burst-to-burst separation histograms from all observed pulse sequences, 
where each sequence is approximately 2000 pulses long. 
The pulse clustering is very clear and reveals a dramatic
periodicity of about 170 pulses between each pulse of a cluster and the pulses
in other clusters. If all bright bursts were equal in length then the height of
the peaks would be equal for nearby separations and slowly decline for large
separations (see Section \ref{BBB_patten}), but we see a striking drop in the level
of these peaks for the two closest bursts to a given burst, followed by a strong
third peak. More distant peaks also have irregular levels, but a general
periodicity of about 170 pulses is maintained. In essence, it is the structure
revealed by this PCF which underlies the bimodal distribution of the burst phase
separations shown in Figure \ref{BBB_lengh_gap_fig}(b) with its second peak at about 500
pulses. 

To understand our result we formed the Fourier spectrum (FFT) of the PCF, which
is shown in the inset diagram of Figure \ref{pcfplots}(a). This indicates two 
separate periodicities corresponding to approximately 170 pulses 
and 270 pulses, with the former dominating. The weighted sum of two sine-waves
with these periodicities is overlaid on the PCF and demonstrates a good match
with the PCF peaks. The 170 pulse and 270 pulse periodicities in the PCF are approximately the
third and the second harmonics respectively of $\approx 500$ pulse periodicity,
similar to the one obtained from Figure \ref{j1738lrf} in Section \ref{null_behavior_sect}. The peaks produced by the dominant periodicity of 170 pulses
are diminished for two successive peaks and then enhanced for the third by the
weaker but significant harmonically-related periodicity of 270 pulses. 
The weaker periodicity is not \emph{precisely} harmonically related to 170 and
thereby produces a progressive difference in the peak levels. What is very
remarkable is that this reproduces very closely the relative magnitude of the
peaks throughout the combined 2000 pulse separation of Figure \ref{pcfplots}(a).
This suggests an emission pattern which maintains coherence over at least 2000
pulses. However, we must caution that these results may or may not 
persist on time-scales longer than our observations. 

The PCF for \pb\ [Figure \ref{pcfplots}(b)] has its first peak at 
around 500 pulsar periods. This is more pronounced than that of \pa\, but much
broader in terms of pulses and it clearly corresponds to the peak found in the
bright phase separations [Figure \ref{BBB_lengh_gap_fig}(d)]. A second peak occurs at 1150
pulses, which is little late to be simply periodic with the first peak, and later
peaks show very little evidence of long-term coherence. We obtained the Fourier
spectra of the PCF,  which is shown in the inset diagram in Figure
\ref{pcfplots}(b), indicating three periodic features at 540, 595 and 490
pulses, with the first dominating. The weighted sum of the three sine-waves is overlaid
on the PCF but, in contrast to \pa, the generated wave loses coherence beyond 
1500 pulses as only the first two peaks are matched. Thus in \pb, three sine
waves are needed to yield just the two leading peaks of the PCF, in stark
contrast to \pa, where two sine waves were enough to match the entire
observation. This suggests that the decomposition into sine waves
has little physical significance in this pulsar.

It is indeed apparent in Figures \ref{spdisplay}(b) and \ref{BBB_lengh_gap_fig}(d) 
that the bursts of \pb\ appear with a wide variety of unpredictable separations. 
Thus the superficial impression of quasi-periodicity is only maintained by the fact that that the
separation of successive pulses is rarely less than 500 pulses, as is indicated
by the PCF. At two and three burst separations, there seems little evidence of
memory operating between bursts and even less of an underlying periodicity. 

\section{Pulse energy modulation in bright phases}
\label{BBB_patten}

\begin{figure}[h!]
\centering
 \subfigure[]{  
    \centering
    \includegraphics[width=6in,height=2in,angle=0,bb=0 0 430 150]{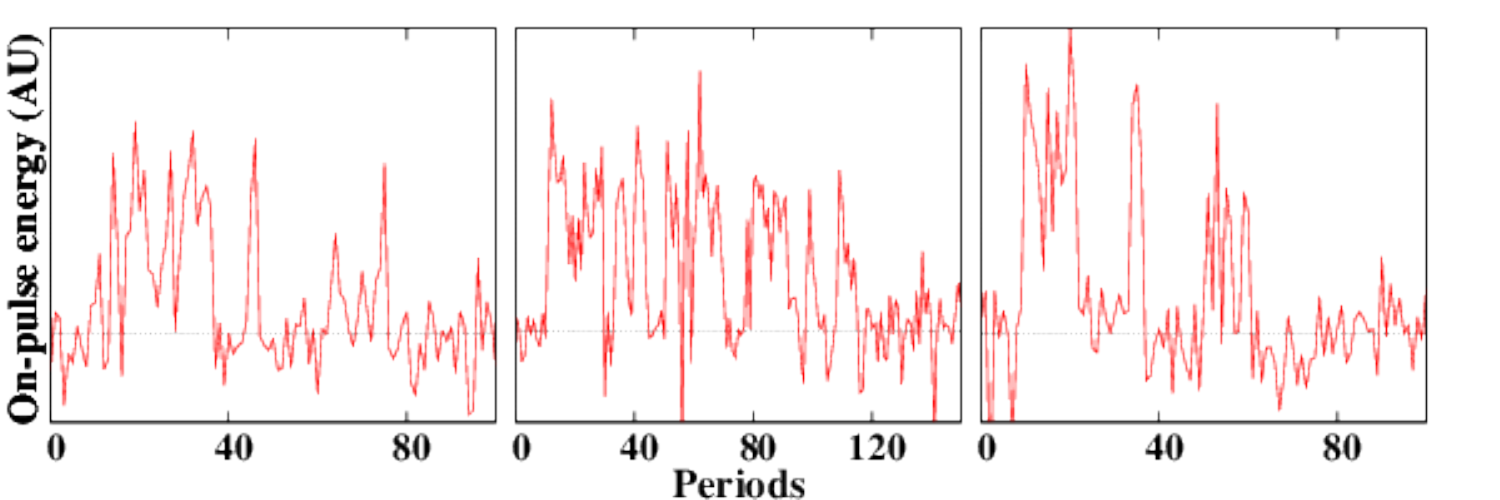}
    \label{j1738BBB}
  }
  \subfigure[]{  
    \centering
    \includegraphics[width=6in,height=2in,angle=0,bb=0 0 430 150]{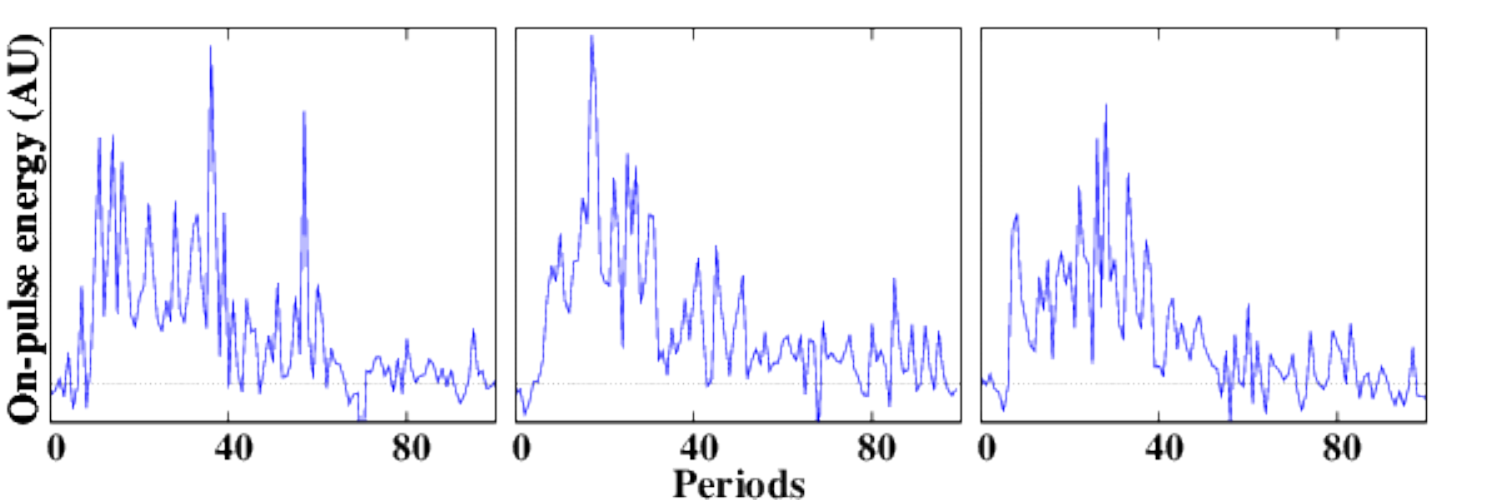}
    \label{j1752BBB}
  }
 \caption[Examples of the on-pulse energy for three bright phases for PSRs \pa\ and \pb]
 {Examples of the on-pulse energy for three bright phases for PSRs (a) \pa\ (b) \pb\ 
 extracted from the GMRT observations. Both  pulsars show gradual decay of 
 the on-pulse energy at the end of the bright phase. 
 Note the short nulls inside the bright phase for \psra. 
 Contrary to that, most  pulses of a bright phase  from \psrb\ are 
 burst pulses.}
 \label{BBB_example}
\end{figure}

\begin{figure}[h!]
  \centering
  \subfigure[]{
    \centering
    \includegraphics[width=4.1in,height=2.5in,bb=0 0 360 250]{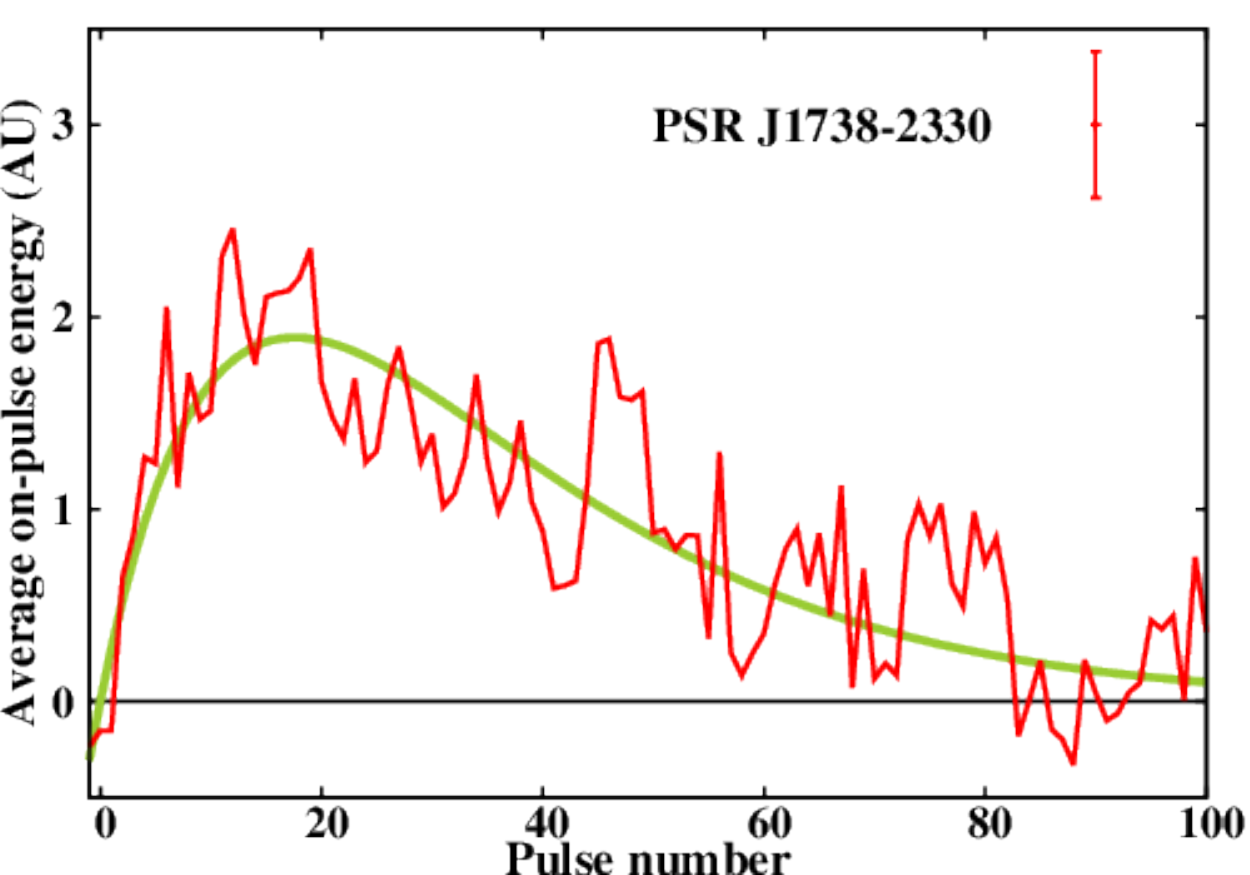}
    \label{j1738burst_fit} 
  }
  \centering
  \subfigure[]{ 
    \centering
    \includegraphics[width=4.1in,height=2.5in,bb=0 0 360 250]{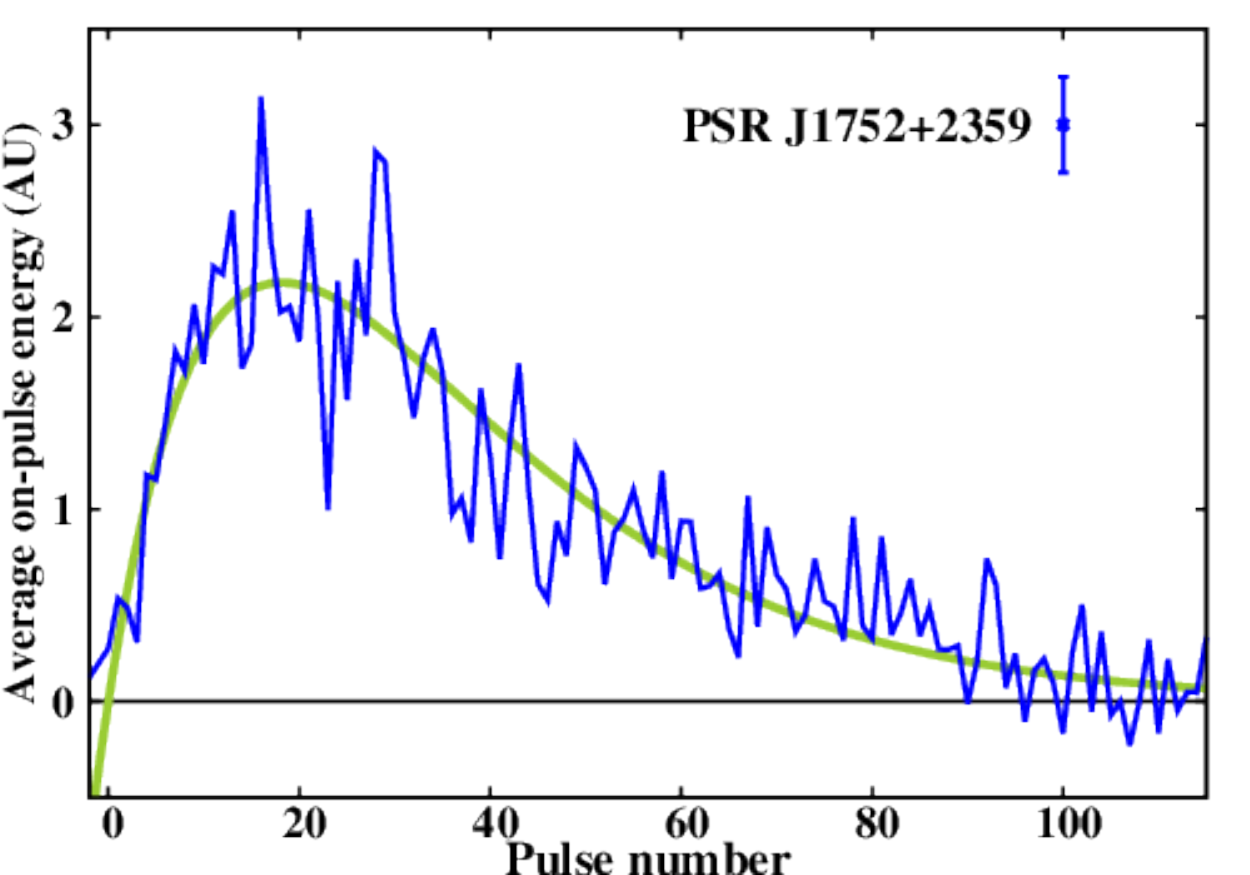}
 % burst_fit_forPaper_j1752.eps: 0x0 pixel, 300dpi, 0.00x0.00 cm, bb=50 50 554 770
    \label{j1752burst_fit_avg}
  }
 \caption[On-pulse energy for 100 pulses, averaged over 12 bright phases]
 {On-pulse energy for 100 to 110 pulses, averaged over 12 bright phases,
 for PSRs (a) \pa\ and (b) \pb\ in the GMRT observations, 
 demonstrating a similar decay in the bright phases of both pulsars.  
 The off-pulse root-mean-square deviations are shown in the top right corner for both
 pulsars. The solid green lines are fitted models given by equation \ref{fx}, 
 with the reduced $\chi^2$ of around 1.2 and 1.9 for PSRs \pa\ and \pb, respectively.}
\label{burst_fit_avg}
\end{figure}

\pa\ exhibits bright phase structures of various lengths, 
consisting of short burst bunches interspersed with 
short nulls, which was also highlighted in Figure \ref{j1738_bunches} 
in Section \ref{null_behavior_sect}. 
The onset of a bright phase is relatively sudden for \pa\ 
with a strong burst pulse, which is followed by a change in the 
emission throughout the bright phase duration. This change 
is manifested by either a reduction in the intensity of 
single pulses or by an increase in the number and/or 
length of short null states, as can be seen in the three 
examples displayed in Figure \ref{BBB_example}(a). 
At the end of every bright phase, the pulsed emission 
clearly goes below the detection threshold 
and produces long null phase or off-phase. We extracted 100 
pulses starting with the first identified burst pulse 
and averaged these over several bright phases to obtain 
averaged bright phase profile. 
A few bright phases in our observations were separated by less than 
100 pulses from the next consecutive bright phase, hence they 
were not included in this analysis. The on-pulse 
energy averaged over 12 bright phases is shown in 
Figure \ref{burst_fit_avg}(a), where a decline in 
the pulse intensity towards the end of the averaged bright phase 
is evident. 

The onset of the bright phase in \pb\ is more gradual, 
spanning typically 5 to 10 pulses. The decline in the intensity 
from its peak is also more striking, as was 
also reported by \cite{lwf+04}. Figure \ref{BBB_example}(b) shows three examples
of the decline in on-pulse energy during a bright phase in this pulsar. Note the absence of
convincing null pulses during the decline. The on-pulse energy for 110 pulses,
averaged from an equal number of bright phases (i.e. 12) as that for 
\pa, is shown in Figure \ref{burst_fit_avg}(b) to illustrate the 
similarity of average bright phase on-pulse energy variations in the two pulsars. 

\subsection{Bright phase modelling}
\label{bbb_modeling}
\begin{figure}
 \centering
 \includegraphics[width=4.2in,height=3in,angle=0,bb=0 0 360 252]{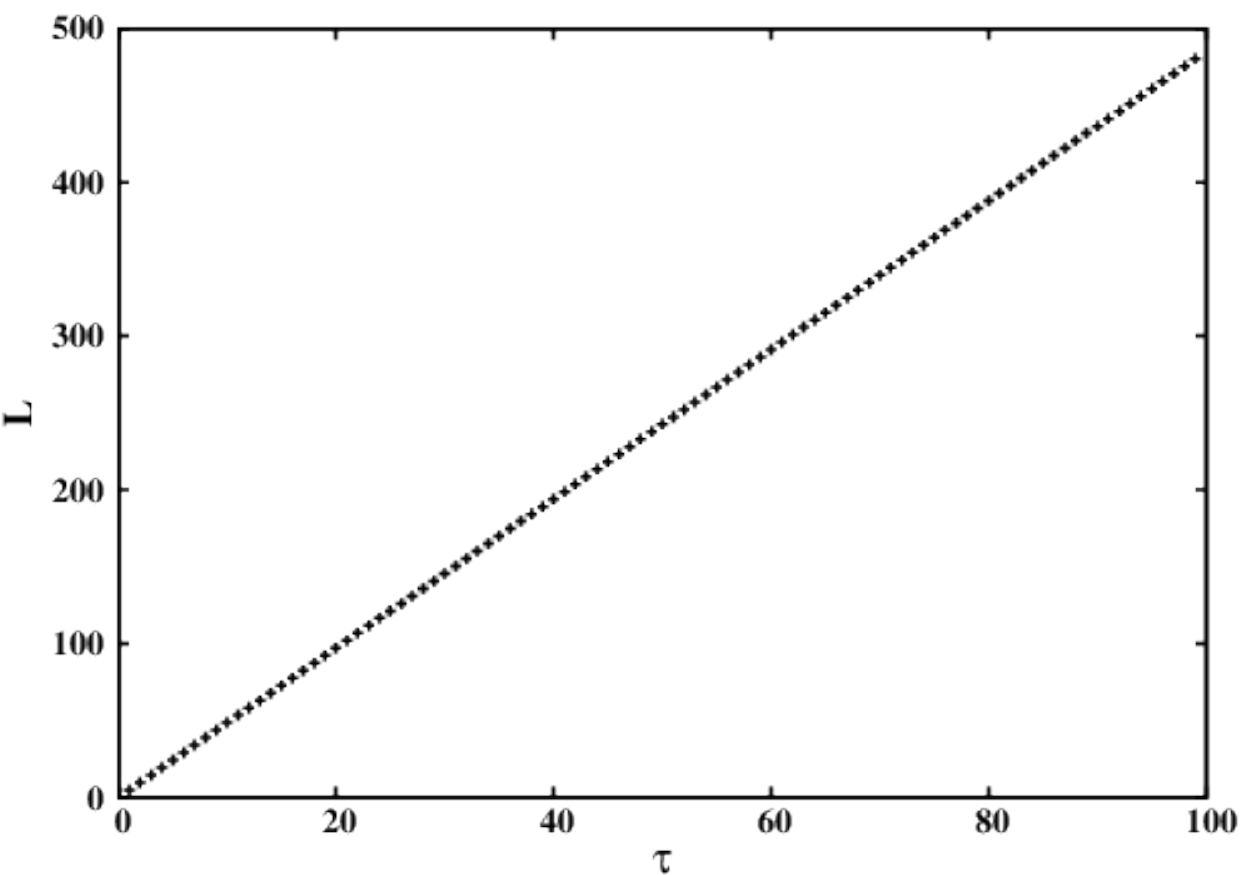}
 % tau_diff_forPaper.eps: 0x0 pixel, 300dpi, 0.00x0.00 cm, bb=50 50 410 302
  \caption{The linear dependence between the obtained 10\% bright phase width (i.e. $L$) and
 the decay parameter (i.e. $\tau$) with a slope of $\sim$ 4.9$\pm$0.05.}
 \label{tau_diff}
\end{figure}

Our analysis of \pb\ clearly shows that on-pulse energy for 
most of the individual bright phases follows the model given 
by equation \ref{fx}. \pa\ also has a similar average on-pulse energy 
variation for its bright phases. 
\begin{equation}
f(x)~=~{\alpha}\cdot{x}\cdot{e^{-(x/\tau)}} 
\label{fx}
\end{equation}
Here, $\alpha$ is a scaling parameter and $\tau$ is the decay 
time-scale. The values for $\alpha$ and $\tau$ were obtained by a least-square-fit 
of function \ref{fx} to the on-pulse energy in a bright 
phase with errors on each energy measurement given by 
off pulse rms.  The length of a given bright phase 
was defined as the difference between 
the two points where $f(x)$ crosses  of 10\% of the $f(x)_{max}$.
It can be shown that $x_{max}$ is given by $\tau$. So the points 
where the $f(x)$ attains 10\% of the peak value (i.e. $x=x_{10}$) 
are given by, 
\begin{equation}
{f(x)}\arrowvert{_{x = {x_{10}}}}~ = ~ 
{\alpha}\cdot{x_{10}}\cdot{e^{-(x_{10}/\tau)}} ~ = ~{0.1}\cdot\frac{\alpha\cdot\tau}{e}. 
\label{f10}
\end{equation}
$f(x)$ attains these values on both sides of the peak position 
and these points can be determined by solution of equation \ref{f10}, 
obtained using numerical methods and the difference between these 
two points was defined as the length of a given bright phase 
(i.e. $L$ =  $x_{10h}$ $-$ $x_{10l}$). It can be seen 
from equation \ref{f10} that values of $x_{10h}$ and $x_{10l}$ do 
not depend upon $\alpha$. To quantify the dependence of the 
bright phase length on $\tau$, we solved equation \ref{f10}  
for a range of $\tau$ values. Figure \ref{tau_diff} shows 
the linear dependence of L on this range of $\tau$, which can also be expressed as,  
\begin{equation}
 L ~ \approx ~ 4.9\times\tau. 
\end{equation}
The error on L is given by, 
\begin{equation}
% \triangle{f(x)}\arrowvert{_{x = {x_{10}}}}~ = ~\frac{0.01}{e}\cdot(\tau\cdot\triangle\alpha~ + ~ \alpha\cdot\triangle\tau)  
\bigtriangleup{L} ~ \approx ~ 4.9\times\bigtriangleup\tau. 
 \label{f10err}
\end{equation}
 
\begin{figure}[h!]
 \centering
\includegraphics[height=3in, width=4.5in,angle=0,bb=0 0 360 252]{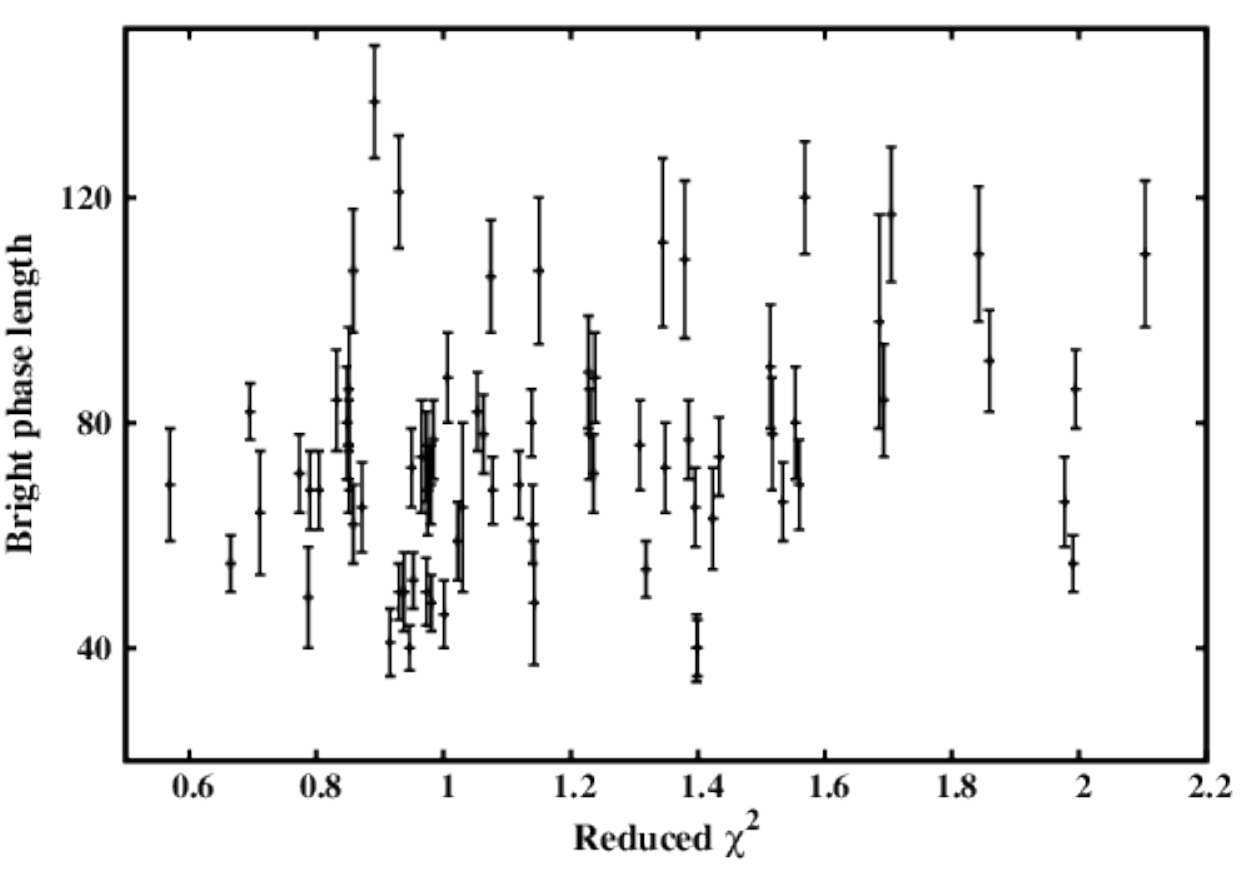}
\caption[Acquired bright phase lengths for \pb]
{Acquired bright phase lengths obtained after the modelling 
83 individual extracted bright phases for \pb\ with their associated errors. 
The bright phase lengths are shown as a function of reduced $\chi^2$ from the individual fit. 
The attained average bright phase length is around 77$\pm$20 periods from these measurements.}
\label{BBB_length_fit}
\end{figure}

The average length of a bright phase derived from the 
above mentioned least-squares fit are 86$\pm$4 pulses 
and 88$\pm$3 pulses for \pa\ and \pb, respectively 
and are consistent with the histograms of bright phase 
lengths discussed in Section \ref{qperiodsection} 
(Figures \ref{j1738BBB_length} and \ref{j1752BBB_length}, which were obtained 
from the visual inspection). The length of the 
individual bright phase for \pb\ was obtained 
in a similar manner. Out of 123 observed bright 
phases in \pb, only 83, with higher S/N burst pulses, were fitted to obtain 
their lengths (shown in Figure \ref{BBB_length_fit}). 
We obtained the reduced $\chi^2$ in the range of 
around 0.6 to 2.2 for these fits. The average 
length of the bright phases from these measurements 
is 77$\pm$20 pulses. 

\subsection{Bright phases of \psrb}
\begin{figure}[h!]
\centering
\begin{center}
\includegraphics[width=5.5in,height=3.2in,angle=0,bb=0 0 359 252]{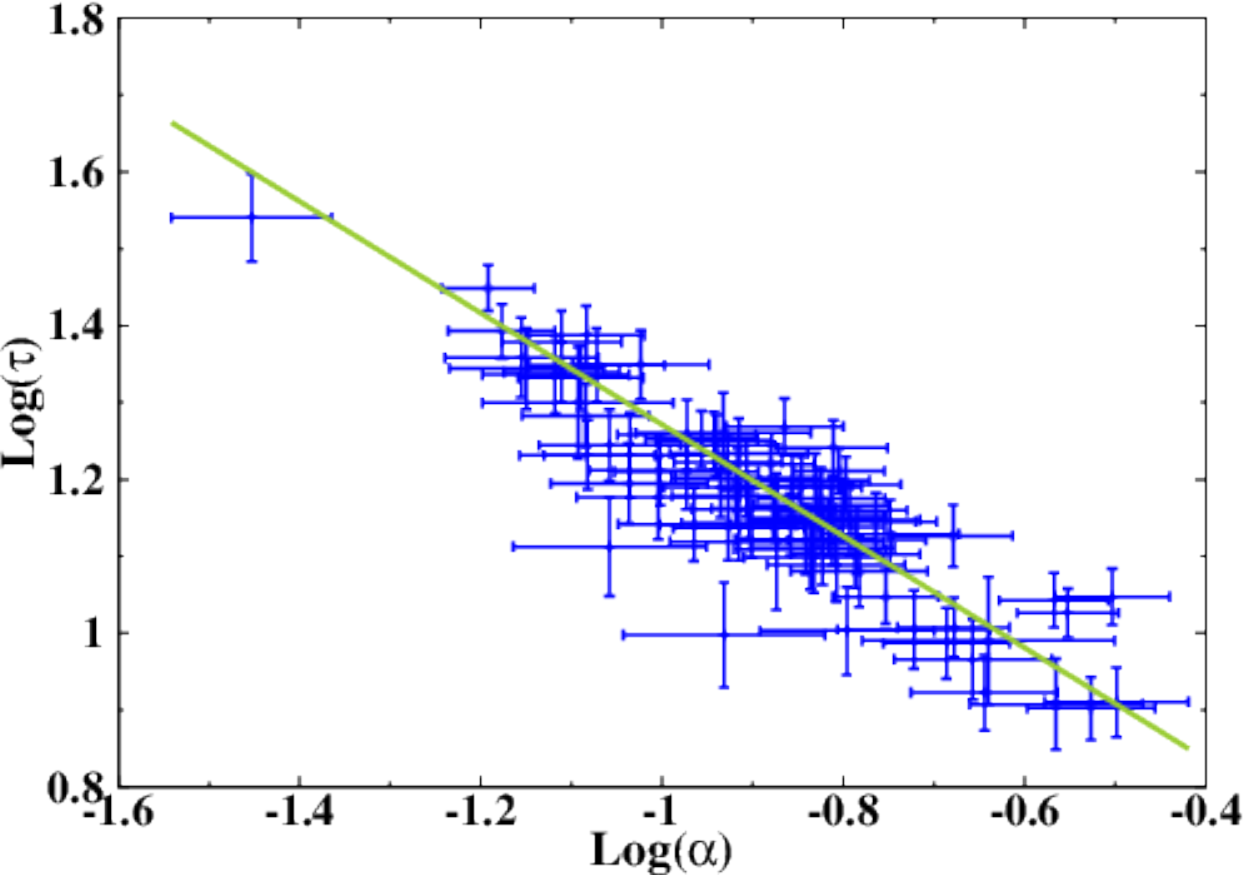}
\caption[The relationship between bright phase parameters for \psrb]
  {The relationship between bright phase parameters for \psrb\ 
  obtained from the GMRT observations. The bright phase on-pulse energy decay time-scale (i.e. $\tau$) 
  as a function of peak of the bright phase on-pulse energy (i.e. $\alpha$) on a log-log scales. 
  The slope was fitted after considering errors in both coordinates. 
  A strong anti-correlation is evident in this diagram. }
  \label{bbb_beta_decay}
\end{center}
\end{figure}

We could investigate the nature of bright phases in \pb\ further  
due to their large number in our $\sim$ 8 hours data, observed with 
the GMRT, as well as  high S/N 30 minute data with the Arecibo
telescope. The on-pulse energy for around 83 observed bright phases 
were fitted to the model given by equation \ref{fx} and 
their respective $\alpha$ and $\tau$  were obtained (shown in Figure \ref{BBB_length_fit}). 
The log-log plot in Figure \ref{bbb_beta_decay} clearly displays a power law 
dependence of  $\tau$ with $\alpha$. The fitted line in the Figure \ref{bbb_beta_decay} 
gives the power law index of around -0.74 $\pm$ 0.04  
incorporating the errors on both axes. The Kendall$\textquoteright$s tau 
rank order correlation between  $\tau$ and 
$\alpha$ is around -0.67 with a very small 
probability ($<10^{-7}$) of random chance. Thus, 
bright phases with large peak intensities decay faster 
for \pb.
\begin{figure}[h!]
 \centering
  \subfigure[]{
  \centering
  \includegraphics[width=4.5in,height=2.7in,angle=0,bb=0 0 360 250]{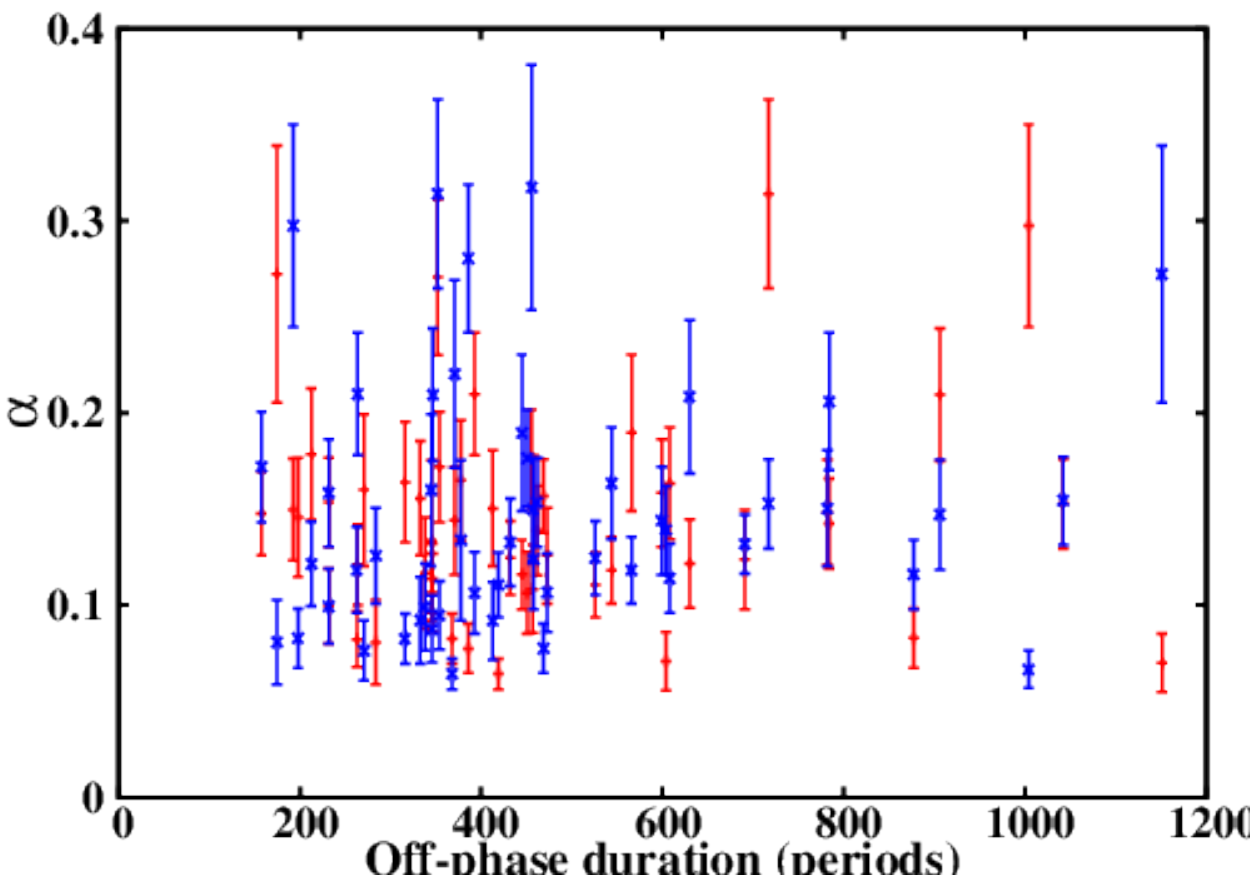}
% Gap_pre_post_A_para_comp.eps: 0x0 pixel, 300dpi, 0.00x0.00 cm, bb=50 50 410 302
  }
  \subfigure[]{
  \centering
  \includegraphics[width=4.5in,height=2.7in,angle=0,bb=0 0 360 250]{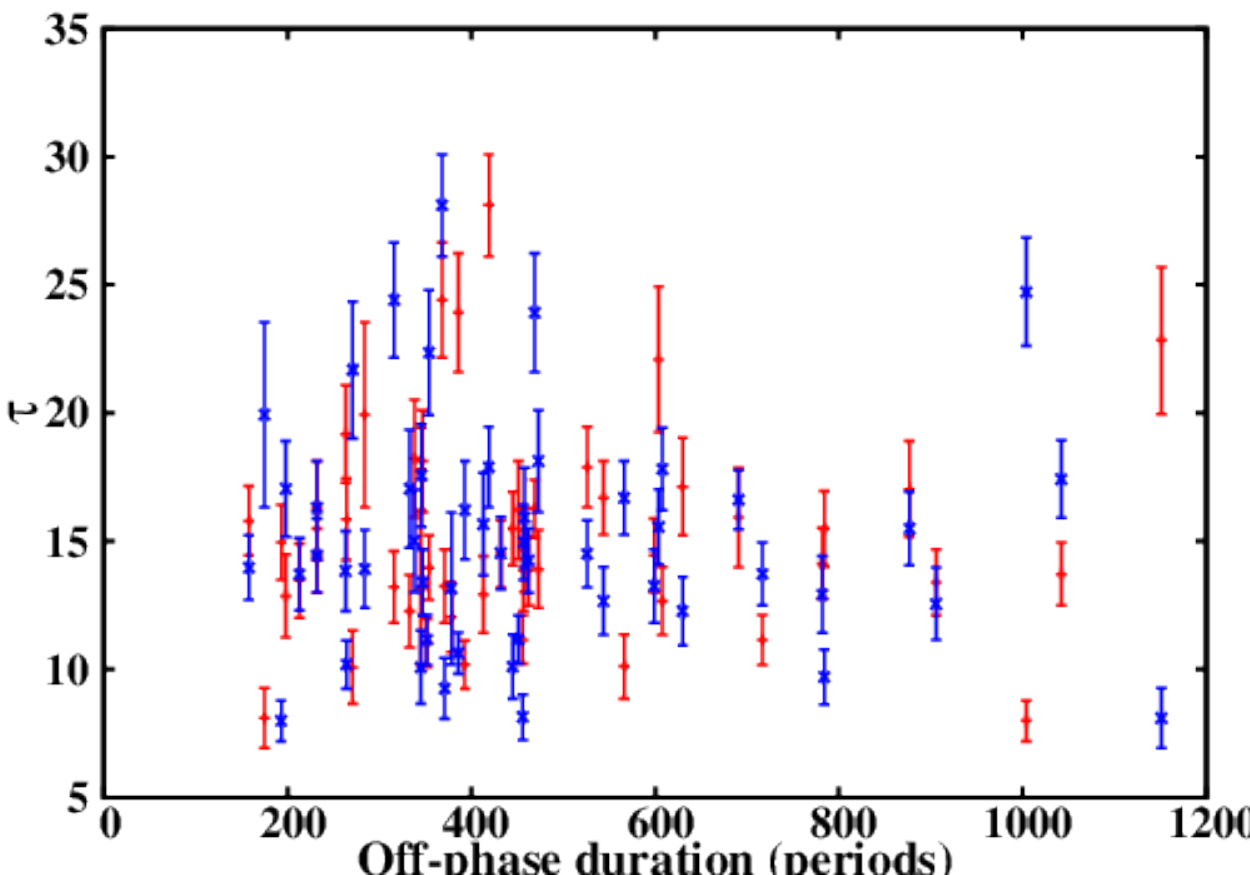}
% Gap_pre_post_T_para_comp.eps: 0x0 pixel, 300dpi, 0.00x0.00 cm, bb=50 50 410 302 
  }
  \caption[The relationship between bright phase parameters and the off-phase durations for \pb]{The 
  relationship between bright phase parameters and the off-phase durations for \pb\ 
  obtained from the GMRT observations. (a) The peak of the bright phase pulse energy and 
  (b) the decay parameter as a function of length of the associated off-phase. 
  The red points and the blue points are parameter values of bright phases before and after 
  the corresponding long off-phase, respectively. The scatter in the plots suggests that both 
  the parameters are uncorrelated with the duration of off phase preceding and succeeding a given bright phase.}
  \label{bbb_para_gap_relation}
\end{figure}

We also investigated the relationship between the separation 
between consecutive bright phases with the  
parameters of bright phase preceding and succeeding the null/off phase  
under consideration. We plotted the $\alpha$ and 
the $\tau$ of a bright phase as a function of the length 
of the off-phase preceding and succeeding it (shown in Figure \ref{bbb_para_gap_relation}). 
The lengths of the long off-phases were estimated 
as number of pulses between the last and the first pulse 
of two consecutive bright phases (for example number of pulses between 
pulse number 2885 and 3421 in Figure \ref{sp_en_oz_combined}).
We did not find any correlation as the parameters showed similar scatter 
for all lengths. Hence, bright phase parameters are independent of the 
length of the off-phase occurring before and after it. 
 
\begin{figure}[h!]
 \centering
 \includegraphics[width=4.7 in,height=3 in,angle=0, bb=0 0 360 250]{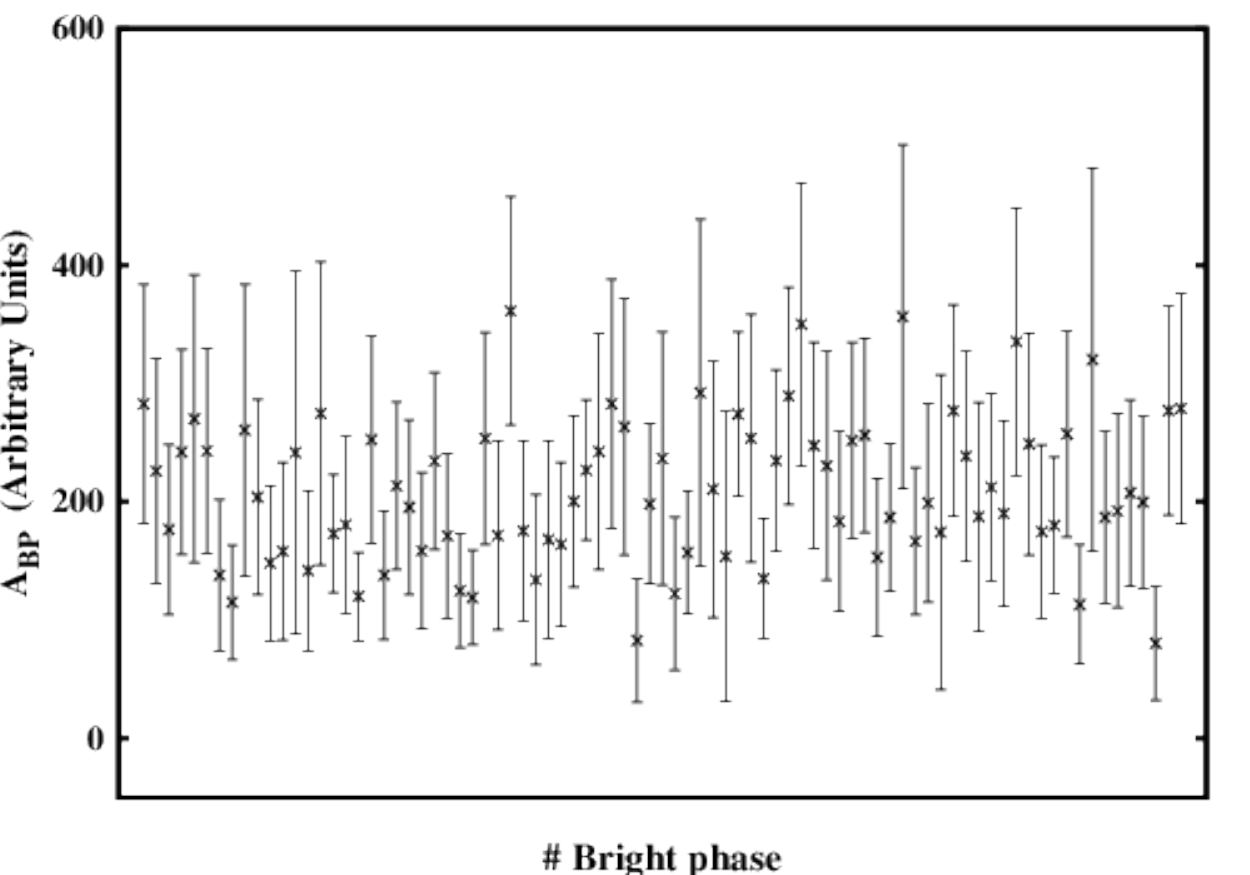}
% Area_under_bbb.ps: 504x720 pixel, 72dpi, 17.78x25.40 cm, bb=0 0 504 720
 \caption[The area under the on-pulse energy modelled curve for the 83 ascertained bright phases]
 {The area under the on-pulse energy modelled curve for the 83 identified bright phases 
 with their respective errors for \pb\ observed from the GMRT. The distribution of the data 
 points displays the spread of the mean is smaller compared 
 to the errors on the individual data points, indicating the average area 
 under every bright phase to be approximately constant.}
\label{area_bbb}
\end{figure}

The strong anti-correlation between $\alpha$ and  
$\tau$ suggests that the area under the on-pulse energy envelope 
for a bright phase for \pb\ is constant. 
As the full pulse energy of a bright phase can be modelled by equation \ref{fx}, 
the area under a given bright phase ($A_{BP}$) was calculated by,  
\begin{equation}
 A_{BP}~=~\int_0^\infty \mathrm{\alpha}\cdot{x}\cdot{e^{-(x/\tau)}}\,\mathrm{d}x 
 =~\alpha{\cdot}\tau^2
\end{equation}
The error in the obtained area can be obtained after propagating the fitting 
error $\bigtriangleup{\alpha}$ and $\bigtriangleup{\tau}$ as, 
\begin{equation}
\bigtriangleup{A_{BP}}~=~\bigtriangleup{\alpha}\cdot{\tau^2}~+~2{\alpha}\cdot{\tau}\cdot{\bigtriangleup{\tau}}.
\end{equation}
These were calculated for the 83 observed bright phases and 
were the same, within errors, for all of them (shown in Figure \ref{area_bbb}). 
This indeed confirms that the total intensity of bright phase is same irrespective 
of its length or peak intensity and consequently, 
the total energy released during a bright phase is likely to be 
approximately constant. 

\section{First and Last bright phase pulse}
\label{first_and_last_bbb_section}
\begin{figure}[h!]
 \centering
 \subfigure[]{
 \includegraphics[width=3.0 in,height=4.5 in,angle=-90,bb=50 50 554 770]{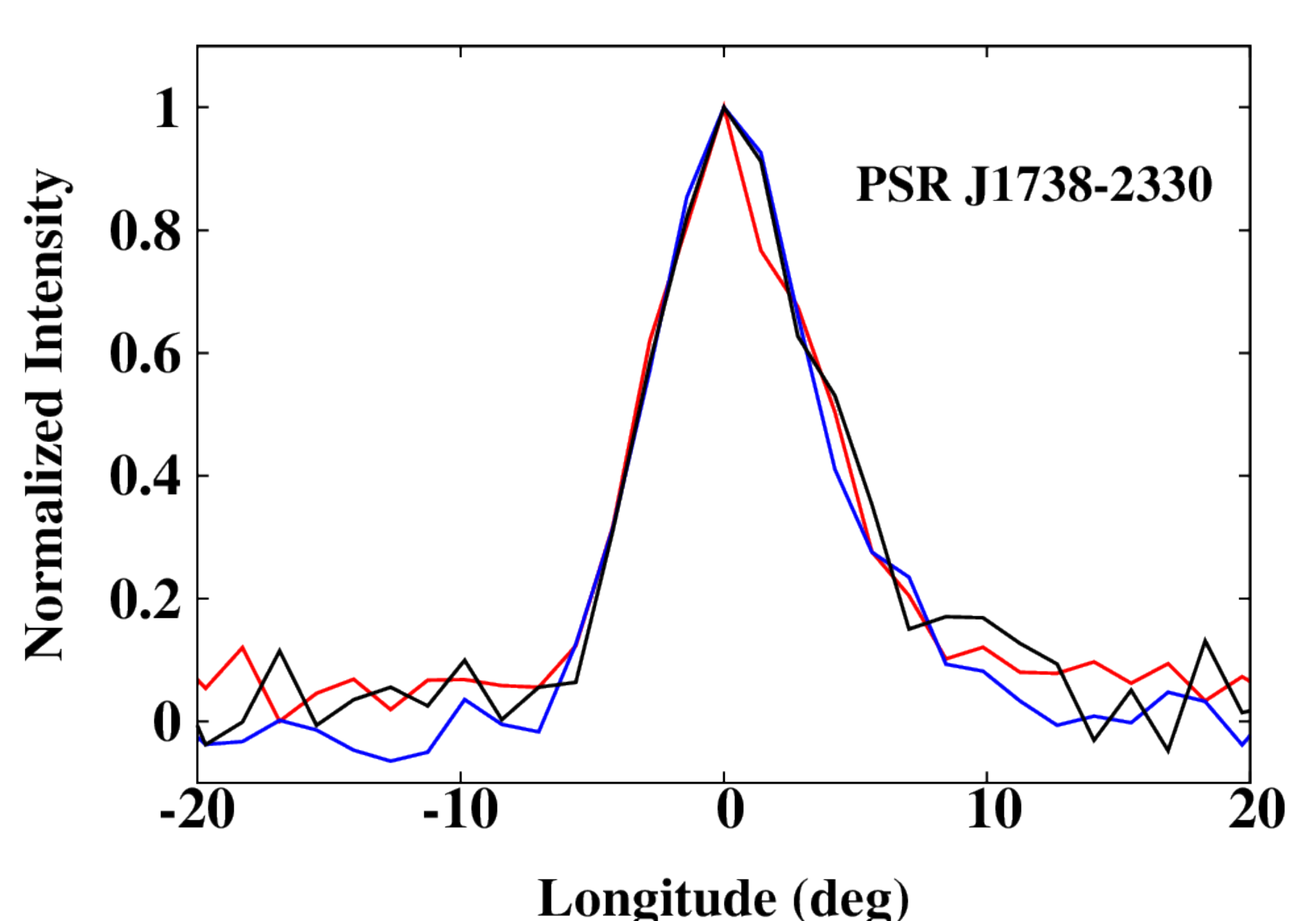}
 % First_Last_Full_profile_j1738.eps: 0x0 pixel, 300dpi, 0.00x0.00 cm, bb=50 50 410 302
 \label{j1738frst_last}
 }
 \subfigure[]{  
 \includegraphics[width=3.0 in,height=4.5 in,angle=-90,bb=50 50 554 770]{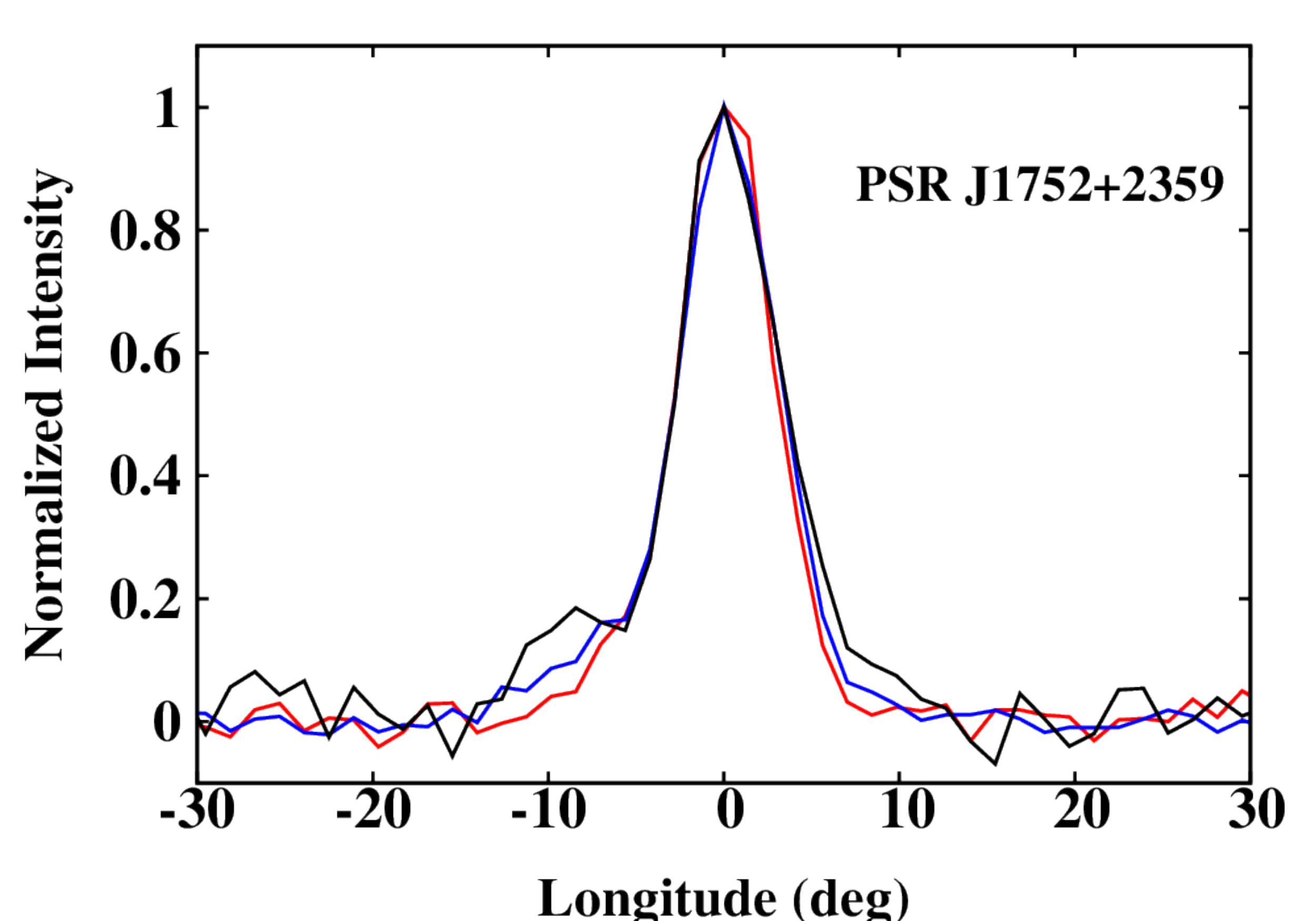}
 % First_Last_Full_profile_J1752.eps: 0x0 pixel, 300dpi, 0.00x0.00 cm, bb=50 50 410 302
 \label{j1752frst_last}
 }
 \caption[The first-bright-phase-pulse profile and the last-bright-phase-pulse profiles]
 {The first-bright-phase-pulse profile (red solid line) and the last-bright-phase-pulse profile 
 (black solid line) obtained from the GMRT observations for PSRs (a) \pa\ and (b) \pb\ 
 plotted against the observed pulse phase. The blue solid lines are the 
 average pulse profiles. All profiles were normalized with their respective peak intensities for 
 comparison.} 
 \label{first_last}
\end{figure}

Compared with \pb, \pa\ emits relatively strong individual pulses with
high S/N during the entire span of its bright phases [as can 
be seen from Figure \ref{BBB_example}(a)]. 
Although the intensity of the \pa's pulses during a bright phase shows 
a decline towards its end, the energy of the last pulse 
is sufficiently above the detection threshold to identify them 
clearly. The first and the last pulse of 18 bright phases were 
combined to form the first-bright-phase-pulse profile and 
the last-bright-phase-pulse profile. Figure \ref{first_last}(a) 
shows these profiles for \pa\ along with its average pulse 
profile. All these profiles look similar. A KS-test comparison,  
carried out between the first-bright-phase-pulse profile and the average pulse 
profile,  indicates similar distributions with 
94\% probability. Similarly, a KS-test comparison 
between the last-bright-phase-pulse profile and the average pulse profile also suggests  
similar distributions with even higher probability of 99\%. 
These results suggest that \pa\ switches 
back into the bright phase mode from the off-phase 
(and vice-versa) without any significant change in the 
emission. However, these results should be treated 
with caution due to the small number of pulses available 
in forming first-bright-phase-pulse profile and last-bright-phase-pulse profile.
 
For \pb, only 114 out of 123 observed bright phases were used 
to obtain the first-bright-phase-pulse as the remaining were affected by RFI. 
The red solid line in Figure \ref{first_last}(b)  
shows this profile. While the profiles look similar, a KS-test 
comparison between the first-bright-phase-pulse profile and the average pulse profile 
rejects the null hypothesis of similar distributions with 
99\% probability. The last pulse for most of the 
bright phases of \pb\ is very difficult to identify  
due to the decline in the pulse energy during this phase. 
Instead, a range of last pulses (10 to 20 pulses) were used for 
each bright phase to form the last-bright-phase-pulse profile. The length of each bright phase  
was determined with its associated error from the least-square-fit 
as explained in Section \ref{bbb_modeling}. 
To obtain the last-bright-phase-pulse profile, we averaged 
a range of pulses within the error bars 
(shown in Figure \ref{BBB_length_fit}) around the 
expected last pulse, obtained by adding the estimated length for every 
bright phase to the pulse number of its first pulse. 
The last-bright-phase-pulse profile averaged from 114 bright phases 
were compared with the average pulse profile 
using the KS-test, which rejected the null hypotheses 
of similar distributions with 99.9\% probability. 
Figure \ref{first_last}(b) shows the last-bright-phase-pulse profile with  
a significant ($>$ 3 times off-pulse rms) 
component preceding the pulse and shoulder emission 
after the pulse. Neither of these features is present in its 
average profile. 
\section{Emission in the off-phase of \psrb}
\label{emission_in_null}
\begin{figure}[h!]
 \centering
 \includegraphics[width=4.2in,height=4.5in,bb=0 0 250 250]{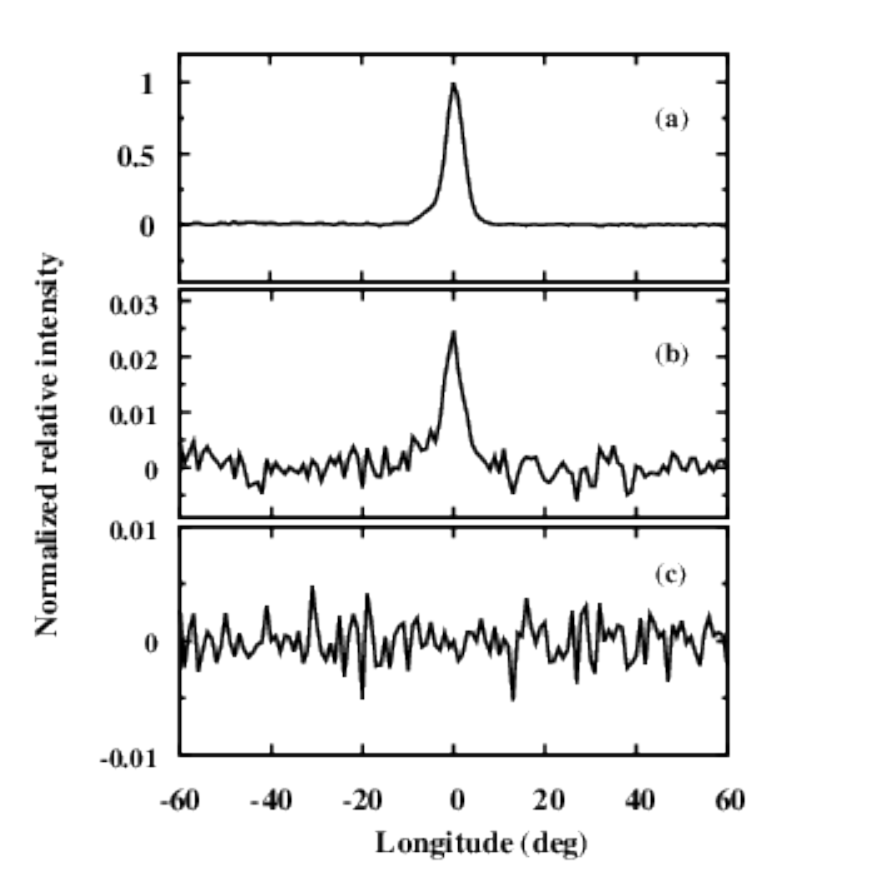}
 % J1752_separated_profile.eps: 0x0 pixel, 300dpi, 0.00x0.00 cm, bb=50 50 302 302
  \caption[Plot of three profiles with the normalized intensities to display emission in the null region]
  {Plot of three profiles with the normalized intensities as a function of observed pulse phase 
  for \psrb\ obtained from the GMRT observations. (a) The pulse
  profile obtained from all pulses in the bright phases (9000 pulses),
  (b) the off-phase pulse profile 
  from the pulses between the bright phases (41500 pulses) and (c) the
  absolute-null-pulse profile from the null pulses remaining  
  after removing the IBPs (38800 pulses). All three profiles 
  are on the similar relative scales after normalizing 
  them with the peak intensity of the bright phase pulse profile.}
 \label{BBB_inter_null_profile}
\end{figure}

In Section \ref{nulls} we noted that, unlike \pa, \pb\ exhibited
intermittent single pulse bursts during the off-phase. Weak emission
during an apparent null phase is not unknown \cite[]{elg+05} and may question
whether weak pulses are sometimes confused `true' nulls. 
We therefore investigated this by forming the off-phase pulse 
profile by averaging all pulses between the 
bright phases. Those parts of the observations (about 15\%), affected by the RFI, 
were excluded from this analysis. All the pulses in identified bright phases, 
amounting to around 9000 pulses, were separated 
from the remaining single pulse observations . The profile obtained 
from these pulses (i.e. bright phase pulse profile) is shown 
in Figure \ref{BBB_inter_null_profile}(a). The remaining 
41,500 pulses, which occurred between the bright phases, were 
integrated to form the off-phase pulse profile. Surprisingly, the 
null-pulse profile showed weak emission  
with a significance of around 20 standard deviations 
[Figure \ref{BBB_inter_null_profile}(b)]. It is therefore important to clarify whether
this emission originates from bright but rare single pulses and/or from
underlying weak emission.

In many nulling pulsars, the burst pulses are strong enough to create a bimodal
intensity distribution and can then be separated by putting a threshold between 
the two peaks in the histogram (as shown in Section \ref{separation_of_null_burst_sect}). 
\pb\ does not show such a bimodal distribution due to the presence of many weak energy 
pulses [Figure \ref{hist_both_fig}(b)]. 
Hence, the use of any threshold will lead to a wrong identification 
of weaker burst pulses as nulls and result in 
a weak emission profile during the off-phase, as seen in Figure \ref{BBB_inter_null_profile}(b).
This could be due to (a) weak emission throughout the off
phase, (b) emission from weak burst pulses in the 
diminishing tail of a bright phase, or (c) emission from a few wrongly
identified individual burst pulses during the off-phase. 

\begin{figure}[h!]
 \centering
 \includegraphics[width=5.5in,height=3.2in,angle=0,bb=0 0 360 252]{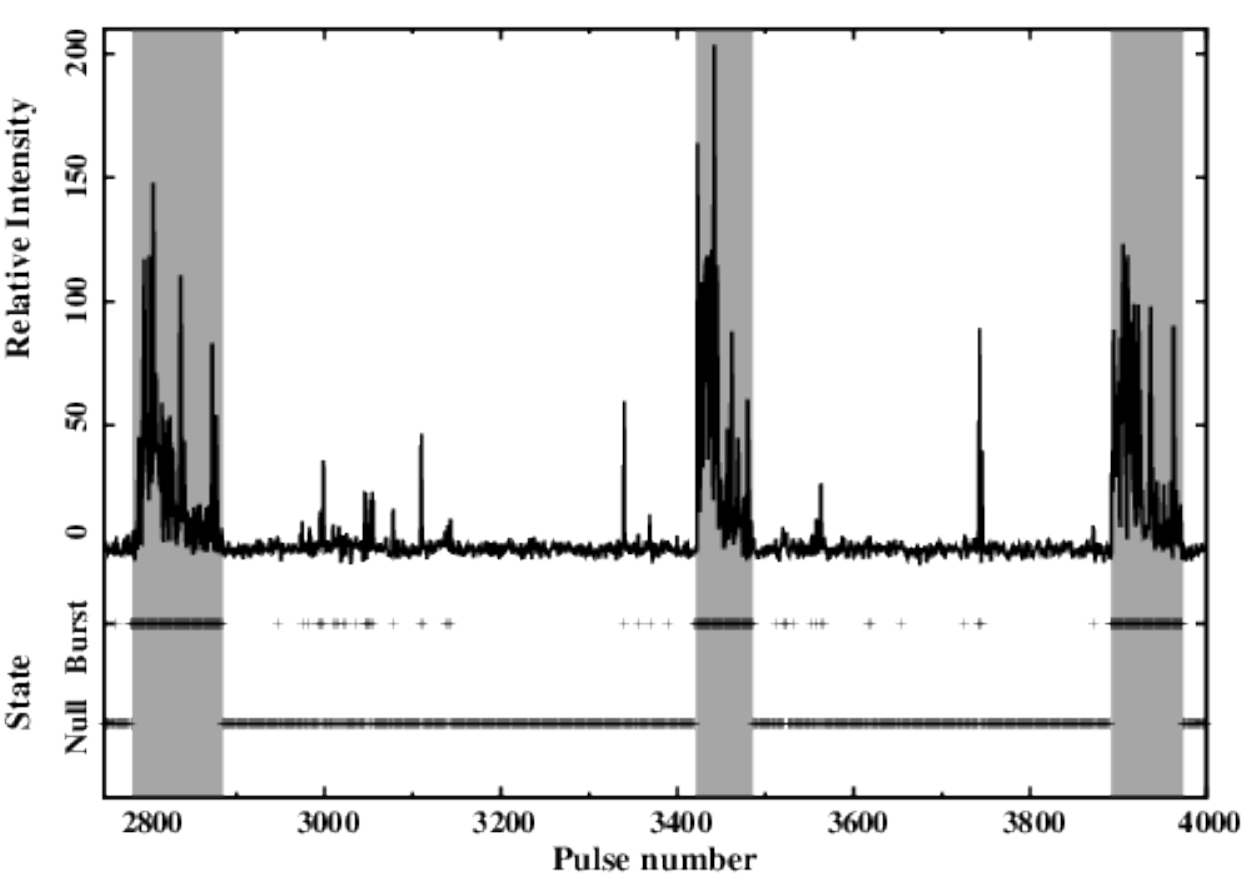}
 % J1752_onenergy_onezero.eps: 0x0 pixel, 300dpi, 0.00x0.00 cm, bb=50 50 410 302
\caption[Plot of on-pulse energy variation of around 1200 pulses of \pb\ observed
at 327-MHz with the Arecibo telescope]{Plot of on-pulse energy variation of 
around 1200 pulses of \pb\ observed at 327-MHz with the Arecibo 
telescope is shown in the top panel. The three
identified bright phases are shown inside the grey shaded regions for clarity. 
Every individual pulse was classified either as a null or as a burst pulse. The
bottom panel shows the null or burst state for the corresponding pulse 
consistent with a random distribution of the IBPs.}
\label{sp_en_oz_combined}
\end{figure}

To distinguish between the above possibilities, we 
used higher S/N observations for this pulsar obtained from the Arecibo telescope 
(see Section \ref{Arecibo_sect} for details). Figure \ref{sp_en_oz_combined} shows a section of the
observations  from the Arecibo observation. The pulse energy plot clearly 
shows three bright phases with about 60 isolated single 
burst pulses occurring during the off-phase between them. 
This strongly supports our earlier suspicion that weak emission,   
seen during the off-phase in this pulsar, is due to these 
isolated burst pulses (i.e. pulses outside the bright phases or IBPs). 

To investigate this further with GMRT observations, all pulses remaining after 
separating the bright phases were arranged in ascending order of their on-pulse
energy. A variable threshold technique, discussed in Section \ref{separation_of_null_burst_sect}, 
was used to separate these weak burst pulses from the null pulses. 
% A threshold was moved from the high energy end towards the low energy end till 
% the pulses below the threshold did not 
% show an average profile with a significant (above 3$\sigma$) 
% component.  All the pulses below this threshold 
% were tagged as null pulses. 
All the separated null pulses were again visually checked 
for wrongly identified nulls due to  presence of low level RFI. 
The average profile, obtained from 
all the pulses tagged as null after this procedure, 
is shown in Figure \ref{BBB_inter_null_profile}(c), 
indicating that tagged pulses are all nulled pulses,  
ruling out the possibility of weak emission throughout the null. 

This method of identifying the IBPs was also used on 
the Arecibo observation. The null or the burst state for each individual 
pulse was identified and is shown in Figure \ref{sp_en_oz_combined}   
in the bottom panel. It can be seen from this figure 
that IBPs are not localised near either the start or the end 
of a bright phase, but are distributed randomly inside the long off
phase. In fact, these pulses may not be confined to off-phases and may occur at
random through all phases since within a burst phase the their numbers would be
small and not detectable. To the best of our knowledge, 
the presence of such pulses has not been reported in a pulsar before.

\begin{figure}[h!]
 \centering
 \includegraphics[width=3.8 in,height=3 in,angle=0,bb=0 0 350 252]{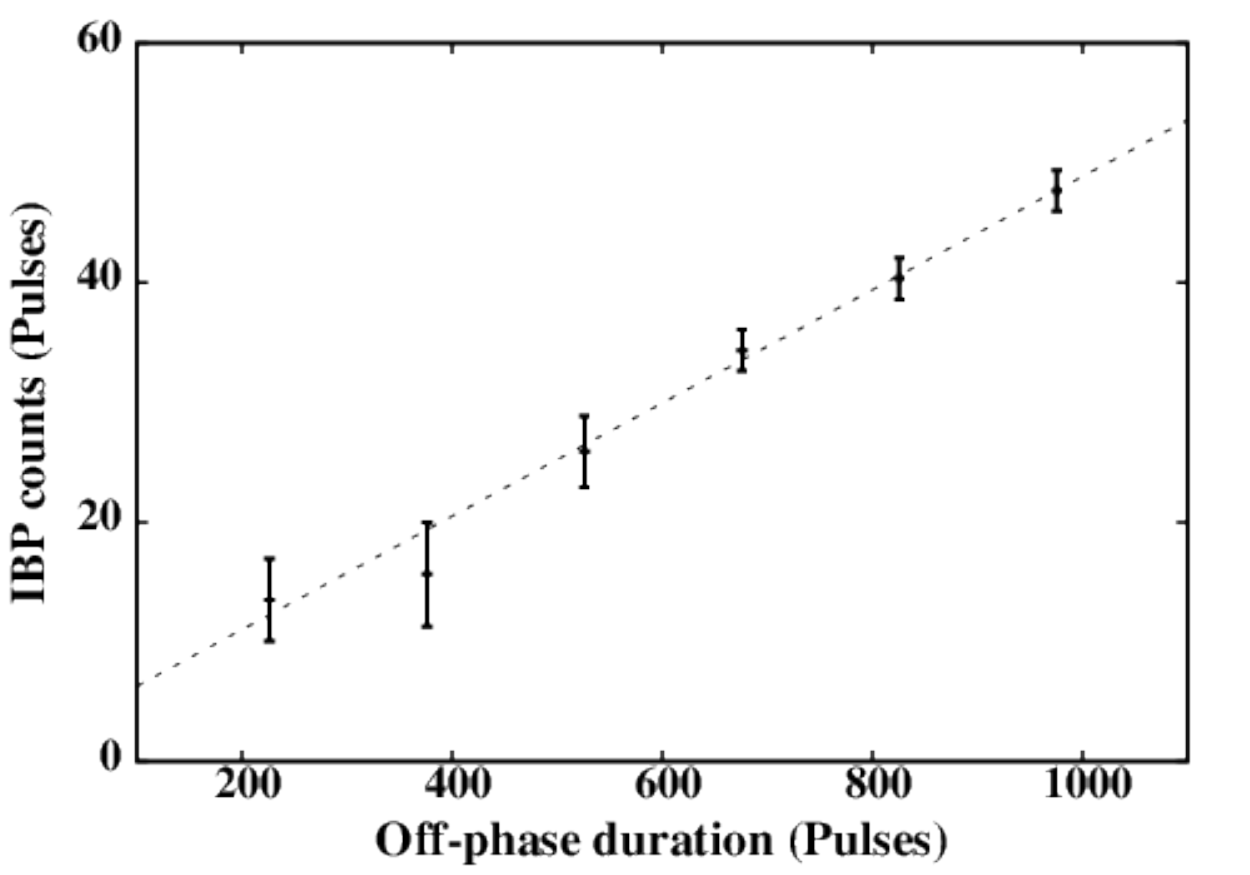}
 % burst_rate_corrected_range_wErr.eps: 0x0 pixel, 300dpi, 0.00x0.00 cm, bb=35 33 524 736
 \caption[Plot of inter-burst pulse (IBP) counts as a function of length of
 off-phase  between two consecutive bright phases as measured from the GMRT 
 observations of \pb]{Plot of inter-burst pulse (IBP) counts as a function of length of
 off-phase  between two consecutive bright phases as measured from the GMRT 
 observations of \pb. A linear relation is evident between the IBP count 
 and the corresponding off-phase length, indicating that the IBP rate is 
 independent of duration of off-phase.}
 \label{burst_rate_hist}
\end{figure}

A random occurrence of IBPs implies that the rate of IBPs 
should remain constant.  The Arecibo observation were short and hence 
statistical analysis of the IBP rate was not possible. 
Using GMRT observations we identified around 2700 such pulses, 
using the variable threshold method described above. 
As discussed in Section \ref{BBB_patten}, the length of every 
bright phase  was derived after fits to equation \ref{fx}. 
An acceptable fit could not be obtained for a few bright phases 
due to either low pulse energy or due the presence of strong RFI. 
We only considered those off-phases bounded at both ends by 
bright phases, with acceptable fits. 
A few such off-phases were also affected by RFI and hence 
not included. The number of identified IBPs were 
counted for each off-phase and the IBP rate estimated using about 53 such off
phases of various lengths. Figure \ref{burst_rate_hist} shows 
a histogram of the number of IBPs for a given off-phase 
between two consecutive bright phases. The error bars on them were obtained 
from Poisson statistics. Figure \ref{burst_rate_hist} clearly 
demonstrates that the number of IBPs 
are linearly correlated with the corresponding off-phase lengths. 
Hence, the IBP rate is independent of the length of a given null phase and could
remain fixed throughout the entire emission of the pulsar. 

\section{Comparison of \pb\ polarization profiles}
\label{Arecibo}

\begin{figure}[h!]
 \centering
 \includegraphics[width=4.5 in,height=3.5 in,angle=0,bb=0 0 350 252]{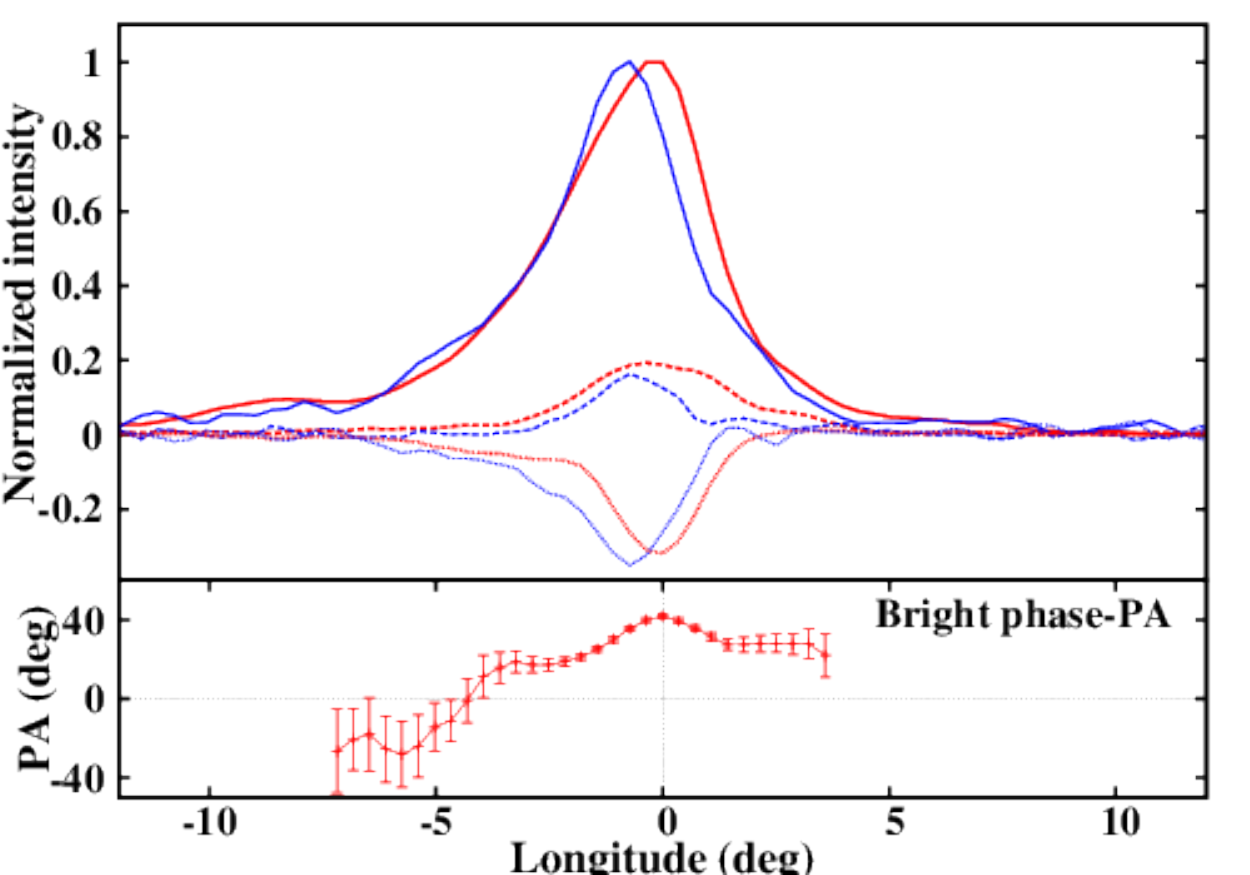}
 \caption[Plot of average polarization profiles
  from a half-hour observation of \pb\ at 327-MHz using the Arecibo
  telescope]{Plot of average polarization profiles
  from a half-hour observation of \pb\ at 327-MHz using the Arecibo
  telescope. The total intensity profile for bright phase pulses (red solid line) and IBPs
  (blue solid line) are both normalised to their peak intensities. The linear
  polarization profiles for bright phase pulses (red dashed line) and IBPs (blue dashed line) 
  are normalized by the peak intensity from their respective total intensity
  profiles. The circular polarization profiles for bright phase pulses (red dotted
  line) and IBPs (blue dotted line) are also normalized in a similar manner. 
  The bottom panel shows the derived position angle (red solid line) as a
  function of pulse longitude for bright phase pulses.  A clear offset is 
  evident in the total intensity and the circular polarization profiles between 
  the profiles for bright phase pulses and the IBPs.}
 \label{pol_comp}
\end{figure}

We analysed the full polarization data from the Arecibo telescope, observed 
with 25-MHz bandwidth at 327 MHz (Rankin, private communication). 
The observed data were calibrated to measure the 
four Strokes parameters (i.e. I,Q,U and V). Each of the Stokes parameter was 
corrected for interstellar Faraday rotation, various instrumental polarization 
effects, and dispersion. The bright phase pulses and the IBPs were separated from the observed  
single pulse data and separate polarization profiles were obtained 
for both. The linear polarization (L) for every phase bin was 
measured after obtaining U and Q parameters for the corresponding 
phase bins by calculating $\sqrt{U^2 + Q^2}$. The off-pulse mean was subtracted from it
to remove the positive bias. Figure \ref{pol_comp} shows the total intensity 
profiles along with the linear and the circular polarization profiles for 
the bright phase pulses and the IBPs. The position angle swing was also measured 
for phase bins where the observed linear polarization was more than 
3 times the off-pulse RMS in the linear polarization profile. 
The bright phase linear polarization profile showed wider range 
of such phase bins. However, the IBP linear polarization 
profile showed very limited number of such phase bins (around 3 to 4), 
hence an acceptable PA swing was not possible to obtain for the IBPs. 
Figure \ref{pol_comp} shows the PA swing for the bright phase pulses calculated by,
\begin{equation}
PA~=~\frac{1}{2}{tan^{-1}\bigg(\frac{U}{Q}\bigg)}, 
\label{PA_eq}
\end{equation}
for each phase bin by the obtained U and Q profiles. 
The error in the PA was calculated as \cite[]{mli04},   
\begin{equation}
  \bigtriangleup{PA}~=~\frac{\sqrt{(U\times\bigtriangleup{U})^2~+~(Q\times\bigtriangleup{Q})^2}}{2\times{L^2}}. 
\end{equation}
Where $\bigtriangleup{U}$ and $\bigtriangleup{Q}$ are the off$-$pulse RMS 
from Stokes U and Q average profiles respectively. 

Figure \ref{pol_comp} shows striking differences between the intensity and
polarization profiles of the bright phase pulses and the IBPs. The total intensity profile 
of the IBPs is clearly shifted to earlier phase with respect to that of
bright phase. A KS-test comparison between these two profiles rejected the null hypothesis of similar 
distributions with 99.9\% probability. The average intensity of the IBPs 
is around 5 to 7 times weaker than the average bright phase pulses. 
A Gaussian function was fitted on both the profiles to estimate the 
position of their peaks.  The offset between the peaks in the total 
intensity profiles for bright phase pulses and the IBPs was estimated  
to be around 0.6$\pm$0.1$^{\circ}$.  The linear polarization 
profile for the bright phase pulses is wider than that for IBPs and 
shows strong linear polarization of around 20\% as compared 
to about 16\% for IBPs. The shift in the position 
of peak intensity between these two profiles 
is not very significant. The circular polarization 
profile of the IBPs is offset from that of the bright phase pulses
by 0.54$\pm$0.07$^{\circ}$. However, in contrast to the overall reduction 
in the pulse energy during the IBPs, the IBP circular polarization 
fraction shows a small increase compared to that for the bright 
phase pulses (circular polarization of 35\% and 32\%, respectively) 
as is also evident from Figure \ref{pol_comp}. 

\section{No Giant pulses from PSR J1752+2359}
\label{gps}
\begin{figure}[h!]
 \centering
\includegraphics[height=2.5in,width=3.5in,bb=0 0 350 252]{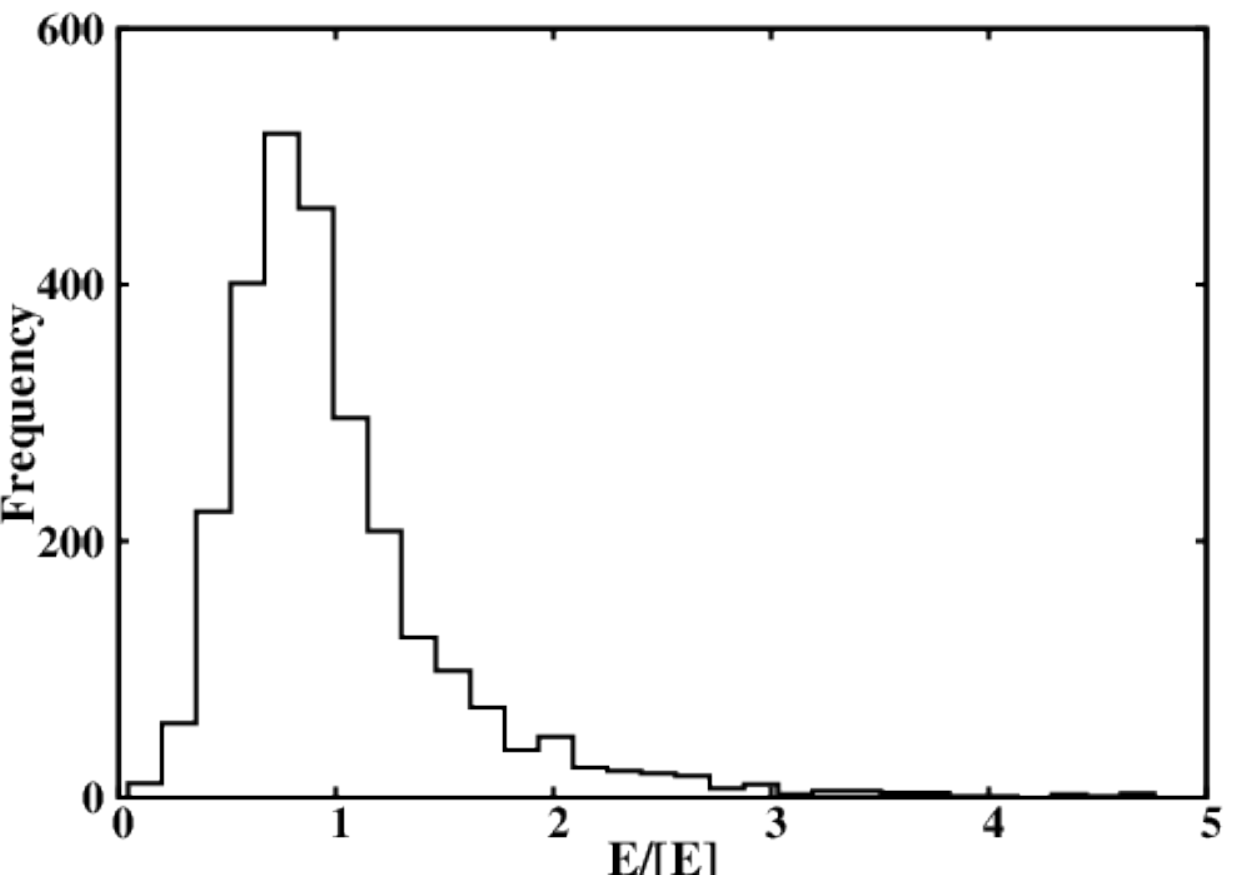}
\caption{The on-pulse energy histogram of pulses with S/N $\geq$ 5 
with the on-pulse energies scaled as explained in the text.}
\label{ESPs}
\end{figure}

PSR J1752+2359 is one of the few pulsars, which showed 
giant pulses (GPs) at 111-MHz. 
The energy of these GPs was reported to be  200 times 
the energy of the mean pulse \cite[]{EK05}.  
Our data show only a gradual distribution 
up to 20 times the mean pulse energy [Figure \ref{hist_both_fig}(b)].
We separated all single pulses with  S/N greater than 5 
and estimated the mean pulse energy from these well defined burst
pulses. The distribution of on-pulse energies of all burst pulses, 
scaled by this mean pulse energy, is shown in the 
Figure \ref{ESPs}. indicating that these are distributed  
only up to 5 times the mean pulse energy, which is significantly lower 
than that reported by \cite{EK05} at 111-MHz.  
In the earlier studies of GPs \cite[]{sr68,kt00,J04,kss11}, it was reported 
that (a) GPs have significantly smaller pulse widths compared to 
average pulses and (b) GPs tend  to occur mostly at the edge of 
the average pulse profile. We separated the strongest pulses to 
compare their widths with that of the average pulses and no 
significant difference was found. While the profile for IBPs 
is slightly narrower and shifted towards the leading edge, their 
intensities are 5 to 7 times weaker than the average bright phase pulses. 
If the IBPs have different spectral behaviour compared to bright phase pulses, 
then it is likely that they may give rise to GPs at 111-MHz. 
However, this needs to be tested with simultaneous observations 
at both frequencies. Hence, it can be concluded from our data  
that PSR J1752+2359 does not show GPs at 325 MHz. 

\section{Conclusions}
\label{conclusion}

% \begin{landscape}
\begin{table}[h!]
{\small
 \centering 
 \begin{center}
  \begin{tabular}[ht]{|m{5.5cm}|m{4.5cm}|m{4.5cm}|}
%   \begin{tabular}[ht]{|L|L|L|}
%  \begin{tabularx}[ht]{\textwidth}{X|X|X}
 \hline 
Comparison    &   \psra\       & \psrb\ 	   \\
 \hline
 \hline
     &                &              \\ 
 NF  &   85.1(2.3)\%  & $<$ 89\%     \\ 
     &                &              \\ 
 Quasi-periodicities & 170 and 270 pulses  & 540 (490,595) pulses \\
     &                &              \\
 Coherence of quasi-periodicities & $>$ 2000 pulses (11 peaks) & 1000 pulses (2 peaks) \\ 
     &                &              \\    
 Variation in on-pulse energy during bright phase  & Exponential decline with flickering nulls & Steady exponential decline \\
     &                &              \\ 
 Length of bright phase & 86 $\pm$ 4  pulses         & 88 $\pm$ 3 pulses \\
     &                &              \\ 
 Separation between bright phases & $\sim$ 170 and 500 pulses    & $\sim$ 570 pulses \\
     &                &              \\ 
 Inter-burst pulses   & No (?)    & Yes; with a fixed random rate \\
     &                &              \\ 
 First and last bright phase pulses & Similar with each other and average profile & Distinct with each other and average profile \\ 
     &                &              \\ 
 \hline
 \end{tabular}
%  \end{tabularx}
 \end{center} 
 \caption{The table of comparison between two pulsars.}
 \label{table_comp}
 \hfill{}
 }
\end{table}

As we have stressed from the start, the superficial similarity in the emission
patterns of  \pa\ and \pb\ hides many differences. \pa\ is the younger
pulsar but with a stronger magnetic field, which has enabled it to spin down to
a period five times longer than the older, but weaker, \pb\ (see Table \ref{paratable}).
Nevertheless, both have arrived at a point where their nulling fractions are
greater than 80\% and their bursts of emission are of similar length 
(measured in pulses) separated by long null sequences which give the overall appearance of
being quasi-periodic (Figure \ref{spdisplay}, Table \ref{table_comp}).

However, detailed examination of the patterns of emission bursts reveals
significant differences. Firstly, the burst separations of the younger pulsar
\pa\ turn out to be underpinned by periodic behaviour. A histogram of the
burst pulse separations (PCF in Figure \ref{pcfplots}) could be modelled by the
superposition of two near-harmonically related sine-waves and matched to the
observations for at least 2000 pulses (i.e 11 burst phases). This is strong
evidence of long-term memory in this pulsar. However, although the coherence of
the sine waves is not lost, individual burst phases often fail to materialise
except in some vestigial form. This can be seen as the effect of the stronger
sine periodicity sometimes countered by the effect of the weaker near-harmonic
(2:3) periodicity.

By contrast, in \pb\ there is no evidence of long-term periodicity. This
pulsar requires a complex wave superposition to reproduce just the first two
burst phase intervals in its pulse-separation histogram  (Figure \ref{pcfplots}b). 
Beyond about 1000 pulses, coherence is lost rapidly and there is no memory of any
periodicity. The impression of quasi-periodicity is maintained since the burst
separations usually range between 300 and 600 pulses [Figure \ref{BBB_lengh_gap_fig}(d)].

Additionally, \pb\ has a striking feature not present in the younger pulsar --
and hitherto not reported in any pulsar. In the long null intervals between the
bursts of \pb\  the nulls are interrupted at random by burst pulses which are
mostly single and relatively weak in intensity (IBPs). These pulses have a
profile significantly shifted with respect to the main profile and different
polarisation properties (Figure \ref{pol_comp}), hinting at a different physical
origin. They may well occur continuously through both burst and off-phases,
providing a diffuse background to the pulsar's more structured emission of
bright and off-phases. 

The bright phases of the pulsars, although similar in duration when measured in
pulses, have different substructures. In \pa, the bursts start suddenly and
as they progress, they are increasingly punctuated by nulls until the off-phase begins
[Figure \ref{BBB_example}(a)]. Thus the onset and the end of the bright phases are clearly
marked, and profiles obtained by integrating first and last burst pulses are
effectively identical to the pulsar's overall profile [Figure \ref{first_last}(a)].
In \pb, the bursts take longer to reach their peak emission, followed by an
exponential decline in intensity. The bright phase fades with weak emission and
it is often difficult to pinpoint its true end [Figure \ref{BBB_example}(b)]. There is
evidence that an additional small leading component appears in the emission
profile as the phase closes [Figure \ref{first_last}(b)].

In both pulsars the progression of the intensity of their burst phases can, when
averaged, be fitted by the same functional form (equation \ref{fx}, see Figure
\ref{burst_fit_avg}). In the case of \pb, we find that individual bright
phases with a higher peak intensity tend to be shorter in length with a
significant anti-correlation ($-$0.7) between these two parameters 
(Figure \ref{bbb_beta_decay}), implying that the output in energy integrated over a
bright phase is constant, at least  at the observing frequency. Furthermore, the
bright phase parameters are independent of the length of the off-phase before
and after the bright phase.

\section{Discussion}
\label{discussion}
\begin{figure}[h]
\begin{center}
 \includegraphics[width=4 in,height=3 in,angle=0,bb=0 0 350 252]{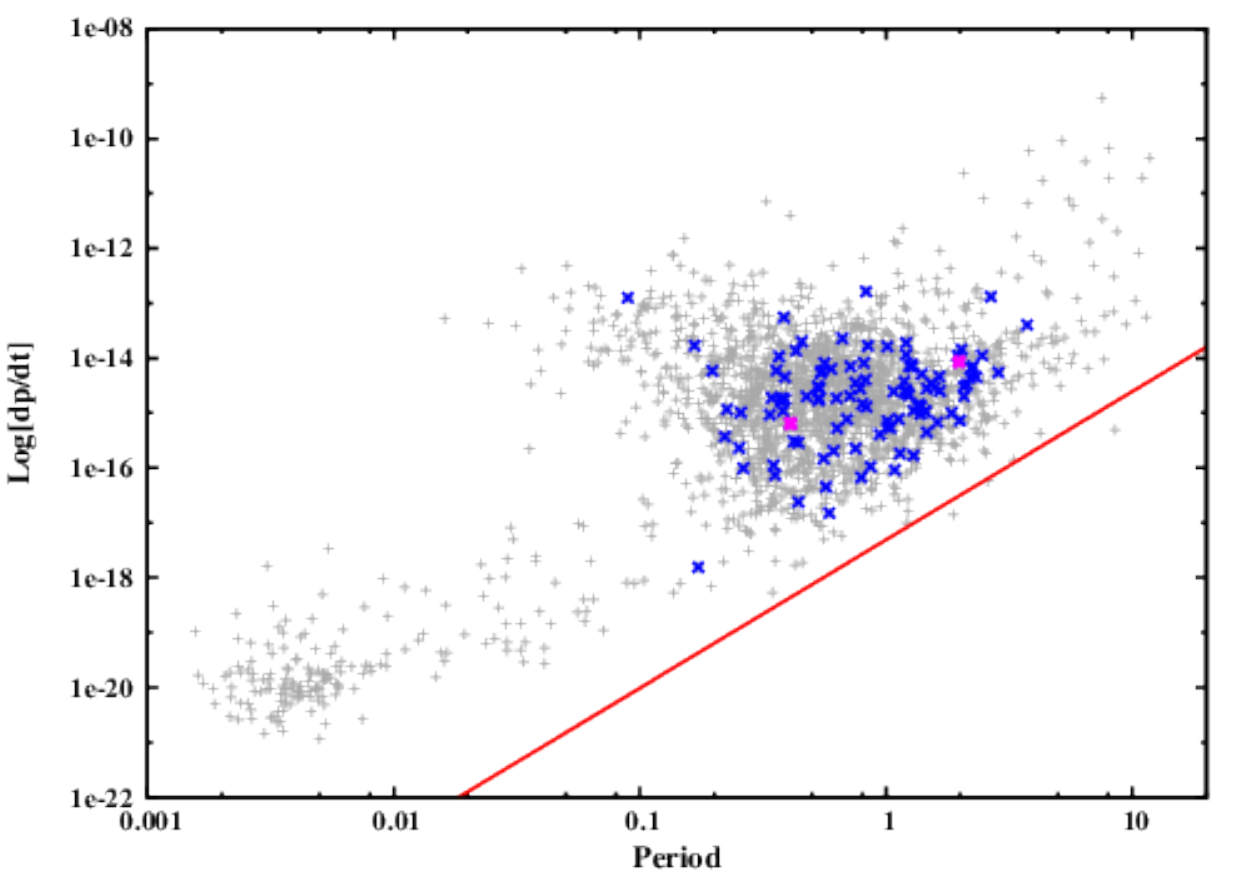}
 % Two_Nulling_PSR_p_pdot.eps: 0x0 pixel, 300dpi, 0.00x0.00 cm, bb=50 50 410 302
  \caption[PSRs \pa\ and \pb\ in the $P-\dot{P}$ diagram]{PSRs \pa\ and \pb\ in the $P-\dot{P}$ diagram. 
    All the pulsars are shown with grey points while all the known nulling pulsars 
    (listed in Table \ref{table_all_null_psr}) are shown with blue points. 
    Both the pulsars, studied in this chapter, are shown with magenta points. The death line, 
    derived using the RS model is shown as red solid line (see Section \ref{spark_sect}).}
   \label{two_null_ppdot}
\end{center}
\end{figure}

As summarized in Section \ref{chronical_obs_sect}, 
it has long been known that no strong correlations exist 
between NF and other pulsar parameters such as period or 
period derivative \cite[]{rit76,ran86,big92a,wmj07}. 
This is also true here since, despite their similar NFs, 
the positions of PSRs \pa\ and \pb\ in the $P-\dot{P}$ diagram 
could hardly be further apart. Furthermore, Figure \ref{two_null_ppdot} 
clearly shows that, both are far from the so-called 
``death line" -- contradicting a simple view that 
pulsars die through progressive increase in NF \cite[]{rit76}. 
However, what high NF pulsars do seem to have in 
common is that their individual null pulses do not appear at 
random with a single fixed probability. 
In fact, their few non-null pulses have a tendency to cluster 
in what we have called ``bright phases", even when these phases are separated by 
hours or even days \cite[]{klo+06,lem+12}.  

The simplest way of generating clusters of bright emission 
is to assume a two-state probability model, such that the 
probability of a null during a bright phase is fixed but 
lower than that during an off-phase [see Section \ref{sect_expted_time_scale} 
and \cite{cor13}]. This represents an elementary Markov (Poisson point) process and 
results in separate exponential distributions for the null and 
burst lengths, so that short nulls/bursts are very common 
and long nulls/bursts much less likely. This model has been 
plausibly applied to the null and burst statistics of several 
pulsars with low to medium NFs [see Appendix B and C and 
Figure 1 of \cite{cor13}]. In all these models burst/null clustering occurs and their 
power spectra show broad red features, but they do not generate quasi-periodic features.

However, in the case of the two pulsars described in this chapter we have become 
convinced that neither of them has null/burst distributions which are exponential 
in form. This is most evident in the burst distribution of \pb\ [Figure \ref{nbhist_both}(c)], 
which appears to be bimodal, and when the IBPs are disregarded [as in Figure \ref{BBB_lengh_gap_fig}(c)] 
a simple Gaussian-like distribution becomes clear. The same is true of the pulsar's 
apparently exponential null distribution in  Figure \ref{nbhist_both}(d), which, 
on neglect of the IBPs, is transformed into a single hump in Figure \ref{BBB_lengh_gap_fig}(d). 
In \pa, the burst/null distributions of Figures \ref{nbhist_both}(a) and \ref{nbhist_both}(b) 
are superficially exponential, but we know them to partly consist of short bursts and 
nulls which arise exclusively during the bright phases, 
again meaning that the distribution is bimodal. When these short nulls/bursts 
are disregarded, as in Figures \ref{BBB_lengh_gap_fig}(a) and \ref{BBB_lengh_gap_fig}(b), 
a single-hump distribution is evident (in fact a double 
hump in the case of Figure \ref{BBB_lengh_gap_fig}(b) for reasons given in Section 3). 
This line of reasoning is similar to that of \cite{kr10}
in their study of the high NF pulsar B1944+17. They suggest that the shortest 
nulls of that pulsar are ``pseudo-nulls", which in fact integrate to a weak profile, 
leaving a hump-like distribution for the remaining nulls.

We are therefore forced to abandon the simple (Poisson point) assumption of a 
separate but fixed null probability for each of the two phases. Instead, we 
see that the probability of the length of time the two pulsars spend in their 
bright and off-phases is dependent on its respective Gaussian-like distribution. 
It is this which gives the pulse sequences their quasi-periodic character, as 
typical bright phase and off-phase alternate with lengths scattered around 
the means of their distributions. 

Within this picture, the burst phases will not occur in precisely periodic sequences, 
but can be expected to gradually lose coherence as their separation varies randomly about a mean.
This can be seen very clearly in the PCF of \pb\ [Figure \ref{pcfplots}(b)], where coherence is 
lost after only two peaks.  In \pa, the typical burst phase separations 
are less and their distribution narrower [Figure \ref{BBB_lengh_gap_fig}(b)], so we might 
expect coherence to persist for more peaks, as is indeed the case [Figure \ref{pcfplots}(a)]. 
However, the degree of coherence in this pulsar's PCF is 
exceptionally high and we have been able to reproduce it well by combining 
just two near-harmonic underlying periodicities. This hints at the presence, 
at least in this pulsar, of forcing periodicities with suitable phase shifts \cite[]{cor13}. 
How this is achieved physically is a matter of speculation. For example, these periodicities may arise from 
an external body \cite[]{csh08} or neutron star oscillations \cite[]{cr04a} 
or near-chaotic switches in the magnetosphere's non-linear system \cite[]{tim10}.

In both pulsars the bright phases themselves show clear evolution and therefore 
have some kind of internal memory (Figures \ref{BBB_example} and \ref{burst_fit_avg}). 
Both diminish in intensity and \pa\ has an 
increasing number of null pulses as the burst proceeds. 
A decline of energy during a burst has been 
noted in several other pulsars. Recently, \cite{lem+12} 
have found that  PSR J1502$-$5653 has long nulls interspersed with weakening 
bursts (their Figure 2) whose peak drifts as the burst develops. 
A similar phenomenon appears to occur in PSR J1819+1305 \cite[]{rw08} 
and \cite{bgg10} have also reported a gradual fall in pulse intensity before 
the onset of null states for PSR B0818$-$41. In other studies, \cite{ysw+12} 
report intermittent long off-phases in PSR B0823+26 and their Figure 1 clearly 
shows a decline in intensity before null onset. In all cases it seems that the 
off-phases do not come out of the blue, so that the burst phase may represent 
a reset or relaxation of the magnetospheric conditions. It is clear that a 
change-of-state occurs when the pulsars move to the off-phase, quite possibly 
involving a global magnetospheric change \cite[]{ckf99,con05,tim10}.

The appearance of random isolated pulses (IBPs) during the off-phase of \pb, 
exhibiting a different integrated profile to the burst profile, has not been 
reported in other long-null pulsars. Its random nature is reminiscent of RRATs 
and the bright single pulses which appear in the weak mode of PSR B0826$-$34 
\cite[]{elg+05} or the RRAT phase of PSR J0941$-$39 \cite[]{bb10}. 
We cannot know if this emission represents an additional property of the stable but 
intermittent off-phase magnetospheric state or whether it is a permanent background 
phenomenon such as accretion \cite[]{w79}, which has a separate physical origin.
Recently, \cite{ysw+12} also reported similar state switching between normal 
nulling pulsar mode and intermittent pulsar mode in PSR B0823+26.

The differences in the statistics and the structure of the bright phases of 
PSRs \pa\ and \pb\ do not necessarily imply that the two pulsars produce nulls 
in a fundamentally different way. It is possible that through having a 
stronger surface magnetic field and a wider light cylinder \pa\ is 
somehow able to maintain near-periodic coherence for longer than \pb\ 
(considerably longer if measured in clock time rather than pulse numbers), 
and possible differences in the unknown inclinations of the pulsars' dipole 
axes to their rotation axes may play a role \cite[]{csh08}. 
Our results suggest that pursuing in detail the ``quasi-periodic"  behaviour of 
any pulsar may well yield valuable physical clues to the nature of its 
magnetosphere and environment.

An earlier study of pulsars with \emph{low} NFs (Section \ref{sect_null_comp}) 
found that the NF percentage was not predictive of a pulsar's subpulse behaviour. 
Our detailed study of PSRs \pa\ and \pb\ has shown that pulsars with very 
similar, but \emph{high}, NFs can also have subtle but important differences 
in emission. 
\clearpage\null\newpage
\chapter[Broadband nulling behaviour]{Simultaneous multi-frequency study of pulse nulling behaviour in two pulsars} 
\graphicspath{{Images/}{Images/}{Images/}}

\section{Introduction}
Pulsar emission at different radio frequencies, 
originates at different locations in the pulsar magnetosphere \cite[]{kom70}. 
This has been discussed with details in Section \ref{emission_hgt_sect}.  
All of the nulling studies, reported in Chapter 4 and 5, 
were carried out at a single observing frequency.  
Moreover, most of them were conducted only for 
a typical observations lasting about an hour. 
There are very a few long simultaneous 
observations of nulling pulsars reported so far 
in the literature. Section \ref{broadband_intro_sect} 
highlights some of the studies that have been 
conducted to scrutinize the broadband behaviour of the nulling phenomena. 
% In a simultaneous single pulse study 
% of two pulsars, PSRs B0329+54 and B1133+16 at 327 and 2695 MHz, 
% \cite{bs78} showed highly correlated pulse energy 
% fluctuations. Simultaneous observations of 
% PSR B0809+74 for about 350 pulses 
% indicated that only 6 out of 9 nulls were simultaneous 
% at 102 and 408 MHz \citep{dls+84}. Contrary to that, 
% \cite{big92a} reported broadband nulling 
% from observations at two frequencies \emph{viz.} 645 and 843 MHz. 
% About half of nulls were reported to occur simultaneously at 325, 610, 1400 
% and 4850 MHz for PSR B1133+16 \citep{bgk+07}. In contrast, 
% simultaneous nulls were reported at 303 and 610 MHz for 
% PSR B0826$-$34 \citep{bgg08}. While the former 
Among the investigated pulsars, PSRs B0809+74 and B1133+16, 
represent a conal cut of pulsar beam \cite[]{rr03,ran93}, 
while PSR B0826$-$34 has an almost aligned pulsar beam \cite[]{bmh+85}. 
Thus, it is not clear if nulling represents a global 
failure of pulse radiation or is due to a shift in pulsar 
beam manifesting as lack of emission at the 
given observation frequency due to the geometry 
of pulse emission. Thus, as suggested in Section \ref{motivation_sect}, 
long, sensitive, and preferably simultaneous observations at multiple frequencies 
of a carefully selected sample of pulsars are motivated by these previous investigations.

In this chapter, we report on long simultaneous 
multi-frequency observations of two pulsars, 
PSRs B0809+74 and B2319+60 to 
investigate the broadband nature of pulse 
nulling. These pulsars were chosen as 
(a) they are strong pulsars allowing an 
easy determination of nulls, 
(b) PSR B0809+74, show prominent drifting and single 
component profile [see Section \ref{null_behavior_sect} and \cite{la83}]
indicating a tangential and peripheral 
line of sight traverse of their emission beam, 
and (c) PSR B2319+60, show long prominent nulls 
and have high NF [about 30\%, see Section \ref{null_behavior_sect} and Figure \ref{sp_2037_2111_2319}(c)]. 
It also shows a multiple component 
profile with drifting in outer components indicative  
a more central line of sight traverse of the emission 
beam \cite[]{ran86,gl98}. Thus, this sample allows us to test the effect 
of pulse nulling as a function of observational 
frequency for different parts of pulsar beam 
and discriminate between a geometric or 
an intrinsic origin for pulse nulling. 
In Section \ref{obsanal}, observations and essential 
time-series alignments between various observed frequencies 
are described. The results on PSR B0809+74 are 
presented in Section \ref{b0809_sect}, while the results 
on PSR B2319+60 are presented in Section \ref{b2319_sect}. 
The conclusions of the study are presented 
in Section \ref{conc_sect} along with a discussion on 
the implications of the results in Section \ref{discussion_sect}.
   
\section{Observations and time-series alignments} 
\label{obsanal}
\begin{table*}[!h]
\footnotesize
\begin{center}
\begin{tabular}{l|c|c|l|c|c}
\tableline\tableline
Pulsar & Period & Dispersion     & Date of       & Number of pulses & Frequencies\\
       &        & Measure        & observations  &              & of observations    \\
\tableline
       &  (s)   &(pc\,cm$^{-3}$) &                &            & (MHz)      \\
\tableline
% PSR B0031$-$07 & 0.942951 & 11.38  & 2011 February 5  &  6776 & 313, 607, 1380\\
PSR B0809+74   & 1.292241 &  6.12  & 2011 February 17 &  10003 & 313, 607, 1380, 4850\\
PSR B2319+60   & 2.256488 & 94.59  & 2011 February 6  &  5126 & 313, 607, 1380, 4850 \\
\tableline
\end{tabular}
\end{center}
\caption{Parameters of the observed pulsars and details of Observations}
\label{tabobs}
\end{table*}
\normalsize

PSRs B0809+74 and B2319+60 were observed simultaneously 
with the Giant Meterwave Radio Telescope (GMRT), 
the Westerbok Synthesis Radio telescope (WSRT) and 
the Effelsberg Radio Telescope. The details of observations 
are discussed in details in Chapter 3 and also 
listed in Table \ref{tabobs}, with exact observing frequency 
for a comparison. 

Given the difference in the longitude of the 
observatories, the total overlap at all frequency was 
smaller than the total duration of observations at 
each telescope. Part of the data during the overlap 
was affected by radio frequency interference (RFI)
at one or the other telescope and was not considered 
for the analysis described below. The remaining number 
of pulses observed simultaneously are indicated 
in Table \ref{tabobs}.

The data from all observatories, as mentioned 
in Section \ref{single_pulse_technique}, were converted into a standard format required for 
SIGPROC\footnote{\url{http://sigproc.sourceforge.net/}} 
analysis package and dedisperesed using the programs 
provided in the package. These were then folded to 
1000 bins across the period using \emph{polycos} obtained 
from the ephemeris of these pulsars using 
TEMPO\footnote{\url{http://tempo.sourceforge.net/}} 
package to obtain a single pulse sequence. 

First, the pulse sequence for the longest data file, 
typically consisting of 6000 pulses, were averaged  
for each frequency to obtain an integrated profile, 
which was used to form a noise-free template, 
after centering the pulse, for the pulsar 
at that frequency. The template at each frequency was 
used to estimate the number of samples to be removed 
from the beginning of each file for these two frequencies 
so that the pulse is centred in a single period and 
time stamps for single pulses were corrected 
by these offsets. The single pulse sequences were then 
aligned by converting these time stamps to solar system 
barycentre (SSB) using TEMPO$^2$. This conversion also takes into 
account the delay at lower frequencies due to dispersion 
in the inter-stellar medium (ISM). Then, the pulses 
corresponding to identical time stamps at SSB across 
all frequencies were extracted from the data. 
As the observations were typically recorded 
in 2$-$3 data files at the GMRT (325 and 610 MHz),
the data recorded at the WSRT and the Effelsberg were split 
into similar number of files with observations 
duration equal to that at the GMRT. 

The single pulse sequences were then visually examined to 
remove any single pulses with excessive Radio Frequency 
Interference (RFI). The number of pulses available 
at all frequencies simultaneously 
after eliminating pulses affected by RFI are indicated 
in Table \ref{tabobs} for each pulsar. Broadband nulling 
behaviour for each pulsar, using these single pulse 
sequences, are discussed in the following sections. 

\section{PSR B0809+74}
\label{b0809_sect}
\begin{figure}[h!]
 \centering
 \includegraphics[width=5.5 in,height=5 in,angle=0,bb=17 -7 803 561]{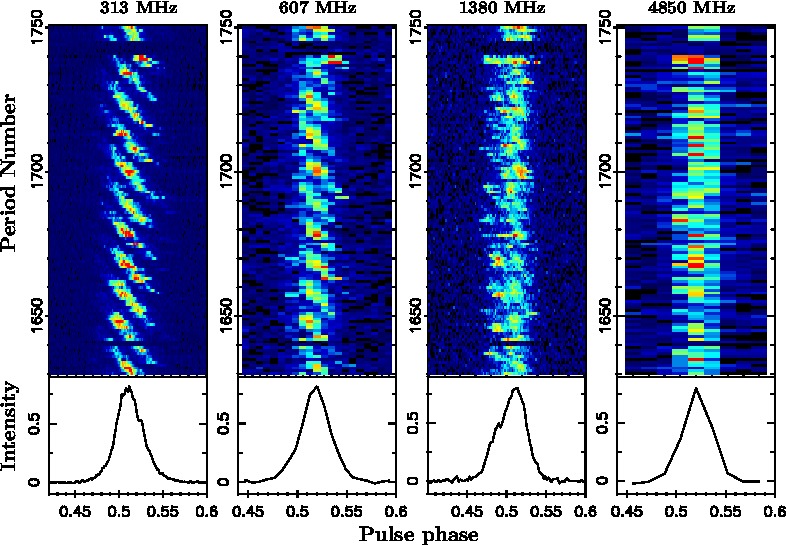}
  % spcomb_label_color.eps: 0x0 pixel, 300dpi, 0.00x0.00 cm, bb=17 -7 803 561
  \caption[Single pulse sequences as a function of pulse number and pulse phase for a subset 
  of data for all frequencies for PSR B0809+74]{Single pulse sequences as a function of 
   pulse number and pulse phase for a subset of data for all frequencies for PSR B0809+74. 
   The sequences were observed simultaneously at all frequencies. The pulse intensities 
   are shown in a color ramp from blue to red. The bottom panel in each plot shows the 
   respective integrated profile.}
  \label{b0809_sp_all_freq}
\end{figure}

A subset of simultaneous pulse sequences, obtained 
at all four frequencies, are shown in Figure \ref{b0809_sp_all_freq}. 
At 1380 MHz, single pulses were detected with highest S/N, 
while they showed relatively low S/N at 4850 MHz. 
The drifting of subpulses is clearly evident 
for single pulses at 313, 607 and 1380 MHz. 
To improve the single pulse S/N at 4850 MHz, 
we added 10 consecutive phase-bins. 
Thus, such drifting behaviour is not possible to 
identify at 4850 MHz in Figure \ref{b0809_sp_all_freq}.
The simultaneity in the absence of emission is also evident 
across all four frequencies in Figure \ref{b0809_sp_all_freq}.
Presented section of single pulses clearly shows two distinct 
null regions near period numbers 1640 and 1740 at each frequency.
To verify and quantify this behaviour further, the NFs, 
the correlation between the nulling pattern 
represented by the one-bit sequence and the distributions 
of null and burst lengths were compared across four frequencies. 

\begin{figure}[h!]
 \begin{center}
 \includegraphics[width=5 in,height=3.5 in,angle=0,bb=0 0 360 252]{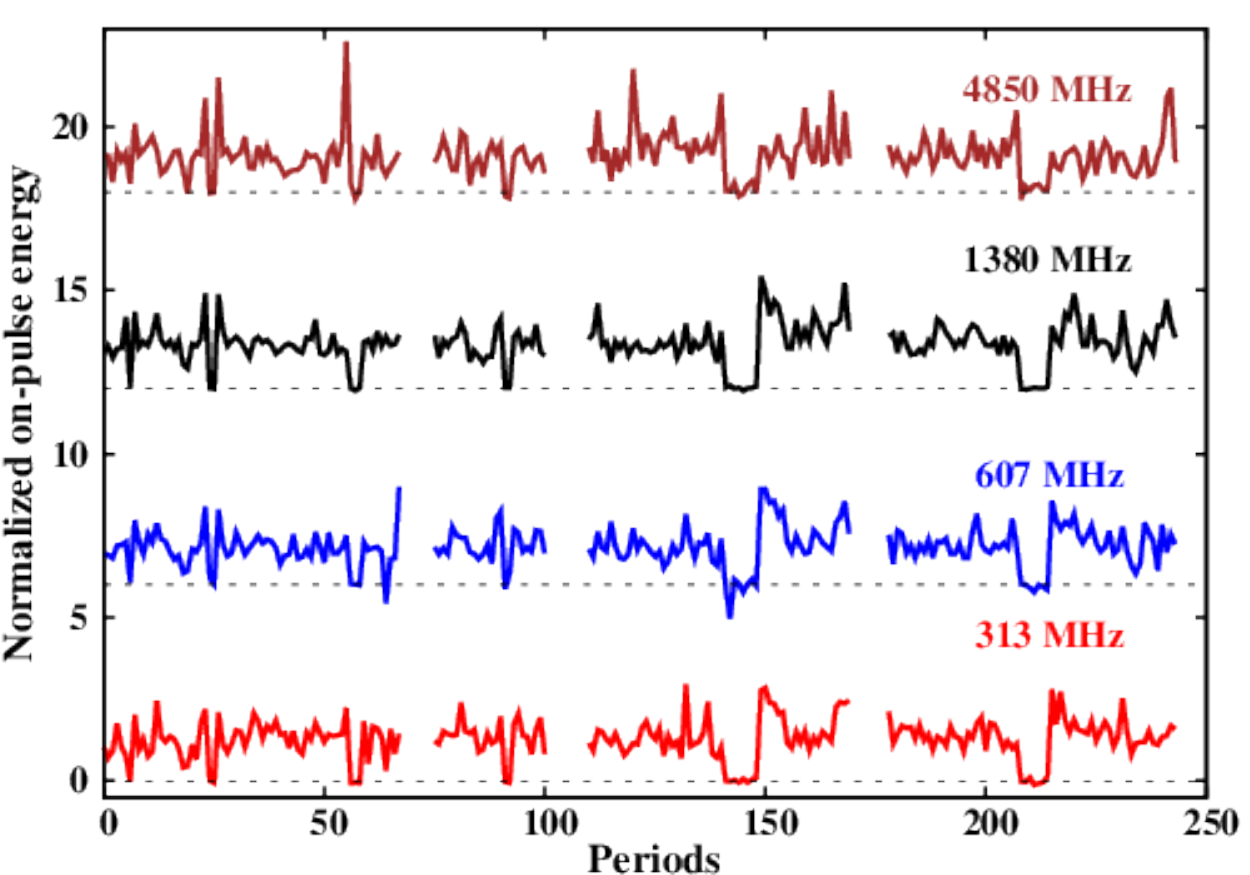}
 % Energy_aligned_all_freq.eps: 0x0 pixel, 300dpi, 0.00x0.00 cm, bb=50 50 410 302
 \caption[On-pulse energy sequences as a function of pulse number for 
  a subset of data from all four frequencies for PSR B0809+74]
  {On-pulse energy sequences as a function of pulse number for 
  a subset of data from all four frequencies for PSR B0809+74. The ordinate 
  present pulse energy in arbitrary units, which is presented with an offset in the 
  vertical direction for each frequency for clarity. The dotted horizontal lines 
  present respective zero pulse energy at a given observing frequency. 
  The period numbers in the abscissa are superficial as four separate null regions, 
  extracted at each frequency, are shown here to display correlated 
  pulse energy fluctuations.}
 \label{b0809_ope_all_freq}
 \end{center}
\end{figure}

First, we obtained the on-pulse energy sequence by selecting 
an appropriate on-pulse window at each frequency from their 
respective integrated profile. Due to the interstellar scintillation, 
the on-pulse energy shows large fluctuations at all frequencies. 
As discussed in Section \ref{scintillation_sect}, this can cause 
the on-pulse energy to plunge around zero pulse energy, 
which makes identification of the null pulses a daunting task. 
Thus, for all those sections, which showed low S/N single pulses at 
any particular frequency (around 2\% of the observed pulses), 
were excluded at all four frequencies. For a few 
different sections, pulse energy was marginally reduced which was  
normalized by subtracting a moving average, or a box average (see Section \ref{scintillation_sect}). 
Figure \ref{b0809_ope_all_freq} shows sections 
of on-pulse energy sequences at all frequencies, after subtracting a box average. 
As PSR B0809+74 exhibit small fraction of null pulses separated by 
large burst phases, to display more simultaneous 
null regions, four different sections of pulse sequences 
around these null states were extracted for display at each frequency. 
The null pulses at each frequency clearly shows 
simultaneous occurrence across all observed frequencies. 
A visual inspection of the entire data broadly confirms this behaviour.
\subsection{NFs comparison}
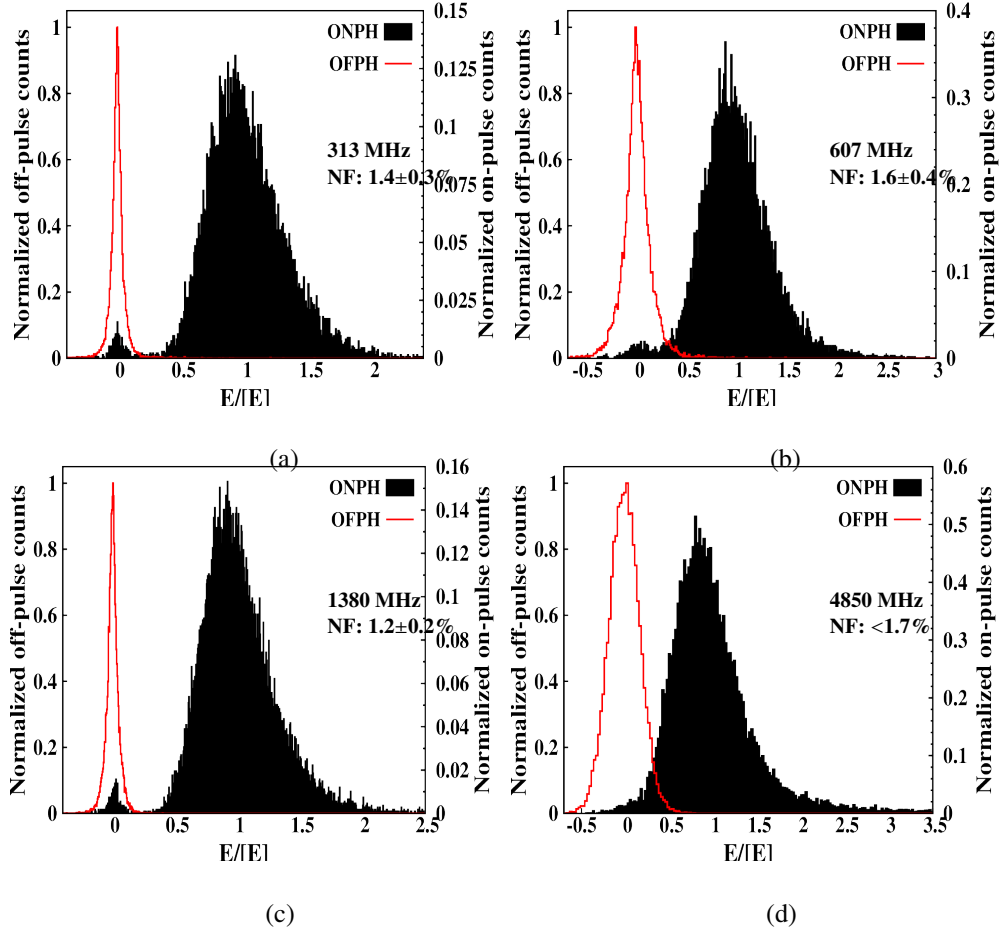
\begin{figure}[h!]
 \centering
 \subfigure[]{
 \includegraphics[width=2.1 in,height=2.5 in,angle=-90,bb=59 53 555 750]{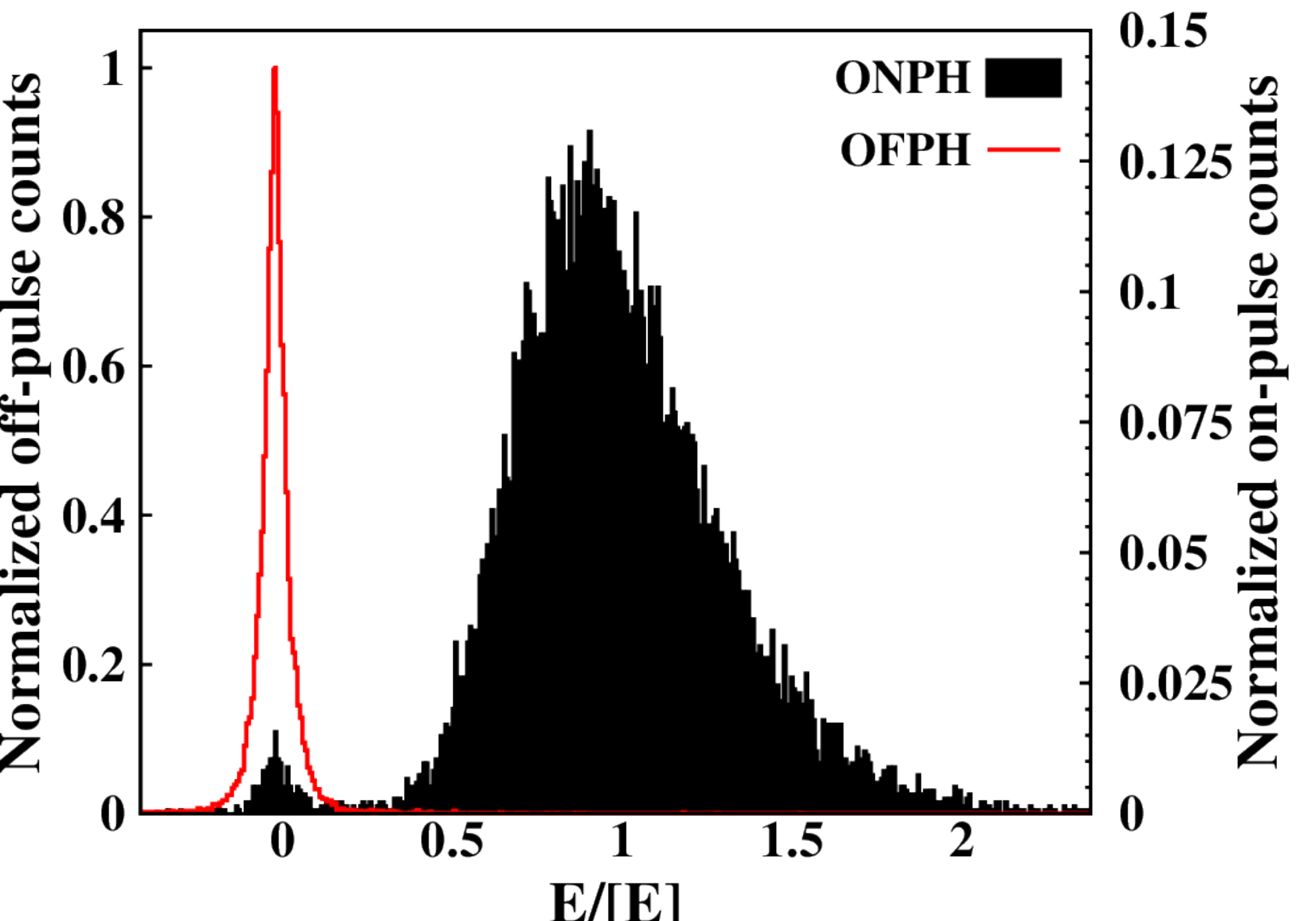}
 % b0809.histogram.325.eps.pdf: 504x720 pixel, 72dpi, 17.78x25.40 cm, bb=0 0 504 720
 }
 \subfigure[]{
 \includegraphics[width=2.1 in,height=2.5 in,angle=-90,bb=59 53 555 750]{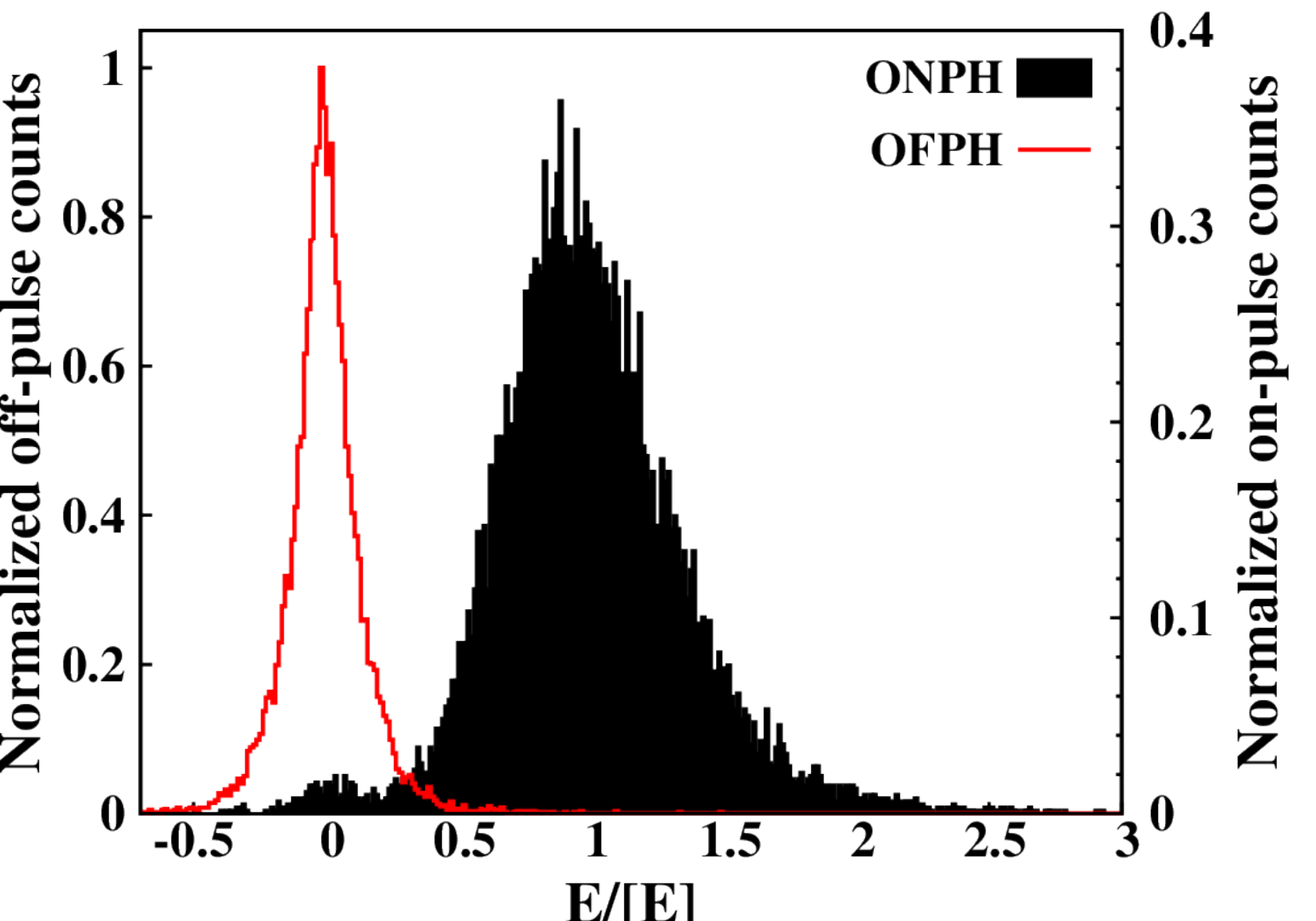}
 }
 \subfigure[]{
 \includegraphics[width=2.1 in,height=2.5 in,angle=-90,bb=59 53 555 750]{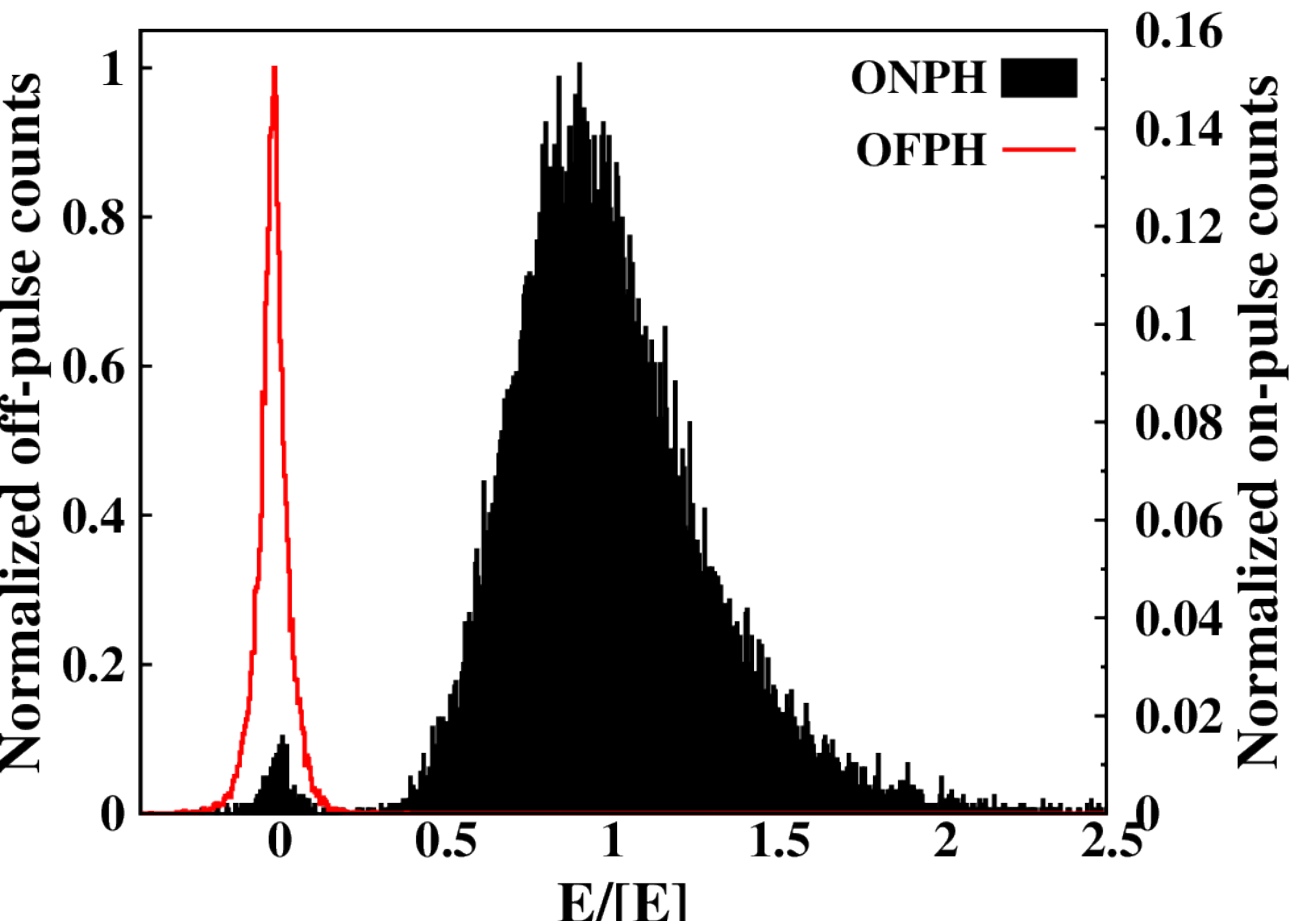}
 }
 \subfigure[]{
 \includegraphics[width=2.1 in,height=2.5 in,angle=-90,bb=59 53 555 750]{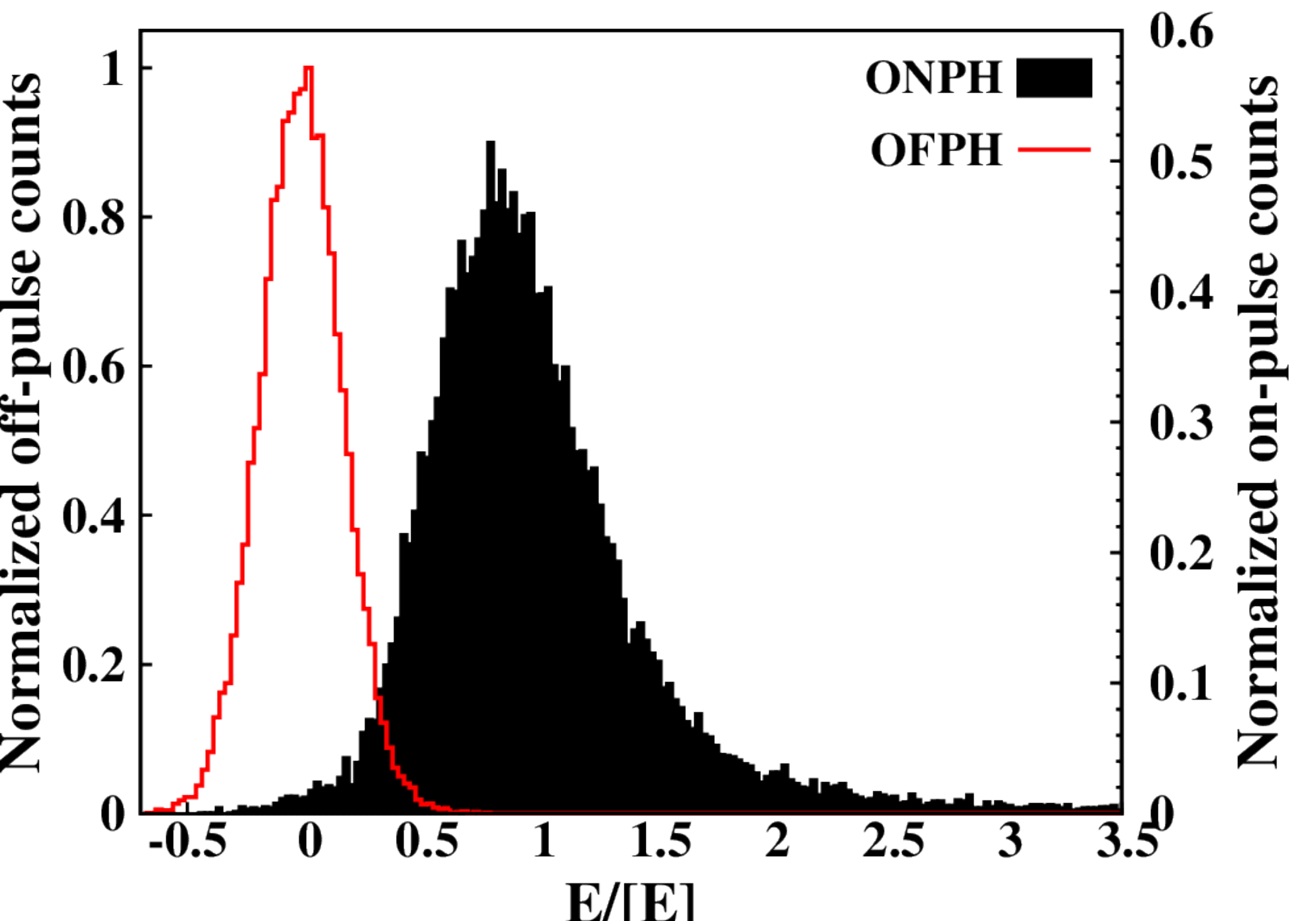}
 }
  \begin{picture}(0,0)
   \put(-270,130){\scriptsize \bf 313 MHz}
   \put(-270,120){\scriptsize \bf NF: 1.4$\pm$0.3\%}
   \put(-80,130){\scriptsize \bf 607 MHz}
   \put(-80,120){\scriptsize \bf NF: 1.6$\pm$0.4\%}
   \put(-270,-40){\scriptsize \bf 1380 MHz}
   \put(-270,-50){\scriptsize \bf NF: 1.2$\pm$0.2\%}
   \put(-80,-40){\scriptsize \bf 4850 MHz}
   \put(-80,-50){\scriptsize \bf NF: $<$1.7\%}
  \end{picture}
 \caption[The on-pulse and the off-pulse energy histograms at all four frequencies for PSR B0809+74]
 {The on-pulse and the off-pulse energy histograms at all four frequencies for PSR B0809+74, 
 i.e. (a) 313 MHz, (b) 607 MHz, (c) 1380 MHz and (d) 4850 MHz. 
 The abscissa presents normalized energy obtained using the block average (see Chapter 3). 
 The ordinate presents normalized counts of occurrence for each energy bin. 
 The OFPHs are shown with the red solid lines while the ONPHs are shown with black filled curve. 
 The counts in both the histograms were normalized by the peak from the corresponding OFPH histogram 
 at each frequency. The obtained NF along with the observed frequency are displayed in the inset texts.
 Total 10003 pulses were used to obtain these histograms at each frequency.}
 \label{b0809_hist_all_freq}
\end{figure}
\begin{figure}[h!]
 \centering
 \includegraphics[width=4 in,height=3 in,angle=0,bb=0 0 360 252]{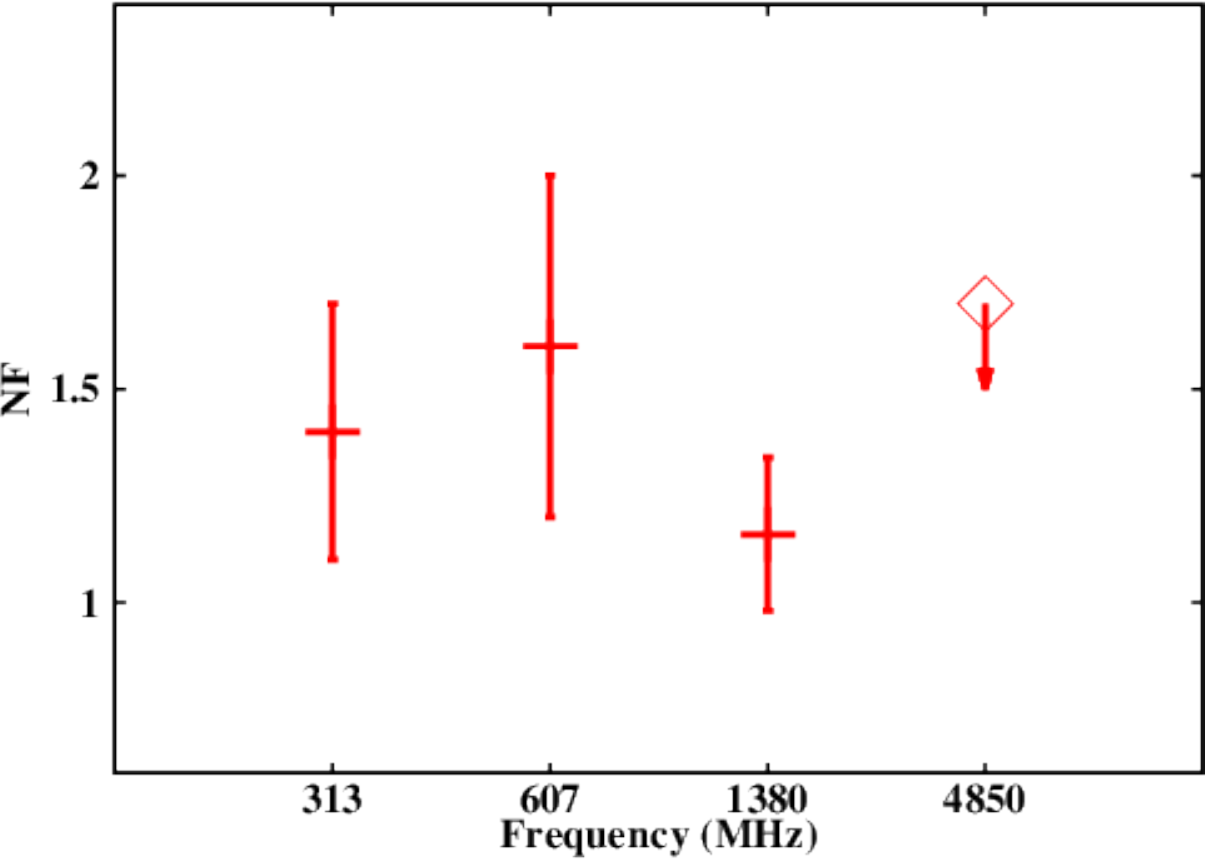}
 % NF_freq.b0809.forthesis.eps: 347x249 pixel, 72dpi, 12.24x8.78 cm, bb=0 0 347 249
 \caption{The obtained NFs at all four observed frequencies for PSR B0809+74.}
 \label{b0809_NF_all_freq}
\end{figure}

As a first test to quantify the similarity of nulling behaviour, obtained NFs at 
all four frequencies were compared. The on-pulse energy sequence and the off-pulse 
energy sequence at each frequency were binned. The obtained 
ONPHs and OFPHs are shown in Figure \ref{b0809_hist_all_freq}. All four ONPHs 
show very small fraction of null pulses. The method to obtain the NF is discussed 
with details in Section \ref{NF_tech_sect}. The ONPHs at 313, 607 and 1380 MHz 
[i.e. Figure \ref{b0809_hist_all_freq}(a),(b) and (c)]
show clear bi-modal distributions of the pulse energy.  Thus, we were able to obtain accurate estimate of the 
NFs at these lower frequencies. However, at 4850 MHz the 
single pulse S/N was not sufficient to produce such clear bi-modal distribution 
in the ONPH. For such cases, consecutive pulses can be integrated to improve 
the S/N and to get better separation between the null and the burst pulse energy 
distributions in the ONPH. This technique has been used for many weak pulsars, 
as discussed in Chapter 4. However, PSR B0809+74 exhibit 
null states with significantly smaller lengths (1 or 2 periods). 
Integrating consecutive pulses will lead to large fraction 
of null pulses merging with the neighbouring burst pulses.
Thus, such sub-integration was not possible to deploy 
for NF estimation. By using the standard Gaussian fit technique 
on the ONPH and OFPH (see Section \ref{NF_tech_sect}), 
only an upper limit on the NF was possible to derive 
due to the mixture of weak burst pulses with the null pulses near the zero 
pulse energy. The obtained NFs are also listed in the inset text of 
Figure \ref{b0809_hist_all_freq} for each frequency. 
Figure \ref{b0809_NF_all_freq} shows these NFs for all frequencies, 
to demonstrate a good match between the obtained 
quantities within the error bars. Although the NF does show a good match 
across a decade of frequencies, as suggested in Section \ref{sect_null_comp}, 
it does not quantify nulling behaviour in full details. Hence, it is 
essential to confirm pulse-to-pulse matching 
of the nulling behaviour in order to check the
true broadband behaviour of this phenomenon. 
% the null length and the burst length distributions for all pairs 
% of observed frequencies. 
\subsection{One-bit sequence comparison}
\label{b0809_onebit_sect}
The null pulses and the burst pulses were identified and separated at all four frequencies. 
For the lower three frequencies, i.e. 313, 607 and 1380 MHz, a threshold was first 
set on the respective ONPH, where the null pulse distribution and the burst pulse 
distribution overlap with each other. Pulses below the threshold were 
tagged as null pulses while pulses above the threshold were tagged 
as burst pulses. A visual inspection was carried out on the separated null 
and burst pulses to check for any misidentified pulses. At the 
highest observing frequency, 4850 MHz, single pulses were weak and hence 
such threshold was not possible to set to separate null and burst pulses. 
Hence, we initially arrange all pulses in the ascending order of their on-pulse energy. 
A threshold was moved from lower energy end towards the higher energy end till 
pulses below the threshold did not form significant profile component (with S/N$\geq$3). 
All the pulses below the threshold were tagged as null pulses 
while pulses above the threshold were tagged as burst pulses. 
This method, also discussed in Section \ref{separation_of_null_burst_sect}, 
allows separation of weak burst pulses from true null pulses. All the separated pulses were carefully 
examined visually to eliminate any possibility of misidentification. 
Using these separated null and burst pulses, one-bit sequences 
were formed for all observed frequencies (see Section \ref{nlh_blh_intro_sect} and Figure \ref{onezero_example}). 
\begin{figure}[h!]
 \centering
 \includegraphics[width=5 in,height=3 in,angle=0,bb=0 0 360 252]{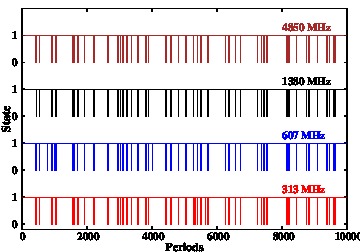}
 % b0809_onezero_all_freq.eps.pdf: 0x0 pixel, 300dpi, 0.00x0.00 cm, bb=50 50 410 302
 \caption[One-zero sequences at four frequencies for PSR B0809+74]
 {The identified one-zero state of the pulsar emission at all four 
 frequencies for PSR B0809+74. These pulses are not contiguous from observations. 
 As explained in the text, different RFI affected regions as well as 
 low S/N pulses were removed identically across all frequencies.} 
 \label{b0809_onezero_all_freq}
\end{figure}

As a second test, one-bit sequences for 
a pair of observing frequencies were compared using a contingency table analysis 
\cite{pftv86}. Figure \ref{b0809_onezero_all_freq} shows these 
sequences for the entire range of simultaneously observed pulses. 
To obtain these sequences, different observed sections 
were patched together at all frequencies to compare 
the synchronicity of null states. A visual comparison 
between various frequency pairs suggests that pulsar 
shows simultaneous switching between the null and the burst 
states at all frequencies, for large fraction of periods. 
To quantify this simultaneity, a correlation between these states for a pair of frequencies can be obtain 
as a 2 $\times$ 2 contingency table. In statistics, contingency table 
presents measure of association between two variables. 
In this chapter, the contingency table was used to measure 
the number of occurrences when the pulsar was in a 
similar emission state at both frequencies under consideration. 
Thus, such contingency table was obtained for each pair 
of frequencies. Table \ref{b0809_cont_table_all_freq} 
shows the different contingency tables for all frequency pairs. 
It can be seen from these tables that around 9 to 11 
pulses show a mismatch for different pairs of frequencies. 
For example, a comparison between one-bit sequences 
obtained at 313 and 1380 MHz indicates around 5 pulses for which 
the pulsar showed absence of emission at 313 MHz but it showed burst emission at 
1380 MHz (see Table \ref{b0809_cont_table_all_freq}). 
Similarly, for 4 pulses, the pulsar showed absence of 
emission at 1380 MHz while it was in a burst state at 313 MHz. 
These are significantly small numbers ($<$0.1\%), 
hence it can be concluded that PSR B0809+74 exhibit 
similar emission states at all pairs of 
frequencies for the duration of our observations. 
\begin{table}[h!]
\centering
\begin{subtable}{}
\centering
\begin{tabular}{cc|c|c|}
\cline{3-4}
 &         & \multicolumn{2}{c|}{607} \\
 \cline{3-4} 
 &         & Null & Burst \\
\cline{1-4}
\multicolumn{1}{|c|}{\multirow{2}{*}[-0.1cm]{\rotatebox{90}{313}}} & Null & 141  & 4 \\
\cline{2-4}     
\multicolumn{1}{|c|}{} & Burst& 7  & 9851 \\
\tableline
\end{tabular}  
\end{subtable}
\vspace{0.1cm}
\begin{subtable}{}
\centering
\begin{tabular}{cc|c|c|}
\cline{3-4}
 &         & \multicolumn{2}{c|}{1380} \\
 \cline{3-4} 
 &         & Null & Burst \\
\cline{1-4}
\multicolumn{1}{|c|}{\multirow{2}{*}[-0.1cm]{\rotatebox{90}{313}}} & Null & 140  & 5 \\
\cline{2-4}     
\multicolumn{1}{|c|}{} & Burst& 4  & 9854 \\
\tableline
\end{tabular}  
\end{subtable}
\vspace{1 cm}
\begin{subtable}{}
\centering
\begin{tabular}{cc|c|c|}
\cline{3-4}
 &         & \multicolumn{2}{c|}{4850} \\
 \cline{3-4} 
 &         & Null & Burst \\
\cline{1-4}
\multicolumn{1}{|c|}{\multirow{2}{*}[-0.1cm]{\rotatebox{90}{313}}} & Null & 140  & 5 \\
\cline{2-4}     
\multicolumn{1}{|c|}{} & Burst& 5  & 9853 \\
\tableline
\end{tabular}  
\end{subtable}
\vspace{0.1cm}
\begin{subtable}{}
\centering
\begin{tabular}{cc|c|c|}
\cline{3-4}
 &         & \multicolumn{2}{c|}{1380} \\
 \cline{3-4} 
 &         & Null & Burst \\
\cline{1-4}
\multicolumn{1}{|c|}{\multirow{2}{*}[-0.1cm]{\rotatebox{90}{607}}} & Null & 141  & 7 \\
\cline{2-4}     
\multicolumn{1}{|c|}{} & Burst& 3  & 9852 \\
\tableline
\end{tabular}  
\end{subtable}
\vspace{1 cm}
\begin{subtable}{}
\centering
\begin{tabular}{cc|c|c|}
\cline{3-4}
 &         & \multicolumn{2}{c|}{4850} \\
 \cline{3-4} 
 &         & Null & Burst \\
\cline{1-4}
\multicolumn{1}{|c|}{\multirow{2}{*}[-0.1cm]{\rotatebox{90}{607}}} & Null & 141  & 7 \\
\cline{2-4}     
\multicolumn{1}{|c|}{} & Burst& 4  & 9851 \\
\tableline
\end{tabular}  
\end{subtable}
\vspace{0.1cm}
\begin{subtable}{}
\centering
\begin{tabular}{cc|c|c|}
\cline{3-4}
 &         & \multicolumn{2}{c|}{4850} \\
 \cline{3-4} 
 &         & Null & Burst \\
\cline{1-4}
\multicolumn{1}{|c|}{\multirow{2}{*}[-0.1cm]{\rotatebox{90}{1380}}} & Null & 140  & 4 \\
\cline{2-4}     
\multicolumn{1}{|c|}{} & Burst& 5  & 9854 \\
\tableline
\end{tabular}  
\end{subtable}
\caption[Contingency tables for different pairs of frequencies for PSR B0809+74]
{Contingency tables for different pairs of frequencies for PSR B0809+74. 
Each frequency is shown with two corresponding emission states (i.e. the null state 
and the burst state). A 2 $\times$ 2 matrix for a given pair of 
frequencies displays number of pulses in four possible conditions.}
\label{b0809_cont_table_all_freq}
\end{table}

A $\phi$ test (Cramer-V) and uncertainty test based on entropy calculations can then be 
used to assess the strength and significance of these correlations. 
Both these tests result in a value between 0 and 1. A value 
close to 1 indicates a strong correlation. Cramer-V for a 
2 $\times$ 2 contingency table, as in our case, is just a measure 
of reduced chi-square. Hence, a value close to 1 indicates a 
very high significance of correlation. Likewise, a value 
close to unity for the uncertainty coefficient calculated 
from entropy arguments indicates a very high 
probability of observing null (burst) pulses at both 
frequencies. See \cite{pftv86} for details about these tests. 
The results of these tests are presented in Table \ref{b0809_cramv_all_freq}. 
Both Cramer-V and the uncertainty coefficients have values very close to 1 for 
PSR B0809+74 indicating a significantly high 
association across all pairs of frequencies. 

\begin{table}[h!]
\begin{center}
\begin{tabular}{|l||c|c|c|}
\tableline
Frequenc & 607  & 1380 & 4850  \\
(MHz)    &      &      &       \\
\tableline
\tableline
313  & 0.96 & 0.97 & 0.96  \\
     & 0.92 & 0.92 & 0.91  \\
607  & $-$  & 0.96 & 0.96  \\
     & $-$  & 0.91 & 0.90  \\ 
1380 &      & $-$  & 0.97  \\
     &      & $-$  & 0.92  \\
\tableline
\end{tabular}
\end{center}
\caption[Estimate of correlation strength and significance for 
one-bit sequences between a pair of frequencies for PSR B0809+74]
{Estimate of correlation strength and significance for 
one-bit sequences between a pair of frequencies for PSR B0809+74. 
The first row in each column for a given 
frequency gives the Cramer$-$V indicating the strength of correlation 
of one-bit sequence associated with the frequency in the column. 
Similarly, the second row in each column for a given frequency gives the 
corresponding uncertainty coefficient derived from entropy 
arguments [See \cite{pftv86}]}
\label{tabcont}
\label{b0809_cramv_all_freq}
\end{table}

\subsection{Nonconcurrent pulses}
\label{b0809_mismatch_sect}
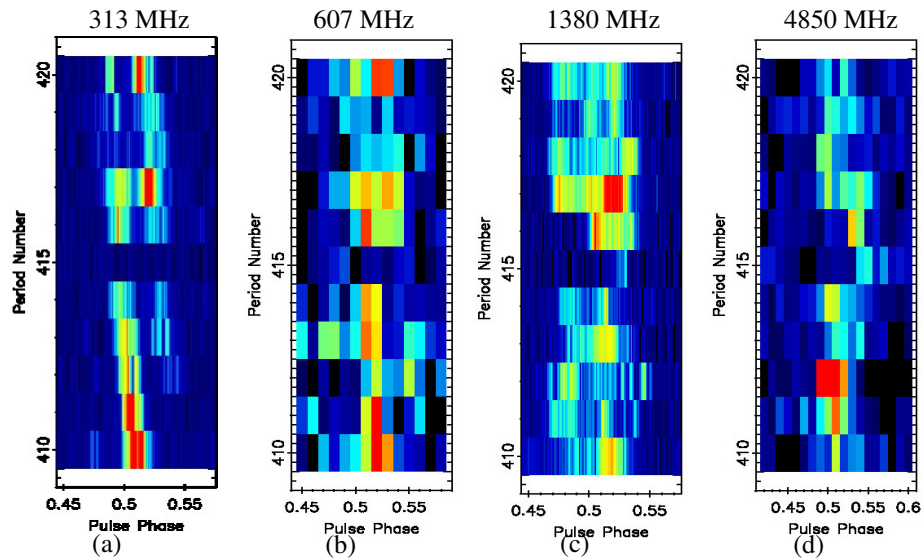
\begin{figure}[h!]
 \vspace{0.2cm}
 \centering
 \subfigure[]{
 \includegraphics[width=2.6 in,height=1.1 in,angle=-90,bb=0 0 496 187]{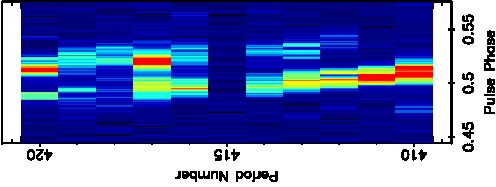}
 % spdisplay_example1.325.eps: 494x183 pixel, 72dpi, 17.43x6.46 cm, bb=52 583 546 766
 }
 \subfigure[]{
 \includegraphics[width=2.6 in,height=1.1 in,angle=-90,bb=0 0 496 187]{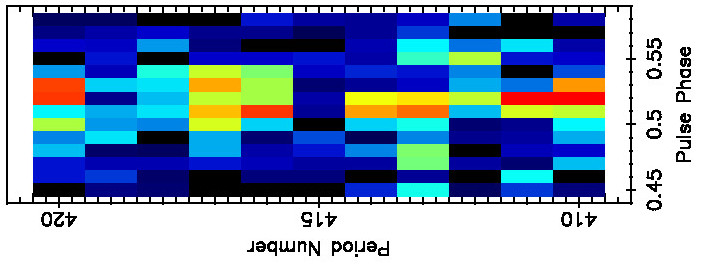}
 }
 \subfigure[]{
 \includegraphics[width=2.6 in,height=1.1 in,angle=-90,bb=0 0 496 187]{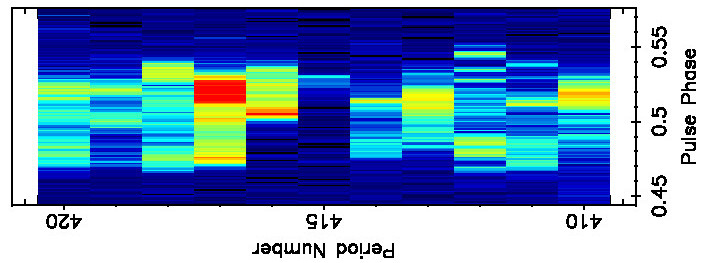}
 % spdisplay_example1.325.eps: 494x183 pixel, 72dpi, 17.43x6.46 cm, bb=52 583 546 766
 }
 \subfigure[]{
 \includegraphics[width=2.6 in,height=1.1 in,angle=-90,bb=0 0 496 187]{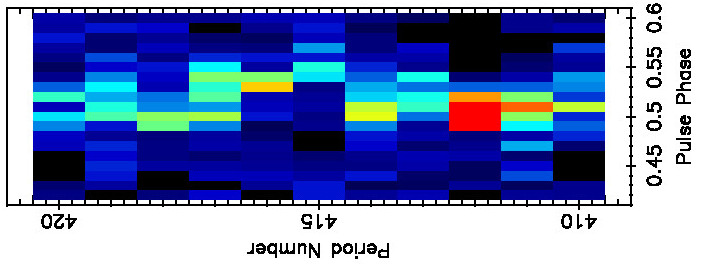}
 }
 \begin{picture}(0,0)
  \put(-320,3){\footnotesize 313 MHz}
  \put(-235,3){\footnotesize 607 MHz}
  \put(-147,3){\footnotesize 1380 MHz}
  \put(-57,3){\footnotesize 4850 MHz}
 \end{picture}
 \caption[First example of simultaneously observed sequence of pulses at four frequencies for PSR B0809+74]
 {Sections of simultaneously observed sequence of pulses at (a) 313 MHz, (b) 607 MHz, 
 (c) 1380 MHz and (d) 4850 MHz for PSR B0809+74. At the pulse number 
 415, pulsar shows clear a null pulse at 313 and 607 MHz 
 while weak burst pulses can be seen at 1380 and 4850 MHz.}
 \label{b0809_sp_mismatch1}
\end{figure}

As can be seen from Table \ref{b0809_cont_table_all_freq}, 
for a small number of pulses, the above association 
does not hold for various pairs. 
In a comparison across all four frequencies 
12 pulses were found to show nonconcurrence in the emission 
states among 10003 compared pulses. 
All such pulses were carefully 
examined to verify their true nature and identify 
their locations. Surprisingly, 7 of the nonconcurrent  
pulses occurred either at the start or at the end of a burst. 
A few examples of such nonconcurrent pulses are shown 
in Figures \ref{b0809_sp_mismatch1}, \ref{b0809_sp_mismatch2} and 
\ref{b0809_sp_mismatch3}. 
As shown in Figure \ref{b0809_sp_mismatch1}, at period number 415, 
pulsar exhibits absence of emission at 313 and 607 MHz 
[Figure \ref{b0809_sp_mismatch1}(a) and (b)] while 
weak burst pulses can be seen at 1380 and 4850 MHz 
[Figure \ref{b0809_sp_mismatch1}(c) and (d)].
As mentioned previously, a deeper examination 
reveals that such nonconcurrent events 
are most likely to occur at the transition stage. 
Figure \ref{b0809_sp_mismatch2} shows example of two such pulses 
where narrow pulsed emission can be seen at 1380 MHz, while no detectable 
emission is present at 313, 607 and 4850 MHz. One of these burst pulses 
is localized right in the middle of the long null state, while the other 
is located near the onset of the burst phase at 1380 MHz. 
On a few occasions, as shown Figure \ref{b0809_sp_mismatch3}, pulsar also exhibits  
absence of emission at higher frequencies (i.e. 607, 1380 and 5100 MHz) 
while a narrow and weak burst pulse can be 
seen at the lowest observing frequency, 313 MHz, 
again near the onset of the burst phase. 
In summary, a comparison across all four frequencies revealed 
about 12 out of 10003 pulses ($\sim$0.1\%) to be 
nonconcurrent null (or burst) state for PSR B0809+74.
 \begin{figure}[h!]
 \centering
 \subfigure[]{
 \includegraphics[width=2.6 in,height=1.1 in,angle=-90,bb=0 0 496 187]{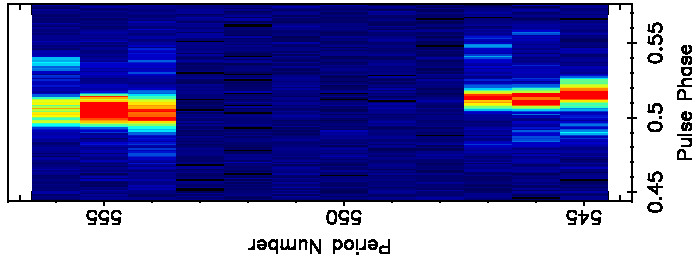}
 }
 \subfigure[]{
 \includegraphics[width=2.6 in,height=1.1 in,angle=-90,bb=0 0 496 187]{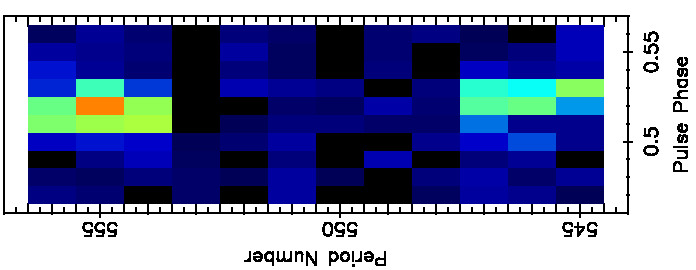}
 }
 \subfigure[]{
 \includegraphics[width=2.6 in,height=1.1 in,angle=-90,bb=0 0 496 187]{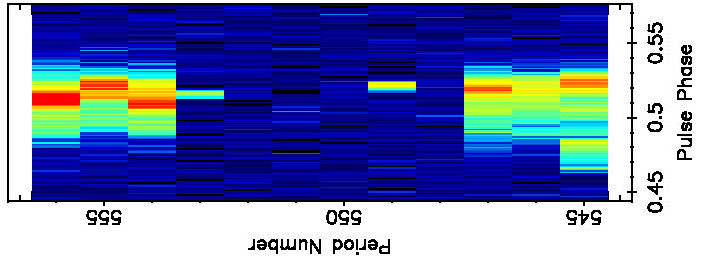}
 }
 \subfigure[]{
 \includegraphics[width=2.6 in,height=1.1 in,angle=-90,bb=0 0 496 187]{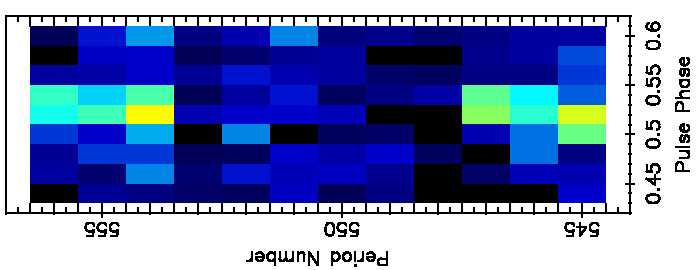}
 }
 \begin{picture}(0,0)
  \put(-320,3){\footnotesize 313 MHz}
  \put(-235,3){\footnotesize 607 MHz}
  \put(-147,3){\footnotesize 1380 MHz}
  \put(-57,3){\footnotesize 4850 MHz}
 \end{picture}
 \caption[Second example of simultaneously observed sequence of pulses at four frequencies for PSR B0809+74]
 {Sections of simultaneously observed sequence of pulses at 
 (a) 313 MHz, (b) 607 MHz, (c) 1380 MHz and (d) 4850 MHz for PSR B0809+74. 
 At the pulse numbers 549 and 553, pulsar shows clear nulls at 313, 607 and 4850 MHz 
 while narrow burst pulses can be seen at 1380 MHz.}
\label{b0809_sp_mismatch2}
\end{figure}
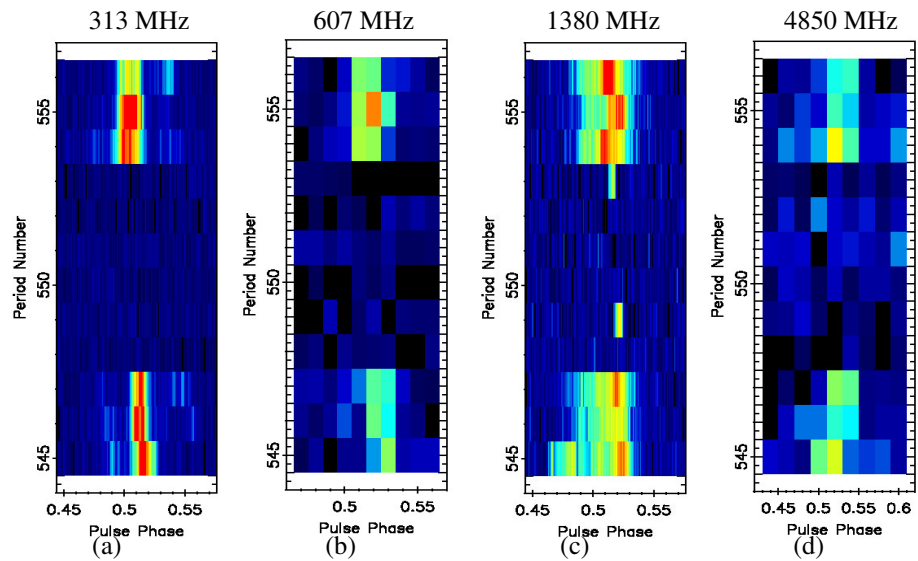
 \begin{figure}[h!]
 \vspace{0.8cm}
 \centering
 \subfigure[]{
 \includegraphics[width=2.6 in,height=1.1 in,angle=-90,bb=0 0 496 187]{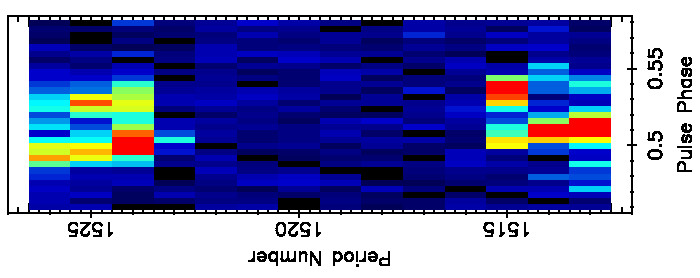}
 }
 \subfigure[]{
 \includegraphics[width=2.6 in,height=1.1 in,angle=-90,bb=0 0 496 187]{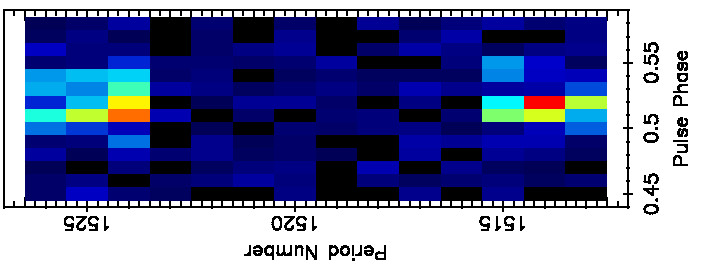}
 }
 \subfigure[]{
 \includegraphics[width=2.6 in,height=1.1 in,angle=-90,bb=0 0 496 187]{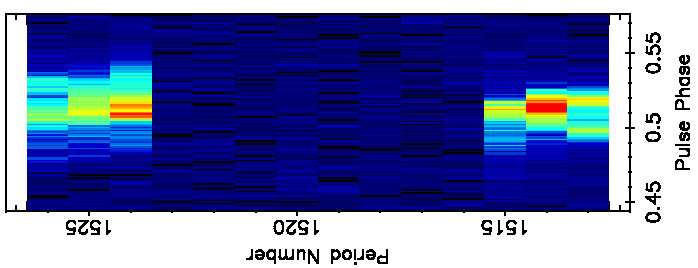}
 }
 \subfigure[]{
 \includegraphics[width=2.6 in,height=1.1 in,angle=-90,bb=0 0 496 187]{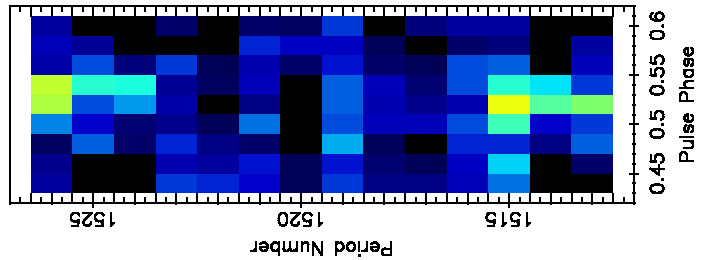}
 }
 \begin{picture}(0,0)
  \put(-320,3){\footnotesize 313 MHz}
  \put(-235,3){\footnotesize 607 MHz}
  \put(-147,3){\footnotesize 1380 MHz}
  \put(-57,3){\footnotesize 4850 MHz}
 \end{picture}
 \caption[Third example of simultaneously observed sequence of pulses at four frequencies for PSR B0809+74]
 {Sections of simultaneously observed sequence of pulses at 
 (a) 313 MHz, (b) 607 MHz, (c) 1380 MHz and (d) 4850 MHz for PSR B0809+74. 
 At the pulse number 1523, pulsar shows clear null pulses at 607, 1380 and 4850 MHz,  
 while a weak burst pulse can be seen at 313 MHz.}
\label{b0809_sp_mismatch3} 
\end{figure}
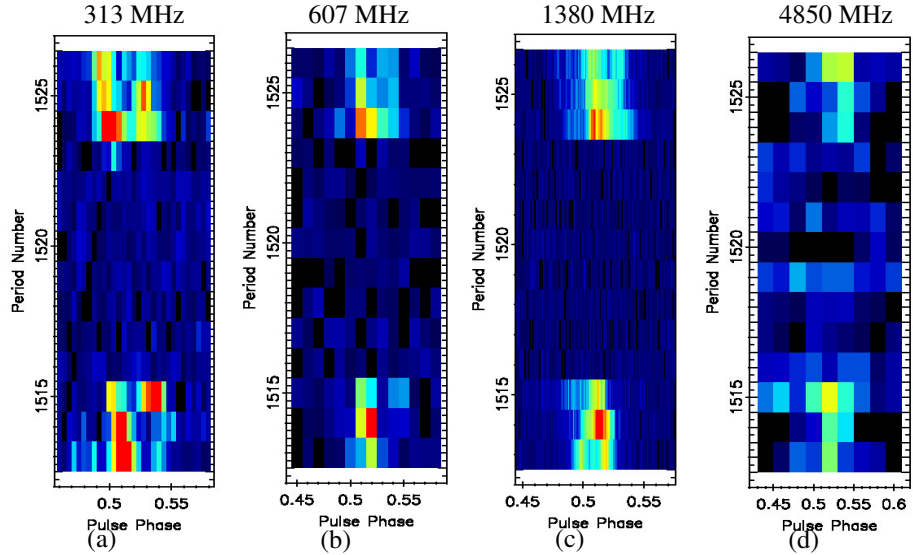
\begin{figure}[h!]
 \vspace{0.8cm}
 \centering
 \subfigure[]{
 \includegraphics[width=2 in,height=2 in,angle=0,bb=0 0 360 250]{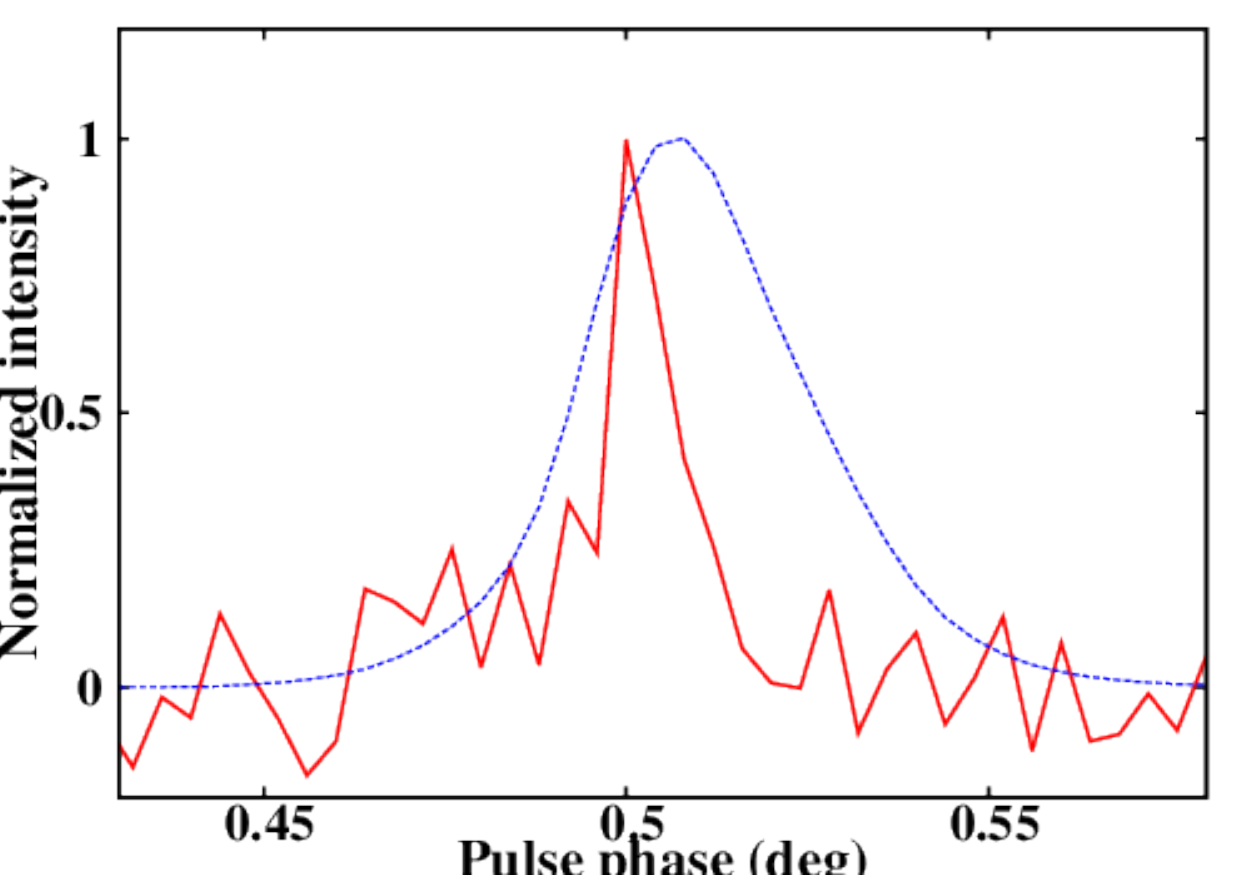}
 % Exclusive_325_burst.eps.pdf: 0x0 pixel, 300dpi, 0.00x0.00 cm, bb=50 50 410 302
 }
 \subfigure[]{
 \includegraphics[width=2 in,height=2 in,angle=0,bb=0 0 360 250]{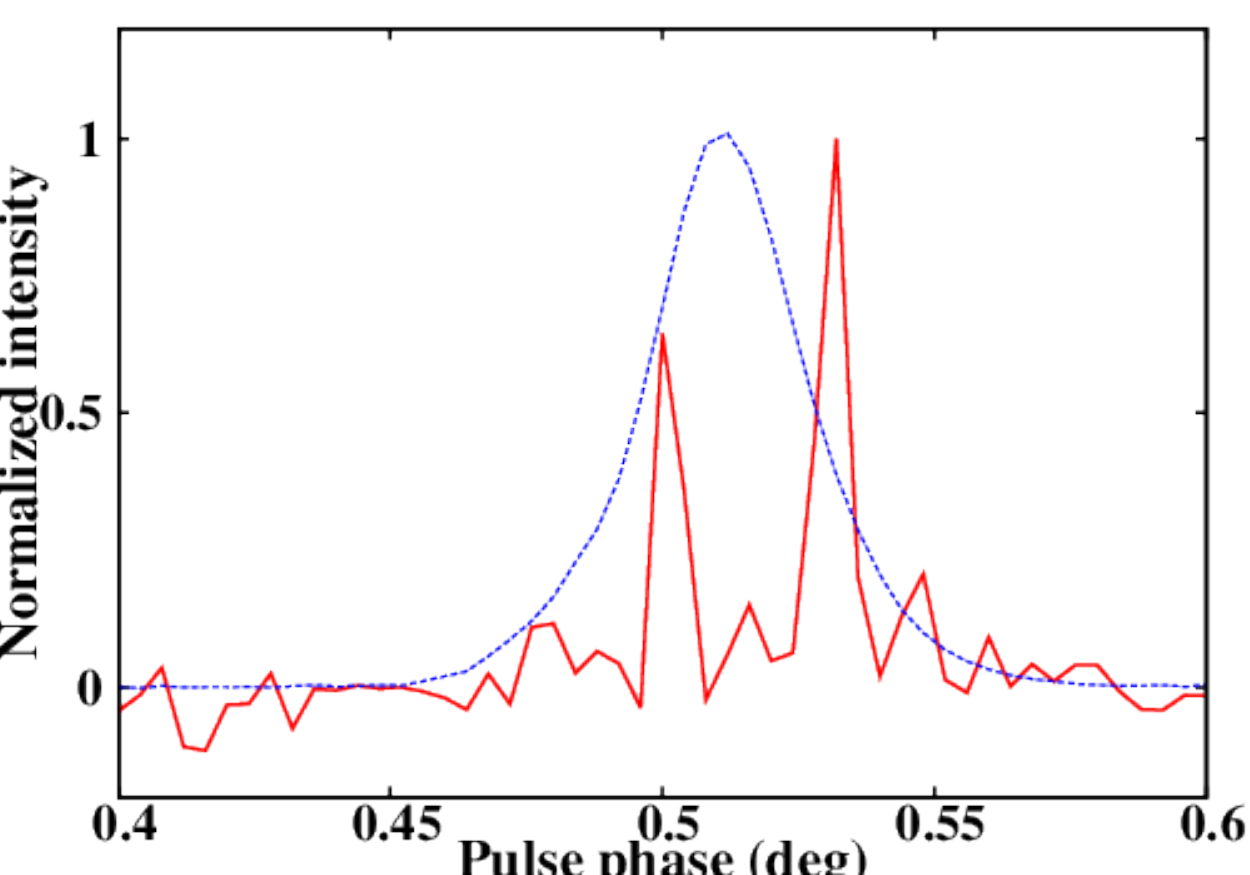}
 % Exclusive_325_burst.eps.pdf: 0x0 pixel, 300dpi, 0.00x0.00 cm, bb=50 50 410 302
 }
 \subfigure[]{
 \includegraphics[width=2 in,height=2 in,angle=0,bb=0 0 360 250]{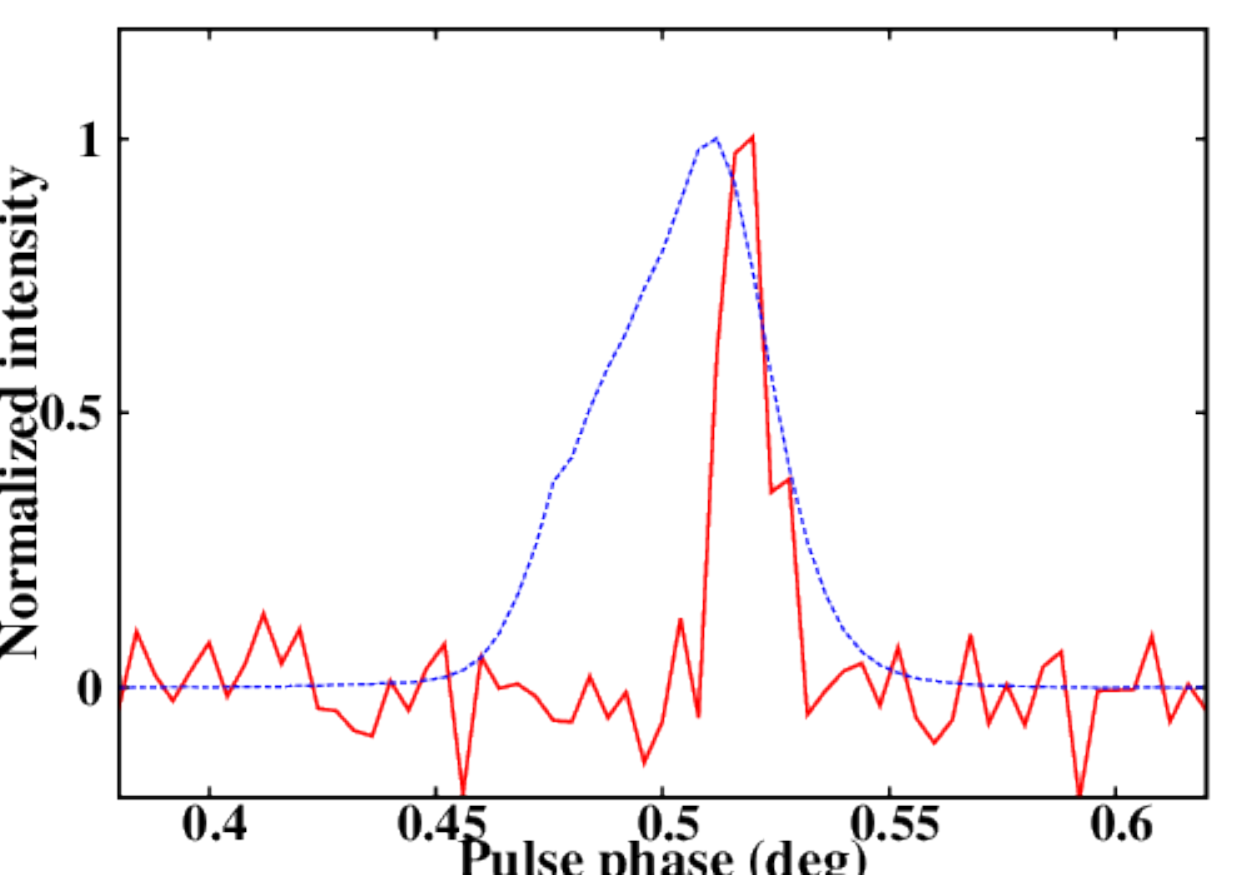}
 % Exclusive_325_burst.eps.pdf: 0x0 pixel, 300dpi, 0.00x0.00 cm, bb=50 50 410 302
 }
 \subfigure[]{
 \includegraphics[width=2 in,height=2 in,angle=0,bb=0 0 360 250]{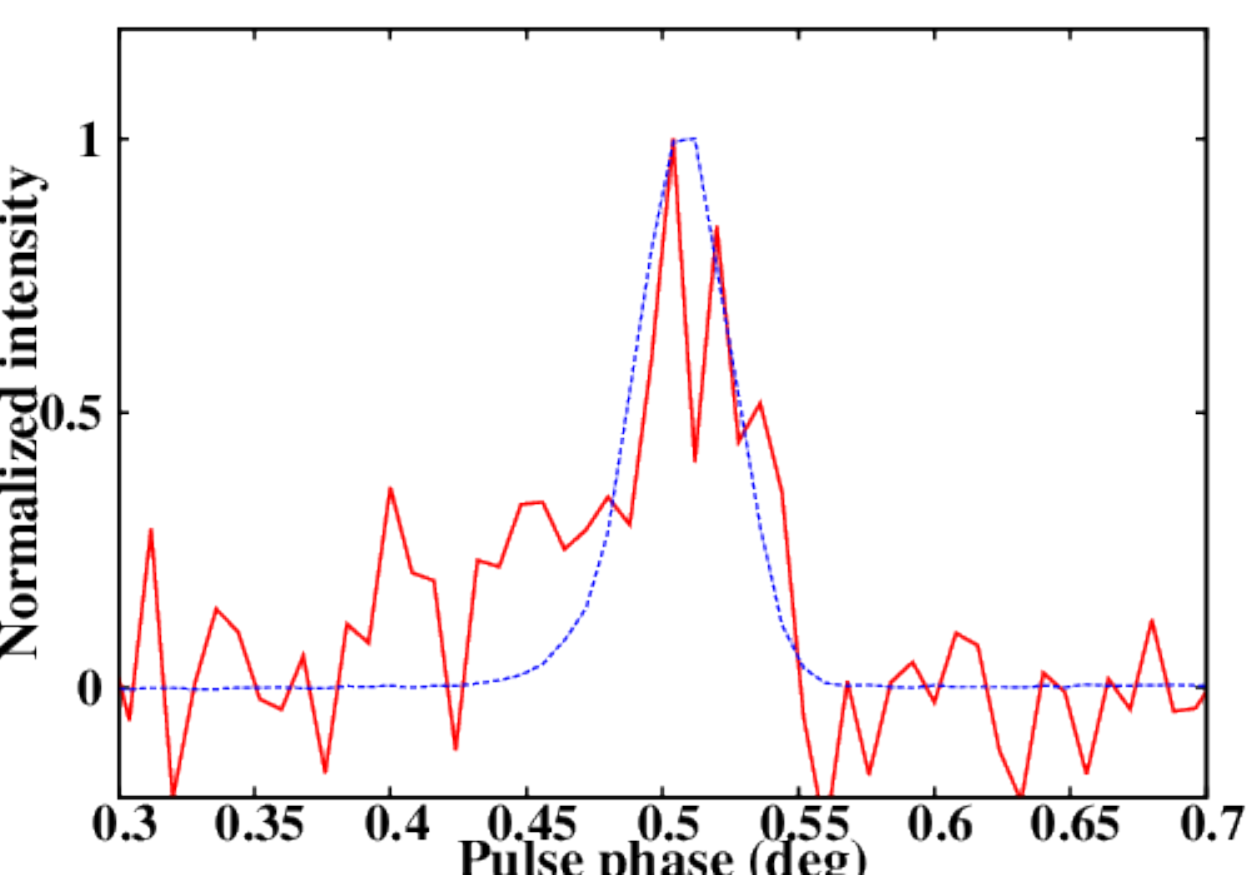}
 % Exclusive_325_burst.eps: 0x0 pixel, 300dpi, 0.00x0.00 cm, bb=50 50 410 302
 }
 \begin{picture}(0,0)
  \put(-280,280){\bf\footnotesize 313 MHz}
  \put(-280,270){\bf\footnotesize (3 pulses)}
  \put(-130,280){\bf\footnotesize 607 MHz}
  \put(-130,270){\bf\footnotesize (2 pulses)}
  \put(-280,120){\bf\footnotesize 1380 MHz}
  \put(-280,110){\bf\footnotesize (3 pulses)}
  \put(-130,120){\bf\footnotesize 4580 MHz}
  \put(-130,110){\bf\footnotesize (4 pulses)}
 \end{picture}
 \caption[Exclusive nonconcurrent burst pulses for PSR B0809+74]
 {Exclusive nonconcurrent burst pulses at (a) 313 MHz, 
 (b) 607 MHz, (c) 1380 MHz and (d) 4850 MHz for PSR B0809+74. 
 For each frequency, all the burst pulses 
 (quoted in the corresponding inset texts), 
 occuring only at a given single frequency, 
 were averaged and shown with a red solid line. 
 The blue dotted line represent the integrated 
 profile at the corresponding frequency.}
 \label{b0809_mismatch_profiles}
\end{figure}
 
As its evident from Figures \ref{b0809_sp_mismatch1}, \ref{b0809_sp_mismatch2} and 
\ref{b0809_sp_mismatch3}, when pulsar displays nonconcurrence in the emission state across 
different frequencies, the burst emission is likely to be weak or narrow. 
To scrutinize this behaviour further, we identified and separated 
all the {\itshape exclusive} nonconcurrent burst pulses.\footnote{Pulses that 
are seen to occur only at a given frequency 
while for the same pulse no emission is seen at other frequencies. Thus, 
the number of such pulses are smaller compared to those quoted in Table \ref{b0809_cont_table_all_freq}.} 
These pulses were combined at each frequency and an average  
pulse shape was obtained, which is shown in Figure \ref{b0809_mismatch_profiles}
for all frequencies. The peak S/N of 
these profiles\footnote{It should be noted that, it is not similar to 
the classical integrated profile, but just the aggregate power from all 
the exclusive pulses.} are around 11, 30, 23 and 8 
for the profiles at 313, 607, 1380 and 4850 MHz, respectively. 
For the highest observed frequency, 4850 MHz, no exclusive 
burst pulse was noticed. Figure \ref{b0809_mismatch_profiles}(d) shows 
average profile from the pulses which showed emission at 4850 MHz while  
no detectable emission was seen at the lowest frequency (i.e. 313 MHz). 
Figure \ref{b0809_mismatch_profiles} also shows comparisons of exclusive 
nonconcurrent pulse profiles with the standard integrated profiles. 
The narrowness of the exclusive nonconcurrent pulse profile can be spotted easily 
from these figures. It should be noted that, PSR B0809+74 exhibit prominent drifting behaviour. 
Thus, most of the individual single pulses consist of two or three narrow subpulses. 
As the number of exclusive nonconcurrent pulses are very small, a broad single component 
smooth profile can not be obtained. This could be a likely reason behind the 
narrowness of the exclusive burst pulse profiles. We also can not reject the possibility of 
detection limits at various frequencies. If the emission is 
narrow and weak at a certain frequency, it is likely to be  
below detection limits at other frequencies. If there is an 
undetectable weak level emission at other frequencies, 
during the occurrence of exclusive burst 
pulse at a certain frequency, averaging these null pulses may 
provide a detectable component at these frequencies. 
However, no detectable emission was seen during the occurrence of 
such nonconcurrent pulses in our data. This scenario can not be truly tested here, 
again due to the small number of exclusive burst pulses and also due to the 
small number of nonconcurrent pulses for different pairs of frequencies. 

% We did not see any 
% correlation across frequencies in these mismatched pulses 
% as they appear to be similar for different pairs of frequencies 
% for all three pulsars. 
\subsection{Null length and burst length comparison}
\begin{figure}[h!]
 \centering
 \subfigure[]{
 \includegraphics[width=2.1 in,height=2 in,angle=0,bb=0 0 360 250]{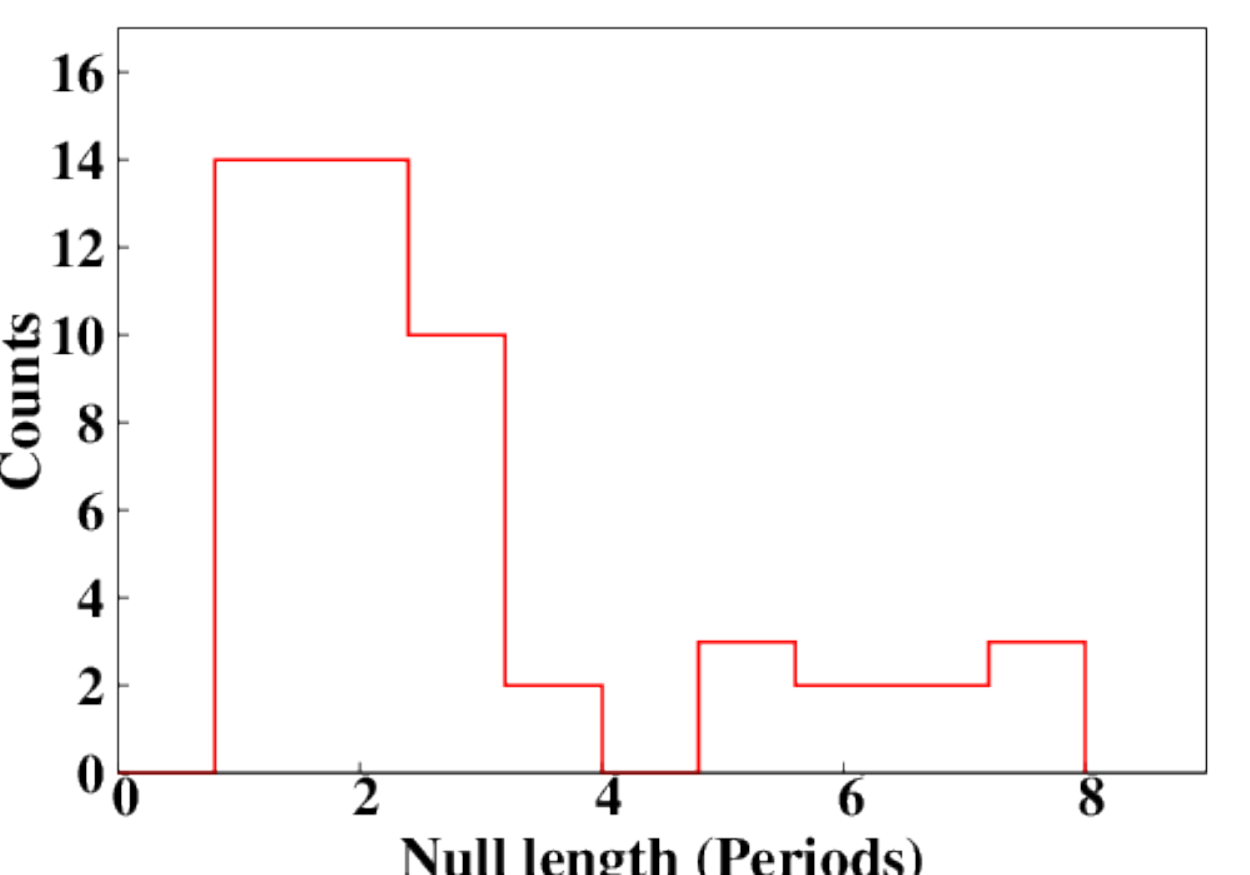}
 % b0809.nlh.325.eps: 360x252 pixel, 72dpi, 12.70x8.89 cm, bb=0 0 360 252
 }
 \subfigure[]{
 \includegraphics[width=2.1 in,height=2 in,angle=0,bb=0 0 360 250]{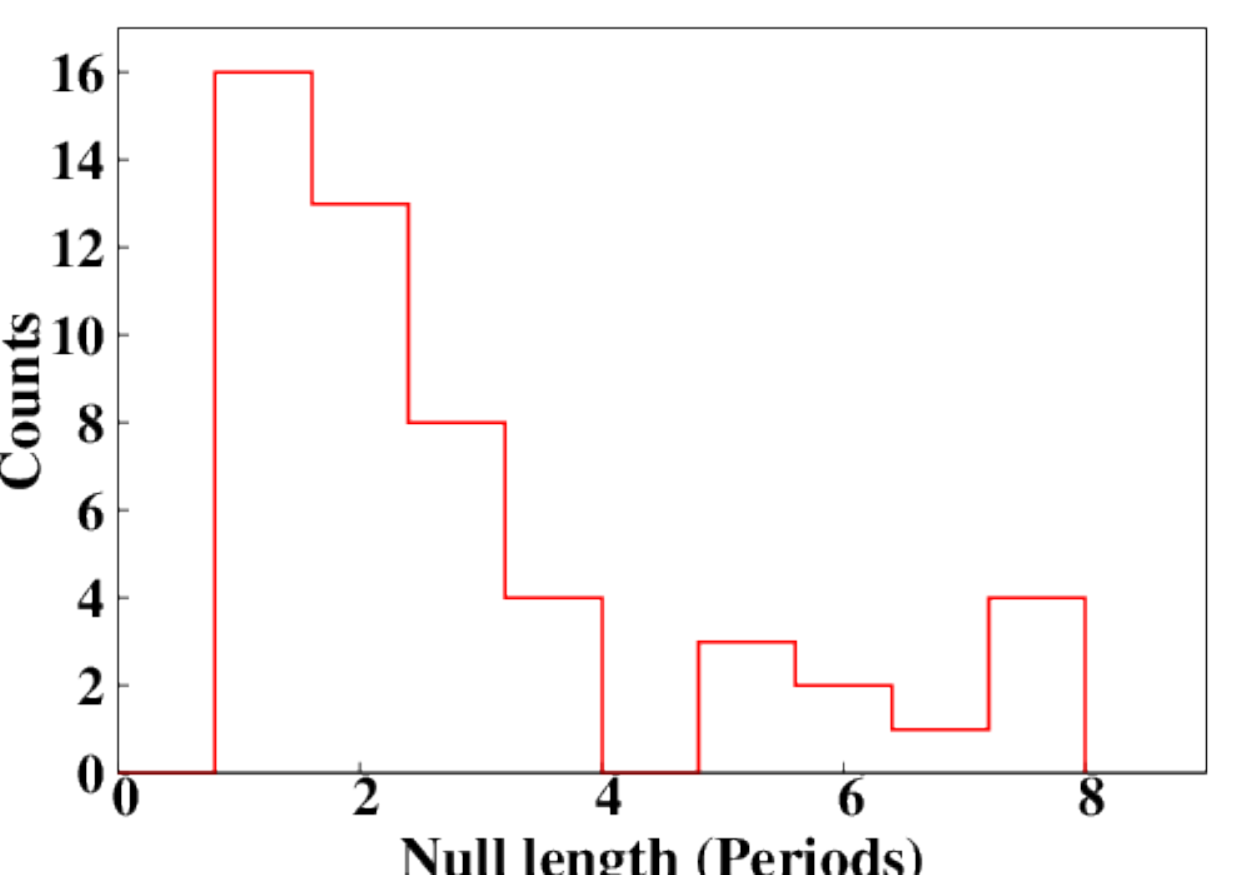}
 % b0809.nlh.325.eps: 360x252 pixel, 72dpi, 12.70x8.89 cm, bb=0 0 360 252
 }
 \subfigure[]{
 \includegraphics[width=2.1 in,height=2 in,angle=0,bb=0 0 360 250]{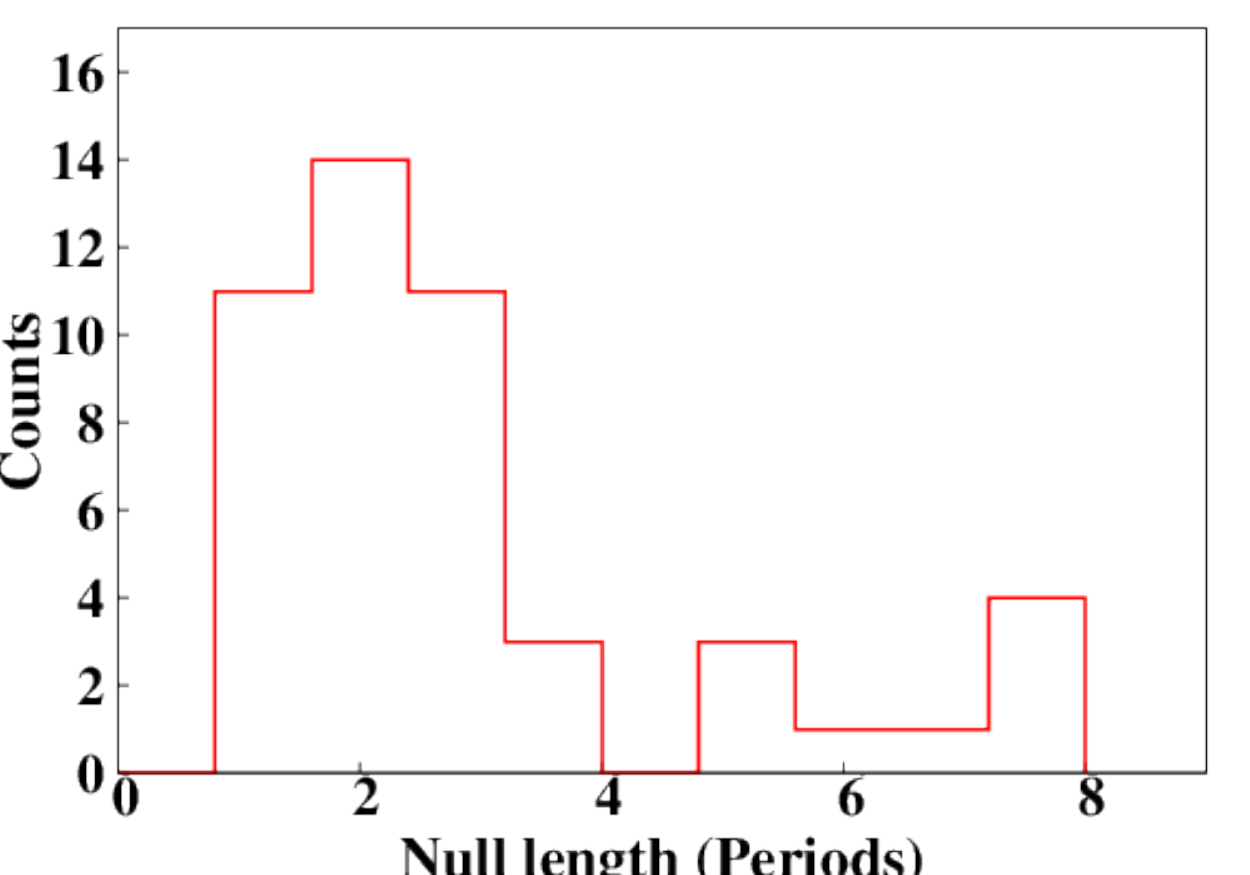}
 % b0809.nlh.325.eps: 360x252 pixel, 72dpi, 12.70x8.89 cm, bb=0 0 360 252
 }
 \subfigure[]{
 \includegraphics[width=2.1 in,height=2 in,angle=0,bb=0 0 360 250]{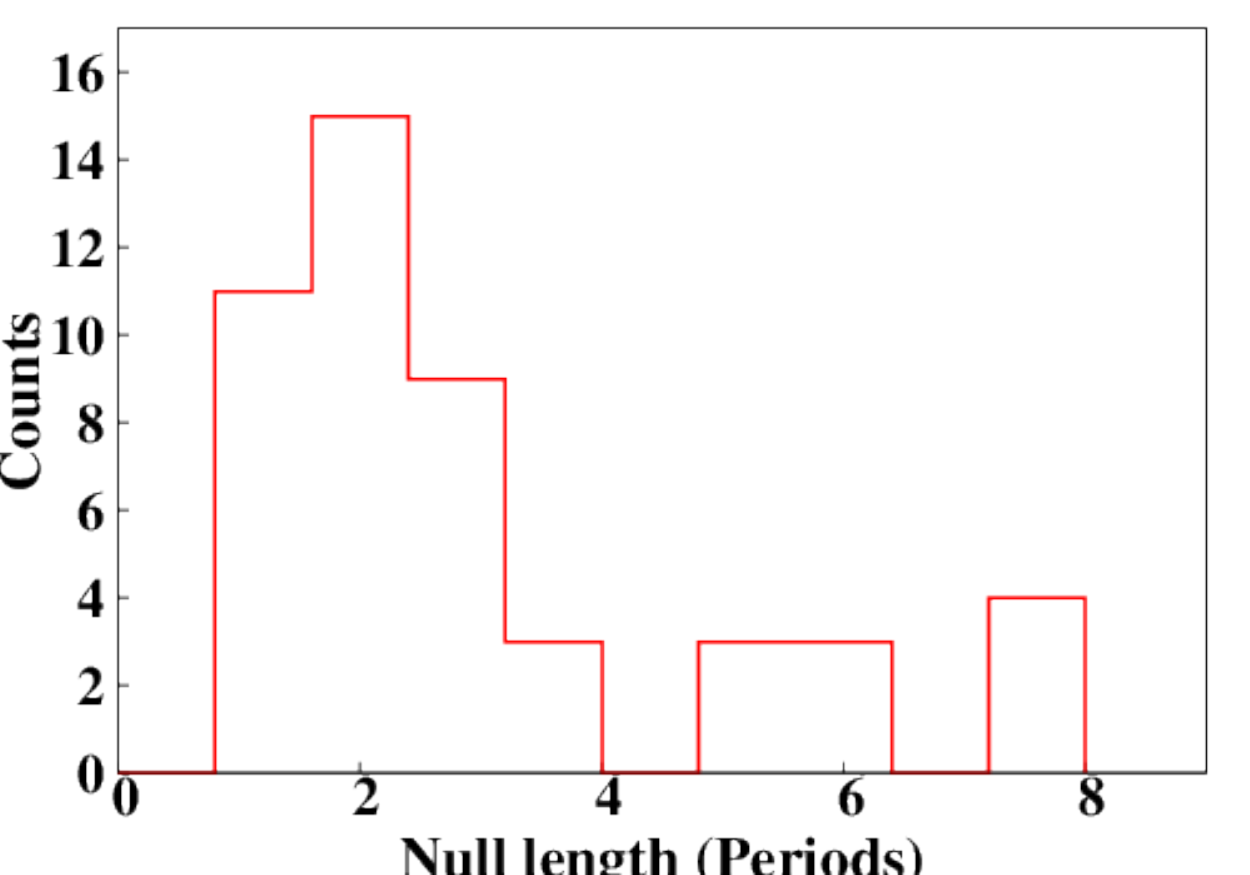}
 % b0809.nlh.325.eps: 360x252 pixel, 72dpi, 12.70x8.89 cm, bb=0 0 360 252
 }
  \begin{picture}(0,0)
  \put(-220,290){\bf\footnotesize 313 MHz}
  \put(-60,290){\bf\footnotesize 607 MHz}
  \put(-220,120){\bf\footnotesize 1380 MHz}
  \put(-60,120){\bf\footnotesize 4850 MHz}
 \end{picture}
 \caption[The obtained NLHs at all four frequencies for PSR B0809+74]
 {The obtained NLHs at (a) 313 MHz, (b) 607 MHz, (c) 1380 MHz and (d) 4850 MHz for PSR B0809+74.
 There are minor differences near the single and double period nulls due to 
 the small number of overall null pulses.}
 \label{b0809_NLH_all_freq}
\end{figure}
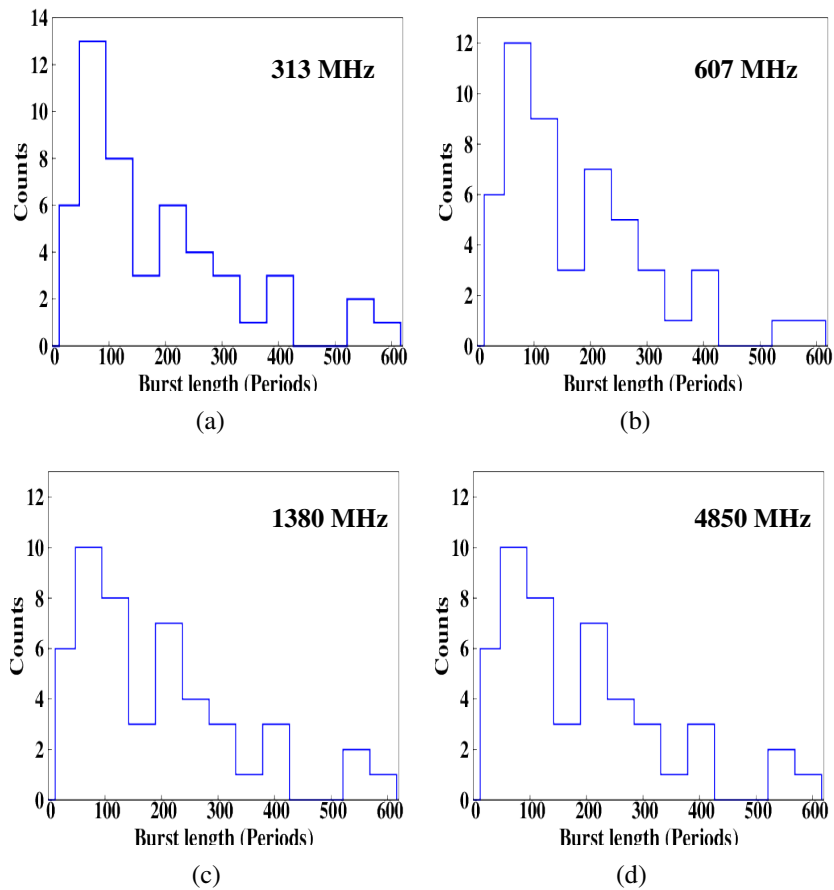
\begin{figure}[h!]
 \centering
 \subfigure[]{
 \includegraphics[width=2.1 in,height=2 in,angle=0,bb=0 0 360 250]{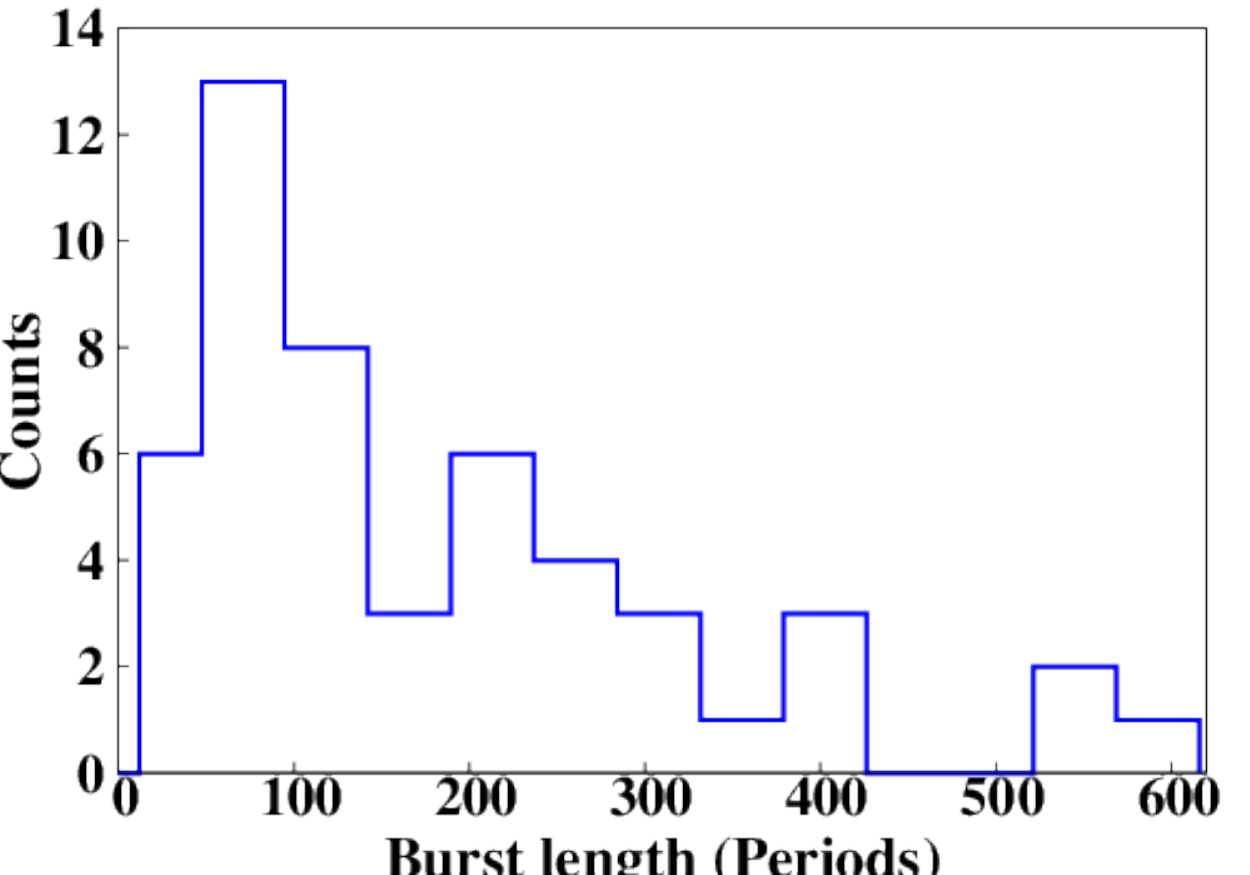}
 % b0809.nlh.325.eps: 360x252 pixel, 72dpi, 12.70x8.89 cm, bb=0 0 360 252
 }
 \subfigure[]{
 \includegraphics[width=2.1 in,height=2 in,angle=0,bb=0 0 360 250]{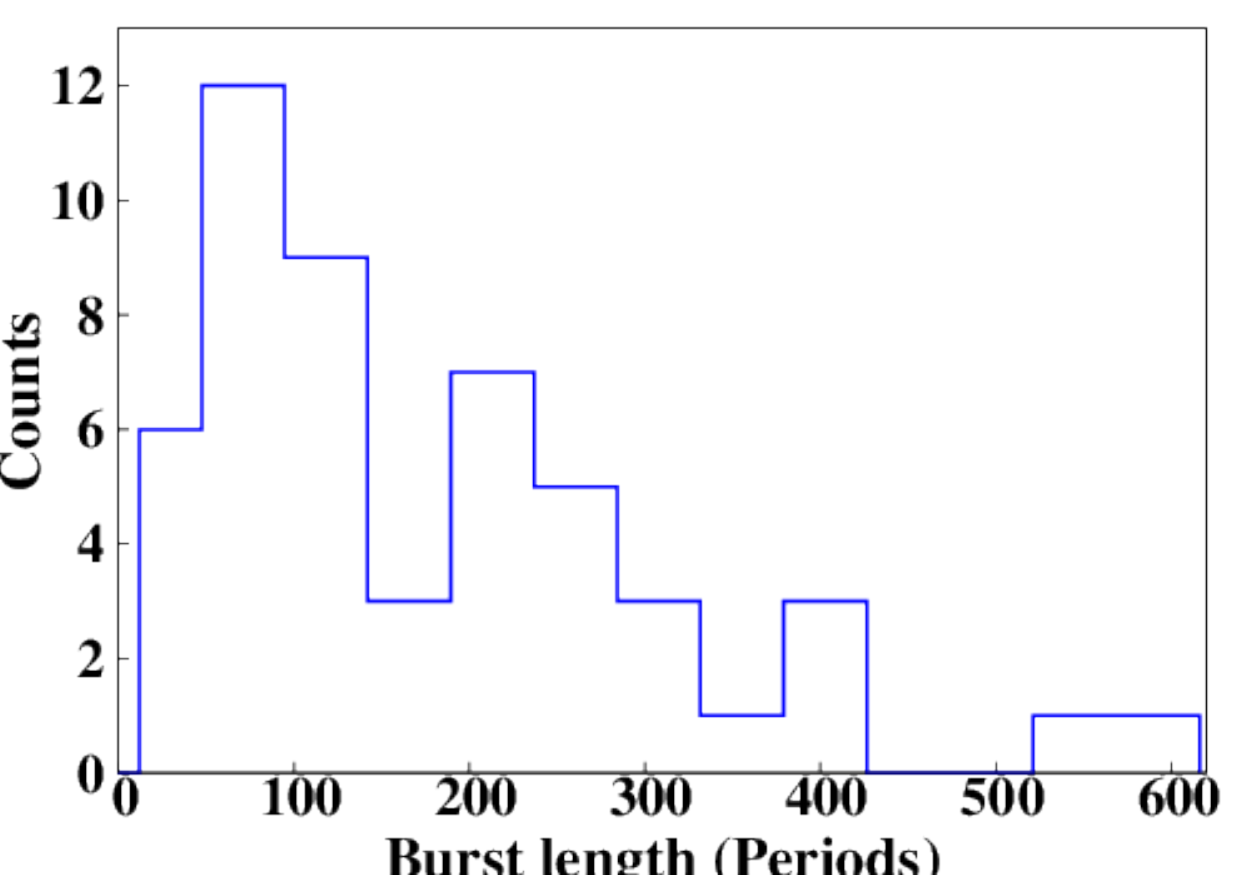}
 % b0809.nlh.325.eps: 360x252 pixel, 72dpi, 12.70x8.89 cm, bb=0 0 360 252
 }
 \subfigure[]{
 \includegraphics[width=2.1 in,height=2 in,angle=0,bb=0 0 360 250]{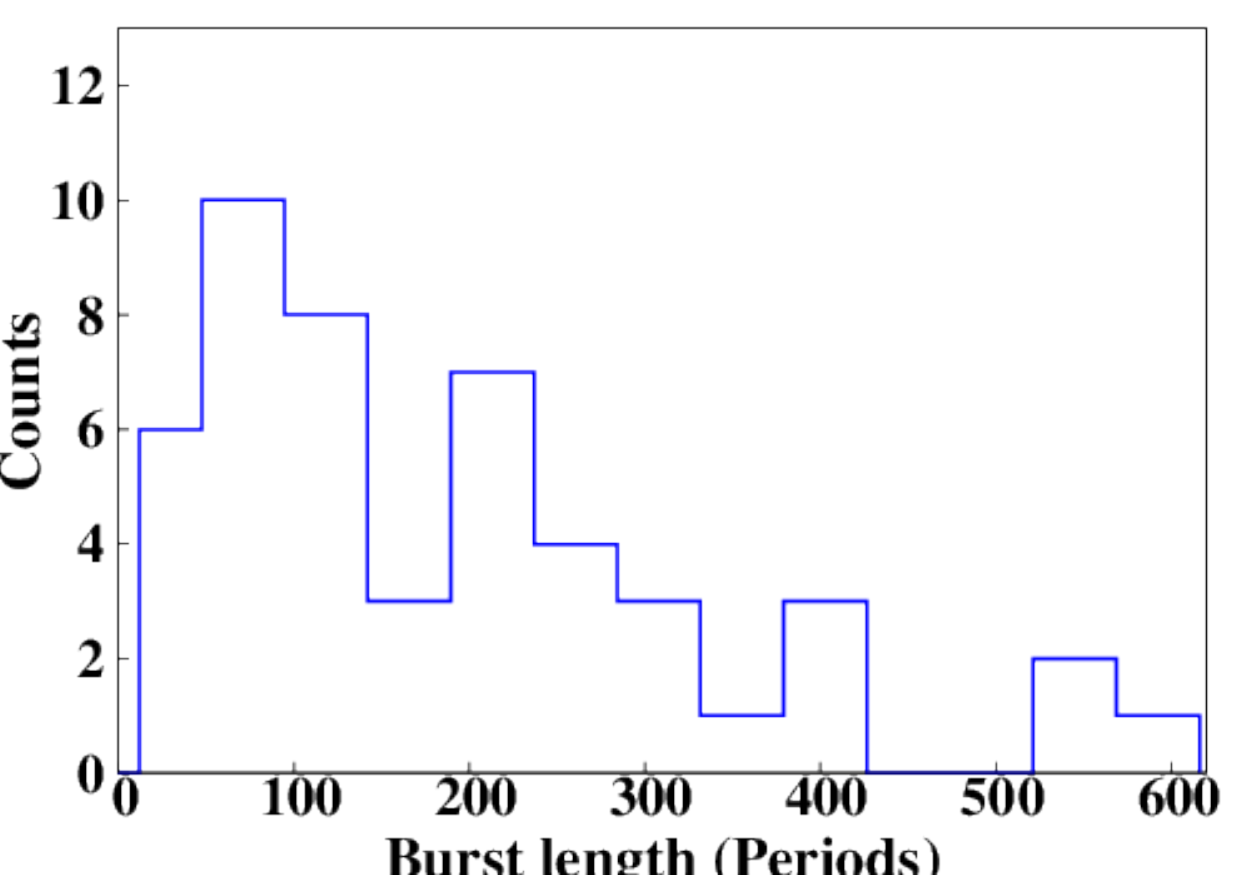}
 % b0809.nlh.325.eps: 360x252 pixel, 72dpi, 12.70x8.89 cm, bb=0 0 360 252
 }
 \subfigure[]{
 \includegraphics[width=2.1 in,height=2 in,angle=0,bb=0 0 360 250]{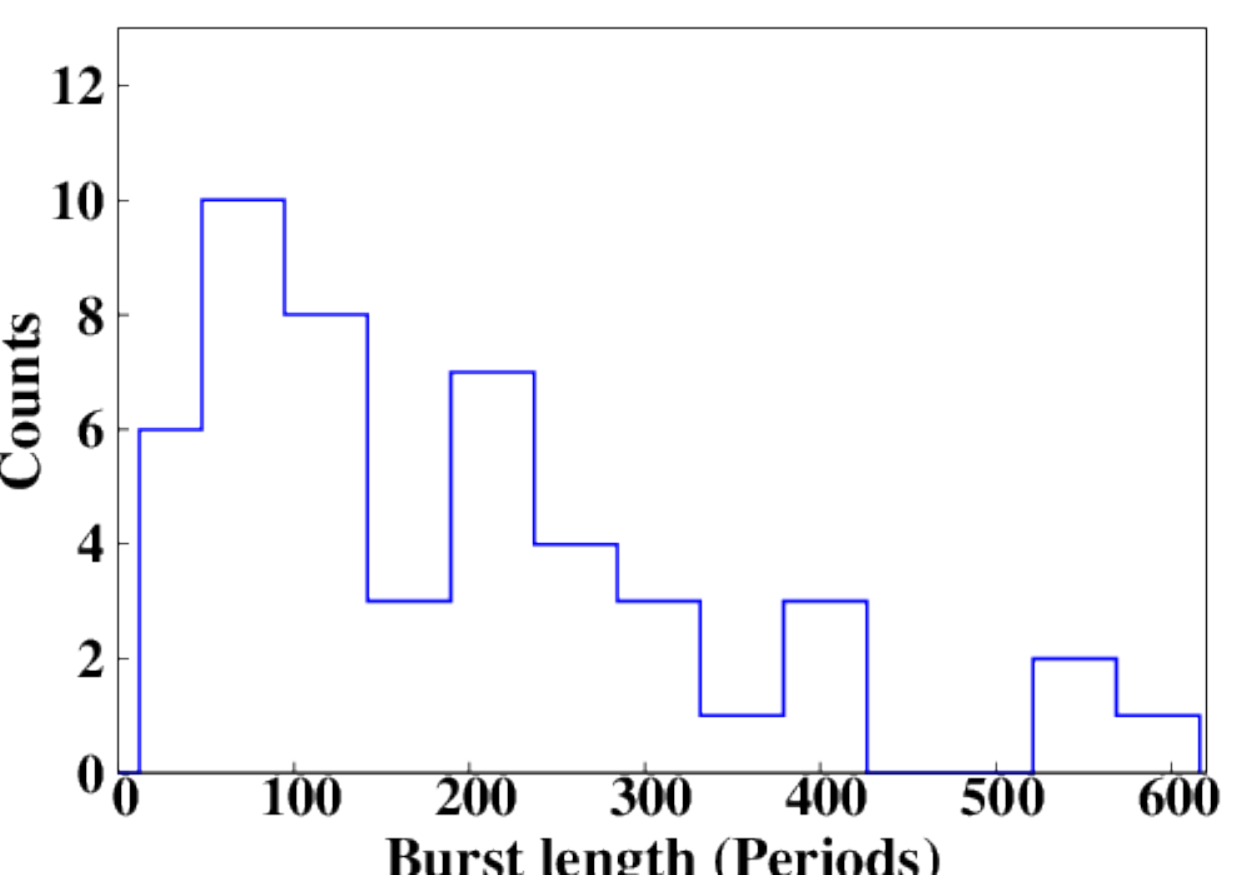}
 % b0809.nlh.325.eps: 360x252 pixel, 72dpi, 12.70x8.89 cm, bb=0 0 360 252
 }
  \begin{picture}(0,0)
  \put(-220,290){\bf\footnotesize 313 MHz}
  \put(-60,290){\bf\footnotesize 607 MHz}
  \put(-220,120){\bf\footnotesize 1380 MHz}
  \put(-60,120){\bf\footnotesize 4850 MHz}
 \end{picture}
 \caption[The obtained BLHs at all four frequencies for PSR B0809+74]
 {The obtained BLHs at (a) 313 MHz, (b) 607 MHz, 
 (c) 1380 MHz and (d) 4850 MHz for PSR B0809+74. 
 Note the remarkable similarity between various frequencies.}
 \label{b0809_BLH_all_freq}
\end{figure}
\begin{figure}[h!]
\begin{center}
 \subfigure[]{
 \includegraphics[width=2.1 in,height=2.1 in,angle=0,bb=0 0 360 250]{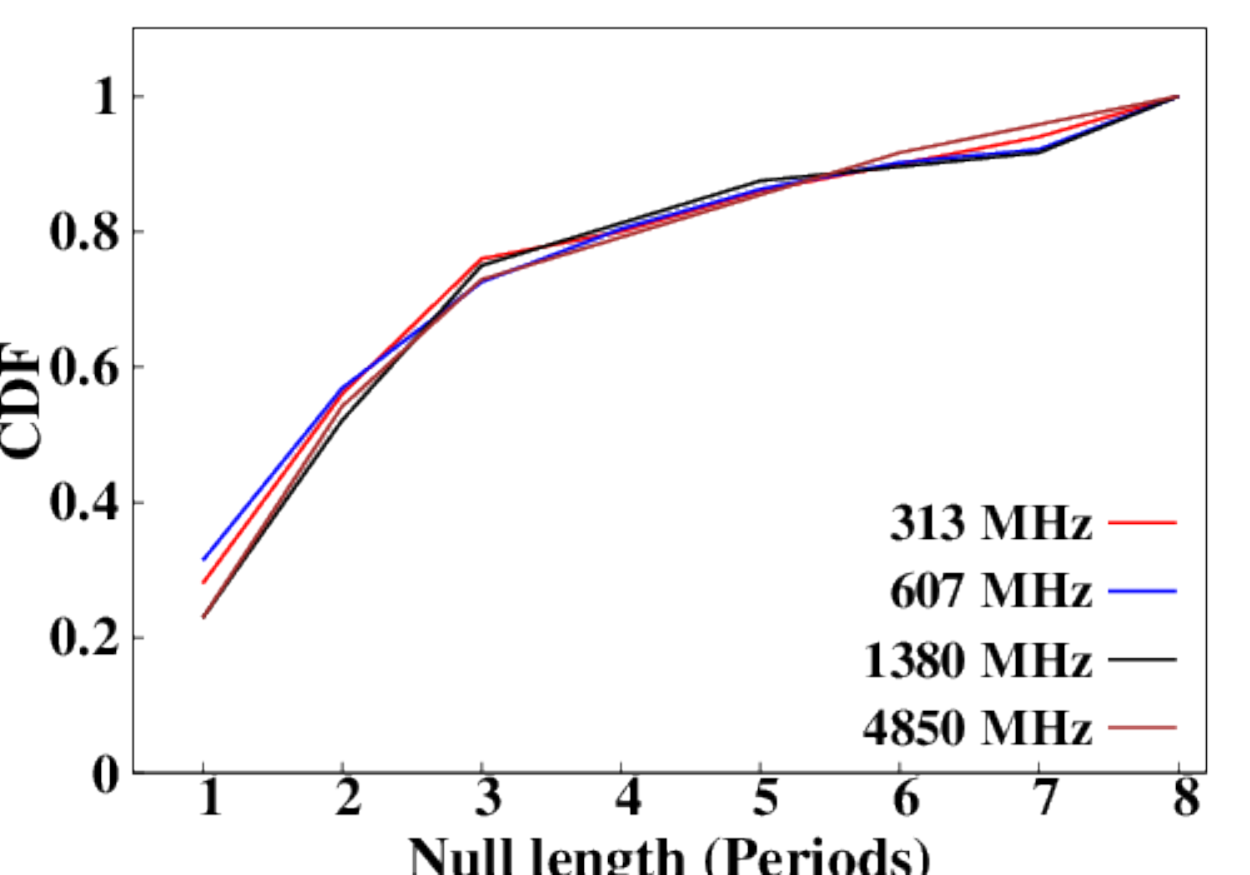}
 % b0809.nlh.cdf.eps: 360x252 pixel, 72dpi, 12.70x8.89 cm, bb=49 46 404 299
 }
 \subfigure[]{
 \includegraphics[width=2.1 in,height=2.1 in,angle=0,bb=0 0 360 250]{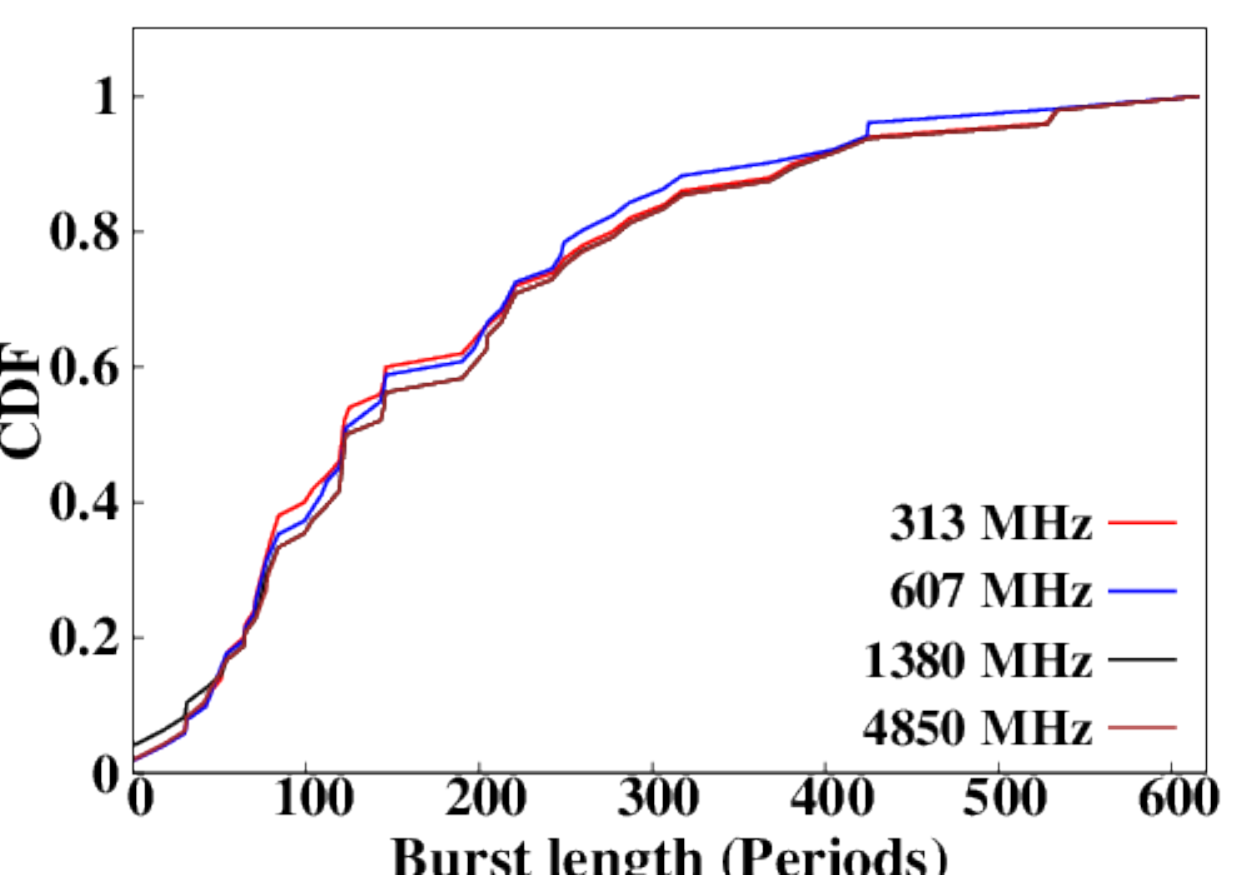}
 % b0809.nlh.cdf.eps: 360x252 pixel, 72dpi, 12.70x8.89 cm, bb=49 46 404 299
 }
\end{center}
 \caption[The obtained CDFs for (a) null length and (b) burst length distributions for PSR B0809+74]
 {The obtained CDFs for (a) null length and (b) burst length distributions for PSR B0809+74 at 
 all four frequencies. Both distributions show noteworthy similarity between all observed frequencies.}
 \label{b0809_cdf_all_freq}
\end{figure}
\begin{table}[h!]
\begin{center}
\begin{tabular}{|l||c|c|c|}
\tableline
Frequency & 607  & 1380 & 4850  \\
(MHz)     &      &      &       \\
\tableline
\tableline
313  & 99.9 & 99.9 & 99.9  \\
     & 99.9 & 99.9 & 99.9  \\
607  & $-$  & 99.9 & 99.9  \\
     & $-$  & 99.9 & 99.9  \\ 
1380 &      & 99.9 & 99.9  \\
     &      & 99.9 & 99.9  \\
\tableline
\end{tabular}
\end{center}
 \caption[Results of a two sample KS-tests on null and burst length 
 distributions between a pair of frequencies with the assumed null 
 hypothesis of the two distributions being different for PSR B0809+74]
 {Results of a two sample KS-tests on null and burst length 
 distributions between a pair of frequencies with the assumed null 
 hypothesis of the two distributions being different for PSR B0809+74. 
 The first row in each column for a given frequency 
 gives the probability of the two distributions being similar for the null 
 length and the second row gives that for the burst length.}
\label{tabnullburst}
\label{b0809_KS_all_freq}
\end{table}

To check the effect of small number of nonconcurrent emission states,
on the overall nulling pattern at various frequencies, 
their NLHs and BLHs were compared. 
As elaborated in Section \ref{b0809_mismatch_sect}, 
more then half of the nonconcurrent pulses are localized 
either at the start or at the end of the burst phase. 
This can be further tested by comparing overall null length and burst 
length distributions, which should not show significant differences. 
The separated null and burst pulses, as discussed 
in Section \ref{b0809_onebit_sect}, were used 
to construct the NLHs and the BLHs for all observed frequencies
(Figures \ref{b0809_NLH_all_freq} and \ref{b0809_BLH_all_freq}). 
Figure \ref{b0809_NLH_all_freq} shows the similarity 
in the null length distribution across all observed frequencies. 
The minor differences seen at the single and double period 
nulls occur due to the a few nonconcurrent pulses. 
These differences are superficial and only due to the small 
number of overall null pulses seen in this pulsar. To check the 
statistical significance of the similarity 
between the null length distributions, we 
carried out a two sample KS-test, assuming a null hypothesis of different 
distribution. Figure \ref{b0809_cdf_all_freq}(a) 
shows CDFs, obtained using the observed null lengths, 
for all four frequencies which presents analogous distributions. 
Similarly, Figure \ref{b0809_BLH_all_freq} shows almost 
identical burst length distributions. To quantify these 
similarities, these distributions were again compared using the  
two sample KS-test, assuming a null hypothesis of different 
distribution. Figure \ref{b0809_cdf_all_freq}(b)
again shows the remarkable similarity between CDFs, 
obtained using the observed burst lengths at all four frequencies. 
Table \ref{b0809_KS_all_freq} summarizes the results on both 
comparisons for each pair of observing frequencies. 
For all pairs of frequencies, the null hypothesis was rejected, 
at a very high significance, confirming the alternate hypothesis 
that these distributions for a pair of frequencies are similar. 
This confirms that nonconcurrent pulses occur near the null to burst 
transitions (or vice-verse). 

Thus, while the nulling patterns for this pulsar is largely broadband, 
deviations from this behaviour is seen in about 0.1 percent 
of pulses, more than half of which occur at the transition from 
null to burst (or vice-verse). 
\newpage
\section{PSR B2319+60}
\label{b2319_sect}
\begin{figure}[h!]
 \centering
 \includegraphics[width=5.5 in,height=5 in,angle=0,bb=17 -7 803 561]{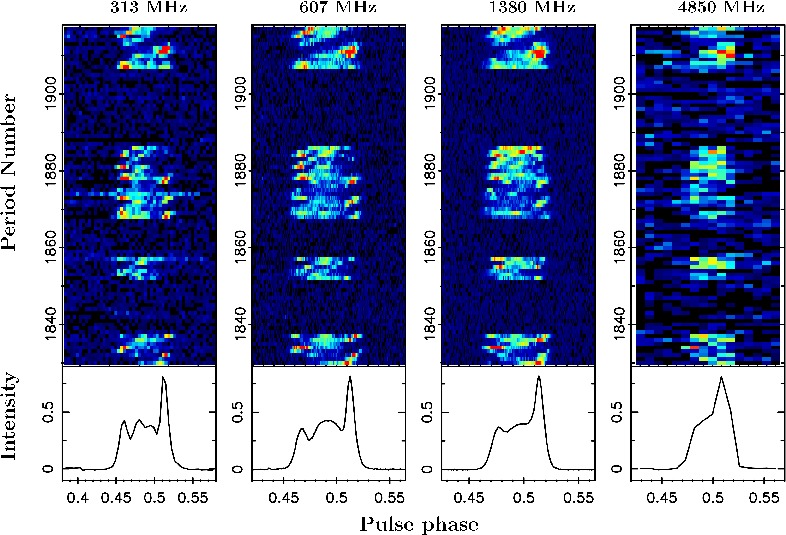}
  % spcomb_label_color.eps: 0x0 pixel, 300dpi, 0.00x0.00 cm, bb=17 -7 803 561
  \caption{Single pulse sequence as a function of 
   pulse number and pulse phase for a subset of data for all frequencies for PSR B2319+64. 
   The sequences were observed simultaneously at all frequencies. The bottom panel in all 
   four plots displays respective integrated profile for the corresponding frequency.}
  \label{b2319_sp_all_freq}
\end{figure}

PSR B2319+60, as also discussed in Section \ref{null_behavior_sect}, 
is a strong mode-changing pulsar with a prominent drifting in the outer 
two conal components. This section discusses the broadband 
behaviour of the nulling phenomenon seen 
in this pulsar. Sections of observed single pulse sequences 
are shown in Figure \ref{b2319_sp_all_freq} at all four 
observed frequencies. A clear drifting feature in the conal 
component can be seen at 313, 607 and 1380 MHz. For the highest 
observed frequency, to improve the single pulse S/N, 
consecutive phase-bins were added, to give 
a total of 125 bins across the profile. 
Thus, such drifting behaviour is 
not possible to spot in Figure \ref{b2319_sp_all_freq}. 
Moreover, Figure \ref{b2319_sp_all_freq} clearly shows three distinct 
null states between period numbers 1837--1851, 
1859--1867 and 1887--1905. The simultaneity in the absence of 
emission is also evident across all 
four frequencies in Figure \ref{b2319_sp_all_freq}. 
To scrutinize and quantify this behaviour 
further, the NFs, the correlation between the nulling pattern 
represented by the one-bit sequence and the distributions 
of null and burst lengths were compared across four frequencies. 

\begin{figure}[h!]
 \begin{center}
 \includegraphics[width=5 in,height=3.5 in,angle=0,bb=0 0 360 250]{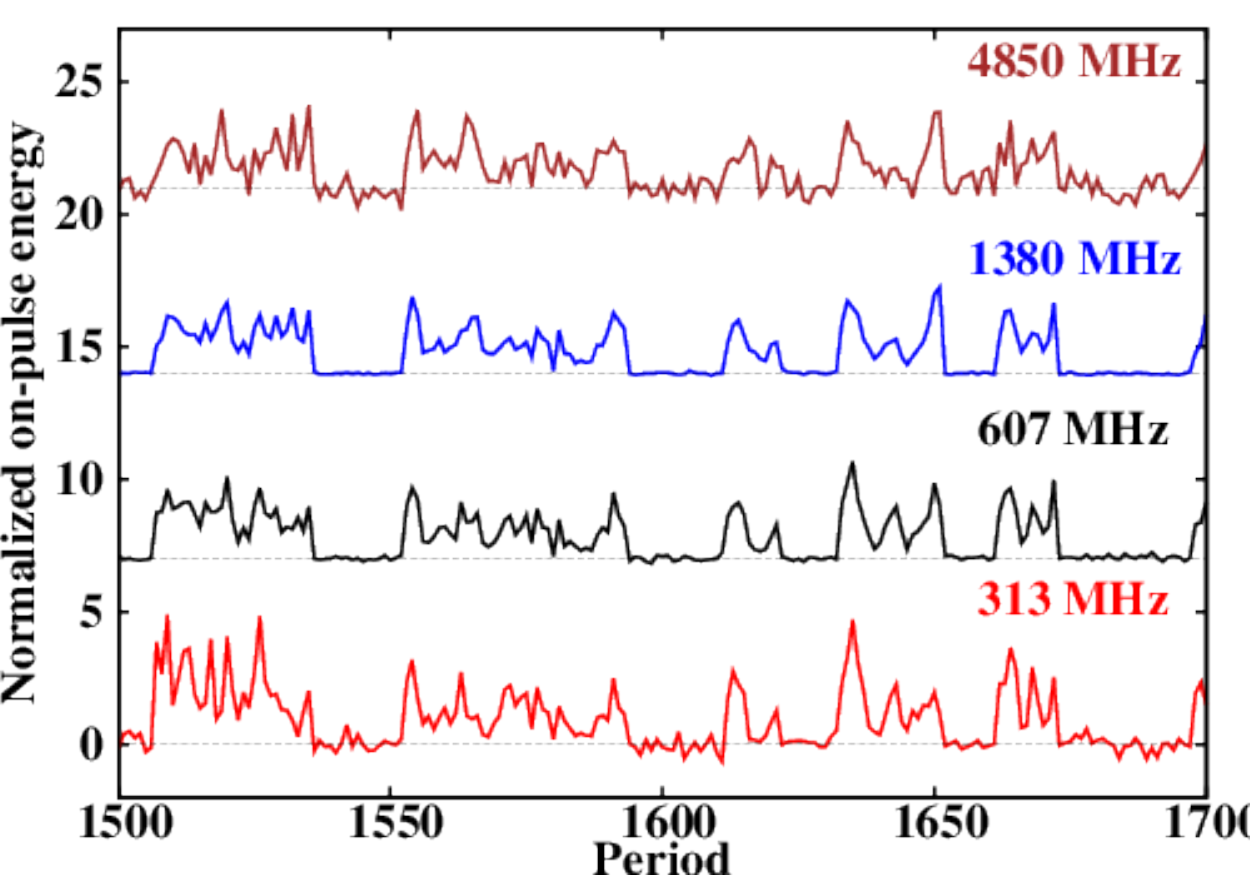}
 % Energy_aligned_all_freq.eps: 0x0 pixel, 300dpi, 0.00x0.00 cm, bb=50 50 410 302
 \caption[On-pulse energy sequences as a function of pulse number for 
  a subset of data at all four frequencies for PSR B2319+60]
 {On-pulse energy sequences as a function of pulse number for 
  a subset of data at all four frequencies for PSR B2319+60. The ordinate 
  present pulse energy in arbitrary units, which is presented with an offset in the 
  vertical direction for each frequency for clarity. The dotted horizontal line 
  presents respective zero pulse energy at a given observing frequency. 
  The abscissa presents the sequence of contiguous period numbers. 
  The simultaneous pulse energy fluctuations is clearly evident across all four frequencies.}
 \label{b2319_ope_all_freq}
 \end{center}
\end{figure}

Similar to PSR B0809+74, we obtained the on-pulse energy sequence by selecting 
an appropriate on-pulse window at each frequency from their 
respective integrated profile. Figure \ref{b2319_ope_all_freq} 
shows sections of concurrent on-pulse energy sequences 
at all four frequencies. The coordinated pulse 
energy fluctuations are also evident from these 
time-series. The on-pulse energy clearly plunges 
to zero level for around six times, depicting 
an occurrence of null state at each frequency. 
These null states clearly show 
simultaneous occurrence across all observed 
frequencies. A visual inspection of 
the entire data broadly confirms this behaviour. 
\subsection{NF comparison}
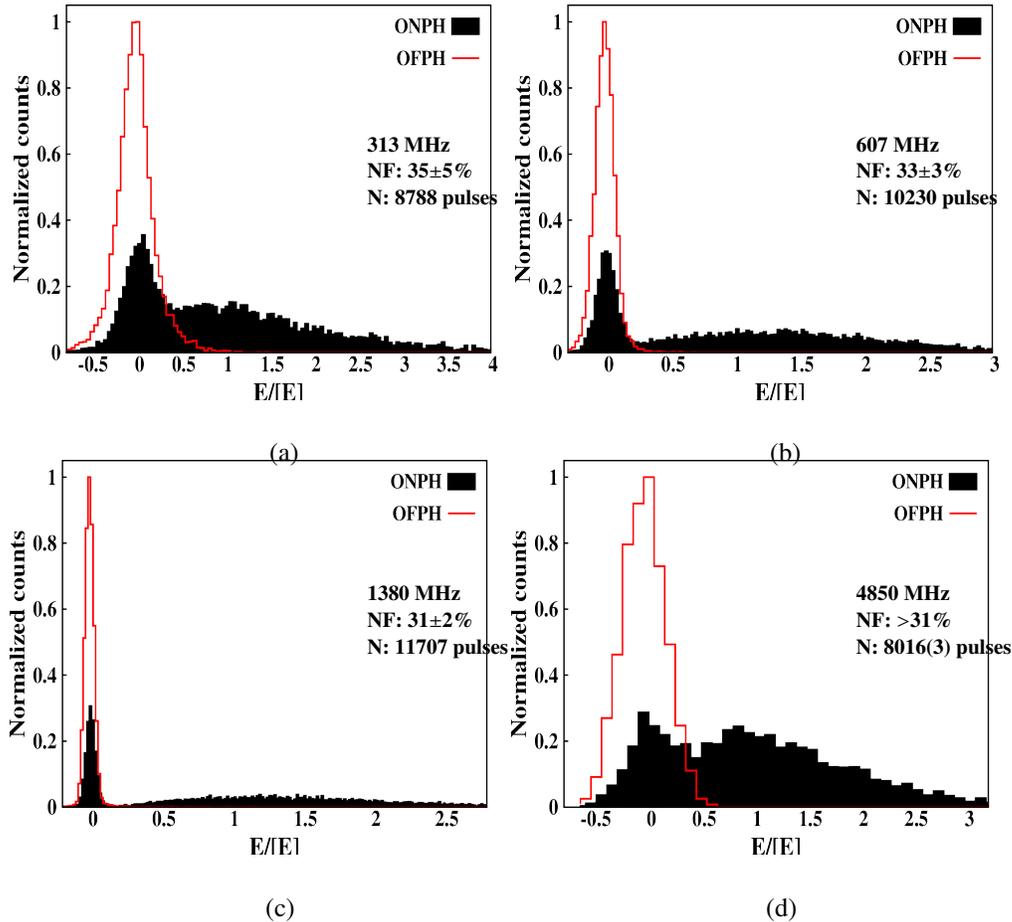
\begin{figure}[h!]
 \centering
 \subfigure[]{
 \includegraphics[width=2.1 in,height=2.5 in,angle=-90,bb=59 53 555 750]{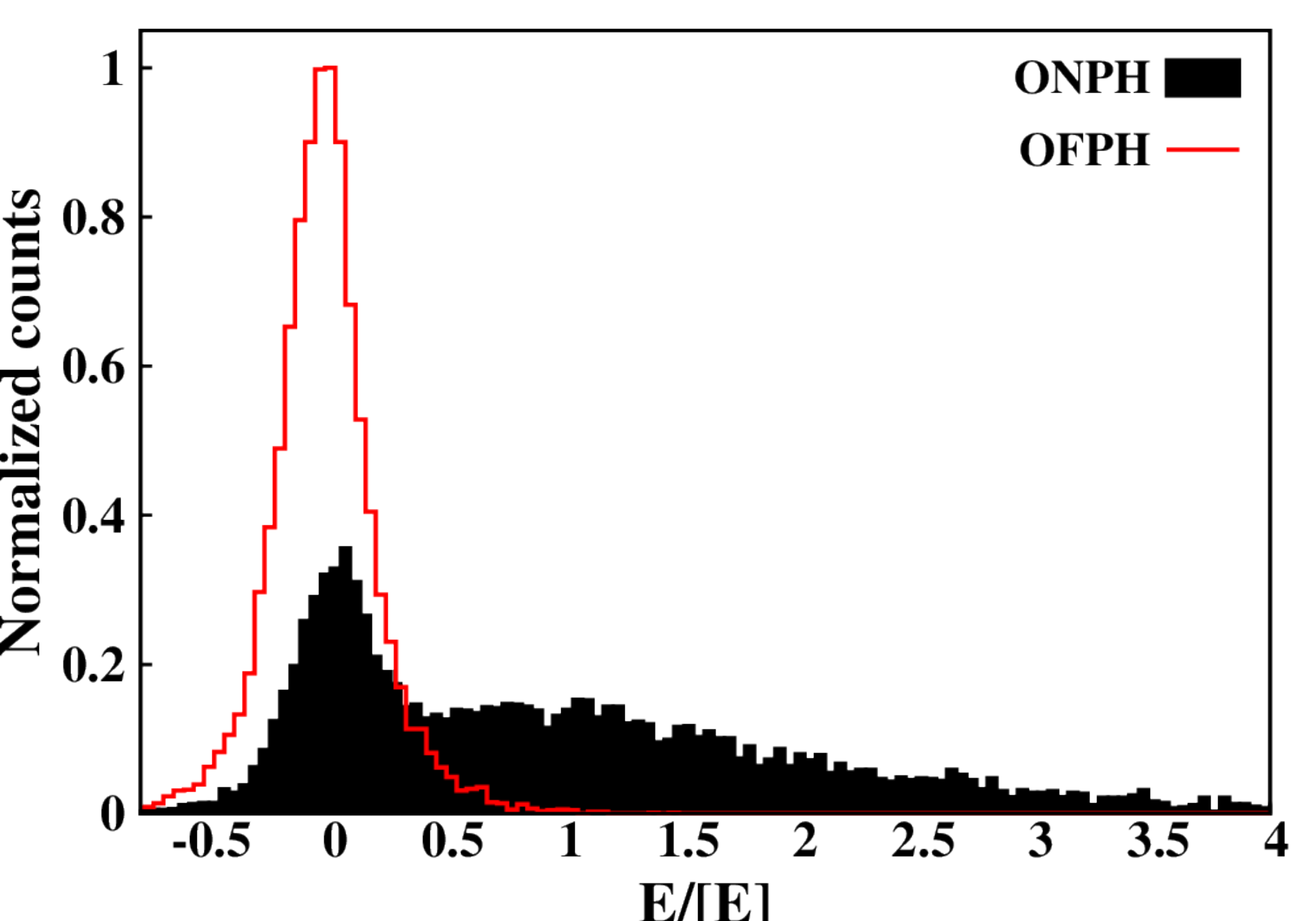}
 % b0809.histogram.325.eps: 504x720 pixel, 72dpi, 17.78x25.40 cm, bb=0 0 504 720
 }
 \subfigure[]{
 \includegraphics[width=2.1 in,height=2.5 in,angle=-90,bb=59 53 555 750]{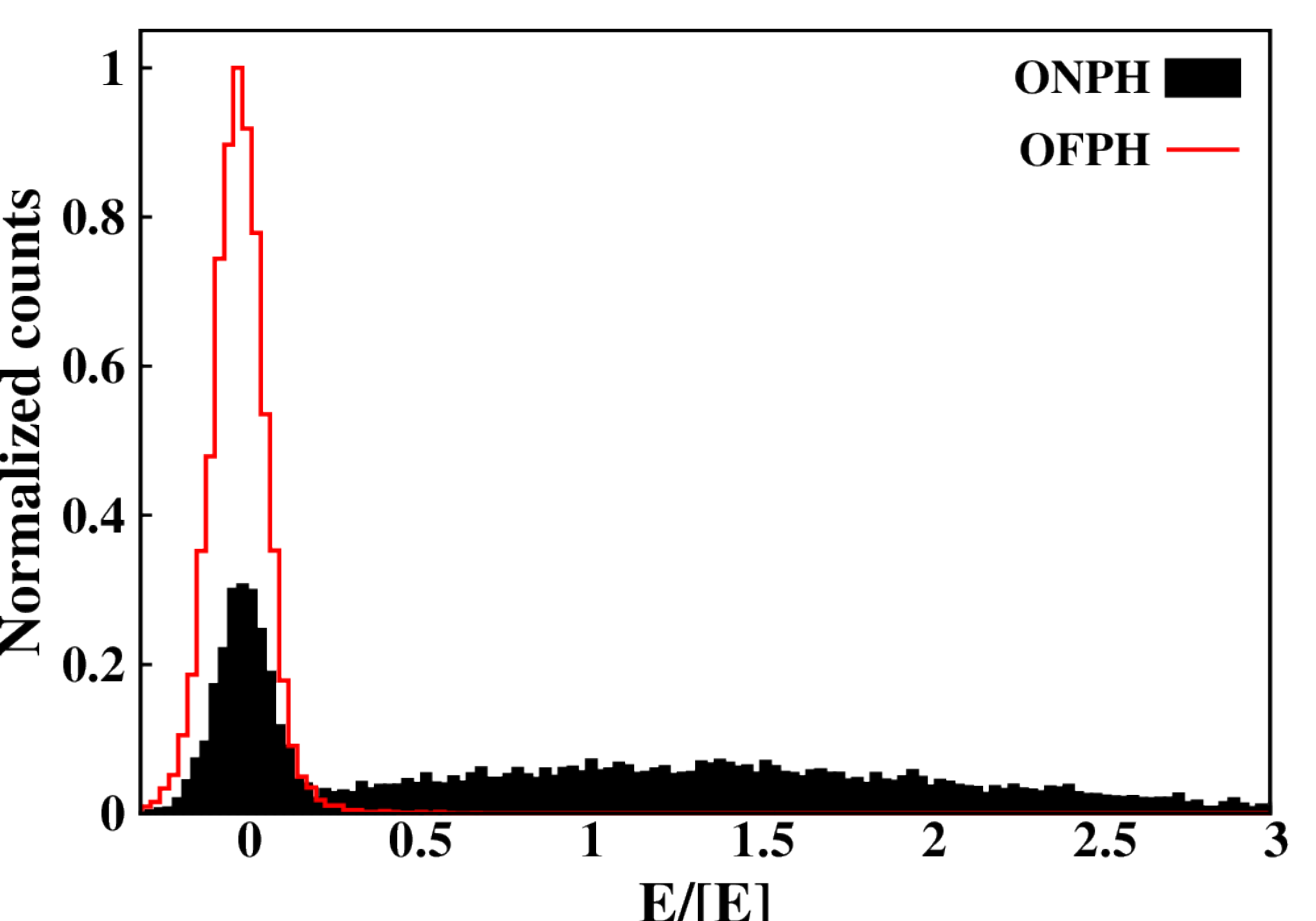}
 }
 \subfigure[]{
 \includegraphics[width=2.1 in,height=2.5 in,angle=-90,bb=59 53 555 750]{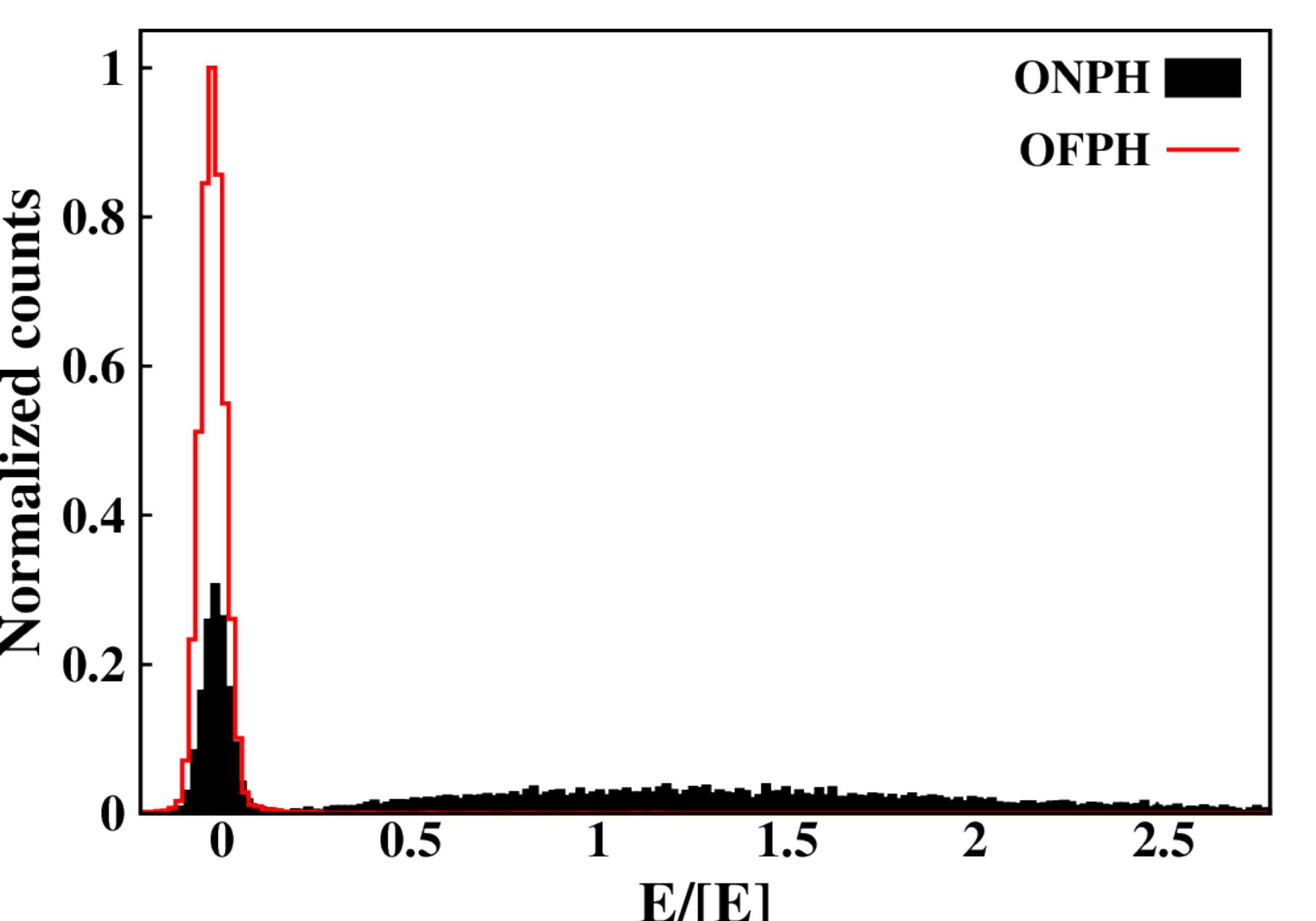}
 }
 \subfigure[]{
 \includegraphics[width=2.1 in,height=2.5 in,angle=-90,bb=59 53 555 750]{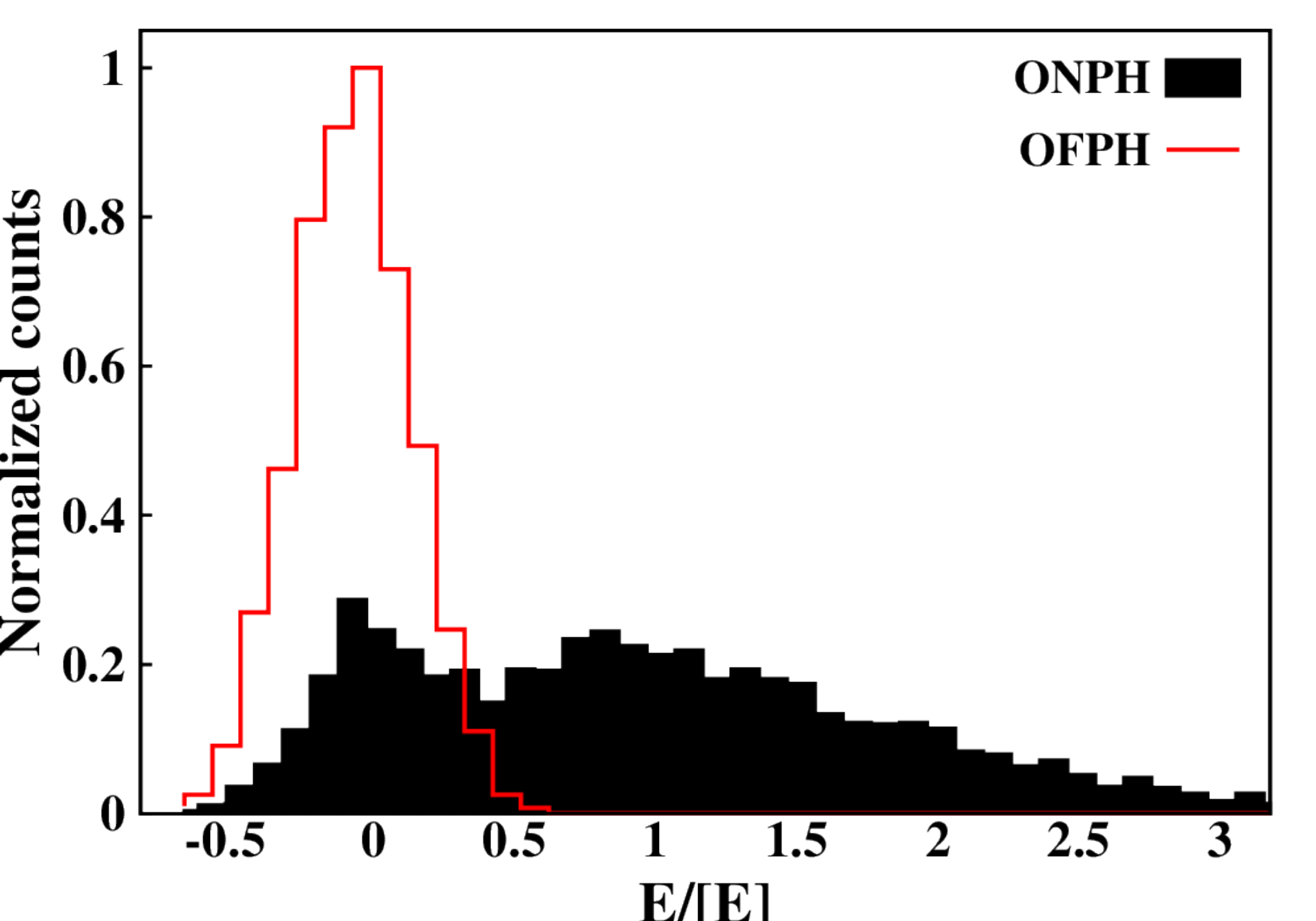}
 }
  \begin{picture}(0,0)
   \put(-255,130){\scriptsize \bf 313 MHz}
   \put(-255,120){\scriptsize \bf NF: 35$\pm$5\%}
   \put(-255,110){\scriptsize \bf N: 8788 pulses}
   \put(-70,130){\scriptsize \bf 607 MHz}
   \put(-70,120){\scriptsize \bf NF: 33$\pm$3\%}
   \put(-70,110){\scriptsize \bf N: 10230 pulses}
   \put(-255,-40){\scriptsize \bf 1380 MHz}
   \put(-255,-50){\scriptsize \bf NF: 31$\pm$2\%}
   \put(-255,-60){\scriptsize \bf N: 11707 pulses}
   \put(-70,-40){\scriptsize \bf 4850 MHz}
   \put(-70,-50){\scriptsize \bf NF: $>$31\%}
   \put(-70,-60){\scriptsize \bf N: 8016(3) pulses}
  \end{picture}
 \caption[The on-pulse and the off-pulse energy histograms at all four 
 frequencies for PSR B2319+60]{The on-pulse and the off-pulse energy 
 histograms at (a) 313 MHz, (b) 607 MHz, (c) 1380 MHz and (d) 4850 MHz for PSR B2319+60.   
 The abscissa presents normalized energy obtained using the block average (see Section \ref{NF_tech_sect}). 
 The ordinate presents normalized count of occurrences for each energy bin. 
 The OFPHs are shown with the red solid lines while the ONPHs are shown with black filled curves. 
 The counts in both the histograms were normalized by the peak from the corresponding OFPH histogram 
 at each frequency. The observed frequency, the NF and the
 number of pulses used during the analysis are displayed in the inset texts.}
 \label{b2319_hist_all_freq}
\end{figure}

A few sections of the data, specially at the lower frequencies, 
showed presence of RFI. These sections were removed at each frequency 
before estimating corresponding NFs. The obtained NFs at all 
four frequencies were compared as the first test to quantify 
the similarity of nulling behaviour. Figure \ref{b2319_hist_all_freq} 
shows the obtained ONPHs and OFPHs for all observed frequencies. 
Each ONPH, at the corresponding observed frequency, point towards 
a significant number of null pulses, presented as a 
separate distribution around the zero mean pulse 
energy. The method to obtain the NF is discussed 
with details in Section \ref{NF_tech_sect}. The number of pulses used 
during these estimation are indicated in the inset texts 
at the respective frequency plots. The ONPHs, obtained 
from the single pulses, at 313, 607 and 1380 MHz 
[i.e. Figure \ref{b2319_hist_all_freq}(a),(b) and (c)]
show clear bi-modal distributions of the pulse energy.  
Thus, we were able to obtain accurate estimate of the 
NFs at these lower frequencies. However, at 4850 MHz, 
as previously mentioned, the single pulse S/N 
was not sufficient to produce such clear bi-modal distribution 
in the ONPH. Hence, 3 consecutive pulses were integrated 
to obtain the ONPH and OFPH. Such integration, as discussed in 
Section \ref{NF_tech_sect}, will lead to a lower estimate of the NF as 
few null pulses will get mixed with the neighbouring burst pulses. 
Figure \ref{b2319_hist_all_freq}(d) shows a lower limit 
on the NF at 4850 MHz due to the sub-integrations. 
The obtained NF is also listed in the 
inset texts of Figure \ref{b2319_hist_all_freq} for each frequency. 
Figure \ref{b2319_hist_all_freq} demonstrate a good match between 
the obtained NFs within the error bars. Although the NF does show a good match 
across a decade of frequencies, as suggested in Section \ref{sect_null_comp}, 
it does not quantify nulling behaviour in full detail. 
Hence, it is essential to confirm pulse-to-pulse matching 
of the nulling behaviour in order to check if this phenomenon 
is truly broadband. 

\begin{figure}[h!]
 \centering
 \includegraphics[width=4 in,height=3 in,angle=0,bb=0 0 360 250]{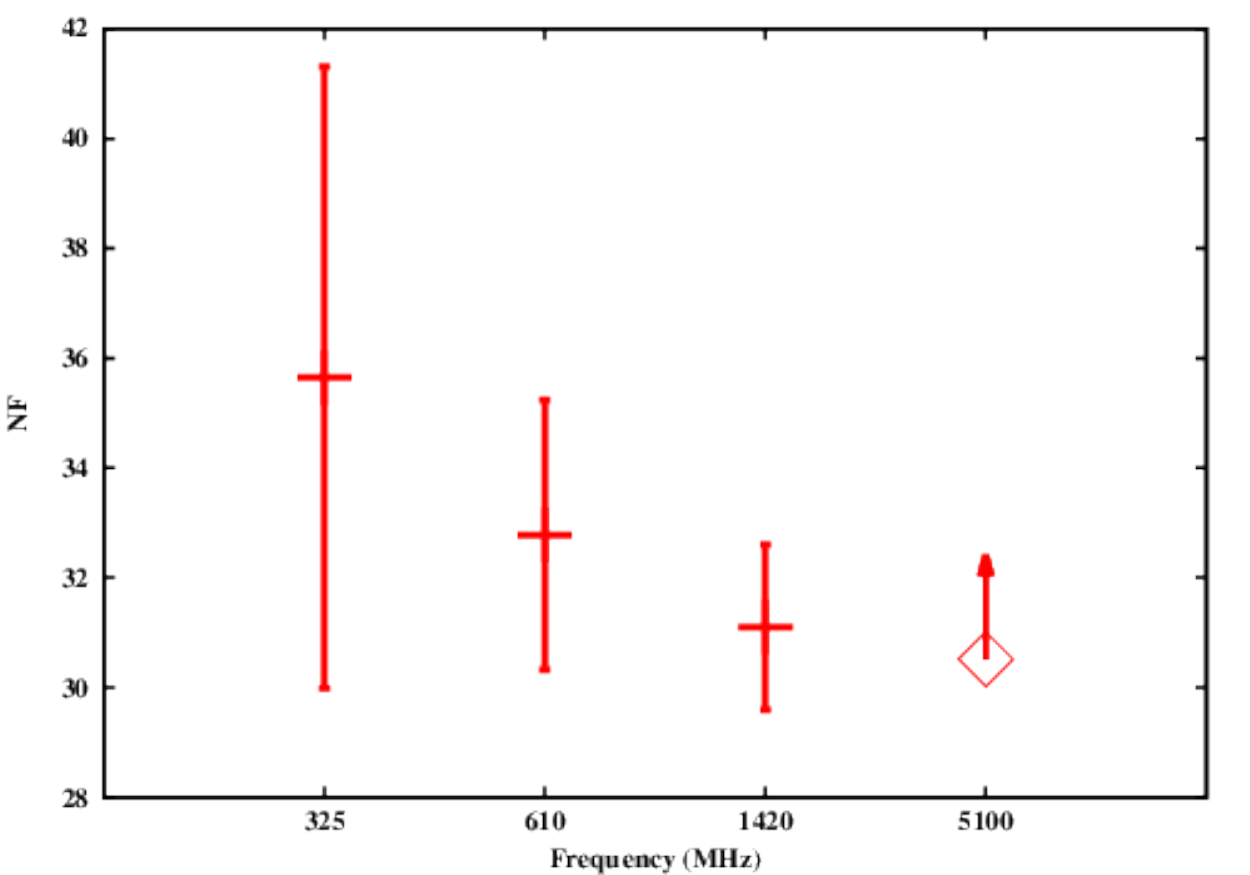}
 % NF_freq.b0809.forthesis.eps: 347x249 pixel, 72dpi, 12.24x8.78 cm, bb=0 0 347 249
 \caption{The obtained NFs at all four observed frequencies for PSR B2319+60.}
 \label{b2319_NF_all_freq}
\end{figure}
\subsection{One-bit sequence comparison}
\label{b2319_onebit_sect}
\begin{figure}[h!]
 \centering
 \includegraphics[width=5 in,height=3 in,angle=0,bb=0 0 360 250]{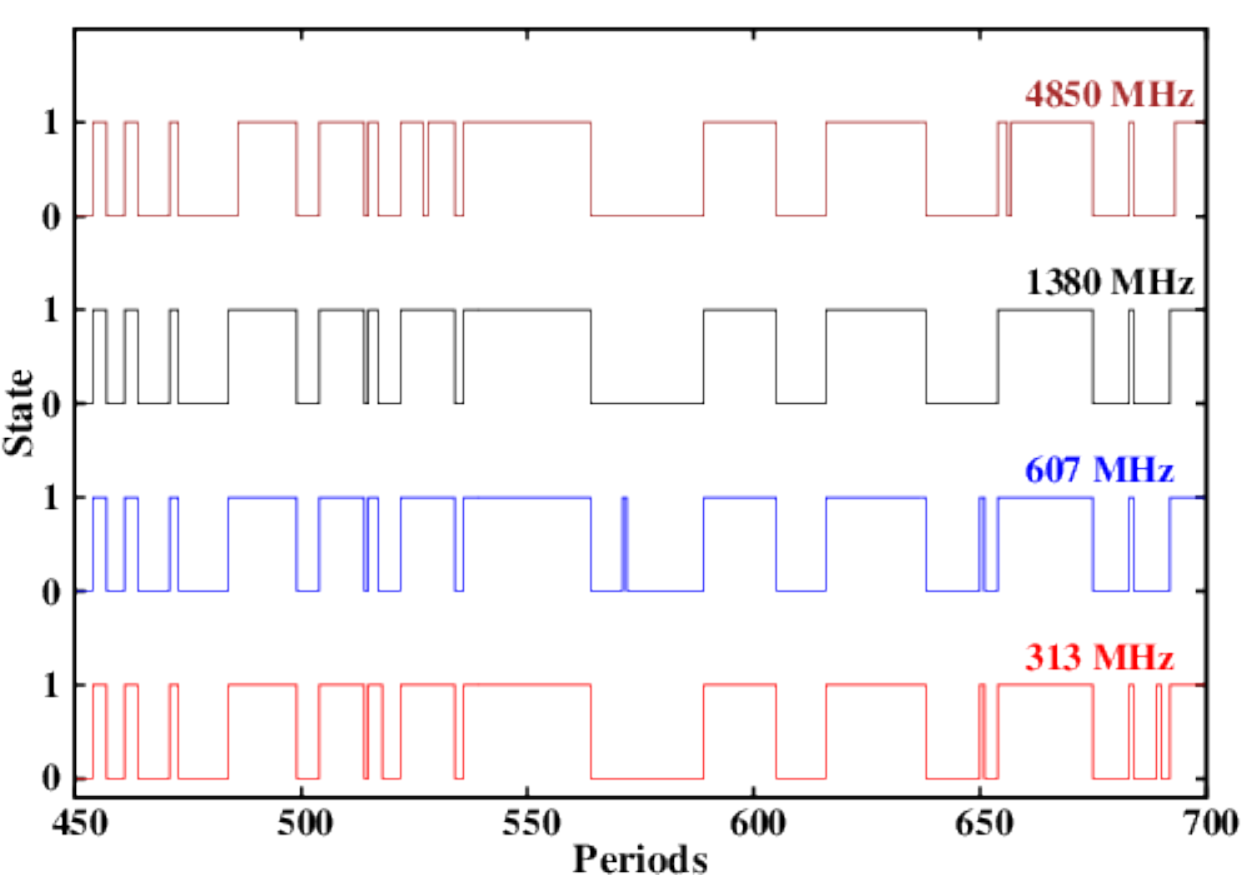}
 % b0809_onezero_all_freq.eps: 0x0 pixel, 300dpi, 0.00x0.00 cm, bb=50 50 410 302
 \caption[One-bit sequences at four frequencies for PSR B2319+60]{The identified one-zero states of 
 the pulsar emission at all four frequencies for PSR B2319+60.} 
 \label{b2319_onezero_all_freq}
\end{figure}

The null pulses and the burst pulses were identified and separated at all four frequencies. 
For the lower three frequencies, i.e. 313, 607 and 1380 MHz, a threshold was first 
set on the respective ONPH, where the null pulse distribution and the burst pulse 
distribution overlapped each other. Pulses below the threshold were 
tagged null pulses while pulses above the threshold were tagged 
as burst pulses. A visual inspection was carried out on the separated null 
and burst pulses to separate any misidentified pulses. At the 
highest observing frequency, 4850 MHz, single pulses were weak and hence 
such threshold was not possible to set to separate null and burst pulses. 
We first obtain the single pulse ONPH and OFPH which showed 
large fraction of null pulses due to the mixture of 
true null pulses with the weak burst pulses near the zero 
pulse energy. A slightly higher threshold was first 
set in such a way that it included all null pulses below it. 
Pulses above this higher threshold were all high S/N burst pulses, 
while pulse below the threshold presented a mixture. 
Pulses, only belonging to this mixture, were arranged in the ascending order of their 
on-pulse energy. A threshold was moved from the lower energy end 
towards the higher energy end till pulses below the threshold 
did not form a significant profile component (with S/N$\geq$3). 
All the pulses below the threshold were tagged as true null pulses 
while pulses above the threshold were tagged as burst pulses (see Section \ref{separation_of_null_burst_sect}). 
All the separated pulses were carefully 
examined visually to eliminate any possibility of misidentification. 
Using these separated null and burst pulses, one-bit sequences 
were formed at all observed frequencies, in which zero represents 
a null and one represents a burst. Figure \ref{b2319_onezero_all_freq} 
shows a section of around 350 pulses with their identified 
emission states at all frequencies. 

The one-bit sequences for a pair of observing 
frequencies were compared using a contingency table analysis 
\cite{pftv86}, as a second test. A visual comparison 
between various frequencies clearly confirms 
that pulsar shows simultaneous switching between the null and the burst 
states at all frequencies. To quantify this simultaneity, the 
correlation between these states for a pair of frequencies can be obtain, 
as a 2 $\times$ 2 contingency table (see \cite{pftv86} and 
section \ref{b0809_onebit_sect} for details). 
Table \ref{b2319_cont_table_all_freq} shows these 
different contingency tables for all frequency pairs. 
It can be seen from these tables that around 14 to 77 
pulses show a mismatch between different pairs of frequencies. 
For example, a comparison of one-bit sequence 
between 313 and 1380 MHz shows around 21 pulses for which 
the pulsar showed absence of emission at 313 MHz 
but it showed burst emission at 1380 MHz 
(see Table \ref{b0809_cont_table_all_freq}). 
Similarly, for 14 pulses, the pulsar showed absence of 
emission at 1380 MHz while it was in a burst state at 313 MHz. 
Total number of pulses which were compared for each pair were 
around 5126 pulse, hence these numbers presents a small fraction 
of nonconcurrence. Hence, it appears that PSR B2319+60 exhibits similar 
emission states at all pairs of frequencies 
for the duration of our observations. 
\begin{table}[h!]
\centering
\begin{subtable}{}
\centering
\begin{tabular}{cc|c|c|}
\cline{3-4}
 &         & \multicolumn{2}{c|}{607} \\
 \cline{3-4} 
 &         & Null & Burst \\
\cline{1-4}
\multicolumn{1}{|c|}{\multirow{2}{*}[-0.1cm]{\rotatebox{90}{313}}} & Null & 1524  & 39 \\
\cline{2-4}     
\multicolumn{1}{|c|}{} & Burst& 14  & 3532 \\
\tableline
\end{tabular}  
\end{subtable}
\vspace{0.1cm}
\begin{subtable}{}
\centering
\begin{tabular}{cc|c|c|}
\cline{3-4}
 &         & \multicolumn{2}{c|}{1380} \\
 \cline{3-4} 
 &         & Null & Burst \\
\cline{1-4}
\multicolumn{1}{|c|}{\multirow{2}{*}[-0.1cm]{\rotatebox{90}{313}}} & Null & 1545  & 21 \\
\cline{2-4}     
\multicolumn{1}{|c|}{} & Burst& 14 & 3522 \\
\tableline
\end{tabular}  
\end{subtable}
\vspace{1 cm}
\begin{subtable}{}
\centering
\begin{tabular}{cc|c|c|}
\cline{3-4}
 &         & \multicolumn{2}{c|}{4850} \\
 \cline{3-4} 
 &         & Null & Burst \\
\cline{1-4}
\multicolumn{1}{|c|}{\multirow{2}{*}[-0.1cm]{\rotatebox{90}{313}}} & Null & 1492  & 60 \\
\cline{2-4}     
\multicolumn{1}{|c|}{} & Burst& 70 & 3461 \\
\tableline
\end{tabular}  
\end{subtable}
\vspace{0.1cm}
\begin{subtable}{}
\centering
\begin{tabular}{cc|c|c|}
\cline{3-4}
 &         & \multicolumn{2}{c|}{1380} \\
 \cline{3-4} 
 &         & Null & Burst \\
\cline{1-4}
\multicolumn{1}{|c|}{\multirow{2}{*}[-0.1cm]{\rotatebox{90}{607}}} & Null & 1526  & 14 \\
\cline{2-4}     
\multicolumn{1}{|c|}{} & Burst & 21  & 3544 \\
\tableline
\end{tabular}  
\end{subtable}
\vspace{1 cm}
\begin{subtable}{}
\centering
\begin{tabular}{cc|c|c|}
\cline{3-4}
 &         & \multicolumn{2}{c|}{4850} \\
 \cline{3-4} 
 &         & Null & Burst \\
\cline{1-4}
\multicolumn{1}{|c|}{\multirow{2}{*}[-0.1cm]{\rotatebox{90}{607}}} & Null & 1485  & 42 \\
\cline{2-4}     
\multicolumn{1}{|c|}{} & Burst & 77  & 3484 \\
\tableline
\end{tabular}  
\end{subtable}
\vspace{0.1cm}
\begin{subtable}{}
\centering
\begin{tabular}{cc|c|c|}
\cline{3-4}
 &         & \multicolumn{2}{c|}{4850} \\
 \cline{3-4} 
 &         & Null & Burst \\
\cline{1-4}
\multicolumn{1}{|c|}{\multirow{2}{*}[-0.1cm]{\rotatebox{90}{1380}}} & Null & 1508  & 37 \\
\cline{2-4}     
\multicolumn{1}{|c|}{} & Burst & 65 & 3487 \\
\tableline
\end{tabular}  
\end{subtable}
\caption[Contingency tables for different pairs of frequency for PSR B2319+60]
{Contingency tables for different pairs of frequency for PSR B2319+60. 
Each frequency is shown with two corresponding emission states (i.e. the null 
state and the burst state). A 2 $\times$ 2 matrix 
for a given pair of frequencies displays number of pulses 
in four possible conditions. It should be noted that total number of pulses 
are not similar for different pairs as a few weak pulses were further avoided, 
only for certain frequency pairs, where a decision regarding their true 
emission state was not possible to make.}
\label{b2319_cont_table_all_freq}
\end{table}

A $\phi$ test (Cramer-V) and uncertainty test based on entropy calculations can then be 
used to assess the strength and significance of these correlations. 
These tests are briefly discussed in Section \ref{b0809_onebit_sect}.  
See \cite{pftv86} for details about these tests. 
The results of these tests are presented in Table \ref{b2319_cramv_all_freq}. 
Both Cramer-V and the uncertainty coefficients have values very close to 1,  
indicating a significantly high association across all 
pairs of frequencies. However, the strength is marginally smaller 
for association between pairs involving 4850 MHz. 
\begin{table}[h!]
\begin{center}
\begin{tabular}{|l||c|c|c|}
\tableline
Frequenc & 607  & 1380 & 4850  \\
(MHz)    &      &      &       \\
\tableline
\tableline
313  & 0.98 & 0.98 & 0.94 \\
     & 0.91 & 0.93 & 0.81 \\
607  & $-$ & 0.98 & 0.94 \\
     & $-$ & 0.94  & 0.83 \\ 
1380 &     & $-$   & 0.95  \\
     &     & $-$   & 0.85 \\
\tableline
\end{tabular}
\end{center}
\caption[Estimate of correlation strength and significance for 
one-bit sequences between a pair of frequencies for PSR B2319+60]
{Estimate of correlation strength and significance for 
one-bit sequences between a pair of frequencies for PSR B2319+60. 
The first row in each column for a given 
frequency gives the Cramer$-$V indicating the strength of correlation 
of one-bit sequence associated with the frequency in the column. 
Similarly, the second row in each column for a given frequency gives the 
corresponding uncertainty coefficient derived from the entropy 
arguments (See \cite{pftv86})}
\label{tabcont}
\label{b2319_cramv_all_freq}
\end{table}
\subsection{Nonconcurrent pulses}
\label{b2319_mismatch_sect}
As can be seen from Table \ref{b2319_cont_table_all_freq}, 
for a small number of pulses, the above association 
does not hold. In a comparison across all
four frequencies, about 158 out of 5126 
pulses ($\sim$ 3\%) do not show concurrent null (or burst) 
state for PSR B2319+60. All such pulses were carefully 
examined to verify their true nature and identify 
their locations. Surprisingly, 82 of these nonconcurrent  
pulses ($\sim$52\% nonconcurrent pulses) occurred either at the start 
or at the end of a burst. A few examples of such nonconcurrent pulses are shown 
in Figures \ref{b2319_sp_mismatch1}, \ref{b2319_sp_mismatch2} and 
\ref{b2319_sp_mismatch3}. As shown in Figure \ref{b2319_sp_mismatch1}, 
pulsar showed absence of emission at 313 and 607 MHz [Figure \ref{b2319_sp_mismatch1}(a) and (b)]
while weak and narrow burst pulses can be seen at 1380 and 4850 MHz 
[Figure \ref{b2319_sp_mismatch1}(c) and (d)].
Figure \ref{b2319_sp_mismatch2} shows example of a pulse 
that is seen to produce detectable emission at 313, 607 and 1380 MHz,
[Figures \ref{b2319_sp_mismatch2}(a),(b) and (c)] while 
no detectable emission is present at 4850 MHz. Similarly, 
during a burst to null transition, as shown in Figure \ref{b2319_sp_mismatch3}, 
a narrow weak pulse can be seen at 1380 MHz [Figure \ref{b2319_sp_mismatch3}(c)] 
while no detectable emission is present at 313, 607 and 4850 MHz [Figure \ref{b2319_sp_mismatch3}(a), (b) and 
(d)]. 

\begin{figure}[h!]
 \vspace{0.2cm}
 \centering
 \subfigure[]{
 \includegraphics[width=2.6 in,height=1.1 in,angle=-90,bb=0 0 496 187]{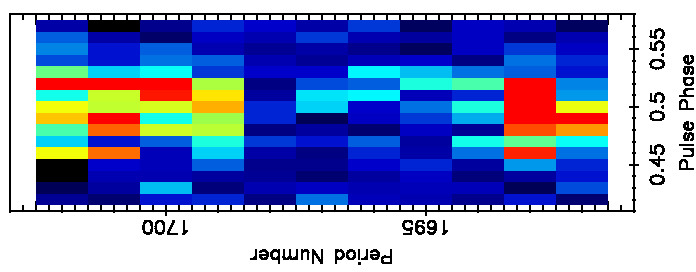}
 % spdisplay_example1.325.eps: 494x183 pixel, 72dpi, 17.43x6.46 cm, bb=52 583 546 766
 }
 \subfigure[]{
 \includegraphics[width=2.6 in,height=1.1 in,angle=-90,bb=0 0 496 187]{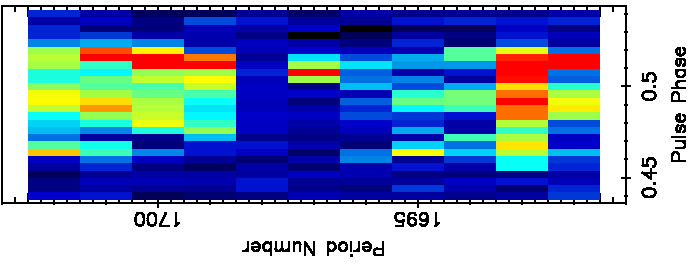}
 }
 \subfigure[]{
 \includegraphics[width=2.6 in,height=1.1 in,angle=-90,bb=0 0 496 187]{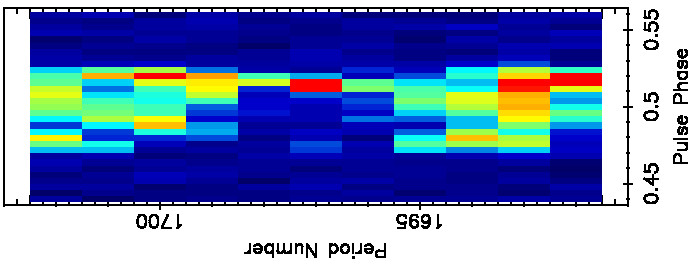}
 % spdisplay_example1.325.eps: 494x183 pixel, 72dpi, 17.43x6.46 cm, bb=52 583 546 766
 }
 \subfigure[]{
 \includegraphics[width=2.6 in,height=1.1 in,angle=-90,bb=0 0 496 187]{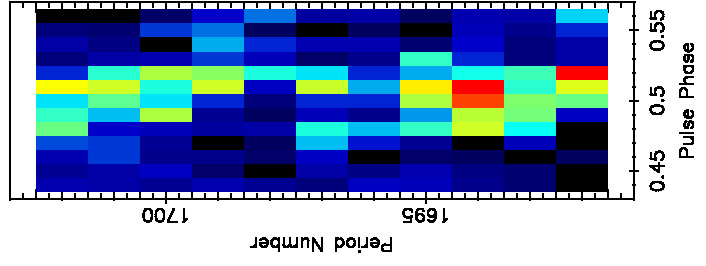}
 }
 \begin{picture}(0,0)
  \put(-320,3){\footnotesize 313 MHz}
  \put(-235,3){\footnotesize 607 MHz}
  \put(-147,3){\footnotesize 1380 MHz}
  \put(-57,3){\footnotesize 4850 MHz}
 \end{picture}
 \caption[First example of simultaneously observed sequence of pulses at four frequencies for PSR B2319+60]
 {Sections of simultaneously observed sequence of pulses at 
 (a) 313 MHz, (b) 607 MHz, (c) 1380 MHz and (d) 4850 MHz for PSR B2319+60. 
 At the pulse number 1698, pulsar shows clear null pulses at 313 and 607 MHz 
 while narrow burst pulses can be seen at 1380 and 4850 MHz.}
 \label{b2319_sp_mismatch1}
\end{figure}
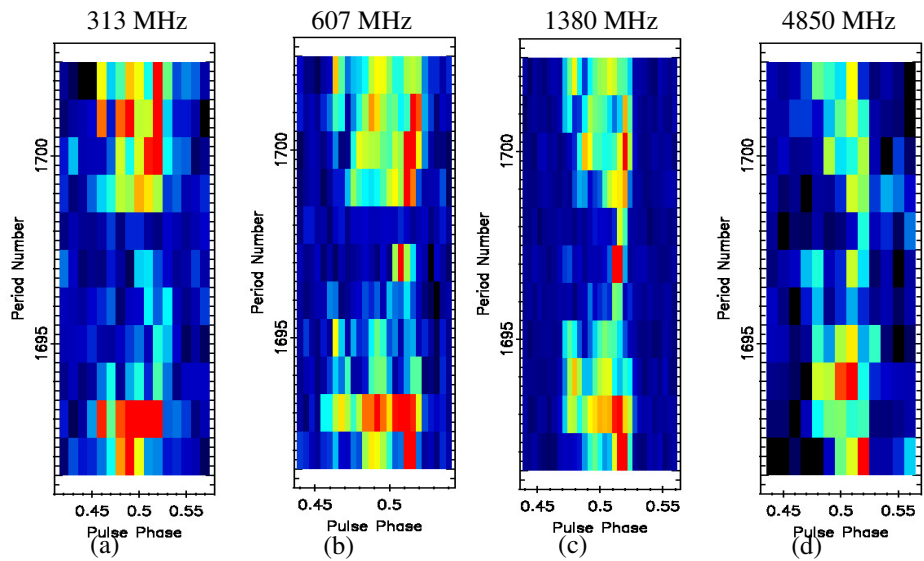
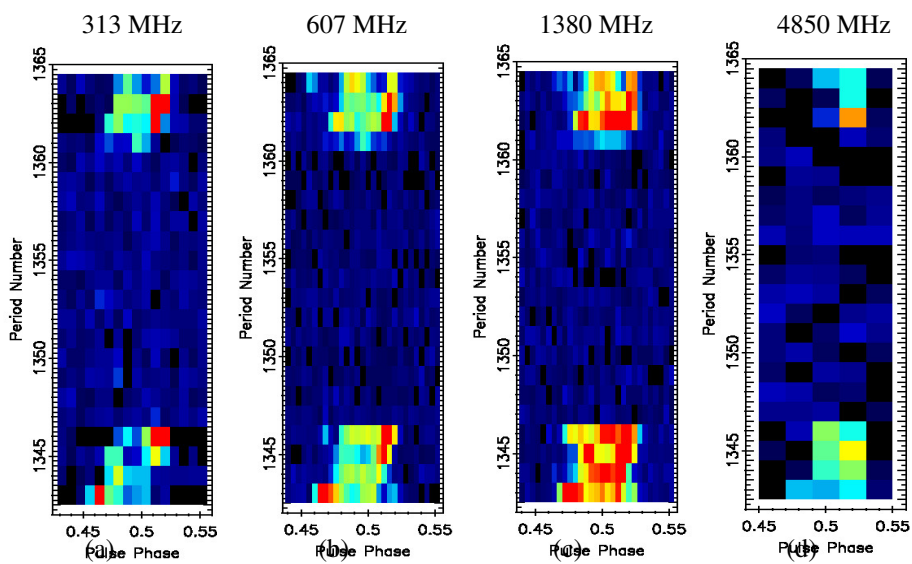
\begin{figure}[h!]
 \vspace{0.2cm}
 \centering
 \subfigure[]{
 \includegraphics[width=2.6 in,height=1.1 in,angle=-90,bb=0 0 496 187]{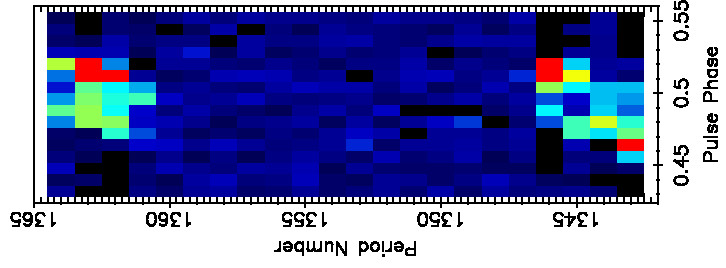}
 % spdisplay_example1.325.eps: 494x183 pixel, 72dpi, 17.43x6.46 cm, bb=52 583 546 766
 }
 \subfigure[]{
 \includegraphics[width=2.6 in,height=1.1 in,angle=-90,bb=0 0 496 187]{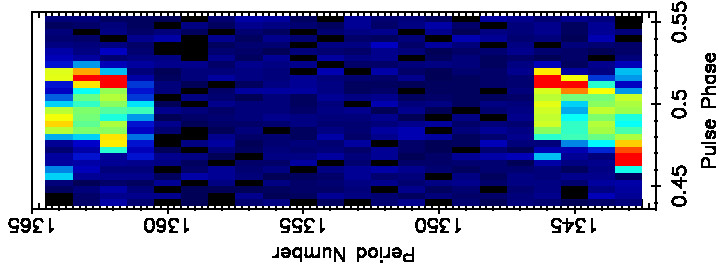}
 }
 \subfigure[]{
 \includegraphics[width=2.6 in,height=1.1 in,angle=-90,bb=0 0 496 187]{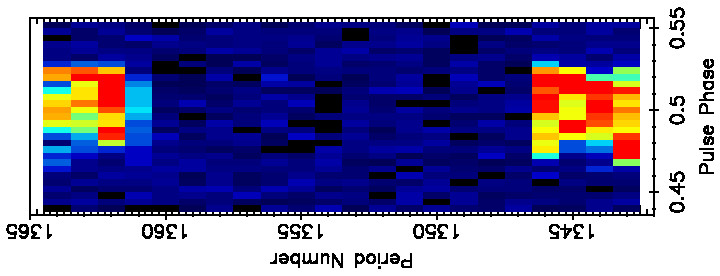}
 % spdisplay_example1.325.eps: 494x183 pixel, 72dpi, 17.43x6.46 cm, bb=52 583 546 766
 }
 \subfigure[]{
 \includegraphics[width=2.6 in,height=1.1 in,angle=-90,bb=0 0 496 187]{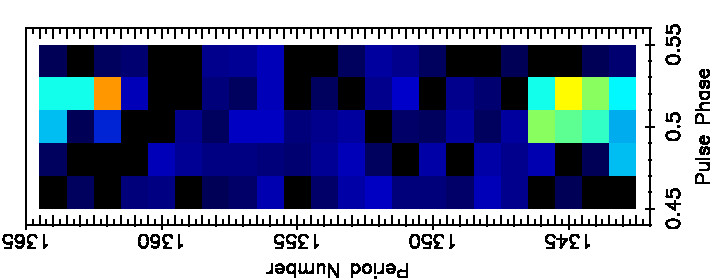}
 }
 \begin{picture}(0,0)
  \put(-320,3){\footnotesize 313 MHz}
  \put(-235,3){\footnotesize 607 MHz}
  \put(-147,3){\footnotesize 1380 MHz}
  \put(-57,3){\footnotesize 4850 MHz}
 \end{picture}
 \caption[Second example of simultaneously observed sequence of pulses at four frequencies for PSR B2319+60]
 {Sections of simultaneously observed sequence of pulses at 
 (a) 313 MHz, (b) 607 MHz, (c) 1380 MHz and (d) 4850 MHz for PSR B2319+60. 
 At the pulse number 1361, pulsar shows clear null pulse at 4850 MHz, while weak burst 
 pulses can be seen 313, 607 and 1380 MHz. However, such descripencies are likely to come from 
 low S/N at 4850 MHz.}
 \label{b2319_sp_mismatch2}
\end{figure}

\begin{figure}[h!]
 \vspace{0.2cm}
 \centering
 \subfigure[]{
 \includegraphics[width=2.6 in,height=1.1 in,angle=-90,bb=0 0 496 187]{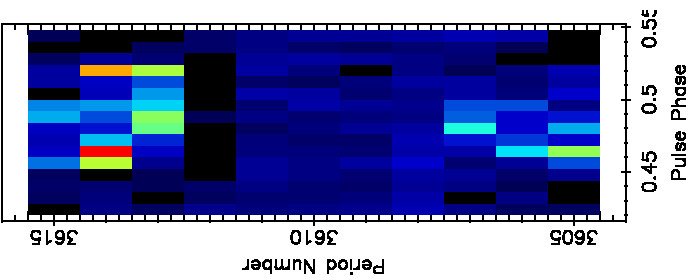}
 % spdisplay_example1.325.eps: 494x183 pixel, 72dpi, 17.43x6.46 cm, bb=52 583 546 766
 }
 \subfigure[]{
 \includegraphics[width=2.6 in,height=1.1 in,angle=-90,bb=0 0 496 187]{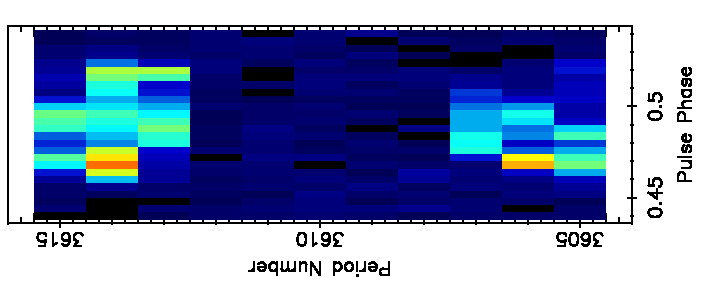}
 }
 \subfigure[]{
 \includegraphics[width=2.6 in,height=1.1 in,angle=-90,bb=0 0 496 187]{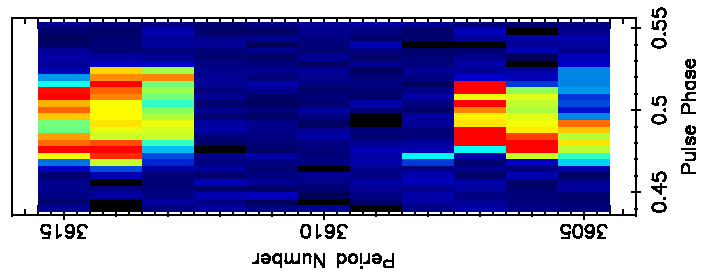}
 % spdisplay_example1.325.eps: 494x183 pixel, 72dpi, 17.43x6.46 cm, bb=52 583 546 766
 }
 \subfigure[]{
 \includegraphics[width=2.6 in,height=1.1 in,angle=-90,bb=0 0 496 187]{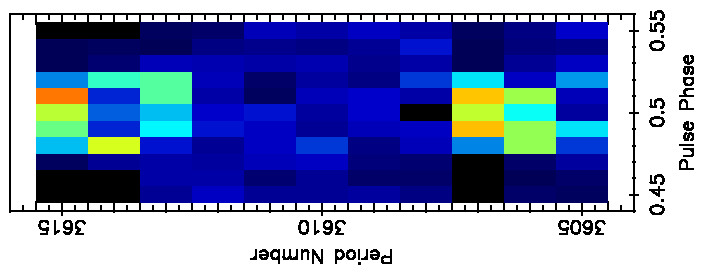}
 }
 \begin{picture}(0,0)
  \put(-320,3){\footnotesize 313 MHz}
  \put(-235,3){\footnotesize 607 MHz}
  \put(-147,3){\footnotesize 1380 MHz}
  \put(-57,3){\footnotesize 4850 MHz}
 \end{picture}
 \caption[Third example of simultaneously observed sequence of pulses at four frequencies for PSR B2319+60]
 {Sections of simultaneously observed sequence of pulses at 
 (a) 313 MHz, (b) 607 MHz, (c) 1380 MHz and (d) 4850 MHz for PSR B2319+60. 
 At the pulse number 3608, pulsar shows clear null pulses 
 at 313, 607 and 4850 MHz while a narrow burst pulse can be seen at 1380 MHz.}
 \label{b2319_sp_mismatch3}
\end{figure}
\begin{figure}[h!]
 \vspace{0.8cm}
 \centering
 \subfigure[]{
 \includegraphics[width=2 in,height=2 in,angle=0,bb=0 0 360 250]{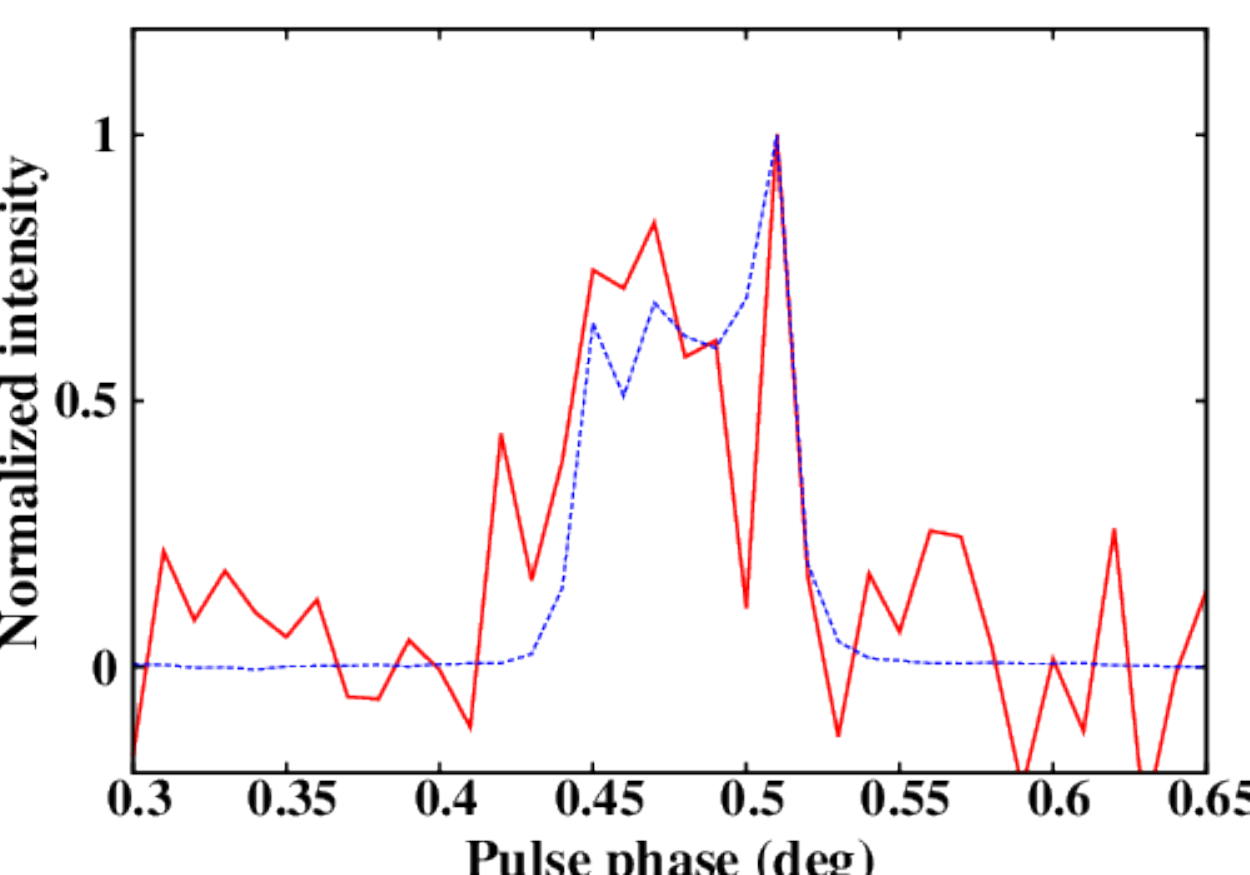}
 % Exclusive_325_burst.eps.pdf: 0x0 pixel, 300dpi, 0.00x0.00 cm, bb=50 50 410 302
 }
 \subfigure[]{
 \includegraphics[width=2 in,height=2 in,angle=0,bb=0 0 360 250]{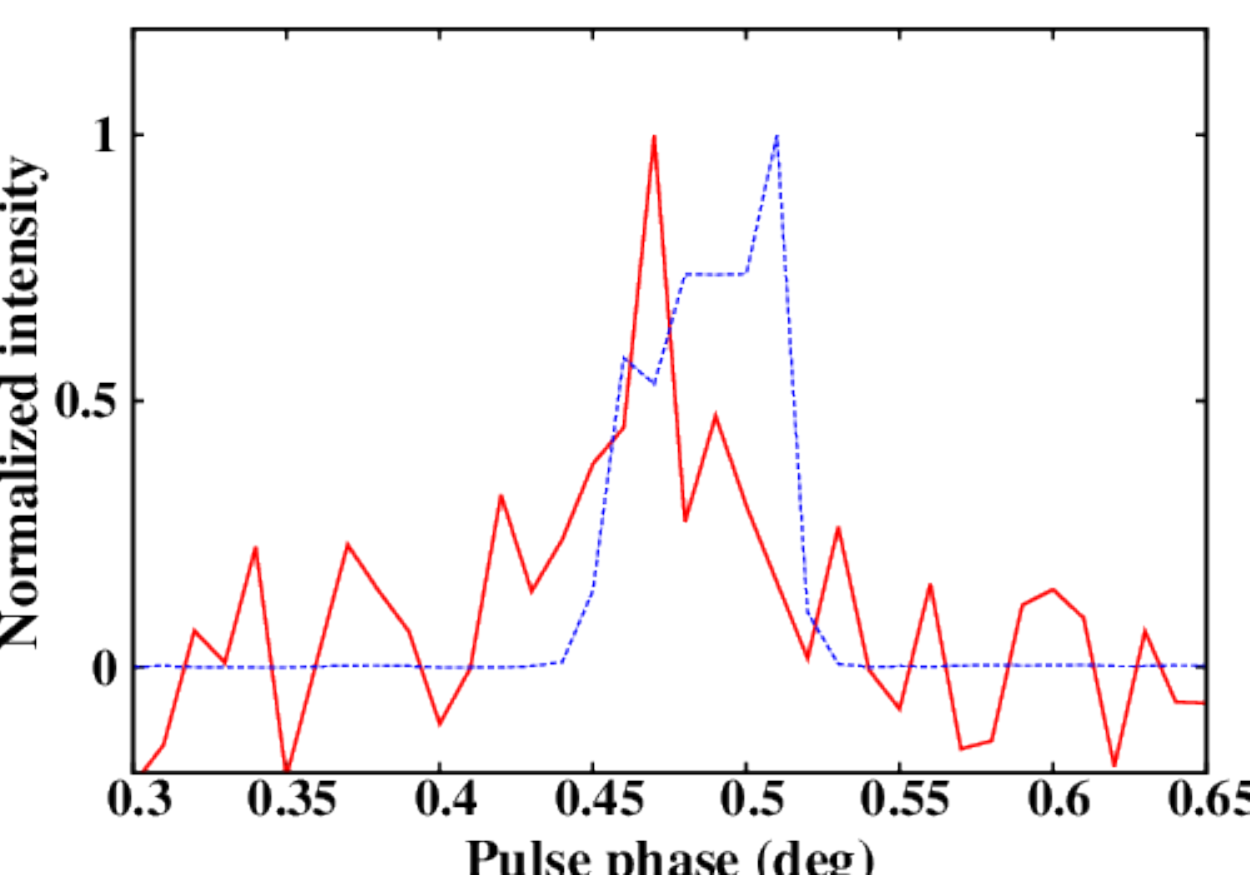}
 % Exclusive_325_burst.eps.pdf: 0x0 pixel, 300dpi, 0.00x0.00 cm, bb=50 50 410 302
 }
 \subfigure[]{
 \includegraphics[width=2 in,height=2 in,angle=0,bb=0 0 360 250]{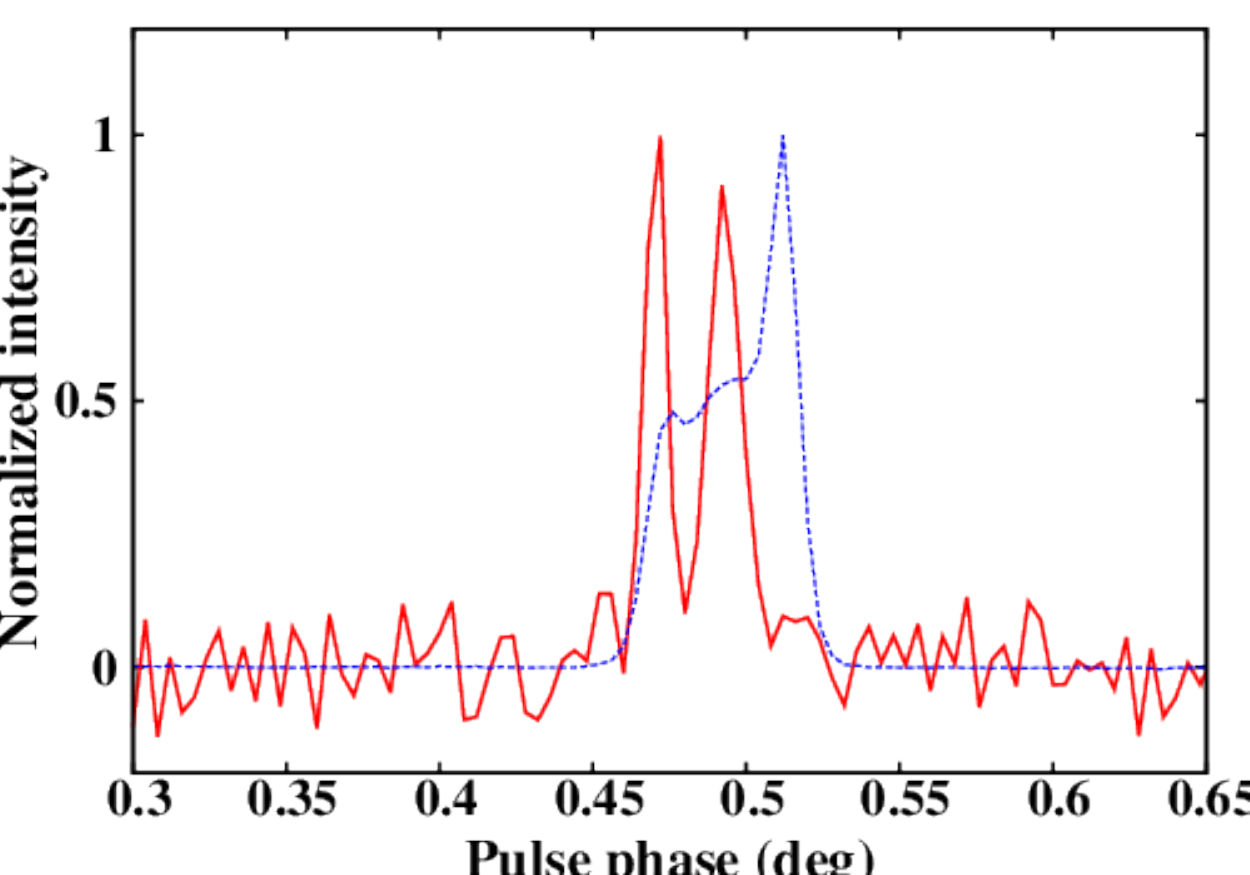}
 % Exclusive_325_burst.eps.pdf: 0x0 pixel, 300dpi, 0.00x0.00 cm, bb=50 50 410 302
 }
 \subfigure[]{
 \includegraphics[width=2 in,height=2 in,angle=0,bb=0 0 360 250]{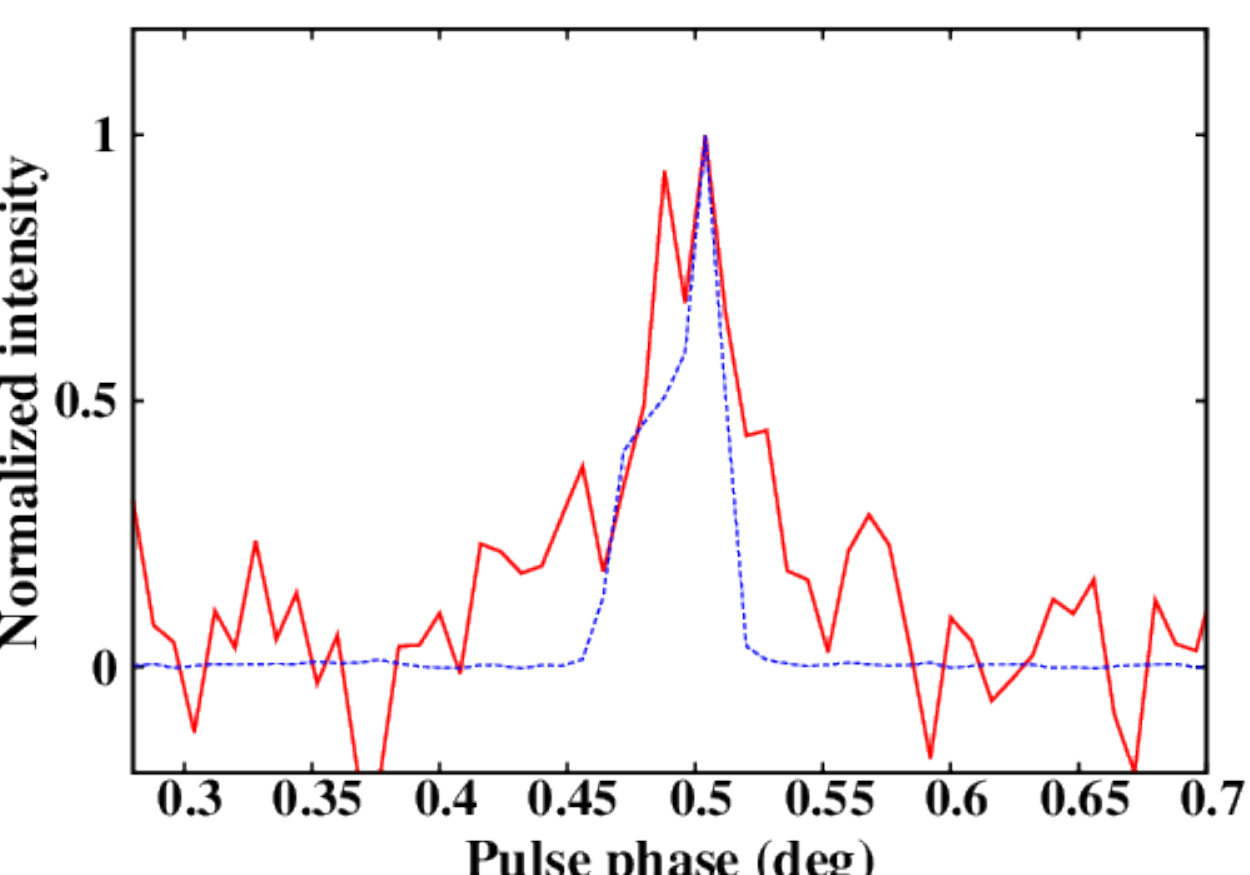}
 % Exclusive_325_burst.eps.pdf: 0x0 pixel, 300dpi, 0.00x0.00 cm, bb=50 50 410 302
 }
 \begin{picture}(0,0)
  \put(-280,320){\bf\footnotesize 313 MHz}
  \put(-280,310){\bf\footnotesize (13 pulses)}
  \put(-130,320){\bf\footnotesize 607 MHz}
  \put(-130,310){\bf\footnotesize (13 pulses)}
  \put(-280,120){\bf\footnotesize 1380 MHz}
  \put(-280,110){\bf\footnotesize (7 pulses)}
  \put(-130,120){\bf\footnotesize 4580 MHz}
  \put(-130,110){\bf\footnotesize (12 pulses)}
 \end{picture}
 \caption[Exclusive nonconcurrent burst pulses from B2319+60]
 {Exclusive nonconcurrent burst pulses at 
 (a) 313 MHz, (b) 607 MHz, (c) 1380 MHz and (d) 4850 MHz for PSR B2319+60. 
 For each frequency, all the burst pulses (quoted in the corresponding inset texts), 
 occuring only at a given single frequency, were combined and the average profile is 
 shown with a red solid line. The blue dotted line represents  
 the integrated profile at the corresponding frequency.}
 \label{b2319_mismatch_profiles}
\end{figure}
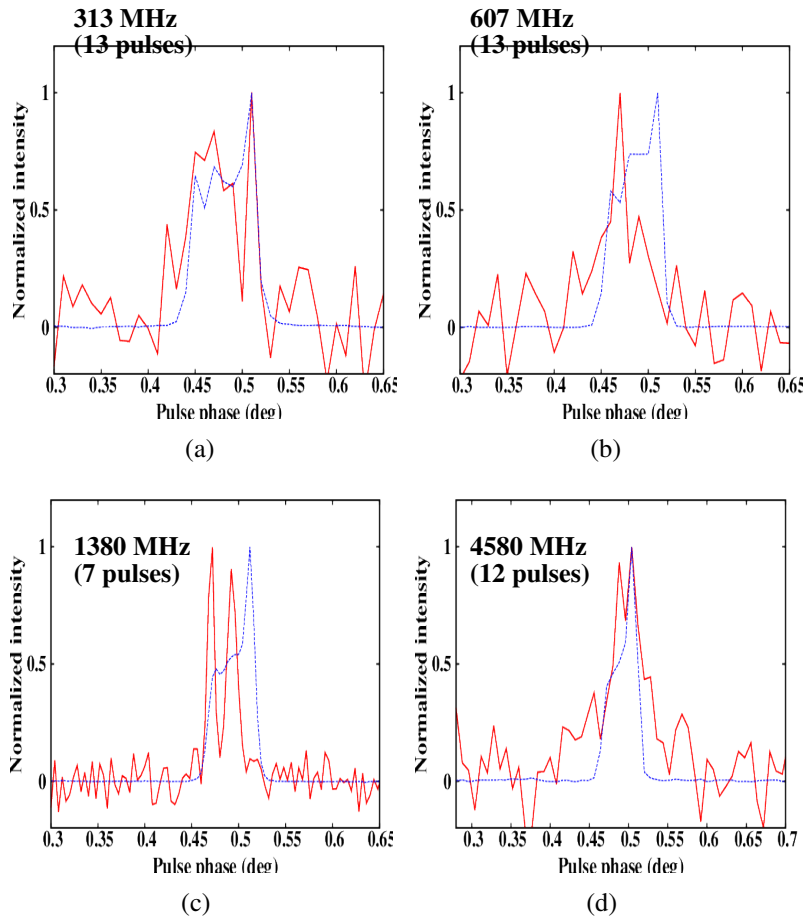

As it is evident from Figures \ref{b2319_sp_mismatch1}, \ref{b2319_sp_mismatch2} and 
\ref{b2319_sp_mismatch3}, when pulsar displays nonconcurrence in the 
emission state across different frequencies, 
the burst emission is likely to be weak (also narrow on a few occasions). 
To scrutinize this behaviour further, we 
identified and separated, all the {\itshape exclusive} 
nonconcurrent burst pulses (similar to those 
mentioned in Section \ref{b0809_mismatch_sect} for PSR B0809+74). 
These pulses were combined and the average 
profile was obtained at each frequency. 
These profiles are shown in Figure \ref{b2319_mismatch_profiles} 
from the exclusive pulses at all frequencies. 
Approximately 7 to 13 pulses were averaged to obtain 
these profiles. It should be noted that these pulses are only 
seen to occur at a given observed frequency while they show absence of 
emission at all other frequencies. Thus, they are smaller in 
numbers compared to number of pulses given in Table \ref{b2319_cont_table_all_freq} 
for different pairs. The peak S/N of these profiles are relatively low  
(i.e. peak S/N of 4.6, 7, 15 and 7.5 at 313, 607, 1380 and 4850 MHz respectively)
compared to PSR B0809+74. 
% However, they are significant to investigate 
% these emission differences. Figure \ref{b2319_mismatch_profiles} also shows 
% comparisons of exclusive nonconcurrent pulse profiles with the integrated profiles. 
% It can be noted from this comparison that nonconcurrence emission is most likely to occur 
% towards the leading edge of the pulse profile. 
% For example, most significant difference seen in Figure \ref{b2319_mismatch_profiles}
% is for the exclusive pulse profile at 1380 MHz [Figure \ref{b2319_mismatch_profiles}(c)]  
% which showed relatively strong emission at the leading edge.  
% This can also be confirmed by Figure \ref{b2319_sp_mismatch3}(c), 
% where an exclusive pulse at 1380 MHz is located near the 
% leading edge of the pulse. 
% However, exclusive pulses 
% seen at 1380 MHz are extremely narrow. They are 
% likely to be below the detection limits at other frequencies. 
% However, it is also possible to speculate that pulsar 
% exhibits mode-changing phenomena, during these exclusive or non-exclusive  
% nonconcurrent pulses. However, this is not possible to confirm, 
% due to their small numbers (similar to the case of PSR B0809+74). 
% 
\subsection{Null length and burst length comparison}
\begin{figure}[h!]
 \centering
 \subfigure[]{
 \includegraphics[width=2.1 in,height=2 in,angle=0,bb=0 0 360 250]{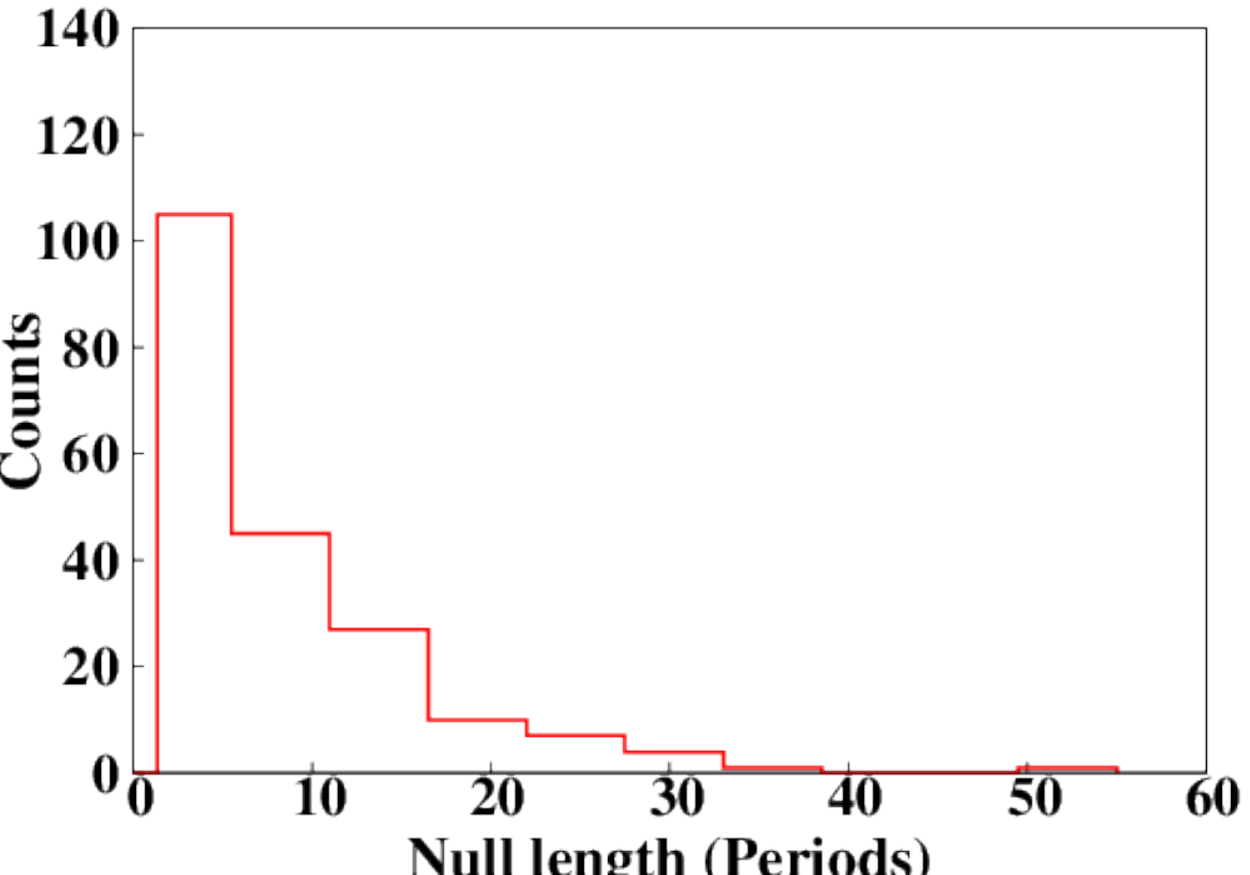}
 % b0809.nlh.325.eps.pdf: 360x252 pixel, 72dpi, 12.70x8.89 cm, bb=0 0 360 252
 }
 \subfigure[]{
 \includegraphics[width=2.1 in,height=2 in,angle=0,bb=0 0 360 250]{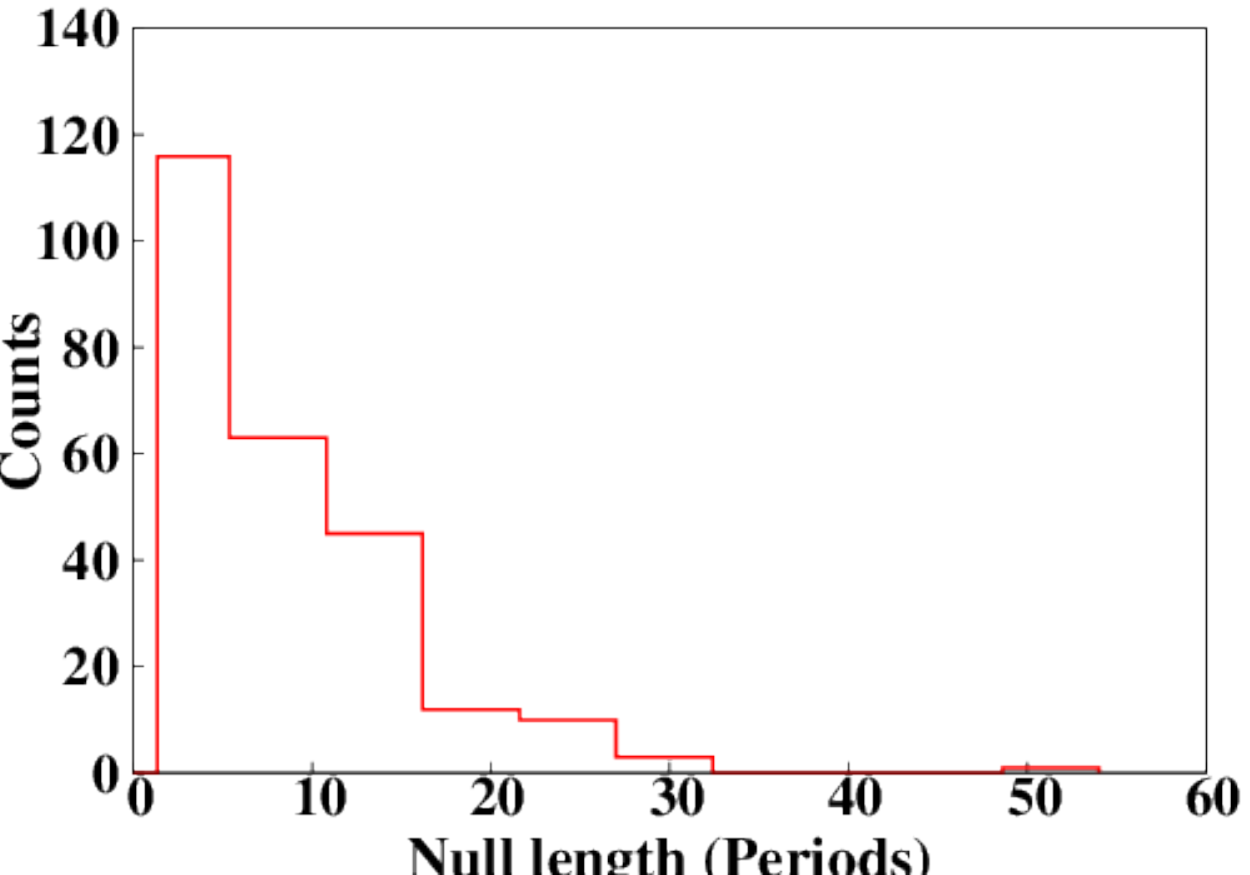}
 % b0809.nlh.325.eps.pdf: 360x252 pixel, 72dpi, 12.70x8.89 cm, bb=0 0 360 252
 }
 \subfigure[]{
 \includegraphics[width=2.1 in,height=2 in,angle=0,bb=0 0 360 250]{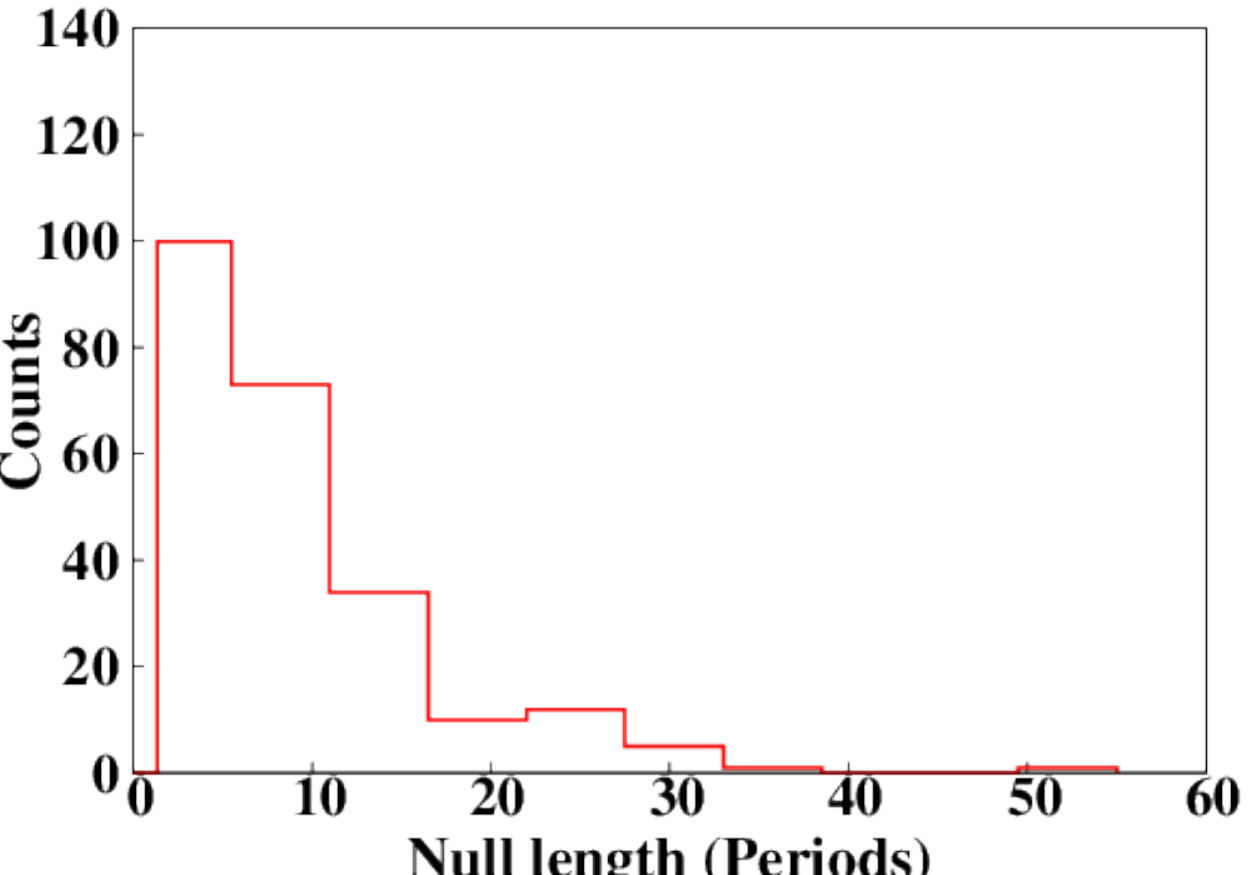}
 % b0809.nlh.325.eps.pdf: 360x252 pixel, 72dpi, 12.70x8.89 cm, bb=0 0 360 252
 }
 \subfigure[]{
 \includegraphics[width=2.1 in,height=2 in,angle=0,bb=0 0 360 250]{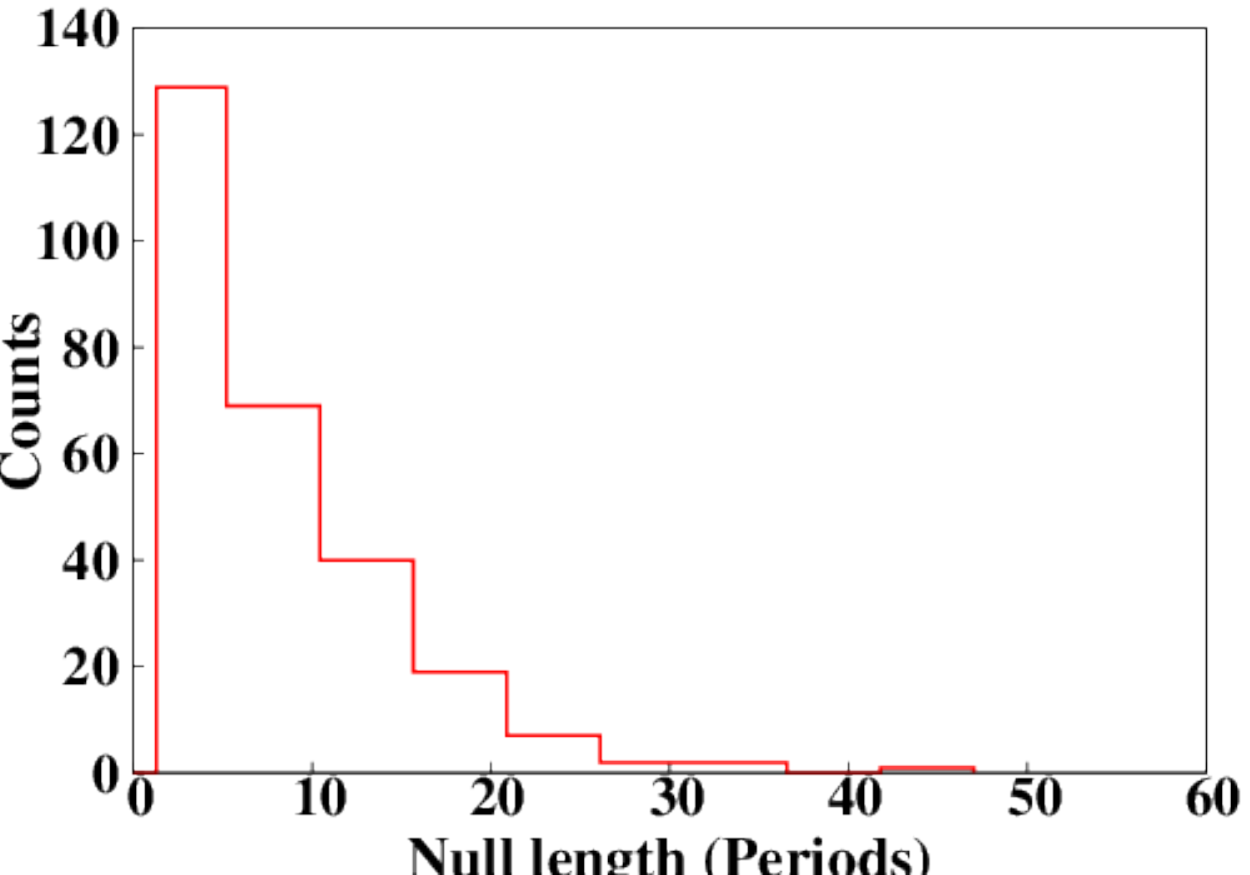}
 % b0809.nlh.325.eps.pdf: 360x252 pixel, 72dpi, 12.70x8.89 cm, bb=0 0 360 252
 }
  \begin{picture}(0,0)
  \put(-220,290){\bf\footnotesize 313 MHz}
  \put(-60,290){\bf\footnotesize 607 MHz}
  \put(-220,120){\bf\footnotesize 1380 MHz}
  \put(-60,120){\bf\footnotesize 4850 MHz}
 \end{picture}
 \caption[The obtained NLHs at all four frequencies for PSR B2319+60]
 {The obtained NLHs at (a) 313 MHz, (b) 607 MHz, 
 (c) 1380 MHz and (d) 4850 MHz for PSR B2319+60. 
 There are minor differences near the single 
 and double period nulls due to the 
 small number of overall null pulses.}
 \label{b2319_NLH_all_freq}
\end{figure}

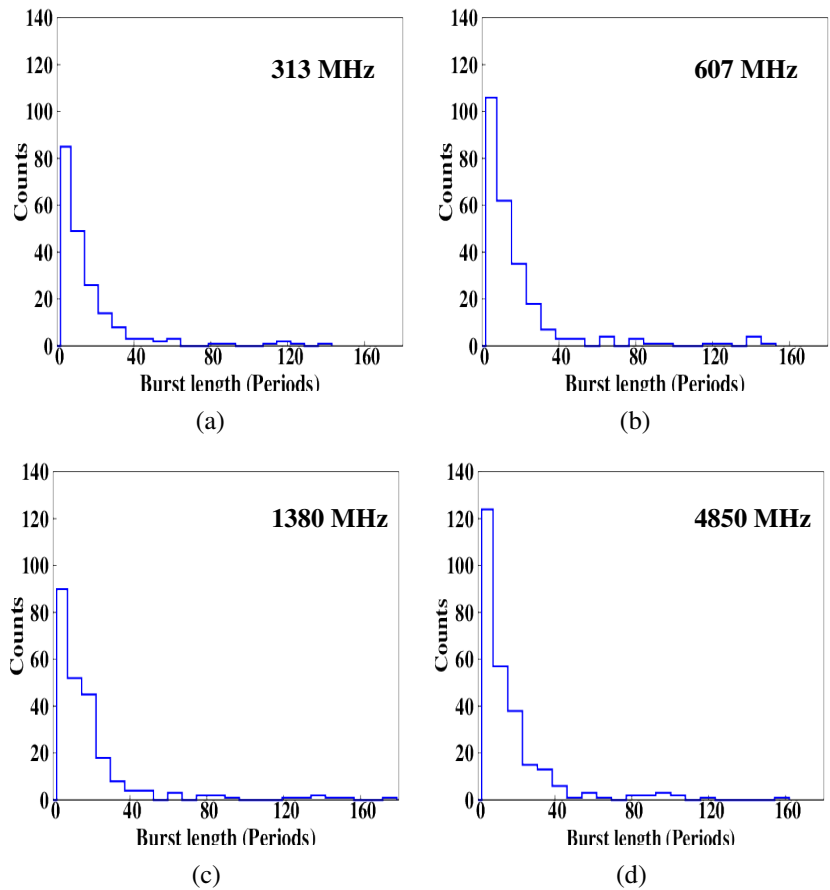
\begin{figure}[h!]
 \centering
 \subfigure[]{
 \includegraphics[width=2.1 in,height=2 in,angle=0,bb=0 0 360 250]{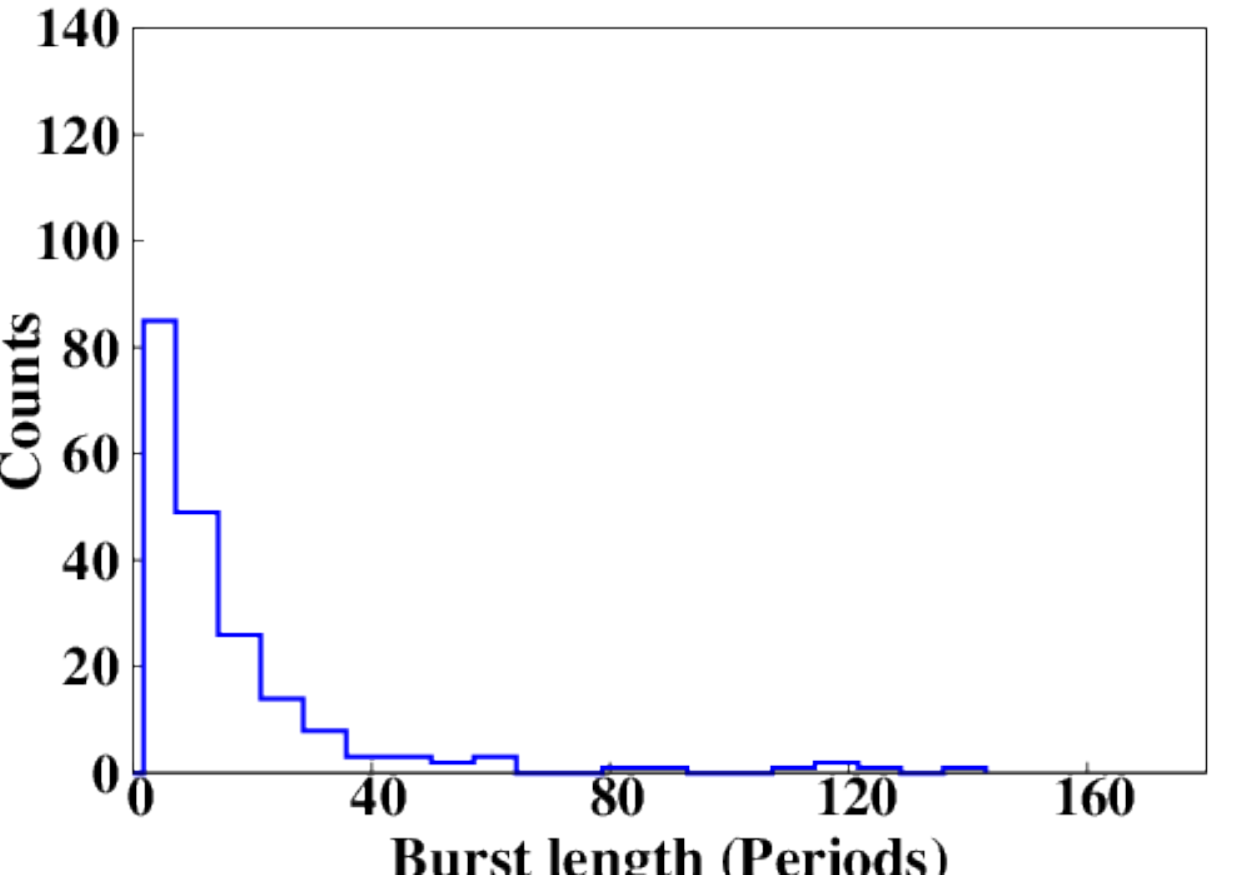}
 % b0809.nlh.325.eps.pdf: 360x252 pixel, 72dpi, 12.70x8.89 cm, bb=0 0 360 252
 }
 \subfigure[]{
 \includegraphics[width=2.1 in,height=2 in,angle=0,bb=0 0 360 250]{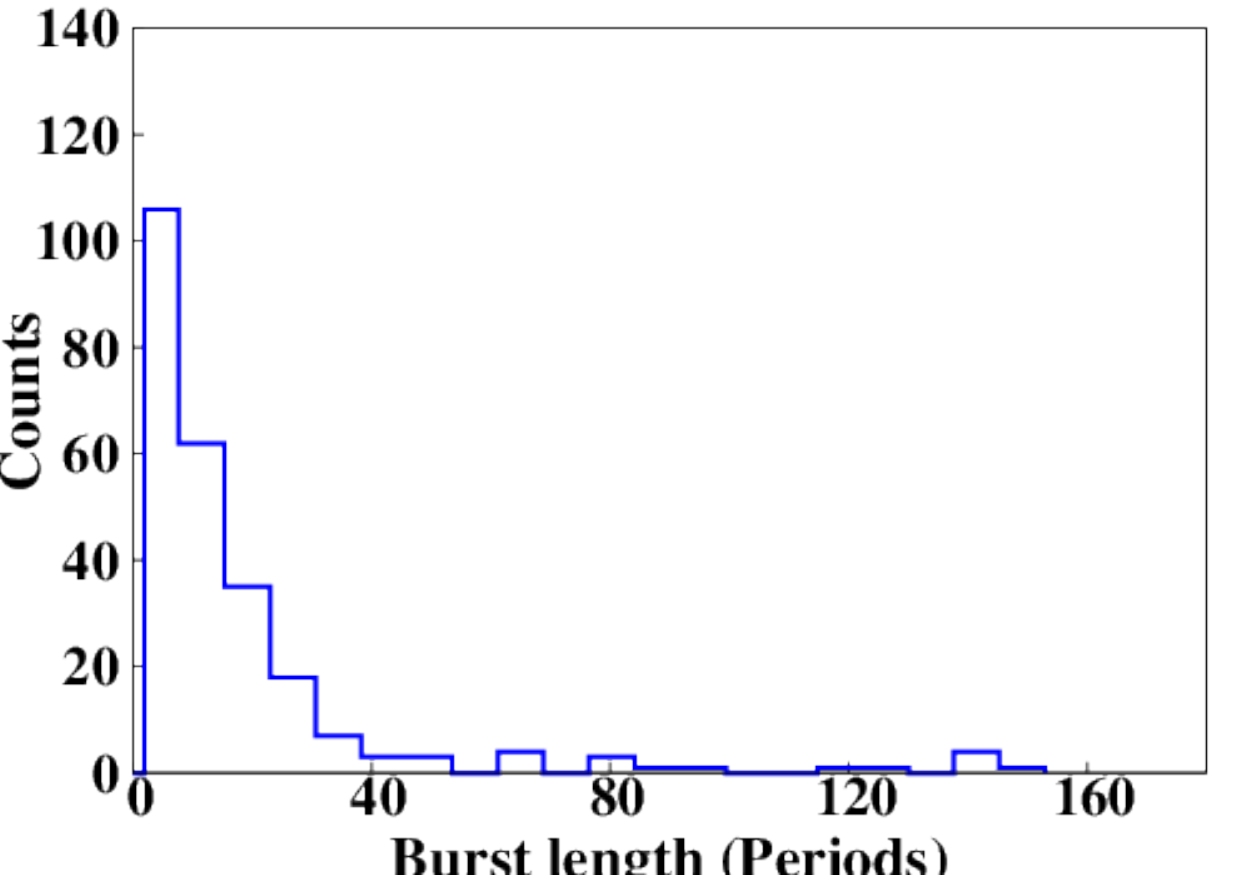}
 % b0809.nlh.325.eps.pdf: 360x252 pixel, 72dpi, 12.70x8.89 cm, bb=0 0 360 252
 }
 \subfigure[]{
 \includegraphics[width=2.1 in,height=2 in,angle=0,bb=0 0 360 250]{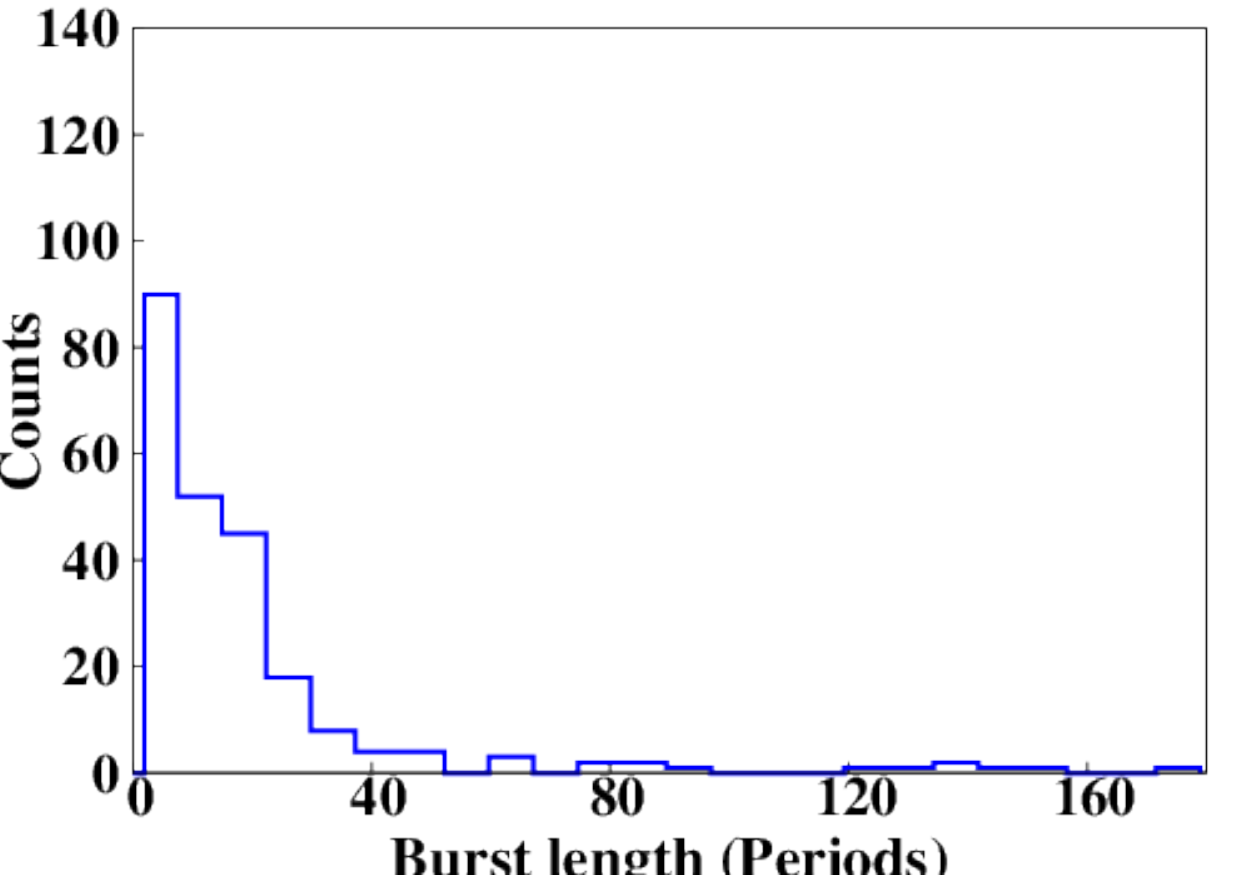}
 % b0809.nlh.325.eps.pdf: 360x252 pixel, 72dpi, 12.70x8.89 cm, bb=0 0 360 252
 }
 \subfigure[]{
 \includegraphics[width=2.1 in,height=2 in,angle=0,bb=0 0 360 250]{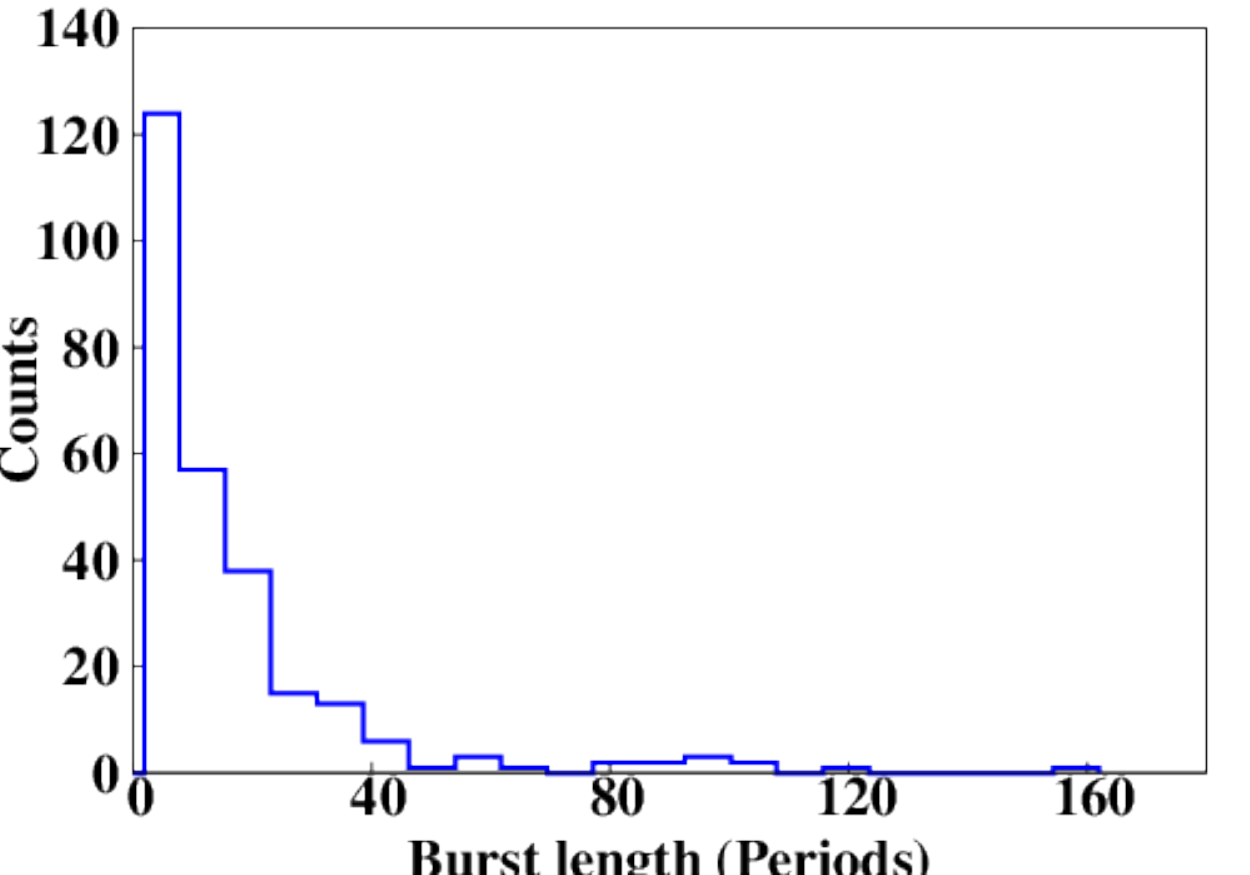}
 % b0809.nlh.325.eps.pdf: 360x252 pixel, 72dpi, 12.70x8.89 cm, bb=0 0 360 252
 }
  \begin{picture}(0,0)
  \put(-220,290){\bf\footnotesize 313 MHz}
  \put(-60,290){\bf\footnotesize 607 MHz}
  \put(-220,120){\bf\footnotesize 1380 MHz}
  \put(-60,120){\bf\footnotesize 4850 MHz}
 \end{picture}
 \caption[The obtained BLHs at all four frequencies for PSR B2319+60]
 {The obtained BLHs at (a) 313 MHz, (b) 607 MHz, 
 (c) 1380 MHz and (d) 4850 MHz for PSR B2319+60. Note the remarkable similarity 
 in the burst length distribution across all observed frequencies.}
 \label{b2319_BLH_all_freq}
\end{figure}

\begin{figure}[h!]
\begin{center}
 \subfigure[]{
 \includegraphics[width=2.1 in,height=2.1 in,angle=0,bb=0 0 360 250]{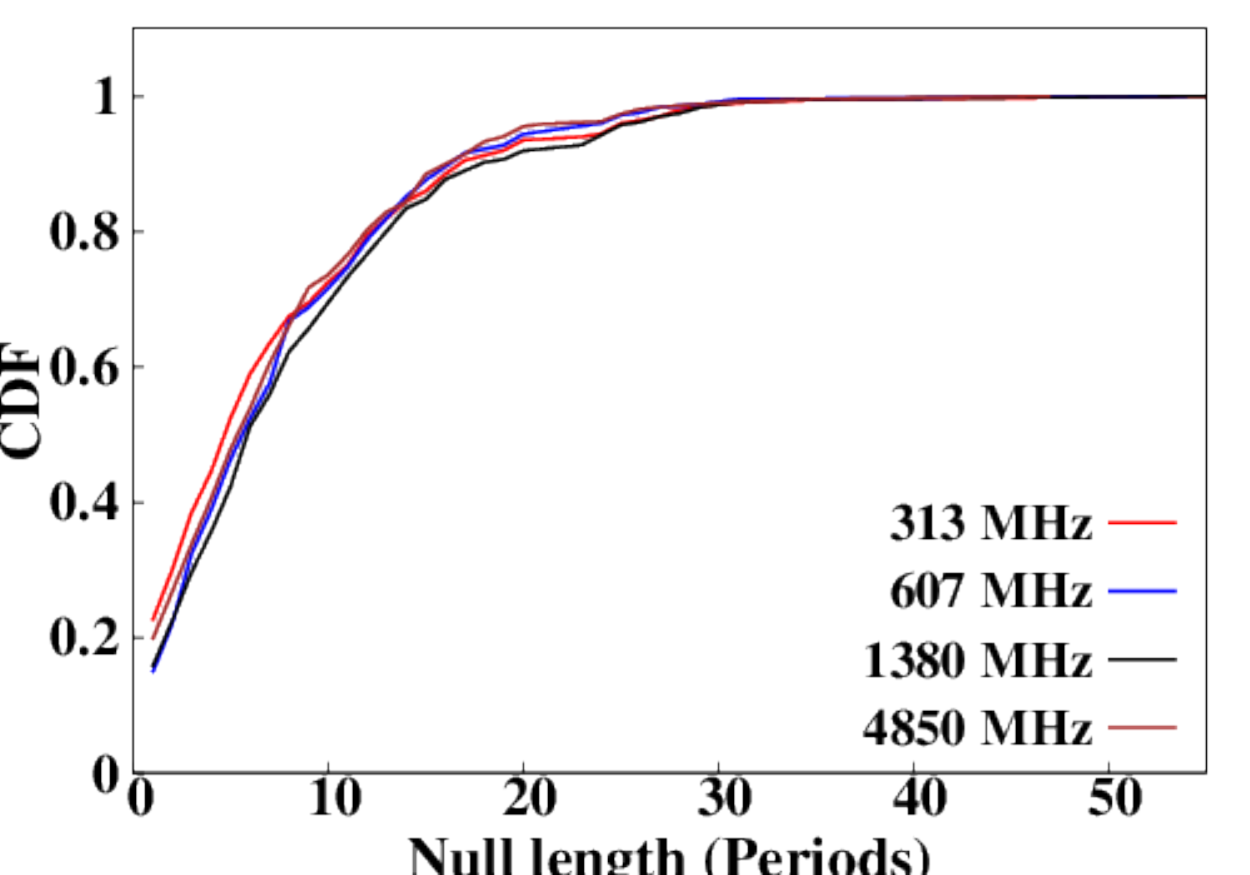}
 % b0809.nlh.cdf.eps.pdf: 360x252 pixel, 72dpi, 12.70x8.89 cm, bb=49 46 404 299
 }
 \subfigure[]{
 \includegraphics[width=2.1 in,height=2.1 in,angle=0,bb=0 0 360 250]{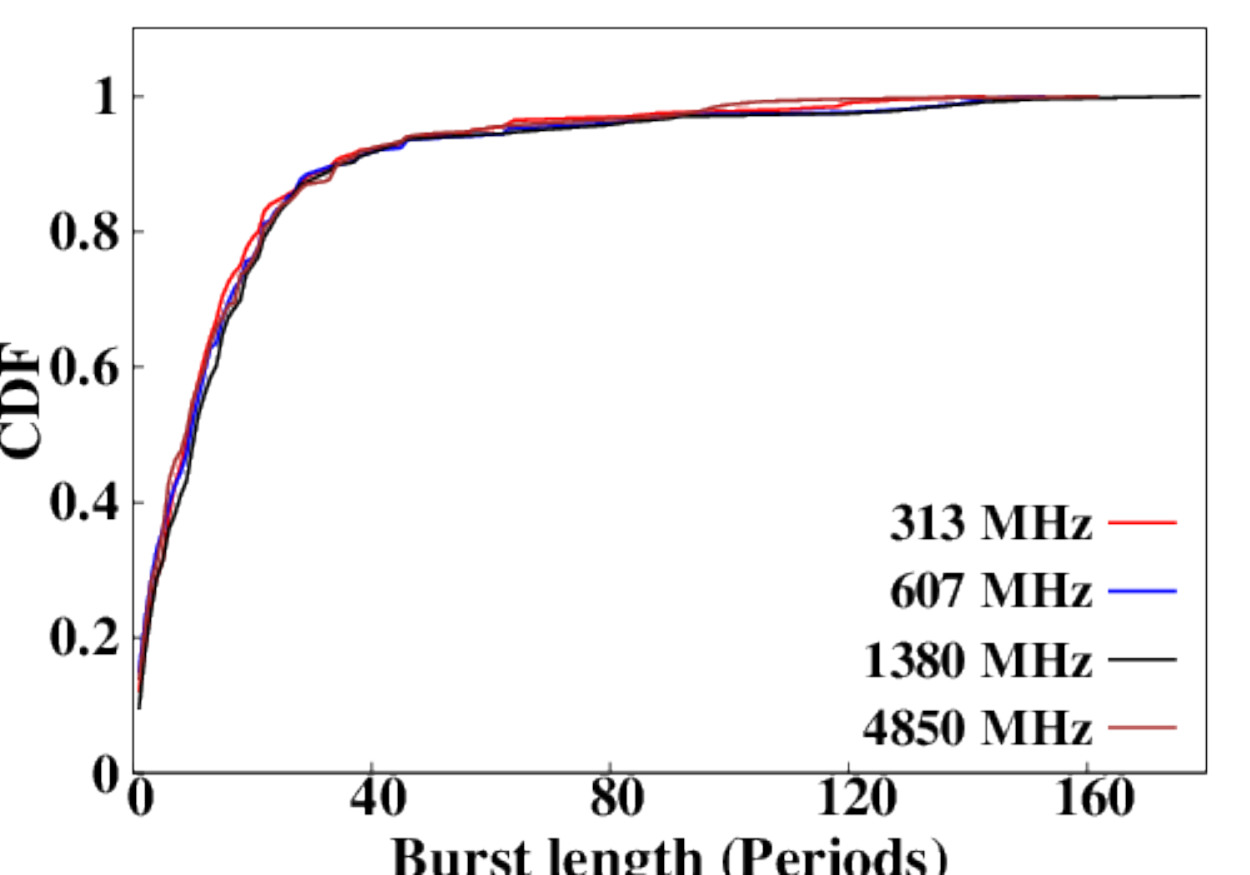}
 % b0809.nlh.cdf.eps.pdf: 360x252 pixel, 72dpi, 12.70x8.89 cm, bb=49 46 404 299
 }
\end{center}
 \caption[The obtained CDFs for null length and burst length distributions for PSR B2319+60]
 {The obtained CDFs of (a) null length and (b) burst 
 length distributions for PSR B2319+60 at 
 all four frequencies. Both distributions show noteworthy 
 similarity between all observed frequencies.}
 \label{b2319_cdf_all_freq}
\end{figure}
\begin{table}[h!]
\begin{center}
\begin{tabular}{|l||c|c|c|}
\tableline
Frequency & 607  & 1380 & 4850  \\
(MHz)     &      &      &       \\
\tableline
\tableline
313  & 99.9 & 99.7 & 99.9  \\
     & 99.9 & 99.9 & 99.9  \\
607  & $-$  & 99.9 & 99.9  \\
     & $-$  & 99.9 & 99.9  \\ 
1380 &      & 99.9 & 99.9  \\
     &      & 99.9 & 99.7  \\
\tableline
\end{tabular}
\end{center}
 \caption[Results of a two sample KS-tests on null and burst length 
 distributions between a pair of frequencies with the assumed null 
 hypothesis of the two distributions being different for PSR B2319+60]
 {Results of a two sample KS-tests on null and burst length 
 distributions between a pair of frequencies with the assumed null 
 hypothesis of the two distributions being different for PSR B2319+60. 
 The first row in each column for a given frequency 
 gives the probability of the two distributions being similar for the null 
 length and the second row gives that for the burst length.}
\label{tabnullburst}
\label{b2319_KS_all_freq}
\end{table}

Similar to PSR B0809+74, to check the effect of nonconcurrent emission states
on the overall nulling pattern at various frequencies, 
their NLHs and BLHs were compared. 
As elaborated in Section \ref{b2319_mismatch_sect}, 
more than half of the nonconcurrent pulses are localized 
either at the start or at the end of the burst phase. 
This can further be tested by comparing the overall null length and burst 
length distributions, which should not show significant differences. 
The separated null and burst pulses, as discussed 
in Section \ref{b2319_onebit_sect}, were used 
to obtain the NLHs and the BLHs for all observed frequencies
(they are shown in Figures \ref{b2319_NLH_all_freq} 
and \ref{b2319_BLH_all_freq}). 
Figure \ref{b2319_NLH_all_freq} shows the NLHs  
with similar exponential decay in the null length distributions 
across all observed frequencies. A two sample KS-test, assuming a 
null hypothesis of different distribution was carried out between 
these distributions. Figure \ref{b2319_cdf_all_freq}(a) 
shows CDFs, obtained using the observed null lengths, 
at all four frequencies which presents remarkably similar distributions. 
Table \ref{b2319_KS_all_freq} shows a high significance ($>$99\%) in 
rejecting the null hypothesis for all pairs of frequencies, 
confirming similar null length distributions at all frequencies.  
Similarly, Figure \ref{b2319_BLH_all_freq} also shows almost 
identical burst length distributions. Figure \ref{b2319_cdf_all_freq}(b)
again shows the remarkable similarity between the CDFs 
obtained using the observed burst lengths at all four frequencies. 
Table \ref{b2319_KS_all_freq} summarizes results on 
a comparison of the burst length distribution, using a two sample 
KS-test conducted for each pair of frequencies. 
Here also the null hypothesis, which assumed different 
burst length distributions, was rejected with a very high significance 
($>$99\%), confirming the alternate hypothesis 
that these distributions for each pair of frequencies are similar. 
This ascertain our results on the location of the nonconcurrent pulses, 
as they do not change the overall length distributions. 
Thus, while the nulling patterns for this pulsar is largely broadband, 
deviations from this behaviour is seen in about 3\%
of pulses, more than half of which occur at the transition from 
null to burst (or vice-verse). 
 
\section{Conclusion}
\label{conc_sect}
A detail study on the simultaneous occurrence of the nulling phenomena
in two pulsars, PSRs B0809+74 and B2319+60, is being reported in this chapter. 
The observations were conducted simultaneously at four different 
frequencies, 313, 607, 1380 and 4850 MHz, from three different 
telescope including the GMRT, the WSRT and the Effelsberg. 
The overlap time for each pulsar was around 6 hours 
between different observatories. 

We obtain single pulses at each frequency and the on-pulse 
energies were compared across all four frequencies, which showed 
remarkable similarity in the pulse energy fluctuations 
(Figures \ref{b0809_ope_all_freq} and \ref{b2319_onezero_all_freq})
for both the pulsars. To quantify these similarities, we obtain the NF 
at each frequency which showed a clear matching, within the error bars, 
at all four frequencies for both the pulsars 
(Figures \ref{b0809_NF_all_freq} and \ref{b2319_NF_all_freq}). 
Similarly, the one-bit sequences were also compared 
using the contingency table analysis. The significance 
of association across frequencies was measured using the 
Cramer-V and the uncertainty coefficient. 
For PSR B0809+74, both the statistical tests showed 
highly significant concurrent behaviour. For PSR B2319+60, 
the significance was marginally lower for the pairs 
involving 4850 MHz. 

We also scrutinize all the nonconcurrent 
pulses for both the pulsar to investigate their true nature. 
For PSR B0809+74, we found that out of 12 nonconcurrent 
pulses, 7 (about 58\%) occurred at the transition point where emission 
state is switching from null to burst (or vice-verse). 
Similarly for PSR B2319+60, we found that out of 158 
nonconcurrent pulses, 82 (about 52\%) occurred at the transition point where 
emission state is switching. Thus, both pulsars showed remarkable 
similarity in the overall broadband behaviour, also accounting the 
fraction of nonconcurrent pulses at the transition points. 
All exclusive pulses were separated and their 
aggregate power was obtained. It can be concluded from these 
exclusive pulses that, during a nonconcurrence event, 
pulsar exhibit weak and narrow emission. Thus, it is likely that 
these pulses occur due to different detection limits at various 
frequencies. A slight difference can be seen for the exclusive 
pulse profiles between both the pulsars. PSR B0809+74 
showed significantly narrow pulses (except at 4850 MHz) 
aligning the overall pulse profile, while PSR B2319+60 showed significantly 
weak pulses shifted towards the leading edge compared to the overall 
pulse profile. However, a strong claim can not be made regarding 
their true shapes, due to their small numbers 
(inset texts of Figures \ref{b0809_mismatch_profiles} and \ref{b2319_mismatch_profiles}). 
Moreover, both the pulsars exhibit prominent drifting behaviour which could easily 
give rise to such differences for individual pulses. Thus, shape of the 
exclusive profiles should be interpreted carefully. 
We also compared the null length and the burst length 
distributions across all observed frequencies which showed similar 
distribution with high significance ($>$99\%) for both the pulsars. 

\section{Discussion}
\label{discussion_sect}
We find that nulling is broadband for PSRs B0809+74 and B2319+60 
across 1:15 frequency range (313 to 4850 MHz). Deviation from  
this behaviour was seen for less than 3\% of pulses, 
most of which occur at the transition phase 
(either at null to burst or vice-verse). If the absence of 
emission during the null state is related 
to the loss of coherence by the secondary particles to produce 
the radio emission [see \cite{fr82} and Section \ref{intrinsic_effects_sect} 
for details], these nonconcurrent pulses represent interesting 
scenario of pulsar magnetospheric state changes. 
The magnetospheric state switching may not be sudden 
and it is likely to posses finite relaxation time to 
switch between different states. For example, PSRs J1752+2359 
and J1738$-$2330, show gradual decay and slow rise in the pulse energy 
before and after the null state respectively (see Chapter 5). 
This kind of gradual changes could manifest itself 
to give rise to differences in the emission state at different 
frequencies near the transition points. These transitional 
states could be chaotic and effect of which can have 
a different frequency dependence for a finite amount of time ($\sim$1 period).

However, the possibility of lower S/N single 
pulses as the prime cause of nonconcurrent pulses can also 
not be rejected.  PSR B0809+74 showed switching between the null 
and the burst state for around 100 times at each frequency 
in our data, out of which only 7 times (7\%) the transition showed 
nonconcurrence. If these pulses are occuring due to real differences 
in emission at various frequencies, then they are likely to 
occur for larger fraction, which was not observed. 
Moreover, the exclusive pulses are narrow 
which makes their detection even more challenging 
at lower S/N single pulses. Thus, its likely that that differences which 
are seen between different frequencies in PSR B0809+74 are 
only due to the limited detection limits at various frequencies. 
Similarly for PSR B2319+60, we observed roughly around 476 
such transitions at each frequency, out of which 82 times (17\%)  
the transition stage showed nonconcurrence across all observed frequencies. 
Moreover, the exclusive pulses seen at different frequencies are 
very weak and again lead to same speculation about their origin, 
which is due to the lower S/N single pulses. Hence, in both pulsars we 
can not reject S/N of the single pulse as the cause of 0.1\% to 3\% differences seen 
in PSRs B0809+74 and B2319+60 respectively. 

These results differ from the ones reported by 
\cite{bgk+07}, in which only 50\% of pulses 
were claimed to be concurrent across similar range of frequencies 
for PSR B1133+16. It is likely that B1133+16 exhibit unique 
frequency dependent nulling behavior while the pulsars studied 
here exhibit frequency independent quenching of pulsed emission. 
One possibility for the seen discrepancy in PSR B1133+16 
can be pointed out as the frequency dependent pulse-to-pulse modulation.
\cite{bsw80} suggested that pulses below critical frequency ($\sim$ 1 GHz) are highly 
correlated while high frequency pulses intermediately correlate with the low frequency pulses. 
As this dependence of the pulse energy between the low and the high frequency pulses varies 
from pulsar to pulsar, one is likely to observe range of non-concurrent behavior. 
Moreover, \cite{bgk+07} also reported lower frequencies to have longer nulls compared to nulls 
at higher frequencies in PSR B1133+16, pointing towards similar deviations at the transition instances. 
The distinct emission seen at higher frequencies for PSR B1133+16 can also be 
consider as a unique mode-changing phenomena in which one magnetospheric state 
does not produce detectable emission at lower frequencies. 

Static geometrical models invoking extinction of sparking region 
are allowed by our results, but ones invoking empty line-of-sight 
are ruled out for these pulsars as the emission will be seen at 
some frequency. Moreover, unexcited emission regions implies some sort of 
periodicity in pulse energy modulation. As concluded in chapter 4, both 
these pulsars exhibit random occurrences of null pulses. 
Thus, empty line-of-sight is unlikely 
to be an explanation. Changes of emission geometry 
as in movement of emission region within the beam also imply 
that emission should be seen at some frequency or other for significant 
fraction of pulses. While this is not ruled out, it is 
unlikely if our results for conal pulsars 
such as PSR B0809+74 are taken together with those 
for a pulsar with more central cut of line of sight such as 
PSR B2319+60. Thus, our results suggest failure 
of emission conditions on a global magnetospheric scale 
\cite[]{ckf99,con05,tim10} as more likely cause of nulling 
seen in these two pulsars. If nulling is another 
form of mode-changing phenomena then our 
results suggest simultaneous occurrence of this phenomena. 
More such observations for number of mode-changing and nulling pulsars, 
covering lower frequencies up to 50 MHz, in future would be of 
great importance. 
\clearpage\null\newpage
\chapter{Conclusions and discussion}
\graphicspath{{Images/}{Images/}{Images/}}

The aim of this thesis was to quantify, model  
and compare nulling behaviour between different classes of pulsars  
to scrutinize the true nature of the nulling phenomenon at multiple frequencies. 
In the previous three chapters, results on different aspects 
of the nulling behaviour with the above mentioned 
core theme are presented. This chapter summarizes the results obtained from each 
of these studies. An attempt has been made towards the end 
of this chapter to examine the possible nulling models 
in view of our results. 

\section{A survey of nulling pulsars using the Giant Meterwave Radio Telescope}
A survey of around 15 nulling pulsars including some newly discovery PKSMB pulsars, 
was conducted using the GMRT at 325 and 610 MHz. Following results were obtained 
from this survey. 
 
\vspace{0.2cm}\hspace{-1cm}
{\bf Subsidiary Results}
\vspace{0.2cm}
\begin{itemize}
 \item Better estimates were obtained on the NFs for four of the observed pulsars, 
       where only an upper/lower limits were previously reported. 
 \item The estimates of reduction in the pulsed emission was also presented for the first time in 11 pulsars. 
 \item NFs values for individual profile components were also presented for two pulsars in the sample \emph{viz.}  
       PSRs B2111+46 and B2020+28.       
 \item Possible mode changing behaviour was suggested by these observations 
       for PSR J1725$-$4043, but this needs to be confirmed with more sensitive observations.       
 \item An interesting quasi-periodic nulling behaviour for PSR J1738$-$2330 was also reported. 
 \item The Wald-Wolfowitz runs test for randomness was carried out to 8 more pulsars to point out the 
       clustering of burst pulses. 
 \item Typical null and burst time-scales were derived for 8 pulsars for the first time.   
\end{itemize}
\vspace{0.2cm}
{\bf Primary Results}
\vspace{0.2cm}
\begin{itemize}
\item It was shown that, nulling patterns differ between 
       PSRs B0809+74, B0818--13, B0835--41 and B2021+51, even though they have similar NFs of around 1\%.      
       Our results confirm that NF does not capture the full detail of the nulling behaviour of a pulsar.       
\item  The durations of the null and the burst states were shown 
       to be modelled by a stochastic Poisson point process suggesting that 
       these transitions occur at random. It was concluded that, the underlying 
       physical process to cause nulls, in the 8 studied pulsars, appears to be 
       random in nature producing nulls and bursts with unpredictable durations. 
\end{itemize}

\section{On the long nulls of PSRs J1738$-$2330 and J1752+2359}
A detailed study of pulse energy modulation in two pulsars, 
PSRs \pa\ and \pb, with similar NFs has been presented 
to compare and contract their nulling behaviour. 

\vspace{0.2cm}\hspace{-1cm}
{\bf Subsidiary Results}
\vspace{0.2cm}
\begin{itemize}
 \item The NFs were estimated to be 85$\pm$2\% and $<$89\% for \pa\ and \pb, respectively.  
 \item Both the pulsars show similar bunching of burst pulses, 
       classified as the bright phases, which are separated by long null states.
       A similar quasi-periodic switching between these two states was observed for both the pulsars. 
  \item Towards the end of each bright phase of \pb, an exponential 
       decline in the pulse energy was reported. \pa\ also showed similar exponential decay 
       along with a flickering emission characterized by short frequent nulls towards 
       the end of each bright phase. 
 \item We modelled the bright phase energy decay for \pb\ and estimated their average durations. 
 \item Unlike J1738$-$2330, the first bright phase pulse profile and the last bright phase pulse profile 
       show striking differences for J1752+2359, hinting differences in 
       transition from null state to burst state and vice-verse between the two pulsars.      
 \item The occurrence rate of the inter-burst pulses (IBP) is random and uncorrelated with the preceding or following 
       bright phase parameters. 
 \item The total intensity and the circular polarization profiles of IBPs are slightly shifted towards the leading side 
       compared to the conventional integrated profile, indicating a change in the emission region. 
 \item We reported absence of any giant pulses in our long observations, although such pulses being 
       reported at low frequencies in \pb.            
\end{itemize}
\vspace{0.2cm}
{\bf Primary Results}
\vspace{0.2cm}
\begin{itemize}
 \item The PCF for \pa\ indicated that the mechanism responsible for bright phases 
       is governed by two quasi-periodic processes with periodicities of 170 and 270 pulses. 
 \item The PCF for \pb\ indicated that the nulling pattern is dominated 
       by 540 pulse quasi-periodicity, which jitters from 490 to 595 pulses.      
 \item It was shown that these processes are not strictly periodic, but retain a memory 
       longer than 2000 pulses for \pa, while the memory of \pb's periodic structure is 
       retained for only about 1000 periods. 
 \item We demonstrated that the area under each bright phase is similar for \pb, suggesting 
       that the energy release during all such events is approximately constant. 
 \item We report, for the first time, peculiar weak burst pulses during the long null phases of \pb, 
       which are similar to emission seen from RRATs. 
 \item These results confirm that even though these two pulsars have similar but significantly high 
       NFs, they show very different nulling behaviour.        
\end{itemize}

\section{Simultaneous multi-frequency study of pulse nulling behaviour in two pulsars}
This study reported a detailed investigation on the simultaneous occurrence of the nulling phenomena
in two pulsars, PSRs B0809+74 and B2319+60 at 313, 607, 1380 and 4850 MHz. 

\vspace{0.2cm}\hspace{-1cm}
{\bf Subsidiary Results}
\vspace{0.2cm}
\begin{itemize}
 \item We obtained single pulses at each frequency and the on-pulse energies were compared 
       across all four frequencies, which showed remarkable similarity in the pulse energy 
       fluctuations for both the pulsars.
 \item We obtained the NF at each frequency which were found to be consistent within the error bars, 
       at all four frequencies for both the pulsars.       
 \item We measured the Cramer-V and the uncertainty coefficient for each pair of frequencies and 
       showed simultaneous nulling behaviour with high significance for both pulsars. 
 \item For B0809+74, we found that out of 12 nonconcurrent pulses, 7 (about 58\%) 
       occurred at the transition point where emission state is switching from null to burst (or vice-verse).        
 \item For B2319+60, we found that out of 158 nonconcurrent pulses, 82 (about 52\%) occurred at 
       the transition point where emission state is switching.       
 \item B0809+74 showed significantly narrow pulses (except at 4850 MHz) aligning the overall pulse profile 
       during it's exclusive emission at only single frequency. 
 \item B2319+60 showed significantly weak exclusive pulses which are shifted towards the 
       leading edge compared to the overall pulse profile.
 \item It was suggested that these exclusive pulses occur due to prominent drifting 
       behaviour seen in both pulsars along with differences in the S/N between different frequencies.  
 \end{itemize}
\vspace{0.2cm}
{\bf Primary Results}
\vspace{0.2cm}
\begin{itemize}
  \item We showed that nulling is a broadband phenomena in these two pulsars, which favors 
  phenomena intrinsic to the pulsar magnetosphere as a more likely cause of nulling. 
  Geometric effects are less likely to give rise to such high concurrent behaviour. 
\end{itemize}

\newpage
\section{Implication of our results}
Following are a few of the implications of our study on the 
overall understanding of the pulsar nulling phenomena. 

\subsection{Quantifying the nulling behaviour}
Nulling behaviour has been classically quantified as the 
fraction of null pulses among the total number of observed pulses. 
\cite{rit76}, who presented a method to estimate 
amount of nulling by the NF, reported a correlation 
between the pulsar period and the NF. \cite{rit76} suggested  
that, pulsars with longer periods tend to null more frequently  
compared to pulsars with smaller periods. 
Pulsar period is directly correlated with the 
age of the pulsar hence, \cite{rit76} suggested 
that pulsar die with increasing fraction of nulls. 
However, as shown in Figure \ref{all_psr_nulling_nf} of Chapter 2, 
such claims can be contested in light of currently known 
sample of nulling pulsars. \cite{ran86} concluded that core single 
pulsars possess small NF compared to other classes.
Investigations by \cite{ran86} contradicted earlier claims by \cite{rit76} 
and suggested that apparent relation between the nulling and the pulsar 
age is due to the profile morphologies. In a given 
profile class, there is no strong correlation 
between the NF and the pulsar age. 
\cite{big92a} reported correlation study between the NF with several 
pulsar parameters, using 72 nulling pulsars. 
The NF-period correlation was highlighted by \cite{big92a}. 
The other correlations which were reported by \cite{big92a} 
includes $\dot{E}, ~\log B_{lc}$ and $\alpha$. However, 
these correlations can be the results of the reported NF-period 
correlation as all these quantities are strongly correlated with the period 
% of the pulsar. Contrary to \cite{ran86}, \cite{wmj07} have claimed 
that there is no correlation between the NF and the profile morphological classes. 
In a detail study, \cite{wmj07} reported that, multicomponent 
profile pulsars tend to have higher NFs but such pulsars are old. 

Thus, NF-period correlation has been tentatively supported by early investigators. 
However, this is not a strong correlation as \psrb, 
which we have studied, has one of the smallest period among the known 
nulling pulsars, but it still exhibit extreme NF. 
Our results on the small NFs as well as on the long NFs pulsars suggest that, 
NF is probably not an ideal parameter to quantify nulling behaviour. 
Thus, our interpretation can be extrapolated to explain it's lack 
of strong correlation with any of the pulsar parameters. 
We have also made an attempt to quantify nulling behaviour by incorporating their length 
distributions to estimate $\tau_n$ and $\tau_b$. 
These quantities needs to be incorporated in order to truly quantify nulling 
behaviour. In similar line of studies, \cite{lr06} made an attempt to correlate 
the longest observed null lengths in 19 pulsars with different pulsar parameters. 
A weak correlation, with pulsar's age, was reported by these comparisons, although strength of this 
correlation was not reported. A visual inspection of Figure 6 in \cite{lr06} 
revels no such correlation. Recently, an attempt has been made by \cite{yhw13} 
to characterize nulling phenomena by incorporating nulling time-scales.  

\subsection{On the randomness of pulsar nulls}
\label{on_randomness_sect}
We have shown extensively in Chapter 4 that, 
nulls do occur randomly with their length distributions originating 
from a stochastic random process, which we modelled as 
a simple Poisson Point Process (PPP). As discussed in Chapter 2, 
for many pulsars, periodic nulling have been reported. 
These analysis were carried by taking a Fourier transform of the 
pulse energy fluctuations \cite[]{hr07,hr09}. 
Moreover, \cite{rr09} reported non-random behaviour in around 14 pulsars 
using a statistical method called Wald–Wolfowitz runs test \cite[]{ww40}. 
Thus, a question can be raised that, \emph{do nulls occur randomly?}

We would like to suggest that, although the runs test is a test of randomness, 
it only describes randomness in the occurrence of single pulses. Many previous studies have 
suggested that burst pulses tend occur in clusters \cite[]{bac70a,rit76}. 
Thus, this clustering is bound to give non-randomness, if the probability 
of random occurrence of null or burst states are estimated for individual pulses
(in the sense of the Wald–Wolfowitz runs test). This should 
not be interpreted as a non-randomness of nulling phenomena. 
We also carried out similar runs test for 8 observed pulsars, in Chapter 4,  
which clearly showed that they fail the runs test with high significance. 
We have also suggested that, if the distribution of nulls 
and bursts follow the stochastic random PPP, the individual pulses 
will fail the runs test due to their excessive clustering. To stretch this point further, 
we generated multiple sequences of null and burst pulses, following the PPP. 
The Poisson point random variable was generated using 
the standard procedures given in \cite{press}. Table \ref{PPP_simul_table} 
shows the random sequences generated from the simulations along 
with their Poisson point variables. Its clearly evident from Table \ref{PPP_simul_table} 
that, runs test will fail for any combination of average null lengths and average burst lengths, 
obtained from the PPP random variable. The null length (or burst length) presents the 
time-scale between the two transitions, first from burst-to-null and second from null-to-burst (or vice-verse). 
Thus, it can be concluded that these transitions occur randomly and they are unpredictable. 
\begin{table}[h!]
\centering
\begin{tabular}{cccc}
\hline
\hline
PSRs  &  Avg. Burst length & Avg. Null length & Z \\ 
      &    (Periods)       &    (Periods)     &   \\
\hline
B0809+74 & 136.43  & 1.47 &  -38.29 \\
B0818-13 & 67.74  & 0.56 & -10.52 \\
B1112+50 & 2.59  &  2.89 & -22.37 \\
B2111+46 &  8.52 & 1.56 & -17.50 \\
\hline
\multicolumn{4}{c}{Simulations} \\
\hline
	   & 100.0 &  10.0 & -89  \\
	   & 100.0 &  20.0 & -94  \\
	   & 50.0 &  0.2 & -15  \\
	   & 40.0 &  5.0 & -76  \\
	   & 40.0 &  40.0 & -94  \\
	   & 20.0 &  4.0 & -68  \\
Simulated  & 10.0 &  6.0 & -71  \\
Pulsar     & 10.0 &  50.0 & -87  \\
	   & 4.0 &  2.0 & -21  \\
	   & 4.0 &  4.0 & -46  \\
	   & 4.0 &  50.0 & -71  \\
	   & 2.0 &  5.0 & -27  \\
	   & 2.0 &  50.0 & -46  \\
	   & 1.0 &  10.0 & -7  \\
	   & 1.0 &  20.0 & -12  \\
\hline
\end{tabular}
\caption[Table of simulated random sequences using the Poisson point random variable]
{Table of simulated random sequences using the Poisson point random variable. 
For both the panels the column gives, pulsar name, average burst length, average 
null length and runs test parameter Z. The top part of the table shows a comparison 
with the average null length and average burst length, in units 
of number pulsar periods, from four of the observed pulsars. 
The bottom panel shows the simulated pulsar with different 
combinations of average null and burst lengths using the Poisson point random 
variables. It should be noted that for all simulated combinations of random variables, 
the runs test fails.}
\label{PPP_simul_table}
\end{table}

In a similar line of studies, we also investigated periodicities in two high 
NF pulsars, \emph{viz.} J1738$-$2330 and J1752+2359. Their quasi-periodicities 
are difficult to model using the above mentioned simple PPP. As suggested in 
Chapter 5, more complicated models have to be invoked to explain the observed 
quasi-periodic fluctuations. Moreover. the observed quasi-periodicities are due to the 
clustering of burst pulses, as typical bright phase and off-phase alternate 
with lengths scattered around the means of Gaussian like distributions. 
However, we have also pointed out that at least for one 
of the pulsar, J1752+2359, the coherence of quasi-periodicity only 
last for around 1000 pulses. 

Recently, \cite{cor13} has extend our work on the randomness in nulling pulsars 
to more classes of pulsars exhibiting mode-changes, drift-rate variations 
and normal-abnormal modes (these classes are introduced in Chapter 2). 
Transitions between different states exhibited by these pulsars 
were modelled by Markov chain (Poisson Point Process is the 
first order Markov chain). For every period, 
the transition probability (q), from previous emission state 
to current emission state can be estimated.  
By obtaining all such probabilities, 
a 2$\times$2 transition matrix (Q) can be formed  
for a nulling pulsar, as shown in Equation \ref{markov_matrix}. 
\begin{equation}
Q ~= ~ \left[ \begin{matrix} q_{NN} & q_{NB} \\ q_{BN} & q_{BB} \end{matrix} \right]
\label{markov_matrix}
\end{equation}
Here, $q_{NN}$ and $q_{NB}$ are null-to-null and null-to-burst probabilities given as, 
\begin{equation}
q_{NN} ~ = ~  \left( 1 ~ - \frac{1}{\tau_N}\right), 
\end{equation}
\begin{equation}
q_{NB} ~ = ~ 1 - q_{NN}. 
\end{equation}
Similarly, $q_{BB}$ and $q_{BN}$ are burst-to-burst and burst-to-null probabilities given as, 
\begin{equation}
q_{BB} ~ = ~  \left( 1 ~ - \frac{1}{\tau_B}\right), 
\end{equation}
\begin{equation}
q_{BN} ~ = ~ 1 - q_{BB}.
\end{equation}
Here,$\tau_N$ and $\tau_B$ are the typical null and burst time-scales, 
defined in Chapter 4. \cite{cor13} also modelled multi-state switching 
which is also seen in many mode-changing pulsars (see Chapter 2 for 
a detailed discussion on multi-state switching). The relevant 
findings, which can also be extrapolated to the quasi-periodicities 
seen in the above mentioned two pulsars, are on the stochastic 
nature of the 35 day quasi-periodicity in the intermittent 
pulsar B1931+24. \cite{cor13} introduced a forcing function 
along with the Markov transition probability, to invoke quasi-periodic 
changes. Such forcing function, $f(t)$, can be given as \cite[]{cor13}, 
\begin{equation}
f(t) ~ = ~ {e}^{-A~cos(2\pi{t}/P_f + \phi)}. 
\end{equation}
Here, A is the amplitude of the forcing function while $P_f$ is the forcing  
periodicity and $\phi$ is the arbitrary phase shift. 
Similarly, such forcing periodicity can also be applied for the 
two high NF pulsars to explain their quasi-periodicities 
using the Markov model. This forcing function can change 
the probabilities of all transitions. 
The transition probability matrix 
in such case can be given as \cite[]{cor13}, 
\begin{equation}
Q ~= ~ \left[ \begin{matrix} 1~ - ~ q_{NB}e^{A_{N}{f(t)}} & q_{NB}e^{A_{N}f(t)} \\ q_{BN}e^{-A_{B}f(t)} & 1 ~ - ~ q_{BN}e^{-A_{B}f(t)} \end{matrix} \right].
\label{markov_matrix_forcing}
\end{equation}
Thus, it can concluded from our studies and by incorporating recent findings 
from \cite{cor13} that, \emph{nulls do occur randomly} as they do not give rise 
to predictable length sequences. For pulsars exhibiting quasi-periodic 
energy fluctuations, forcing functions can be invoked to model 
their transition to follow the Markov chain model. 
As mentioned in Chapter 5, such forcing functions 
may arise from an external body \cite[]{csh08} or neutron star oscillations \cite[]{cr04a} 
or near-chaotic switches in the magnetosphere's non-linear system \cite[]{tim10}. 

\subsection{On the global magnetospheric state changes}
Recently, \cite{wmj07} have suggested that 
nulling is an extreme form of mode-changing phenomena. 
Our discovery of weak emission modes in PSRs J1725$-$4043 
and J1752+2359 support this claim that nulls and profile modes 
can be interpreted as different extremes of similar 
phenomena. Our results on the simultaneous observations of 
two nulling pulsars, point towards the 
global failure of emission processes at all frequencies. 
Thus, it can be speculated that nulls represent 
a different stable magnetospheric state \cite[]{tim10} during 
which the emission is ceased. Similar conclusions were also derived 
for the intermittent pulsar, PSR B1931+24 \cite[]{klo+06}.
Moreover, \cite{lhk+10} reported correlated changes in the pulse 
profiles with the changes in the $\dot{P}$, pointing towards 
global magnetospheric changes. Recent simultaneous 
radio and X-ray observations of PSR B0943+10 
also revealed such global changes \cite[]{hhk+13}.

\section{Implication for possible nulling mechanism}
In this section, an attempt has been made to 
characterize different observations in order to compare 
them with different proposed models of pulsar nulling. 

According to \cite{che81}, temperature changes on the surface may lead to 
changes in the ion flow from the polar cap. Such ion flow may hinder 
the growth of polar cap potential. \cite{cor81,cor83} listed various time-scales for the 
occurrence of different observed phenomena. In which, 
he suspects the temperature fluctuations of the order of few milliseconds 
to several seconds (Figure 3 of \cite{cor81}). Such time-scales 
do match with the derived typical null and burst time-scale 
in our study (Table \ref{tabtau} in Chapter 4). 
Thus, such temperature variations are likely 
to cause nulling in pulsars we studied. However, nulling 
observed in the intermittent pulsars, which occurs with the 
time-scale ranging from days to weeks, are difficult to explain by 
this rapid temperature fluctuations. 

\cite{zqlh97} also considered three gap discharge mechanisms, 
operating at different temperature ranges. These models 
include CR and two different types of ICS processes. The resonant 
ICS mode in which the cross section of interaction 
between the high energy particle and thermal photon is maximum, 
while during the thermal ICS mode where the temperature are high enough 
to produce sufficient thermal photons. These authors argue that 
the normal pulsars lie in the range where the gap discharge process
can switch between these different models. 
They suggested two different critical temperature ranges as T$_{1}$ and T$_{2}$ with T$_{1}$ $<$ T$_{2}$. 
Above T$_{2}$, the thermal ICS could be active while below T$_{1}$, CR 
is a more viable mechanism. Most of the normal pulsars possess 
temperature in the range between T$_{1}$ and T$_{2}$. Thus, they are likely to 
switch rapidly between different gap discharge models due to 
the thermal fluctuations. This was proposed 
as an explanation for mode-changing phenomena. As nulling is an extreme 
form of mode-changing phenomena, as discussed earlier, this can be extrapolated 
to explain nulling. For some of these pulsars, due to the temperature fluctuations, 
the gap properties and the Lorentz factor of the primary particles can be 
altered on a sudden instant. This can cause inefficiency in the two-stream instability 
and can lead to loss of coherence and eventually to absence of emission. 
\cite{zqh97b} also extrapolated the mode-changing model given 
by \cite{zqlh97} and suggested similar interpretation for PSR B1055$-$52. 

\cite{zx06} reported precession as the cause of extreme nulls. 
If precession is causing the beam to shift, then one is expected 
to get gradual changes in the pulse width as the pulsar 
goes into the null state from the emitting burst state. 
In \pb, we see evidence of slightly narrow profile 
for the first periods of the bright phases while the last pulses are much wider 
and also have an extra component. Moreover, \pb\ also exhibit weak single burst pulses 
during the long null states. For \pa, we do not see any evidence 
of profile variations across the burst state. 
Thus, precession can explain the extreme nulls but our 
results are not consistent with the proposed model. 
Moreover, \cite{gle90} also reported deformation of the emission beam due to 
the misalignment between the rotation and the neutron star symmetry axes. 
Such models can also be rejected due to the absence of significant profile 
variations during the bright phase in two of the pulsars we studied. 

\cite{fr82} has proposed a model of pulse nulling as a break
in the two stream instabilities which occurs during a steady polar gap discharge. 
A brief discussion regarding this model is given in Section \ref{intrinsic_effects_sect}. 
Two stream instabilities are produce due to the propagation of secondary particles with different 
momenta. These secondary particles are generated from the primary 
particles with high gamma, generated at initial gap discharge. A situation can occur
for long period pulsars in which gap discharges at roughly the same rate as the potential drop
would increase before the sparking. In which, Polar cap attains a steady discharge state which does
not produce high gamma primary particles for the consecutive two stream instabilities and
bunching. In the absence of two stream instability, the coherent radio emission will be ceased. 
It can be speculated that, this steady state is reached after a build up from the residual potential, 
left over after each gap discharge during normal non-steady condition (non-null state). Once the
build up of the residual potential reaches the maximum gap height potential, pulsar attains a steady state
which does not emit radio emission (null state). When the residual potential discharges, pulsar
can resume its normal interrupted sparking action which give rise to primary particles 
with high gamma for the consecutive bunching and coherence. The charging and discharging 
of this residual potential on top of normal gap discharge has an inherent 
memory associated with them. The time-scale of this  
memory mechanism should be higher for long period and low surface magnetic field
pulsars. We reported nulling time-scales from the PPP model in 8 pulsars. 
This modelling reveals that there is an inherent memory which 
constrains the transitions to occur in accord with the PPP distributions. 
However, the charging and discharging of the gap potential, on top 
of the left over potential, needs to be simulated in 
order to scrutinize this model for our pulsars. Moreover, correlation 
with the long period and low surface magnetic field can also not 
be tested for such a small sample. However, this appears to be a more viable 
mechanism to cause nulls. 

In Section \ref{geometric_effects_sect}, various geometric phenomena which can give to pulsar nulls were 
also discussed. The phenomena which is related to the missing line-of-sight can be scrutinize 
in light of our results. One of the main prediction of this model is the 
periodic occurrence of null pulses. However, we have conclusively shown 
that occurrence of nulls are completely random (see above section \ref{on_randomness_sect}). 
For all pulsars we have studied, a stochastic random process 
can be invoked to explain their length distributions. Moreover, 
missing line-of-sight will also give rise to large mismatch between 
observations at different frequencies. As the emission at different frequencies, originate 
at different heights from the surface, the structure of the emission beams are also 
likely to be significantly different. The higher frequencies originates from the much lower 
heights in the conical emission beam, with emission sub-beams closely spaced (see Figure \ref{radiation_beam_height}). 
Thus, the likelihood of an empty sight line is significantly lower at higher frequencies 
and this can give rise to high fraction of nonconcurrent pulses between the occurrence 
nulls at different frequencies. However, contrary to that, we have observed 
highly significant concurrent emission behaviour, with only around 1 to 3\% mismatch 
for different frequency pairs. Thus, our results on the randomness of pulsar nulls, 
simultaneously occurring at large range of radio frequencies, do not support 
empty line-of-sight as a likely mechanism in the sample of pulsars we studied. 

\newpage
\section{Future work}
Our study can be extended to incorporate more classes of pulsar in order 
to understand the magnetospheric state switching phenomena. The sample 
of pulsars we have studied are very small for which this state switching 
is shown to be random and global. This samples needs to be enhanced in order 
support these claims. Thus, following future work can be suggested 
from this thesis which can help to unravel the mechanism operating behind 
the nulling phenomena.
\begin{itemize}
 \item The sample of pulsars, with null and burst length distributions, needs 
 to be enhanced by observing several relatively strong nulling pulsars using 
 the GMRT. 
 \item Similarly, simultaneous observations of more nulling pulsars needs to be 
 conducted to conclusively point out global state switching. Such pulsar should include 
 more central line cut to scrutinize the concurrent emission behaviour of the 
 core component. 
 \item The Pair correlation function analysis applied for the high NF pulsars 
 should also be applied to intermittent pulsars, in order to investigate 
 their coherence time-scales. 
 \item The $\dot{P}$ variations can be measured by separating the bright 
 phases and the IBPs for PSR J1752+2359 by monitoring it for a few months. 
 Both these $\dot{P}$ values can be compared to  estimate the changes in 
 the rotation during these different phases in order to scrutinize pair production. 
 \item PSR J1738$-$2330 needs to be observed for a continues stretch of around 8 hours 
 from instruments like the Ooty radio telescope or the Parkes radio telescope to 
 investigate the coherence length of the measured quasi-periodicity. 
 \item Estimation on the NFs for the individual profile components should be extended 
 for more number of pulsars in order to truly understand the phenomena. 
 \item In order to truly scrutinize the global changes during the nulls, simultaneous 
 radio and X-ray observations should conducted for a few strong nulling pulsars with 
 long nulls. 
\end{itemize}
\newpage
\section{Summary}
In light of the results reported in this thesis, following interpretations 
about the nulling phenomena can be suggested. 
\begin{enumerate}
 \item NF is not an ideal parameter to quantify nulling behaviour, confirmed firmly 
  by comparing low NF pulsars as well as high NF pulsars.  
 \item Nulling occurs randomly with unpredictable length durations. 
  Quasi-periodicities seen in the high NF pulsars can also be explained 
  by the Markov models with a forcing function.  
 \item Nulling is an extreme form of mode-changing phenomena which 
  occurs on a global magnetospheric scale. 
 \item Geometric reasons are less favoured as a likely cause of nulling phenomena 
  due the randomness and broadband behaviour reported in this thesis.  
\end{enumerate}
   \backmatter 
   \appendix   
   \begin{appendices}
   \clearpage\null\newpage
  \chapter[Appendix A : Pulse energy histogram]{Appendix A \\ 
Pulse energy Histograms}
% \appendixname{Pulse energy Histograms}
\graphicspath{{Images/}{Images/}{Images/}}

\label{Appendix_PEH}
% \enlargethispage{2cm}

This appendix lists the on-pulse and the off-pulse energy histograms of 
all the pulsars discussed in Chapter 4. 
The on-pulse and the off-pulse energy were binned in various number of bins, 
which are mentioned in the caption for each pulsar. 
The bin sizes are kept similar for both the histograms, 
the ONPH and the OFPH for each pulsar. 
In all figures, the name of the pulsar is listed at the top.
The abscissa presents the normalized energy obtained 
using the block average as discussed in Chapter 3. 
The ordinate presents the normalized counts of occurrence 
for each energy bin. The OFPHs are shown with the red solid 
lines while the ONPHs are shown with black filled curve. 
The counts in both the histograms were normalized 
by the peak from the corresponding OFPH histogram 
for each pulsar. As highlighted in Chapter 3, 
this normalization does not provide probability distribution 
of the pulse energy. It was only adopted here to easily 
discern the NFs from these plots. The obtained NF, 
observed frequency along with the number 
of pulses used during the analysis are displayed in the inset texts.
For weak pulsars, the number of sub-integrated pulses 
are shown in the parentheses beside the total number of pulses in 
the inset texts. 

\setcounter{figure}{0} \renewcommand{\thefigure}{A.\arabic{figure}} 
\setcounter{equation}{0} \renewcommand{\theequation}{A.\arabic{equation}} 

\begin{figure}[h!]
\centering
\includegraphics[width=2.8in, height=5in, angle=-90,bb=0 0 504 720]{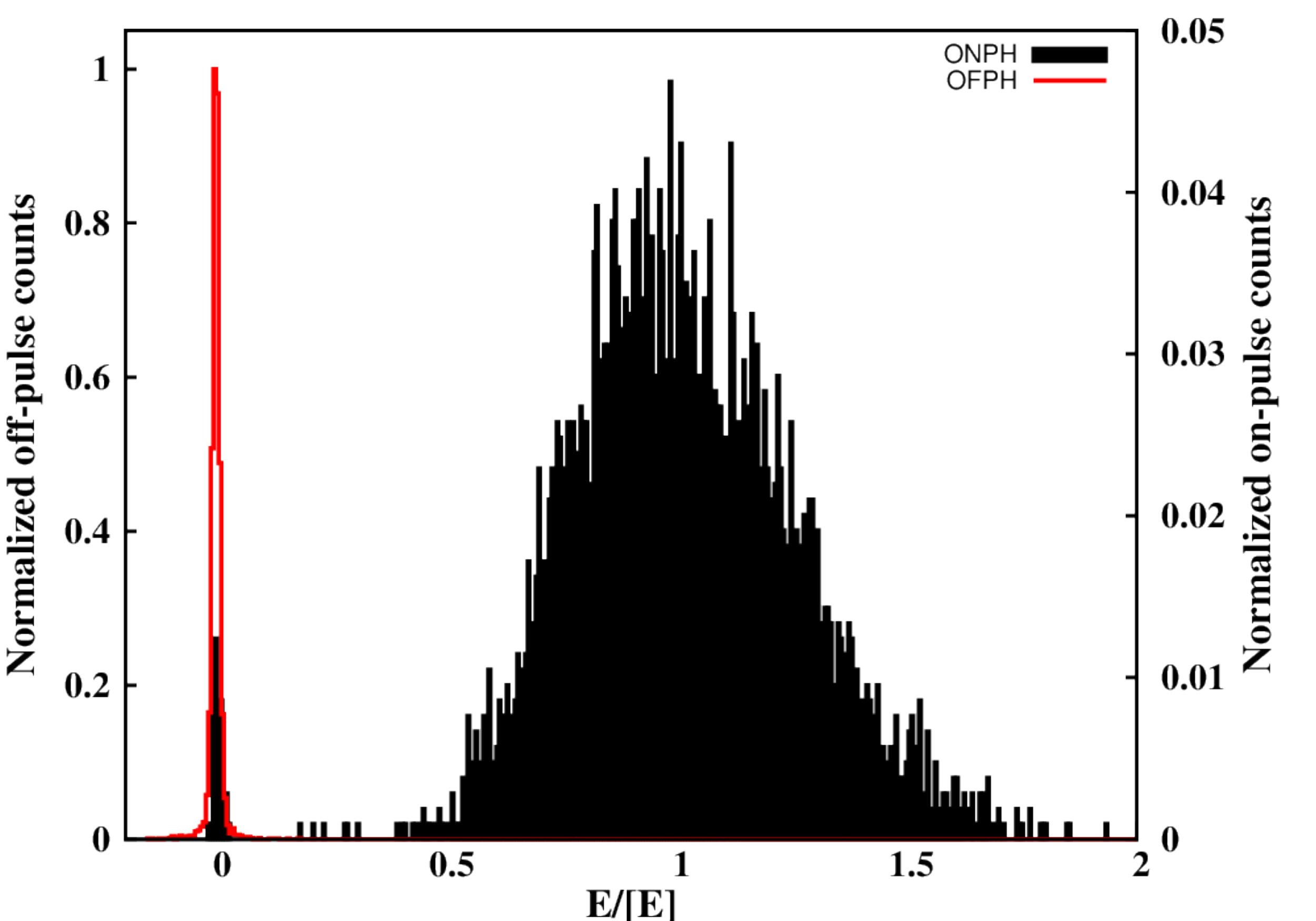}
\begin{picture}(0,0)
 \put(-220,0){\bf PSR B0809+74}
 \put(-115,-40){\scriptsize \bf Freq : 325 MHz}
 \put(-115,-50){\scriptsize \bf N : 13766 pulses}
 \put(-115,-60){\scriptsize \bf NF : 1.0$\pm$0.4\%}
\end{picture}
\caption[The ONPH and the OFPH for PSR B0809+74 observed at 325 MHz]
{The ONPH and the OFPH for PSR B0809+74 observed at 325 MHz. The on-pulse and the off-pulse 
energies were binned in around 500 bins. Note the clear 
bimodal distribution of the ONPH due to the small fraction of null pulses.}
\label{NF_b0809}
\end{figure}
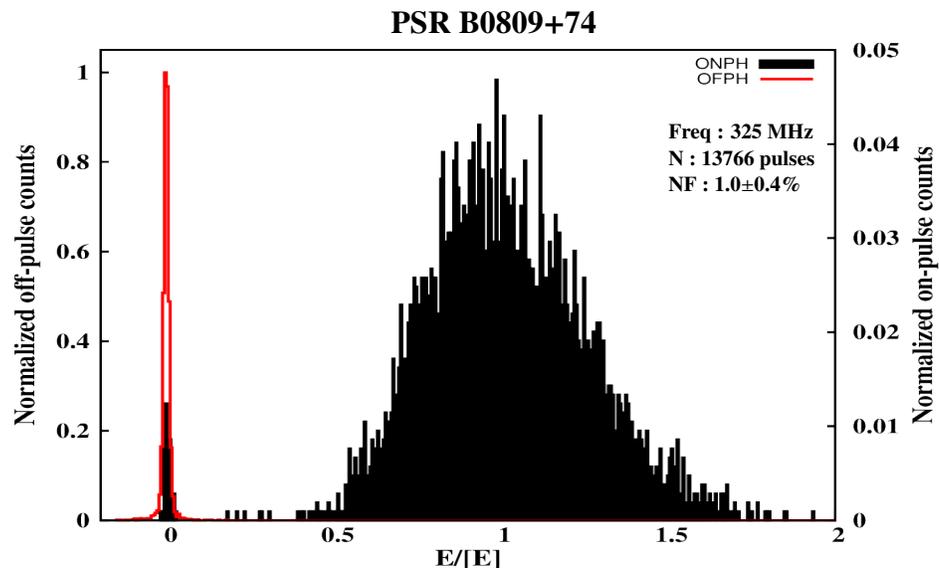
\begin{figure}[h!]
\centering
\includegraphics[width=3.4in, height=5.2in, angle=-90,bb=0 0 504 720]{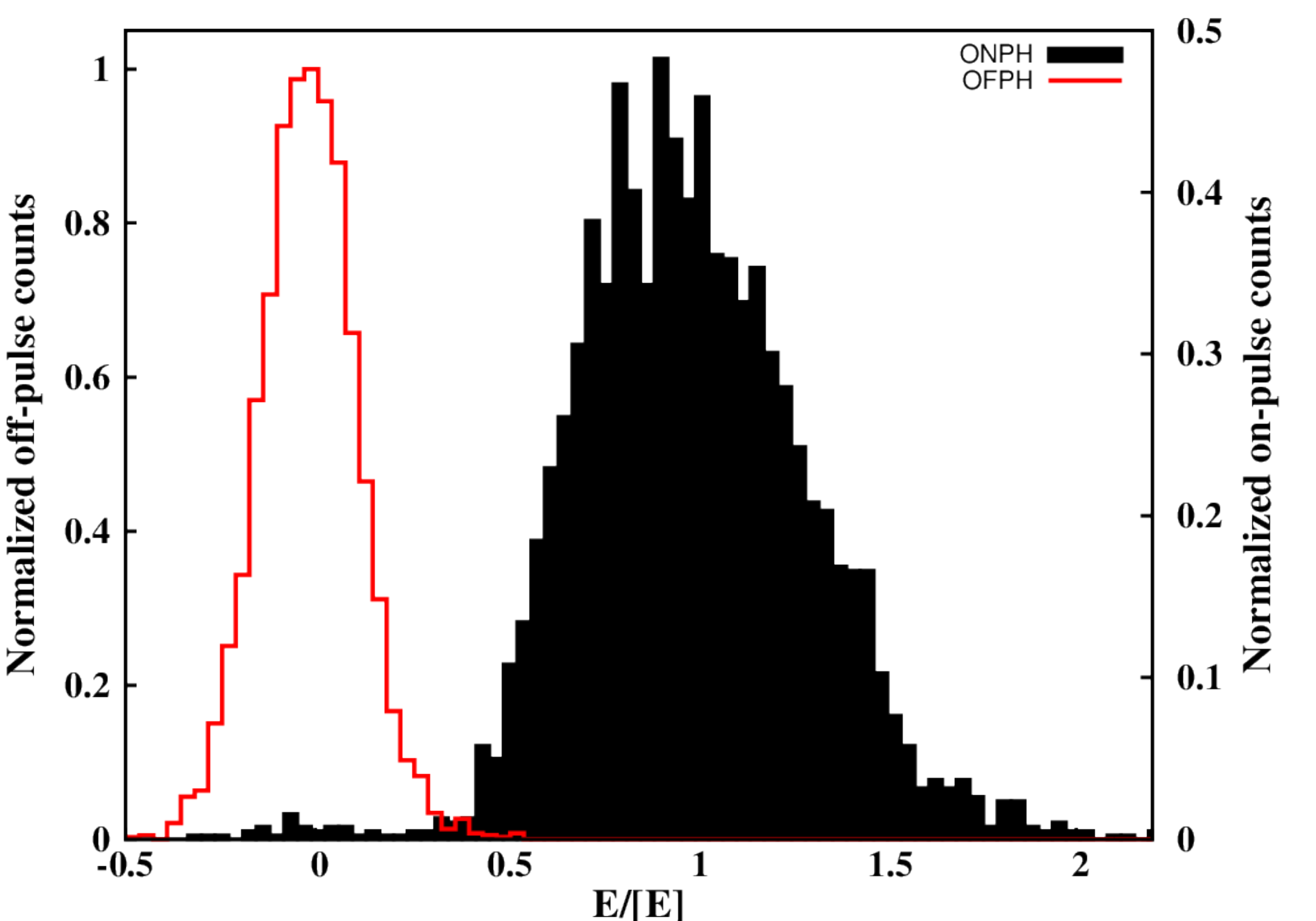}
\begin{picture}(0,0)
 \put(-220,0){\bf PSR B0818$-$13}
 \put(-110,-40){\scriptsize \bf Freq : 610 MHz}
 \put(-110,-50){\scriptsize \bf N : 3341 pulses}
 \put(-110,-60){\scriptsize \bf NF : 0.9$\pm$1.8\%}
\end{picture}
\caption[The ONPH and the OFPH for PSR B0818$-$13 observed at 610 MHz]
{The ONPH and the OFPH for PSR B0818$-$13 observed at 610 MHz. The on-pulse and the off-pulse 
energies were binned in around 100 bins. The ONPH does 
not show very clear separation of null and burst pulses 
due to the weak burst pulses.}
\label{NF_b0818}
\end{figure}
\begin{figure}[h!]
 \centering
\includegraphics[width=3.4in, height=5.2in, angle=-90,bb=0 0 504 720]{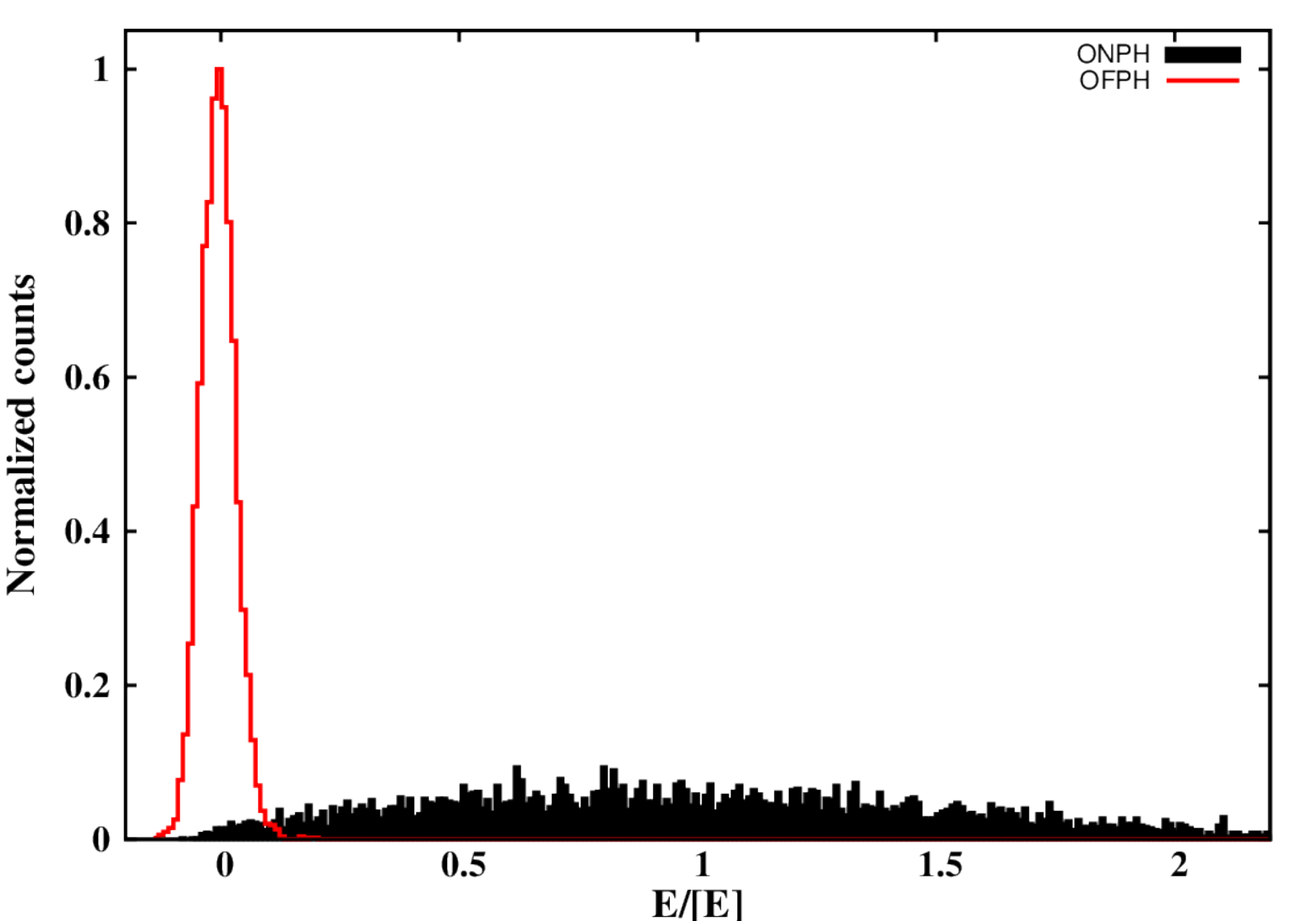}
\begin{picture}(0,0)
 \put(-220,0){\bf PSR B0835$-$41}
 \put(-80,-40){\scriptsize \bf Freq : 610 MHz}
 \put(-80,-50){\scriptsize \bf N : 3335 pulses}
 \put(-80,-60){\scriptsize \bf NF : 1.7$\pm$1.2\%}
\end{picture}
\caption[The ONPH and the OFPH for PSR B0835$-$41 observed at 610 MHz]
{The ONPH and the OFPH for PSR B0835$-$41 observed at 610 MHz. The on-pulse and the off-pulse 
energies were binned in around 300 bins. The ONPH does 
not show very clear separation of null and burst pulses 
due to the weak burst pulses.}
\label{NF_b0835}
\end{figure}

\vspace{0.2cm}

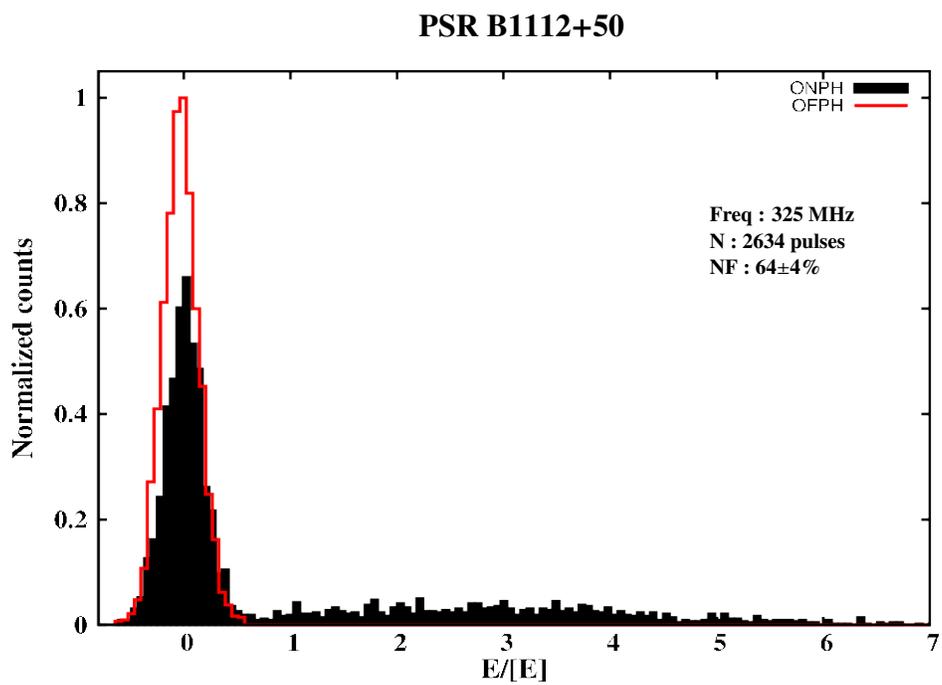
\begin{figure}[h!]
 \centering
\includegraphics[width=3.3in, height=5.0in, angle=-90,bb=0 0 504 720]{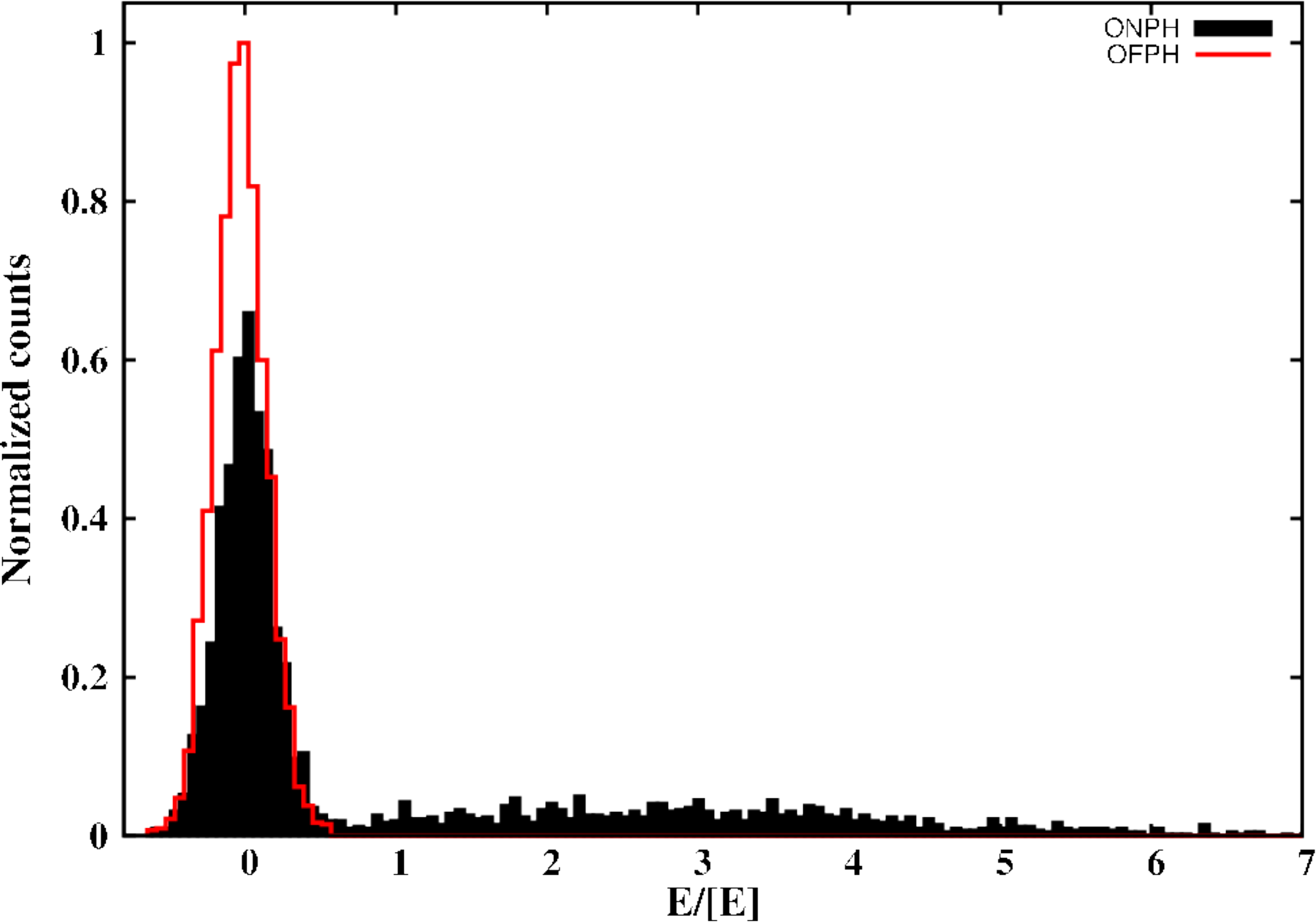}
\caption[The ONPH and the OFPH for PSR B1112+50  observed at 325 MHz]
{The ONPH and the OFPH for PSR B1112+50 observed at 325 MHz. The on-pulse and the off-pulse 
energies were binned in around 200 bins. The ONPH does 
show a very clear bimodal distribution due to the mixture of 
null pulses and burst pulses.}
\begin{picture}(370,60)
 \put(160,360){\bf PSR B1112+50}
 \put(270,290){\scriptsize \bf Freq : 325 MHz}
 \put(270,280){\scriptsize \bf N : 2634 pulses}
 \put(270,270){\scriptsize \bf NF : 64$\pm$4\%}
\end{picture} 
\label{NF_b1112}
\end{figure}

\begin{figure}[h!]
 \centering
\includegraphics[width=3.4in, height=5.2in, angle=-90,bb=0 0 504 720]{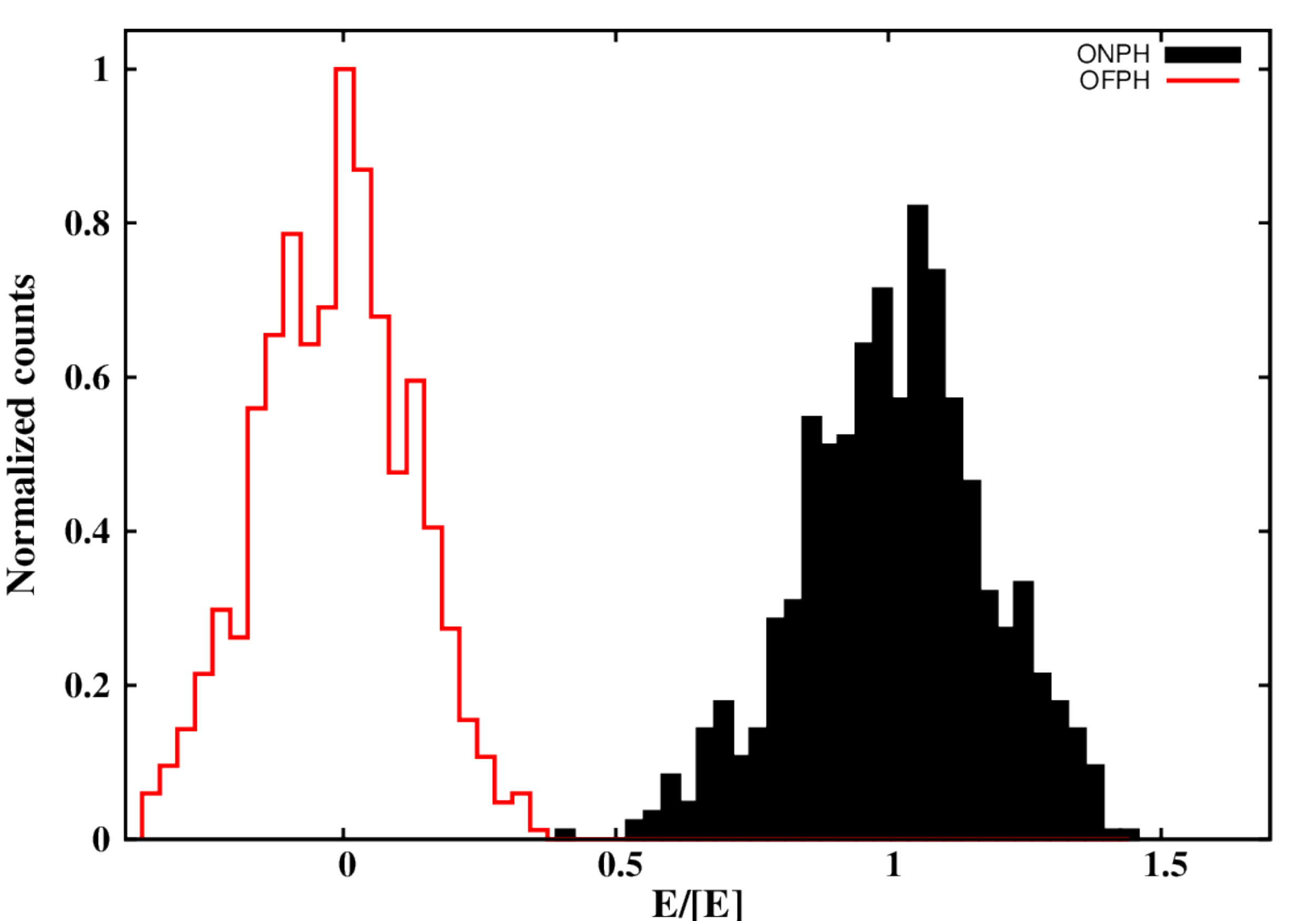}
\caption[The ONPH and the OFPH for PSR J1639$-$4359 observed at 610 MHz]
{The ONPH and the OFPH for PSR J1639$-$4359 observed at 610 MHz. The on-pulse and the off-pulse 
energies were binned in around 100 bins. To obtain these histograms, successive 5 pulses 
were sub-integrated (see Section \ref{sect_j1639}). The NF, displayed in the inset texts, 
was estimated by arranging single pulses in the ascending order of their 
on-pulse energy and separating low energy single pulses. These histograms are 
presented here to demonstrate absence of null pulses among the sub-integrated pulses.}
\begin{picture}(370,60)
 \put(160,400){\bf PSR J1639$-$4359}
 \put(290,320){\scriptsize \bf Freq : 610 MHz}
 \put(290,310){\scriptsize \bf N : 13034 (16) pulses}
 \put(290,300){\scriptsize \bf NF : $\leq$ 0.1\%}
\end{picture}
\label{NF_j1639}
\end{figure}

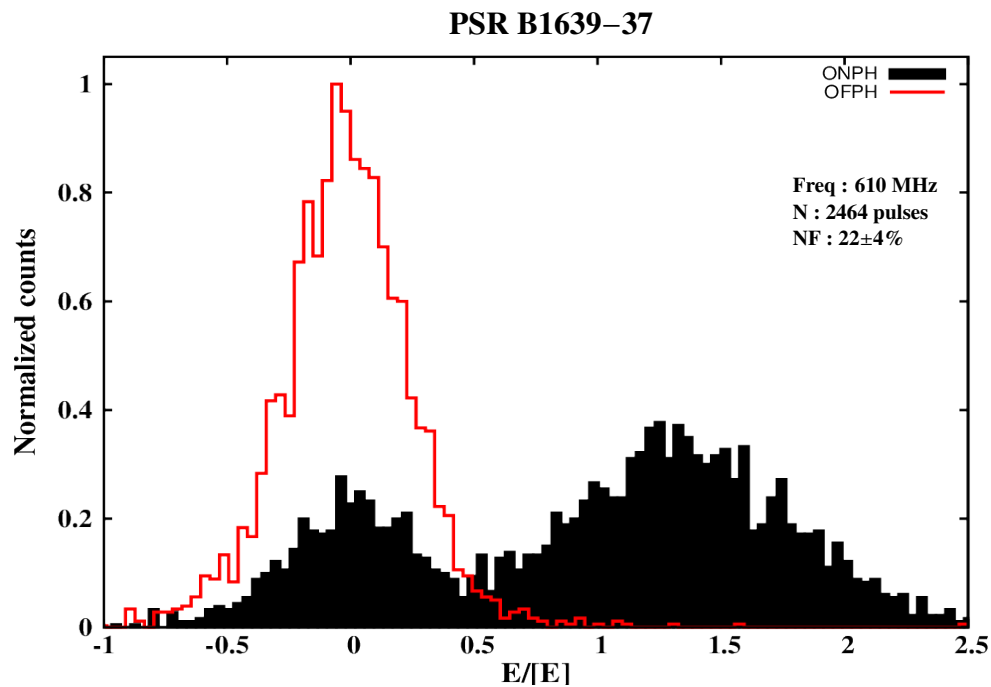
\begin{figure}[h!]
 \centering
 \includegraphics[width=3.4in, height=5.2in, angle=-90,bb=0 0 504 720]{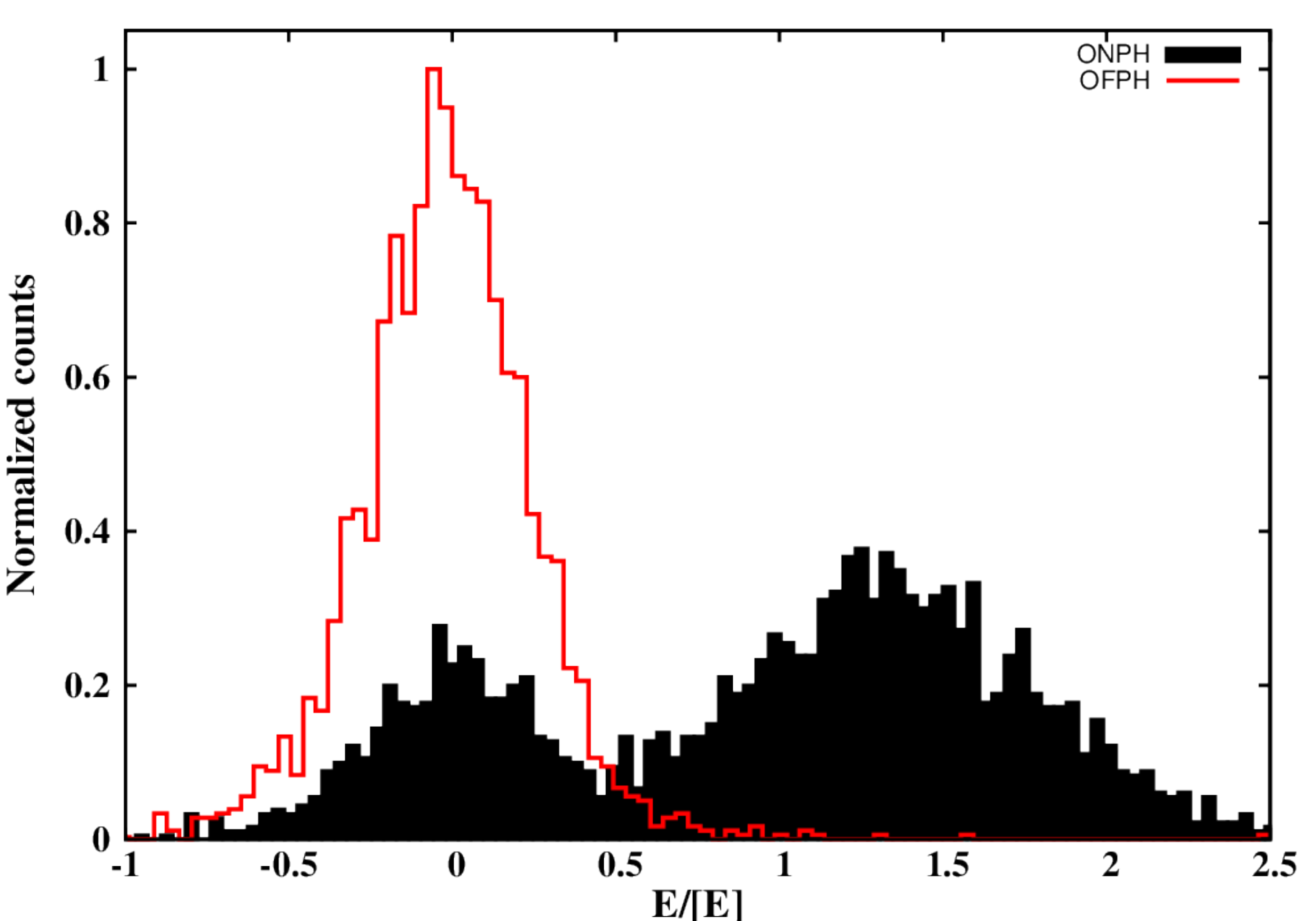}
 % j1701.histogram.eps: 479x744 pixel, 72dpi, 16.90x26.25 cm, bb=0 0 479 744
\caption[The ONPH and the OFPH for PSR B1658$-$37 observed at 610 MHz]
{The ONPH and the OFPH for PSR B1658$-$37 observed at 610 MHz. The on-pulse and the off-pulse 
energies were binned in around 200 bins. The ONPH does show very 
clear bimodal distribution originating from the null pulses and 
relatively stronger burst pulses.}
\begin{picture}(0,0)
 \put(-20,320){\bf PSR B1639$-$37}
 \put(110,260){\scriptsize \bf Freq : 610 MHz}
 \put(110,250){\scriptsize \bf N : 2464 pulses}
 \put(110,240){\scriptsize \bf NF : 22$\pm$4\%}
\end{picture} 
\label{NF_j1701}
\end{figure}

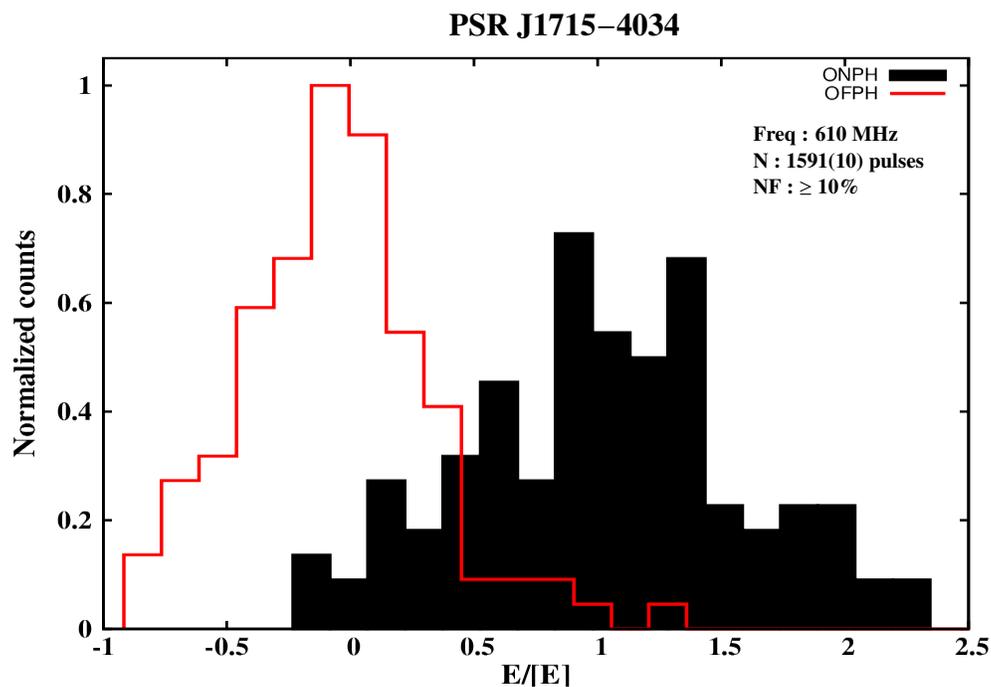
\begin{figure}[h!]
 \centering
 \includegraphics[width=3.4in,height=5.2in, angle=-90,bb=0 0 504 720]{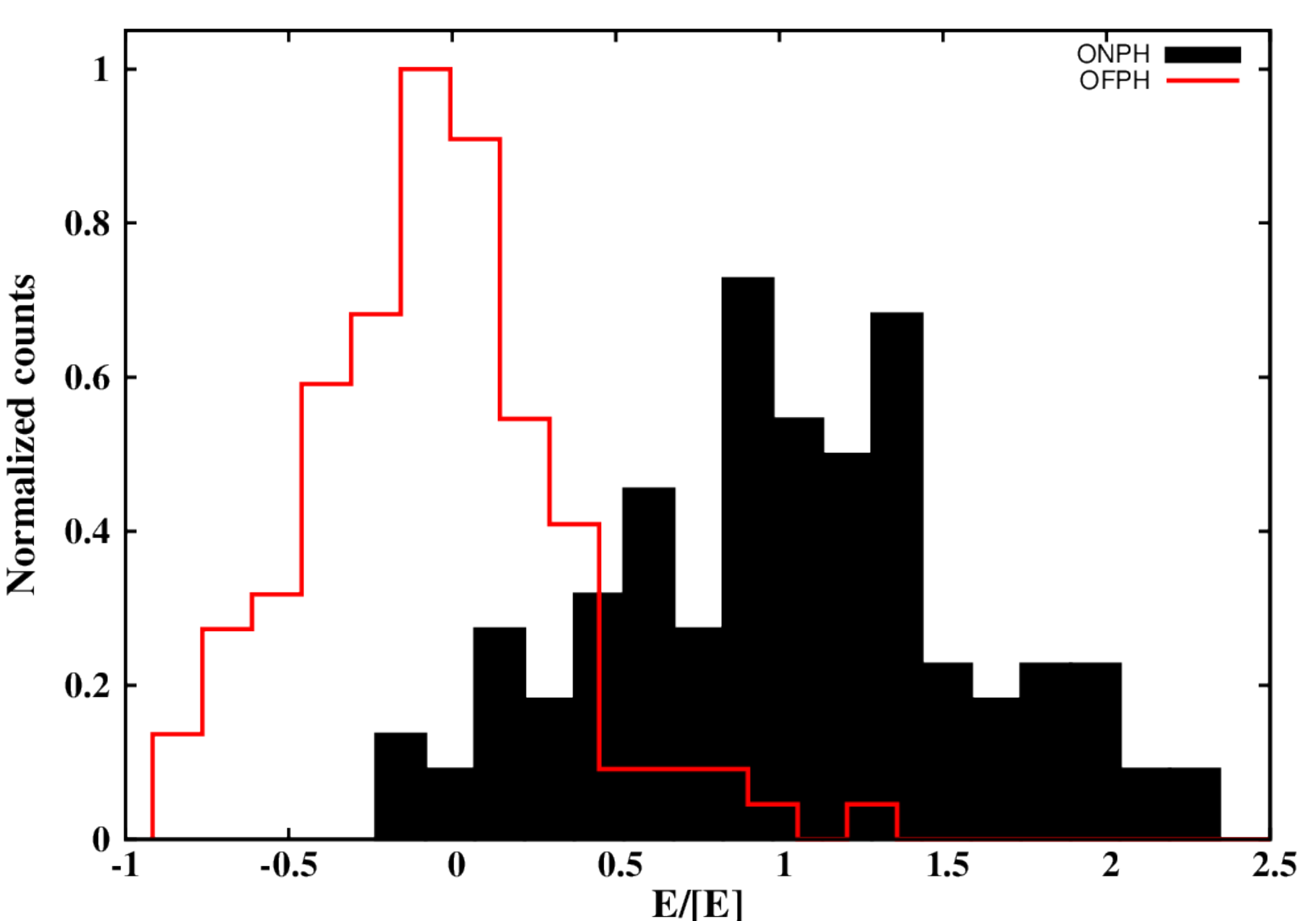} 
 % j1701.histogram.eps.pdf: 479x744 pixel, 72dpi, 16.90x26.25 cm, bb=0 0 479 744
 \caption[The ONPH and the OFPH for PSR J1715$-$4034 observed at 610 MHz]
 {The ONPH and the OFPH for PSR J1715$-$4034 observed at 610 MHz. 
 The on-pulse and the off-pulse energies were binned in around 25 bins. 
 To obtain these histograms, successive 10 pulses 
 were sub-integrated (see Section \ref{sect_j1715}). Thus, NF shown 
 in the inset text, is only a lower limit on the true NF.}
\begin{picture}(0,0)
 \put(-20,320){\bf PSR J1715$-$4034}
 \put(95,280){\scriptsize \bf Freq : 610 MHz}
 \put(95,270){\scriptsize \bf N : 1591(10) pulses}
 \put(95,260){\scriptsize \bf NF : $\geq$ 10\%}
\end{picture}
 \label{NF_j1715}
\end{figure}

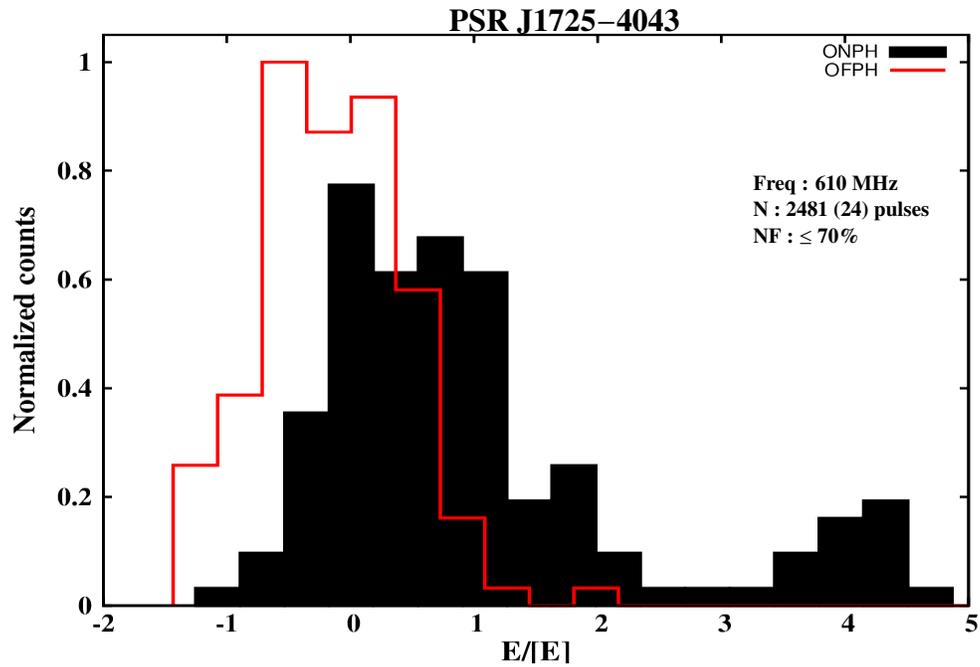
\begin{figure}[h!]
 \centering
 \includegraphics[width=3.4in,height=5.2in, angle=-90,bb=0 0 504 720]{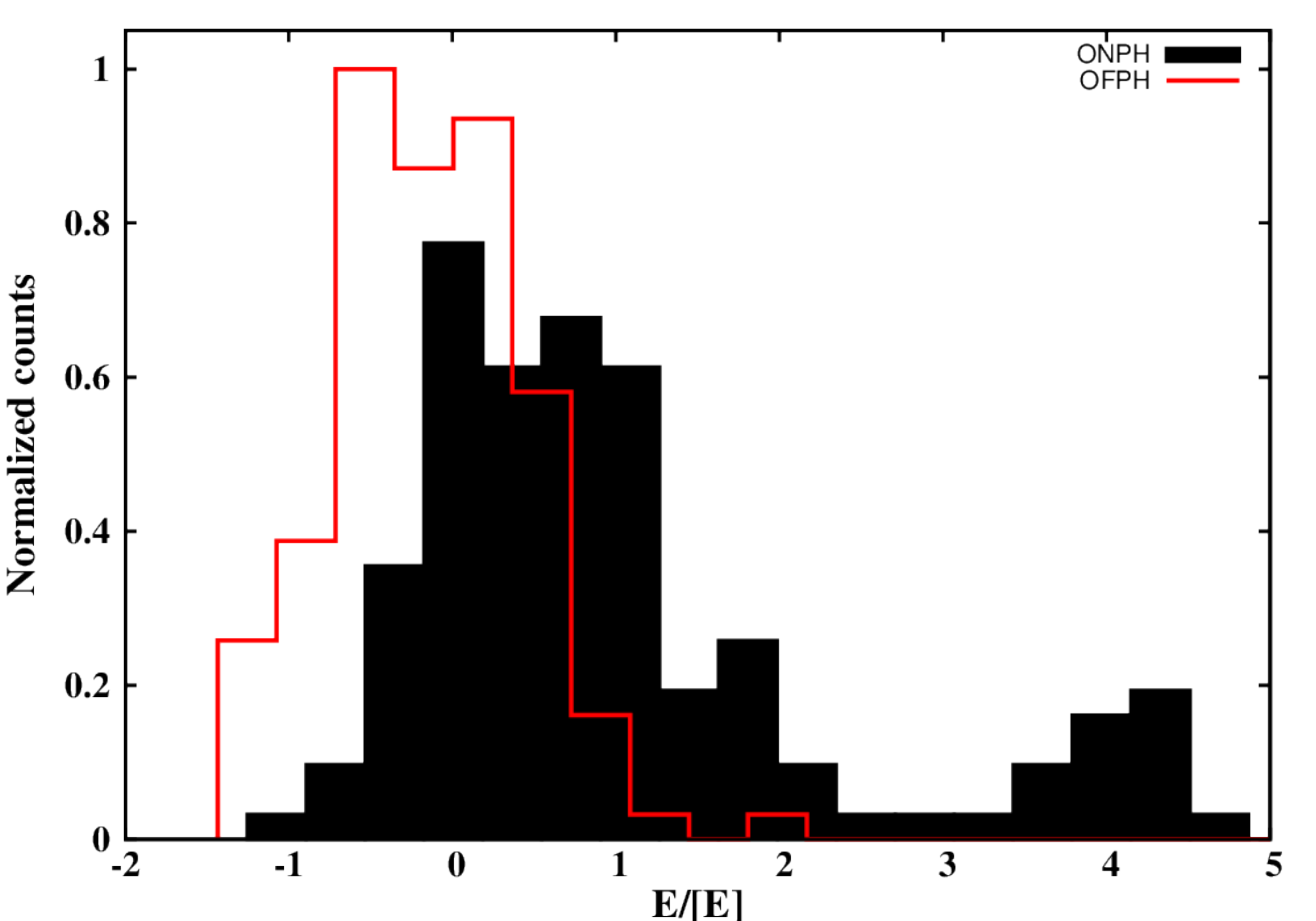}
 \caption[The ONPH and the OFPH for PSR J1725$-$4043 observed at 610 MHz]
 {The ONPH and the OFPH for PSR J1725$-$4043 observed at 610 MHz. 
 The on-pulse and the off-pulse energy were binned in around 20 bins. 
 To obtain these histograms, successive 24 pulses 
 were sub-integrated (see Section \ref{sect_j1715}). 
 The ONPH shows mixture of true null 
 pulses and weak Mode B pulses near the zero pulse energy. 
 The on-pulse energy from strong Mode A pulses clearly shows 
 distribution away from the zero pulse energy.}
 \begin{picture}(0,0)
 \put(-20,340){\bf PSR J1725$-$4043}
 \put(95,280){\scriptsize \bf Freq : 610 MHz}
 \put(95,270){\scriptsize \bf N : 2481 (24) pulses}
 \put(95,260){\scriptsize \bf NF : $\leq$ 70\%}
\end{picture}
 \label{NF_j1725}
\end{figure}

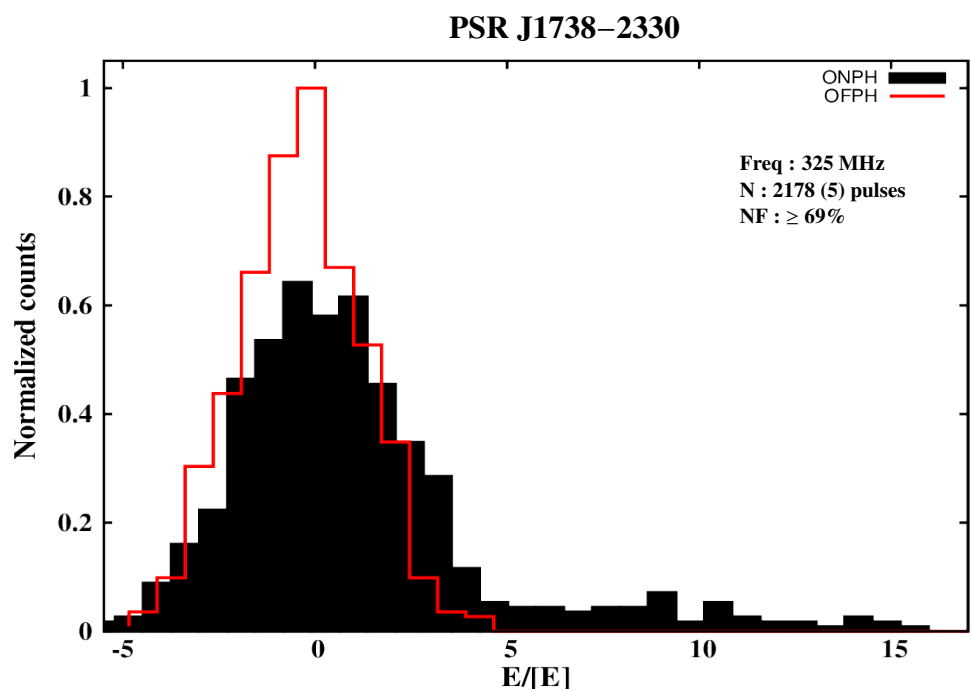
\begin{figure}[h!]
 \centering
 \includegraphics[width=3.4in,height=5.2in, angle=-90,bb=0 0 504 720]{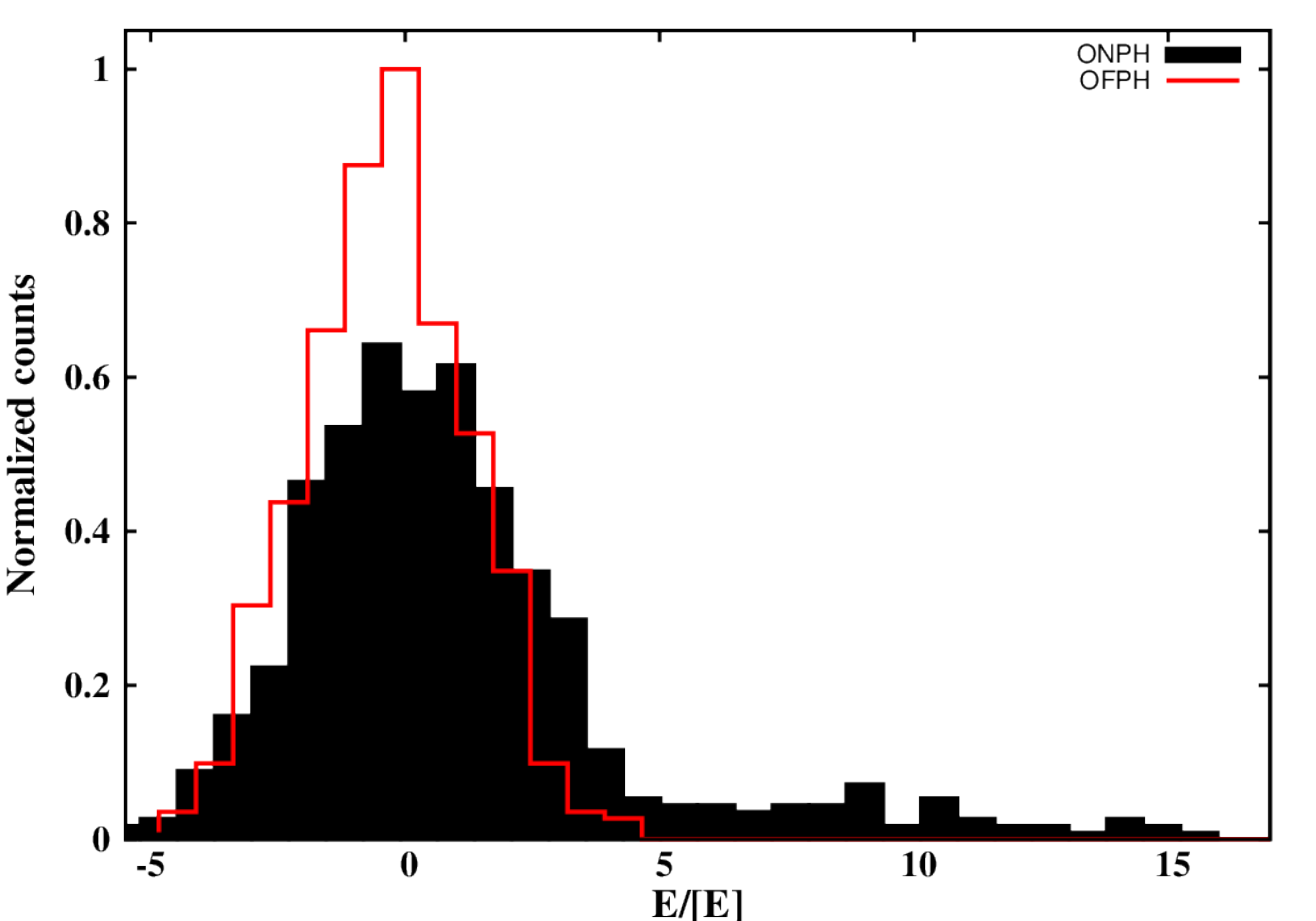}
 \caption[The ONPH and the OFPH for PSR J1738$-$2330 observed at 325 MHz]
 {The ONPH and the OFPH for PSR J1738$-$2330 observed at 325 MHz. 
 The on-pulse and the off-pulse energies were binned in around 40 bins. 
 To obtain these histograms, successive 5 pulses were 
 sub-integrated (see Section \ref{sect_j1715}). 
 The ONPH shows a large fraction of null pulses.}
 \begin{picture}(0,0)
 \put(-20,320){\bf PSR J1738$-$2330}
 \put(90,270){\scriptsize \bf Freq : 325 MHz}
 \put(90,260){\scriptsize \bf N : 2178 (5) pulses}
 \put(90,250){\scriptsize \bf NF : $\geq$ 69\%}
\end{picture}
\label{NF_j1738}
\end{figure}

\begin{figure}[h!]
\centering 
\includegraphics[width=3.4in,height=5.2in, angle=-90,bb=0 0 504 720]{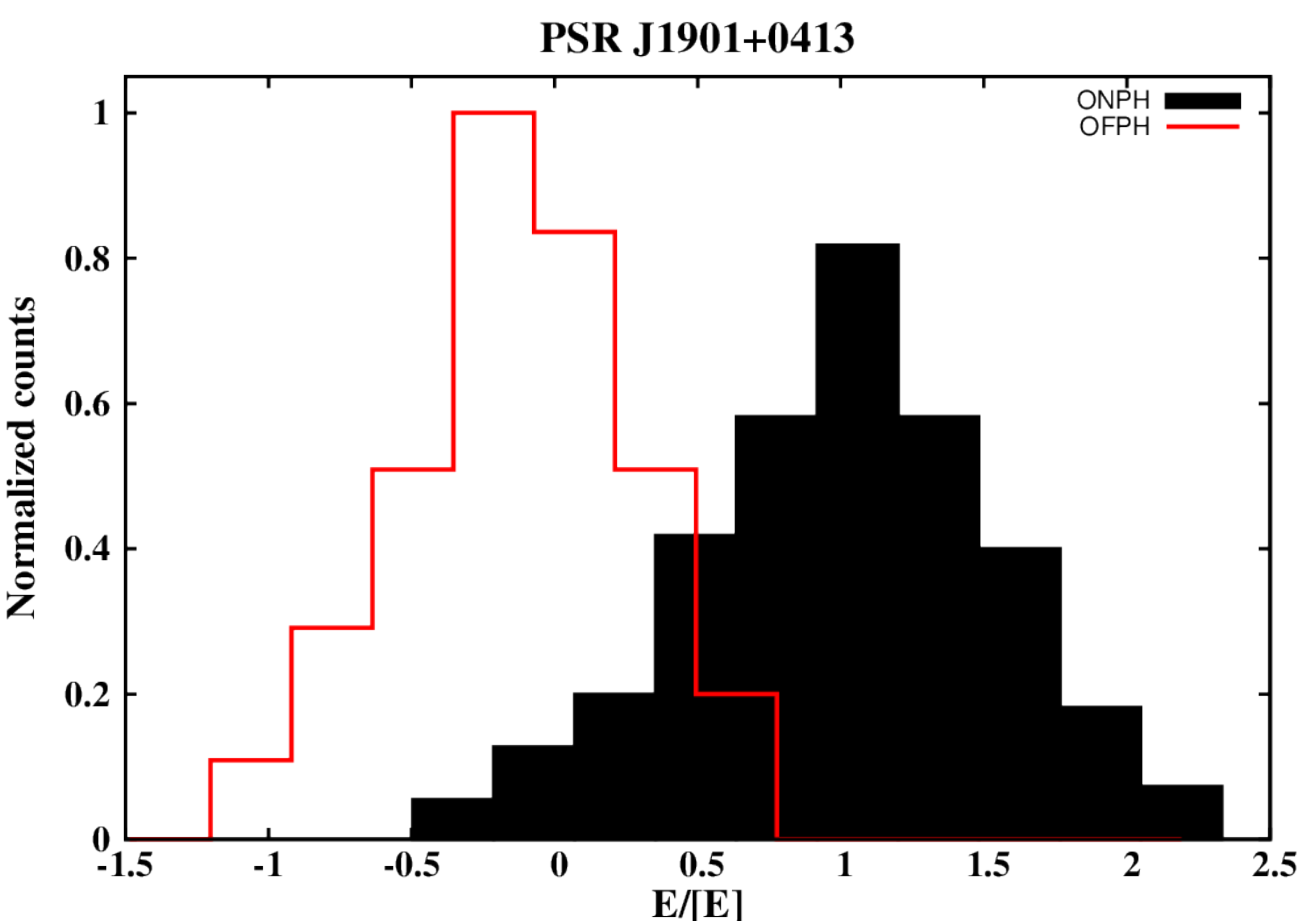}
\caption[The ONPH and the OFPH for PSR J1901+0413 observed at 610 MHz]
{The ONPH and the OFPH for PSR J1901+0413 observed at 610 MHz. The on-pulse and the off-pulse 
energies were binned in around 15 bins. To obtain these histograms, successive 10 pulses 
were sub-integrated (see Section \ref{sect_j1639}). The NF, displayed in the inset texts, 
was estimated by arranging single pulses in the ascending order of their 
on-pulse energy and separating low energy single pulses. These histograms are 
presented here to demonstrate small fraction of null pulses among the sub-integrated pulses.}
\begin{picture}(0,0)
 \put(105,290){\scriptsize \bf Freq : 610 MHz}
 \put(105,280){\scriptsize \bf N : 2605 (10) pulses}
 \put(105,270){\scriptsize \bf NF : $\leq$ 6\%}
\end{picture}
\label{NF_j1901}
\end{figure}

\begin{figure}[h!]
\centering
\includegraphics[width=3.4in,height=5.2in, angle=-90,bb=0 0 504 720]{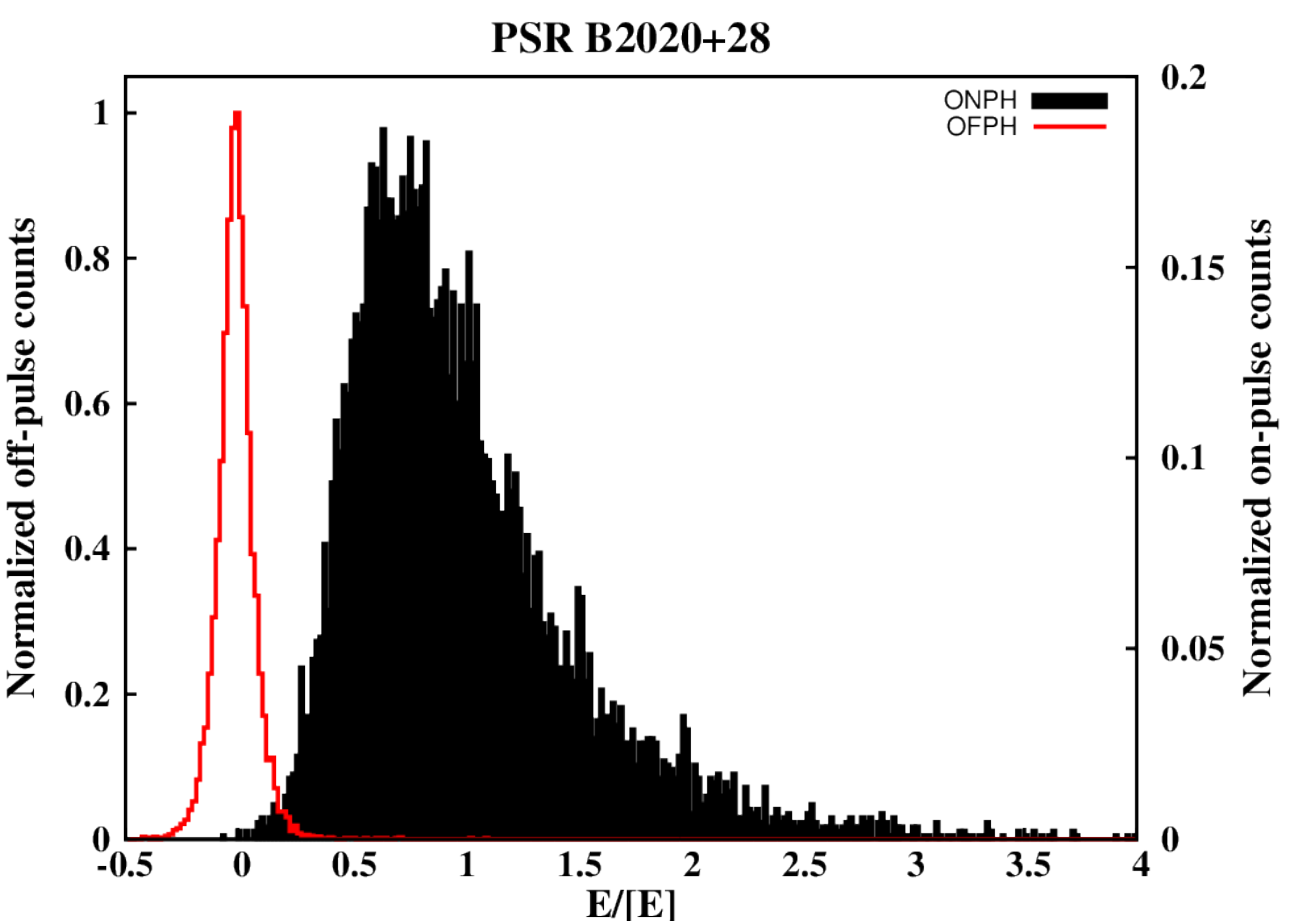}
\caption[The ONPH and the OFPH for PSR B2020+28 observed at 610 MHz]
{The ONPH and the OFPH for PSR B2020+28 observed at 610 MHz. The on-pulse and the off-pulse 
energies were binned in around 300 bins. The obtained NF along with the number 
of pulses used during the analysis are displayed in the inset texts.}
\begin{picture}(0,0)
 \put(80,250){\scriptsize \bf Freq : 610 MHz}
 \put(80,240){\scriptsize \bf N : 8039 pulses}
 \put(80,230){\scriptsize \bf NF : 0.2$\pm$1.6\%}
\end{picture}
\label{NF_b2020}
\end{figure}

\begin{figure}[h!]
\centering
\includegraphics[width=3.4in,height=5.2in, angle=-90,bb=0 0 504 720]{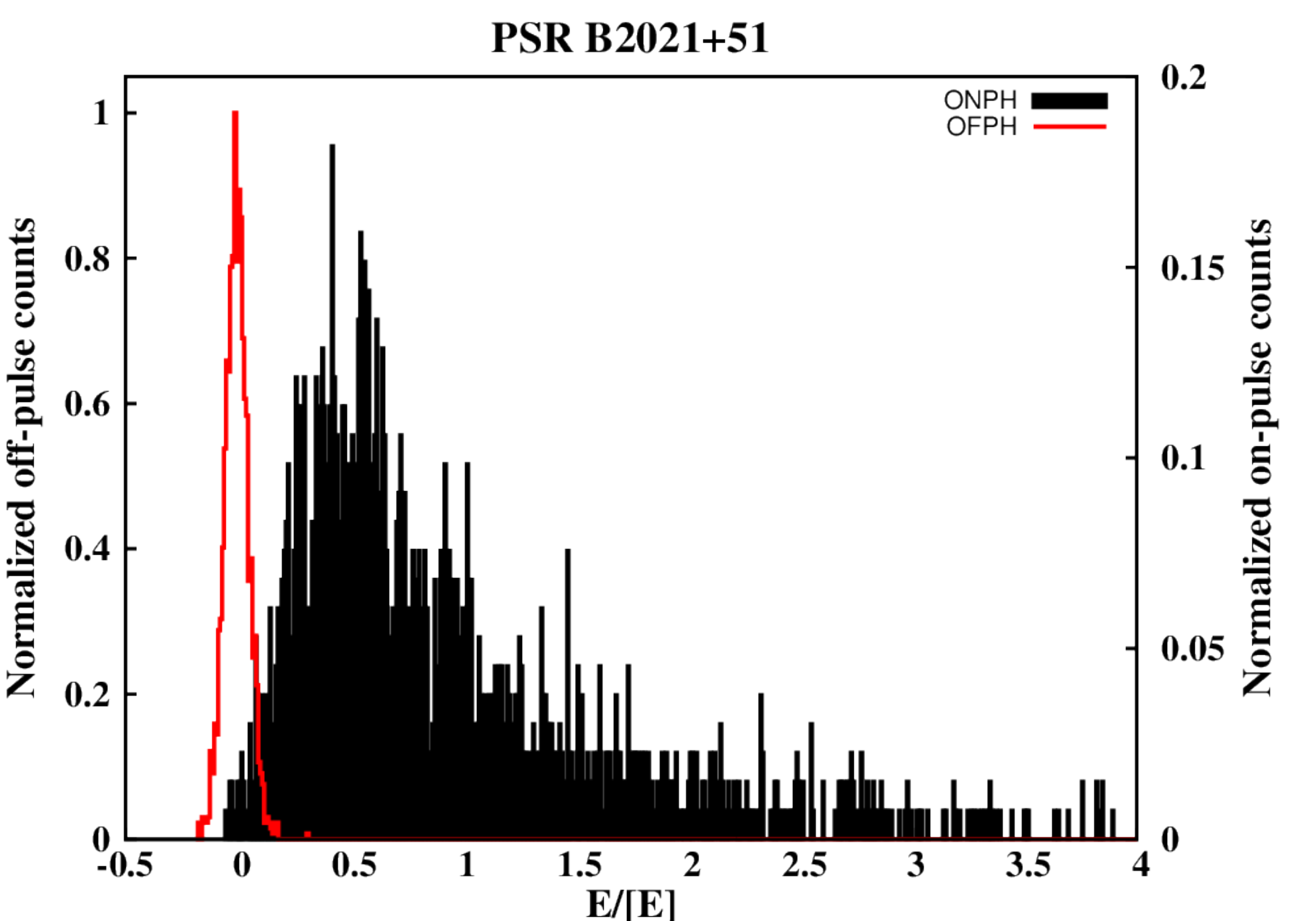}
\caption[The ONPH and the OFPH for PSR B2021+51 observed at 610 MHz]
{The ONPH and the OFPH for PSR B2021+51 observed at 610 MHz. The on-pulse and the off-pulse 
energies were binned in around 800 bins. The obtained NF along with the number 
of pulses used during the analysis are displayed in the inset texts.}
\begin{picture}(0,0)
 \put(80,250){\scriptsize \bf Freq : 610 MHz}
 \put(80,240){\scriptsize \bf N : 1326 pulses}
 \put(80,230){\scriptsize \bf NF : 1.4$\pm$0.7\%}
\end{picture}
\label{NF_b2021}
\end{figure}

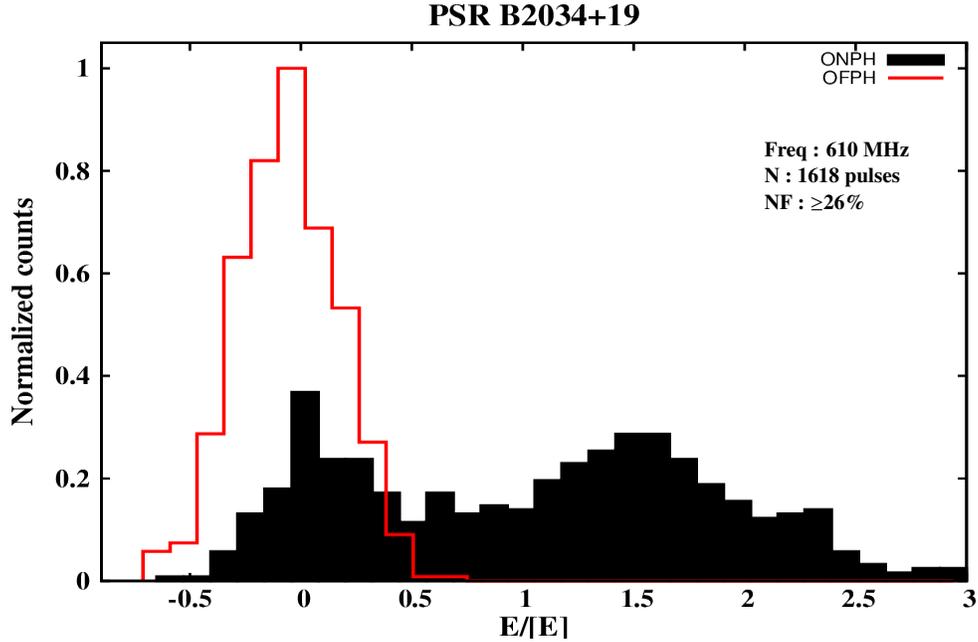
\begin{figure}[h!]
\centering
\includegraphics[width=3.4in,height=5.2in, angle=-90,bb=0 0 504 720]{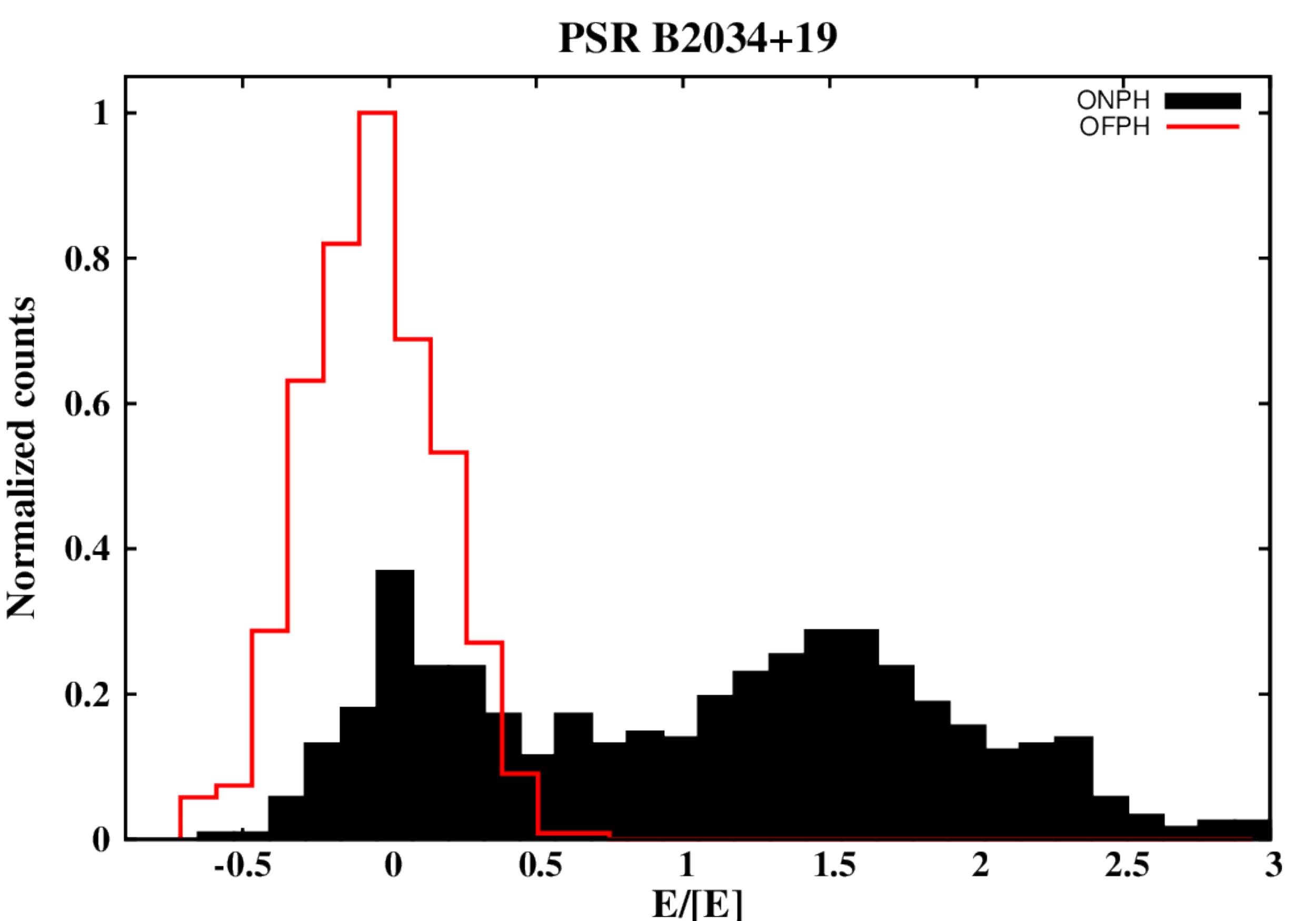}
\caption[The ONPH and the OFPH for PSR B2034+19 observed at 610 MHz]
{The ONPH and the OFPH for PSR B2034+19 observed at 610 MHz. The on-pulse and the off-pulse 
energies were binned in around 30 bins. To obtain these histograms, successive 3 pulses 
were sub-integrated (see Section \ref{sect_j1639}). The ONPH shows clear bi-modal 
distribution of the on-pulse energy in the sub-integrated data originating 
from the null pulses and burst pulses.}
\begin{picture}(0,0)
 \put(100,270){\scriptsize \bf Freq : 610 MHz}
 \put(100,260){\scriptsize \bf N : 1618 pulses}
 \put(100,250){\scriptsize \bf NF : $\geq$26\%}
\end{picture}
\label{NF_b2037}
\end{figure}

\begin{figure}[h!]
\centering
\includegraphics[width=3.4in,height=5.2in, angle=-90,bb=0 0 504 720]{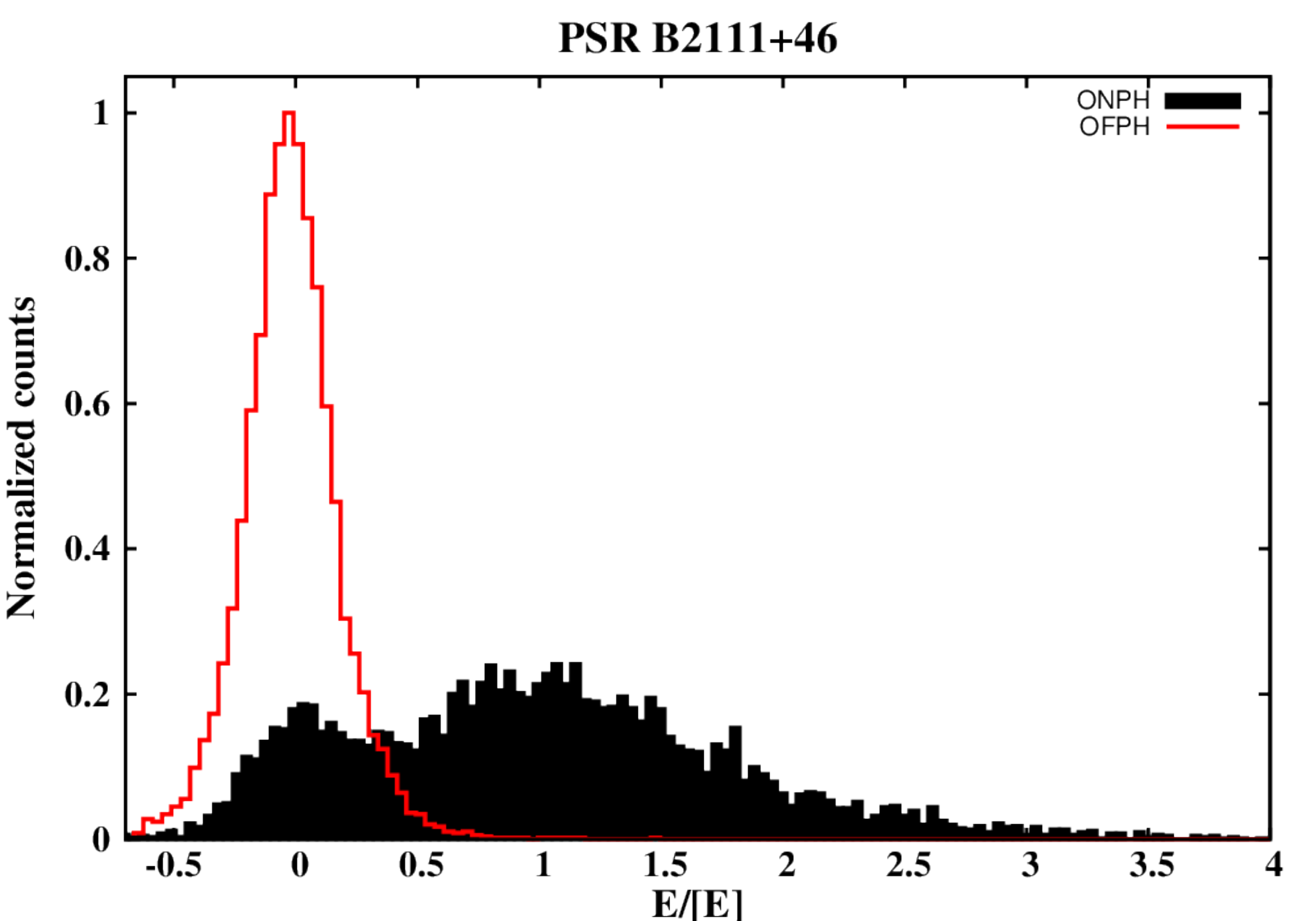}
\caption[The ONPH and the OFPH for PSR B2111+46 observed at 610 MHz]
{The ONPH and the OFPH for PSR B2111+46 observed at 610 MHz. The on-pulse and the off-pulse 
energies were binned in around 200 bins. The obtained NF along with the number 
of pulses used during the analysis are displayed in the inset texts.}
\begin{picture}(0,0)
 \put(100,250){\scriptsize \bf Freq : 610 MHz}
 \put(100,240){\scriptsize \bf N : 6208 pulses}
 \put(100,230){\scriptsize \bf NF : 21$\pm$4\%}
\end{picture}
\label{NF_b2111}
\end{figure}

\begin{figure}[h!]
\centering
\includegraphics[width=3.4in,height=5.2in, angle=-90,bb=0 0 504 720]{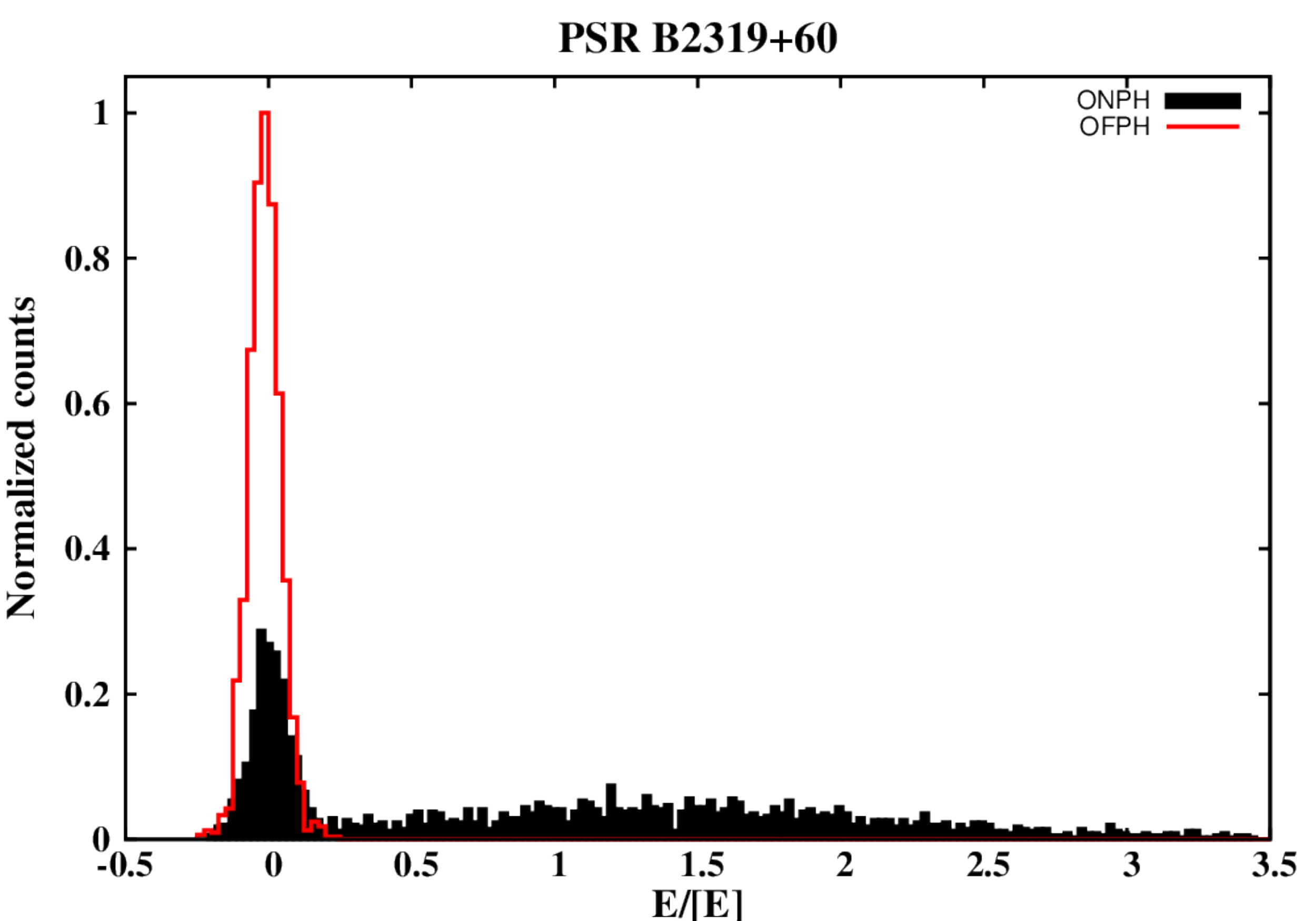}
\caption[The ONPH and the OFPH for PSR B2319+60 observed at 610 MHz]
{The ONPH and the OFPH for PSR B2319+60 observed at 610 MHz. The on-pulse and the off-pulse 
energies were binned in around 200 bins. The obtained NF along with the number 
of pulses used during the analysis are displayed in the inset texts.}
\begin{picture}(0,0)
 \put(100,250){\scriptsize \bf Freq : 610 MHz}
 \put(100,240){\scriptsize \bf N : 1795 pulses}
 \put(100,230){\scriptsize \bf NF : 29$\pm$1\%}
\end{picture}
\label{NF_b2319}
\end{figure}
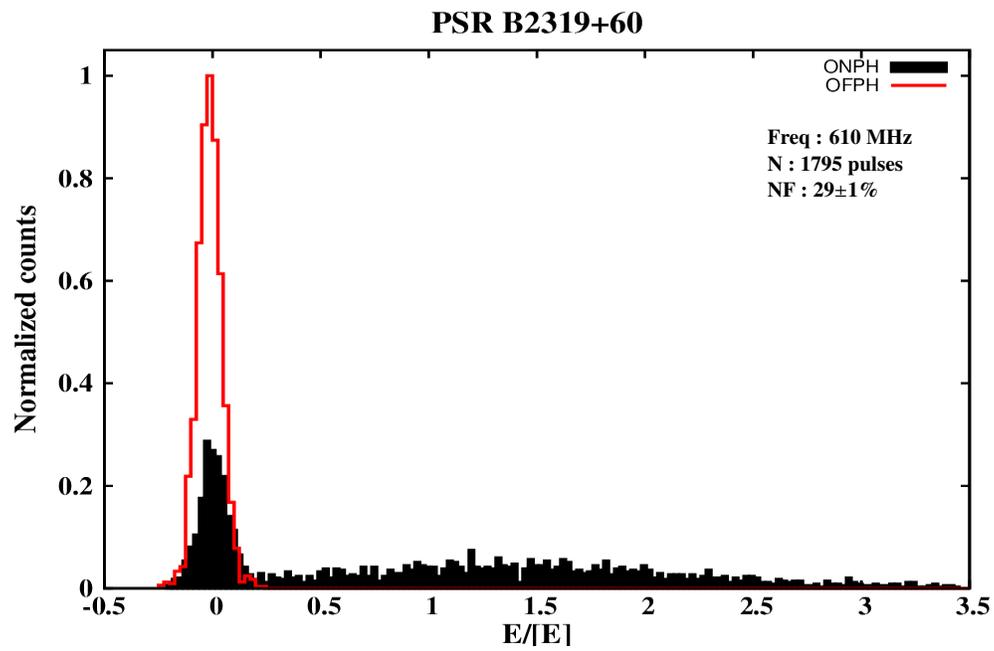

\chapter[Appendix B : Null length and Burst length histograms]{Appendix B \\ Null length and Burst length histograms}
\label{Appendix_NLH_BLH}
This Appendix lists all the null length and the burst length histograms 
obtained for pulsars studied in Chapter 4. The NLHs  
are shown in the left hand panel with red solid line while the BLHs are 
shown on the right hand panel with blue sold line.  
The abscissa present length of contiguous null or burst phases 
in units of pulsar periods. The ordinate presents the normalized 
counts as counts on every bin were normalized by the total number 
of observed contiguous null/burst lengths (mentioned in the caption) 
for each pulsar. The pulsar names are given at the top of the plot. 

\setcounter{figure}{0} \renewcommand{\thefigure}{B.\arabic{figure}} 
\setcounter{equation}{0} \renewcommand{\theequation}{B.\arabic{equation}}

\begin{figure}[h!]
\centering
\includegraphics[width=3.5in,height=6.5in,angle=-90,bb=0 0 504 720]{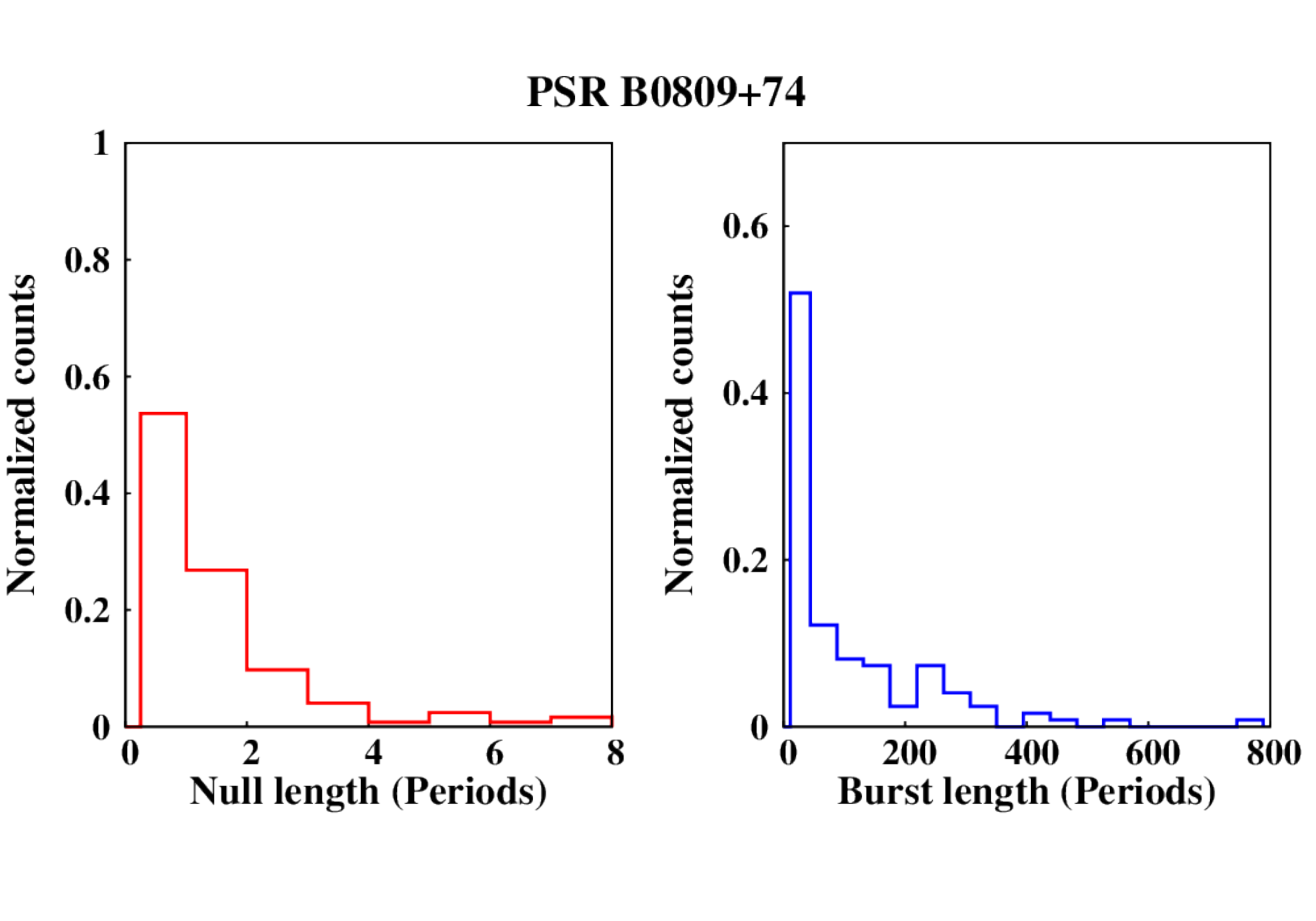}
\caption{The NLH and BLH for PSR B0809+74. Total 73 null lengths and 73 burst lengths were included to construct 
these histograms.}
\label{nlh_blh_b0809}
\end{figure}

\begin{figure}[h!]
\centering
\includegraphics[width=3.5in,height=6.5in,angle=-90,bb=0 0 504 720]{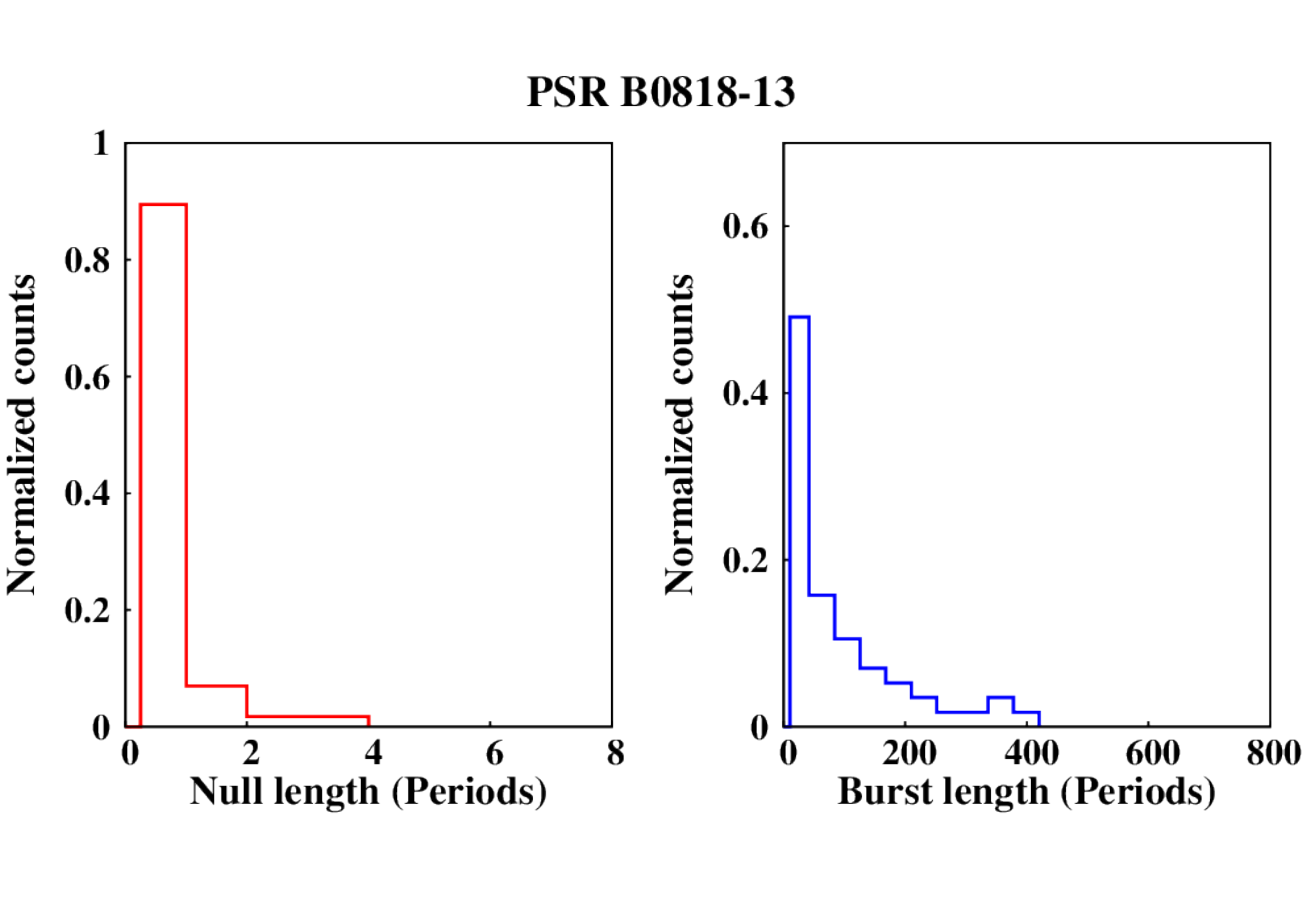}
\caption{The NLH and BLH for PSR B0818$-$13. Total 57 null lengths and 57 burst lengths were included to construct 
these histograms.}
\label{nlh_blh_b0818}
\end{figure}

\begin{figure}[h]
 \centering
 \includegraphics[width=3.5in,height=6.5in,angle=-90,bb=0 0 504 720]{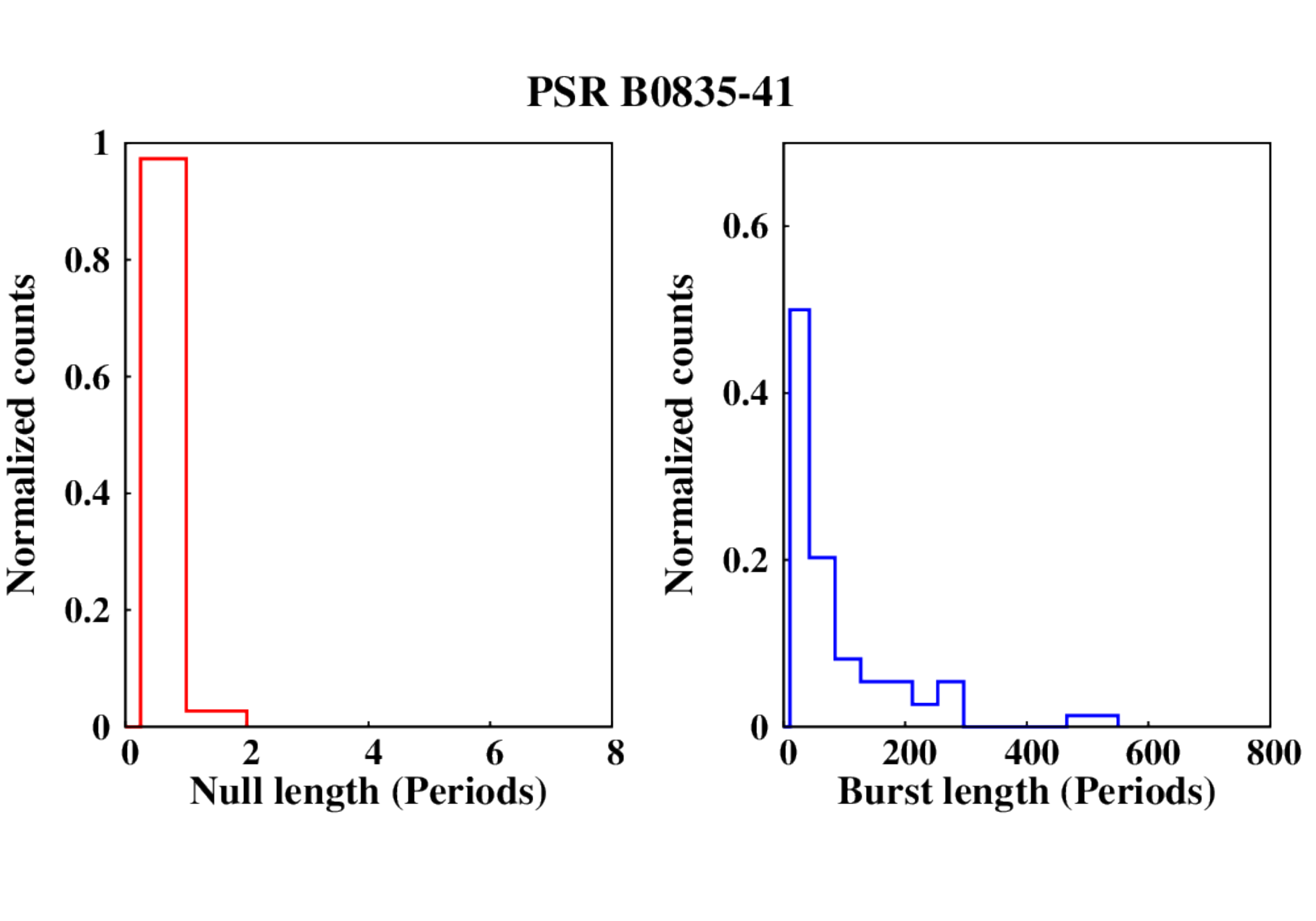}
 % B0837.nlhblh.eps.pdf: 504x720 pixel, 72dpi, 17.78x25.40 cm, bb=0 0 504 720
 \caption{The NLH and BLH for PSR B0835$-$41. Total 74 null lengths and 74 burst lengths were included to construct 
these histograms.}
 \label{nlh_blh_b0835}
\end{figure}

\begin{figure}[h!]
\centering
\includegraphics[width=3.5in,height=6.5in,angle=-90,bb=0 0 504 720]{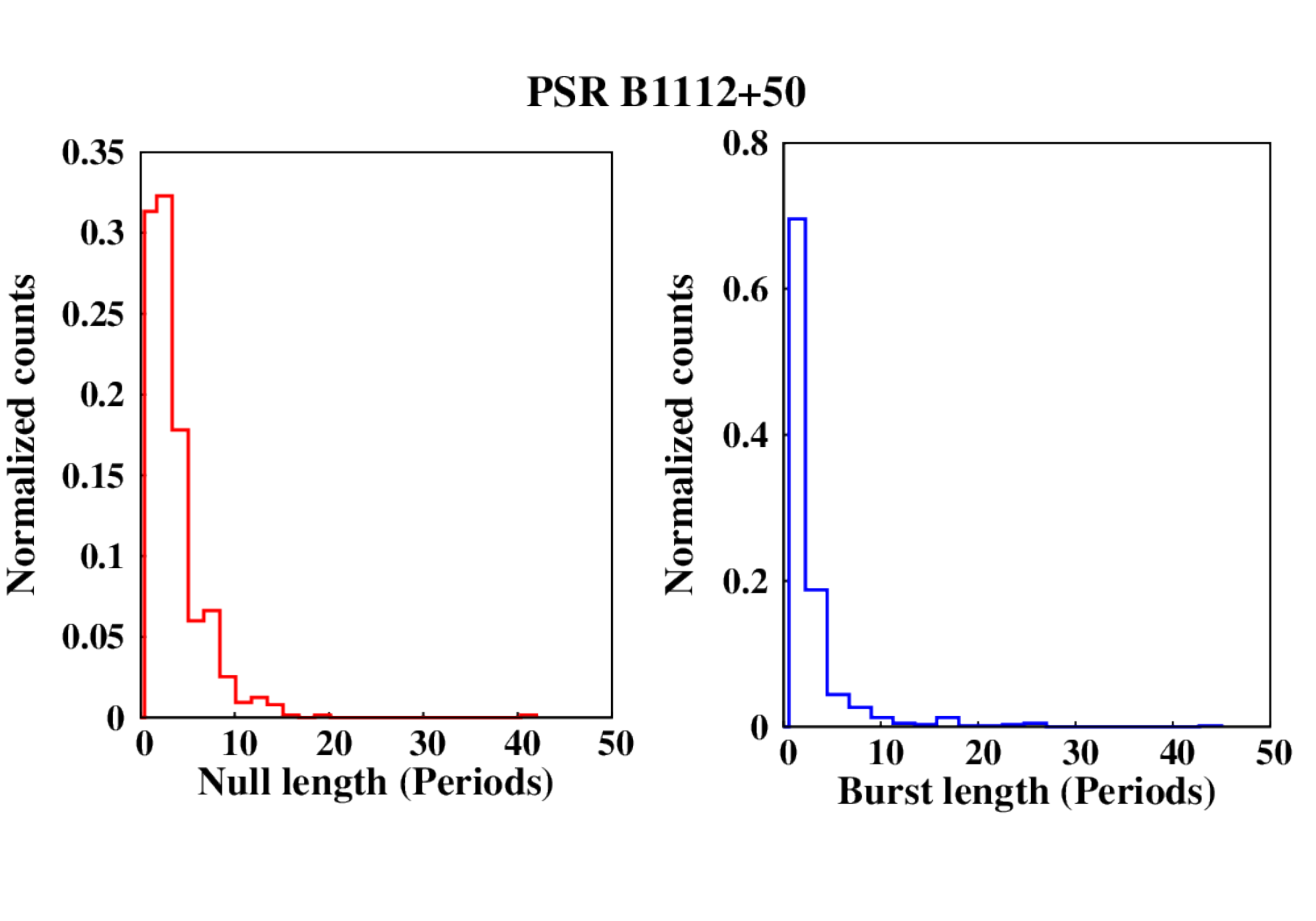}
\caption{The NLH and BLH for PSR B1112+50. Total 635 null lengths and 635 burst lengths were included to construct 
these histograms.}
\label{nlh_blh_b1112}
\end{figure}

\begin{figure}[h!]
\centering
\includegraphics[width=3.5in,height=6.5in,angle=-90,bb=0 0 504 720]{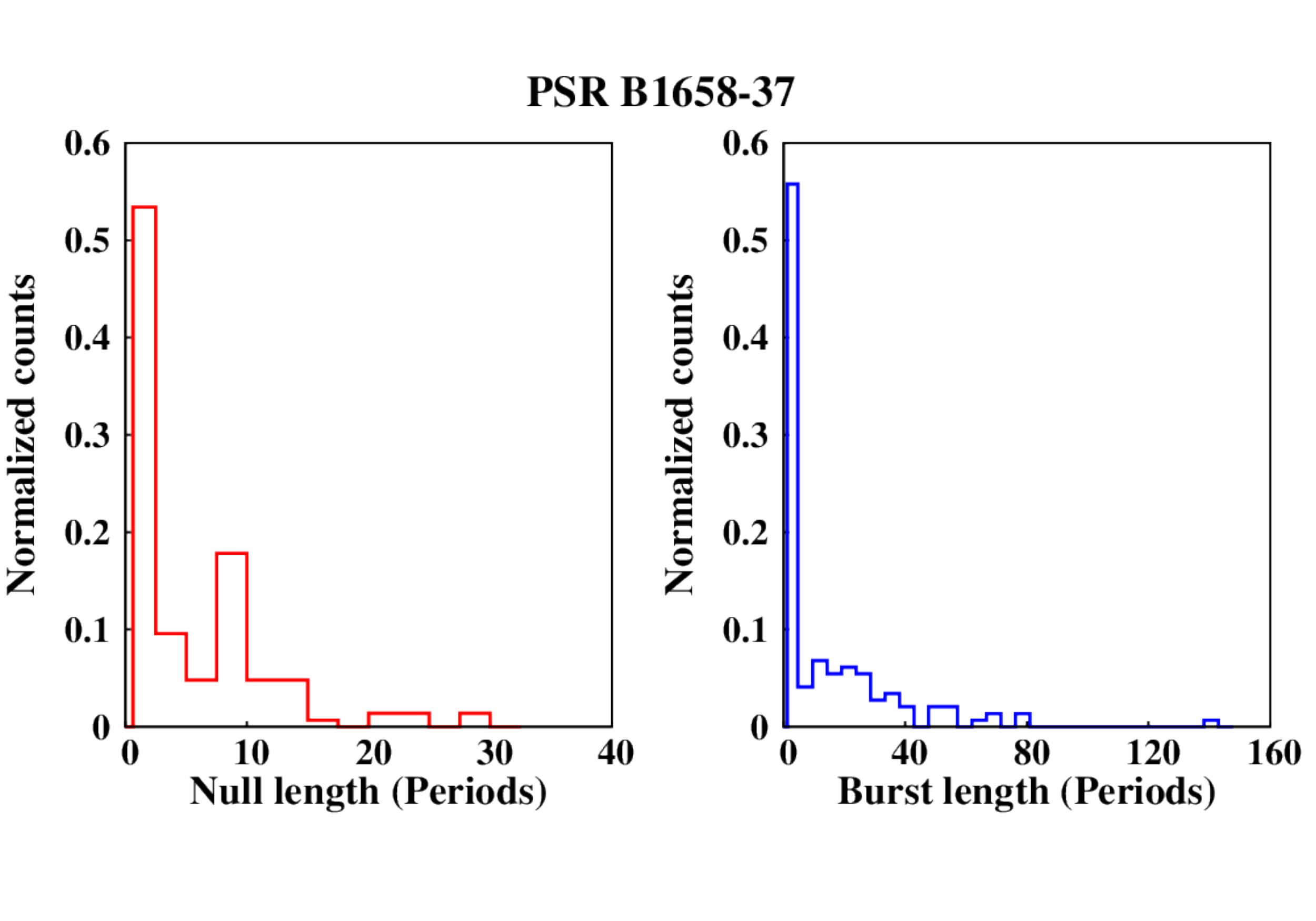}
\caption{The NLH and BLH for PSR B1658$-$37. Total 73 null lengths and 73 burst lengths were included to construct 
these histograms.}
\label{nlh_blh_b1701}
\end{figure}

\begin{figure}[h!]
\centering
\includegraphics[width=3.5in,height=6.5in,angle=-90,bb=0 0 504 720]{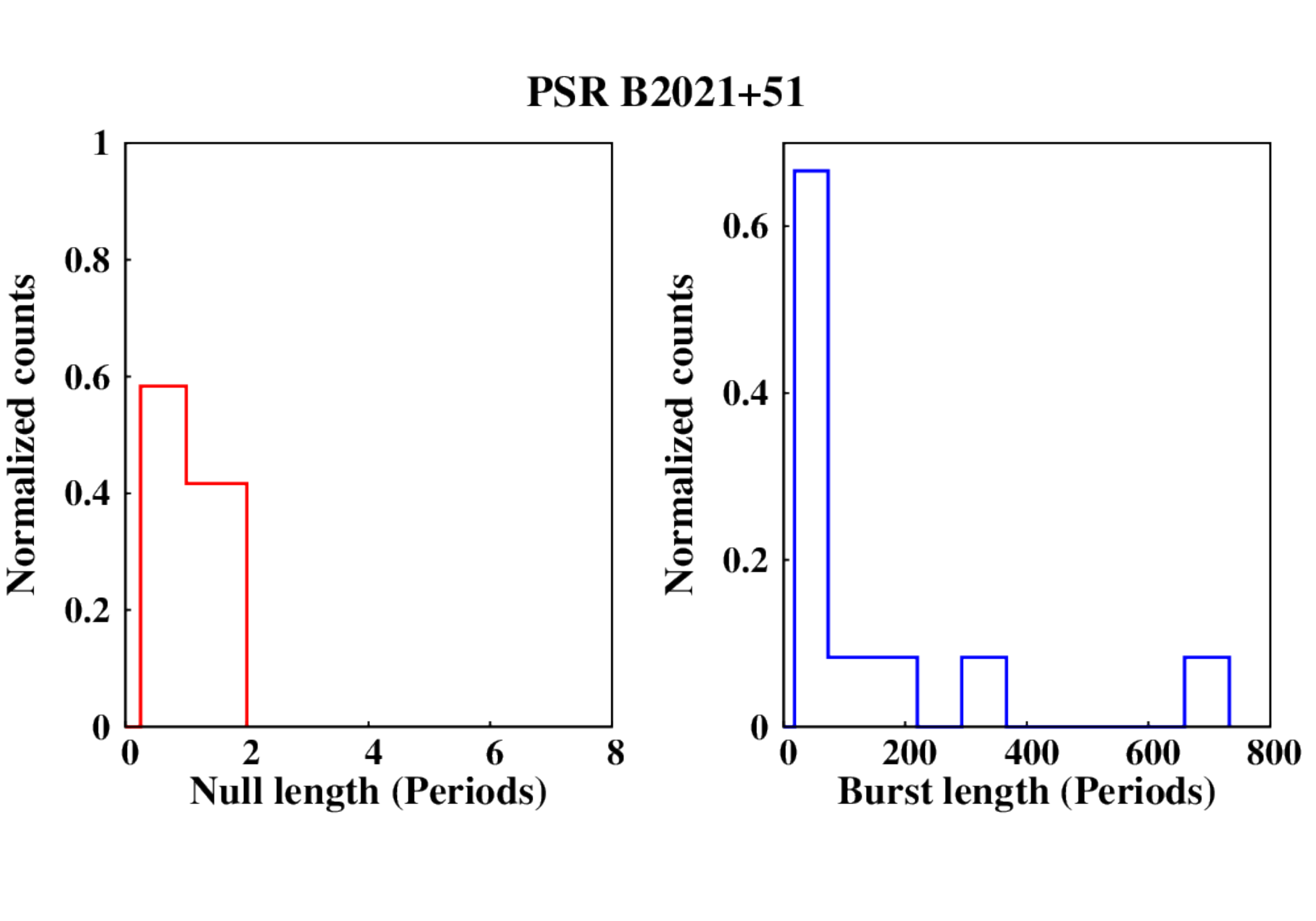}
\caption{The NLH and BLH for PSR B2021+51. Total 12 null lengths and 12 burst lengths were included to construct 
these histograms.}
\label{nlh_blh_b2021}
\end{figure}

\begin{figure}[h!]
\centering
\includegraphics[width=3.5in,height=6.5in,angle=-90,bb=0 0 504 720]{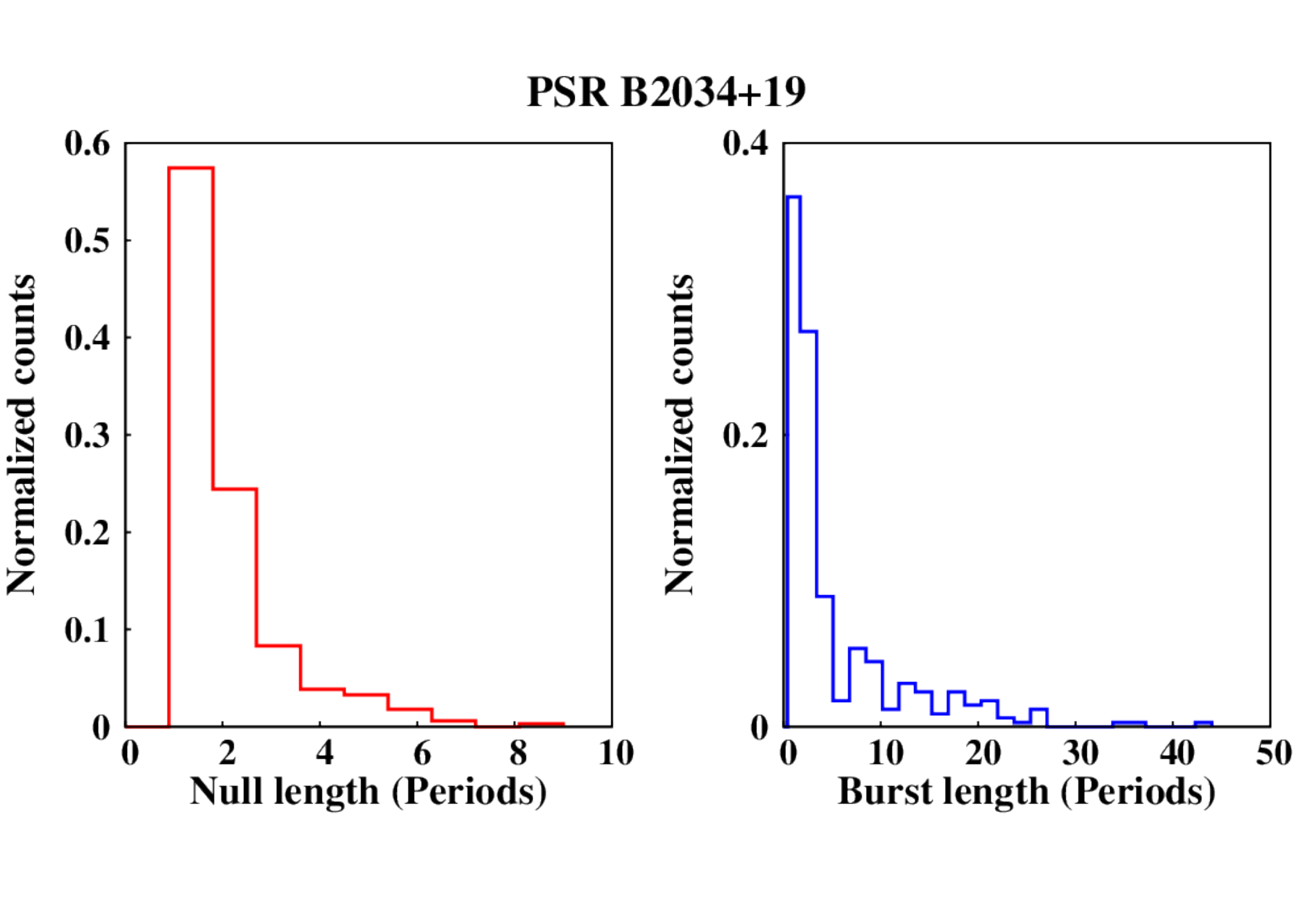}
\caption{The NLH and BLH for PSR B2034+19. Total 336 null lengths and 336 burst lengths were included to construct 
these histograms.}
\label{nlh_blh_b2037}
\end{figure}

\begin{figure}[h!]
\centering
\includegraphics[width=3.5in,height=6.5in,angle=-90,bb=0 0 504 720]{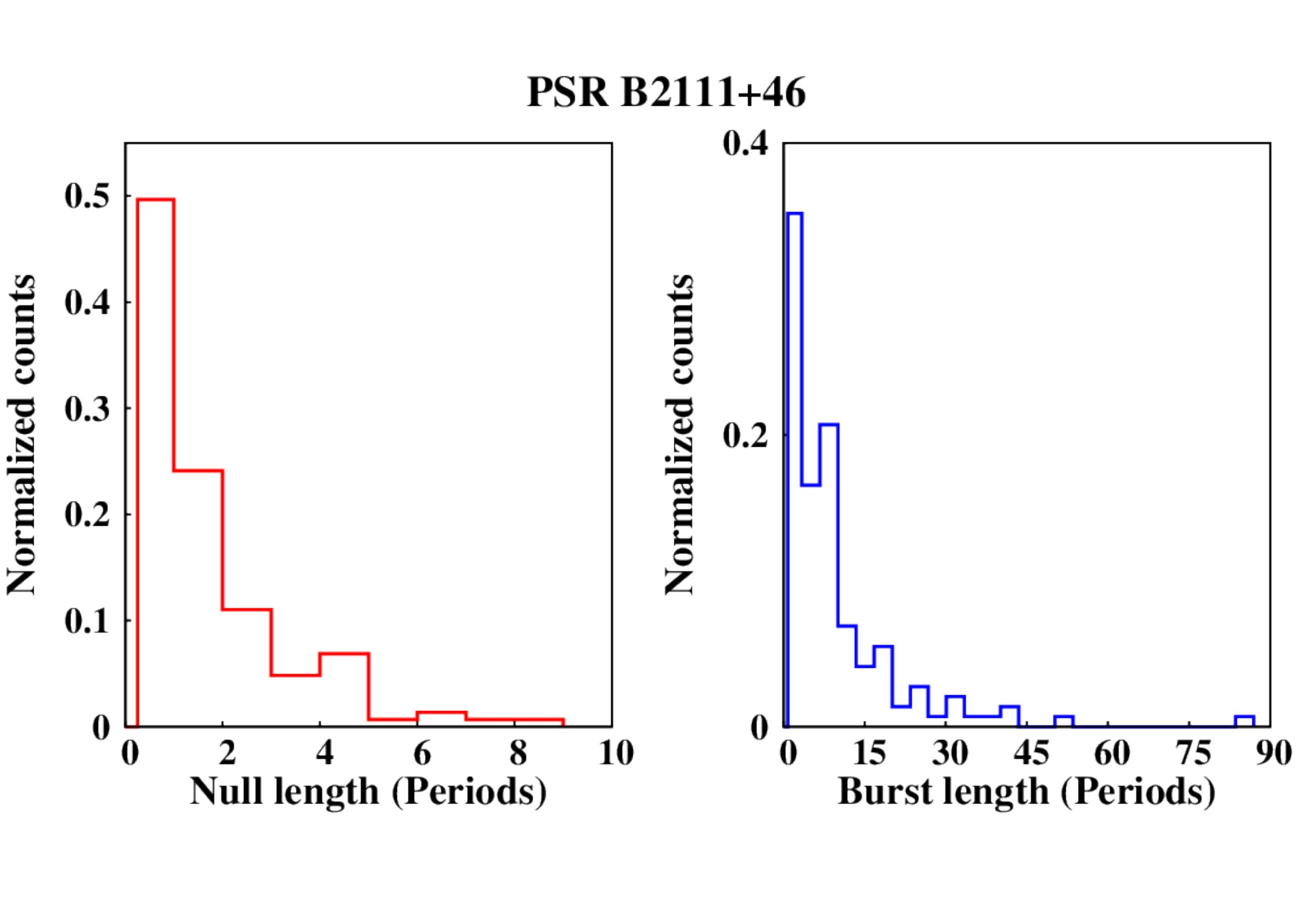}
\caption{The NLH and BLH for PSR B2111+46. Total 145 null lengths and 145 burst lengths were included to construct 
these histograms.} 
\label{nlh_blh_b2111}
\end{figure}

\begin{figure}[h!]
\centering
\includegraphics[width=3.5in,height=6.5in,angle=-90,bb=0 0 504 720]{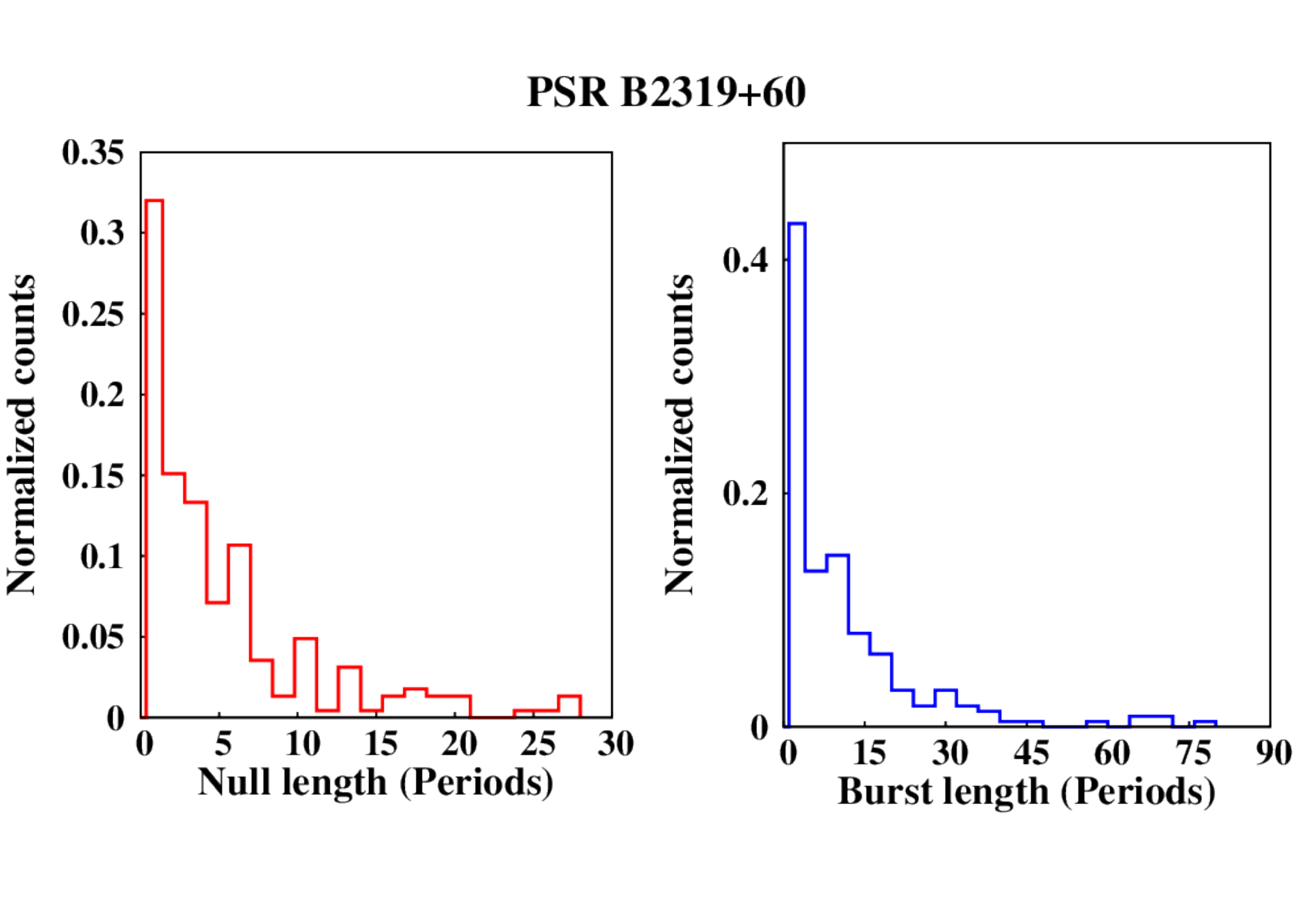}
\caption{The NLH and BLH for PSR B2319+60. Total 225 null lengths and 225 burst lengths were included to construct 
these histograms.}
\label{nlh_blh_b2319}
\end{figure}
 \clearpage\null\newpage
\chapter[Appendix C : Cumulative Distribution Function]{Appendix C \\ Cumulative Distribution Function}
\label{Appendix_CDF}
\setcounter{figure}{0} \renewcommand{\thefigure}{C.\arabic{figure}} 
This appendix displays the Cumulative distribution functions (CDF), as discussed in 
Chapter 4, of eight pulsars for which the NLH and the BLH were possible to obtain. 
PSR B1658$-$37 did not show a suitable match with the 
fitted model, hence it was not included in the following 
plots. Moreover, for PSRs B0837$-$41 and B2021+51 the null length 
distribution only showed single and double period nulls, hence 
they were also excluded in the following plots. For each pulsars, 
both the null length and burst length fits to the Poisson point process, 
given by the following equation, are shown. 
\setcounter{equation}{0} \renewcommand{\theequation}{C.\arabic{equation}}
\begin{equation}
F(x)  = 1  - \exp{(-x/\tau)}
\label{poiscdf_apndx}
\end{equation}
Each plot shows the obtained CDFs (blue line)
using the observed null/burst length distributions along with the 
fitted models (red line). After obtaining the least-square-fit, 
the decay parameter, $\tau$, was obtained for each CDF. 
These fitted parameters, $\tau_n$ and $\tau_b$ in units of pulsar period ($P_1$), 
are also shown for each pulsar. The fitted model was 
checked by carrying out a two-sample KS-test between 
the observed CDF and the obtained CDF from the simulated data, 
as discussed in Section \ref{sect_expted_time_scale}. 
The rejection probability of the null hypothesis of 
similar distributions is also shown for each 
pulsar in the inset texts. 
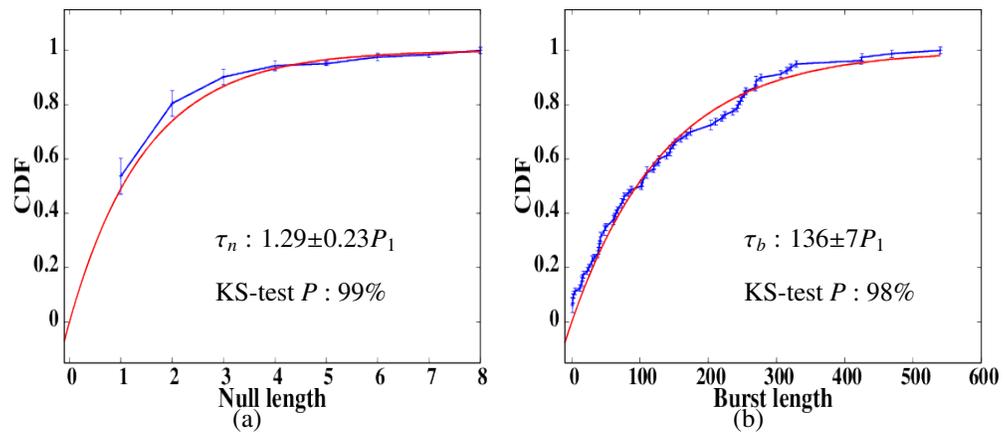
\begin{figure}[H]
 \centering
 \subfigure[]{
 \includegraphics[width=2.5in,height=2in,angle=0,bb=0 20 360 252]{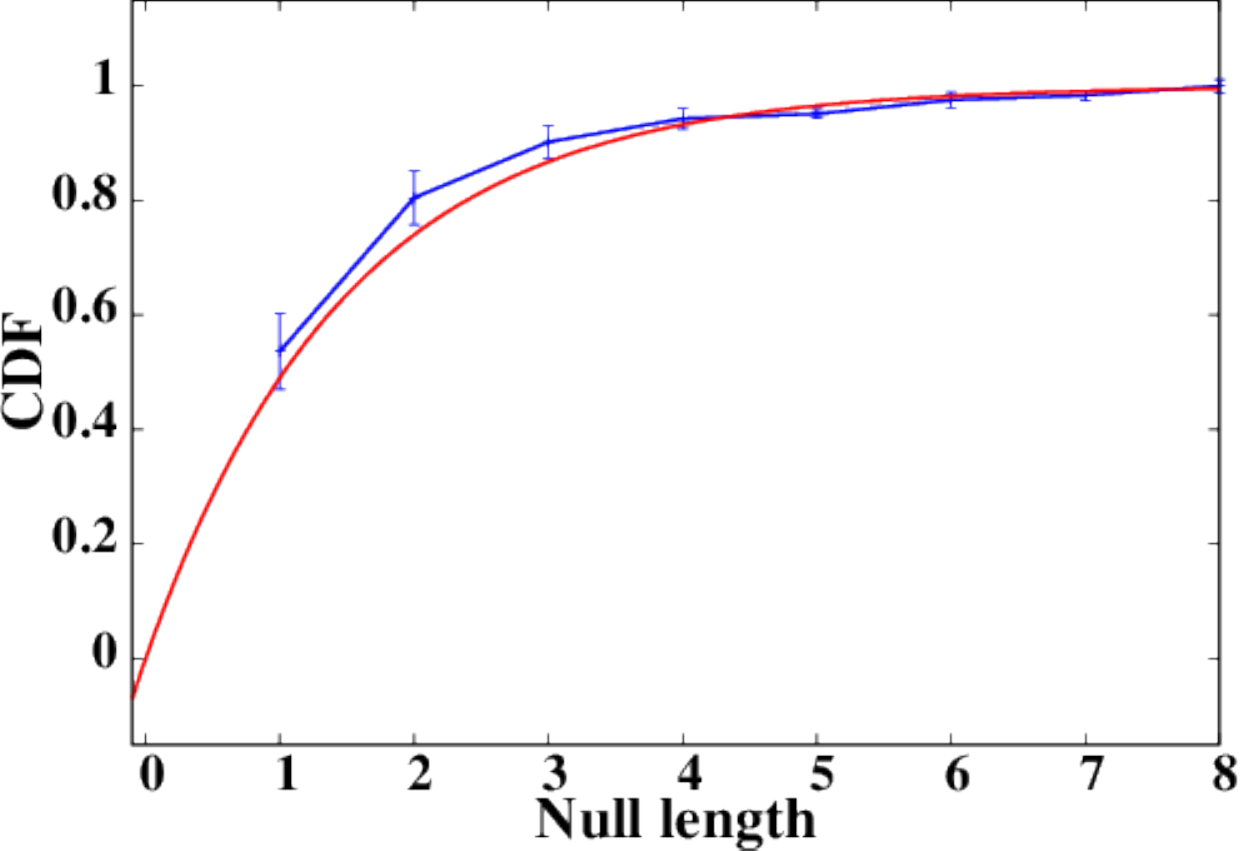}
 % b0809_spdisplay_forThesis.eps.pdf: 0x0 pixel, 300dpi, 0.00x0.00 cm, bb=503 202 1 1
 }
 \subfigure[]{
 \includegraphics[width=2.5in,height=2in,angle=0,bb=0 20 360 252]{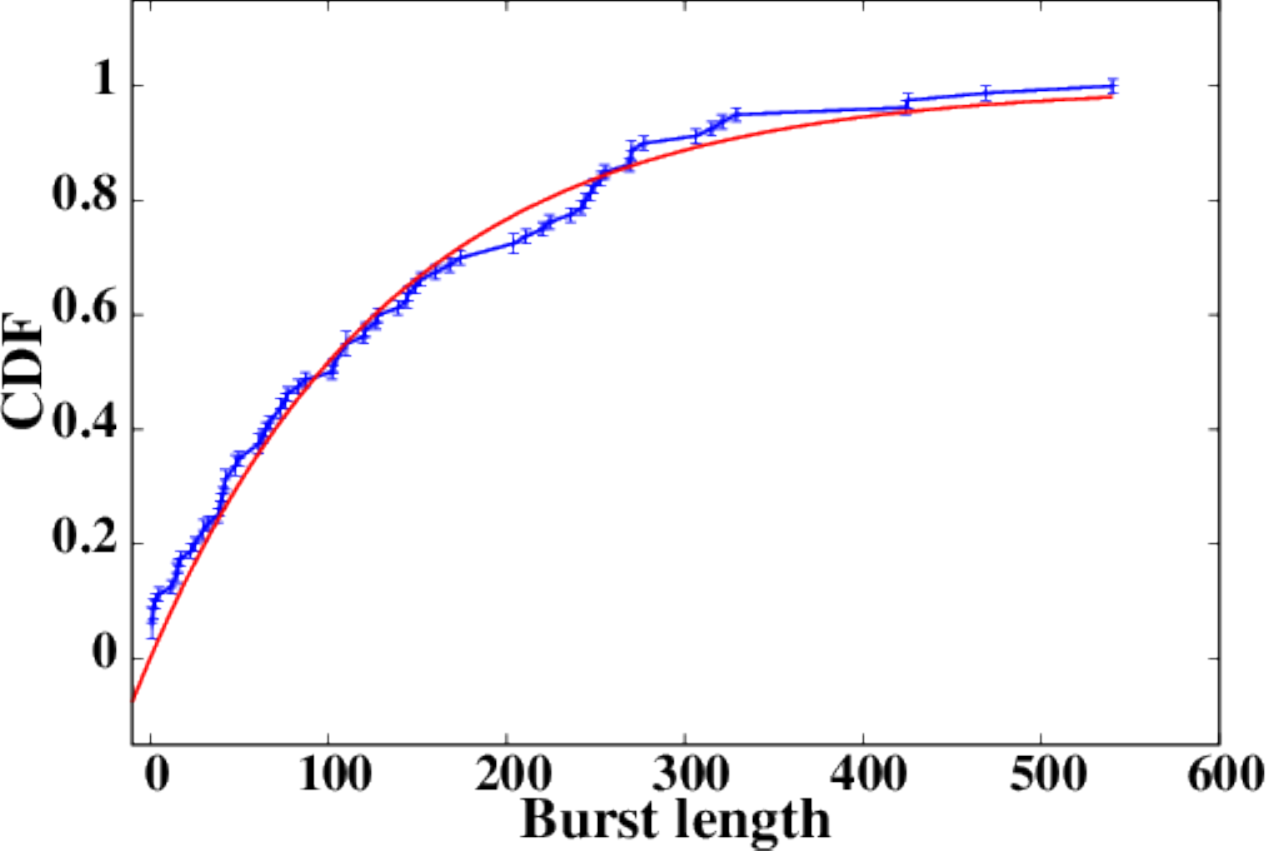}
 % b0809_spdisplay_forThesis.eps.pdf: 0x0 pixel, 300dpi, 0.00x0.00 cm, bb=503 202 1 1
 }
 \begin{picture}(0,0)
  \put(-300,50){\footnotesize $\tau_n$ : 1.29$\pm$0.23$P_1$} 
  \put(-300,30){\footnotesize KS-test \emph{P} : 99\%}
  \put(-100,50){\footnotesize $\tau_b$ : 136$\pm$7$P_1$} 
  \put(-100,30){\footnotesize KS-test \emph{P} : 98\%}
 \end{picture}
 \caption{Null length and burst length CDFs for PSR B0809+74}
 \label{cdf_b0809} 
\end{figure}
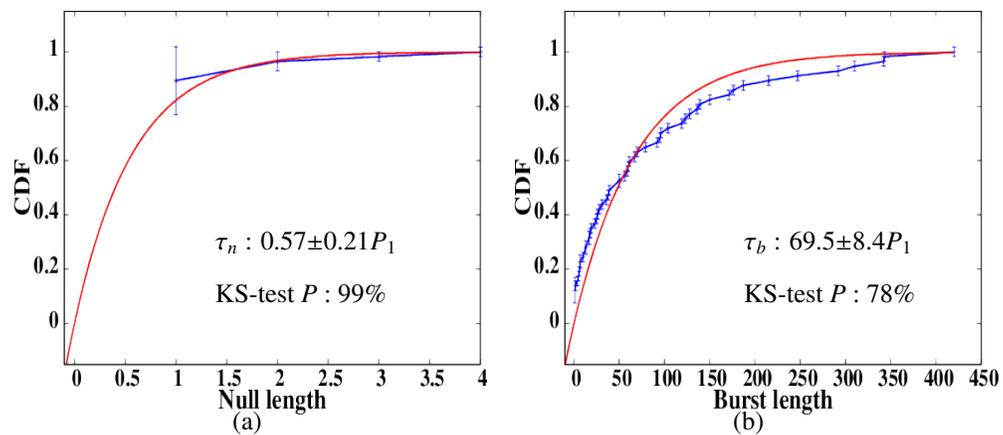
\begin{figure}[H]
 \centering
 \subfigure[]{
 \includegraphics[width=2.5in,height=2in,angle=0,bb=0 20 360 252]{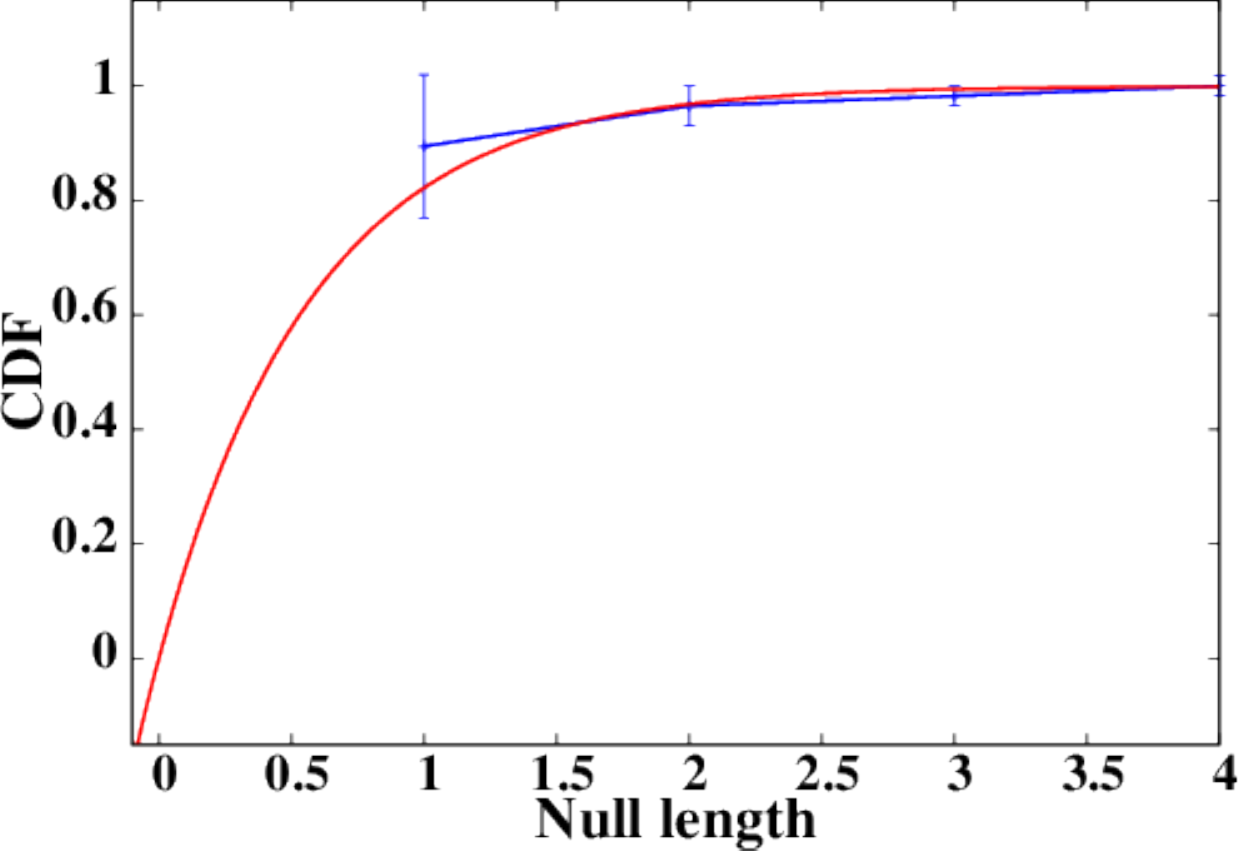}
 % b0809_spdisplay_forThesis.eps.pdf: 0x0 pixel, 300dpi, 0.00x0.00 cm, bb=503 202 1 1
 }
 \subfigure[]{
 \includegraphics[width=2.5in,height=2in,angle=0,bb=0 20 360 252]{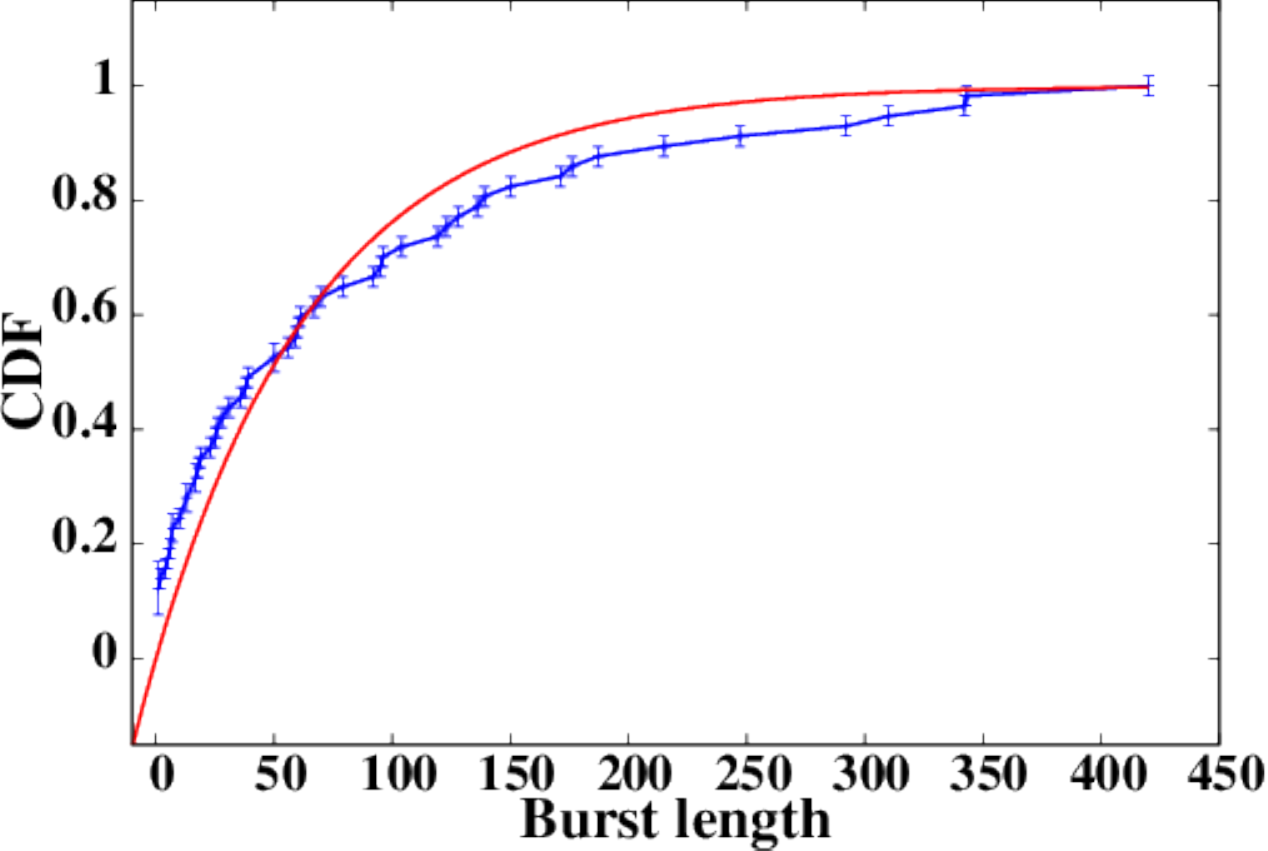}
 % b0809_spdisplay_forThesis.eps.pdf: 0x0 pixel, 300dpi, 0.00x0.00 cm, bb=503 202 1 1
 }
 \begin{picture}(0,0)
  \put(-300,50){\footnotesize $\tau_n$ : 0.57$\pm$0.21$P_1$} 
  \put(-300,30){\footnotesize KS-test \emph{P} : 99\%}
  \put(-100,50){\footnotesize $\tau_b$ : 69.5$\pm$8.4$P_1$} 
  \put(-100,30){\footnotesize KS-test \emph{P} : 78\%}
 \end{picture}
 \caption{Null length and burst length CDFs for PSR B0818$-$13}
 \label{cdf_b0818} 
\end{figure}
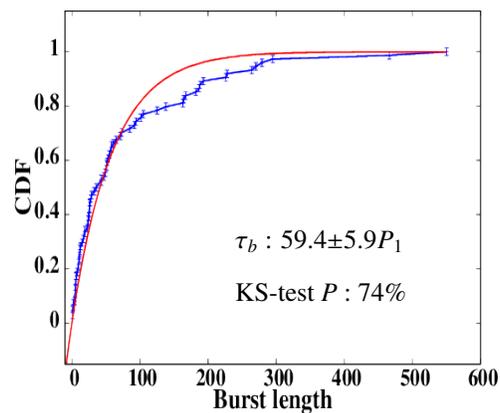
\begin{figure}[H]
 \centering
 \includegraphics[width=2.5in,height=2in,angle=0,bb=0 20 360 252]{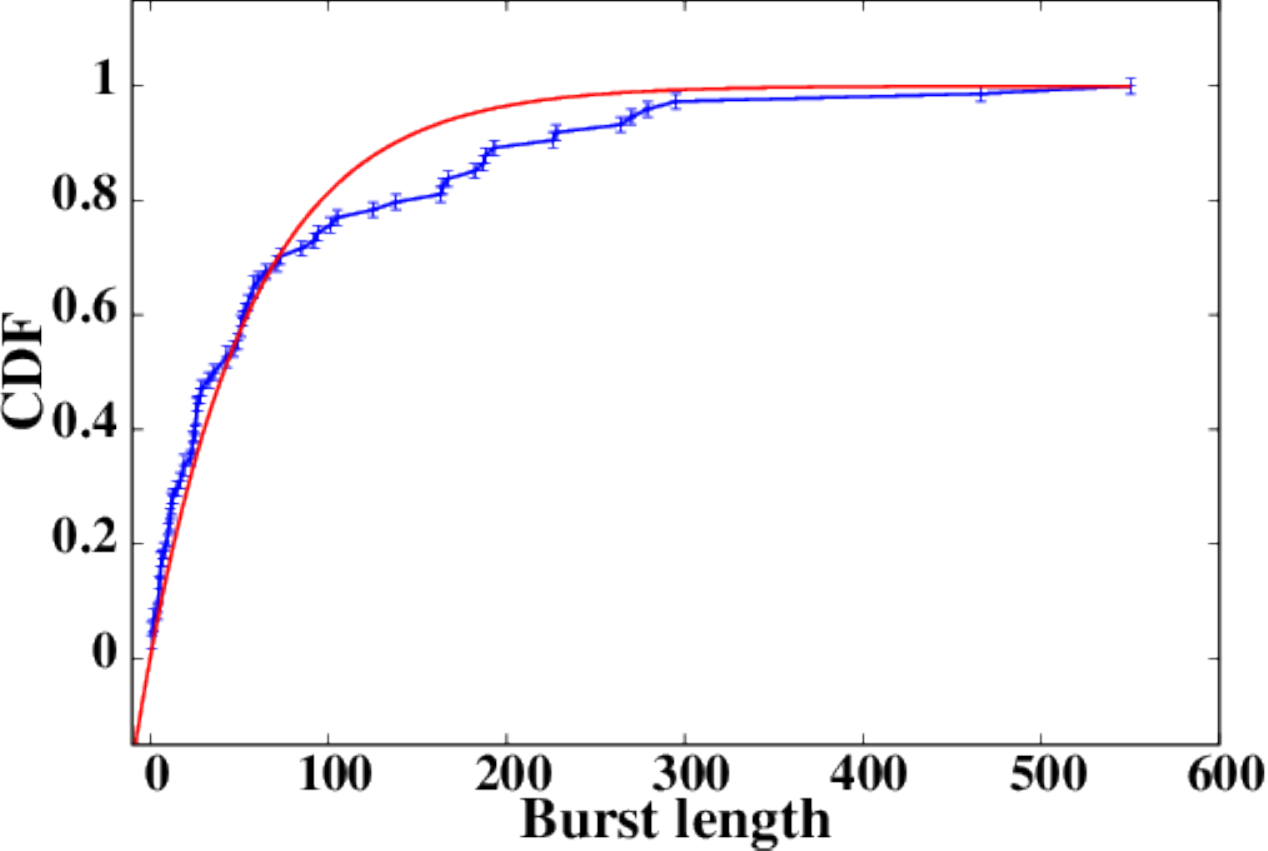}
 % b0809_spdisplay_forThesis.eps.pdf: 0x0 pixel, 300dpi, 0.00x0.00 cm, bb=503 202 1 1
 \begin{picture}(0,0)
  \put(-100,50){\footnotesize $\tau_b$ : 59.4$\pm$5.9$P_1$} 
  \put(-100,30){\footnotesize KS-test \emph{P} : 74\%}
 \end{picture}
 \caption{Null length and burst length CDFs for PSR B0837$-$41}
 \label{cdf_b0837} 
\end{figure}
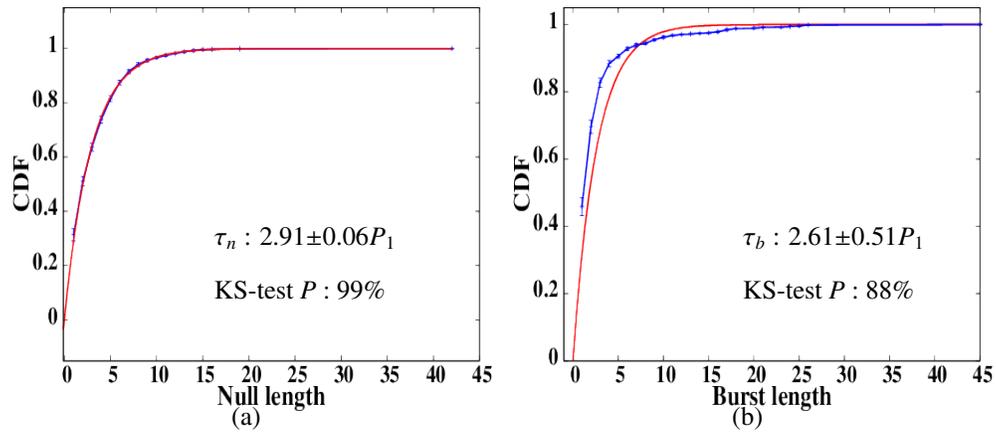
\begin{figure}[H]
 \centering 
 \subfigure[]{
 \centering
 \includegraphics[width=2.5in,height=2in,angle=0,bb=0 20 360 252]{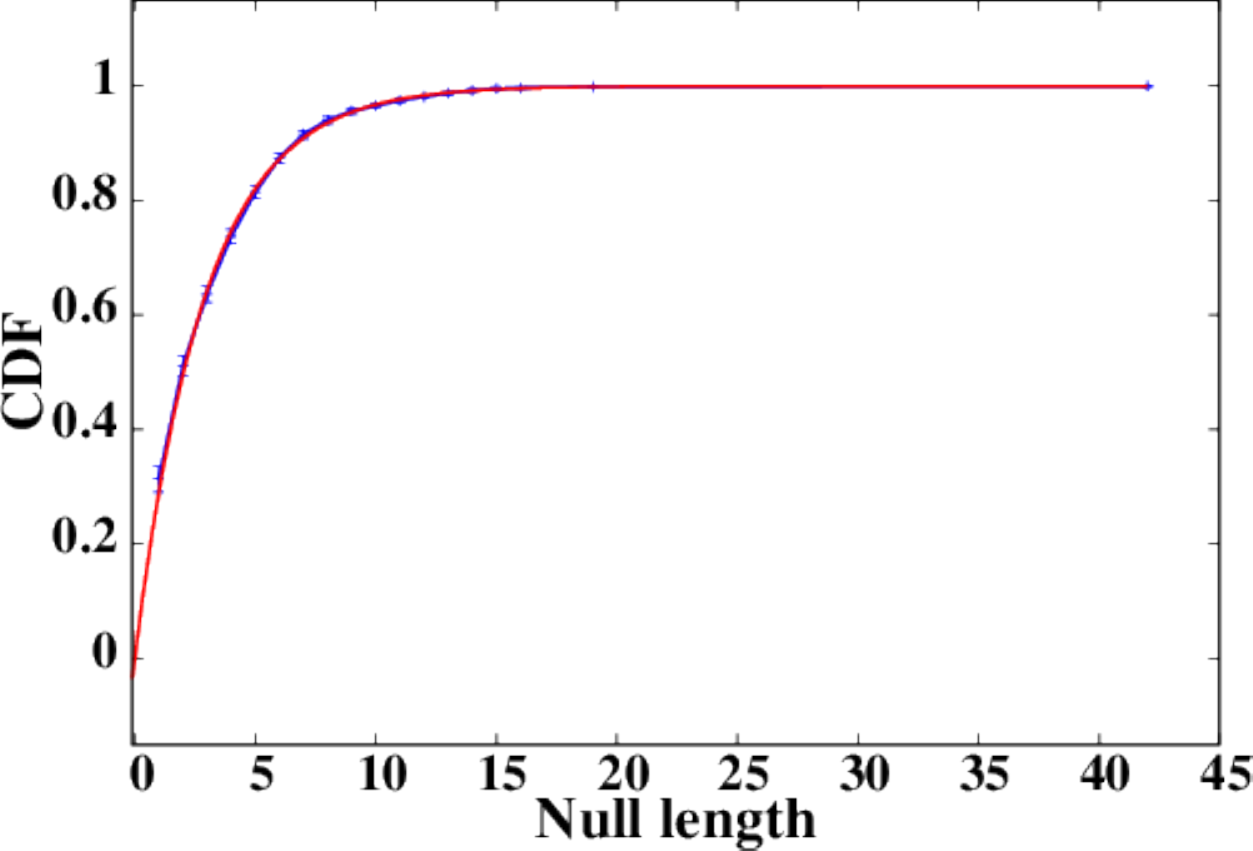}
 % b0809_spdisplay_forThesis.eps.pdf: 0x0 pixel, 300dpi, 0.00x0.00 cm, bb=503 202 1 1
 }
 \subfigure[]{
 \includegraphics[width=2.5in,height=2in,angle=0,bb=0 20 360 252]{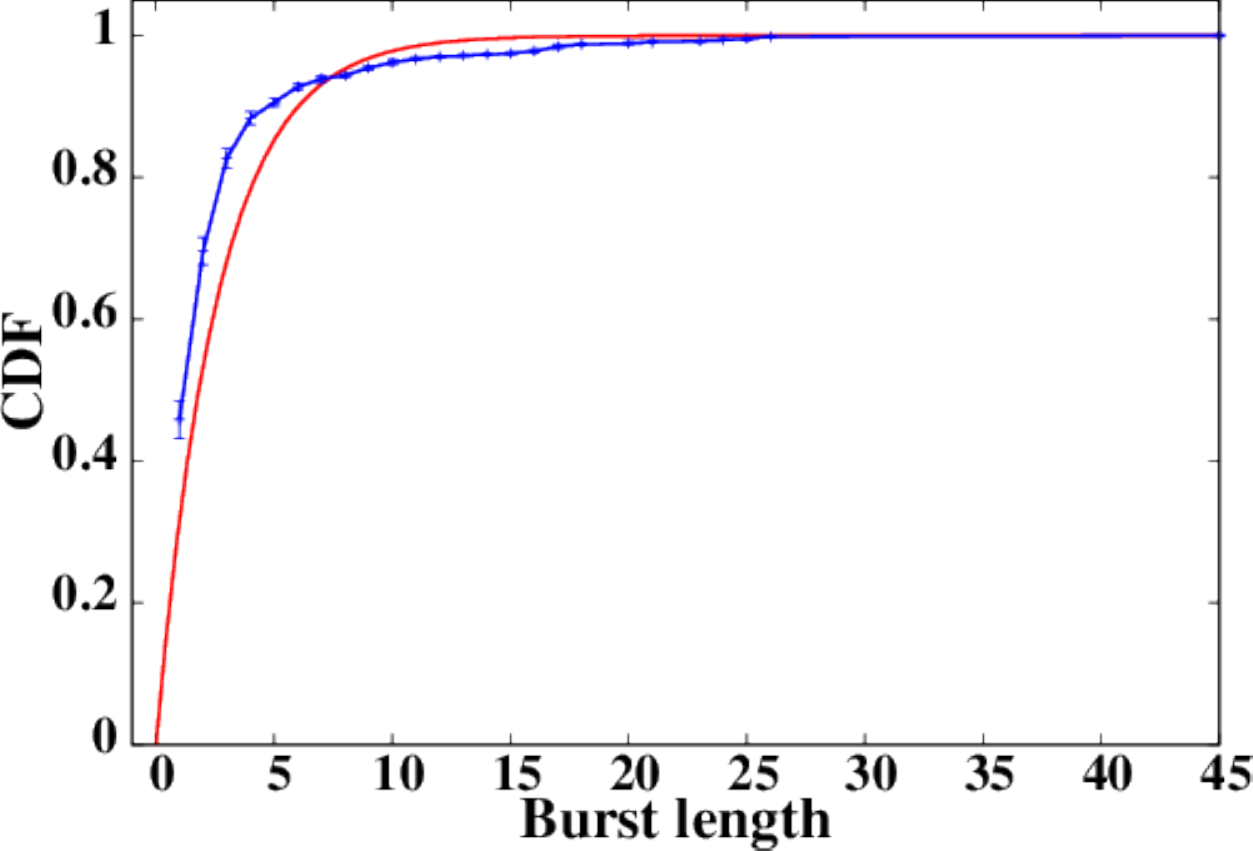}
 % b0809_spdisplay_forThesis.eps.pdf: 0x0 pixel, 300dpi, 0.00x0.00 cm, bb=503 202 1 1
 }
 \begin{picture}(0,0)
  \put(-300,50){\footnotesize $\tau_n$ : 2.91$\pm$0.06$P_1$} 
  \put(-300,30){\footnotesize KS-test \emph{P} : 99\%}
  \put(-100,50){\footnotesize $\tau_b$ : 2.61$\pm$0.51$P_1$} 
  \put(-100,30){\footnotesize KS-test \emph{P} : 88\%}
 \end{picture}
 \caption{Null length and burst length CDFs for PSR B1112+50}
 \label{cdf_b1112} 
\end{figure}
\begin{figure}[H]
 \centering  
 \subfigure[]{
 \includegraphics[width=2.5in,height=2in,angle=0,bb=0 20 360 252]{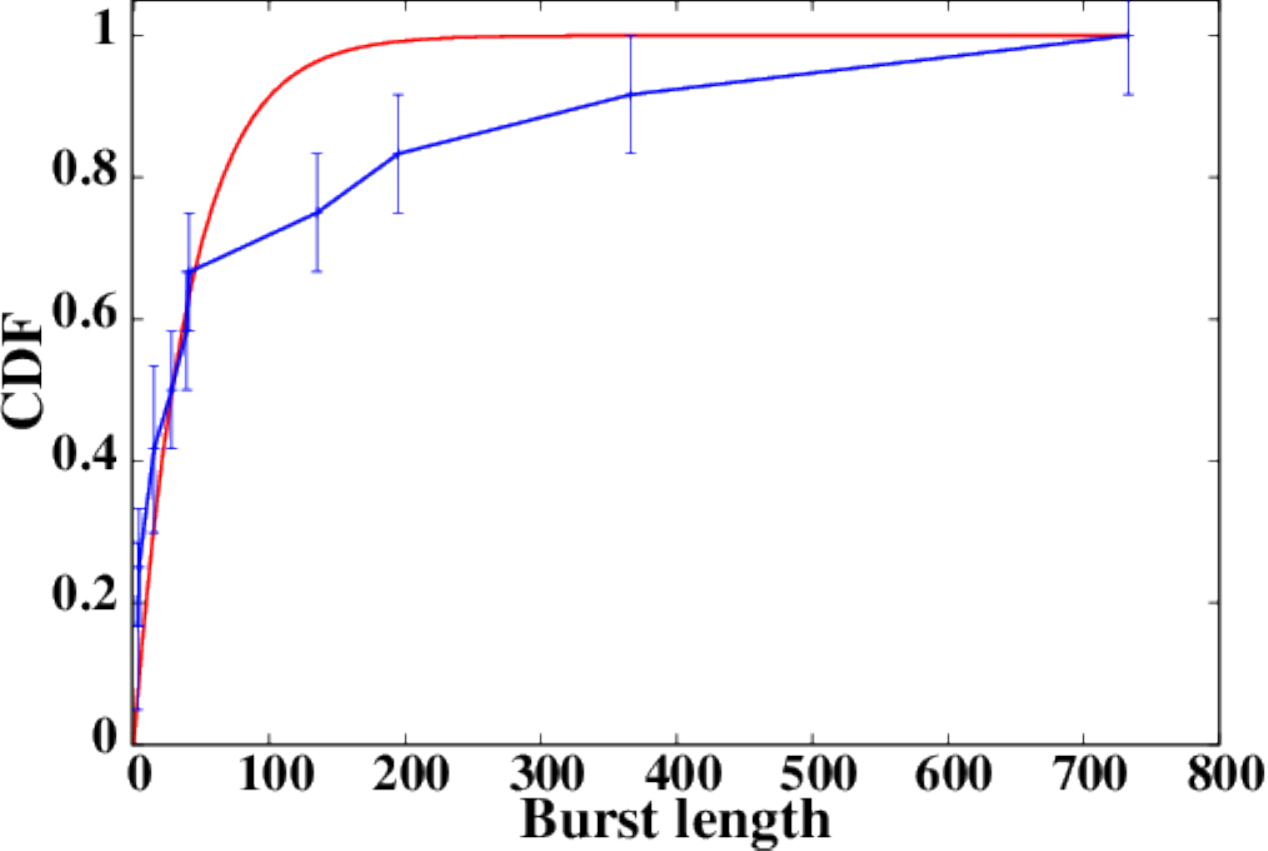}
 % b0809_spdisplay_forThesis.eps.pdf: 0x0 pixel, 300dpi, 0.00x0.00 cm, bb=503 202 1 1
 }
  \begin{picture}(0,0)
  \put(-100,50){\footnotesize $\tau_b$ : 40$\pm$20$P_1$} 
  \put(-100,30){\footnotesize KS-test \emph{P} : 95\%}
 \end{picture}
 \caption{Null length and burst length CDFs for PSR B2021+51}
 \label{cdf_b2021} 
\end{figure}
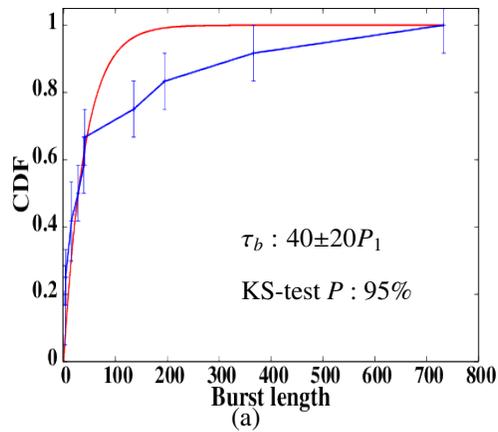
\begin{figure}[H]
 \centering  
 \subfigure[]{
 \centering
 \includegraphics[width=2.5in,height=2in,angle=0,bb=0 20 360 252]{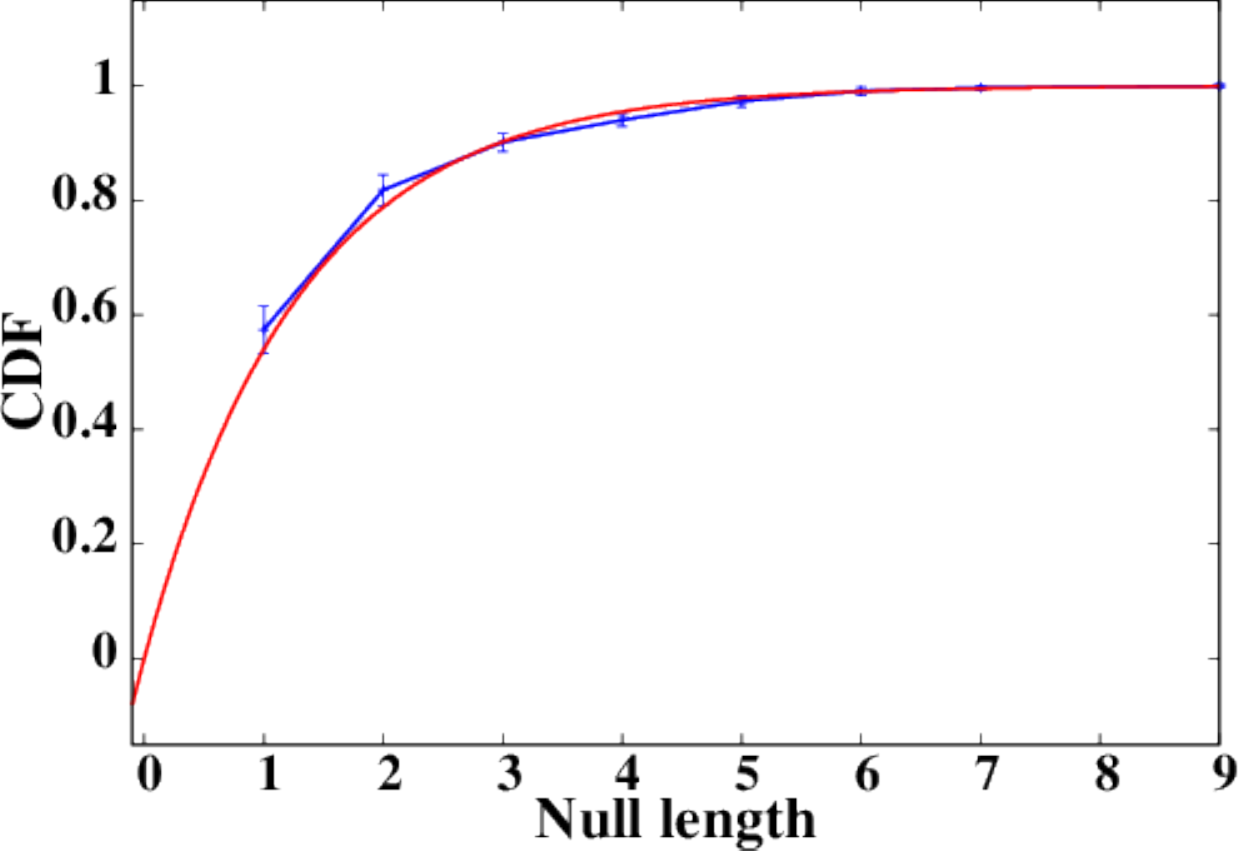}
 % b0809_spdisplay_forThesis.eps.pdf: 0x0 pixel, 300dpi, 0.00x0.00 cm, bb=503 202 1 1
 }
 \subfigure[]{
 \includegraphics[width=2.5in,height=2in,angle=0,bb=0 20 360 252]{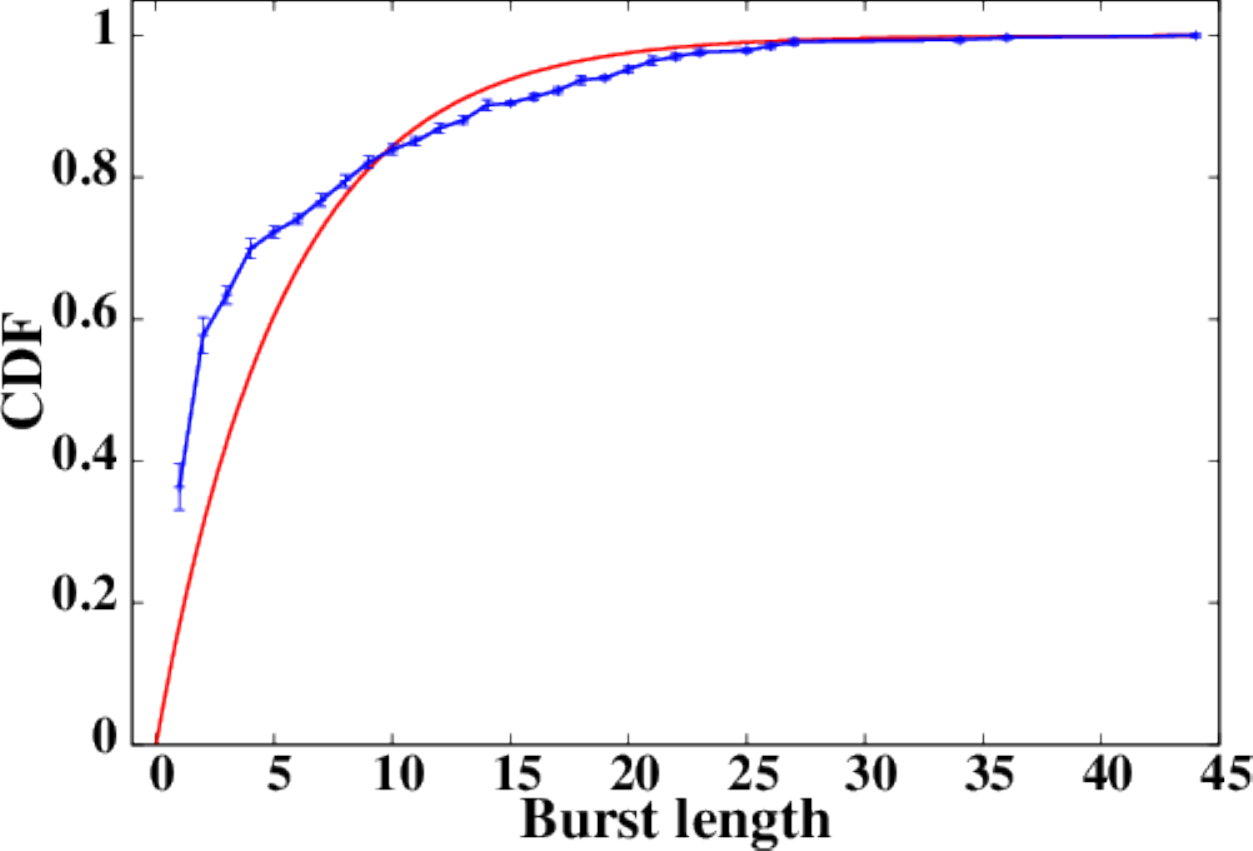}
 % b0809_spdisplay_forThesis.eps.pdf: 0x0 pixel, 300dpi, 0.00x0.00 cm, bb=503 202 1 1
 }
  \begin{picture}(0,0)
  \put(-300,50){\footnotesize $\tau_n$ : 1.3$\pm$0.1$P_1$} 
  \put(-300,30){\footnotesize KS-test \emph{P} : 99\%}
  \put(-100,50){\footnotesize $\tau_b$ : 5.4$\pm$0.7$P_1$} 
  \put(-100,30){\footnotesize KS-test \emph{P} : 22\%}
 \end{picture}
 \caption{Null length and burst length CDFs for PSR B2034+19}
 \label{cdf_b2037} 
\end{figure}
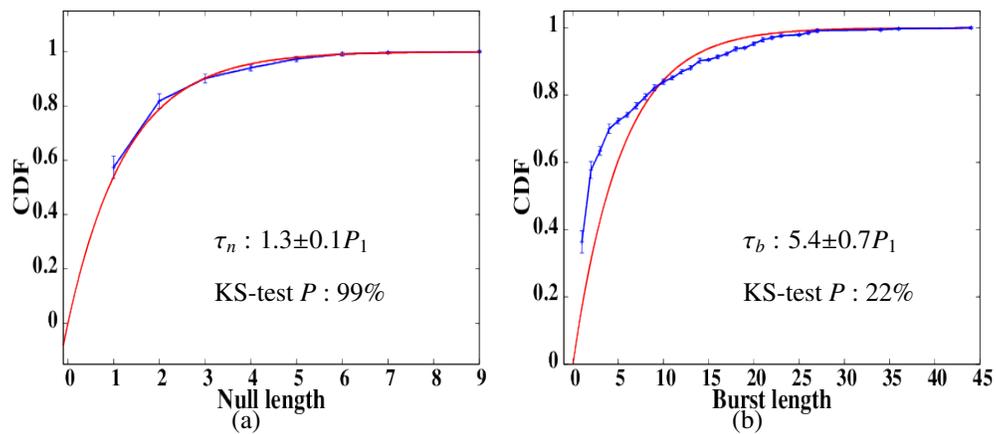
\begin{figure}[H]
 \centering  
 \subfigure[]{
 \centering
 \includegraphics[width=2.5in,height=2in,angle=0,bb=0 20 360 252]{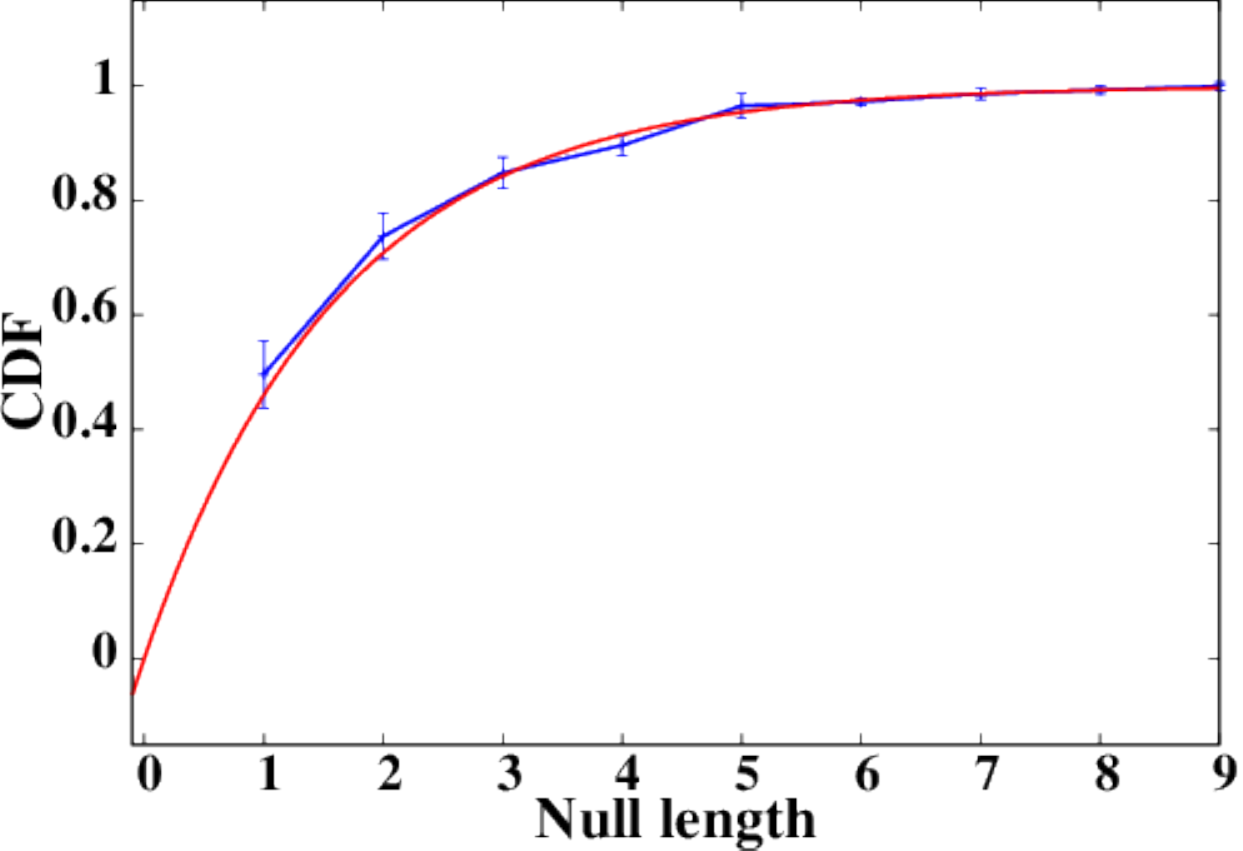}
 % b0809_spdisplay_forThesis.eps.pdf: 0x0 pixel, 300dpi, 0.00x0.00 cm, bb=503 202 1 1
 }
 \subfigure[]{
 \includegraphics[width=2.5in,height=2in,angle=0,bb=0 20 360 252]{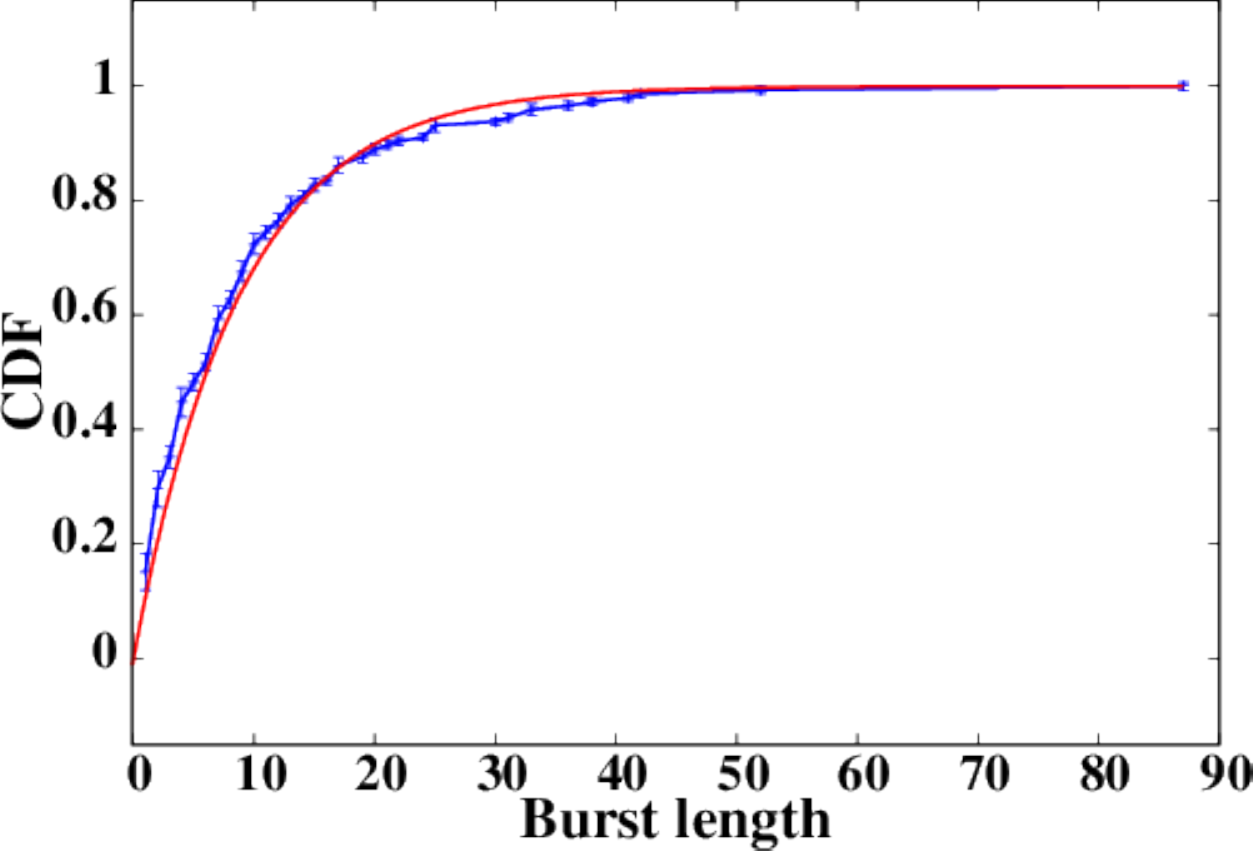}
 % b0809_spdisplay_forThesis.eps.pdf: 0x0 pixel, 300dpi, 0.00x0.00 cm, bb=503 202 1 1
 }
 \begin{picture}(0,0)
  \put(-300,50){\footnotesize $\tau_n$ : 1.62$\pm$0.11$P_1$} 
  \put(-300,30){\footnotesize KS-test \emph{P} : 99\%}
  \put(-100,50){\footnotesize $\tau_b$ : 8.7$\pm$0.5$P_1$} 
  \put(-100,30){\footnotesize KS-test \emph{P} : 99\%}
 \end{picture}
 \caption{Null length and burst length CDFs for PSR B2111+46}
 \label{cdf_b2111} 
\end{figure}
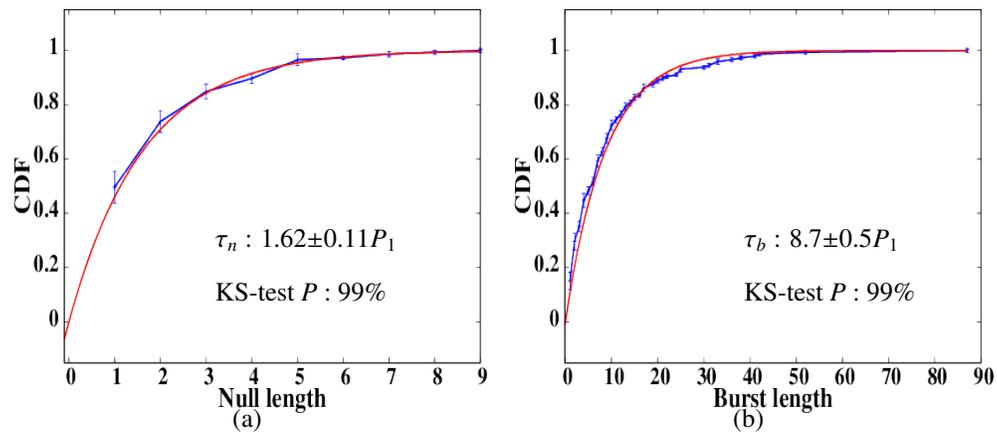
\begin{figure}[H]
 \centering  
 \subfigure[]{
 \centering
 \includegraphics[width=2.5in,height=2in,angle=0,bb=0 20 360 252]{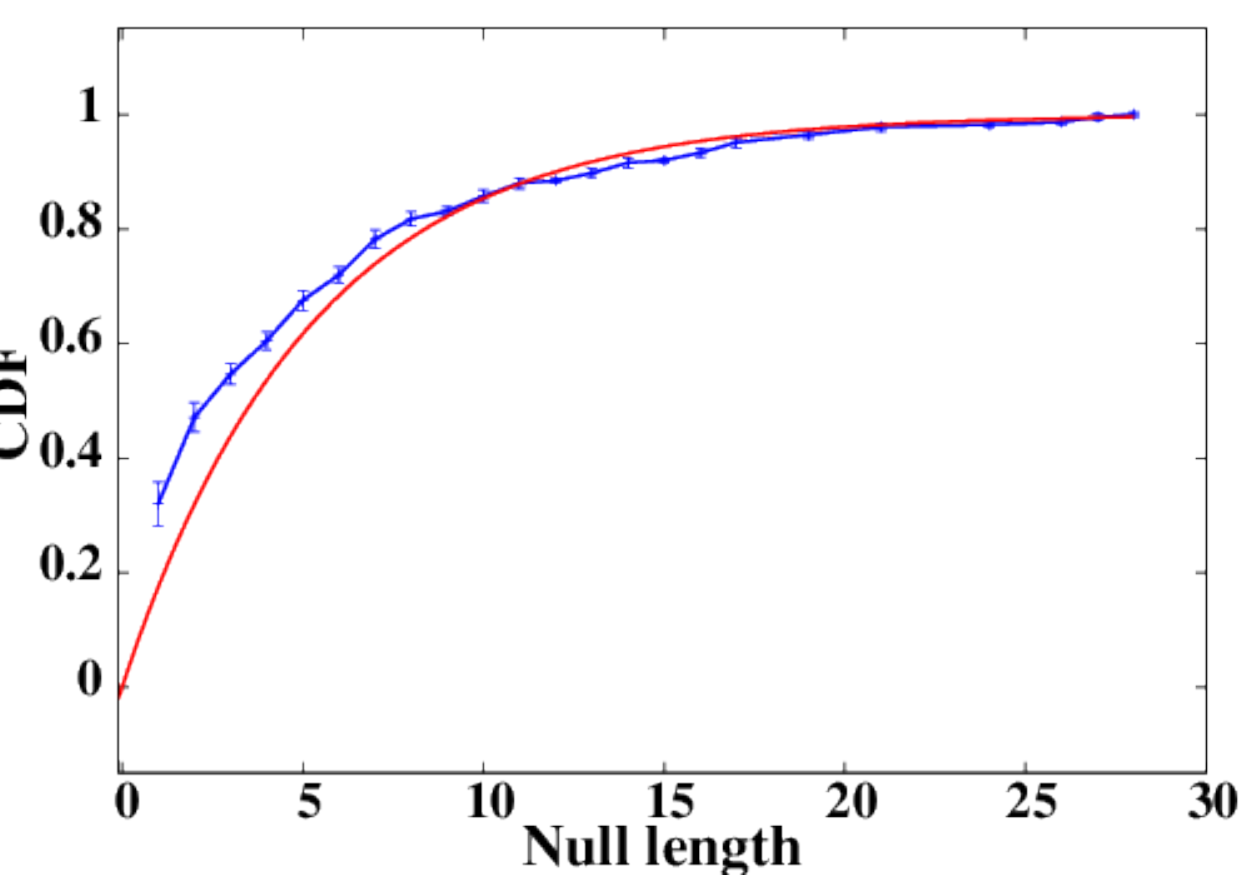}
 % b0809_spdisplay_forThesis.eps.pdf: 0x0 pixel, 300dpi, 0.00x0.00 cm, bb=503 202 1 1
 }
 \subfigure[]{
 \includegraphics[width=2.5in,height=2in,angle=0,bb=0 20 360 252]{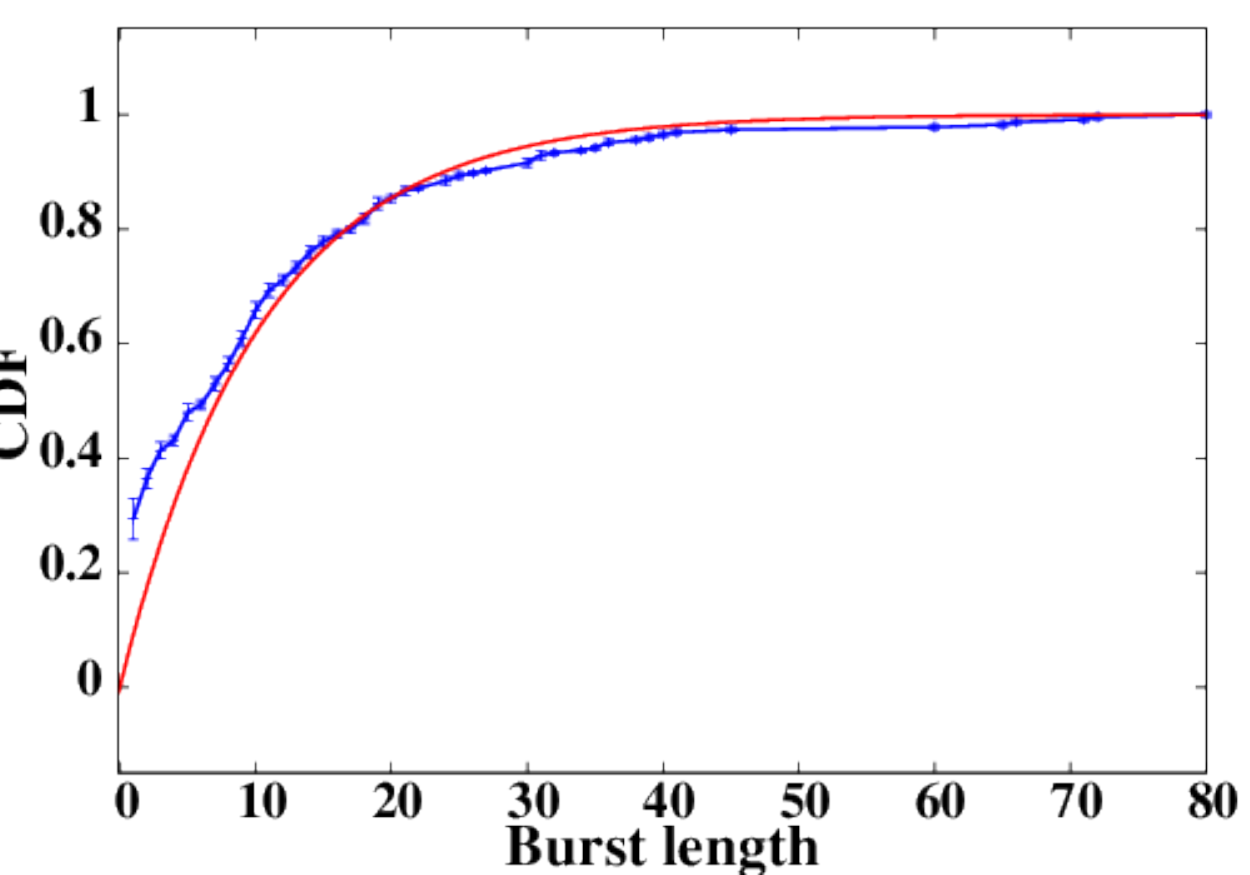}
 % b0809_spdisplay_forThesis.eps.pdf: 0x0 pixel, 300dpi, 0.00x0.00 cm, bb=503 202 1 1
 }
 \begin{picture}(0,0)
  \put(-300,50){\footnotesize $\tau_n$ : 5.2$\pm$0.4$P_1$} 
  \put(-300,30){\footnotesize KS-test \emph{P} : 94\%}
  \put(-100,50){\footnotesize $\tau_b$ : 10.4$\pm$0.7$P_1$} 
  \put(-100,30){\footnotesize KS-test \emph{P} : 33\%}
 \end{picture}
 \caption{Null length and burst length CDFs for PSR B2319+60} 
 \label{cdf_b2319} 
\end{figure}

\clearpage\null\newpage

\chapter[Appendix D : The Pair Correlation Function]{Appendix D \\ The Pair Correlation Function}
\graphicspath{{Images/}{Images/}{Images/}}

\setcounter{figure}{0} \renewcommand{\thefigure}{D.\arabic{figure}} 
\setcounter{equation}{0} \renewcommand{\theequation}{D.\arabic{equation}} 

\label{apppcf}
% 1.introduction
The Pair Correlation Function ({\itshape PCF}) is a probability density
function 
(also known as the radial distribution function or pair separation function) for
the clustering 
of certain objects or events in space and/or time coordinates \cite[]{plu85}  
and is useful for measuring the degree of packing. We have 
used a one-dimensional PCF, which 
identifies the clustering of events (the bursts pulses of a bright 
phase in our case), in the time series data. A brief  
description is provided here as this 
seems to be the first time such a technique is applied 
the clustering of burst pulses in pulsar astronomy. 

% 2.basic derivation
The PCF for a series M pulses 
with N burst pulses  can be derived as follows.  
The pulse index of these burst pulses are 
\begin{equation}
 p_i~ or ~ p_j~=~p_1,~p_2,~...,~p_N.
\end{equation}
Then,  PCF is defined as 
\begin{equation}
%  g(p)~=~\frac{1}{(N^2 -N)}\sum\limits_{i=1}^{N}\sum\limits_{j\neq{i}}^{N}\langle{\delta(p~-~\arrowvert{p_j-p_i}\arrowvert)}\rangle.
 g(p)~=~G\cdot\sum\limits_{i=1}^{N}\sum\limits_{j\neq{i}}^{N}\langle{\delta(p~-~\arrowvert{p_j-p_i}\arrowvert)}\rangle.
 \label{pcfeq}
\end{equation}
where G is a scaling parameter and $\delta$ is the Kronecker delta function.
A normalized binning of PCF, g(p), provides the probability  
of occurrence of certain separations between burst pulses. If a 
pulsar exhibits bunching of burst pulses and periodic  
occurrence of these bunches, the PCF shows prominent peaks 
around repeatedly occurring separations and their 
harmonics. A simple way to detect such periodicities is to 
obtain the Fourier spectra  of the PCF.

\begin{figure}
 \centering 
 \includegraphics[width=3.5 in,height=2.5 in,angle=0,bb=0 0 350 250]{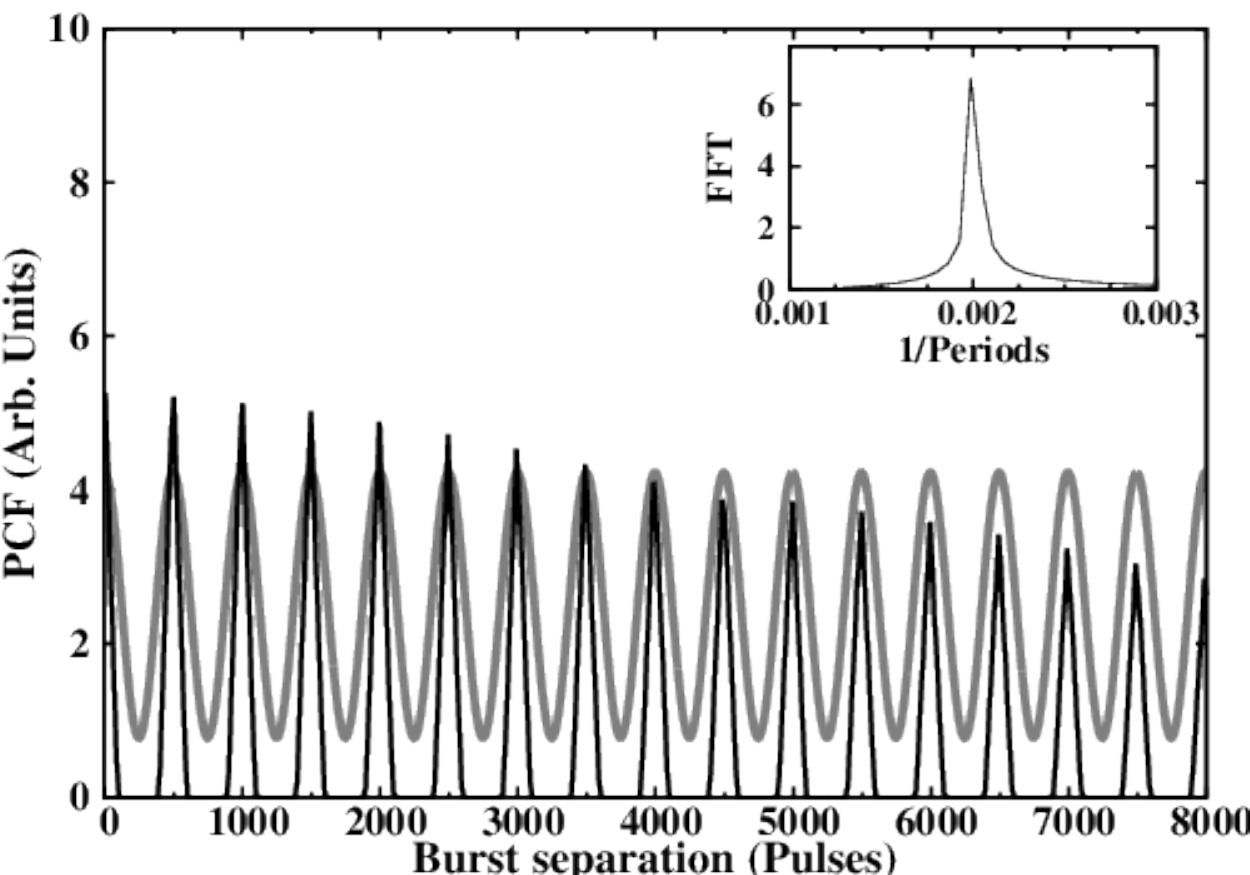}
 % Apndx_PCF_FFT.eps: 0x0 pixel, 300dpi, 0.00x0.00 cm, bb=50 50 410 302
  \caption[PCF from the simulated pulse sequence]
  {PCF of a pulsar whose nulls form clusters of 100 pulses separated
  by 400 periods. It is a histogram of the separations between the simulated
  burst pulses in units of pulse periods. Note the gradual 
  reduction in amplitude for large separations due to the finite 
  length of the simulated data. The inset figure shows the 
  Fourier spectra of the PCF. A sine-wave with a 500 pulse periodicity 
  is overlaid along with the PCF in grey colour for clarity.} 
  \label{Apndx_pcf}
\end{figure}

% 3.figure obtaining procedure
Fig. \ref{Apndx_pcf} shows an example of  the PCF obtained 
for 8000 pulses of a pulsar nulling with a precise periodicity. The pulsar was 
simulated 
by  repeated occurrence of 100 burst pulses 
separated by 400 null pulses, so the NF was 80 \%. 
%We first identified all the burst pulse indexes. 
%These were used to measure the separation between 
%all pairs of burst pulses as a function of separation between them. 
%The obtained separation between all burst pulses (not only 
%the consecutive) in units of periods were binned 
%to get a smooth PCF. 
The periodicity is clearly visible both in the PCF 
and in the Fourier spectra in the inset figure. The peaks in PCF are at 500
pulses, broadened by the 100 pulse spread.

% 3.importance 
The PCF measures not only the periodicity of the clustering but also how long
its coherence persists. If coherence in the bunching 
is lost, the PCF would not show peaks beyond a particular length. 
This makes it superior to a simple Fourier analysis of 
the pulse energy modulation since a PCF emphasises short-lived periodic
features 
as well as providing information regarding the coherence 
length. An additional advantage over conventional Fourier analysis 
is that observations from different sessions can be combined. The maximum coherence 
length measurable in this case will come from the session with 
the longest pulse sequence among all the  observing sessions. 
Hence, the PCF is a useful technique to scrutinize 
periodic pulse energy fluctuations. 
\end{appendices}
   
% The ref are only coming for the \input.. they disappear for \include   
\newpage
\bibliographystyle{mn2e.bst}
%\bibliography{Ar}
% \bibliography{Arxiv_thesis_combined.bbl}
\bibliography{psrrefs,modpsrrefs,mybib,modrefs}

\begin{thebibliography}{}

\bibitem[\protect\citeauthoryear{{Akujor} \& {Okeke}}{{Akujor} \&
  {Okeke}}{1982}]{ao82}
{Akujor} C.~E.,  {Okeke} P.~N.,  1982, \apss, 85, 325

\bibitem[\protect\citeauthoryear{Arons}{Arons}{1981}]{aro81}
Arons J.,  1981, ApJ, 248, 1099

\bibitem[\protect\citeauthoryear{{Arons}}{{Arons}}{1983}]{aro83c}
{Arons} J.,  1983, in {Burns} M.~L.,  {Harding} A.~K.,   {Ramaty} R.,  eds,
  Positron-Electron Pairs in Astrophysics Vol.~101 of American Institute of
  Physics Conference Series, {Electron positron pairs in radio pulsars}.
pp 163--193

\bibitem[\protect\citeauthoryear{Arons}{Arons}{1983}]{aro83b}
Arons J.,  1983, ApJ, 266, 215

\bibitem[\protect\citeauthoryear{Arons \& Barnard}{Arons \&
  Barnard}{1986}]{ab86}
Arons J.,  Barnard J.~J.,  1986, ApJ, 302, 120

\bibitem[\protect\citeauthoryear{Arons \& Scharlemann}{Arons \&
  Scharlemann}{1979}]{as79}
Arons J.,  Scharlemann E.~T.,  1979, ApJ, 231, 854

\bibitem[\protect\citeauthoryear{Asseo, Pelletier \& Sol}{Asseo
  et~al.}{1990}]{aps90}
Asseo E.,  Pelletier G.,    Sol H.,  1990, MNRAS, 247, 529

\bibitem[\protect\citeauthoryear{{Baade} \& {Zwicky}}{{Baade} \&
  {Zwicky}}{1934}]{bz34}
{Baade} W.,  {Zwicky} F.,  1934, Physical Review, 46, 76

\bibitem[\protect\citeauthoryear{{Baars} \& {Hooghoudt}}{{Baars} \&
  {Hooghoudt}}{1974}]{bh74}
{Baars} J.~W.~M.,  {Hooghoudt} B.~G.,  1974, A\&A, 31, 323

\bibitem[\protect\citeauthoryear{Backer}{Backer}{1970a}]{bac70c}
Backer D.~C.,  1970a, Nat., 228, 752

\bibitem[\protect\citeauthoryear{Backer}{Backer}{1970b}]{bac70a}
Backer D.~C.,  1970b, Nat., 228, 1297

\bibitem[\protect\citeauthoryear{Backer}{Backer}{1970c}]{bac70}
Backer D.~C.,  1970c, Nat., 228, 42

\bibitem[\protect\citeauthoryear{Backer}{Backer}{1976}]{bac76}
Backer D.~C.,  1976, ApJ, 209, 895

\bibitem[\protect\citeauthoryear{{Backus}, {Mitra} \& {Rankin}}{{Backus}
  et~al.}{2010}]{bmr10}
{Backus} I.,  {Mitra} D.,    {Rankin} J.~M.,  2010, \mnras, 404, 30

\bibitem[\protect\citeauthoryear{Backus}{Backus}{1981}]{bac81}
Backus P.~R.,  1981, PhD thesis, The University of Massachusetts

\bibitem[\protect\citeauthoryear{{Bartel} \& {Sieber}}{{Bartel} \&
  {Sieber}}{1978}]{bs78}
{Bartel} N.,  {Sieber} W.,  1978, \aap, 70, 307

\bibitem[\protect\citeauthoryear{Bartel, Sieber \& Wolszczan}{Bartel
  et~al.}{1980}]{bsw80}
Bartel N.,  Sieber W.,    Wolszczan A.,  1980, 90, 58

\bibitem[\protect\citeauthoryear{Baym}{Baym}{1991}]{bay91}
Baym G.,  1991, in {Ventura} J.,  {Pines} D.,  eds, NATO ASIC Proc. 344:
  Neutron Stars Vol.~334, {The High Density Interiors of Neutron Stars}.
p.~21

\bibitem[\protect\citeauthoryear{{Bell-Burnell}}{{Bell-Burnell}}{1977}]{bell77}
{Bell-Burnell} S.~J.,  1977, in {Papagiannis} M.~D.,  ed., Eighth Texas
  Symposium on Relativistic Astrophysics Vol.~302 of Annals of the New York
  Academy of Sciences, {Petit Four}.
p.~685

\bibitem[\protect\citeauthoryear{Benford \& Buschauer}{Benford \&
  Buschauer}{1977}]{bb77}
Benford G.,  Buschauer R.,  1977, MNRAS, 179, 189

\bibitem[\protect\citeauthoryear{{Beskin}, {Gurevich} \& {Istomin}}{{Beskin}
  et~al.}{1988}]{bgi88}
{Beskin} V.~S.,  {Gurevich} A.~V.,    {Istomin} I.~N.,  1988, \apss, 146, 205

\bibitem[\protect\citeauthoryear{{Bhat}, {Gupta}, {Kramer}, {Karastergiou},
  {Lyne} \& {Johnston}}{{Bhat} et~al.}{2007}]{bgk+07}
{Bhat} N.~D.~R.,  {Gupta} Y.,  {Kramer} M.,  {Karastergiou} A.,  {Lyne} A.~G.,
    {Johnston} S.,  2007, A\&A, 462, 257

\bibitem[\protect\citeauthoryear{{Bhattacharyya}, {Gupta} \&
  {Gil}}{{Bhattacharyya} et~al.}{2008}]{bgg08}
{Bhattacharyya} B.,  {Gupta} Y.,    {Gil} J.,  2008, MNRAS, 383, 1538

\bibitem[\protect\citeauthoryear{{Bhattacharyya}, {Gupta} \&
  {Gil}}{{Bhattacharyya} et~al.}{2010}]{bgg10}
{Bhattacharyya} B.,  {Gupta} Y.,    {Gil} J.,  2010, \mnras, 408, 407

\bibitem[\protect\citeauthoryear{Biggs}{Biggs}{1992}]{big92a}
Biggs J.~D.,  1992, ApJ, 394, 574

\bibitem[\protect\citeauthoryear{Biggs, McCulloch, Hamilton, Manchester \&
  Lyne}{Biggs et~al.}{1985}]{bmh+85}
Biggs J.~D.,  McCulloch P.~M.,  Hamilton P.~A.,  Manchester R.~N.,    Lyne
  A.~G.,  1985, MNRAS, 215, 281

\bibitem[\protect\citeauthoryear{Blandford}{Blandford}{1975}]{bla75}
Blandford R.~D.,  1975, MNRAS, 170, 551

\bibitem[\protect\citeauthoryear{{Burke-Spolaor} \& {Bailes}}{{Burke-Spolaor}
  \& {Bailes}}{2010}]{bb10}
{Burke-Spolaor} S.,  {Bailes} M.,  2010, \mnras, 402, 855

\bibitem[\protect\citeauthoryear{{Burke-Spolaor}, {Bailes}, {Johnston},
  {Bates}, {Bhat}, {Burgay}, {D'Amico}, {Jameson}, {Keith}, {Kramer}, {Levin},
  {Milia}, {Possenti}, {Stappers} \& {van Straten}}{{Burke-Spolaor}
  et~al.}{2011}]{bbj+11}
{Burke-Spolaor} S.,  {Bailes} M.,  {Johnston} S.,  {Bates} S.~D.,  {Bhat}
  N.~D.~R.,  {Burgay} M.,  {D'Amico} N.,  {Jameson} A.,  {Keith} M.~J.,
  {Kramer} M.,  {Levin} L.,  {Milia} S.,  {Possenti} A.,  {Stappers} B.,
  {van Straten} W.,  2011, \mnras, 416, 2465

\bibitem[\protect\citeauthoryear{{Burke-Spolaor}, {Johnston}, {Bailes},
  {Bates}, {Bhat}, {Burgay}, {Champion}, {D'Amico}, {Keith}, {Kramer}, {Levin},
  {Milia}, {Possenti}, {Stappers} \& {van Straten}}{{Burke-Spolaor}
  et~al.}{2012}]{bjb+12}
{Burke-Spolaor} S.,  {Johnston} S.,  {Bailes} M.,  {Bates} S.~D.,  {Bhat}
  N.~D.~R.,  {Burgay} M.,  {Champion} D.~J.,  {D'Amico} N.,  {Keith} M.~J.,
  {Kramer} M.,  {Levin} L.,  {Milia} S.,  {Possenti} A.,  {Stappers} B.,
  {van Straten} W.,  2012, \mnras, 423, 1351

\bibitem[\protect\citeauthoryear{{Cadez}, {Galicic} \& {Calvani}}{{Cadez}
  et~al.}{1997}]{cgc97}
{Cadez} A.,  {Galicic} M.,    {Calvani} M.,  1997, \aap, 324, 1005

\bibitem[\protect\citeauthoryear{{Camilo}, {Ransom}, {Chatterjee}, {Johnston}
  \& {Demorest}}{{Camilo} et~al.}{2012}]{crc+12}
{Camilo} F.,  {Ransom} S.~M.,  {Chatterjee} S.,  {Johnston} S.,    {Demorest}
  P.,  2012, \apj, 746, 63

\bibitem[\protect\citeauthoryear{Chen \& Ruderman}{Chen \&
  Ruderman}{1993}]{cr93}
Chen K.,  Ruderman M.,  1993, ApJ, 408, 179

\bibitem[\protect\citeauthoryear{{Cheng}}{{Cheng}}{1981}]{che81}
{Cheng} A.~F.,  1981, in {Sieber} W.,  {Wielebinski} R.,  eds, Pulsars: 13
  Years of Research on Neutron Stars Vol.~95 of IAU Symposium, {Observational
  consequences of polar CAP theories}.
pp 99--101

\bibitem[\protect\citeauthoryear{Cheng \& Ruderman}{Cheng \&
  Ruderman}{1977}]{cr77}
Cheng A.~F.,  Ruderman M.,  1977, ApJ, 212, 800

\bibitem[\protect\citeauthoryear{{Clemens} \& {Rosen}}{{Clemens} \&
  {Rosen}}{2004}]{cr04a}
{Clemens} J.~C.,  {Rosen} R.,  2004, ApJ, 609, 340

\bibitem[\protect\citeauthoryear{Cole}{Cole}{1970}]{col70a}
Cole T.~W.,  1970, Nat., 227, 788

\bibitem[\protect\citeauthoryear{{Contopoulos}}{{Contopoulos}}{2005}]{con05}
{Contopoulos} I.,  2005, A\&A, 442, 579

\bibitem[\protect\citeauthoryear{{Contopoulos}, {Kazanas} \&
  {Fendt}}{{Contopoulos} et~al.}{1999}]{ckf99}
{Contopoulos} I.,  {Kazanas} D.,    {Fendt} C.,  1999, ApJ, 511, 351

\bibitem[\protect\citeauthoryear{Cordes}{Cordes}{1978}]{cor78}
Cordes J.~M.,  1978, ApJ, 222, 1006

\bibitem[\protect\citeauthoryear{Cordes}{Cordes}{1981}]{cor81}
Cordes J.~M.,  1981, in Sieber W.,  Wielebinski R.,  eds, Pulsars, {IAU}
  {S}ymposium 95 Radio observational constraints on pulsar emission mechanisms.
Reidel, Dordrecht, p.~115

\bibitem[\protect\citeauthoryear{{Cordes}}{{Cordes}}{1983}]{cor83}
{Cordes} J.~M.,  1983, in {Burns} M.~L.,  {Harding} A.~K.,   {Ramaty} R.,  eds,
  Positron-Electron Pairs in Astrophysics Vol.~101 of American Institute of
  Physics Conference Series, {Radio pulsars - Intensity, polarization, and
  rotation fluctuations}.
pp 98--112

\bibitem[\protect\citeauthoryear{{Cordes}}{{Cordes}}{2013}]{cor13}
{Cordes} J.~M.,  2013, \apj, 775, 47

\bibitem[\protect\citeauthoryear{{Cordes}, {Bhat}, {Hankins}, {McLaughlin} \&
  {Kern}}{{Cordes} et~al.}{2004}]{cbh+04}
{Cordes} J.~M.,  {Bhat} N.~D.~R.,  {Hankins} T.~H.,  {McLaughlin} M.~A.,
  {Kern} J.,  2004, ApJ, 612, 375

\bibitem[\protect\citeauthoryear{{Cordes} \& {Lazio}}{{Cordes} \&
  {Lazio}}{2002}]{NE2001}
{Cordes} J.~M.,  {Lazio} T.~J.~W.,  2002, ArXiv Astrophysics e-prints

\bibitem[\protect\citeauthoryear{{Cordes} \& {Shannon}}{{Cordes} \&
  {Shannon}}{2008}]{csh08}
{Cordes} J.~M.,  {Shannon} R.~M.,  2008, \apj, 682, 1152

\bibitem[\protect\citeauthoryear{{Crawford} \& {Lorimer}}{{Crawford} \&
  {Lorimer}}{2007}]{cl07}
{Crawford} F.,  {Lorimer} D.~R.,  2007, ATNF Proposal, p.~1104

\bibitem[\protect\citeauthoryear{{Daugherty} \& {Harding}}{{Daugherty} \&
  {Harding}}{1986}]{dh86}
{Daugherty} J.~K.,  {Harding} A.~K.,  1986, ApJ, 309, 362

\bibitem[\protect\citeauthoryear{Davies, Lyne, Smith, Izvekova, Kuzmin \&
  Shitov}{Davies et~al.}{1984}]{dls+84}
Davies J.~G.,  Lyne A.~G.,  Smith F.~G.,  Izvekova V.~A.,  Kuzmin A.~D.,
  Shitov Y.~P.,  1984, MNRAS, 211, 57

\bibitem[\protect\citeauthoryear{Deich, Cordes, Hankins \& Rankin}{Deich
  et~al.}{1986}]{dchr86}
Deich W. T.~S.,  Cordes J.~M.,  Hankins T.~H.,    Rankin J.~M.,  1986, ApJ,
  300, 540

\bibitem[\protect\citeauthoryear{{Deneva}, {Cordes}, {McLaughlin}, {Nice},
  {Lorimer}, {Crawford}, {Bhat}, {Camilo}, {Champion}, {Freire}, {Edel},
  {Kondratiev} \& {Hessels}}{{Deneva} et~al.}{2009}]{dcm+09}
{Deneva} J.~S.,  {Cordes} J.~M.,  {McLaughlin} M.~A.,  {Nice} D.~J.,  {Lorimer}
  D.~R.,  {Crawford} F.,  {Bhat} N.~D.~R.,  {Camilo} F.,  {Champion} D.~J.,
  {Freire} P.~C.~C.,  {Edel} S.,  {Kondratiev} V.~I.,    {Hessels} J.~W.~T.,
  2009, \apj, 703, 2259

\bibitem[\protect\citeauthoryear{{Deshpande} \& {Rankin}}{{Deshpande} \&
  {Rankin}}{1999}]{dr99}
{Deshpande} A.~A.,  {Rankin} J.~M.,  1999, ApJ, 524, 1008

\bibitem[\protect\citeauthoryear{Deshpande \& Rankin}{Deshpande \&
  Rankin}{2001}]{dr01}
Deshpande A.~A.,  Rankin J.~M.,  2001, MNRAS, 322, 438

\bibitem[\protect\citeauthoryear{{Deutsch}}{{Deutsch}}{1955}]{deu55}
{Deutsch} A.~J.,  1955, Annales d'Astrophysique, 18, 1

\bibitem[\protect\citeauthoryear{Dewey, Taylor, Maguire \& Stokes}{Dewey
  et~al.}{1988}]{dtms88}
Dewey R.~J.,  Taylor J.~H.,  Maguire C.~M.,    Stokes G.~H.,  1988, ApJ, 332,
  762

\bibitem[\protect\citeauthoryear{{Dhillon}, {Keane}, {Marsh}, {Stappers},
  {Copperwheat}, {Hickman}, {Jordan}, {Kerry}, {Kramer}, {Littlefair}, {Lyne},
  {Mignani} \& {Shearer}}{{Dhillon} et~al.}{2011}]{dkm+11}
{Dhillon} V.~S.,  {Keane} E.~F.,  {Marsh} T.~R.,  {Stappers} B.~W.,
  {Copperwheat} C.~M.,  {Hickman} R.~D.~G.,  {Jordan} C.~A.,  {Kerry} P.,
  {Kramer} M.,  {Littlefair} S.~P.,  {Lyne} A.~G.,  {Mignani} R.~P.,
  {Shearer} A.,  2011, \mnras, 414, 3627

\bibitem[\protect\citeauthoryear{Drake \& Craft}{Drake \& Craft}{1968}]{dc68}
Drake F.~D.,  Craft H.~D.,  1968, Nat., 220, 231

\bibitem[\protect\citeauthoryear{Durdin, Large, Little, Manchester, Lyne \&
  Taylor}{Durdin et~al.}{1979}]{dll+79}
Durdin J.~M.,  Large M.~I.,  Little A.~G.,  Manchester R.~N.,  Lyne A.~G.,
  Taylor J.~H.,  1979, \mnras, 186, 39P

\bibitem[\protect\citeauthoryear{{Dyks}, {Zhang} \& {Gil}}{{Dyks}
  et~al.}{2005}]{dzg05}
{Dyks} J.,  {Zhang} B.,    {Gil} J.,  2005, \apjl, 626, L45

\bibitem[\protect\citeauthoryear{{Ershov} \& {Kuzmin}}{{Ershov} \&
  {Kuzmin}}{2005}]{EK05}
{Ershov} A.~A.,  {Kuzmin} A.~D.,  2005, A\&A, 443, 593

\bibitem[\protect\citeauthoryear{{Esamdin}, {Lyne}, {Graham-Smith}, {Kramer},
  {Manchester} \& {Wu}}{{Esamdin} et~al.}{2005}]{elg+05}
{Esamdin} A.,  {Lyne} A.~G.,  {Graham-Smith} F.,  {Kramer} M.,  {Manchester}
  R.~N.,    {Wu} X.,  2005, MNRAS, 356, 59

\bibitem[\protect\citeauthoryear{{Faulkner}, {Stairs}, {Kramer}, {Lyne},
  {Hobbs}, {Possenti}, {Lorimer}, {Manchester}, {McLaughlin}, {D'Amico},
  {Camilo} \& {Burgay}}{{Faulkner} et~al.}{2004}]{fsk+04}
{Faulkner} A.~J.,  {Stairs} I.~H.,  {Kramer} M.,  {Lyne} A.~G.,  {Hobbs} G.,
  {Possenti} A.,  {Lorimer} D.~R.,  {Manchester} R.~N.,  {McLaughlin} M.~A.,
  {D'Amico} N.,  {Camilo} F.,    {Burgay} M.,  2004, MNRAS, 355, 147

\bibitem[\protect\citeauthoryear{Filippenko \& Radhakrishnan}{Filippenko \&
  Radhakrishnan}{1982}]{fr82}
Filippenko A.~V.,  Radhakrishnan V.,  1982, ApJ, 263, 828

\bibitem[\protect\citeauthoryear{{Filippenko}, {Readhead} \&
  {Ewing}}{{Filippenko} et~al.}{1983}]{fre83}
{Filippenko} A.~V.,  {Readhead} A.~C.~S.,    {Ewing} M.~S.,  1983, in {Burns}
  M.~L.,  {Harding} A.~K.,   {Ramaty} R.,  eds, Positron-Electron Pairs in
  Astrophysics Vol.~101 of American Institute of Physics Conference Series,
  {The effect of nulls on the drifting subpulses in PSR 0809+74}.
pp 113--117

\bibitem[\protect\citeauthoryear{Foster, Cadwell, Wolszczan \& Anderson}{Foster
  et~al.}{1995}]{fcwa95}
Foster R.~S.,  Cadwell B.~J.,  Wolszczan A.,    Anderson S.~B.,  1995, ApJ,
  454, 826

\bibitem[\protect\citeauthoryear{{Gangadhara} \& {Gupta}}{{Gangadhara} \&
  {Gupta}}{2001}]{gg01}
{Gangadhara} R.~T.,  {Gupta} Y.,  2001, ApJ, 555, 31

\bibitem[\protect\citeauthoryear{{Ghosh}}{{Ghosh}}{2007}]{gho07}
{Ghosh} P.,  2007, {Rotation and Accretion Powered Pulsars}.
World Scientific Publishing Co.

\bibitem[\protect\citeauthoryear{Gil, Jessner, Kijak, Kramer, Malofeev, Malov,
  Seiradakis, Sieber \& Wielebinski}{Gil et~al.}{1994}]{gjk+94}
Gil J.~A.,  Jessner A.,  Kijak J.,  Kramer M.,  Malofeev V.,  Malov I.,
  Seiradakis J.~H.,  Sieber W.,    Wielebinski R.,  1994, A\&A, 282, 45

\bibitem[\protect\citeauthoryear{Glendenning}{Glendenning}{1990}]{gle90}
Glendenning N.~K.,  1990, ApJ, 359, 186

\bibitem[\protect\citeauthoryear{{Gold}}{{Gold}}{1968}]{gold68}
{Gold} T.,  1968, \nat, 218, 731

\bibitem[\protect\citeauthoryear{Gold}{Gold}{1968}]{gol68}
Gold T.,  1968, Nat., 218, 731

\bibitem[\protect\citeauthoryear{Goldreich \& Julian}{Goldreich \&
  Julian}{1969}]{gj69}
Goldreich P.,  Julian W.~H.,  1969, ApJ, 157, 869

\bibitem[\protect\citeauthoryear{Gomez-Gonzalez \& Guelin}{Gomez-Gonzalez \&
  Guelin}{1974}]{gg74}
Gomez-Gonzalez J.,  Guelin M.,  1974, 32, 441

\bibitem[\protect\citeauthoryear{Gould \& Lyne}{Gould \& Lyne}{1998}]{gl98}
Gould D.~M.,  Lyne A.~G.,  1998, MNRAS, 301, 235

\bibitem[\protect\citeauthoryear{{Guseinov} \& {Iusifov}}{{Guseinov} \&
  {Iusifov}}{1983}]{gi83}
{Guseinov} O.~K.,  {Iusifov} I.~M.,  1983, \apss, 94, 249

\bibitem[\protect\citeauthoryear{{Hankins}, {Kern}, {Weatherall} \&
  {Eilek}}{{Hankins} et~al.}{2003}]{hkw+03}
{Hankins} T.~H.,  {Kern} J.~S.,  {Weatherall} J.~C.,    {Eilek} J.~A.,  2003,
  \nat, 422, 141

\bibitem[\protect\citeauthoryear{{Herfindal} \& {Rankin}}{{Herfindal} \&
  {Rankin}}{2007a}]{hr07}
{Herfindal} J.~L.,  {Rankin} J.~M.,  2007a, \mnras, 380, 430

\bibitem[\protect\citeauthoryear{{Herfindal} \& {Rankin}}{{Herfindal} \&
  {Rankin}}{2007b}]{hjr07}
{Herfindal} J.~L.,  {Rankin} J.~M.,  2007b, MNRAS, 380, 430

\bibitem[\protect\citeauthoryear{{Herfindal} \& {Rankin}}{{Herfindal} \&
  {Rankin}}{2009}]{hr09}
{Herfindal} J.~L.,  {Rankin} J.~M.,  2009, \mnras, 393, 1391

\bibitem[\protect\citeauthoryear{{Hermsen et al.}}{{Hermsen et
  al.}}{2013}]{hhk+13}
{Hermsen et al.} 2013, Science, 339, 436

\bibitem[\protect\citeauthoryear{{Hessels}, {Ransom}, {Stairs}, {Freire},
  {Kaspi} \& {Camilo}}{{Hessels} et~al.}{2006}]{hrs+06}
{Hessels} J.~W.~T.,  {Ransom} S.~M.,  {Stairs} I.~H.,  {Freire} P.~C.~C.,
  {Kaspi} V.~M.,    {Camilo} F.,  2006, 311, 1901

\bibitem[\protect\citeauthoryear{Hewish, Bell, Pilkington, Scott \&
  Collins}{Hewish et~al.}{1968}]{hbp+68}
Hewish A.,  Bell S.~J.,  Pilkington J. D.~H.,  Scott P.~F.,    Collins R.~A.,
  1968, Nat., 217, 709

\bibitem[\protect\citeauthoryear{{Heyl} \& {Hernquist}}{{Heyl} \&
  {Hernquist}}{2002}]{hh02}
{Heyl} J.~S.,  {Hernquist} L.,  2002, ApJ, 567, 510

\bibitem[\protect\citeauthoryear{{Hobbs}, {Faulkner}, {Stairs}, {Camilo},
  {Manchester}, {Lyne}, {Kramer}, {D'Amico}, {Kaspi}, {Possenti}, {McLaughlin},
  {Lorimer}, {Burgay}, {Joshi} \& {Crawford}}{{Hobbs} et~al.}{2004}]{hfs+04}
{Hobbs} G.,  {Faulkner} A.,  {Stairs} I.~H.,  {Camilo} F.,  {Manchester} R.~N.,
   {Lyne} A.~G.,  {Kramer} M.,  {D'Amico} N.,  {Kaspi} V.~M.,  {Possenti} A.,
  {McLaughlin} M.~A.,  {Lorimer} D.~R.,  {Burgay} M.,  {Joshi} B.~C.,
  {Crawford} F.,  2004, MNRAS, 352, 1439

\bibitem[\protect\citeauthoryear{Hobbs, Lyne, Kramer, Martin \& Jordan}{Hobbs
  et~al.}{2004}]{hlk+04}
Hobbs G.,  Lyne A.~G.,  Kramer M.,  Martin C.~E.,    Jordan C.,  2004, MNRAS,
  353, 1311

\bibitem[\protect\citeauthoryear{{Honnappa}, {Lewandowski}, {Kijak},
  {Deshpande}, {Gil}, {Maron} \& {Jessner}}{{Honnappa} et~al.}{2012}]{hlk+12}
{Honnappa} S.,  {Lewandowski} W.,  {Kijak} J.,  {Deshpande} A.~A.,  {Gil} J.,
  {Maron} O.,    {Jessner} A.,  2012, \mnras, 421, 1996

\bibitem[\protect\citeauthoryear{{Jackson}}{{Jackson}}{1975}]{jackson}
{Jackson} J.~D.,  1975, {Classical electrodynamics}

\bibitem[\protect\citeauthoryear{{Janssen} \& {van Leeuwen}}{{Janssen} \& {van
  Leeuwen}}{2004}]{jv04}
{Janssen} G.~H.,  {van Leeuwen} J.,  2004, A\&A, 425, 255

\bibitem[\protect\citeauthoryear{Johnston, Manchester, Lyne, Kaspi \&
  D'Amico}{Johnston et~al.}{1995}]{jml+95}
Johnston S.,  Manchester R.~N.,  Lyne A.~G.,  Kaspi V.~M.,    D'Amico N.,
  1995, 293, 795

\bibitem[\protect\citeauthoryear{Jones \& Anderson}{Jones \&
  Anderson}{2001}]{ja01}
Jones D.~I.,  Anderson N.,  2001, MNRAS, 324, 811

\bibitem[\protect\citeauthoryear{{Joshi}}{{Joshi}}{2013}]{joshi13}
{Joshi} B.~C.,  2013, International Journal of Modern Physics D, 22, 41008

\bibitem[\protect\citeauthoryear{{Joshi}, {Kramer}, {Lyne}, {McLaughlin} \&
  {Stairs}}{{Joshi} et~al.}{2004}]{J04}
{Joshi} B.~C.,  {Kramer} M.,  {Lyne} A.~G.,  {McLaughlin} M.~A.,    {Stairs}
  I.~H.,  2004, in {F.~Camilo \& B.~M.~Gaensler} ed., Young Neutron Stars and
  Their Environments Vol.~218 of IAU Symposium, {Giant Pulses in Millisecond
  Pulsars}.
pp 319--+

\bibitem[\protect\citeauthoryear{{Joshi}, {McLaughlin}, {Lyne}, {Ludovici},
  {Pawar}, {Faulkner}, {Lorimer}, {Kramer} \& {Davies}}{{Joshi}
  et~al.}{2009}]{jml+09}
{Joshi} B.~C.,  {McLaughlin} M.~A.,  {Lyne} A.~G.,  {Ludovici} D.~A.,  {Pawar}
  N.~A.,  {Faulkner} A.~J.,  {Lorimer} D.~R.,  {Kramer} M.,    {Davies} M.~L.,
  2009, \mnras, 398, 943

\bibitem[\protect\citeauthoryear{{Joshi} \& {Vivekanand}}{{Joshi} \&
  {Vivekanand}}{2000}]{jv00}
{Joshi} B.~C.,  {Vivekanand} M.,  2000, \mnras, 316, 716

\bibitem[\protect\citeauthoryear{{Karuppusamy}, {Stappers} \& {van
  Straten}}{{Karuppusamy} et~al.}{2008}]{ksv08}
{Karuppusamy} R.,  {Stappers} B.,    {van Straten} W.,  2008, PASP, 120, 191

\bibitem[\protect\citeauthoryear{{Karuppusamy}, {Stappers} \&
  {Serylak}}{{Karuppusamy} et~al.}{2011}]{kss11}
{Karuppusamy} R.,  {Stappers} B.~W.,    {Serylak} M.,  2011, A\&A, 525, A55

\bibitem[\protect\citeauthoryear{{Kazbegi}, {Machabeli}, {Melikidze} \&
  {Shukre}}{{Kazbegi} et~al.}{1996}]{kmms96}
{Kazbegi} A.,  {Machabeli} G.,  {Melikidze} G.,    {Shukre} C.,  1996, \aap,
  309, 515

\bibitem[\protect\citeauthoryear{Kazbegi, Machabeli \& Melikidze}{Kazbegi
  et~al.}{1991}]{kmm91}
Kazbegi A.~Z.,  Machabeli G.~Z.,    Melikidze G.~I.,  1991, MNRAS, 253, 377

\bibitem[\protect\citeauthoryear{{Keane}, {Kramer}, {Lyne}, {Stappers} \&
  {McLaughlin}}{{Keane} et~al.}{2011}]{kkl+11}
{Keane} E.~F.,  {Kramer} M.,  {Lyne} A.~G.,  {Stappers} B.~W.,    {McLaughlin}
  M.~A.,  2011, \mnras, 415, 3065

\bibitem[\protect\citeauthoryear{{Keane}, {Ludovici}, {Eatough}, {Kramer},
  {Lyne}, {McLaughlin} \& {Stappers}}{{Keane} et~al.}{2010}]{kle+10}
{Keane} E.~F.,  {Ludovici} D.~A.,  {Eatough} R.~P.,  {Kramer} M.,  {Lyne}
  A.~G.,  {McLaughlin} M.~A.,    {Stappers} B.~W.,  2010, \mnras, 401, 1057

\bibitem[\protect\citeauthoryear{{Keane} \& {McLaughlin}}{{Keane} \&
  {McLaughlin}}{2011}]{km11}
{Keane} E.~F.,  {McLaughlin} M.~A.,  2011, Bulletin of the Astronomical Society
  of India, 39, 333

\bibitem[\protect\citeauthoryear{Kijak \& Gil}{Kijak \& Gil}{2003}]{kg03a}
Kijak J.,  Gil J.,  2003, ApJ, 397, 969

\bibitem[\protect\citeauthoryear{Kijal \& Gil}{Kijal \& Gil}{1998}]{kg98}
Kijal J.,  Gil J.,  1998, MNRAS, 299, 855

\bibitem[\protect\citeauthoryear{Kinkhabwala \& Thorsett}{Kinkhabwala \&
  Thorsett}{2000}]{kt00}
Kinkhabwala A.,  Thorsett S.~E.,  2000, ApJ, 535, 365

\bibitem[\protect\citeauthoryear{{Kloumann} \& {Rankin}}{{Kloumann} \&
  {Rankin}}{2010}]{kr10}
{Kloumann} I.~M.,  {Rankin} J.~M.,  2010, MNRAS, 408, 40

\bibitem[\protect\citeauthoryear{Komesaroff}{Komesaroff}{1970}]{kom70}
Komesaroff M.~M.,  1970, Nat., 225, 612

\bibitem[\protect\citeauthoryear{{Kramer}, {Bell}, {Manchester}, {Lyne},
  {Camilo}, {Stairs}, {D'Amico}, {Kaspi}, {Hobbs}, {Morris}, {Crawford},
  {Possenti}, {Joshi}, {McLaughlin}, {Lorimer} \& {Faulkner}}{{Kramer}
  et~al.}{2003}]{kbm+03}
{Kramer} M.,  {Bell} J.~F.,  {Manchester} R.~N.,  {Lyne} A.~G.,  {Camilo} F.,
  {Stairs} I.~H.,  {D'Amico} N.,  {Kaspi} V.~M.,  {Hobbs} G.,  {Morris} D.~J.,
  {Crawford} F.,  {Possenti} A.,  {Joshi} B.~C.,  {McLaughlin} M.~A.,
  {Lorimer} D.~R.,    {Faulkner} A.~J.,  2003, MNRAS, 342, 1299

\bibitem[\protect\citeauthoryear{{Kramer}, {Lyne}, {O'Brien}, {Jordan} \&
  {Lorimer}}{{Kramer} et~al.}{2006}]{klo+06}
{Kramer} M.,  {Lyne} A.~G.,  {O'Brien} J.~T.,  {Jordan} C.~A.,    {Lorimer}
  D.~R.,  2006, Science, 312, 549

\bibitem[\protect\citeauthoryear{Krause-Polstorff \& Michel}{Krause-Polstorff
  \& Michel}{1985}]{km85}
Krause-Polstorff J.,  Michel F.~C.,  1985, MNRAS, 213, 43P

\bibitem[\protect\citeauthoryear{{Kuzmin} \& {Ershov}}{{Kuzmin} \&
  {Ershov}}{2004}]{ke04}
{Kuzmin} A.~D.,  {Ershov} A.~A.,  2004, A\&A, 427, 575

\bibitem[\protect\citeauthoryear{{Lattimer} \& {Prakash}}{{Lattimer} \&
  {Prakash}}{2001}]{lp01}
{Lattimer} J.~M.,  {Prakash} M.,  2001, ApJ, 550, 426

\bibitem[\protect\citeauthoryear{{Lazaridis} \& {Seiradakis}}{{Lazaridis} \&
  {Seiradakis}}{2006}]{lr06}
{Lazaridis} K.,  {Seiradakis} J.~H.,  2006, in {Solomos} N.,  ed., Recent
  Advances in Astronomy and Astrophysics Vol.~848 of American Institute of
  Physics Conference Series, {Pulsar Nulling Quantitative Analysis}.
pp 309--315

\bibitem[\protect\citeauthoryear{{Lewandowski}, {Wolszczan}, {Feiler},
  {Konacki} \& {Soltysi'nski}}{{Lewandowski} et~al.}{2004}]{lwf+04}
{Lewandowski} W.,  {Wolszczan} A.,  {Feiler} G.,  {Konacki} M.,
  {Soltysi'nski} T.,  2004, ApJ, 600, 905

\bibitem[\protect\citeauthoryear{{Li}, {Esamdin}, {Manchester}, {Qian} \&
  {Niu}}{{Li} et~al.}{2012}]{lem+12}
{Li} J.,  {Esamdin} A.,  {Manchester} R.~N.,  {Qian} M.~F.,    {Niu} H.~B.,
  2012, MNRAS, 425, 1294

\bibitem[\protect\citeauthoryear{{Li} \& {Wang}}{{Li} \& {Wang}}{1995}]{lw95}
{Li} X.-d.,  {Wang} Z.-r.,  1995, ChA\&A, 19, 302

\bibitem[\protect\citeauthoryear{{Lorimer}}{{Lorimer}}{2008}]{lor08}
{Lorimer} D.~R.,  2008, Living Reviews in Relativity, 11, 8

\bibitem[\protect\citeauthoryear{{Lorimer}, {Camilo} \& {Xilouris}}{{Lorimer}
  et~al.}{2002}]{lcx02}
{Lorimer} D.~R.,  {Camilo} F.,    {Xilouris} K.~M.,  2002, ApJ, 123, 1750

\bibitem[\protect\citeauthoryear{{Lorimer}, {Faulkner}, {Lyne}, {Manchester},
  {Kramer}, {McLaughlin}, {Hobbs}, {Possenti}, {Stairs}, {Camilo}, {Burgay},
  {D'Amico}, {Corongiu} \& {Crawford}}{{Lorimer} et~al.}{2006}]{lfl+06}
{Lorimer} D.~R.,  {Faulkner} A.~J.,  {Lyne} A.~G.,  {Manchester} R.~N.,
  {Kramer} M.,  {McLaughlin} M.~A.,  {Hobbs} G.,  {Possenti} A.,  {Stairs}
  I.~H.,  {Camilo} F.,  {Burgay} M.,  {D'Amico} N.,  {Corongiu} A.,
  {Crawford} F.,  2006, \mnras, 372, 777

\bibitem[\protect\citeauthoryear{{Lorimer}, {Jessner}, {Seiradakis}, {Lyne},
  {D'Amico}, {Athanasopoulos}, {Xilouris}, {Kramer} \& {Wielebinski}}{{Lorimer}
  et~al.}{1998}]{ljs+98}
{Lorimer} D.~R.,  {Jessner} A.,  {Seiradakis} J.~H.,  {Lyne} A.~G.,  {D'Amico}
  N.,  {Athanasopoulos} A.,  {Xilouris} K.~M.,  {Kramer} M.,    {Wielebinski}
  R.,  1998, \aaps, 128, 541

\bibitem[\protect\citeauthoryear{{Lorimer} \& {Kramer}}{{Lorimer} \&
  {Kramer}}{2004}]{handbook}
{Lorimer} D.~R.,  {Kramer} M.,  2004, {Handbook of Pulsar Astronomy}

\bibitem[\protect\citeauthoryear{{Lorimer}, {Lyne}, {McLaughlin}, {Kramer},
  {Pavlov} \& {Chang}}{{Lorimer} et~al.}{2012}]{llm+12}
{Lorimer} D.~R.,  {Lyne} A.~G.,  {McLaughlin} M.~A.,  {Kramer} M.,  {Pavlov}
  G.~G.,    {Chang} C.,  2012, \apj, 758, 141

\bibitem[\protect\citeauthoryear{Lorimer, Yates, Lyne \& Gould}{Lorimer
  et~al.}{1995}]{lylg95}
Lorimer D.~R.,  Yates J.~A.,  Lyne A.~G.,    Gould D.~M.,  1995, MNRAS, 273,
  411

\bibitem[\protect\citeauthoryear{{Lyne}, {Hobbs}, {Kramer}, {Stairs} \&
  {Stappers}}{{Lyne} et~al.}{2010}]{lhk+10}
{Lyne} A.,  {Hobbs} G.,  {Kramer} M.,  {Stairs} I.,    {Stappers} B.,  2010,
  Science, 329, 408

\bibitem[\protect\citeauthoryear{{Lyne}}{{Lyne}}{2009}]{lyn09}
{Lyne} A.~G.,  2009, in {Becker} W.,  ed., Astrophysics and Space Science
  Library Vol.~357 of Astrophysics and Space Science Library, {Intermittent
  Pulsars}.
p.~67

\bibitem[\protect\citeauthoryear{Lyne \& Ashworth}{Lyne \&
  Ashworth}{1983}]{la83}
Lyne A.~G.,  Ashworth M.,  1983, MNRAS, 204, 519

\bibitem[\protect\citeauthoryear{Lyne \& Manchester}{Lyne \&
  Manchester}{1988}]{lm88}
Lyne A.~G.,  Manchester R.~N.,  1988, MNRAS, 234, 477

\bibitem[\protect\citeauthoryear{Lyne \& Rickett}{Lyne \& Rickett}{1968}]{lr68}
Lyne A.~G.,  Rickett B.~J.,  1968, Nat., 218, 326

\bibitem[\protect\citeauthoryear{{Manchester}, {Hobbs}, {Teoh} \&
  {Hobbs}}{{Manchester} et~al.}{2005}]{mhth05}
{Manchester} R.~N.,  {Hobbs} G.~B.,  {Teoh} A.,    {Hobbs} M.,  2005, \aj, 129,
  1993

\bibitem[\protect\citeauthoryear{Manchester, Lyne, Camilo, Bell, Kaspi,
  D'Amico, McKay, Crawford, Stairs, Possenti, Morris \& Sheppard}{Manchester
  et~al.}{2001}]{mlc+01}
Manchester R.~N.,  Lyne A.~G.,  Camilo F.,  Bell J.~F.,  Kaspi V.~M.,  D'Amico
  N.,  McKay N. P.~F.,  Crawford F.,  Stairs I.~H.,  Possenti A.,  Morris
  D.~J.,    Sheppard D.~C.,  2001, MNRAS, 328, 17

\bibitem[\protect\citeauthoryear{Manchester, Lyne, D'Amico, Bailes, Johnston,
  Lorimer, Harrison, Nicastro \& Bell}{Manchester et~al.}{1996}]{mld+96}
Manchester R.~N.,  Lyne A.~G.,  D'Amico N.,  Bailes M.,  Johnston S.,  Lorimer
  D.~R.,  Harrison P.~A.,  Nicastro L.,    Bell J.~F.,  1996, MNRAS, 279, 1235

\bibitem[\protect\citeauthoryear{Manchester \& Taylor}{Manchester \&
  Taylor}{1977}]{mt77}
Manchester R.~N.,  Taylor J.~H.,  1977, Pulsars.
Freeman, San Francisco

\bibitem[\protect\citeauthoryear{{Maron}, {Kijak}, {Kramer} \&
  {Wielebinski}}{{Maron} et~al.}{2000}]{mkk+00}
{Maron} O.,  {Kijak} J.,  {Kramer} M.,    {Wielebinski} R.,  2000, \aaps, 147,
  195

\bibitem[\protect\citeauthoryear{{McLaughlin}, {Lyne}, {Lorimer}, {Kramer},
  {Faulkner}, {Manchester}, {Cordes}, {Camilo}, {Possenti}, {Stairs}, {Hobbs},
  {D'Amico}, {Burgay} \& {O'Brien}}{{McLaughlin} et~al.}{2006}]{mll+06}
{McLaughlin} M.~A.,  {Lyne} A.~G.,  {Lorimer} D.~R.,  {Kramer} M.,  {Faulkner}
  A.~J.,  {Manchester} R.~N.,  {Cordes} J.~M.,  {Camilo} F.,  {Possenti} A.,
  {Stairs} I.~H.,  {Hobbs} G.,  {D'Amico} N.,  {Burgay} M.,    {O'Brien} J.~T.,
   2006, Nat., 439, 817

\bibitem[\protect\citeauthoryear{{McLaughlin}, {Rea}, {Gaensler}, {Chatterjee},
  {Camilo}, {Kramer}, {Lorimer}, {Lyne}, {Israel} \& {Possenti}}{{McLaughlin}
  et~al.}{2007}]{mrg+07}
{McLaughlin} M.~A.,  {Rea} N.,  {Gaensler} B.~M.,  {Chatterjee} S.,  {Camilo}
  F.,  {Kramer} M.,  {Lorimer} D.~R.,  {Lyne} A.~G.,  {Israel} G.~L.,
  {Possenti} A.,  2007, \apj, 670, 1307

\bibitem[\protect\citeauthoryear{{Melikidze}, {Gil} \& {Pataraya}}{{Melikidze}
  et~al.}{2000}]{mgp00}
{Melikidze} G.~I.,  {Gil} J.~A.,    {Pataraya} A.~D.,  2000, \apj, 544, 1081

\bibitem[\protect\citeauthoryear{Melrose}{Melrose}{1978}]{mel78}
Melrose D.~B.,  1978, ApJ, 225, 557

\bibitem[\protect\citeauthoryear{{Melrose}}{{Melrose}}{1992}]{mel92b}
{Melrose} D.~B.,  1992, Royal Society of London Philosophical Transactions
  Series A, 341, 105

\bibitem[\protect\citeauthoryear{Michel}{Michel}{1991}]{mic91}
Michel F.~C.,  1991, {Theory of Neutron Star Magnetospheres}.
{University of Chicago Press}, Chicago

\bibitem[\protect\citeauthoryear{{Miller}, {McLaughlin}, {Rea}, {Keane},
  {Lyne}, {Kramer}, {Manchester} \& {Lazaridis}}{{Miller}
  et~al.}{2011}]{mmr+11}
{Miller} J.,  {McLaughlin} M.,  {Rea} N.,  {Keane} E.,  {Lyne} A.,  {Kramer}
  M.,  {Manchester} R.,    {Lazaridis} K.,  2011, in {Burgay} M.,  {D'Amico}
  N.,  {Esposito} P.,  {Pellizzoni} A.,   {Possenti} A.,  eds, American
  Institute of Physics Conference Series Vol.~1357 of American Institute of
  Physics Conference Series, {Multiwavelength Studies of Rotating Radio
  Transients}.
pp 161--164

\bibitem[\protect\citeauthoryear{{Mitra} \& {Li}}{{Mitra} \&
  {Li}}{2004}]{mli04}
{Mitra} D.,  {Li} X.~H.,  2004, AAp, 421, 215

\bibitem[\protect\citeauthoryear{Mitra \& Rankin}{Mitra \&
  Rankin}{2002}]{mr02a}
Mitra D.,  Rankin J.~M.,  2002, ApJ, pp 322--336

\bibitem[\protect\citeauthoryear{{Morris}, {Hobbs}, {Lyne}, {Stairs}, {Camilo},
  {Manchester}, {Possenti}, {Bell}, {Kaspi}, {Amico}, {McKay}, {Crawford} \&
  {Kramer}}{{Morris} et~al.}{2002}]{mhl+02}
{Morris} D.~J.,  {Hobbs} G.,  {Lyne} A.~G.,  {Stairs} I.~H.,  {Camilo} F.,
  {Manchester} R.~N.,  {Possenti} A.,  {Bell} J.~F.,  {Kaspi} V.~M.,  {Amico}
  N.~D.,  {McKay} N.~P.~F.,  {Crawford} F.,    {Kramer} M.,  2002, MNRAS, 335,
  275

\bibitem[\protect\citeauthoryear{Narayan \& Vivekanand}{Narayan \&
  Vivekanand}{1982}]{nv82}
Narayan R.,  Vivekanand M.,  1982, A\&A, 113, L3

\bibitem[\protect\citeauthoryear{{Navarro}, {Anderson} \& {Freire}}{{Navarro}
  et~al.}{2003}]{naf03}
{Navarro} J.,  {Anderson} S.~B.,    {Freire} P.~C.,  2003, ApJ, 594, 943

\bibitem[\protect\citeauthoryear{{Nowakowski}, {Usowicz}, {Kepa} \&
  {Wolszczan}}{{Nowakowski} et~al.}{1982}]{nuk+82}
{Nowakowski} L.,  {Usowicz} J.,  {Kepa} A.,    {Wolszczan} A.,  1982, \aap,
  116, 158

\bibitem[\protect\citeauthoryear{{Okeke} \& {Akujor}}{{Okeke} \&
  {Akujor}}{1982}]{oa82}
{Okeke} P.~N.,  {Akujor} C.~E.,  1982, \apss, 84, 243

\bibitem[\protect\citeauthoryear{Oppenheimer \& Volkoff}{Oppenheimer \&
  Volkoff}{1939}]{ov39}
Oppenheimer J.~R.,  Volkoff G.,  1939, Physical Review, 55, 374

\bibitem[\protect\citeauthoryear{Oster \& Sieber}{Oster \&
  Sieber}{1976}]{os76a}
Oster L.,  Sieber W.,  1976, ApJ, 203, 233

\bibitem[\protect\citeauthoryear{Ostriker \& Gunn}{Ostriker \&
  Gunn}{1969}]{og69}
Ostriker J.~P.,  Gunn J.~E.,  1969, ApJ, 157, 1395

\bibitem[\protect\citeauthoryear{Pacini}{Pacini}{1967}]{pac67}
Pacini F.,  1967, Nat., 216, 567

\bibitem[\protect\citeauthoryear{Pacini}{Pacini}{1968}]{pac68}
Pacini F.,  1968, Nat., 219, 145

\bibitem[\protect\citeauthoryear{Page}{Page}{1973}]{pag73}
Page C.~G.,  1973, MNRAS, 163

\bibitem[\protect\citeauthoryear{{Palliyaguru}, {McLaughlin}, {Keane},
  {Kramer}, {Lyne}, {Lorimer}, {Manchester}, {Camilo} \&
  {Stairs}}{{Palliyaguru} et~al.}{2011}]{pmk+11}
{Palliyaguru} N.~T.,  {McLaughlin} M.~A.,  {Keane} E.~F.,  {Kramer} M.,  {Lyne}
  A.~G.,  {Lorimer} D.~R.,  {Manchester} R.~N.,  {Camilo} F.,    {Stairs}
  I.~H.,  2011, \mnras, 417, 1871

\bibitem[\protect\citeauthoryear{{Papoulis} \& {Pillai}}{{Papoulis} \&
  {Pillai}}{2002}]{pp02}
{Papoulis} A.,  {Pillai} S.~U.,  2002, {Probability, Random Variables, and
  Stochastic Processes, Fourth Edition}.
McGraw-Hill Higher Education

\bibitem[\protect\citeauthoryear{{Pimbley} \& {Lu}}{{Pimbley} \&
  {Lu}}{1985}]{plu85}
{Pimbley} J.~M.,  {Lu} T.~M.,  1985, J. Appl. Phys, 57, 1121

\bibitem[\protect\citeauthoryear{{Prabhu}}{{Prabhu}}{1997}]{pr97}
{Prabhu} T.,  1997, MSc Thesis, IISc Bangalore, India

\bibitem[\protect\citeauthoryear{Press, Flannery, Teukolsky \&
  Vetterling}{Press et~al.}{1986}]{pftv86}
Press W.~H.,  Flannery B.~P.,  Teukolsky S.~A.,    Vetterling W.~T.,  1986,
  Numerical Recipes: {T}he Art of Scientific Computing.
Cambridge University Press, Cambridge

\bibitem[\protect\citeauthoryear{{Press}, {Teukolsky}, {Vetterling} \&
  {Flannery}}{{Press} et~al.}{1992}]{press}
{Press} W.~H.,  {Teukolsky} S.~A.,  {Vetterling} W.~T.,    {Flannery} B.~P.,
  1992, {Numerical recipes in C. The art of scientific computing}

\bibitem[\protect\citeauthoryear{{Radhakrishnan} \& {Cooke}}{{Radhakrishnan} \&
  {Cooke}}{1969}]{rc69}
{Radhakrishnan} V.,  {Cooke} D.~J.,  1969, ApJL, 3, 225

\bibitem[\protect\citeauthoryear{Rankin}{Rankin}{1983}]{ran83a}
Rankin J.~M.,  1983, ApJ, 274, 359

\bibitem[\protect\citeauthoryear{Rankin}{Rankin}{1986}]{ran86}
Rankin J.~M.,  1986, ApJ, 301, 901

\bibitem[\protect\citeauthoryear{Rankin}{Rankin}{1990}]{ran90}
Rankin J.~M.,  1990, ApJ, 352, 247

\bibitem[\protect\citeauthoryear{Rankin}{Rankin}{1993}]{ran93}
Rankin J.~M.,  1993, ApJ, 405, 285

\bibitem[\protect\citeauthoryear{{Rankin} \& {Ramachandran}}{{Rankin} \&
  {Ramachandran}}{2003}]{rr03}
{Rankin} J.~M.,  {Ramachandran} R.,  2003, ApJ, 590, 411

\bibitem[\protect\citeauthoryear{Rankin, Stinebring \& Weisberg}{Rankin
  et~al.}{1989}]{rsw89}
Rankin J.~M.,  Stinebring D.~R.,    Weisberg J.~M.,  1989, ApJ, 346, 869

\bibitem[\protect\citeauthoryear{{Rankin} \& {Wright}}{{Rankin} \&
  {Wright}}{2007}]{wr07}
{Rankin} J.~M.,  {Wright} G.~A.~E.,  2007, \mnras, 379, 507

\bibitem[\protect\citeauthoryear{{Rankin} \& {Wright}}{{Rankin} \&
  {Wright}}{2008}]{rw08}
{Rankin} J.~M.,  {Wright} G.~A.~E.,  2008, \mnras, 385, 1923

\bibitem[\protect\citeauthoryear{{Rankin}, {Wright} \& {Brown}}{{Rankin}
  et~al.}{2013}]{rwb13}
{Rankin} J.~M.,  {Wright} G.~A.~E.,    {Brown} A.~M.,  2013, \mnras, 433, 445

\bibitem[\protect\citeauthoryear{{Redman} \& {Rankin}}{{Redman} \&
  {Rankin}}{2009}]{rr09}
{Redman} S.~L.,  {Rankin} J.~M.,  2009, MNRAS, 395, 1529

\bibitem[\protect\citeauthoryear{{Redman}, {Wright} \& {Rankin}}{{Redman}
  et~al.}{2005}]{rwr05}
{Redman} S.~L.,  {Wright} G.~A.~E.,    {Rankin} J.~M.,  2005, MNRAS, 357, 859

\bibitem[\protect\citeauthoryear{{Reynolds}, {Borkowski}, {Gaensler}, {Rea},
  {McLaughlin}, {Possenti}, {Israel}, {Burgay}, {Camilo}, {Chatterjee},
  {Kramer}, {Lyne} \& {Stairs}}{{Reynolds} et~al.}{2006}]{rbg+06}
{Reynolds} S.~P.,  {Borkowski} K.~J.,  {Gaensler} B.~M.,  {Rea} N.,
  {McLaughlin} M.,  {Possenti} A.,  {Israel} G.,  {Burgay} M.,  {Camilo} F.,
  {Chatterjee} S.,  {Kramer} M.,  {Lyne} A.,    {Stairs} I.,  2006, \apjl, 639,
  L71

\bibitem[\protect\citeauthoryear{Rickett}{Rickett}{1969}]{ric69}
Rickett B.~J.,  1969, Nat., 221, 158

\bibitem[\protect\citeauthoryear{Ritchings}{Ritchings}{1976}]{rit76}
Ritchings R.~T.,  1976, MNRAS, 176, 249

\bibitem[\protect\citeauthoryear{Romani \& Johnston}{Romani \&
  Johnston}{2001}]{rj01}
Romani R.,  Johnston S.,  2001, ApJ, 557, L93

\bibitem[\protect\citeauthoryear{{Rosen} \& {Cameron}}{{Rosen} \&
  {Cameron}}{1972}]{rc72}
{Rosen} L.~C.,  {Cameron} A.~G.~W.,  1972, \apss, 15, 137

\bibitem[\protect\citeauthoryear{{Roy}, {Gupta}, {Pen}, {Peterson}, {Kudale} \&
  {Kodilkar}}{{Roy} et~al.}{2010}]{rgp+00}
{Roy} J.,  {Gupta} Y.,  {Pen} U.-L.,  {Peterson} J.~B.,  {Kudale} S.,
  {Kodilkar} J.,  2010, Experimental Astronomy, 28, 25

\bibitem[\protect\citeauthoryear{Ruderman}{Ruderman}{1974}]{rud74}
Ruderman M.~A.,  1974, Science, 184, 1079

\bibitem[\protect\citeauthoryear{Ruderman \& Sutherland}{Ruderman \&
  Sutherland}{1975}]{rs75}
Ruderman M.~A.,  Sutherland P.~G.,  1975, ApJ, 196, 51

\bibitem[\protect\citeauthoryear{Scheuer}{Scheuer}{1968}]{sch68}
Scheuer P. A.~G.,  1968, Nat., 218, 920

\bibitem[\protect\citeauthoryear{{Shitov}, {Kuzmin}, {Dumskii} \&
  {Losovsky}}{{Shitov} et~al.}{2009}]{skd+09}
{Shitov} Y.~P.,  {Kuzmin} A.~D.,  {Dumskii} D.~V.,    {Losovsky} B.~Y.,  2009,
  Astronomy Reports, 53, 561

\bibitem[\protect\citeauthoryear{{Spitkovsky}}{{Spitkovsky}}{2004}]{spi04}
{Spitkovsky} A.,  2004, in {Camilo} F.,  {Gaensler} B.~M.,  eds, Young Neutron
  Stars and Their Environments Vol.~218 of IAU Symposium, {Electrodynamics of
  Pulsar Magnetospheres}.
p.~357

\bibitem[\protect\citeauthoryear{Staelin \& Reifenstein {III}}{Staelin \&
  Reifenstein}{1968}]{sr68}
Staelin D.~H.,  Reifenstein {III} E.~C.,  1968, Science, 162, 1481

\bibitem[\protect\citeauthoryear{Sturrock}{Sturrock}{1971}]{stu71}
Sturrock P.~A.,  1971, ApJ, 164, 529

\bibitem[\protect\citeauthoryear{{Surnis}, {Joshi}, {McLaughlin} \&
  {Gajjar}}{{Surnis} et~al.}{2013}]{sjm+13}
{Surnis} M.~P.,  {Joshi} B.~C.,  {McLaughlin} M.~A.,    {Gajjar} V.,  2013, in
  IAU Symposium Vol.~291 of IAU Symposium, {Discovery of an intermittent
  pulsar: PSR J1839+15}.
pp 508--510

\bibitem[\protect\citeauthoryear{{Swarup}, {Ananthakrishnan}, {Kapahi}, {Rao},
  {Subrahmanya} \& {Kulkarni}}{{Swarup} et~al.}{1991}]{sak+91}
{Swarup} G.,  {Ananthakrishnan} S.,  {Kapahi} V.~K.,  {Rao} A.~P.,
  {Subrahmanya} C.~R.,    {Kulkarni} V.~K.,  1991, Current Science, 60, 95

\bibitem[\protect\citeauthoryear{Taylor \& Huguenin}{Taylor \&
  Huguenin}{1971}]{th71}
Taylor J.~H.,  Huguenin G.~R.,  1971, ApJ, 167, 273

\bibitem[\protect\citeauthoryear{Taylor \& Manchester}{Taylor \&
  Manchester}{1977}]{tm77}
Taylor J.~H.,  Manchester R.~N.,  1977, ApJ, 215, 885

\bibitem[\protect\citeauthoryear{{Timokhin}}{{Timokhin}}{2006}]{tim06}
{Timokhin} A.~N.,  2006, \mnras, 368, 1055

\bibitem[\protect\citeauthoryear{{Timokhin}}{{Timokhin}}{2010}]{tim10}
{Timokhin} A.~N.,  2010, MNRAS, 408, L41

\bibitem[\protect\citeauthoryear{Unwin, Readhead, Wilkinson \& Ewing}{Unwin
  et~al.}{1978}]{urwe78}
Unwin S.~C.,  Readhead A. C.~S.,  Wilkinson P.~N.,    Ewing M.~S.,  1978,
  MNRAS, 182, 711

\bibitem[\protect\citeauthoryear{Ursov \& Usov}{Ursov \& Usov}{1988}]{uu88}
Ursov V.~N.,  Usov V.~V.,  1988, \apss, 140, 325

\bibitem[\protect\citeauthoryear{Usov}{Usov}{1987}]{uso87}
Usov V.~V.,  1987, ApJ, 320, 333

\bibitem[\protect\citeauthoryear{{Usov}}{{Usov}}{2002}]{uso02}
{Usov} V.~V.,  2002, in {Becker} W.,  {Lesch} H.,   {Tr{\"u}mper} J.,  eds,
  Neutron Stars, Pulsars, and Supernova Remnants {Two-stream Instability in
  Pulsar Magnetospheres}.
p.~240

\bibitem[\protect\citeauthoryear{{van Leeuwen}, {Kouwenhoven}, {Ramachandran},
  {Rankin} \& {Stappers}}{{van Leeuwen} et~al.}{2002}]{vkr+02}
{van Leeuwen} A.~G.~J.,  {Kouwenhoven} M.~L.~A.,  {Ramachandran} R.,  {Rankin}
  J.~M.,    {Stappers} B.~W.,  2002, A\&A, 387, 169

\bibitem[\protect\citeauthoryear{{van Leeuwen}, {Stappers}, {Ramachandran} \&
  {Rankin}}{{van Leeuwen} et~al.}{2003}]{vsrr03}
{van Leeuwen} A.~G.~J.,  {Stappers} B.~W.,  {Ramachandran} R.,    {Rankin}
  J.~M.,  2003, A\&A, 399, 223

\bibitem[\protect\citeauthoryear{{Vivekanand}}{{Vivekanand}}{1995}]{viv95}
{Vivekanand} M.,  1995, {MNRAS}, 274, 785

\bibitem[\protect\citeauthoryear{Vivekanand \& Joshi}{Vivekanand \&
  Joshi}{1997}]{vj97}
Vivekanand M.,  Joshi B.~C.,  1997, ApJ, 477, 431

\bibitem[\protect\citeauthoryear{Vivekanand \& Radhakrishnan}{Vivekanand \&
  Radhakrishnan}{1980}]{vr80}
Vivekanand M.,  Radhakrishnan V.,  1980, Journal of Astrophysics and Astronomy,
  1, 119

\bibitem[\protect\citeauthoryear{{Wald} \& {Wolfowitz}}{{Wald} \&
  {Wolfowitz}}{1940}]{ww40}
{Wald} A.,  {Wolfowitz} J.,  1940, Ann. Math. Stat., 11, 147

\bibitem[\protect\citeauthoryear{{Wang}, {Manchester} \& {Johnston}}{{Wang}
  et~al.}{2007}]{wmj07}
{Wang} N.,  {Manchester} R.~N.,    {Johnston} S.,  2007, MNRAS, 377, 1383

\bibitem[\protect\citeauthoryear{Wang, Manchester, Zhang, Wu, Yusup, Lyne,
  Cheng \& Chen}{Wang et~al.}{2001}]{wmz+01}
Wang N.,  Manchester R.~N.,  Zhang J.,  Wu X.~J.,  Yusup A.,  Lyne A.~G.,
  Cheng K.~S.,    Chen M.~Z.,  2001, MNRAS, 328, 855

\bibitem[\protect\citeauthoryear{{Wang} \& {Chu}}{{Wang} \& {Chu}}{1981}]{wc81}
{Wang} Z.~r.,  {Chu} Y.,  1981, ChA\&A, 5, 329

\bibitem[\protect\citeauthoryear{Weisberg, Armstrong, Backus, Cordes, Boriakoff
  \& Ferguson}{Weisberg et~al.}{1986}]{wab+86}
Weisberg J.~M.,  Armstrong B.~K.,  Backus P.~R.,  Cordes J.~M.,  Boriakoff V.,
    Ferguson D.~C.,  1986, \aj, 92, 621

\bibitem[\protect\citeauthoryear{Weisberg, Romani \& Taylor}{Weisberg
  et~al.}{1989}]{wrt89}
Weisberg J.~M.,  Romani R.~W.,    Taylor J.~H.,  1989, ApJ, 347, 1030

\bibitem[\protect\citeauthoryear{{Weisberg} \& {Taylor}}{{Weisberg} \&
  {Taylor}}{2005}]{wt05}
{Weisberg} J.~M.,  {Taylor} J.~H.,  2005, in {Rasio} F.~A.,  {Stairs} I.~H.,
  eds, Binary Radio Pulsars Vol.~328 of Astronomical Society of the Pacific
  Conference Series, {The Relativistic Binary Pulsar B1913+16: Thirty Years of
  Observations and Analysis}.
p.~25

\bibitem[\protect\citeauthoryear{{Weltevrede}}{{Weltevrede}}{2007}]{wel07}
{Weltevrede} P.,  2007, PhD thesis, University of Amsterdam

\bibitem[\protect\citeauthoryear{{Weltevrede}, {Edwards} \&
  {Stappers}}{{Weltevrede} et~al.}{2006}]{wes06}
{Weltevrede} P.,  {Edwards} R.~T.,    {Stappers} B.~W.,  2006, A\&A, 445, 243

\bibitem[\protect\citeauthoryear{{Weltevrede}, {Stappers} \&
  {Edwards}}{{Weltevrede} et~al.}{2007}]{wse07}
{Weltevrede} P.,  {Stappers} B.~W.,    {Edwards} R.~T.,  2007, \aap, 469, 607

\bibitem[\protect\citeauthoryear{{Wheaton}, {Doty}, {Primini}, {Cooke},
  {Dobson}, {Goldman}, {Hecht}, {Howe}, {Hoffman} \& {Scheepmaker}}{{Wheaton}
  et~al.}{1979}]{wdp+79}
{Wheaton} W.~A.,  {Doty} J.~P.,  {Primini} F.~A.,  {Cooke} B.~A.,  {Dobson}
  C.~A.,  {Goldman} A.,  {Hecht} M.,  {Howe} S.~K.,  {Hoffman} J.~A.,
  {Scheepmaker} A.,  1979, Nat., 282, 240

\bibitem[\protect\citeauthoryear{White, Swank \& Holt}{White
  et~al.}{1983}]{wsh83}
White N.~E.,  Swank J.~H.,    Holt S.~S.,  1983, ApJ, 270, 711

\bibitem[\protect\citeauthoryear{{Wolszczan}}{{Wolszczan}}{1980}]{wol80}
{Wolszczan} A.,  1980, \aap, 86, 7

\bibitem[\protect\citeauthoryear{{Wright}, {Weltevrede}, {Rankin} \&
  {Herfindal}}{{Wright} et~al.}{2012}]{wwr+12}
{Wright} G.,  {Weltevrede} P.,  {Rankin} J.,    {Herfindal} J.,  2012, in
  {Lewandowski} W.,  {Maron} O.,   {Kijak} J.,  eds, Electromagnetic Radiation
  from Pulsars and Magnetars Vol.~466 of Astronomical Society of the Pacific
  Conference Series, {The Null Patterns of Pulsars J1649+2533 and B2310+42}.
p.~87

\bibitem[\protect\citeauthoryear{Wright \& Fowler}{Wright \&
  Fowler}{1981}]{wf81}
Wright G.~A.,  Fowler L.~A.,  1981, A\&A, 101, 356

\bibitem[\protect\citeauthoryear{{Wright}}{{Wright}}{1979}]{w79}
{Wright} G.~A.~E.,  1979, \nat, 280, 40

\bibitem[\protect\citeauthoryear{{Wright}, {Sieber} \& {Wolszczan}}{{Wright}
  et~al.}{1986}]{wsw86}
{Wright} G.~A.~E.,  {Sieber} W.,    {Wolszczan} A.,  1986, A\&A, 160, 402

\bibitem[\protect\citeauthoryear{{Wu} \& {Gil}}{{Wu} \& {Gil}}{1995}]{wg95}
{Wu} X.,  {Gil} J.~A.,  1995, Acta Astrophysica Sinica, 15, 40

\bibitem[\protect\citeauthoryear{{Yang}, {Han} \& {Wang}}{{Yang}
  et~al.}{2013}]{yhw13}
{Yang} A.~Y.,  {Han} J.~L.,    {Wang} N.,  2013, ArXiv e-prints

\bibitem[\protect\citeauthoryear{{Young}, {Stappers}, {Weltevrede}, {Lyne} \&
  {Kramer}}{{Young} et~al.}{2012}]{ysw+12}
{Young} N.~J.,  {Stappers} B.~W.,  {Weltevrede} P.,  {Lyne} A.~G.,    {Kramer}
  M.,  2012, MNRAS, 427, 114

\bibitem[\protect\citeauthoryear{{Zhang} \& {Qiao}}{{Zhang} \&
  {Qiao}}{1996}]{zq96}
{Zhang} B.,  {Qiao} G.~J.,  1996, A\&A, 310, 135

\bibitem[\protect\citeauthoryear{{Zhang}, {Qiao} \& {Han}}{{Zhang}
  et~al.}{1997}]{zqh97b}
{Zhang} B.,  {Qiao} G.~J.,    {Han} J.~L.,  1997, \apj, 491, 891

\bibitem[\protect\citeauthoryear{Zhang, Qiao, Lin \& Han}{Zhang
  et~al.}{1997}]{zqlh97}
Zhang B.,  Qiao G.~J.,  Lin W.~P.,    Han J.~L.,  1997, ApJ, 478, 313

\bibitem[\protect\citeauthoryear{{Zhang}, {Qiao}, {Han}, {Lee} \&
  {Wang}}{{Zhang} et~al.}{2007}]{zqh+07}
{Zhang} H.,  {Qiao} G.~J.,  {Han} J.~L.,  {Lee} K.~J.,    {Wang} H.~G.,  2007,
  \aap, 465, 525

\bibitem[\protect\citeauthoryear{{Zhu} \& {Xu}}{{Zhu} \& {Xu}}{2006}]{zx06}
{Zhu} W.~W.,  {Xu} R.~X.,  2006, \mnras, 365, L16

\end{thebibliography}
%\bibliography{/home/vishal/My_paper/Bibfiles/psrrefs,/home/vishal/My_paper/Bibfiles/modpsrrefs,/home/vishal/My_paper/Bibfiles/mybib,/home/vishal/My_paper/Bibfiles/modrefs}
\end{document}